\documentclass[showpacs,reprint,superscriptaddress,prb]{revtex4-2}

\usepackage{amsmath}
\usepackage{amssymb}
\usepackage{graphicx}
\usepackage{tikz}
\usepackage{slashed}
\usepackage{multirow}
\usepackage{hhline}
\usepackage{subcaption}
\usepackage{caption}
\usepackage{float}
\usepackage{xcolor}
\usepackage{tikz-feynhand}
\usepackage{hyperref}
\usepackage{ulem}
\usepackage{array}
\usepackage{comment}

\makeatletter
\def\l@subsection#1#2{}
\def\l@subsubsection#1#2{}
\makeatother

\begin{document}

\title{Effects of general non-magnetic quenched disorder on a spin-density-wave quantum critical metallic system in two spatial dimension}
\author{Iksu Jang}
\email{iksujang@mx.nthu.edu.tw}
\affiliation{Department of Physics, Pohang University of Science and Technology, Pohang, Gyeongbuk 790-784, Korea}

\affiliation{Department of Physics, National Tsing Hua University, Hsinchu 30013, Taiwan}
\author{Ki-Seok Kim}
\email{tkfkd@postech.ac.kr}
\affiliation{Department of Physics, Pohang University of Science and Technology, Pohang, Gyeongbuk 790-784, Korea}
\affiliation{Asia Pacific Center for Theoretical Physics (APCTP), Pohang, Gyeongbuk 37673, Korea}
\date{\today}

\begin{abstract}
We investigate the effects of general non-magnetic quenched disorder on a two-dimensional spin-density-wave (SDW) quantum critical metallic system using a renormalization group (RG) method. We consider (i) all possible scattering channels by a random charge potential for fermion fields and (ii) a random mass term for a SDW boson order parameter as effects of the non-magnetic quenched disorder. From the one-loop analysis, we find a weakly disordered non-Fermi liquid metallic fixed point (random mass disordered non-Fermi liquid fixed point) when only the random boson mass vertex is considered. However, in the general case where all disorder vertices are considered, it turns out that there is no stable fixed point and the low-energy RG flows are governed by the large random charge potential vertices especially channels in a `Direct' category. Focusing on the physical meanings of the low-energy RG flows, we provide a detailed explanation of the one-loop results. Beyond the one-loop level, we first discuss partial two-loop corrections to the random charge potential vertices. Furthermore, we examine the possibility of different low-energy RG flows compared to that of the one-loop results by considering the two-loop corrections to the random boson mass vertex and, discuss low energy properties in relation to the random singlet phase. Related to physical properties, we calculate anomalous dimensions of the four superconducting channels in the one-loop level. 

\end{abstract}

\maketitle
\section{Introduction}

Two-dimensional quantum criticality with a Fermi surface is regarded as a strong coupling problem beyond the perturbative approach \cite{LeeLargeN}. More specifically, in the high-temperature regime where we will refer to as a $T>T^*$ regime, a mean-field type theory the so-called Hertz-Moriya-Millis (HMM) theory \cite{Hertz,Moriya,Millis} works well as it is shown by the sign-free quantum Monte-Carlo method \cite{SchattnerQMC,LedererQMC} recently. However in the vicinity of the quantum critical point referred to as a $T<T^*$ regime, the HMM theory (mean-field theory) breaks down \cite{AbanovChubukovHMM,LiuQMC}. In addition, the conventional large-$N$ approach is also proven to break down in the low energy limit by Lee \cite{LeeLargeN}, Metlitsky and Sachdev \cite{Metlitsky1,Metlitsky2}. In other words, the perturbative non-Fermi liquid state described by the HMM theory evolves into a strongly coupled non-Fermi liquid phase across the energy scale $T^*$.

To understand the strongly coupled non-Fermi liquid states, there have been many theoretical studies both in analytical and numerical ways. Analytically, several methods involving the new expansion parameters \cite{Mross,DaliLee,SurLee,LeeReview} and the modified large-$N$ \cite{Damia} approach are used to investigate the strongly coupled non-Fermi liquid states in a controllable way. Numerically, as mentioned previously, sign-free quantum-Monte-Carlo methods have been developed to study the many quantum critical systems \cite{SchattnerQMC,LedererQMC,LiuQMC,SachedevBerg,berg}. Recently, Schlief et al. \cite{SchLee,LuntsLee} obtain exact anomalous dimensions analytically in a spin-density wave (SDW) quantum critical system.

However, most of these works do not take account of effects of disorder. In real systems such as cuprate, an existence of randomness is unavoidable considering a quantum critical point is tuned by doping. 


Previously, Kirkpatrick and Belitz \cite{KirkpatricBelitz} studied the effect of the disorder in the SDW quantum criticality by introducing the random boson mass effect in the HMM theory. In their study, they found a disordered stable fixed point so called `Long-Range Ordered’ phase. Additionally, they identify the instability of the fixed point, due to the oscillating renormalization group (RG) flows and two-loop corrections, which leads the RG-flow to `Random singlet’ fixed point.  However it is necessary to consider the fermion fields without integrating out them since the HMM theory breaks down due to non-local corrections coming from the integration of the gap-less fermion fields \cite{AbanovChubukovHMM}. Therefore the results obtained by Kirkpatric and Belitz may not be applicable to understanding the effect of the disorder on the low energy properties of the quantum critical metallic systems.

 Recently, Halbinger and Punk \cite{Punk} studied the effect of the random charge potential in the SDW quantum critical metallic systems based on the effective field theory without integrating out the fermion fields \cite{SurLee}. Using the one-loop RG analysis, they figured out that there was no stable fixed point due to the random charge potential and the low energy property is governed by the random charge potential effect. Based on this one-loop result and the previous study by Kirkpatrick and Belitz, they speculate the low energy phase is given by `random single phase’.  Although this study provides a fruitful result about an effect of the random charge potential on the SDW quantum critical metallic systems, there are some loopholes. First of all, the random boson mass effect is not considered. From the perspective of the renormalization group method, the random boson mass vertex is the most relevant vertex. Therefore physically the random boson mass effect should be considered. Second, only one scattering channel within a hot spot by the random charge potential is considered. Generally it is possible to have several different scattering channels among total 8 hot spots by the random charge potential. In addition, Umklapp scatterings are possible due to the commensurable nesting vector. As a result, it is necessary to deal with these loopholes for a  more physically reliable model. 
 	 
In our study, we investigate effects of the general non-magnetic disorder in the SDW quantum critical metallic systems with a physically more reliable setting by dealing with these loopholes. Here we consider both the random mass and the random charge potential effects.  We introduce all scattering channels, by the random charge potential, which are classified into three categories: `Direct', `Exchange' and `Umklapp'. Using the two regularization methods, we apply the RG method. 

From the one-loop RG analysis, we find that there exists a stable `random mass disordered non-Fermi liquid' fixed point when the random mass effect is considered only. However we discover that there is no stable fixed point in a general case considering both random mass and random charge potential effects. Practically, low energy RG flows of the general case converge to that obtained by Halbinger and Punk \cite{Punk}. Although the one-loop RG analysis shows the similar results to the previous study, we discover that the physical interpretation identifying the low energy phase as a `random singlet phase’ by Halbinger and Punk needs more care based on the analysis using dimensionless parameters considering the random boson mass effect. 

In addition to the one-loop results, we consider the partial two-loop corrections. From the partial two-loop corrections, we observe that two phase spaces, named as a Random Charge Potential Dominant (RCPD) and a Random Boson Mass Dominant (RBMD) phase spaces, are possible in the low energy limit depending on the relative strength between the random charge potential effect and the random boson effect. The low energy RG flow in the RCPD phase space is basically same to that of the general one-loop case while the RG flow in the RBMD phase space shows completely different behavior. We find that the physical properties of the RBMD phase space are more consistent with that of the `random singlet’ phase rather than the RCPD phase space or the general one-loop case which was speculated as a `random singlet phase' in the previous study \cite{Punk}.

Regarding physical properties, we calculate anomalous dimensions of four superconducting channels in the one-loop level \cite{SurLee,MandalSC}. From this study, we discover that the $d$-wave zero momentum superconducting channel, which is experimentally relevant, is suppressed by the disorder in the setting where the all interactions and disorder effects are considered in equal footing. 
Our result suggest an importance of the disorder in understanding a discrepancy between the theoretical results in clean systems \cite{SCMross,SchLee} showing that superconducting instability develops at a higher temperature than the temperature that the non-fermi liquid state appears and the conventional phase diagram showing the non-Fermi liquid phase appears at the higher temperature before the superconducting dome develops.

Our paper is organized as follows. In Sec. \ref{sec:EffTheory}, we introduce a effective theoretical model for a two-dimensional disordered SDW quantum critical metallic system and two regularization methods. In Sec. \ref{sec:RGanalysis}, we discuss RG results in detail. We focus on the physical interpretation of the RG-flows and discuss the resulting low energy properties. In Sec. \ref{sec:PhysicalProperties}, we discuss the effects of the disorder on the superconducting instability channels within the one-loop results. In Sec. \ref{sec:SummaryConclusion}, we give a brief summary of our results. Then, we discuss some limitations and technical difficulties in our research. Finally, we consider two possible alternative approaches to this research direction. One is based on SYK-type models \cite{FuSachdev,Song,Debanjan1,Debanjan2,Patel1,Patel2,GuoSachdev,Kim,Esterlis1,Esterlis2,Wang1,Wang2,GeorgesReview}, and the other is an approach starting from the clean non-Fermi liquid fixed point directly \cite{Nosov}.

\section{Effective Field Theory} \label{sec:EffTheory}

\subsection{Spin-Density-Wave hot spot model}

\def\a{6} 
\def\r{2.5} 
\def\b{2.4}

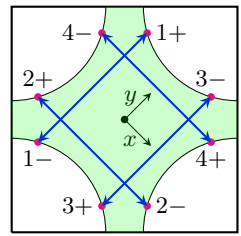
\begin{figure}
	\centering
	\begin{tikzpicture}[scale=0.5,>=stealth]
		\draw[style=thick] (0,0)--(\a,0)--(\a,\a)--(0,\a)--(0,0);
		\draw[style=thin] (0,\r) arc (90:0:\r);
		\draw[style=thin] (\a,\r) arc (90:180:\r);
		\draw[style=thin] (\a,\a-\r) arc (-90:-180:\r);
		\draw[style=thin] (0,\a-\r) arc (-90:0:\r);
		
		\fill[color=black] ({\a/2},{\a/2}) circle (0.1);
		
		\fill[color=magenta] (\a-\b,{\a-sqrt((\r^2-\b^2))}) circle (0.1) node[black,right]{$1+$}; 
		\fill[color=magenta] ({sqrt((\r^2-\b^2))},\b) circle (0.1) node[black,below]{$1-$}; 
		
		\fill[color=magenta] ({sqrt((\r^2-\b^2))},\a-\b) circle (0.1) node[black,above]{$2+$}; 
		\fill[color=magenta] (\a-\b,{sqrt((\r^2-\b^2))}) circle (0.1) node[black,right]{$2-$}; 
		
		\fill[color=magenta] (\b,{sqrt((\r^2-\b^2))}) circle (0.1) node[black,left]{$3+$}; 
		\fill[color=magenta] ({\a-sqrt((\r^2-\b^2))},\a-\b) circle (0.1) node[black,above]{$3-$}; 
		
		\fill[color=magenta] ({\a-sqrt((\r^2-\b^2))},\b) circle (0.1) node[black,below]{$4+$}; 
		\fill[color=magenta] (\b,{\a-sqrt((\r^2-\b^2))}) circle (0.1) node[black,left]{$4-$}; 
		
		\draw[thick,<->,color=blue] (\a-\b,{\a-sqrt((\r^2-\b^2))})--({sqrt((\r^2-\b^2))},\b);
		\draw[thick,<->,color=blue] ({sqrt((\r^2-\b^2))},\a-\b)-- (\a-\b,{sqrt((\r^2-\b^2))});
		\draw[thick,<->,color=blue]  (\b,{sqrt((\r^2-\b^2))})--({\a-sqrt((\r^2-\b^2))},\a-\b);
		\draw[thick,<->,color=blue] ({\a-sqrt((\r^2-\b^2))},\b)--(\b,{\a-sqrt((\r^2-\b^2))});
		
		\draw[thin,->] (\a/2,\a/2)--({\a/2+1/sqrt(2)},{\a/2-1/sqrt(2)}) node[pos=0.8,left]{$x$};
		\draw[thin,->] (\a/2,\a/2)--({\a/2+1/sqrt(2)},{\a/2+1/sqrt(2)}) node[pos=0.8,left] {$y$};
		
		\fill[color=green, opacity=0.2] (0,\r)--(0,\a-\r) to [out=0,in=-90] (\r,\a)--(\a-\r,\a) to [out=-90, in=-180] (\a,\a-\r)--(\a,\r) to [out=180, in=90] (\a-\r,0)--(\r,0) to [out=90,in=0] (0,\r);
	\end{tikzpicture}
	\caption{The two-dimensional Fermi surface for an electron-doped cuprate system. Magenta dots with an index $n=1,2,3,4$ and $m=\pm$ denote hot spots connected by SDW nesting vectors ($\mathbf{Q}_i\in \{(\pi,\pi),(-\pi,-\pi),(\pi,-\pi),(-\pi,\pi)\}$).} \label{fig:SDWFS}
\end{figure}

\begin{figure}
	\centering
	\includegraphics[scale=0.25]{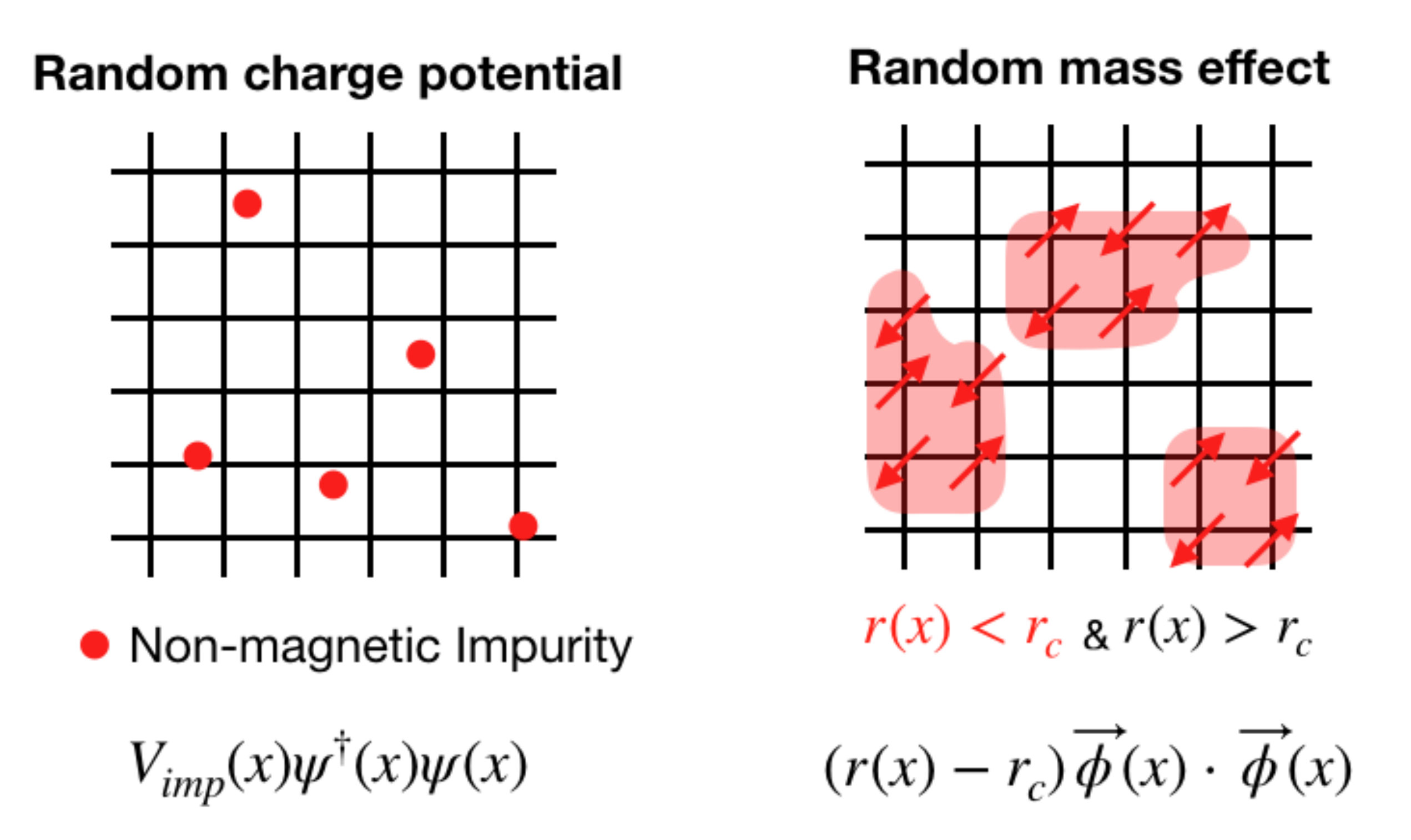}
	\caption{Two types of disorder effects.} \label{fig:DisorderEffects}
\end{figure}

For an electron-doped cuprate system, a Fermi surface structure near two-dimensional SDW quantum criticality and its low energy effective field theory is given as follows \cite{PhysRevLett.84.5608}:
\begin{align}
	S&=\sum_{n=1}^{4}\sum_{m=\pm}\sum_{\sigma}\int dk\psi_{n,\sigma}^{(m)*}(k)[ik_0+e_n^{m}(\mathbf{k})]\psi_{n,\sigma}^{(m)}(k)\nonumber\\
	&+\frac{1}{2}\int dq [q_0^2+c^2|\mathbf{q}|^2]\vec{\phi}(q)\cdot\vec{\phi}(-q)+g_0\sum_{n=1}^4\sum_{\sigma}\int dk \int dq \nonumber\\
	&\times [\vec{\phi}(q)\cdot\psi_{n,\sigma}^{(+)*}(k+q)\vec{\tau}_{\sigma,\sigma'}\psi_{n,\sigma'}^{(-)}(k)+c.c.]\nonumber \\
	&+u_0\int dk_1\int dk_2\int dq[\vec{\phi}(k_1+q)\cdot\vec{\phi}(k_2-q)]\nonumber\\
	&\times [\vec{\phi}(-k_1)\cdot\vec{\phi}(-k_2)] . \label{eq:CleanSDWAction}
\end{align}
Here, $\psi_{n,\sigma}^{(m)}$ represents an electron field living in a hot spot denoted by $n=1,2,3,4$ and $m=\pm$, shown in Fig. \ref{fig:SDWFS}. The short-handed integral expression means $\int dk=\int \frac{dk_0}{2\pi}\int\frac{d^2\mathbf{k}}{(2\pi)^2}$. Dispersions of hot spot electrons are given by $e_{1}^{\pm}(\mathbf{k})=vk_x\pm k_y$, $e_2^{\pm}(\mathbf{k})=vk_y\mp k_x$, $e_3^{\pm}(\mathbf{k})=-vk_x\mp k_y$ and $e_4^{\pm}(\mathbf{k})=-vk_y\pm k_x$ where $v$ is a perpendicular component of the Fermi velocity to the nesting vector $\vec{Q}$. $\vec{\phi}(q)$ is a Fourier transformed SDW order parameter in momentum space. The SDW order-parameter field has a `relativistic' dispersion with its group velocity $c$. $\vec{\tau}$ represent Pauli matrices acting on the electron spin.

The first and second terms in Eq. \eqref{eq:CleanSDWAction} represent the kinetic energy of electrons and boson order parameters, respectively. The third term represents an anti-ferromagnetic interaction between the SDW order parameter and the electron field, refereed to as Yukawa interaction with a strength $g_{0}$. The last term describes a $\phi^4$-like self-interaction between the spin-density order parameters $\vec{\phi}$ with a strength $u_{0}$. 

\subsection{Non-magnetic quenched disorders}

 We consider two non-magnetic disorder effects; random charge potential and random boson mass effects described in Fig. \ref{fig:DisorderEffects} with following actions
\begin{gather}
	S_{rCP}=\int d\tau \int d^2x\; V(x)\psi_{\sigma}^\dagger(\tau,x)\psi_{\sigma}(\tau,x),\\
	S_{rBM}=\int d\tau \int d^2x\; m^2(x)\vec{\phi}(\tau,x)\cdot\vec{\phi}(\tau,x)
\end{gather}
where $V(x)$ is a random charge potential by nonmagnetic impurities and $m^2(x)$ is a random mass.


In the clean SDW quantum criticality, only hot-spot electrons have been considered for non-Fermi liquid physics. However, cold-spot electrons also should be considered for the SDW non-Fermi liquid physics since scattering processes between cold and hot spots are possible due to the random charge potential. In spite of this possibility, we focus on scattering processes between hot spots only in the present study and ignore possible scattering processes between hot and cold spots. This is because we are mainly interested in the thermodynamic properties of critical electrons on hot spots, responsible for the non-Fermi liquid physics. However, we admit that the cold-spot dynamics have to be considered in order to understand transport properties \cite{RevModPhys.77.579}. Unfortunately, this is beyond our current work.

As a result, an effective action for disorder scattering is given by
\begin{align}
	S_{dis}&=\int d\tau \int d^2x \Big[V(x)\psi_{\sigma,hot}^{\dagger}(\tau,x)\psi_{\sigma,hot}(\tau,x)\nonumber\\
	&+ m^2(x)\vec{\phi}(\tau,x)\cdot\vec{\phi}(\tau,x)\Big] . \label{eq:Disorder}
\end{align}
Here, we consider the Gaussian distribution function for both disorders given by $\mathcal{P}_{rCP}[V(x)]= \mathcal{N}_{V} \exp\Big(-\int d^2 x\frac{V^2(x)}{2\Gamma}\Big)$, $
	\mathcal{P}_{rBM}[m^2(x)]= \mathcal{N}_{m^{2}} \exp\Big(-\int d^2x\frac{m^2(x)m^2(x)}{2\Gamma_M} \Big)$ where
$\mathcal{N}_{V}$ and $\mathcal{N}_{m^{2}}$ are normalization constants.

After disorder averaging process using `replica trick' \cite{zinn2002quantum}, we obtain a following disorder-averaged effective action 
\begin{align}
	&S_{eff}=\sum_{a=1}^R\Bigg[\sum_{n=1}^4\sum_{m=\pm}\sum_{\sigma}\int dk \psi_{a,n,\sigma}^{(m)*}(k)[ik_0+e_{n}^{m}(k)]\nonumber\\
	&\times \psi_{a,n,\sigma}(k)+\frac{1}{2}\int dq[q_0^2+c^2|\mathbf{q}|^2]\vec{\phi}_a(q)\cdot\vec{\phi}_a(-q)\nonumber\\
	&+g_0\sum_{n=1}^4\sum_{\sigma}\int dx \vec{\phi}_a(x)\cdot\psi_{a,n,\sigma}^{(+)*}(x)\vec{\tau}_{\sigma,\sigma'}\psi_{a,n,\sigma'}^{(-)}(x)\nonumber\\
	&+u_0\int dx[\vec{\phi}_a(x)\cdot\vec{\phi}_a(x)]^2\Bigg]-\sum_{a,b=1}^R\int d\tau\int d\tau'\int d^2\mathbf{x}\nonumber\\
	&\times \Bigg[\sum_{\sigma,\sigma'}\frac{\Gamma^{m_1,m_2,m_3,m_4}_{n_1,n_2,n_3,n_4}}{2} \psi_{a,n_1,\sigma}^{(m_1)*}(\tau,\mathbf{x}) \psi_{a,n_2,\sigma}^{(m_2)}(\tau,\mathbf{x})\nonumber\\
	&\times  \psi_{b,n_3,\sigma'}^{(m_3)*}(\tau',\mathbf{x})\psi_{b,n_4,\sigma'}^{(m_4)}(\tau',\mathbf{x}) +\frac{\Gamma_M}{2}[\vec{\phi}_a(\tau,\mathbf{x})\cdot\vec{\phi}_a(\tau,\mathbf{x})] \nonumber\\
	&\times [\vec{\phi}_b(\tau',\mathbf{x})\cdot\vec{\phi}_b(\tau',\mathbf{x})]\Bigg] , \label{eq:effAction}
\end{align}
where $\int dk=\int \frac{dk_0}{2\pi}\int\frac{d^2\mathbf{k}}{(2\pi)^2}$, $\int dx=\int d\tau\int d^2\mathbf{x}$ and the Gaussian integrals with respect to both distribution functions have been performed. Here, $a$ and $b$ are replica indices, and $\Gamma_{n_1,n_2,n_3,n_4}^{m_1,m_2,m_3,m_4}$ is a coupling constant of an impurity scattering process between hot-spot electrons with indices $(n_1,m_1),(n_2,m_2),(n_3,m_3),(n_4,m_4)$ depicted in Fig. \ref{fig:SDWFS}.

\subsection{Classification of fermion random charge potential vertices} \label{subsec:ClassificationOfDiosrderVertices}

All possible scattering channels due to the fermion random charge potential are divided into two groups; normal and Umklapp processes. In the case of normal process, the total momentum is conserved while that modulo reciprocal lattice vectors $\mathbf{G}$ is conserved in Umklapp process. 
 From a tight-binding modeling of the system, an action of the random charge potential vertices is given by
\begin{align}
	S_{TB-dis}&=-\sum_{a,b=1}^R\sum_{\sigma,\sigma'}\frac{\Gamma_{i_1i_2i_3i_4}}{2}\int \frac{d\omega}{2\pi}\int \frac{d\omega'}{2\pi}\prod_{i=1}^4\int \frac{d^2\mathbf{k}_i}{(2\pi)^2} \nonumber\\
	& \psi^*_{a,\sigma}(\omega,\mathbf{k}_F^{(i_1)}+\mathbf{k}_1)\psi_{a,\sigma}(\omega,\mathbf{k}_F^{(i_2)}+\mathbf{k}_2)\nonumber\\
	&\times \psi^*_{b,\sigma'}(\omega',\mathbf{k}_F^{(i_3)}+\mathbf{k}_3) \psi_{b,\sigma'}(\omega',\mathbf{k}_F^{(i_4)}+\mathbf{k}_4)\nonumber\\
	&\times \sum_{\mathbf{G}}\delta(\mathbf{k}_F^{(i_1)}+\mathbf{k}_F^{(i_2)}-\mathbf{k}_F^{(i_3)}-\mathbf{k}_F^{(i_4)}+\mathbf{G})\nonumber\\
	&\times\delta(\mathbf{k}_1+\mathbf{k}_3-\mathbf{k}_2-\mathbf{k}_4) ,
	\label{eq:TBdisAction}
\end{align}
where $i_r=(n_r,m_r)$ is a hot-spot index depicted in Fig. \ref{fig:SDWFS} and $\mathbf{k}_F^{(i_r)}$ is a Fermi wave vector from the center of the First Brillioun zone to the hot spot with index $(n_r,m_r)$. The random charge potential vertices in the effective action $S_{eff}$ of Eq. \eqref{eq:effAction} can be reconstructed from the above action $S_{TB-dis}$ (Eq. \eqref{eq:TBdisAction}), using the following correspondence: $\psi_{a,\sigma}(\omega,\mathbf{k}_F^{(i_r)}+\mathbf{k}_r)=\psi_{a,n_r,\sigma}^{m_r}(\omega,\mathbf{k}_r)$ and $\Gamma_{i_1i_2i_3i_4}=\Gamma_{n_{1},n_{2},n_{3},n_{4}}^{m_{1},m_{2},m_{3},m_{4}}$.

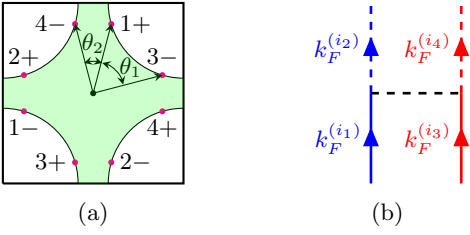
\begin{figure}
	\centering
	\begin{subfigure}{0.2\textwidth}
		\begin{tikzpicture}[scale=0.4,>=stealth]
			\draw[style=thick] (0,0)--(\a,0)--(\a,\a)--(0,\a)--(0,0);
			\draw[style=thin] (0,\r) arc (90:0:\r);
			\draw[style=thin] (\a,\r) arc (90:180:\r);
			\draw[style=thin] (\a,\a-\r) arc (-90:-180:\r);
			\draw[style=thin] (0,\a-\r) arc (-90:0:\r);
			
			\fill[color=magenta] (\a-\b,{\a-sqrt((\r^2-\b^2))}) circle (0.1) node[black,right]{$1+$}; 
			\fill[color=magenta] ({sqrt((\r^2-\b^2))},\b) circle (0.1) node[black,below]{$1-$}; 
			
			\fill[color=magenta] ({sqrt((\r^2-\b^2))},\a-\b) circle (0.1) node[black,above]{$2+$}; 
			\fill[color=magenta] (\a-\b,{sqrt((\r^2-\b^2))}) circle (0.1) node[black,right]{$2-$}; 
			
			\fill[color=magenta] (\b,{sqrt((\r^2-\b^2))}) circle (0.1) node[black,left]{$3+$}; 
			\fill[color=magenta] ({\a-sqrt((\r^2-\b^2))},\a-\b) circle (0.1) node[black,above]{$3-$}; 
			
			\fill[color=magenta] ({\a-sqrt((\r^2-\b^2))},\b) circle (0.1) node[black,below]{$4+$}; 
			\fill[color=magenta] (\b,{\a-sqrt((\r^2-\b^2))}) circle (0.1) node[black,left]{$4-$}; 
			
			\fill[color=black] ({\a/2},{\a/2}) circle (0.1);
			
			\draw[thin, ->] (\a/2,\a/2)--(\a-\b,{\a-sqrt((\r^2-\b^2))}); 
			\draw[thin, ->] (\a/2,\a/2)--(\b,{\a-sqrt((\r^2-\b^2))}); 
			\draw[thin,->] (\a/2,\a/2)--({\a-sqrt((\r^2-\b^2))},\a-\b); 
			
			\draw[<->] (\a/2+\a/4-\b/2,{\a/2+\a/4-sqrt((\r^2-\b^2)/2)}) arc (74:106:1.1) node[pos=0.5,above] {$\theta_2$};
			\draw[<->] (\a/2+\a/4-\b/2,{\a/2+\a/4-sqrt((\r^2-\b^2)/2)}) arc (74:14:1.) node[pos=0.3,right] {$\theta_1$};
			
			\fill[color=green, opacity=0.2] (0,\r)--(0,\a-\r) to [out=0,in=-90] (\r,\a)--(\a-\r,\a) to [out=-90, in=-180] (\a,\a-\r)--(\a,\r) to [out=180, in=90] (\a-\r,0)--(\r,0) to [out=90,in=0] (0,\r);
		\end{tikzpicture}
		\caption{}\label{fig:AnglesInFS}
	\end{subfigure}
	~
	\begin{subfigure}{0.2\textwidth}
		\begin{tikzpicture}[scale=1.2]
			\begin{feynhand}
				\vertex (a1) at (0,0);
				\vertex (a2) at (1,0);
				\vertex (b1) at (0,1);
				\vertex (b2) at (1,1);
				\vertex (c1) at (0,2);
				\vertex (c2) at (1,2);
				
				\propag[fer,blue] (a1) to [edge label=$k_F^{(i_1)}$](b1);
				\propag[fer,red,very thick] (a2) to [edge label=$k_F^{(i_3)}$] (b2);
				\propag[fer,blue,dashed,very thick] (b1) to [edge label=$k_F^{(i_2)}$](c1);
				\propag[fer,red,dashed,very thick] (b2) to [edge label=$k_F^{(i_4)}$](c2);
				\propag[sca,very thick] (b1) to (b2);
				
			\end{feynhand}
		\end{tikzpicture}
		\caption{}\label{fig:FeynmanDiagramOfScatteringTerm}
	\end{subfigure}
	\caption{Schematic figures for the classification of scattering channels by the random charge potential: (a) Angles $\theta_1$ and $\theta_2$ in the Fermi surface and (b) Feynman diagram representation of a disorder scattering channel given by $\psi^*_{a,\sigma}(\omega,\mathbf{k}_F^{(i_1)}+\mathbf{k}_1) \psi_{a,\sigma}(\omega,\mathbf{k}_F^{(i_2)}+\mathbf{k}_2) \psi^*_{b,\sigma'}(\omega,\mathbf{k}_F^{(i_3)}+\mathbf{k}_3) \psi^*_{b,\sigma'}(\omega,\mathbf{k}_F^{(i_4)}+\mathbf{k}_4)$.} \label{fig:AnglesInFSandFeynmanRepresentation}
\end{figure}

In $S_{TB-dis}$ (Eq. \eqref{eq:TBdisAction}), the terms with $\mathbf{G}=0$ correspond to normal scattering process ($S_{normal}$) and those with $\mathbf{G} \neq 0$ do to Umklapp one ($S_{umklapp}$). Both normal and Umklapp disorder-scattering processes consist of various scattering channels. 
We classify all possible scattering channels. Here, we use angles in the Fermi surface and Feynman diagram representations depicted in Fig. \ref{fig:AnglesInFSandFeynmanRepresentation} to specify scattering processes pictorially.

\subsubsection{Normal processes}

In the normal process, the total momentum is conserved as $\mathbf{k}_F^{(i_1)}+\mathbf{k}_F^{(i_3)}=\mathbf{k}_F^{(i_2)}+\mathbf{k}_F^{(i_4)}$. Based on this momentum conservation with the Fermi-surface geometry in Fig. \ref{fig:SDWFS}, all the possible normal processes are classified into the following three classes:
(i) $\mathbf{k}_F^{(i_1)}=\mathbf{k}_F^{(i_2)}, \mathbf{k}_F^{(i_3)}=\mathbf{k}_F^{(i_4)}(\mathbf{k}_F^{(i_1)}+\mathbf{k}_F^{(i_3)}\neq 0)$, (ii) $\mathbf{k}_F^{(i_1)}=\mathbf{k}_F^{(i_4)},\mathbf{k}_F^{(i_3)}=\mathbf{k}_F^{(i_2)}(\mathbf{k}_F^{(i_1)}+\mathbf{k}_F^{(i_3)}\neq 0)$ and (iii) $\mathbf{k}_F^{(i_1)}+\mathbf{k}_F^{(i_3)}=\mathbf{k}_F^{(i_2)}+\mathbf{k}_F^{(i_4)}=0$.
These are nothing but the forward (direct), exchange, and Cooper channels described in the Fermi liquid theory \cite{ShankarRMP}, respectively. Each class can be further classified based on an scattering angle. All the possible scattering channels of the normal processes are described in Fig. \ref{fig:NormalScatteringChannels}.

\subsubsection{Umklapp processes}

We first classify Umklapp scattering channels into two sub-classes based on the magnitude of a reciprocal lattice vector $\mathbf{G}$; (i) $|\mathbf{G}|=\frac{\pi}{a}$ and (ii) $|\mathbf{G}|=\frac{\sqrt{2}\pi}{a}$. Since other scattering channels with bigger reciprocal lattice vectors ($|\mathbf{G}|>\frac{\sqrt{2}\pi}{a}$) are the same as those of either (i) or (ii) in this system, these two classes are the complete set. Each sub-class, based on the magnitude of $\mathbf{G}$, is further classified into more specific channels. The resulting all possible scattering channels of the Umklapp processes are given in Fig. \ref{fig:UmklappScatteringChannels}.

As a result of the classification, all disorder scattering processes between hot spots is modeled with total 27 scattering channels: 18 scattering channels of the normal process and 9 scattering channels of the Umklapp process.  Explicit expressions of the random charge potential vertices in action forms are given in the Supplementary Material \footnote{In the Supplementary Material, explicit forms of the random charge potential vertices and details of the two loop calculations are given.}.
	
\begin{widetext}
	
	\begin{figure}
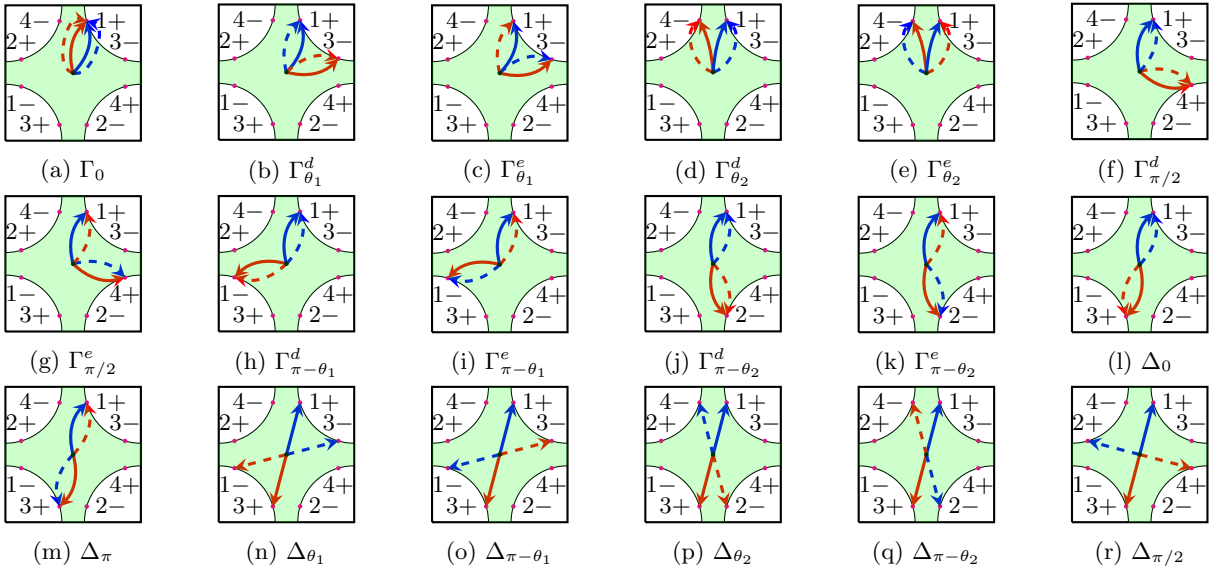

		\centering
		\begin{subfigure}{0.14\textwidth}
			\centering

			\caption{$\Gamma_0$ }
		\end{subfigure}
		~
		\begin{subfigure}{0.14\textwidth}
			\centering
			%
			\caption{$\Gamma_{\theta_1}^d$}
		\end{subfigure}
		~
		\begin{subfigure}{0.14\textwidth}
			\centering
			%
			\caption{$\Gamma_{\theta_1}^e$}
		\end{subfigure}
		~
		\begin{subfigure}{0.14\textwidth}
			\centering
			%
			\caption{$\Gamma_{\theta_2}^d$}
		\end{subfigure}
		~
		\begin{subfigure}{0.14\textwidth}
			\centering
			%
			\caption{$\Gamma_{\theta_2}^e$}
		\end{subfigure}
		~
		\begin{subfigure}{0.14\textwidth}
			\centering
			%
			\caption{$\Gamma_{\pi/2}^d$}
		\end{subfigure}
		~
		\begin{subfigure}{0.14\textwidth}
			\centering
			%
			\caption{$\Gamma_{\pi/2}^e$}
		\end{subfigure}
		~
		\begin{subfigure}{0.14\textwidth}
			\centering
			%
			\caption{$\Gamma_{\pi-\theta_1}^d$}
		\end{subfigure}
		~
		\begin{subfigure}{0.14\textwidth}
			\centering
			%
			\caption{$\Gamma_{\pi-\theta_1}^e$}
		\end{subfigure}
		~
		\begin{subfigure}{0.14\textwidth}
			\centering
			%
			\caption{$\Gamma_{\pi-\theta_2}^d$}
		\end{subfigure}
		~
		\begin{subfigure}{0.14\textwidth}
			\centering
			%
			\caption{$\Gamma_{\pi-\theta_2}^e$}
		\end{subfigure}
		~
		\begin{subfigure}{0.14\textwidth}
			\centering
			%
			\caption{$\Delta_{0}$}
		\end{subfigure}
		~
		\begin{subfigure}{0.14\textwidth}
			\centering
			%
			\caption{$\Delta_{\pi}$}
		\end{subfigure}
		~
		\begin{subfigure}{0.14\textwidth}
			\centering
			%
			\caption{$\Delta_{\theta_1}$}
		\end{subfigure}
		~
		\begin{subfigure}{0.14\textwidth}
			\centering
			%
			\caption{$\Delta_{\pi-\theta_1}$}
		\end{subfigure}
		~
		\begin{subfigure}{0.14\textwidth}
			\centering
			%
			\caption{$\Delta_{\theta_2}$}
		\end{subfigure}
		~
		\begin{subfigure}{0.14\textwidth}
			\centering
			%
			\caption{$\Delta_{\pi-\theta_2}$}
		\end{subfigure}
		~
		\begin{subfigure}{0.14\textwidth}
			\centering
			%
			\caption{$\Delta_{\pi/2}$}
		\end{subfigure}
		\caption{All channels of the normal processes. Descriptions of the line (dashed) arrows and angles are given in Fig. \ref{fig:AnglesInFSandFeynmanRepresentation}. Here, $\Gamma_{\theta}^{d/e}$ ($\Delta_{\theta}$) denotes a random charge potential vertex coupling constant of (i) Forward / (ii) Exchange ((iii) Cooper) scattering channel with an angle $\theta$ between two Fermi wave vectors; $\mathbf{k}_F^{(i_1)},\; \mathbf{k}_F^{(i_3)} ~ (\mathbf{k}_F^{(i_1)},\; \mathbf{k}_F^{(i_2)})$. In the case of $\Gamma_0$, we do not have to specify whether it is forward or exchange scattering channel: they are same when $\theta$ is zero.} \label{fig:NormalScatteringChannels}
	\end{figure}

	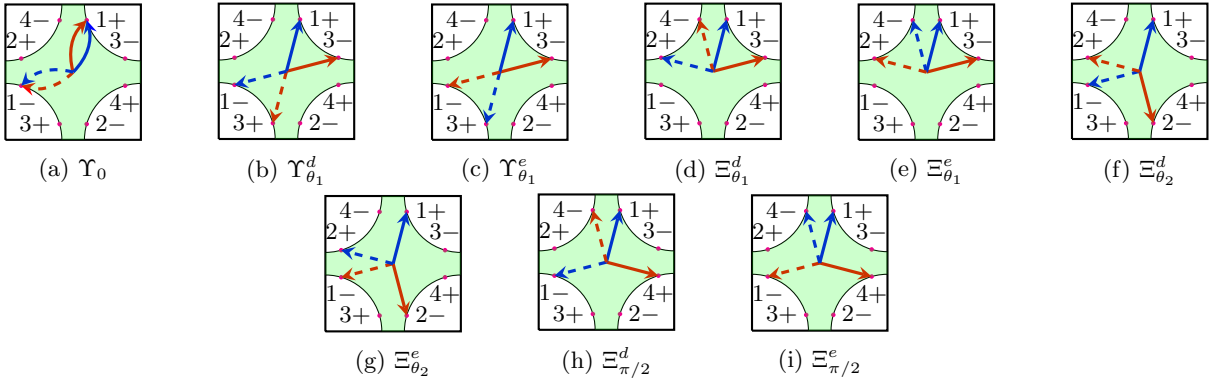
\begin{figure}
		\centering
		\begin{subfigure}{0.14\textwidth}
			\centering
			\begin{tikzpicture}[scale=0.3,>=stealth]
				\draw[style=thick] (0,0)--(\a,0)--(\a,\a)--(0,\a)--(0,0);
				\draw[style=thin] (0,\r) arc (90:0:\r);
				\draw[style=thin] (\a,\r) arc (90:180:\r);
				\draw[style=thin] (\a,\a-\r) arc (-90:-180:\r);
				\draw[style=thin] (0,\a-\r) arc (-90:0:\r);
				
				\fill[color=magenta] (\a-\b,{\a-sqrt((\r^2-\b^2))}) circle (0.1) node[black,right]{$1+$}; 
				\fill[color=magenta] ({sqrt((\r^2-\b^2))},\b) circle (0.1) node[black,below]{$1-$}; 
				
				\fill[color=magenta] ({sqrt((\r^2-\b^2))},\a-\b) circle (0.1) node[black,above]{$2+$}; 
				\fill[color=magenta] (\a-\b,{sqrt((\r^2-\b^2))}) circle (0.1) node[black,right]{$2-$}; 
				
				\fill[color=magenta] (\b,{sqrt((\r^2-\b^2))}) circle (0.1) node[black,left]{$3+$}; 
				\fill[color=magenta] ({\a-sqrt((\r^2-\b^2))},\a-\b) circle (0.1) node[black,above]{$3-$}; 
				
				\fill[color=magenta] ({\a-sqrt((\r^2-\b^2))},\b) circle (0.1) node[black,below]{$4+$}; 
				\fill[color=magenta] (\b,{\a-sqrt((\r^2-\b^2))}) circle (0.1) node[black,left]{$4-$}; 
				
				\draw[very thick, ->,color=red] (\a/2,\a/2) to [bend left](\a-\b,{\a-sqrt((\r^2-\b^2))}); 
				\draw[very thick, ->,color=blue] (\a/2,\a/2) to [bend right](\a-\b,{\a-sqrt((\r^2-\b^2))}); 
				
				\draw[very thick, ->,color=red,dashed] (\a/2,\a/2) to [bend left]({sqrt((\r^2-\b^2))},\b); 
				\draw[very thick, ->,color=blue,dashed] (\a/2,\a/2) to [bend right]({sqrt((\r^2-\b^2))},\b); 
				
				\fill[color=green, opacity=0.2] (0,\r)--(0,\a-\r) to [out=0,in=-90] (\r,\a)--(\a-\r,\a) to [out=-90, in=-180] (\a,\a-\r)--(\a,\r) to [out=180, in=90] (\a-\r,0)--(\r,0) to [out=90,in=0] (0,\r);
			\end{tikzpicture}
			\caption{$\Upsilon_0$} 
		\end{subfigure}
		~
		\begin{subfigure}{0.14\textwidth}
			\centering
			\begin{tikzpicture}[scale=0.3,>=stealth]
				\draw[style=thick] (0,0)--(\a,0)--(\a,\a)--(0,\a)--(0,0);
				\draw[style=thin] (0,\r) arc (90:0:\r);
				\draw[style=thin] (\a,\r) arc (90:180:\r);
				\draw[style=thin] (\a,\a-\r) arc (-90:-180:\r);
				\draw[style=thin] (0,\a-\r) arc (-90:0:\r);
				
				\fill[color=magenta] (\a-\b,{\a-sqrt((\r^2-\b^2))}) circle (0.1) node[black,right]{$1+$}; 
				\fill[color=magenta] ({sqrt((\r^2-\b^2))},\b) circle (0.1) node[black,below]{$1-$}; 
				
				\fill[color=magenta] ({sqrt((\r^2-\b^2))},\a-\b) circle (0.1) node[black,above]{$2+$}; 
				\fill[color=magenta] (\a-\b,{sqrt((\r^2-\b^2))}) circle (0.1) node[black,right]{$2-$}; 
				
				\fill[color=magenta] (\b,{sqrt((\r^2-\b^2))}) circle (0.1) node[black,left]{$3+$}; 
				\fill[color=magenta] ({\a-sqrt((\r^2-\b^2))},\a-\b) circle (0.1) node[black,above]{$3-$}; 
				
				\fill[color=magenta] ({\a-sqrt((\r^2-\b^2))},\b) circle (0.1) node[black,below]{$4+$}; 
				\fill[color=magenta] (\b,{\a-sqrt((\r^2-\b^2))}) circle (0.1) node[black,left]{$4-$}; 
				
				\draw[very thick, ->,color=red] (\a/2,\a/2) to ({\a-sqrt((\r^2-\b^2))},\a-\b); 
				\draw[very thick, ->,color=blue] (\a/2,\a/2) to (\a-\b,{\a-sqrt((\r^2-\b^2))}); 
				
				\draw[very thick, ->,color=red,dashed] (\a/2,\a/2) to (\b,{sqrt((\r^2-\b^2))}); 
				\draw[very thick, ->,color=blue,dashed] (\a/2,\a/2) to ({sqrt((\r^2-\b^2))},\b); 
				
				\fill[color=green, opacity=0.2] (0,\r)--(0,\a-\r) to [out=0,in=-90] (\r,\a)--(\a-\r,\a) to [out=-90, in=-180] (\a,\a-\r)--(\a,\r) to [out=180, in=90] (\a-\r,0)--(\r,0) to [out=90,in=0] (0,\r);
			\end{tikzpicture}
			\caption{$\Upsilon_{\theta_1}^d$ }
		\end{subfigure}
		~
		\begin{subfigure}{0.14\textwidth}
			\centering
			\begin{tikzpicture}[scale=0.3,>=stealth]
				\draw[style=thick] (0,0)--(\a,0)--(\a,\a)--(0,\a)--(0,0);
				\draw[style=thin] (0,\r) arc (90:0:\r);
				\draw[style=thin] (\a,\r) arc (90:180:\r);
				\draw[style=thin] (\a,\a-\r) arc (-90:-180:\r);
				\draw[style=thin] (0,\a-\r) arc (-90:0:\r);
				
				\fill[color=magenta] (\a-\b,{\a-sqrt((\r^2-\b^2))}) circle (0.1) node[black,right]{$1+$}; 
				\fill[color=magenta] ({sqrt((\r^2-\b^2))},\b) circle (0.1) node[black,below]{$1-$}; 
				
				\fill[color=magenta] ({sqrt((\r^2-\b^2))},\a-\b) circle (0.1) node[black,above]{$2+$}; 
				\fill[color=magenta] (\a-\b,{sqrt((\r^2-\b^2))}) circle (0.1) node[black,right]{$2-$}; 
				
				\fill[color=magenta] (\b,{sqrt((\r^2-\b^2))}) circle (0.1) node[black,left]{$3+$}; 
				\fill[color=magenta] ({\a-sqrt((\r^2-\b^2))},\a-\b) circle (0.1) node[black,above]{$3-$}; 
				
				\fill[color=magenta] ({\a-sqrt((\r^2-\b^2))},\b) circle (0.1) node[black,below]{$4+$}; 
				\fill[color=magenta] (\b,{\a-sqrt((\r^2-\b^2))}) circle (0.1) node[black,left]{$4-$}; 

				\draw[very thick, ->,color=red] (\a/2,\a/2) to ({\a-sqrt((\r^2-\b^2))},\a-\b); 
				\draw[very thick, ->,color=blue] (\a/2,\a/2) to (\a-\b,{\a-sqrt((\r^2-\b^2))}); 
				
				\draw[very thick, ->,color=blue,dashed] (\a/2,\a/2) to (\b,{sqrt((\r^2-\b^2))}); 
				\draw[very thick, ->,color=red,dashed] (\a/2,\a/2) to ({sqrt((\r^2-\b^2))},\b); 
				
				\fill[color=green, opacity=0.2] (0,\r)--(0,\a-\r) to [out=0,in=-90] (\r,\a)--(\a-\r,\a) to [out=-90, in=-180] (\a,\a-\r)--(\a,\r) to [out=180, in=90] (\a-\r,0)--(\r,0) to [out=90,in=0] (0,\r);
			\end{tikzpicture}
			\caption{$\Upsilon_{\theta_1}^e$} 
		\end{subfigure}
		~
		\begin{subfigure}{0.14\textwidth}
			\centering
			\begin{tikzpicture}[scale=0.3,>=stealth]
				\draw[style=thick] (0,0)--(\a,0)--(\a,\a)--(0,\a)--(0,0);
				\draw[style=thin] (0,\r) arc (90:0:\r);
				\draw[style=thin] (\a,\r) arc (90:180:\r);
				\draw[style=thin] (\a,\a-\r) arc (-90:-180:\r);
				\draw[style=thin] (0,\a-\r) arc (-90:0:\r);
				
				\fill[color=magenta] (\a-\b,{\a-sqrt((\r^2-\b^2))}) circle (0.1) node[black,right]{$1+$}; 
				\fill[color=magenta] ({sqrt((\r^2-\b^2))},\b) circle (0.1) node[black,below]{$1-$}; 
				
				\fill[color=magenta] ({sqrt((\r^2-\b^2))},\a-\b) circle (0.1) node[black,above]{$2+$}; 
				\fill[color=magenta] (\a-\b,{sqrt((\r^2-\b^2))}) circle (0.1) node[black,right]{$2-$}; 
				
				\fill[color=magenta] (\b,{sqrt((\r^2-\b^2))}) circle (0.1) node[black,left]{$3+$}; 
				\fill[color=magenta] ({\a-sqrt((\r^2-\b^2))},\a-\b) circle (0.1) node[black,above]{$3-$}; 
				
				\fill[color=magenta] ({\a-sqrt((\r^2-\b^2))},\b) circle (0.1) node[black,below]{$4+$}; 
				\fill[color=magenta] (\b,{\a-sqrt((\r^2-\b^2))}) circle (0.1) node[black,left]{$4-$}; 
				
				\draw[very thick, ->,color=red] (\a/2,\a/2) to ({\a-sqrt((\r^2-\b^2))},\a-\b); 
				\draw[very thick, ->,color=blue] (\a/2,\a/2) to (\a-\b,{\a-sqrt((\r^2-\b^2))}); 
				
				\draw[very thick, ->,color=red,dashed] (\a/2,\a/2) to (\b,{\a-sqrt((\r^2-\b^2))}); 
				\draw[very thick, ->,color=blue,dashed] (\a/2,\a/2) to ({sqrt((\r^2-\b^2))},\a-\b); 
				
				\fill[color=green, opacity=0.2] (0,\r)--(0,\a-\r) to [out=0,in=-90] (\r,\a)--(\a-\r,\a) to [out=-90, in=-180] (\a,\a-\r)--(\a,\r) to [out=180, in=90] (\a-\r,0)--(\r,0) to [out=90,in=0] (0,\r);
			\end{tikzpicture}
			\caption{$\Xi_{\theta_1}^d$} 
		\end{subfigure}
		~
		\begin{subfigure}{0.14\textwidth}
			\centering
			\begin{tikzpicture}[scale=0.3,>=stealth]
				\draw[style=thick] (0,0)--(\a,0)--(\a,\a)--(0,\a)--(0,0);
				\draw[style=thin] (0,\r) arc (90:0:\r);
				\draw[style=thin] (\a,\r) arc (90:180:\r);
				\draw[style=thin] (\a,\a-\r) arc (-90:-180:\r);
				\draw[style=thin] (0,\a-\r) arc (-90:0:\r);
				
				\fill[color=magenta] (\a-\b,{\a-sqrt((\r^2-\b^2))}) circle (0.1) node[black,right]{$1+$}; 
				\fill[color=magenta] ({sqrt((\r^2-\b^2))},\b) circle (0.1) node[black,below]{$1-$}; 
				
				\fill[color=magenta] ({sqrt((\r^2-\b^2))},\a-\b) circle (0.1) node[black,above]{$2+$}; 
				\fill[color=magenta] (\a-\b,{sqrt((\r^2-\b^2))}) circle (0.1) node[black,right]{$2-$}; 
				
				\fill[color=magenta] (\b,{sqrt((\r^2-\b^2))}) circle (0.1) node[black,left]{$3+$}; 
				\fill[color=magenta] ({\a-sqrt((\r^2-\b^2))},\a-\b) circle (0.1) node[black,above]{$3-$}; 
				
				\fill[color=magenta] ({\a-sqrt((\r^2-\b^2))},\b) circle (0.1) node[black,below]{$4+$}; 
				\fill[color=magenta] (\b,{\a-sqrt((\r^2-\b^2))}) circle (0.1) node[black,left]{$4-$}; 
				
				\draw[very thick, ->,color=red] (\a/2,\a/2) to ({\a-sqrt((\r^2-\b^2))},\a-\b); 
				\draw[very thick, ->,color=blue] (\a/2,\a/2) to (\a-\b,{\a-sqrt((\r^2-\b^2))}); 
				
				\draw[very thick, ->,color=blue,dashed] (\a/2,\a/2) to (\b,{\a-sqrt((\r^2-\b^2))}); 
				\draw[very thick, ->,color=red,dashed] (\a/2,\a/2) to ({sqrt((\r^2-\b^2))},\a-\b); 
				
				\fill[color=green, opacity=0.2] (0,\r)--(0,\a-\r) to [out=0,in=-90] (\r,\a)--(\a-\r,\a) to [out=-90, in=-180] (\a,\a-\r)--(\a,\r) to [out=180, in=90] (\a-\r,0)--(\r,0) to [out=90,in=0] (0,\r);
			\end{tikzpicture}
			\caption{$\Xi_{\theta_1}^e$} 
		\end{subfigure}
		~
		\begin{subfigure}{0.14\textwidth}
			\centering
			\begin{tikzpicture}[scale=0.3,>=stealth]
				\draw[style=thick] (0,0)--(\a,0)--(\a,\a)--(0,\a)--(0,0);
				\draw[style=thin] (0,\r) arc (90:0:\r);
				\draw[style=thin] (\a,\r) arc (90:180:\r);
				\draw[style=thin] (\a,\a-\r) arc (-90:-180:\r);
				\draw[style=thin] (0,\a-\r) arc (-90:0:\r);
				
				\fill[color=magenta] (\a-\b,{\a-sqrt((\r^2-\b^2))}) circle (0.1) node[black,right]{$1+$}; 
				\fill[color=magenta] ({sqrt((\r^2-\b^2))},\b) circle (0.1) node[black,below]{$1-$}; 
				
				\fill[color=magenta] ({sqrt((\r^2-\b^2))},\a-\b) circle (0.1) node[black,above]{$2+$}; 
				\fill[color=magenta] (\a-\b,{sqrt((\r^2-\b^2))}) circle (0.1) node[black,right]{$2-$}; 
				
				\fill[color=magenta] (\b,{sqrt((\r^2-\b^2))}) circle (0.1) node[black,left]{$3+$}; 
				\fill[color=magenta] ({\a-sqrt((\r^2-\b^2))},\a-\b) circle (0.1) node[black,above]{$3-$}; 
				
				\fill[color=magenta] ({\a-sqrt((\r^2-\b^2))},\b) circle (0.1) node[black,below]{$4+$}; 
				\fill[color=magenta] (\b,{\a-sqrt((\r^2-\b^2))}) circle (0.1) node[black,left]{$4-$}; 
				
				\draw[very thick, ->,color=red] (\a/2,\a/2) to (\a-\b,{sqrt((\r^2-\b^2))}); 
				\draw[very thick, ->,color=blue] (\a/2,\a/2) to (\a-\b,{\a-sqrt((\r^2-\b^2))}); 
				
				\draw[very thick, ->,color=blue,dashed] (\a/2,\a/2) to ({sqrt((\r^2-\b^2))},\b); 
				\draw[very thick, ->,color=red,dashed] (\a/2,\a/2) to ({sqrt((\r^2-\b^2))},\a-\b); 
				
				\fill[color=green, opacity=0.2] (0,\r)--(0,\a-\r) to [out=0,in=-90] (\r,\a)--(\a-\r,\a) to [out=-90, in=-180] (\a,\a-\r)--(\a,\r) to [out=180, in=90] (\a-\r,0)--(\r,0) to [out=90,in=0] (0,\r);
			\end{tikzpicture}
			\caption{$\Xi_{\theta_2}^d$}
		\end{subfigure}
		~
		\begin{subfigure}{0.14\textwidth}
			\centering
			\begin{tikzpicture}[scale=0.3,>=stealth]
				\draw[style=thick] (0,0)--(\a,0)--(\a,\a)--(0,\a)--(0,0);
				\draw[style=thin] (0,\r) arc (90:0:\r);
				\draw[style=thin] (\a,\r) arc (90:180:\r);
				\draw[style=thin] (\a,\a-\r) arc (-90:-180:\r);
				\draw[style=thin] (0,\a-\r) arc (-90:0:\r);
				
				\fill[color=magenta] (\a-\b,{\a-sqrt((\r^2-\b^2))}) circle (0.1) node[black,right]{$1+$}; 
				\fill[color=magenta] ({sqrt((\r^2-\b^2))},\b) circle (0.1) node[black,below]{$1-$}; 
				
				\fill[color=magenta] ({sqrt((\r^2-\b^2))},\a-\b) circle (0.1) node[black,above]{$2+$}; 
				\fill[color=magenta] (\a-\b,{sqrt((\r^2-\b^2))}) circle (0.1) node[black,right]{$2-$}; 
				
				\fill[color=magenta] (\b,{sqrt((\r^2-\b^2))}) circle (0.1) node[black,left]{$3+$}; 
				\fill[color=magenta] ({\a-sqrt((\r^2-\b^2))},\a-\b) circle (0.1) node[black,above]{$3-$}; 
				
				\fill[color=magenta] ({\a-sqrt((\r^2-\b^2))},\b) circle (0.1) node[black,below]{$4+$}; 
				\fill[color=magenta] (\b,{\a-sqrt((\r^2-\b^2))}) circle (0.1) node[black,left]{$4-$}; 
				
				\draw[very thick, ->,color=red] (\a/2,\a/2) to (\a-\b,{sqrt((\r^2-\b^2))}); 
				\draw[very thick, ->,color=blue] (\a/2,\a/2) to (\a-\b,{\a-sqrt((\r^2-\b^2))}); 
				
				\draw[very thick, ->,color=red,dashed] (\a/2,\a/2) to ({sqrt((\r^2-\b^2))},\b); 
				\draw[very thick, ->,color=blue,dashed] (\a/2,\a/2) to ({sqrt((\r^2-\b^2))},\a-\b); 
				
				\fill[color=green, opacity=0.2] (0,\r)--(0,\a-\r) to [out=0,in=-90] (\r,\a)--(\a-\r,\a) to [out=-90, in=-180] (\a,\a-\r)--(\a,\r) to [out=180, in=90] (\a-\r,0)--(\r,0) to [out=90,in=0] (0,\r);
			\end{tikzpicture}
			\caption{$\Xi_{\theta_2}^e$}
		\end{subfigure}
		~
		\begin{subfigure}{0.14\textwidth}
			\centering
			\begin{tikzpicture}[scale=0.3,>=stealth]
				\draw[style=thick] (0,0)--(\a,0)--(\a,\a)--(0,\a)--(0,0);
				\draw[style=thin] (0,\r) arc (90:0:\r);
				\draw[style=thin] (\a,\r) arc (90:180:\r);
				\draw[style=thin] (\a,\a-\r) arc (-90:-180:\r);
				\draw[style=thin] (0,\a-\r) arc (-90:0:\r);
				
				\fill[color=magenta] (\a-\b,{\a-sqrt((\r^2-\b^2))}) circle (0.1) node[black,right]{$1+$}; 
				\fill[color=magenta] ({sqrt((\r^2-\b^2))},\b) circle (0.1) node[black,below]{$1-$}; 
				
				\fill[color=magenta] ({sqrt((\r^2-\b^2))},\a-\b) circle (0.1) node[black,above]{$2+$}; 
				\fill[color=magenta] (\a-\b,{sqrt((\r^2-\b^2))}) circle (0.1) node[black,right]{$2-$}; 
				
				\fill[color=magenta] (\b,{sqrt((\r^2-\b^2))}) circle (0.1) node[black,left]{$3+$}; 
				\fill[color=magenta] ({\a-sqrt((\r^2-\b^2))},\a-\b) circle (0.1) node[black,above]{$3-$}; 
				
				\fill[color=magenta] ({\a-sqrt((\r^2-\b^2))},\b) circle (0.1) node[black,below]{$4+$}; 
				\fill[color=magenta] (\b,{\a-sqrt((\r^2-\b^2))}) circle (0.1) node[black,left]{$4-$}; 
				
				\draw[very thick, ->,color=red] (\a/2,\a/2) to ({\a-sqrt((\r^2-\b^2))},\b); 
				\draw[very thick, ->,color=blue] (\a/2,\a/2) to (\a-\b,{\a-sqrt((\r^2-\b^2))}); 
				
				\draw[very thick, ->,color=blue,dashed] (\a/2,\a/2) to ({sqrt((\r^2-\b^2))},\b); 
				\draw[very thick, ->,color=red,dashed] (\a/2,\a/2) to (\b,{\a-sqrt((\r^2-\b^2))}); 
				
				\fill[color=green, opacity=0.2] (0,\r)--(0,\a-\r) to [out=0,in=-90] (\r,\a)--(\a-\r,\a) to [out=-90, in=-180] (\a,\a-\r)--(\a,\r) to [out=180, in=90] (\a-\r,0)--(\r,0) to [out=90,in=0] (0,\r);
			\end{tikzpicture}
			\caption{$\Xi_{\pi/2}^d$}
		\end{subfigure}
		~
		\begin{subfigure}{0.14\textwidth}
			\centering
			\begin{tikzpicture}[scale=0.3,>=stealth]
				\draw[style=thick] (0,0)--(\a,0)--(\a,\a)--(0,\a)--(0,0);
				\draw[style=thin] (0,\r) arc (90:0:\r);
				\draw[style=thin] (\a,\r) arc (90:180:\r);
				\draw[style=thin] (\a,\a-\r) arc (-90:-180:\r);
				\draw[style=thin] (0,\a-\r) arc (-90:0:\r);
				
				\fill[color=magenta] (\a-\b,{\a-sqrt((\r^2-\b^2))}) circle (0.1) node[black,right]{$1+$}; 
				\fill[color=magenta] ({sqrt((\r^2-\b^2))},\b) circle (0.1) node[black,below]{$1-$}; 
				
				\fill[color=magenta] ({sqrt((\r^2-\b^2))},\a-\b) circle (0.1) node[black,above]{$2+$}; 
				\fill[color=magenta] (\a-\b,{sqrt((\r^2-\b^2))}) circle (0.1) node[black,right]{$2-$}; 
				
				\fill[color=magenta] (\b,{sqrt((\r^2-\b^2))}) circle (0.1) node[black,left]{$3+$}; 
				\fill[color=magenta] ({\a-sqrt((\r^2-\b^2))},\a-\b) circle (0.1) node[black,above]{$3-$}; 
				
				\fill[color=magenta] ({\a-sqrt((\r^2-\b^2))},\b) circle (0.1) node[black,below]{$4+$}; 
				\fill[color=magenta] (\b,{\a-sqrt((\r^2-\b^2))}) circle (0.1) node[black,left]{$4-$}; 
				
				\draw[very thick, ->,color=red] (\a/2,\a/2) to ({\a-sqrt((\r^2-\b^2))},\b); 
				\draw[very thick, ->,color=blue] (\a/2,\a/2) to (\a-\b,{\a-sqrt((\r^2-\b^2))}); 
				
				\draw[very thick, ->,color=red,dashed] (\a/2,\a/2) to ({sqrt((\r^2-\b^2))},\b); 
				\draw[very thick, ->,color=blue,dashed] (\a/2,\a/2) to (\b,{\a-sqrt((\r^2-\b^2))}); 
				
				\fill[color=green, opacity=0.2] (0,\r)--(0,\a-\r) to [out=0,in=-90] (\r,\a)--(\a-\r,\a) to [out=-90, in=-180] (\a,\a-\r)--(\a,\r) to [out=180, in=90] (\a-\r,0)--(\r,0) to [out=90,in=0] (0,\r);
			\end{tikzpicture}
			\caption{$\Xi_{\pi/2}^e$}
		\end{subfigure}
		\caption{All channels of the Umklapp processes. (i) $|\mathbf{G}|=\frac{\pi}{a}$: (e), (f), (g), (h), (i), (j) and (ii) $|\mathbf{G}|=\frac{\sqrt{2}\pi}{a}$: (a), (b), (c), (d). Descriptions of the line (dashed) arrows and angles are given in Fig. \ref{fig:AnglesInFSandFeynmanRepresentation}. Here, $\Upsilon_{\theta}^{d/e}$ ($\Xi_{\theta}^{d/e}$) denotes a random charge potential vertex coupling constant of a direct/exchange channel with a reciprocal lattice vector $|\mathbf{G}|=\frac{\sqrt{2}\pi}{a}$ ($|\mathbf{G}|=\frac{\pi}{a}$), and an angle between two Fermi wave vectors of incoming fermions ($\mathbf{k}_F^{(i_1)},\; \mathbf{k}_{F}^{(i_3)}$) is given by $\theta$. 
			One can think terms with `direct' and `exchange' notations of the Umklapp processes as the terms used to label some scattering channels and the ir counterpart scattering channels with exchanged out-going Fermi wave vectors ($\mathbf{k}_F^{(i_2)},\; \mathbf{k}_F^{(i_4)}$). In the case of $\Upsilon_0$, we do not have to specify wether it is direct or exchange channel: they are same when $\theta$ is zero.} \label{fig:UmklappScatteringChannels}
	\end{figure}
\end{widetext}

\subsection{Regularized effective action} \label{sec:Regularization}

To re-sum quantum corrections in the perturbative RG analysis, it is necessary to regularize UV divergences. Here, there are four types of scattering vertices, (i) Yukawa coupling, (ii) boson self-interaction, (iii) fermion random charge potential, and (iv) boson random mass, denoted by ($g$, $u$, $\Gamma$, $\Gamma_M$), respectively, in the effective action $S_{eff}$ (Eq. \eqref{eq:effAction}). When there exist several types of massless degrees of freedom, it is not easy to find a single regularization scheme, setting all the interaction vertices to be marginal \cite{YANG2021168462}. This is called generic scale invariance \cite{RevModPhys.77.579}. In this study, there are two kinds of gapless fluctuations, hot-spot electrons on the Fermi surface and critical SDW order-parameter fluctuations, in the presence of four types of scattering vertices, as mentioned above. To regularize quantum fluctuations involved with hot-spot Fermi-surface electrons, we use a co-dimensional regularization method 
\cite{DaliLee,SurLee}. We find that the co-dimensional regularization scheme fails to regularize quantum fluctuations from the random boson mass vertex. This will be clarified in the tree-level scaling analysis of the next section. In this respect, we introduce another regularization scheme, so-called `correlated random mass probability regularization' to regularize the quantum fluctuation from the random boson mass vertex.

\subsubsection{Correlated random boson mass probability regularization} \label{sec:CorrelatedRandomBosonMassRegularization}

A conventional regularization scheme for quantum fluctuations from the random boson mass vertex is to change the scaling dimension of time \cite{zinn2002quantum}. Unfortunately, it turns out that quantum fluctuations from both the Yukawa interaction and the random boson mass vertex can not be regularized at the same time using this regularization scheme. Therefore we introduce another regularization scheme \cite{Moon} which changes the Gaussian distribution function to a following non-local form: $\mathcal{P}_{rBM,\alpha}[m^2(x)]=e^{-\int\frac{d^2q}{(2\pi)^2}\frac{m^2(q)m^2(-q)}{2\Gamma_M|\vec{q}|^{\alpha}}}$. Here, $|\vec{q}|^\alpha$ is introduced into the distribution function of which $\alpha$ is bigger than zero. If the $\alpha$ is set to zero, it is reduced to the conventional Gaussian distribution function. One can think the introduction of $|\vec{q}|^{\alpha}$ as a change of the variance from $\Gamma_M$ to $\Gamma_M|\vec{q}|^\alpha \equiv \Gamma_{M,\alpha}(\vec{q})$ for the $m^2(q)$ field. Physically, it means that the variance of the random boson mass is reduced as the momentum decreases. As a result, the magnitude of the variance of the random boson mass is reduced in the low energy regime (small $|q|$).

A disorder-averaged effective action with the changed probability distribution is given by
\begin{align}
	S_{rBM}^{reg.} &= \sum_{a,b=1}^R\int\frac{d\omega}{2\pi}\frac{d\omega'}{2\pi}\prod_{i=1}^4\frac{d^2\mathbf{k}_i}{(2\pi)^2}\frac{\Gamma_M|\mathbf{k}_1+\mathbf{k}_2|^\alpha}{2}\nonumber\\
	&\vec{\phi}^a(\omega,\mathbf{k}_1)\cdot\vec{\phi}^a(\omega,\mathbf{k}_2)\vec{\phi}^b(\omega',\mathbf{k}_3)\cdot\vec{\phi}^b(\omega',\mathbf{k}_4)(2\pi)^2\nonumber\\
	&\delta(\mathbf{k}_1+\mathbf{k}_2+\mathbf{k}_3+\mathbf{k}_4) .\label{eq:rBMregulAction}
\end{align}
The non-local structure of this scattering vertex does not allow self-renormalizations from quantum fluctuations, distinguished from other local interaction vertices. This is explicitly shown in the one-loop calculation, given in Appendix \ref{Appendix:OneLoopRBMVertex}. This is a general drawback of regularization schemes introducing action of the non-local forms \cite{Mross,DaliLee}.

\begin{widetext}
	\subsubsection{Co-dimensional regularization}
	Since the co-dimensional regularization technique is well explained in the previous studies \cite{DaliLee,SurLee}, here we only present results of the co-dimensional regularization. The resulting regularized effective action is given by 
	\begin{align}
		S_{eff}&=\sum_{a=1}^R\Bigg[\sum_{i_f=1}^{N_f}\sum_{n=1}^4\sum_{\sigma=1}^{N_c}\int dk \Bar{\Psi}^a_{n,\sigma,i_f}(k)[i\gamma_0k_0+i\mathbf{\Gamma}_\perp\cdot\mathbf{K}_\perp+i\gamma_{d-1}\epsilon_n(\mathbf{k})]\Psi^a_{n,\sigma,i_f}(k)\nonumber\\
		&+\frac{1}{4}\int dq [q_0^2+c_\perp^2 |\mathbf{Q}_\perp|^2+c^2|\mathbf{q}|^2]Tr[\Phi^a(-q)\Phi^a(q)]+\frac{ig}{\sqrt{N_f}}\sum_{i_f=1}^{N_f}\sum_{n=1}^4\sum_{\sigma,\sigma'=1}^{N_c}\int dx \bar{\Psi}^a_{\bar{n},\sigma,i_f}(x)\Phi_{\sigma,\sigma'}^a(x)\gamma_{d-1}\Psi^a_{n,\sigma',i_f}(x)\nonumber\\
		&+\frac{1}{4}\int dx \Big(u_1Tr[\Phi^a(x)\Phi^a(x)] Tr[\Phi^a(x)\Phi^a(x)]+u_2  Tr[\Phi^a(x)\Phi^a(x) \Phi^a(x)\Phi^a(x)]\Big)\Bigg]-\sum_{a,b=1}^R\int \frac{d\omega}{2\pi}\int \frac{d\omega'}{2\pi}\prod_{i=1}^4 \int \frac{d^{d}\mathbf{k}_i}{(2\pi)^d}\nonumber\\
		&\times (2\pi)^d\Big[\sum_{i_f,j_f=1}^{N_f}\sum_{i=1}^{27}\sum_{\sigma,\sigma'=1}^{N_c}\frac{\Gamma_i}{2N_f}\delta(\mathbf{k}_1+\mathbf{k}_3-\mathbf{k}_2-\mathbf{k}_4) \Big([\bar{\Psi}^a_{n,\sigma,i_f}(\omega,\mathbf{k}_1) \mathcal{M}^i_{nm}\Psi^a_{m,\sigma,i_f}(\omega,\mathbf{k}_2)]\nonumber\\
		&\times [\bar{\Psi}^b_{k,\sigma',j_f}(\omega',\mathbf{k}_3) \tilde{\mathcal{M}}^i_{kl}\Psi^b_{l,\sigma',j_f}(\omega',\mathbf{k}_4)]+\cdots\Big)+\frac{\Gamma_M\Big(|\mathbf{K}_{1,\perp}+\mathbf{K}_{2,\perp}|^\alpha+\kappa |\vec{k}_1+\vec{k}_2|^{\alpha} \Big)}{8}\nonumber\\
		&\times Tr[\Phi^a(\omega,\mathbf{k}_1)\Phi^a(\omega,\mathbf{k}_2)] Tr[\Phi^b(\omega',\mathbf{k}_3)\Phi^b(\omega',\mathbf{k}_4)] \delta(\mathbf{k}_1+\mathbf{k}_2+\mathbf{k}_3+\mathbf{k}_4)\Big] , \label{eq:effActionRegularized}
	\end{align}
\end{widetext}
where $a$,$b$ are replica indices, $\sigma$ and $i_f$ are a spin and flavor index respectively. $\int dx=\int d\tau\int d x_{d-1}\int dx_{d}\int d^{d-2}d\mathbf{X}_\perp$, $\int dk = \int \frac{dk_0}{2\pi} \int \frac{dk_{d-1}}{2\pi} \int \frac{dk_d}{2\pi} \int \frac{d^{d-2}\mathbf{K}_\perp}{(2\pi)^{d-2}}$. $\Psi_{i,\sigma}$ is a two-component Fermion spinor field given by $\Psi_{1,\sigma}^a=({\psi_{a,1,\sigma}^{(+)}},\psi_{a,3,\sigma}^{(+)})^T $, $\Psi_{2,\sigma}^a=({\psi_{a,2,\sigma}^{(+)}},\psi_{a,4,\sigma}^{(+)})^T $, $\Psi_{3,\sigma}^a=({\psi_{a,1,\sigma}^{(-)}},-\psi_{a,3,\sigma}^{(-)})^T$ and $\Psi_{4,\sigma}^a=({\psi_{a,2,\sigma}^{(-)}},-\psi_{a,4,\sigma}^{(-)})^T $. $\Phi^a(q)=\sum_{i=1}^{N_c^2-1}\phi_i^a(q)\tau^i$ is an SU($N_{c}$) matrix order-parameter field. The flavor number of fermions and the number of spins are increased from 1 to $N_f$ and $2$ to $N_c$ respectively. Due to the increased number of the spin, a single boson interaction vertex $u_0$ is generalized to the two boson interaction vertices given by $u_1$ and $u_2$ \cite{SurLee}. The Yukawa coupling constant $g$ and the random charge potential coupling constant $\Gamma_i$ are changed to $\frac{g}{\sqrt{N_f}}$ and $\frac{\Gamma_i}{N_f}$ respectively to make the effective action the order of $\mathcal{O}(N_f)$. Here, a symbol $\perp$ denotes the co-dimension of the Fermi-surface. Explicit forms of the regularized random charge potential vertices are given in the Supplementary Material \cite{Note1}. In the random boson mass vertex, the factor $|\vec{q}|^{\alpha}$ in Eq. \eqref{eq:rBMregulAction} is translated to $\Big(|\mathbf{Q}_\perp|^\alpha+\kappa |\vec{q}|^\alpha\Big)$  due to lack of the Lorentz symmetry between $\mathbf{Q}_\perp$ and $\vec{q}$. Since $\kappa$ is not renormalized, we set it to 1 in the remaining context.

\subsubsection{Translation symmetry breaking in the co-dimensional regularization} \label{sec:TranslationSymBreakingInCoDimRegularization}

Before proceeding to the RG analysis, let us discuss the effects of translation symmetry breaking, due to the co-dimensional regularization, on the random charge potential vertices. As explained in Ref. \cite{SurLee}, translation symmetry is explicitly broken in the co-dimensional regularized effective action $S_{eff}$ (Eq. \eqref{eq:effActionRegularized}). Indeed, $S_{eff}$ (Eq. \eqref{eq:effActionRegularized}) has a $p_{2k_F,z}-$wave charge density wave order in three spatial dimensions, which corresponds to the upper critical dimension of the two-dimensional SDW quantum critical system. See the Supplementary Material \cite{Note1} for details of the translation symmetry breaking by the co-dimensional regularization. It turns out that the translation symmetry breaking by the co-dimensional regularization causes the following problem in dealing with the disordered system: \textit{Random charge potential vertices which do not exist in the original effective action (Eq. \eqref{eq:effAction}) can be generated in the co-dimensional regularized effective action (Eq. \eqref{eq:effActionRegularized}) by loop corrections.} This is a general problem we face in considering disorders using the co-dimensional regularization or the dimensional regularization. In this paper, we assume that these terms are fine-tuned to zero.

\section{Renormalization Group Analysis} \label{sec:RGanalysis}

We perform the perturbative RG analysis for the co-dimensional regularized effective action $S_{eff}$ (Eq. \eqref{eq:effActionRegularized}) in the high-energy scheme. We refer to Appendix \ref{Appendix:RGSetting} for details of the RG setting. Here, we focus on the results of the RG analysis.

\subsection{Classical scaling}
\begin{figure}
	\centering
	\includegraphics[scale=0.2]{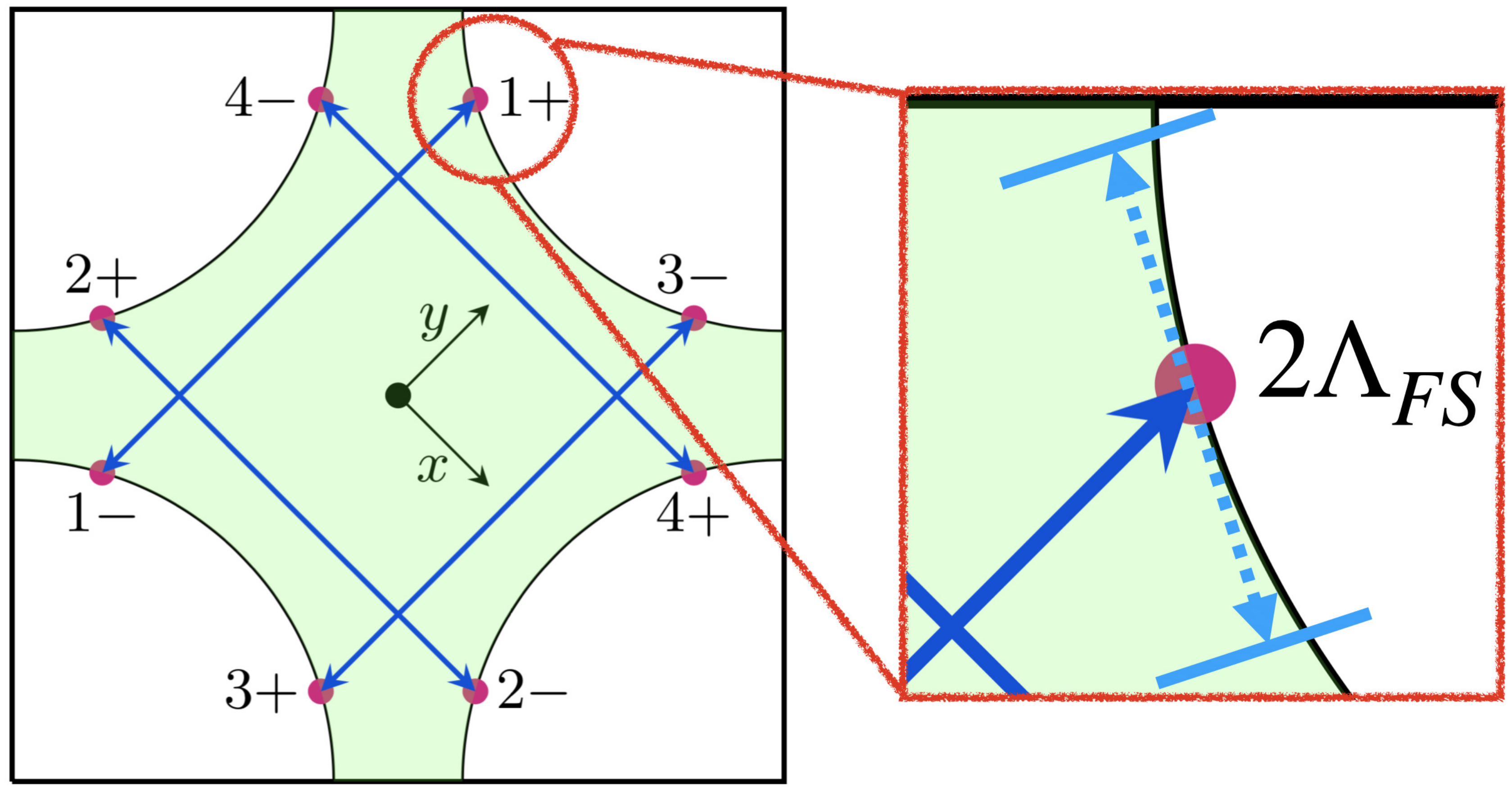}
	\caption{$\Lambda_{FS}$ is a size of a hot-spot.} \label{fig:SizeOfHotSpot}
\end{figure}

First, we start from the tree-level scaling analysis. Demanding the regularized effective action $S_{eff}$ (Eq. \eqref{eq:effActionRegularized}) to be dimensionless, we obtain classical scaling dimensions of the coupling constants and fields as follows
\begin{gather*}
	[\Psi_{n,\sigma,i_f}^a]=-\frac{d+2}{2},\;[\phi_i^a]=-\frac{d+3}{2},\; [g]=\frac{3-d}{2},\\ [u_1]=[u_2]=3-d,\; [\Gamma_i]=2-d,\; [\Gamma_M]=4-d-\alpha .
\end{gather*}
where we used $[k_0]=[\mathbf{K}_\perp]=[k_{d-1}]=[k_{d}]=1$. Here, $[\mathcal{O}]$ denotes the scaling dimension of a parameter $\mathcal{O}$. $\Gamma_i \in (\Gamma_0,\cdots,\Xi_{\pi/2}^e)$ is a coupling constant of the random charge potential vertices. Setting $d=3-\epsilon$ and $\alpha=1-\bar{\epsilon}$, the scaling dimensions of the coupling constants are given by $[g]=\frac{\epsilon}{2}$, $[u_1]=[u_2]=\epsilon$, $[\Gamma_i]=-1+\epsilon$ and $[\Gamma_M]=\epsilon+\bar{\epsilon}$. Three types of coupling constants, Yukawa $g$, boson self-interactions $u_1$ ($u_2$), and boson random mass $\Gamma_M$ are marginal at $d=3$ and $\alpha=1$. On the other hand, random charge potential vertices $\Gamma_i$ for hot-spot fermions are irrelevant at $d=3$ and $\alpha=1$. Therefore, it seems that all random charge potential vertices can be ignored in the low energy limit. However, it turns out that a real expansion parameter is not $\Gamma_i$ but $\bar{\Gamma}_i=\Gamma_i\Lambda_{FS}$, where $\Lambda_{FS}$ is a size of the hot spots, shown in Fig. \ref{fig:SizeOfHotSpot}.

To discuss the origin of $\Lambda_{FS}$ in the expansion parameter $\bar{\Gamma}_i$, let us first recall the Shankar's RG approach \cite{ShankarRMP} to the Fermi liquid state. In the conventional RG analysis with a Fermi surface, directions parallel to the Fermi surface are not scaled and only the direction perpendicular to the Fermi surface is scaled. This is because the Fermi energy changes only along the perpendicular direction, and high-energy fermion fields are integrated out to give renormalization effects. As a result, the size of the Fermi surface does not change in the RG process, and the scaling dimensions of coupling constants are determined by the scaling dimension of a momentum coordinate perpendicular to the Fermi surface. However, momentum coordinates parallel to the Fermi surface is also scaled in the hot spot model. This is due to the existence of critical boson excitations coupled to the Fermi surface. In the case of boson fields, energy increases in all directions (both parallel and perpendicular to the Fermi surface), which differs from that of fermion fields. Therefore, loop corrections involving boson fields can give UV divergences responsible for renormalization of the effective action even by integrals of momentums parallel to the Fermi surface. On the other hand, the integral of the parallel momentum does not contribute to the UV divergence in dealing with the random charge potential vertices as in the RG approach to the Fermi liquid state. 
As a result, Feynman-diagram calculations involving the random charge potential vertices give 
$(\Gamma_i\Lambda_{FS})^{n_{\Gamma}}\frac{1}{\epsilon}=(\bar{\Gamma}_i)^{n_{\Gamma}}\frac{1}{\epsilon}$ instead of $(\Gamma_i)^{n_{\Gamma}}\frac{1}{\epsilon}$ where $n_{\Gamma}$ is a number of the random charge potential vertices in a given diagram. The proof of the new expansion parameter $\bar{\Gamma}_i$ for loops giving the log-divergence is given in Appendix \ref{Appendix:ProofOfNewExpansionParameter}. Since the tree-level scaling dimensions of momentum coordinate variables, $k_0$, $\mathbf{K}_\perp$, $k_{d-1}$, and $k_d$ are set to 1, the dimension of $\Lambda_{FS}$ is also given by 1. This results in the scaling dimension of the new expansion parameter $\bar{\Gamma}_i$ is given by $\epsilon$ ($\because [\bar{\Gamma}_i]=[\Gamma_i]+[\Lambda_{FS}]=\epsilon$) which is marginal at $d=3-\epsilon$. As a consequence, all the coupling constants of the interaction vertices given by $g$, $u_1$, $u_2$, $\bar{\Gamma}_i$, and $\Gamma_M$ are marginal near $d=3$ and $\alpha=1$ and  perturbative RG analysis is applicable. From now on, we denote $\bar{\Gamma}_i$ as $\Gamma_i$ for notational simplicity.

\subsection{Renormalization group results in the one-loop level} \label{sec:OneLoopRG}
	\begin{figure}
		\centering
		\begin{subfigure}[h]{0.15\textwidth}
			\centering
			\begin{tikzpicture}[scale=0.8]
				\begin{feynhand}
					\vertex (a) at (0,0); \vertex (b) at (0.7,0); \vertex (c) at (2,0); \vertex (d) at (2.7,0);
					\propag[fermion] (a) to (b); \propag[fermion] (b) to (c); \propag[fermion] (c) to (d);
					\propag[boson] (b) to [out=90, in=90,looseness=1.5](c);
				\end{feynhand}
			\end{tikzpicture}
			\caption{}\label{fig:SDWFermionSelfEnergy}
		\end{subfigure}
		~
		\begin{subfigure}[h]{0.15\textwidth}
			\centering
			\begin{tikzpicture}[scale=0.8]
				\begin{feynhand}
					\vertex (a) at (0,0); \vertex (b) at (0.7,0); \vertex (c) at (2,0); \vertex (d) at (2.7,0);
					\propag[boson] (a) to (b); \propag[fermion] (b) to [out=50,in=130,looseness=1.5](c); \propag[fermion] (c) to [out=230, in=-50,looseness=1.5](b); \propag[boson] (c) to (d);
				\end{feynhand}
			\end{tikzpicture}
			\caption{}\label{fig:SDWBosonSelfEnergy}
		\end{subfigure}
		~
		\begin{subfigure}[h]{0.15\textwidth}
			\centering
			\begin{tikzpicture}[scale=0.7]
				\begin{feynhand}
					\vertex (a) at (0,0); \vertex (b) at (0.5,0); \vertex (c) at (1.5,0.5); \vertex (d) at (2,1);
					\vertex (e) at (1.5,-0.5); \vertex (f) at (2,-1);
					\propag[boson] (a) to (b); \propag[fermion] (b) to (c); \propag[fermion] (c) to (d); \propag[fermion] (f) to (e); \propag[fermion] (e) to (b); \propag[boson] (c) to (e);
				\end{feynhand}
			\end{tikzpicture}
			\caption{}\label{fig:SDWBFVertexCorrection}
		\end{subfigure}
		~
		\begin{subfigure}[h]{0.25\textwidth}
			\centering
			\begin{tikzpicture}
				\begin{feynhand}
					\vertex (a) at (0,0.5); \vertex (b) at (0.5,0); \vertex (c) at (0,-0.5); \vertex (d) at (1.5,0);
					\vertex (e) at (2,0.5); \vertex (f) at (2,-0.5);
					\propag[boson] (a) to (b); \propag[boson] (b) to (c); \propag[boson] (b) to [out=50,in=130](d); \propag[boson] (d) to [out=230, in=-50](b); \propag[boson] (f) to (d); \propag[boson] (e) to (d);
				\end{feynhand}
			\end{tikzpicture}
			~
			\begin{tikzpicture}[scale=0.7]
				\begin{feynhand}
					\vertex (a) at (-1,1); \vertex (b) at (1,1); \vertex (c) at (1,-1); \vertex (d) at (-1,-1);
					\vertex (e) at (-0.5,0.5); \vertex (f) at (0.5,0.5); \vertex (g) at (0.5,-0.5); \vertex (h) at (-0.5,-0.5);
					\propag[boson] (a) to (e); \propag[boson] (b) to (f); \propag[boson] (c) to (g); \propag[boson] (d) to (h); \propag[fermion] (e) to (f); \propag[fermion] (f) to (g); \propag[fermion] (g) to (h); \propag[fermion] (h) to (e);
				\end{feynhand}
			\end{tikzpicture}
			\caption{}\label{fig:SDWBSICorrection}
		\end{subfigure}
		\caption{One-loop Feynman diagrams in the clean case; (a) Fermion self energy, (b) Boson self energy, (c) Yukawa vertex correction and (d) Boson self-interaction vertex correction. Here, the real (wavy) line represents an electron (boson) propagator.} \label{fig:SDWCleanOneLoopDiagrams}
	\end{figure}
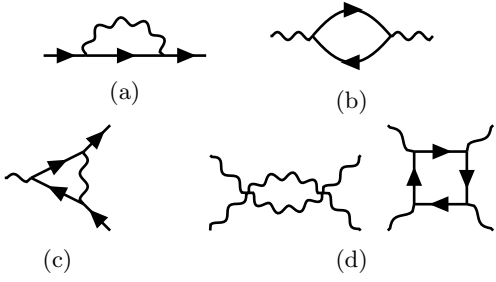

	We review the one-loop results of the clean system \cite{SurLee} first. One-loop Feynman diagrams for the clean system are given in Fig. \ref{fig:SDWCleanOneLoopDiagrams}. Ref. \cite{SurLee} found a nontrivial low-energy fixed point, specified by $c^*=v^*=g^*=u_1^*=u_2^*=0$. Nature of this fixed point is clarified, considering dimensionless parameters given by $w=\frac{v}{c}$, $\lambda=\frac{g^2}{v}$, $\kappa_1=\frac{u_1}{c^2}$, and $\kappa_2=\frac{u_2}{c^2}$ which have following fixed point values: $w^*=\frac{N_cN_f}{N_c^2-1}$, $\lambda^*=\frac{4\pi(N_c^2+N_cN_f-1)}{N_c^2+N_cN_f-3}\epsilon$, $\kappa_1^*=\kappa_2^*=0$ \cite{SurLee}.
	
		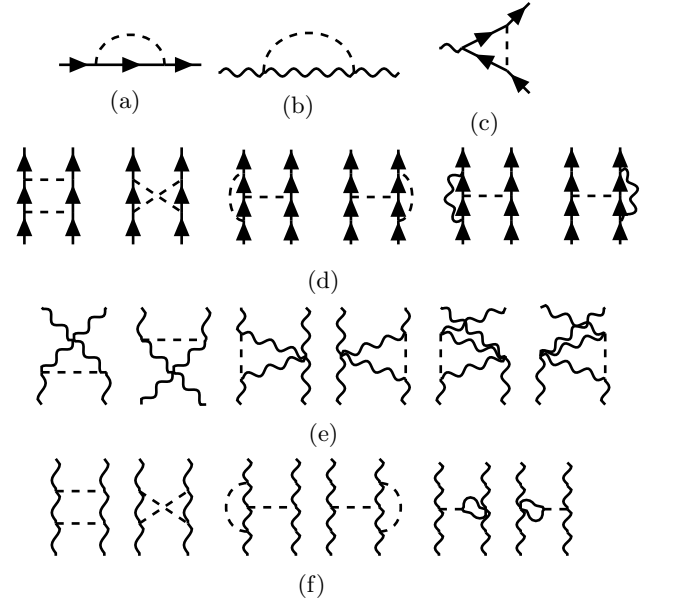
\begin{figure}
		\centering
		\begin{subfigure}[h]{0.1\textwidth}
			\centering
			\begin{tikzpicture}[scale=0.7]
				\begin{feynhand}
					\vertex (a) at (0,0); \vertex (b) at (0.7,0); \vertex (c) at (2,0); \vertex (d) at (2.7,0);
					\propag[fermion] (a) to (b); \propag[fermion] (b) to (c); \propag[fermion] (c) to (d);
					\propag[scalar] (b) to [out=90, in=90,looseness=1.5](c);
				\end{feynhand}
			\end{tikzpicture}
			\caption{}\label{fig:SDWDisorderFermionSelfEnergy}
		\end{subfigure}
		~
		\begin{subfigure}[h]{0.12\textwidth}
			\begin{tikzpicture}[scale=0.6]
				\begin{feynhand}
					\vertex (a) at (-2,0); \vertex (b) at (-1,0); \vertex (c) at (1,0); \vertex (d) at (2,0);
					\propag[boson] (a) to (b); \propag[boson] (b) to (c); \propag[boson] (c) to (d); \propag[sca] (b) to [out=90, in=90, looseness=1.5](c);
				\end{feynhand}
			\end{tikzpicture}
			\caption{}\label{fig:SDWDisorderBosonSelfEnergy}
		\end{subfigure}
		~
		\begin{subfigure}[h]{0.12\textwidth}
			\centering
			\begin{tikzpicture}[scale=0.6]
				\begin{feynhand}
					\vertex (a) at (0,0); \vertex (b) at (0.5,0); \vertex (c) at (1.5,0.5); \vertex (d) at (2,1);
					\vertex (e) at (1.5,-0.5); \vertex (f) at (2,-1);
					\propag[boson] (a) to (b); \propag[fermion] (b) to (c); \propag[fermion] (c) to (d); \propag[fermion] (f) to (e); \propag[fermion] (e) to (b); \propag[scalar] (c) to (e);
				\end{feynhand}
			\end{tikzpicture}
			\caption{}\label{fig:SDWDisorderBFVertexCorrection}
		\end{subfigure}
		~
		\begin{subfigure}[h]{0.5\textwidth}
			\centering
			\begin{tikzpicture}[scale=0.43]
				\begin{feynhand}
					\vertex (a1) at (0,0); \vertex (a2) at (0,1); \vertex (a3) at (0,2); \vertex (a4) at (0,3);
					\vertex (b1) at (1.5,0); \vertex (b2) at (1.5,1); \vertex (b3) at (1.5,2); \vertex (b4) at (1.5,3);
					\propag[fer] (a1) to (a2); \propag[fer] (a2) to (a3); \propag[fer] (a3) to (a4);
					\propag[fer] (b1) to (b2); \propag[fer] (b2) to (b3); \propag[fer] (b3) to (b4);
					\propag[sca] (a2) to (b2); \propag[sca] (a3) to (b3);
					\node at (-0.3,0); \node at (1.8,0);
				\end{feynhand}
			\end{tikzpicture}
			~
			\begin{tikzpicture}[scale=0.43]
				\begin{feynhand}
					\vertex (a1) at (0,0); \vertex (a2) at (0,1); \vertex (a3) at (0,2); \vertex (a4) at (0,3);
					\vertex (b1) at (1.5,0); \vertex (b2) at (1.5,1); \vertex (b3) at (1.5,2); \vertex (b4) at (1.5,3);
					\propag[fer] (a1) to (a2); \propag[fer] (a2) to (a3); \propag[fer] (a3) to (a4);
					\propag[fer] (b1) to (b2); \propag[fer] (b2) to (b3); \propag[fer] (b3) to (b4);
					\propag[sca] (a2) to (b3); \propag[sca] (a3) to (b2);
					\node at (-0.3,0); \node at (1.8,0);
				\end{feynhand}
			\end{tikzpicture}
			~
			\begin{tikzpicture}[scale=0.42]
				\begin{feynhand}
					\vertex (a1) at (0,0); \vertex (a2) at (0,3/4); \vertex (a3) at (0,6/4); \vertex (a4) at (0, 9/4); \vertex(a5) at (0, 3);
					\vertex (b1) at (1.5,0); \vertex (b2) at (1.5,3/4); \vertex (b3) at (1.5,6/4); \vertex (b4) at (1.5,9/4); \vertex(b5) at (1.5, 3);
					\propag[fer] (a1) to (a2); \propag[fer] (a2) to (a3); \propag[fer] (a3) to (a4); \propag[fer] (a4) to (a5);
					\propag[fer] (b1) to (b2); \propag[fer] (b2) to (b3); \propag[fer] (b3) to (b4); \propag[fer] (b4) to (b5);
					\propag[sca] (a3) to (b3); \propag[sca] (a2) to [out=180, in=180](a4);
					\node at (-0.3,0); \node at (1.8,0);
				\end{feynhand}
			\end{tikzpicture}
			~
			\begin{tikzpicture}[scale=0.42]
				\begin{feynhand}
					\vertex (a1) at (0,0); \vertex (a2) at (0,3/4); \vertex (a3) at (0,6/4); \vertex (a4) at (0, 9/4); \vertex(a5) at (0, 3);
					\vertex (b1) at (1.5,0); \vertex (b2) at (1.5,3/4); \vertex (b3) at (1.5,6/4); \vertex (b4) at (1.5,9/4); \vertex(b5) at (1.5, 3);
					\propag[fer] (a1) to (a2); \propag[fer] (a2) to (a3); \propag[fer] (a3) to (a4); \propag[fer] (a4) to (a5);
					\propag[fer] (b1) to (b2); \propag[fer] (b2) to (b3); \propag[fer] (b3) to (b4); \propag[fer] (b4) to (b5);
					\propag[sca] (a3) to (b3); \propag[sca] (b2) to [out=0, in=0](b4);
					\node at (-0.3,0); \node at (1.8,0);
				\end{feynhand}
			\end{tikzpicture}
			~
			\begin{tikzpicture}[scale=0.43]
				\begin{feynhand}
					\vertex (a1) at (0,0); \vertex (a2) at (0,3/4); \vertex (a3) at (0,6/4); \vertex (a4) at (0, 9/4); \vertex(a5) at (0, 3);
					\vertex (b1) at (1.5,0); \vertex (b2) at (1.5,3/4); \vertex (b3) at (1.5,6/4); \vertex (b4) at (1.5,9/4); \vertex(b5) at (1.5, 3);
					\propag[fer] (a1) to (a2); \propag[fer] (a2) to (a3); \propag[fer] (a3) to (a4); \propag[fer] (a4) to (a5);
					\propag[fer] (b1) to (b2); \propag[fer] (b2) to (b3); \propag[fer] (b3) to (b4); \propag[fer] (b4) to (b5);
					\propag[sca] (a3) to (b3); \propag[boson] (a2) to [out=180, in=180](a4);
					\node at (-0.3,0); \node at (1.8,0);
				\end{feynhand}
			\end{tikzpicture}
			~
			\begin{tikzpicture}[scale=0.43]
				\begin{feynhand}
					\vertex (a1) at (0,0); \vertex (a2) at (0,3/4); \vertex (a3) at (0,6/4); \vertex (a4) at (0, 9/4); \vertex(a5) at (0, 3);
					\vertex (b1) at (1.5,0); \vertex (b2) at (1.5,3/4); \vertex (b3) at (1.5,6/4); \vertex (b4) at (1.5,9/4); \vertex(b5) at (1.5, 3);
					\propag[fer] (a1) to (a2); \propag[fer] (a2) to (a3); \propag[fer] (a3) to (a4); \propag[fer] (a4) to (a5);
					\propag[fer] (b1) to (b2); \propag[fer] (b2) to (b3); \propag[fer] (b3) to (b4); \propag[fer] (b4) to (b5);
					\propag[sca] (a3) to (b3); \propag[boson] (b2) to [out=0, in=0](b4);
					\node at (-0.3,0); \node at (1.8,0);
				\end{feynhand}
			\end{tikzpicture}
			\caption{}\label{fig:SDWDisorderRCP}
		\end{subfigure}
		\\
		\begin{subfigure}[h]{0.5\textwidth}
			\centering
			\begin{tikzpicture}[scale=0.43]
				\begin{feynhand}
					\vertex (a) at (-1,1); \vertex (b) at (1,1); \vertex (c) at (0,0); \vertex (d) at (-1,-1); \vertex (e) at (1,-1); \vertex (f) at (-1,-2); \vertex (g) at (1,-2);
					\propag[boson] (a) to (c); \propag[boson] (b) to (c); \propag[boson] (c) to (d); \propag[boson] (c) to (e);
					\propag[boson] (d) to (f); \propag[boson] (e) to (g);
					\propag[sca] (d) to (e);
				\end{feynhand}
			\end{tikzpicture}
			~
			\begin{tikzpicture}[scale=0.43]
				\begin{feynhand}
					\vertex (a) at (-1,1); \vertex (b) at (1,1); \vertex (c) at (0,0); \vertex (d) at (-1,-1); \vertex (e) at (1,-1); \vertex (f) at (-1,2); \vertex (g) at (1,2);
					\propag[boson] (a) to (c); \propag[boson] (b) to (c); \propag[boson] (c) to (d); \propag[boson] (c) to (e);
					\propag[boson] (a) to (f); \propag[boson] (b) to (g);
					\propag[sca] (a) to (b);
				\end{feynhand}
			\end{tikzpicture}
			~
			\begin{tikzpicture}[scale=0.43]
				\begin{feynhand}
					\vertex (a) at (-1,1.5); \vertex (b) at (1,1.5); \vertex (c) at (-1,0.7); \vertex (d) at (-1,-0.7);  \vertex(e) at (1, 0);
					\vertex (f) at (-1,-1.5); \vertex (g) at (1,-1.5);
					\propag[boson] (a) to (c); \propag[sca] (c) to (d); \propag[boson] (d) to (f);
					\propag[boson] (c) to (e); \propag[boson] (d) to (e);   \propag[boson] (e) to (b); \propag[boson] (e) to (g);
				\end{feynhand}
			\end{tikzpicture}
			~
			\begin{tikzpicture}[scale=0.43]
				\begin{feynhand}
					\vertex (a) at (1,1.5); \vertex (b) at (-1,1.5); \vertex (c) at (1,0.7); \vertex (d) at (1,-0.7);  \vertex (e) at (-1, 0);
					\vertex (f) at (1,-1.5); \vertex (g) at (-1,-1.5);
					\propag[boson] (a) to (c); \propag[sca] (c) to (d); \propag[boson] (d) to (f);
					\propag[boson] (c) to (e); \propag[boson] (d) to (e);   \propag[boson] (e) to (b); \propag[boson] (e) to (g);
				\end{feynhand}
			\end{tikzpicture}
			~
			\begin{tikzpicture}[scale=0.43]
				\begin{feynhand}
					\vertex (a) at (-1,1.5); \vertex (b) at (1,1.5); \vertex (c) at (-1,0.7); \vertex (d) at (-1,-0.7);  \vertex (e) at (1, 0);
					\vertex (f) at (-1,-1.5); \vertex (g) at (1,-1.5);
					\propag[boson] (a) to (e); \propag[sca] (c) to (d); \propag[boson] (d) to (f);
					\propag[boson] (c) to (e); \propag[boson] (d) to (e);   \propag[boson] (c) to (b); \propag[boson] (e) to (g);
				\end{feynhand}
			\end{tikzpicture}
			~
			\begin{tikzpicture}[scale=0.43]
				\begin{feynhand}
					\vertex (a) at (1,1.5); \vertex (b) at (-1,1.5); \vertex (c) at (1,0.7); \vertex (d) at (1,-0.7);  \vertex (e) at (-1, 0);
					\vertex (f) at (1,-1.5); \vertex (g) at (-1,-1.5);
					\propag[boson] (a) to (e); \propag[sca] (c) to (d); \propag[boson] (d) to (f);
					\propag[boson] (c) to (e); \propag[boson] (d) to (e);   \propag[boson] (c) to (b); \propag[boson] (e) to (g);
				\end{feynhand}
			\end{tikzpicture}
			\caption{}\label{fig:SDWDisorderBSI}
		\end{subfigure}
		~
		\begin{subfigure}[h]{0.39\textwidth}
			\centering
			\begin{tikzpicture}[scale=0.43]
				\begin{feynhand}
					\vertex (a1) at (0,0); \vertex (a2) at (0,1); \vertex (a3) at (0,2); \vertex (a4) at (0,3);
					\vertex (b1) at (1.5,0); \vertex (b2) at (1.5,1); \vertex (b3) at (1.5,2); \vertex (b4) at (1.5,3);
					\propag[boson] (a1) to (a2); \propag[boson] (a2) to (a3); \propag[boson] (a3) to (a4);
					\propag[boson] (b1) to (b2); \propag[boson] (b2) to (b3); \propag[boson] (b3) to (b4);
					\propag[sca] (a2) to (b2); \propag[sca] (a3) to (b3);
				\end{feynhand}
			\end{tikzpicture}
			~
			\begin{tikzpicture}[scale=0.43]
				\begin{feynhand}
					\vertex (a1) at (0,0); \vertex (a2) at (0,1); \vertex (a3) at (0,2); \vertex (a4) at (0,3);
					\vertex (b1) at (1.5,0); \vertex (b2) at (1.5,1); \vertex (b3) at (1.5,2); \vertex (b4) at (1.5,3);
					\propag[boson] (a1) to (a2); \propag[boson] (a2) to (a3); \propag[boson] (a3) to (a4);
					\propag[boson] (b1) to (b2); \propag[boson] (b2) to (b3); \propag[boson] (b3) to (b4);
					\propag[sca] (a2) to (b3); \propag[sca] (a3) to (b2);
				\end{feynhand}
			\end{tikzpicture}
			~
			\begin{tikzpicture}[scale=0.43]
				\begin{feynhand}
					\vertex (a1) at (0,0); \vertex (a2) at (0,3/4); \vertex (a3) at (0,6/4); \vertex (a4) at (0, 9/4); \vertex(a5) at (0, 3);
					\vertex (b1) at (1.5,0); \vertex (b2) at (1.5,3/4); \vertex (b3) at (1.5,6/4); \vertex (b4) at (1.5,9/4); \vertex(b5) at (1.5, 3);
					\propag[boson] (a1) to (a2); \propag[boson] (a2) to (a3); \propag[boson] (a3) to (a4); \propag[boson] (a4) to (a5);
					\propag[boson] (b1) to (b2); \propag[boson] (b2) to (b3); \propag[boson] (b3) to (b4); \propag[boson] (b4) to (b5);
					\propag[sca] (a3) to (b3); \propag[sca, looseness=1.5] (a2) to [out=180, in=180](a4);
				\end{feynhand}
			\end{tikzpicture}
			~
			\begin{tikzpicture}[scale=0.43]
				\begin{feynhand}
					\vertex (a1) at (0,0); \vertex (a2) at (0,3/4); \vertex (a3) at (0,6/4); \vertex (a4) at (0, 9/4); \vertex(a5) at (0, 3);
					\vertex (b1) at (1.5,0); \vertex (b2) at (1.5,3/4); \vertex (b3) at (1.5,6/4); \vertex (b4) at (1.5,9/4); \vertex(b5) at (1.5, 3);
					\propag[boson] (a1) to (a2); \propag[boson] (a2) to (a3); \propag[boson] (a3) to (a4); \propag[boson] (a4) to (a5);
					\propag[boson] (b1) to (b2); \propag[boson] (b2) to(b3); \propag[boson] (b3) to (b4); \propag[boson] (b4) to (b5);
					\propag[sca] (a3) to (b3); \propag[sca, looseness=1.5] (b2) to [out=0, in=0](b4);
				\end{feynhand}
			\end{tikzpicture}
			~
			\begin{tikzpicture}[scale=0.31]
				\begin{feynhand}
					\vertex (a1) at (0,0); \vertex (a2) at (0,1); \vertex (a3) at (0,2); \vertex (a4) at (0, 3); \vertex(a5) at (0, 4);
					\vertex (b1) at (2,0); \vertex (b2) at (2,1); \vertex (b3) at (2,2); \vertex (c) at (1,2); \vertex (b4) at (2,3); \vertex(b5) at (2, 4);
					\propag[boson] (a1) to (a2); \propag[boson] (a2) to (a3); \propag[boson] (a3) to (a4); \propag[boson] (a4) to (a5);
					\propag[boson] (b1) to (b2); \propag[boson] (b2) to(b3); \propag[boson] (b3) to (b4); \propag[boson] (b4) to (b5);
					\propag[sca] (a3) to (c); \propag[boson, looseness=1.5] (c) to [out=60, in=120](b3); \propag[boson, looseness=1.5] (c) to [out=-60, in=-120](b3);
				\end{feynhand}
			\end{tikzpicture}
			~
			\begin{tikzpicture}[scale=0.31]
				\begin{feynhand}
					\vertex (a1) at (0,0) ; \vertex (a2) at (0,1); \vertex (a3) at (0,2); \vertex (a4) at (0, 3); \vertex(a5) at (0, 4) ;
					\vertex (b1) at (2,0); \vertex (b2) at (2,1); \vertex (b3) at (2,2); \vertex (c) at (1,2); \vertex (b4) at (2,3); \vertex(b5) at (2, 4);
					\propag[boson] (a1) to (a2); \propag[boson] (a2) to (a3); \propag[boson] (a3) to (a4); \propag[boson] (a4) to (a5);
					\propag[boson] (b1) to (b2); \propag[boson] (b2) to(b3); \propag[boson] (b3) to (b4); \propag[boson] (b4) to (b5);
					\propag[sca] (b3) to (c); \propag[boson, looseness=1.5] (c) to [out=120, in=60](a3); \propag[boson, looseness=1.5] (c) to [out=-120, in=-60](a3);
				\end{feynhand}
			\end{tikzpicture}
			\caption{}\label{fig:SDWDisorderRBM}
		\end{subfigure}
		\caption{Additional one-loop Feynman diagrams in the disordered case; (a) Fermion self energy, (b) Boson self energy, (c) Yukawa vertex correction, (d) Random charge-potential vertex corrections, (e) Boson self-interaction vertex corrections, and (f) Random mass vertex corrections. Here, the dashed line results from disorder scattering.} \label{fig:SDWDisorderOneLoopDiagrams}
	\end{figure}

This clean two-dimensional SDW quantum critical point becomes unstable when disorder effects are introduced. In the presence of disorder effects, additional one-loop Feynman diagrams have to be considered, given in Fig. \ref{fig:SDWDisorderOneLoopDiagrams}. Based on the RG setting given in Appendix \ref{Appendix:RGSetting}, we find all one-loop counter terms. See Appendix \ref{Appendix:OneLoopCounterTerms} and \ref{Appendix:CalOfOneLoopFeynmanDiagrams} for details of calculations. The (i) two types of dynamical critical exponents,
 $z_\tau$ for $\omega$ scaling and $z_\perp$ for $\mathbf{K}_\perp$ scaling, (ii) anomalous dimensions of fermions $\eta_{\psi}$ and bosons $\eta_\phi$, are given by
 \begin{align}
 	z_\perp&=\Big[1-\frac{N_c^2-1}{4\pi^2 N_cN_f}\frac{g^2}{c}[h_2(c,c_\perp,v)-h_3(c,c_\perp,v)]\Big]^{-1},\label{eq:zperp}\\
 	z_{\tau}&=z_\perp\Big[1+\frac{N_c^2-1}{4\pi^2 N_cN_f}\frac{g^2}{c}[h_1(c,c_\perp,v)-h_2(c,c_\perp,v)]\nonumber\\
 	&+F_{dis}(\{\Gamma_i,v\})\Big],\label{eq:ztau}\\
	\eta_\psi&=\frac{z_\perp}{2}\Big[\frac{N_c^2-1}{4\pi^2 N_cN_f}\frac{g^2}{c}[h_2(c,c_\perp,v)-h_3(c,c_\perp,v)]\epsilon\nonumber\\
	&-\frac{N_c^2-1}{4\pi^2 N_cN_f}\frac{g^2}{c}[h_1(c,c_\perp,v)+h_2(c,c_\perp,v)-3h_3(c,c_\perp,v)]\nonumber\\
	&-F_{dis}(\{\Gamma_i\},v)\Big]\label{eq:etapsi},\\
	\eta_{\phi}&=\frac{z_\perp}{2}\Big[\frac{N_c^2-1}{4\pi^2 N_cN_f}\frac{g^2}{c}[h_2(c,c_\perp,v)-h_3(c,c_\perp,v)]\epsilon\nonumber\\
	&-\frac{N_c^2-1}{4\pi^2 N_cN_f}\frac{g^2}{c}[3h_1(c,c_\perp,v)+h_2(c,c_\perp,v)-4h_3(c,c_\perp,v)]\nonumber\\
	&+\frac{g^2}{4\pi v}-3F_{dis}(\{\Gamma_i\},v)\Big]+\frac{\Gamma_M}{2\pi^2 c^2c_\perp^2}\Big(1+\frac{\pi}{2}\frac{c_\perp}{c}\kappa\Big)\frac{z_\perp\epsilon+\bar{\epsilon}}{\epsilon+\bar{\epsilon}}\label{eq:etaphi},
\end{align}
and (iii) one-loop beta functions($\beta_{c}$, $\beta_{c_\perp}$, $\beta_v$, $\beta_g$, $\beta_{u_1}$, $\beta_{u_2}$, $\beta_{\Gamma_i}$, and $\beta_{\Gamma_M}$) are given by 

\begin{widetext}

			\begin{align}
		\beta_v&=v z_\perp\frac{N_c^2-1}{2\pi^2N_cN_f}\frac{g^2}{c}h_3(c,c_\perp,v),\label{eq:betav}\\
		\beta_{c}&=z_\perp\frac{c}{2}\Big[\frac{g^2}{4\pi v}-\frac{N_c^2-1}{2\pi^2 N_c N_f}\frac{g^2}{c}[h_1(c,c_\perp,v)-h_3(c,c_\perp,v)]-2 F_{dis}(\{\Gamma_i\},v)\Big]+\frac{\Gamma_M}{2\pi^2 cc_\perp^2}\Big(1+\frac{3\pi}{4}\frac{c_\perp}{c}\kappa\Big)\frac{\epsilon z_\perp+\bar{\epsilon}}{\epsilon+\bar{\epsilon}},\label{eq:betac}\\
		\beta_{c_\perp}&=z_\perp\frac{c_\perp}{2}\Big[\frac{g^2}{4\pi v}\Big(1-\frac{1}{c_\perp^2}\Big)-\frac{N_c^2-1}{2\pi^2 N_cN_f}\frac{g^2}{c}[h_1(c,c_\perp,v)-h_2(c,c_\perp,v)]-2F_{dis}(\{\Gamma_i,v\})\Big]\nonumber\\
		&+\frac{\Gamma_M}{\pi^2c^2 c_\perp}\Big(1+\frac{\pi}{4}\frac{c_\perp}{c}\kappa\Big)\frac{z_\perp \epsilon+\bar{\epsilon}}{\epsilon+\bar{\epsilon}}\label{eq:betacperp},\\
		\beta_{g}&=z_\perp\frac{g}{2}\Big[-\epsilon+\frac{g^2}{4\pi v}-\frac{1}{4\pi^3N_cN_f}\frac{g^2}{c}\Big(h_4(c,c_\perp,v)+\pi(N_c^2-1)[h_1(c,c_\perp,v)-h_2(c,c_\perp,v)-2h_3(c,c_\perp,v)]\Big)
	\nonumber\\
	&-F_{dis}(\{\Gamma_i\},v)+2G_{dis}(\{\Gamma_i\},v)\Big]+\frac{g\Gamma_M}{2\pi^2 c^2c_\perp^2}\Big(1+\frac{\pi}{2}\frac{c_\perp}{c}\kappa\Big)\frac{z_\perp \epsilon+\bar{\epsilon}}{\epsilon+\bar{\epsilon}}\label{eq:betag},\\
		\beta_{u_1}&=z_\perp u_1\Big[-\epsilon+\frac{g^2}{2 \pi v }-\frac{N_c^2-1}{4 \pi^2 N_cN_f}\frac{g^2}{c} [3h_1(c,c_\perp,v)-h_2(c,c_\perp,v)-2h_3(c,c_\perp,v)]-3F_{dis}(\{\Gamma_i,v\})\nonumber\\
		&+\frac{1}{2\pi^2 c^2c_\perp}\Big((N_c^2+7)u_1+\frac{2u_2(2N_c^2-3)}{N_c}+3\frac{3+N_c^2}{N_c^2} \frac{u_2^2}{u_1}\Big)\Big]-\frac{4u_1\Gamma_M}{\pi^2 c^2c_\perp^2}\Big(1+\frac{\pi}{2}\frac{c_\perp}{c}\kappa\Big)\Big(1-\frac{3}{2}\frac{u_2}{N_cu_1}\Big)\frac{z_\perp \epsilon+\bar{\epsilon}}{\epsilon+\bar{\epsilon}}\label{eq:betau1},\\
		\beta_{u_2}&=z_\perp u_2\Big[-\epsilon+\frac{g^2}{2\pi v}-\frac{N_c^2-1}{4\pi^2N_cN_f} \frac{g^2}{c}[3h_1(c,c_\perp,v)-h_2(c,c_\perp,v)-2h_3(c,c_\perp,v)]+ \frac{1}{\pi^2c^2c_\perp}\Big(6u_1+\frac{N_c^2-9}{N_c}u_2\Big)
		\nonumber\\
		&-3F_{dis}(\{\Gamma_i,v\})\Big]-\frac{4u_2\Gamma_M}{\pi^2c^2c_\perp^2}\Big(1+\frac{\pi}{2}\frac{c_\perp}{c}\kappa\Big)\frac{z_\perp\epsilon+\bar{\epsilon}}{\epsilon+\bar{\epsilon}}\label{eq:betau2},\\
		\beta_{\Gamma_i}&=z_\perp\Gamma_i\Big[-\epsilon+\frac{N_c^2-1}{4\pi^2N_cN_f}\frac{g^2}{c}[h_2(c,c_\perp,v)+h_3(c,c_\perp,v)]+A_{\Gamma_i}^{(1)}\Big]\label{eq:betaGamma},
			\\
		\beta_{\Gamma_M}&=-(z_\perp \epsilon+\bar{\epsilon})\Gamma_M+z_\perp\Gamma_M\Big[\frac{g^2}{2\pi v}-\frac{N_c^2-1}{4\pi^2 N_cN_f}\frac{g^2}{c}[4h_1(c,c_\perp,v)-h_2(c,c_\perp,v)-3h_3(c,c_\perp,v)]-4F_{dis}(\{\Gamma_i,v\})\nonumber\\
		&+\frac{N_c^2+1}{\pi^2c_\perp c^2 }\Big(u_1+\frac{1}{N_c}u_2\Big)\Big], \label{eq:betaGammaM}
	\end{align}
\end{widetext}
where functions $F_{dis}(\{\Gamma_i\},v)$, $G_{dis}(\{\Gamma_i\},v)$ and  $h_i(c,c_\perp,v)(i=1,2,3,4)$ are given in the Appendix \ref{Appendix:OneLoopCounterTerms}.
	In the beta functions of random charge potential vertices, $A_{\Gamma_i}^{(1)}$ is a coefficient of the $\frac{1}{\epsilon}$-pole in the random charge potential vertex counter term $A_{\Gamma_i}$, given by $A_{\Gamma_i}^{(1)}=\lim_{\epsilon\rightarrow 0}\epsilon A_{\Gamma_i}$. Here, we choose the high-energy convention for signs of the beta functions. As a result, the coupling constant $\alpha$ increases when $\beta_\alpha < 0$ while it decreases when $\beta_\alpha > 0$ in the low energy limit.
	
We point out that there is no contribution of $\Gamma_M^2$-term in the beta function $\beta_{\Gamma_M}$ (Eq. \eqref{eq:betaGammaM}). The absence of the $\Gamma_M^2$-term is due to the non-local structure of the regularized effective action related to the random boson mass.

Before proceeding further, let us introduce ways to simplify the analysis of the beta functions. 

First, the number of the random charge potential vertices is reduced from 27 to 15. The original random charge potential vertices of 27-channels can be simplified into 15 groups that are classified into three categories; `Direct', `Exchange', and `Umklapp'. The resulting categories and groups are given as follows:

	\begin{align*}
		\text{Direct:}&\left\{\begin{array}{l}\Gamma_{G1}^d=\{\Gamma_0,\; \Delta_0\},\; 	\Gamma_{G2}^d=\{\Gamma_{\theta_1}^d,\; \Gamma_{\pi-\theta_1}^d\}\\
			\Gamma_{G3}^d=\{\Gamma_{\theta_2}^d,\; \Gamma_{\pi-\theta_2}^d\},\; 	\Gamma_{G4}^d=\{\Gamma_{\pi/2}^d\}
		\end{array}
		\right\},
		\\
		\text{Exchange:}&\left\{\begin{array}{l}\Gamma_{G5}^e=\{\Gamma_{\theta_1}^e,\; \Delta_{\theta_1}\},\; \Gamma_{G6}^e=\{\Gamma_{\pi-\theta_1}^e,\; \Delta_{\pi-\theta_1}\}\\
			\Gamma_{G7}^e=\{\Gamma_{\theta_2}^e,\; \Delta_{\theta_2}\},\; \Gamma_{G8}^e=\{\Gamma_{\pi/2}^e,\; \Delta_{\pi/2}\}\\
			\Gamma_{G9}^e=\{\Gamma_{\pi-\theta_2}^e,\; \Delta_{\pi-\theta_2}\},\; \Gamma_{G10}^e=\{\Delta_{\pi}\}\end{array}
		\right\},
		\\
		\text{Umklapp:}&\left\{\begin{array}{l}\Gamma_{G11}^u=\{\Upsilon_{0},\; \Upsilon_{\theta_1}^d\},\;\Gamma_{G12}^u=\{\Xi_{\theta_2}^d,\; \Xi_{\pi/2}^d\}\\
			\Gamma_{G13}^u=\{\Xi_{\theta_1}^d,\; \Xi_{\theta_2}^e\},\; \Gamma_{G14}^u=\{\Upsilon_{\theta_1}^e\}\\
			\Gamma_{G15}^u=\{\Xi_{\theta_1}^e,\; \Xi_{\pi/2}^e\}\end{array}\right\}
	\end{align*}

If the initial values of disorder scattering channels are set to be the same within the same group, the RG flows are the same within the same group. Instead of the original 27 disorder channels, these 15 groups are used in RG analysis for simplicity. Therefore we provide counter terms of the random charge potential vertices in terms of these reduced channels only (Appendix \ref{Appendix:OneLoopCounterTerms}).

Secondly, in the RG analysis, we use the following dimensionless parameters given by $s=\frac{c_\perp}{c}$, $w=\frac{v}{c}$, $\lambda=\frac{g^2}{v}$, $\kappa_i=\frac{u_i}{c^2c_\perp}$, $\gamma_M=\frac{\Gamma_M}{c^2c_\perp^2}$ to analyze RG-flows. These parameters show relative magnitudes of interaction and disorder with respect to kinetic energy. In this respect, the dimensionless parameters play the role of actual expansion parameters in the RG analysis, analogous to the Fine structure constant $\alpha=\frac{1}{4\pi\epsilon_0}\frac{e^2}{\hbar c}$ in quantum electrodynamics. Therefore, RG flows of the dimensionless parameters are more important to identify the low energy properties than that of the original parameters. As a result, we use beta functions of the dimensionless parameters in the RG analysis. Rewritten RG beta functions in terms of dimensionless parameters are given as follows
\begin{widetext}
	\begin{align}
	\beta_{c_\perp}&=z_\perp\frac{c_\perp}{2}\Big[\frac{\lambda}{4\pi}\Big(1-\frac{1}{c_\perp^2}\Big)-\frac{N_c^2-1}{2\pi^2 N_cN_f}w\lambda [h_1(c,c_\perp,v)-h_2(c,c_\perp,v)]-2F_{dis}(\{ \Gamma_i,v\}) \Big]+\frac{c_\perp\gamma_M}{\pi^2}\Big(1+\frac{\pi}{4}\kappa s \Big)\frac{z_\perp\epsilon+\bar{\epsilon}}{\epsilon+\bar{\epsilon}},\label{eq:relaBetacperp}\\
	\beta_w&=w\Bigg[z_\perp\Bigg\{-\frac{\lambda}{8\pi}+\frac{N_c^2-1}{4\pi^2N_cN_f}\lambda w\Big(h_1(c,c_\perp,v)+h_3(c,c_\perp,v)\Big)+F_{dis}(\{\Gamma_i,v\})\Bigg\}-\frac{\gamma_M}{2\pi^2}\Big(1+\frac{3\pi}{4}\kappa s\Big)\frac{z_\perp \epsilon+\bar{\epsilon}}{\epsilon+\bar{\epsilon}}\Bigg]\label{eq:betaw},\\
		\beta_{\lambda}&=\lambda \Bigg[z_\perp\Bigg\{-\epsilon+\frac{\lambda}{4\pi}-\frac{\lambda w}{4\pi^3 N_cN_f}\Big(h_4(c,c_\perp,v)+\pi(N_c^2-1)[h_1(c,c_\perp,v)-h_2(c,c_\perp,v)]\Big)-F_{dis}(\{\Gamma_i,v\})\nonumber\\
		&+2G_{dis}(\{\Gamma_i,v\})\Bigg\}+\frac{\gamma_M}{\pi^2}\Big(1+\frac{\pi}{2}\kappa s\Big)\frac{z_\perp\epsilon+\bar{\epsilon}}{\epsilon+\bar{\epsilon}}\Bigg]\label{eq:betalambda},\\
		\beta_{\kappa_1}&=\kappa_1\Bigg[z_\perp\Bigg\{-\epsilon+\frac{\lambda}{8\pi}\Big(1+\frac{1}{c_\perp^2}\Big)+\frac{1}{2\pi^2}\Big((N_c^2+7)\kappa_1+\frac{2(2N_c^2-3)}{N_c}\kappa_2+\frac{3(3+N_c^2)}{N_c^2}\frac{\kappa_2^2}{\kappa_1}\Big)\Bigg\}\nonumber\\
		&-\frac{\gamma_M}{\pi^2}\Bigg\{3(2+\pi\kappa s)-6\Big(1+\frac{\pi}{2}\kappa s\Big)\frac{1}{N_c}\frac{\kappa_2}{\kappa_1}\Bigg\}\frac{z_\perp\epsilon+\bar{\epsilon}}{\epsilon+\bar{\epsilon}}\Bigg]\label{eq:betakappa1},\\
		\beta_{\kappa_2}&=\kappa_2\Bigg[z_\perp\Bigg\{-\epsilon+\frac{\lambda}{8\pi}\Big(1+\frac{1}{c_\perp^2}\Big)+\frac{1}{\pi^2}\Big(6\kappa_1+\frac{N_c^2-9}{N_c}\kappa_2\Big)\Bigg\}-\frac{\gamma_M}{\pi^2}3(2+\pi\kappa s)\frac{z_\perp\epsilon+\bar{\epsilon}}{\epsilon+\bar{\epsilon}}\Bigg]\label{eq:betakappa2},\\
		\beta_{\Gamma_{i}}&=z_\perp\Gamma_i\Big[-\epsilon+\frac{N_c^2-1}{4\pi^2N_cN_f}w\lambda[h_2(c,c_\perp,v)+h_3(c,c_\perp,v)]+A_{\Gamma_i}^{(1)}\Big]\label{eq:betaGamma2}\\
		\beta_{\gamma_M}&=\gamma_M\Bigg[-(z_\perp \epsilon+\bar{\epsilon})+z_\perp\Bigg\{\frac{\lambda}{4\pi}\frac{1}{c_\perp^2}-\frac{N_c^2-1}{4\pi^2 N_cN_f}\lambda w\Big(h_2(c,c_\perp,v)-h_3(c,c_\perp,v)\Big)+\frac{N_c^2+1}{\pi^2 }\Big(\kappa_1+\frac{1}{N_c}\kappa_2\Big)\Bigg\}\nonumber\\
		&-\frac{\gamma_M}{\pi^2}\Big(3+\frac{5\pi}{4}\kappa s\Big)\frac{z_\perp\epsilon+\bar{\epsilon}}{\epsilon+\bar{\epsilon}}\Bigg]\label{eq:betagammaM}.
	\end{align}
\end{widetext}

Thirdly, we consider the case of $N_c = 2$ and $N_f = 1$ for `physical relevance' in the analysis. Since the boson self-interaction vertex $u_2$ is identical to the $u_1$ vertex in the case of $N_c$ less than 4 \cite{SurLee}, we keep $u_2$ or $\kappa_2$ to be zero in the remaining parts.

Finally, to present RG-analysis in a systematic way, we first consider three limiting cases where one of parameter among the Yukawa interaction vertex ($\lambda$), the random charge-potential vertex ($\Gamma_i$), and the random boson mass vertex ($\gamma_M$) is set to zero: (i) No Yukawa interaction case (No-YI case: $\lambda=0$, $\Gamma_i \neq 0$, $\gamma_M \neq 0$), (ii) No random charge potential case (No-rCP case: $\lambda \neq 0$, $\Gamma_i=0$, $\gamma_M \neq 0$), and (iii) No random-boson mass case (No-rBM case: $\lambda \neq 0$, $\Gamma_i \neq 0$, $\gamma_M=0$). Based on the analysis of limiting cases, we discuss the general case with all ingredients.

\begin{figure}
	\begin{subfigure}{0.48\textwidth}
		\includegraphics[scale=0.23]{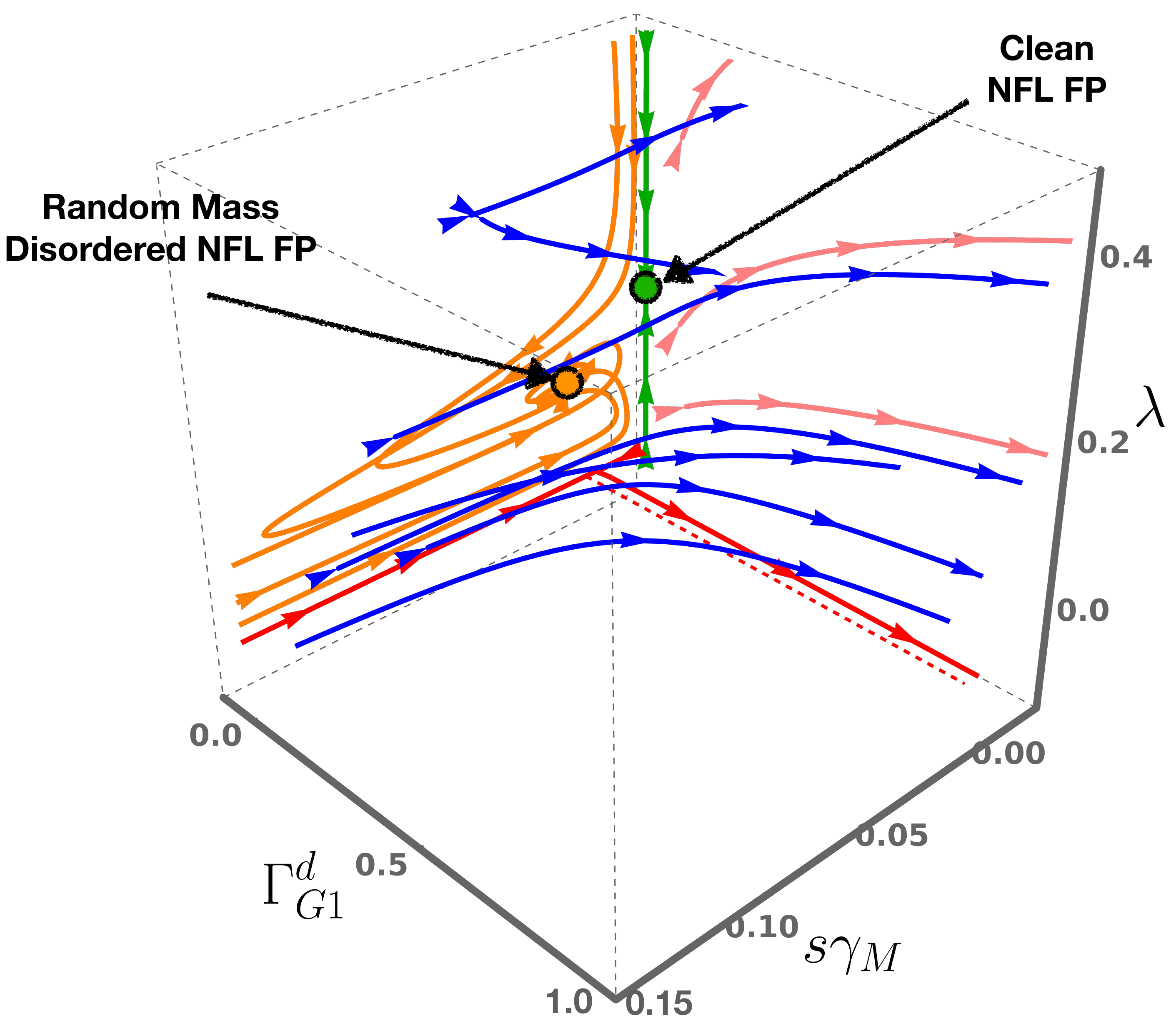}
		\caption{RG flows in the parameter space of $(\lambda,\Gamma_{G1}^d,s\gamma_M)$.}
	\end{subfigure}
	~
	\begin{subfigure}{0.48\textwidth}
		\includegraphics[scale=0.22]{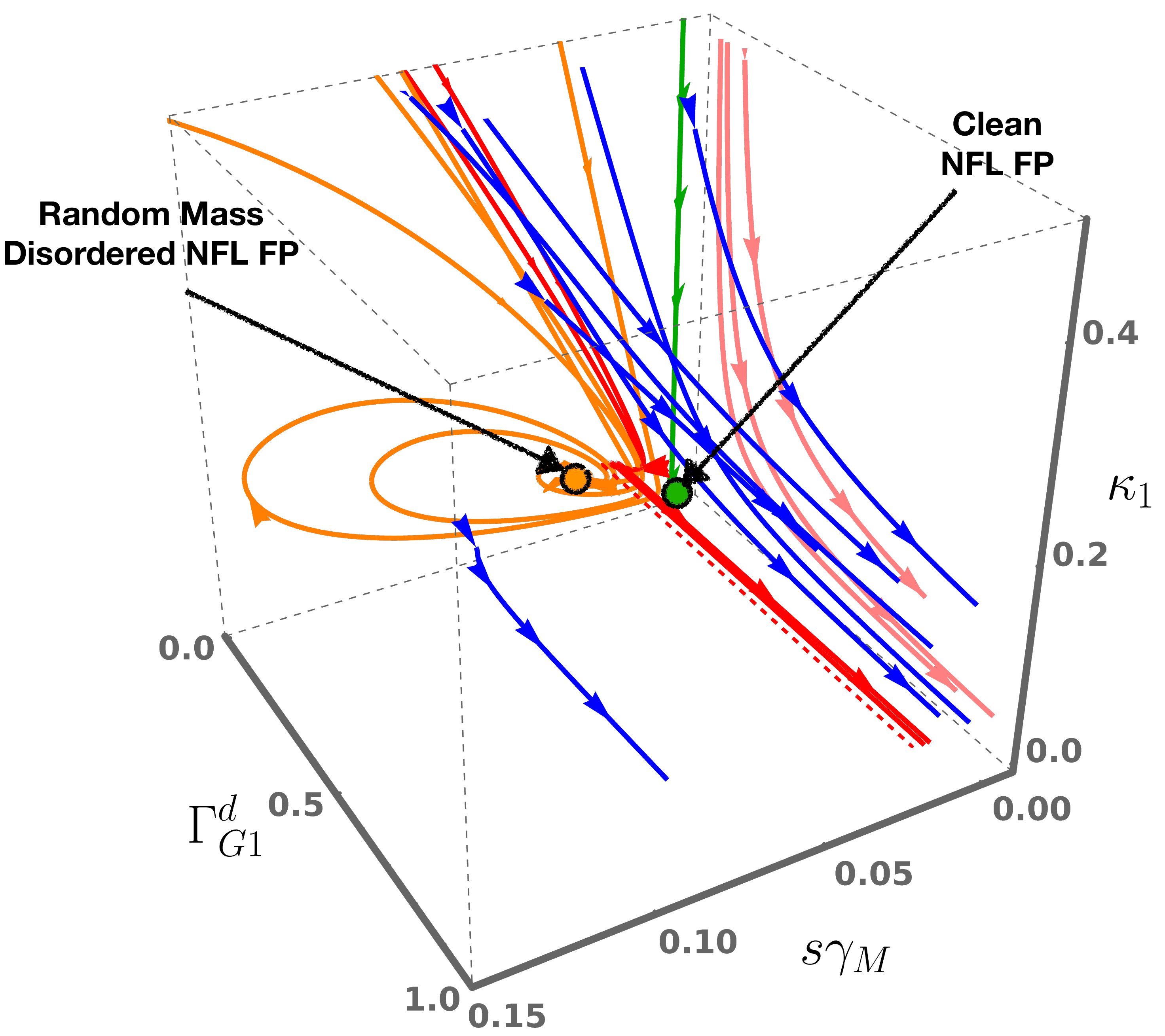}
		\caption{RG flows in the parameter space of $(\kappa_1,\Gamma_{G1}^d,s\gamma_M)$.}
	\end{subfigure}
	\caption{RG flows with several initial conditions. Green, orange, red, and pink colored flow lines correspond to those of four limiting cases: (0) Green: Clean Case ($\Gamma_i=\gamma_M=0$), (i) Orange: No random charge potential ($\Gamma_i=0$), (ii) Red: No Yukawa interaction ($\lambda=0$), and (iii) Pink: No random boson mass ($\gamma_M=0$). Blue colored flow lines correspond to the RG flows of the general case ($\lambda\neq0$, $\Gamma_i\neq0$, and $\gamma_M\neq0$). There are two unstable fixed points; Clean Non-Fermi Liquid Fixed Point (green dot) and Random Mass Disordered Non-Fermi Liquid Fixed Point (orange dot). The dashed red line denotes an unstable fixed line with fixed values of $\kappa_1$ and $s\gamma_M$. Detailed values of these fixed points and lines are given in Appendix \ref{Appendix:DetailedAnalysisOneLoopBetaFunctions}. For this numerical plots, we used $\epsilon=\bar{\epsilon}=0.01$, $v_c=0.05$, $N_f=1$, $N_c=2$, and $\kappa=1$.} \label{fig:3dFlow}
\end{figure}

Figure \ref{fig:3dFlow} shows the three-dimensional RG flows in two parameter spaces, (i) ($\Gamma_{G1}^d,s\gamma_M,\lambda$) and (ii) $(\Gamma_{G1}^d,s\gamma_M,\kappa_1)$. 
Green-colored, red-colored, orange-colored, and pink-colored lines in these figures represent the RG flows of the clean case, No-YI case, No-rCP case, and No-rBM case, respectively. Blue-colored ones represent the RG flows of the general case. The green dot, orange dot, and red dashed lines show fixed points or a line of each case. In Table \ref{table:OneLoopAnalysisSumUp}, the summary of the fixed points and RG flows is given. In the remaining parts of the section, our presentation focuses on physical aspects of these fixed points and low-energy RG flows instead of technical details of the analysis. For technical details of analysis about RG flows and fixed points, we would like to refer to Appendix \ref{Appendix:DetailedAnalysisOneLoopBetaFunctions}.

\renewcommand{\arraystretch}{1.2}
\begin{table}
	\scalebox{0.9}{
		\begin{tabular}{|c|c|c|c|}
			\hline
			Case & Fixed point & RG flow & Remarks \\
			\hhline{====}
			\parbox[c]{2.cm}{No-YI\\ (red line) } & \parbox[c]{2.cm}{$(s\gamma_M)^*$, $\kappa_1^*$\\(red dashed line, Appendix \ref{Appendix:DetailedAnalysisOneLoopBetaFunctions:NoYICase})}  &\parbox[c]{2.5cm}{\begin{gather*}
					(s\gamma_M)\rightarrow (s\gamma_M)^*,\\ \kappa_1\rightarrow \kappa_1^*,\\ \Gamma_{G1}^d \nearrow, 
			\end{gather*}}& \parbox[c]{2.cm}{Oscillating flow} \\
			\hline
			\parbox[c]{2.cm}{No-rCP\\ (orange line) }& \parbox[c]{2.cm}{$(s\gamma_M)^*$, $\kappa_1^*$,\; $\lambda^*$ \\(orange dot, Appendix \ref{Appendix:DetailedAnalysisOneLoopBetaFunctions:NorCPCase})}  & \parbox[c]{2.5cm}{\begin{gather*}
					(s\gamma_M)\rightarrow (s\gamma_M)^*,\\ \kappa_1\rightarrow \kappa_1^*,\\ \lambda \rightarrow \lambda^*
			\end{gather*}}& \parbox[c]{2.cm}{Oscillating flow, Ref. \cite{KirkpatricBelitz}}\\
			\hline
			\parbox[c]{2.cm}{No-rBM\\ (pink line) } & No & \parbox[c]{2.5cm}{\begin{gather*}
					\kappa_1\searrow,\; \lambda \nearrow, \\ \Gamma_{G1}^d\nearrow 
			\end{gather*}} & Ref. \cite{Punk} \\
			\hline
			\parbox[c]{2.cm}{General\\ (blue line) } & No &   \parbox[c]{2.5cm}{\begin{gather*}
					\kappa_1\searrow,\; \lambda \nearrow, \\ \Gamma_{G1}^d\nearrow ,\; s\gamma_M\searrow
			\end{gather*}}  & \parbox[c]{2.cm}{same to the No-rBM case} \\
			\hline
		\end{tabular}
	}
	\caption{Summary of the one-loop RG flows} \label{table:OneLoopAnalysisSumUp}
\end{table}

\subsubsection{(i) No Yukawa interaction (No-YI) case $\&$ (ii) No random charge potential (No-rCP) case} \label{sec:NoYIAndNorCPCase}

First, we consider two limiting cases; (i) No-YI case and (ii) No-rCP case at the same time since they share some common points.

In the No-YI case (red lines in Fig. \ref{fig:3dFlow}), boson and fermion fields are decoupled from each other. Then, the boson sector is reduced to the model studied by Kirkpatrick and Belitz \cite{KirkpatricBelitz}. One difference is that the effect of the Landau damping on the boson dynamics disappears in the No-YI case. There is a stable fixed point specified with finite values of the effective random boson mass vertex ($s\gamma_M$) and the effective boson self-interaction vertex ($\kappa_1$), represented by the red dashed lines in Fig \ref{fig:3dFlow}. Here, we use the same name `Long-range-ordered' fixed point as that of Ref. \cite{KirkpatricBelitz}. Additionally, oscillating patterns of the RG flows \cite{KirkpatricBelitz} are observed as shown in Fig. \ref{fig:2dFlowNoInterCase} (Appendix \ref{Appendix:DetailedAnalysisOneLoopBetaFunctions:NoYICase}). This oscillating pattern originates from the interplay between the random boson mass vertex ($\gamma_M$) and the boson self-interaction vertex ($\kappa_1$).  See Appendix \ref{Appendix:DetailedAnalysisOneLoopBetaFunctions:NoYICase} for technical details.

In the No-rCP case (orange lines in Fig. \ref{fig:3dFlow}), we obtain similar low energy behaviors of $\kappa_1$ and $s\gamma_M$ to the No-YI case. There is a stable fixed point specified with finite values of $\lambda$, $s\gamma_M$, and $\kappa_1$ with an oscillating pattern as shown in Fig. \ref{fig:3dFlow} and in Fig. \ref{fig:2dBetaFunctionNoRCPCase} of Appendix \ref{Appendix:DetailedAnalysisOneLoopBetaFunctions:NorCPCase}. As in the No-YI case, the interplay between $\gamma_M$ and $\kappa_1$ causes the oscillating pattern. However, unlike the No-Yi case, fermion and boson fields are coupled to each other through a finite Yukawa interaction in the low energy limit. Since the Yukawa interaction is taken into account in this case, the fixed point might be more closely related to that discussed in Ref. \cite{KirkpatricBelitz}, where the Landau damping term is considered in the dispersion of the boson dynamics. We call this fixed point `random-mass disordered non-Fermi liquid’ fixed point as shown in Fig. \ref{fig:3dFlow}.

To clarify the role of the random boson mass vertex on the clean system, we compare the low energy physics of the No-rCP case and the clean case. See Appendix \ref{Appendix:DetailedAnalysisOneLoopBetaFunctions:NorCPCase} for detailed analysis on the beta functions of the No-rCP case. With a setting $\epsilon=\bar{\epsilon}=0.01 ,\; N_c=2,\; N_f=1$, fixed points of the clean and the No-rCP case are specified by following numerical values of the parameters:
\begin{align*}
	\text{Clean case: }&c_\perp^*=1,\; w^*\approx 0.67,\; \lambda^*=0.21,\; \kappa_1^*=0.\\
	\text{No-rCP case: }&c_\perp^{*}=0.856,\; w^*=1.035,\; \lambda^*=0.144,\\ &\kappa_1^*=0.055,\; (s\gamma_M)^*=0.03.
\end{align*}
Compared to the clean case, the boson velocity along the co-dimensional direction ($c_\perp$) and the effective Yukawa interaction parameter ($\lambda$) decrease while the ratio between boson and fermion velocity ($w=v/c$) and the effective boson self-interaction parameter ($\kappa_1$) increase. Mathematically, it can be understood from the beta functions. In $\beta_{c_\perp}$ (Eq. \eqref{eq:relaBetacperp}) and $\beta_{\lambda}$ (Eq. \eqref{eq:betalambda}), there are terms proportional to $\gamma_M$ with a positive sign. This means that the random boson mass vertex gives screening to these variables. In contrast, there are terms proportional to $\gamma_M$ with a negative sign in the beta functions $\beta_{w}$ (Eq. \eqref{eq:betaw}) and $\beta_{\kappa_1}$ (Eq. \eqref{eq:betakappa1}) which give the anti-screening. As a result, the fixed point values of $c_\perp$ and $\lambda$ are reduced while those of $w$ and $\kappa_1$ are enhanced compared to the clean case.

The above discussion can be translated into more physical terms. 

First, the random boson mass vertex leads the boson dynamics to be localized. As a result, boson velocities $c$ and $c_\perp$ are reduced and the velocity ratio $w=v/c$ is enhanced. 

Second, the reduction of the effective Yukawa interaction ($\lambda$) can be explained by the reduced correlation of boson fields due to the random boson mass fluctuations. The reduced correlation between boson fields in the low energy limit is reflected in the increasing anomalous dimension of boson fields $\eta_{\phi}$ in Eq. \eqref{eq:etaphi} by the random boson mass vertex. Here the increasing anomalous dimension of boson fields can be physically interpreted as loss of the quasi-particleness of boson field in the low energy limit. Since boson fields mediate the Yukawa interaction between fermion fields, the reduced correlation of boson fields leads to the reduced effective Yukawa interaction. This physical interpretation is confirmed by the term proportional to the $\Gamma_M$ coming from the anomalous dimension of boson ($\eta_\phi$) with a positive sign in the beta function of the Yukawa interaction ($g$). 

Finally, let us consider the effective boson self-interaction parameter ($\kappa_1$). Enhancement of $\kappa_1$ in the No-rCP case is due to two factors. The first one is the reduction of boson velocities ($c$ and $c_\perp$), discussed previously. This results in a reduction of the boson kinetic energy and leads to enhancement of $\kappa_1$, which is the ratio between the boson self-interaction energy and the boson kinetic energy. The second factor is that the boson self-interaction $u_1$ gets an anti-screening effect from one-loop corrections by the random boson mass vertex, shown in Fig. \ref{fig:SDWDisorderBSI}.

\subsubsection{(iii) No random boson mass (No-rBM) case $\&$ (iv) General case} \label{sec:NorBMandGeneralCase}

Detailed analysis on the general case in Appendix \ref{Appendix:DetailedAnalysisOneLoopBetaFunctions:GeneralCase} shows that the random boson mass vertex becomes irrelevant in the low energy limit within the phase space where the one-loop RG analysis remains to be valid. However, it turns out that the oscillating RG flows observed in the No-Yi and No-rCP cases reduce the one-loop valid phase space and lead the RG flow to the phase space where the random boson mass vertex becomes relevant. In this respect, we will discuss possible low energy physics involved with the random boson mass vertex beyond the one-loop RG analysis in section \ref{sec:TwoLoopResultsAndpossibleTwoDisorderedPhase}. Here, we presume the irrelevance of the random boson mass vertex in the one-loop RG analysis and discuss the role of random charge potential vertices in the two-dimensional clean SDW quantum criticality.

\begin{figure}[h]
	\begin{subfigure}{0.45\textwidth}
		\includegraphics[width=0.8\textwidth]{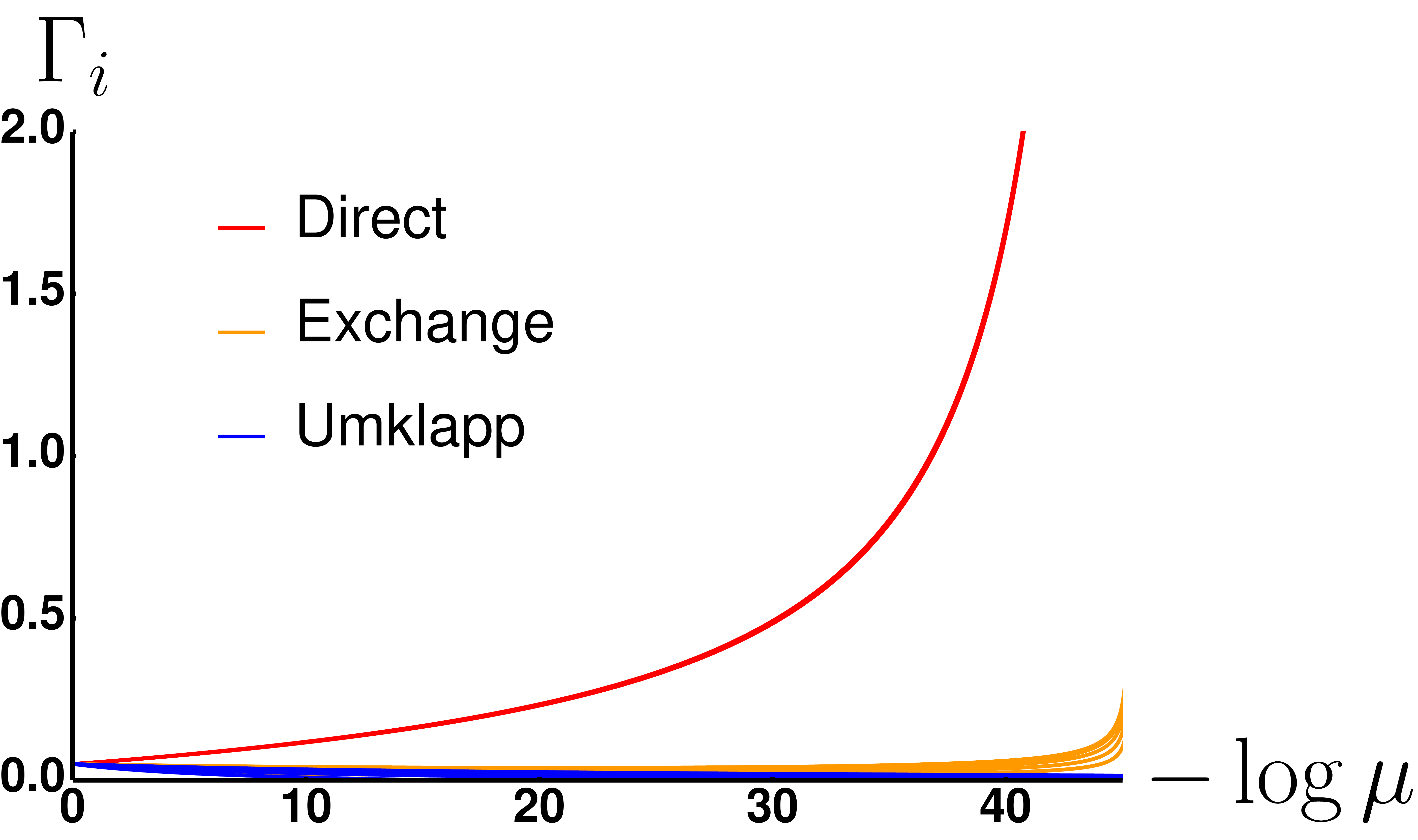}
		\caption{RG flows of all scattering channels}\label{subfig:RCPFlowGamma1}
	\end{subfigure}
	~
	\begin{subfigure}{0.45\textwidth}
		\includegraphics[width=0.8\textwidth]{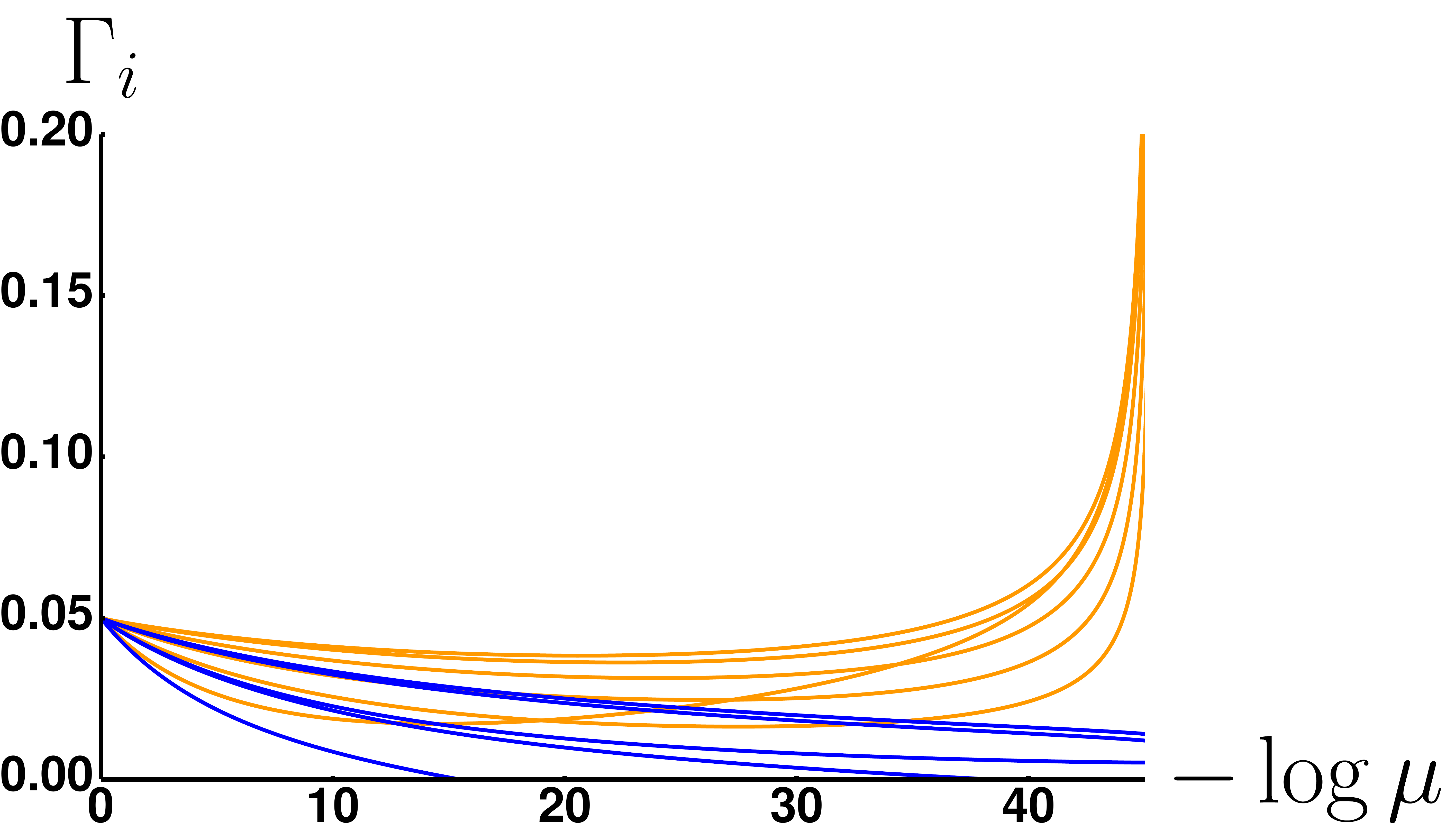}
		\caption{RG flows of scattering channels in the `Exchange' and the `Umklapp' categories only}\label{subfig:RCPFlowGamma2}
	\end{subfigure}
	\caption{RG flows of all random charge potential vertices $\{ \Gamma_{i}\}$ in the general case. Here, red, orange and, blue-colored lines denote scattering channels in the `Direct', the `Exchange', and the `Umkalpp' categories, respectively. We used $\epsilon=\bar{\epsilon}=0.01$, $v_c=0.05$, $N_f=1$, $N_c=2$, and $\kappa=1$.} \label{fig:RCPFlowGamma}
\end{figure}

First, we consider low energy behaviors of random charge potential vertices. Figure \ref{fig:RCPFlowGamma} shows the RG flows of the 15 groups of random charge potential vertices, which are classified into three categories (`Direct', `Exchange', `Umklapp'). The scattering channels in the `Direct' category rapidly increase and become the most dominant channels in the low energy limit as shown in Fig. \ref{subfig:RCPFlowGamma1}. In the case of the channels in the `Exchange' and `Umklapp' categories, they first show decreasing RG flows. However, the channels in the `Exchange' category change the flows and increase in the low energy limit while those in the `Umklapp' category keep decreasing as shown in Fig. \ref{subfig:RCPFlowGamma2}. These different low-energy RG flows between the channels in the different categories is understood mainly from the sign of the term $A_{\Gamma_i}^{(1)}$ in the beta function $\beta_{\Gamma_i}$ (Eq. \eqref{eq:betaGamma}). Here, $A_{\Gamma_i}^{(1)}$ is a counter term of the random charge potential vertex. For the channels in the `Direct' and `Exchange', the signs of the $A_{\Gamma_i}^{(1)}$ terms are negative which means that they give rise to an anti-screening effect to $\Gamma_{Direct/Exchange}$. On the other hand, the signs of the $A_{\Gamma_i}^{(1)}$ terms are positive for the channels in the Umklapp category which results in the screening of $\Gamma_{Umklapp}$. See Appendix \ref{Appendix:OneLoopCounterTerms} for details of one-loop counterterms ($A_{\Gamma_i}^{(1)}$) of the random charge vertices. To understand the reason for the different RG flows between the `Direct' and the `Exchange' categories, we need to consider the second term proportional to $\frac{g^2}{c}$ in the Eq. \eqref{eq:betaGamma}. This term comes from the wave function renormalization constant of the fermion fields ($Z_{\psi}$) and gives a screening effect to the $\Gamma_i$. Physically, it can be interpreted as a reduction of the ballistic scattering by the charge impurity due to loss of the quasi-particleness by the Yukawa interaction. However, for channels in the `Direct' category, the second term ($\sim\frac{g^2}{c}$) is almost canceled by one-loop corrections in $A_{\Gamma_{Direct}}^{(1)}$ involving both the Yukawa and the random charge potential vertices due to the Ward identity while this does not happen for the channels in the `Exchange' category.   Moreover, there are more channels to give anti-screening in the case of the `Direct' category, compared to the case of the `Exchange' category. As a result, scattering channels in the `Direct' category increase much faster than those in the `Exchange' category. 

Now we discuss the low energy behaviors of other parameters; $c_\perp$, $w$, $\lambda$, and $\kappa_1$. We find that the effective Yukawa interaction ($\lambda$) converges to a value $4 \pi F_{dis}(\{\Gamma_i,v\})$ in the low energy limit. Since the term $F_{dis}(\{\Gamma_i,v\})$ is proportional to $\Gamma_{G1}^d$, it is a rapidly increasing function in the low energy limit. Therefore the effective Yukawa potential ($\lambda$) increases in the low energy limit as shown in Fig. \ref{fig:2dBetaFunctionNoRBMCase} (Appendix \ref{Appendix:DetailedAnalysisOneLoopBetaFunctions:NorBMCase}) and Fig. \ref{fig:2dBetaFunctionGeneralCase} (Appendix \ref{Appendix:DetailedAnalysisOneLoopBetaFunctions:GeneralCase}). This low energy behavior of the effective Yukawa interaction determines the low energy behaviors of $c$, $w$, and $\kappa_1$. Substituting $\lambda$ with $4 \pi F_{dis}(\{\Gamma_i,v\})$ into the beta functions of $\beta_{c_\perp}$ (Eq. \eqref{eq:relaBetacperp}), $\beta_w$ (Eq. \eqref{eq:betaw}), and $\beta_{\kappa_1}$ (Eq. \eqref{eq:betakappa1}), it is found that $\beta_c$ gets strong anti-screening while both $\beta_w$ and $\beta_{\kappa_1}$ obtain strong screening effects by random charge potential vertices. As a result, $c_\perp$ increases rapidly while both $w$ and $\kappa_1$ decrease in the low energy limit. For more detailed analysis, see Appendix \ref{Appendix:DetailedAnalysisOneLoopBetaFunctions:GeneralCase}. A summary of the one-loop RG flows of the parameters for the general case is given in Table \ref{table:SummaryOfRGFlowGeneralCase}.

\renewcommand{\arraystretch}{1.5}
\begin{table}
		\begin{tabular}{|>{\centering\arraybackslash}p{8cm}|}
			\hline
			 One-loop RG flows of the general case \\
			\hhline{=}
			$\Gamma_{Direct/Exchange}\nearrow, \; \Gamma_{Umklapp}\searrow$\\
			\hline
			 $\gamma_M\searrow, \; s\gamma_M\searrow$\\
			\hline
			$c_\perp \nearrow,\; w\searrow$\\
			\hline
			$\lambda\rightarrow 4\pi F_{dis}(\{\Gamma_i,v\})\nearrow,\; \kappa_1\searrow$\\
			\hline
		\end{tabular}
	\caption{Summary of One-loop RG-flows in the general case. Here $\Gamma_{Direct/Exchange/Umklapp}$ denote scattering vertices of three types of random charge potentials, `Direct', `Exchange', and `Umklapp', respectively.} \label{table:SummaryOfRGFlowGeneralCase}
\end{table}
Since $\gamma_M$ and $s\gamma_M$ decrease in the low energy limit, the RG flows for the general case become the same as that for the No-rBM case where $\gamma_M$ and $s\gamma_M$ are set to be zero. These low-energy RG flows are the almost same to that obtained in Ref. \cite{Punk}.

To understand these results in more physical terms, we first define `physical' velocities of fermion and boson, which take into account renormalized scaling dimensions of the time ($z_{\tau}$) and the extended spatial coordinate ($z_{\perp}$). Here, we consider a ratio given by $\frac{\text{velocity} \times \text{time}}{\text{distance}}$ to define the physical velocities. If this ratio becomes larger in the low energy limit, we interpret it as physical velocity is increasing. We employ a fact that time and spatial coordinates are transformed as follows under the RG transformation in defining the physical velocities:
$\tau\rightarrow b^{-z_\tau}\tau$, $\mathbf{X}_\perp\rightarrow b^{-z_\perp}\mathbf{X}_\perp$, $x_{d-1}\rightarrow b^{-1}x_{d-1}$, and $x_d\rightarrow b^{-1}x_d$ where $b$ is a scaling parameter larger than 1.

First, let us consider fermion fields. There is only one velocity parameter $v$ which is a perpendicular component to the nesting vector $\mathbf{Q}$. Therefore we introduce two more velocities of $V_\perp$ and $v_{||}$ which are parallel components to the $\mathbf{X}_\perp$ direction and the nesting vector $\mathbf{Q}$, respectively. In our setting, $V_\perp$ and $v_{||}$ are set to 1 and their values are not changed under the RG transformation. Then, the ratios $\frac{\text{velocity} \times \text{time}}{\text{distance}}$ of the fermion fields evolve under the RG transformation as follows:
$\frac{V_\perp|\tau|}{|\mathbf{X}_\perp|}\rightarrow  b^{z_\perp-z_\tau} \frac{V_\perp|\tau|}{|\mathbf{X}_\perp|}$, $\frac{v_{||}|\tau|}{|x_{||}|}\rightarrow b^{1-z_\tau}\frac{v_{||}|\tau|}{|x_{||}|}$ and $\frac{v|\tau|}{|x_\perp|}\rightarrow b^{dim[v]+1-z_{\tau}}\frac{v|\tau|}{|x_\perp|}$ where $dim[v]$ is a dimension of $v$ given by $dim[v]=-\frac{\beta_v}{v}$. From the above transformations of the $\frac{\text{velocity} \times \text{time}}{\text{distance}}$ ratios, we can define following physical velocities $\tilde{V}_\perp$, $\tilde{v}_{||}$, and $\tilde{v}$: $\tilde{V}_\perp=b^{z_\perp-z_\tau}V_\perp$, $\tilde{v}_{||}=b^{1-z_\tau}v_{||}$ and $\tilde{v}=b^{1-z_\tau}v$. Here, both scaling dimensions $z_\tau$ and $z_\perp$ are introduced into the definitions of the physical velocities. From the above definitions, it is straightforward to know the scaling dimensions of the physical velocities, given by $dim[\tilde{V}_\perp]=z_\perp-z_\tau$, $dim[\tilde{v}_{||}]=1-z_\tau$ and $dim[\tilde{v}]=dim[v]+1-z_\tau$ where $dim[v]=-\frac{\beta_v}{v}$. By using the one-loop results of the $z_\tau$ (Eq. \eqref{eq:ztau}), $z_\perp$ (Eq. \eqref{eq:zperp}),  and $\beta_v$ (Eq. \eqref{eq:betav}), we obtain one-loop scaling dimensions of the fermion physical velocities as follows
\begin{subequations}\label{eq:DimOfPhysicalVelocitiesFermion2}
	\begin{align}
		dim[\tilde{V}_\perp]&=-z_\perp\Big[\frac{N_c^2-1}{4\pi^2 N_cN_f}w\lambda[h_1(c,c_\perp,v)-h_2(c,c_\perp,v)]\nonumber\\
		&+F_{dis}(\{\Gamma_i\},v)\Big]\\
		dim[\tilde{v}_{||}]&=-z_\perp\Big[\frac{N_c^2-1}{4\pi^2N_cN_f}w\lambda[h_1(c,c_\perp,v)-h_3(c,c_\perp,v)]\nonumber\\
		&+F_{dis}(\{\Gamma_i,v\})\Big]\\
		dim[\tilde{v}]&=-z_\perp\Big[\frac{N_c^2-1}{4\pi^2N_cN_f}w\lambda[h_1(c,c_\perp,v)+h_3(c,c_\perp,v)]\nonumber\\
		&+F_{dis}(\{\Gamma_i,v\})\Big] .
	\end{align}
\end{subequations}

Next, we consider boson fields. There are two velocity parameters, $c_\perp$ and $c$. Following the same procedure as the above, we obtain physical velocities of the boson fields as follows $\tilde{c}_\perp=b^{z_\perp-z_\tau}c_\perp$, $\tilde{c}=b^{1-z_\tau}c$. Resulting scaling dimensions of the physical velocities are given by $dim[\tilde{c}_\perp]=dim[c_\perp]+z_\perp-z_\tau$ and $dim[\tilde{c}]=dim[c]+1-z_\tau$ where $dim[c_\perp]=-\frac{\beta_{c_\perp}}{c_\perp}$ and $dim[c]=-\frac{\beta_c}{c}$. Using the one-loop of the $z_\tau$ (Eq. \eqref{eq:ztau}), $z_\perp$ (Eq. \eqref{eq:zperp}), $\beta_c$ (Eq. \eqref{eq:betac}), and $\beta_{c_\perp}$ (Eq. \eqref{eq:betacperp}), we find the following one-loop scaling dimensions of the boson physical velocities
\begin{subequations}\label{eq:DimOfPhysicalVelocitiesBoson2}
	\begin{align}
		dim[\tilde{c}_\perp]&=-\frac{z_\perp}{2}\frac{\lambda}{4\pi }\Big(1-\frac{1}{c_\perp^2}\Big)-\gamma_M\Big(1+\frac{\pi}{4}s\kappa\Big)\frac{z_\perp\epsilon+\bar{\epsilon}}{\epsilon+\bar{\epsilon}},\\
		dim[\tilde{c}]&=-\frac{z_\perp}{2}\frac{\lambda}{4\pi }-\frac{\gamma_M}{2\pi^2}\Big(1+\frac{3\pi}{4}s\kappa\Big)\frac{z_\perp \epsilon+\bar{\epsilon}}{\epsilon+\bar{\epsilon}} .
	\end{align}
\end{subequations}

Now we are ready to interpret the one-loop RG flows for the general case (Table  \ref{table:SummaryOfRGFlowGeneralCase}) in physical terms, based on the one-loop scaling dimensions of physical velocities for fermions (Eq. \eqref{eq:DimOfPhysicalVelocitiesFermion2}) and bosons (Eq. \eqref{eq:DimOfPhysicalVelocitiesBoson2}). 

First, let us discuss the implications of the one-loop RG flows on the physical velocities of boson and fermion fields. The physical velocities of fermions ($\tilde{v}$, $\tilde{v}_{||}$, $\tilde{V}_{\perp}$) (Eqs. \eqref{eq:DimOfPhysicalVelocitiesFermion2}) decrease rapidly in the low energy limit due to the term $F_{dis}(\{,\Gamma_i,v\})$ which is proportional to the $\Gamma_{G1}^d$. Physically, it means that fermion fields become strongly localized in the low energy limit by the random charge impurity. The physical velocities of bosons ($\tilde{c}_{\perp}$, $\tilde{c}_{||}$) (Eq. \eqref{eq:DimOfPhysicalVelocitiesBoson2}) also decrease in the low energy limit. However, it is mainly due to Yukawa interaction ($\lambda$), which is converging to the value of $4\pi F_{dis}(\Gamma_i,v)$, rather than random mass vertex ($\gamma_M$) since the random boson mass becomes irrelevant in the low energy limit. As a result, both fermion and boson fields become localized in the low energy limit but by different mechanisms. However, a more important thing in understanding the low energy physics is the fact that which field between fermion and boson is more localized. It is determined from ratios between the physical velocities of fermions and those of bosons. One of them is $\frac{\tilde{v}}{\tilde{c}}$ which turns out to be same as $w=\frac{v}{c}$. In the general case, $w$ decreases at low energies, which means that the Fermion field is more localized than the Boson field. As discussed above it is because that the fermion field becomes localized by the random charge potential, which is the most dominant vertex in the general case, directly while the boson field gets effect from the random charge potential indirectly through the Yukawa interaction.

Next, the enhancement of the effective Yukawa interaction $\lambda$ in the general case is understood in the following way. Here, the effective Yukawa interaction is a ratio between the interaction energy and the kinetic energy of fermions. Naively, we can think that reduced physical velocities of fermions cause the enhanced effective Yukawa interactions since the kinetic energy of fermions is reduced. However as we discuss in section \ref{sec:NoYIAndNorCPCase}, the correlation of the boson fields or more specifically wave function renormalization constant of the boson field also needs to be considered in understanding the low-energy RG-flows of the effective Yukawa interaction ($\lambda$). In this case, the reduction of the fermion kinetic energy due to random charge-potential vertices is larger than the reduction of the boson correlations due to the Yukawa interaction. Therefore it results in the enhancement of the effective Yukawa interaction. However, there can appear an opposite case which will be discussed in the next section \ref{sec:TwoLoopResultsAndpossibleTwoDisorderedPhase}. 

Finally, let us discuss the physical meaning of the decreasing effective boson self-interaction $\kappa_1$ in the low energy limit. The effective boson self-interaction parameter $\kappa_1$ is defined as a ratio between the boson self-interaction energy and the boson kinetic energy. Therefore it seems that $\kappa_1$ would increase since the boson kinetic energy decreases in the low energy limit. However, as pointed out above in the case of the effective Yukawa interaction $\lambda$, quasi-particleness, more directly, the wave-function renormalization of the boson field ($Z_\phi$) also has to be considered. In the anomalous dimension of boson fields $\eta_\phi$ (Eq. \eqref{eq:etaphi}), a term proportional to the effective Yukawa interaction $\lambda$ becomes dominant in the low energy limit, which results from the boson self-energy diagram (Fig. \ref{fig:SDWBosonSelfEnergy}). This is the `Landau damping' effect. Due to the Landau damping, the boson self-interaction acquires screening as explicitly shown by the term $\frac{g^2}{2\pi v}$ in $\beta_{u_1}$ (Eq. \eqref{eq:betau1}). Additionally, the boson self-interaction $u_1$ screens itself by one-loop corrections (Fig. \ref{fig:SDWBSICorrection}). Considering all these effects on $\kappa_1$, we find that $\kappa_1$ decreases: The reduction of the interaction energy $u_1$ by both the Landau damping and the self-screening is larger than the reduction of the kinetic energy in this case.

\subsubsection{Physical interpretations of the one-loop low energy state}
Before proceeding to the next section, we would like to give a comment on the physical interpretation given in Ref. \cite{Punk}. In the previous study, the authors speculated that the low energy physics of the No-rBM case would be a random singlet phase based on the fact that $u_1$ diverges in the low energy limit. However, as pointed out before, dimensionless parameters ($w$, $\lambda$, $\kappa_i$, $\gamma_M$) should be considered in determining the low energy physics rather than the original parameters ($c$, $c_\perp$, $v$, $g$, $u_i$, $\Gamma_M$). Therefore the dimensionless variable $\kappa_1$, the ratio between the boson self-interaction energy and the boson kinetic energy, needs to be considered to determine the low energy physics rather than the $u_1$ variable. In the No-rBM phase, $\kappa_1$ decreases in the low energy limit while $u_1$ diverges. In this respect, we speculate that the low energy state of boson fields is governed by the boson kinetic energy rather than the boson self-interaction energy. As a result, it is unlikely for boson fields to form a random-singlet phase. Additionally, we show that the effective random boson mass vertex $\gamma_M$ or $s\gamma_M$ decreases in the general case. It also supports that the ground state would not be the random singlet phase. In the next section \ref{sec:TwoLoopResultsAndpossibleTwoDisorderedPhase}, we discuss the possibility of the random singlet phase when two-loop corrections are considered.

\subsection{Incomplete two-loop RG analysis and possibility of another strongly disordered phase} \label{sec:TwoLoopResultsAndpossibleTwoDisorderedPhase}

In the one-loop results, we have found that there is no stable fixed point in the presence of the random charge potential vertices. Therefore, to have stable fixed points, we need to consider two-loop Feynman diagrams which can give rise to a screening of the random charge vertices. However, there are too many Feynman diagrams in the two-loop level compared to the one-loop case. As a result, here we only consider two-loop Feynman diagrams composed of random charge potential vertices. Details of our partial two-loop calculations and results can be found in the Supplementary Material \cite{Note1}. Unfortunately, it turns out that there is still no stable fixed point and the strengths of random charge potential vertices rather increase by the partial two-loop diagrams. This further enhancement of the random charge potential vertices seems to be consistent with the fact that the random charge disorder cannot be screened by the disorder scattering itself in two dimensional systems: Anderson localization \cite{RevModPhys.57.287}. To find a stable fixed point, it appears that we should consider other two-loop Feynman diagrams coming from combinations of Yukawa interactions and random charge potential vertices. 

Up to now, we have focused on a way to screen the random charge potential vertices since one-loop results are broken down by the run-away flow of the random charge potential vertices. However, there is another non-trivial way that one-loop analysis breaks down by the interplay between the random boson mass vertex ($\gamma_M$) and the effective boson self-interaction vertex ($\kappa_1$). 
From the one-loop analysis, we have seen there appears an oscillating RG flow of $\gamma_M$ and $\kappa_1$ in the section section \ref{sec:NoYIAndNorCPCase}. Due to this oscillating behavior, one-loop RG flow become more unstable with respect to the higher order corrections \cite{KirkpatricBelitz}. As a result, we can argue that there is a phase space where the one-loop analysis breaks down in the early stage of the RG flow even before the random charge potential vertices become large enough to break the one-loop results. 
See Appendix \ref{Appendix:BreakdownOfOneLoopByOscillatingRG} for the detailed argument. 

Since the one-loop result breaks down by the random boson vertex and boson self-interaction vertex, we should consider two-loop corrections to the random boson mass and boson self-interaction vertices. Since the two-loop results of the random boson mass and boson self-interaction vertices are already known \cite{KirkpatricBelitz}, here we use the results to discuss how these two-loop corrections change the low energy physics. Let us suppose there is a parameter region with finite $ (s\gamma_M, \kappa_1)$, where both $\gamma_M$ and $\kappa_1$ keep increasing in the low energy limit due to such two-loop order quantum corrections \cite{KirkpatricBelitz}. With this presumption, we can argue that there is a parameter region where both $\frac{\Gamma_{G1}^d}{\gamma_M}$ and $\lambda$ decrease while both $w$ and $\kappa_1$ increase in the low energy limit. See Appendix \ref{Appendix:StabilityOfRBMDphase} for the detailed analysis. 

\begin{figure}
	\includegraphics[width=0.28\textwidth]{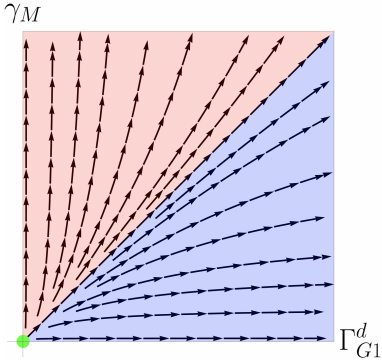}
	\caption{Schematic phase diagram. Arrows show RG flows as lowering an energy scale. The red colored region is an RBMD (`random boson mass dominant') phase space while the blue colored region is an RCPD (`random charge potential dominant') phase space. The green point corresponds to the clean case. Here, the `Interacting LRO' phase is not presented for simplicity.} \label{fig:SchematicPhaseDiagram}
\end{figure}

Increasing $w$ means that boson fields become more localized than fermion fields due to stronger random boson mass vertex than the random charge potential vertices. More localized boson fields result in the enhancement of the effective boson self-interaction $\kappa_1$. Additionally, since the reduction of the correlation between the boson fields is larger than the reduction of the fermion kinetic energy, the effective Yukawa interaction $\lambda$ decreases. Based on the \cite{KirkpatricBelitz}, we suspect that this low-energy RG flow arrives at a fixed point which might correspond to the random-singlet-like phase. The physical picture of the random-singlet phase seems to be consistent with the possible RG flows discussed here. For example, reduction of the effective Yukawa interaction ($\lambda$) can be understood by decoupling between boson and fermion fields due to the singlet formation of the boson fields. This low-energy RG flow is different from that of the No-rBM and the general case discussed in Sec. \ref{sec:NorBMandGeneralCase}. It seems that these two different low-energy RG flows are determined by which disorder effects between the random charge potential and the random boson mass are more dominant. To distinguish these two different cases, we call two low-energy phase spaces that have different RG flows a `random charge potential dominant' (RCPD)  and a `random boson mass dominant' (RBMD) phase space, respectively. Figure \ref{fig:SchematicPhaseDiagram} shows a schematic phase diagram in the parameter space of the random charge potential and the random boson mass. A blue-colored region corresponds to the RCPD phase space while a red-colored region corresponds to the RBMD phase space. Low-energy RG flows of these two cases are summarized in Table \ref{table:TwoDisorderedPhasesSumUp}.

\renewcommand{\arraystretch}{1.2}
\begin{table}
	\scalebox{0.9}{
		\begin{tabular}{|c|c|c|}
			\hline
			& Low-energy RG flows & Remarks \\
			\hhline{===}
			\parbox[c]{2.5cm}{RCPD phase space} & \parbox[c]{2.5cm}{\begin{gather*}
					\Gamma_{G1}^d/\gamma_M, \lambda:\;  \nearrow,  \\ w,\kappa_1:\; \searrow
			\end{gather*}}& \parbox[c]{2.5cm}{No-rBM case\\ General case} \\
			\cline{2-3}\cline{1-1}
			\parbox[c]{2.5cm}{RBMD phase space}&    \parbox[c]{2.5cm}{\begin{gather*}
					\Gamma_{G1}^d/\gamma_M, \lambda:\;  \searrow ,  \\ w,\kappa_1:\; \nearrow
			\end{gather*}}& \parbox[c]{2.5cm}{random-singlet \\ like phase}\\
			\hline
		\end{tabular}
	}
	\caption{Summary of the RG flows in two different phase spaces in the low energy limit: `random charge potential dominant' (RCPD) and `random boson mass dominant' (RBMD) phase spaces. Here, $\Gamma_{G1}^d$ ($\gamma_M$) are the parameters of random charge potential vertices (effective random boson mass vertex), $w$ is the ratio of $\tilde{v}$ to $\tilde{c}$, and $\lambda$ $(\kappa_1)$ is the effective interaction parameter of the Yukawa interaction (boson self-interaction).} \label{table:TwoDisorderedPhasesSumUp}
\end{table}

We have discussed two different low-energy RG flows originating from the competition between two non-magnetic quenched disorder effects in the presence of the Yukawa interaction. However, this is not a complete result. As we discussed at first in this section, if the additional two-loop corrections, that can give rise to screening to the random charge potential vertices, are considered, there can be a stable fixed point within a finite phase space region. 
Outside of this finite phase space region, we expect that the above two different RG flows would appear. 


\section{Physical properties} \label{sec:PhysicalProperties}

Based on our one-loop RG analysis, 
we check out superconducting instabilities for four pairing channels in the low energy limit by calculating the anomalous dimensions of one-loop order.

\subsection{Superconducting instabilities}

\def\a{6} 
\def\r{2.5} 
\def\b{2.4}

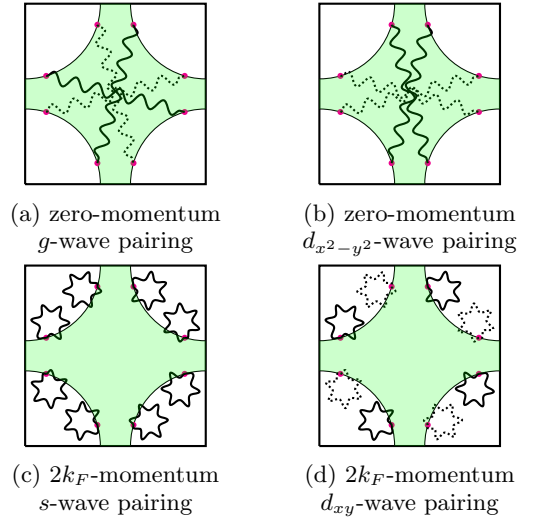
\begin{figure}
	\centering
	\begin{subfigure}{0.2\textwidth}
		\begin{tikzpicture}[scale=0.4,>=stealth]
			\draw[style=thick] (0,0)--(\a,0)--(\a,\a)--(0,\a)--(0,0);
			\draw[style=thin] (0,\r) arc (90:0:\r);
			\draw[style=thin] (\a,\r) arc (90:180:\r);
			\draw[style=thin] (\a,\a-\r) arc (-90:-180:\r);
			\draw[style=thin] (0,\a-\r) arc (-90:0:\r);

			\fill[color=magenta] (\a-\b,{\a-sqrt((\r^2-\b^2))}) circle (0.1); 
			\fill[color=magenta] ({sqrt((\r^2-\b^2))},\b) circle (0.1); 
			
			\fill[color=magenta] ({sqrt((\r^2-\b^2))},\a-\b) circle (0.1); 
			\fill[color=magenta] (\a-\b,{sqrt((\r^2-\b^2))}) circle (0.1); 
			
			\fill[color=magenta] (\b,{sqrt((\r^2-\b^2))}) circle (0.1); 
			\fill[color=magenta] ({\a-sqrt((\r^2-\b^2))},\a-\b) circle (0.1); 
			
			\fill[color=magenta] ({\a-sqrt((\r^2-\b^2))},\b) circle (0.1); 
			\fill[color=magenta] (\b,{\a-sqrt((\r^2-\b^2))}) circle (0.1); 

			\draw[decorate,decoration={snake,amplitude=.8mm,segment length=3mm,post length=0mm},thick] (\a-\b,{\a-sqrt((\r^2-\b^2))})--(\b,{sqrt((\r^2-\b^2))});
			
			\draw[decorate,decoration={snake,amplitude=.8mm,segment length=3mm,post length=0mm},thick,densely dotted] (\b,{\a-sqrt((\r^2-\b^2))})--(\a-\b,{sqrt((\r^2-\b^2))});
			
			\draw[decorate,decoration={snake,amplitude=.8mm,segment length=3mm,post length=0mm},thick] ({sqrt((\r^2-\b^2))},\a-\b)--({\a-sqrt((\r^2-\b^2))},\b);
			
			\draw[decorate,decoration={snake,amplitude=.8mm,segment length=3mm,post length=0mm},thick,densely dotted] ({sqrt((\r^2-\b^2))},\b)--({\a-sqrt((\r^2-\b^2))},\a-\b);

			\fill[color=green, opacity=0.2] (0,\r)--(0,\a-\r) to [out=0,in=-90] (\r,\a)--(\a-\r,\a) to [out=-90, in=-180] (\a,\a-\r)--(\a,\r) to [out=180, in=90] (\a-\r,0)--(\r,0) to [out=90,in=0] (0,\r);
		\end{tikzpicture}
		\caption{zero-momentum $g$-wave pairing}
	\end{subfigure}
	~
	\begin{subfigure}{0.2\textwidth}
		\begin{tikzpicture}[scale=0.4,>=stealth]
			\draw[style=thick] (0,0)--(\a,0)--(\a,\a)--(0,\a)--(0,0);
			\draw[style=thin] (0,\r) arc (90:0:\r);
			\draw[style=thin] (\a,\r) arc (90:180:\r);
			\draw[style=thin] (\a,\a-\r) arc (-90:-180:\r);
			\draw[style=thin] (0,\a-\r) arc (-90:0:\r);

			\fill[color=magenta] (\a-\b,{\a-sqrt((\r^2-\b^2))}) circle (0.1); 
			\fill[color=magenta] ({sqrt((\r^2-\b^2))},\b) circle (0.1); 
			
			\fill[color=magenta] ({sqrt((\r^2-\b^2))},\a-\b) circle (0.1); 
			\fill[color=magenta] (\a-\b,{sqrt((\r^2-\b^2))}) circle (0.1); 
			
			\fill[color=magenta] (\b,{sqrt((\r^2-\b^2))}) circle (0.1); 
			\fill[color=magenta] ({\a-sqrt((\r^2-\b^2))},\a-\b) circle (0.1); 
			
			\fill[color=magenta] ({\a-sqrt((\r^2-\b^2))},\b) circle (0.1); 
			\fill[color=magenta] (\b,{\a-sqrt((\r^2-\b^2))}) circle (0.1); 

			\draw[decorate,decoration={snake,amplitude=.8mm,segment length=3mm,post length=0mm},thick] (\a-\b,{\a-sqrt((\r^2-\b^2))})--(\b,{sqrt((\r^2-\b^2))});
			
			\draw[decorate,decoration={snake,amplitude=.8mm,segment length=3mm,post length=0mm},thick] (\b,{\a-sqrt((\r^2-\b^2))})--(\a-\b,{sqrt((\r^2-\b^2))});
			
			\draw[decorate,decoration={snake,amplitude=.8mm,segment length=3mm,post length=0mm},thick,densely dotted] ({sqrt((\r^2-\b^2))},\a-\b)--({\a-sqrt((\r^2-\b^2))},\b);
			
			\draw[decorate,decoration={snake,amplitude=.8mm,segment length=3mm,post length=0mm},thick,densely dotted] ({sqrt((\r^2-\b^2))},\b)--({\a-sqrt((\r^2-\b^2))},\a-\b);

			\fill[color=green, opacity=0.2] (0,\r)--(0,\a-\r) to [out=0,in=-90] (\r,\a)--(\a-\r,\a) to [out=-90, in=-180] (\a,\a-\r)--(\a,\r) to [out=180, in=90] (\a-\r,0)--(\r,0) to [out=90,in=0] (0,\r);
		\end{tikzpicture}
		\caption{zero-momentum $d_{x^2-y^2}$-wave pairing}
	\end{subfigure}
	\\
	\begin{subfigure}{0.2\textwidth}
		\begin{tikzpicture}[scale=0.4,>=stealth]
			\draw[style=thick] (0,0)--(\a,0)--(\a,\a)--(0,\a)--(0,0);
			\draw[style=thin] (0,\r) arc (90:0:\r);
			\draw[style=thin] (\a,\r) arc (90:180:\r);
			\draw[style=thin] (\a,\a-\r) arc (-90:-180:\r);
			\draw[style=thin] (0,\a-\r) arc (-90:0:\r);

			\fill[color=magenta] (\a-\b,{\a-sqrt((\r^2-\b^2))}) circle (0.1); 
			\fill[color=magenta] ({sqrt((\r^2-\b^2))},\b) circle (0.1); 
			
			\fill[color=magenta] ({sqrt((\r^2-\b^2))},\a-\b) circle (0.1); 
			\fill[color=magenta] (\a-\b,{sqrt((\r^2-\b^2))}) circle (0.1); 
			
			\fill[color=magenta] (\b,{sqrt((\r^2-\b^2))}) circle (0.1); 
			\fill[color=magenta] ({\a-sqrt((\r^2-\b^2))},\a-\b) circle (0.1); 
			
			\fill[color=magenta] ({\a-sqrt((\r^2-\b^2))},\b) circle (0.1); 
			\fill[color=magenta] (\b,{\a-sqrt((\r^2-\b^2))}) circle (0.1); 

			\draw[decorate,decoration={snake,amplitude=.5mm,segment length=2mm},thick] (\a-\b,{\a-sqrt((\r^2-\b^2))}) arc (180:360+180:0.5cm);
			
			\draw[decorate,decoration={snake,amplitude=.5mm,segment length=2mm},thick] (\b,{\a-sqrt((\r^2-\b^2))}) arc (360:0:0.5cm);
			
			\draw[decorate,decoration={snake,amplitude=.5mm,segment length=2mm},thick] ({sqrt((\r^2-\b^2))},\a-\b) arc (-90:270:0.5cm);
			
			\draw[decorate,decoration={snake,amplitude=.5mm,segment length=2mm},thick] ({sqrt((\r^2-\b^2))},\b) arc (-270:-270-360:0.5cm);
			
			\draw[decorate,decoration={snake,amplitude=.5mm,segment length=2mm},thick] (\b,{sqrt((\r^2-\b^2))}) arc (0:360:0.5cm);
			
			\draw[decorate,decoration={snake,amplitude=.5mm,segment length=2mm},thick] (\a-\b,{sqrt((\r^2-\b^2))}) arc (180:-180:0.5cm);
			
			\draw[decorate,decoration={snake,amplitude=.5mm,segment length=2mm},thick] ({\a-sqrt((\r^2-\b^2))},\b) arc (90:450:0.5cm);
			
			\draw[decorate,decoration={snake,amplitude=.5mm,segment length=2mm},thick] ({\a-sqrt((\r^2-\b^2))},\a-\b) arc (-90:-450:0.5cm);

			\fill[color=green, opacity=0.2] (0,\r)--(0,\a-\r) to [out=0,in=-90] (\r,\a)--(\a-\r,\a) to [out=-90, in=-180] (\a,\a-\r)--(\a,\r) to [out=180, in=90] (\a-\r,0)--(\r,0) to [out=90,in=0] (0,\r);
		\end{tikzpicture}
		\caption{$2k_F$-momentum $s$-wave pairing}
	\end{subfigure}
	~
	\begin{subfigure}{0.2\textwidth}
		\begin{tikzpicture}[scale=0.4,>=stealth]
			\draw[style=thick] (0,0)--(\a,0)--(\a,\a)--(0,\a)--(0,0);
			\draw[style=thin] (0,\r) arc (90:0:\r);
			\draw[style=thin] (\a,\r) arc (90:180:\r);
			\draw[style=thin] (\a,\a-\r) arc (-90:-180:\r);
			\draw[style=thin] (0,\a-\r) arc (-90:0:\r);

			\fill[color=magenta] (\a-\b,{\a-sqrt((\r^2-\b^2))}) circle (0.1); 
			\fill[color=magenta] ({sqrt((\r^2-\b^2))},\b) circle (0.1); 
			
			\fill[color=magenta] ({sqrt((\r^2-\b^2))},\a-\b) circle (0.1); 
			\fill[color=magenta] (\a-\b,{sqrt((\r^2-\b^2))}) circle (0.1); 
			
			\fill[color=magenta] (\b,{sqrt((\r^2-\b^2))}) circle (0.1); 
			\fill[color=magenta] ({\a-sqrt((\r^2-\b^2))},\a-\b) circle (0.1); 
			
			\fill[color=magenta] ({\a-sqrt((\r^2-\b^2))},\b) circle (0.1); 
			\fill[color=magenta] (\b,{\a-sqrt((\r^2-\b^2))}) circle (0.1); 

			\draw[decorate,decoration={snake,amplitude=.5mm,segment length=2mm},thick] (\a-\b,{\a-sqrt((\r^2-\b^2))}) arc (180:360+180:0.5cm);
			
			\draw[decorate,decoration={snake,amplitude=.5mm,segment length=2mm},thick,densely dotted] (\b,{\a-sqrt((\r^2-\b^2))}) arc (360:0:0.5cm);
			
			\draw[decorate,decoration={snake,amplitude=.5mm,segment length=2mm},thick] ({sqrt((\r^2-\b^2))},\a-\b) arc (-90:270:0.5cm);
			
			\draw[decorate,decoration={snake,amplitude=.5mm,segment length=2mm},thick,densely dotted] ({sqrt((\r^2-\b^2))},\b) arc (-270:-270-360:0.5cm);
			
			\draw[decorate,decoration={snake,amplitude=.5mm,segment length=2mm},thick] (\b,{sqrt((\r^2-\b^2))}) arc (0:360:0.5cm);
			
			\draw[decorate,decoration={snake,amplitude=.5mm,segment length=2mm},thick,densely dotted] (\a-\b,{sqrt((\r^2-\b^2))}) arc (180:-180:0.5cm);
			
			\draw[decorate,decoration={snake,amplitude=.5mm,segment length=2mm},thick] ({\a-sqrt((\r^2-\b^2))},\b) arc (90:450:0.5cm);
			
			\draw[decorate,decoration={snake,amplitude=.5mm,segment length=2mm},thick,densely dotted] ({\a-sqrt((\r^2-\b^2))},\a-\b) arc (-90:-450:0.5cm);

			\fill[color=green, opacity=0.2] (0,\r)--(0,\a-\r) to [out=0,in=-90] (\r,\a)--(\a-\r,\a) to [out=-90, in=-180] (\a,\a-\r)--(\a,\r) to [out=180, in=90] (\a-\r,0)--(\r,0) to [out=90,in=0] (0,\r);
		\end{tikzpicture}
		\caption{$2k_F$-momentum $d_{xy}$-wave pairing}
	\end{subfigure}
	\caption{Schematic figures for four types of superconducting instability channels. Dotted lines denote the different sign compared to plain lines.} \label{Fig:SCchannels}
\end{figure}

\begin{figure}
	\begin{subfigure}{0.23\textwidth}
		\begin{tikzpicture}[scale=0.7]
			\begin{feynhand}
				\vertex (a) at (0,0);
				\vertex (b) at (1,0);
				\vertex [dot] (c) at (2,0) {};
				\vertex (d) at (3,0);
				\vertex (e) at (4,0);
				\propag[fer] (a) to [edge label=$n$](b);
				\propag[fer] (b) to [edge label=$\bar{n}$](c);
				\propag[fer] (d) to [edge label=$\bar{n}$](c);
				\propag[fer] (e) to [edge label=$n$](d);
				\propag[boson] (b) to [out=90, in=90, looseness=1.5](d);
			\end{feynhand}
		\end{tikzpicture}
		\caption{ From the Yukawa vertex} \label{Fig:SConeLoop1}
	\end{subfigure}
	~
	\begin{subfigure}{0.23\textwidth}
		\begin{tikzpicture}[scale=0.7]
			\begin{feynhand}
				\vertex (a) at (0,0);
				\vertex (b) at (1,0);
				\vertex [dot] (c) at (2,0) {};
				\vertex (d) at (3,0);
				\vertex (e) at (4,0);
				\node at (1,-0.5) {$\mathcal{M}$};
				\node at (3,-0.5) {$\mathcal{\tilde{M}}$};
				\propag[fer] (a) to [edge label=$n$](b);
				\propag[fer] (b) to [edge label=$m$](c);
				\propag[fer] (d) to [edge label=$m$](c);
				\propag[fer] (e) to [edge label=$n$](d);
				\propag[sca] (b) to [out=90, in=90, looseness=1.5](d);
			\end{feynhand}
		\end{tikzpicture}
		\caption{From the random charge potential vertex} \label{Fig:SConeLoop2}
	\end{subfigure}
	\caption{One-loop Feynman diagrams for superconducting instabilities.} \label{Fig:SConeLoop}
\end{figure}
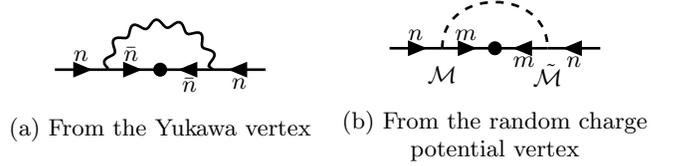

Here, we consider four spin-singlet superconducting channels, identified as most relevant ones in the previous study \cite{SurLee}: Two zero-momentum channels ($g$ and $d_{x^2-y^2}$) and two $2k_F$-momentum channels ($s$ and $d_{xy}$) shown in Fig. \ref{Fig:SCchannels}, where explicit forms of these superconducting channels are given in Appendix \ref{Appendix:SCinstabilityCal}. There are two one-loop Feynman diagrams depicted in Fig. \ref{Fig:SConeLoop} which contribute to the anomalous dimensions of the superconducting channels.

Here, we only present the results. For details of the calculations, see Appendix \ref{Appendix:SCinstabilityCal}. From the results (Eqs. \eqref{eq:betafunctionOfDelta1}$\sim$\eqref{eq:betafunctionOfDelta4}) in Appendix \ref{Appendix:SCinstabilityCal}, the RG beta functions of the four superconducting vertices are given as follows
\begin{align}
	\beta_{\Delta_{0,i}}&=-\Big[1+\frac{1}{4\pi^2}\frac{N_c+1}{N_cN_f}\frac{g^2}{c}\Big(\frac{f^{(\gamma_{d-1})}_{SC}(c,c_\perp,v)}{2\pi}\nonumber\\
	&-(N_c-1)h_3(c,c_\perp,v)\Big)\Big]\Delta_{0,i}\nonumber\\
	&= -(1+\gamma_{\Delta_{0,i}})\Delta_{0,i}\label{eq:betaDelta0}\\
	\beta_{\Delta_{2k_F,i}}&=-\Big[1+\frac{1}{4\pi^2}\frac{N_c+1}{N_cN_f}\frac{g^2}{c}\Big(\frac{f^{(\hat{1})}_{SC}(c,c_\perp,v)}{2\pi}\nonumber\\
	&-(N_c-1)h_3(c,c_\perp,v)\Big)+\frac{\Gamma_0+\Upsilon_{0}}{1+v^2}\Big]\Delta_{2k_F,i}\nonumber\\
	&= -(1+\gamma_{\Delta_{2k_F,i}})\Delta_{2k_F,i} , \label{eq:betaDelta2kF}
\end{align}
where superconducting order parameters are $\Delta_{0,i}=\{\Delta_{0,g},\Delta_{0,d_{x^2-y^2}}\}$ and $\Delta_{2k_F,i}=\{\Delta_{2k_F,d_{xy}},\Delta_{2k_F,s}\}$. The function $f^{(\hat{\Omega})}_{SC}(c,c_\perp,v)$ is given by 
\begin{align*}
	&f^{(\hat{\Omega})}_{SC}(c,c_\perp,v)=\frac{\pi}{\sqrt{1+v^2}}\int_0^1dx\; x(1-x)^{-1/2}\nonumber\\
	&\times \Big(x+(1-x)c_\perp^2\Big)^{-1/2}\Big(x+\frac{c^2}{1+v^2}(1-x)\Big)^{-1/2} \nonumber\\
	&\Big[\frac{1}{x+\frac{c^2}{1+v^2}(1-x)}+Sign(\hat{\Omega})\Big(-1+\frac{1}{x+(1-x)c_\perp^2}\Big)\Big].\\
	&\Big(Sign(\hat{\Omega})=\Big\{\begin{array}{ll}1 &\text{ when }\hat{\Omega}=\hat{1}\\ -1&\text{ when }\hat{\Omega}=\gamma_{d-1} \end{array}\Big) . \nonumber
\end{align*}

In the case of two zero-momentum superconducting channels ($\Delta_{0,i}$), there is a correction from the boson-fermion interaction only (Eq. \eqref{eq:betaDelta0}), reproducing a similar result to that in Ref. \cite{SurLee}. However, for the two $2k_F$-momentum superconducting channels ($\Delta_{2k_F,i}$) we find that there are additional corrections from the random charge potential vertices (Eq. \eqref{eq:betaDelta2kF}). Two random charge potential vertices $\Gamma_0$ (Direct) and $\Upsilon_0$ (Umklapp) enhance the superconductivity of the $2k_F$-momentum pairing channels. As a result, the $2k_F$-momentum superconducting instability channels are more favorable to develop, compared to the zero-momentum superconducting channels in the presence of random charge potential fluctuations.

\begin{figure}
	\includegraphics[width=0.4\textwidth]{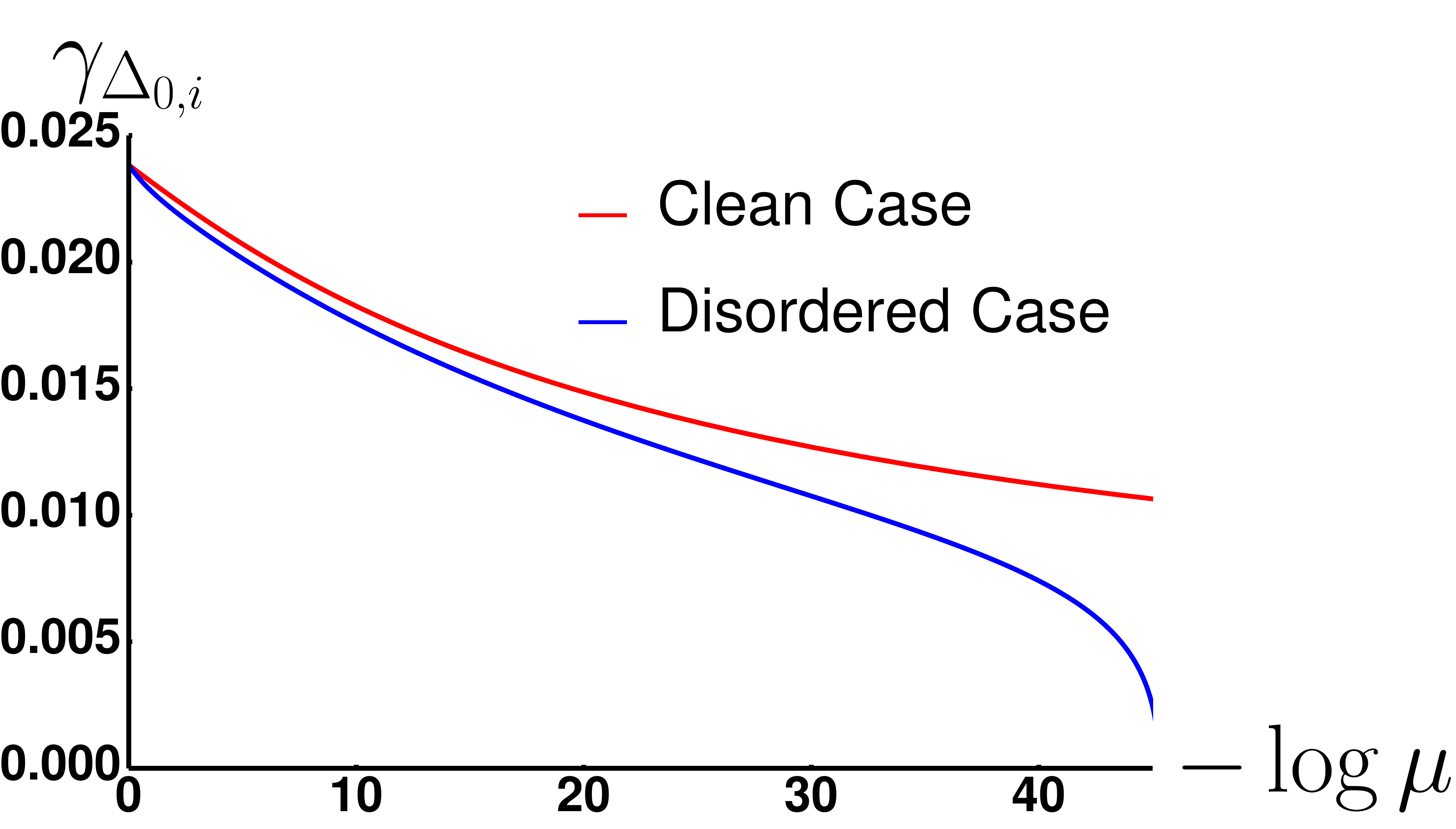}
	\caption{The anomalous scaling dimension of two zero-momentum superconducting instability channels ($\Delta_{0,g}$ and $\Delta_{0,d_{x^2-y^2}}$) for the clean case (red colored line) and the disordered case (blue colored line with $\Gamma_i\neq 0,\; \Gamma_M\neq 0$). Here, we used $\epsilon=\bar{\epsilon}=0.01$, $N_f=1$, $N_c=2$, $v_c=0.05$, and $\kappa=1$.} \label{fig:DimSCInstability}
\end{figure}

From now on, we focus on how the zero-momentum superconducting channel is affected by disorder scattering. Despite there are no direct contributions from the disorder vertices for the zero-momentum superconducting channels ($\Delta_{0,i}$), there are indirect contributions through the Yukawa coupling $g$, the boson velocity $c$, $f_{SC}^{(\hat{\Omega})}(c,c_\perp,v)$, and $h_3(c,c_\perp,v)$ from the disorder effects. Therefore, we calculate the anomalous dimensions of the zero-momentum superconducting channel ($\gamma_{\Delta_{0,i}}$) for the clean case ($\Gamma_i=\Gamma_M=0$) and the general case ($\Gamma_i\neq0,\; \Gamma_M\neq 0$) numerically and compare these results. The numerical result of the anomalous dimensions $\gamma_{\Delta_{0,i}}$ for both the clean and disordered cases is shown in Fig. \ref{fig:DimSCInstability}. The anomalous dimension of the zero-momentum superconducting channels is suppressed in the disordered case, compared to the clean case, which means that the superconducting instability is suppressed by disorder effects. 

It has been theoretically argued that superconductivity (especially $d$-wave superconductivity) is developed at the temperature before electrons near hot spots lose their coherence in the two-dimensional clean SDW quantum critical metallic system \cite{SchLee} and in the two-dimensional nematic Ising quantum critical metallic system \cite{SCMross}. These results are different from conventional phase diagrams of quantum critical systems which usually show the non-Fermi liquid state above the superconducting dome. However, according to our results, $d$-wave superconductivity can be suppressed by the effects of disorders. Additionally, from the discussion in section \ref{sec:RGanalysis}, disorder effects enhance the effective Yukawa interaction in the low energy limit. As a result, the present calculation shows a possibility that the non-Fermi liquid state appears before the superconducting dome in the presence of the disorder which is contrary to the results obtained in the clean cases \cite{SchLee,SCMross}. 


\section{Conclusion} \label{sec:SummaryConclusion}

In this paper, we have investigated the effects of the general non-magnetic quenched disorder on the two-dimensional spin-density-wave(SDW) quantum critical metallic system using a perturbative renormalization group method. As effects of the non-magnetic quenched disorder, we considered (i) a random charge potential for fermions and (ii) a random mass term for a SDW boson order parameter. Particularly, we have taken into account all possible scattering channels of the random charge potential vertex among hot spots in a 2d SDW metallic Fermi surface and classified them into three categories; `Direct', `Exchange' and `Umklapp'. To control strong quantum fluctuations in two spatial dimensions, we used two regularization methods at the same time in the RG analysis: One is a co-dimensional regularization technique for controlling two-dimensional Fermi-surface fluctuations, and the other is a nonlocal-correlated random mass probability method for controlling the random boson mass fluctuation. 

From the one-loop results, we found a weakly-disordered 2d SDW quantum critical NFL fixed point (Random Mass Disordered non-Fermi liquid phase) in the one-loop RG analysis when only random boson mass vertex is considered. However, scattering channels from the random charge potential destabilize this dirty SDW NFL fixed point to have a run-away RG flow in the general case. This low energy run-away RG-flow is driven by both large random charge potential vertices of the `Direct' category and the effective Yukawa interaction. On the other hand, random charge potential vertices of the `Umklapp' category, the effective boson self-interaction, and the effective random boson mass vertex, become irrelevant in the low energy limit. We have discussed these low-energy RG flows in more physical terms, based on the dimensionless parameters and physical velocities. Furthermore, we performed a two-loop level RG analysis to find a stable 2d SDW disordered NFL fixed point in the general case. To find a stable fixed point, we examined a screening of the random charge potential vertices by the two-loop diagrams. Unfortunately, we finished our two-loop level RG analysis only for renormalization effects from random charge potential vertices due to its complexity. It turns out that these two-loop diagrams do not screen the random charge potential vertices in the `Direct' category, but rather increase them. We speculate that the two-loop diagrams involving the Yukawa interaction should be considered for the screening of the random charge potential vertices. Despite a stable fixed point is not found for the general case, we revealed the possibility of another disordered phase space by considering the two-loop corrections to the random boson mass vertex based on the one-loop analysis and the work done by Kirkpatrick and Belitz \cite{KirkpatricBelitz}. In another disordered phase space, we have found that the random boson mass vertex is the most dominant vertex which is different from the general case in the one-loop results where the random charge potential vertices are the most dominant. As a result, we called these two different cases a `random charge potential dominant' (RCPD) phase space and a `random boson mass dominant' (RBMD) phase space, respectively. Additionally, we have discussed the physical properties of the `RBMD' phase space in relation to the random singlet phase. 

Regarding physical properties, we 
discussed anomalous dimensions of four superconducting channels considered by Sur and Lee \cite{SurLee}. We found that 
a zero-momentum $d-$wave superconducting channel is suppressed due to effects of disorders while the superconducting channels with $2k_F$-momentum are enhanced. From the result of the zero momentum $d-$wave superconducting channel, we argued about how the disorder affects can change a phase diagram of a clean non-fermi liquid state with a superconducting dome discussed in Refs. \cite{SchLee,SCMross}

Now, we point out some limitations and technical difficulties in this research direction, regarding both interaction and disorder in strongly coupled quantum critical metallic systems:

\begin{itemize}
	\item In our research, we considered only hot spot fermions. However, we have to take into account fermions at cold regions in the presence of disorder effects. Especially, to investigate transport phenomena, it is essential to consider all the fermions at the Fermi surface.
	
	\item Related to the first point, the patch construction for non-Fermi liquid physics near quantum criticality is not justified in the presence of disorder effects.
	
	\item There are some artifacts coming from our regularization methods. The co-dimensional regularization method generates new random charge potential vertices beyond the original lattice model, which originates from explicit translation symmetry breaking. The correlated random boson mass probability regularization method does not allow physically possible loop corrections to the random boson mass vertex ($\Gamma_M$) due to its non-local nature. These are the cost of making the interaction and disorder vertices marginal.
	
	\item When we take into account not only Yukawa interaction but also effects of disorders perturbatively, there are too many Feynman diagrams we need to consider. It is an incredibly difficult task to calculate all relevant diagrams beyond the one-loop level. Additionally, it is difficult to justify or figure out the controllability of the results.
\end{itemize}

To overcome these limitations and technical difficulties, alternative approaches can be considered to study the effects of disorders on non-fermi liquid systems. Most popular approaches recently are based on SYK-like models \cite{Debanjan1,Debanjan2,Patel1,Patel2,Esterlis1,Esterlis2,Wang1,Wang2,GeorgesReview}. In these SYK-like models, the effects of disorders are simplified by considering all-to-all scattering processes and employing large numbers of fermion and boson species ($N$, $M$). In the large $N$, $M$ limit, it is possible to obtain fully self-consistent Green's functions. Another approach less popular but promising is to consider disorder effects at the non-Fermi liquid fixed point directly \cite{Nosov}. Since the starting point is the non-fermi liquid fixed point instead of the clean fixed point, interaction effects are already incorporated in a controllable way. Only disorder effects need to be considered. The main obstacle to this approach is that it is not easy to identify the clean non-fermi liquid fixed in a controllable way. In the case of the SDW quantum critical metallic system, Schlief et al. \cite{SchLee} proposed a clean non-fermi liquid fixed point including all possible relevant diagrams. Therefore it would be interesting to investigate the effects of disorders on the clean non-fermi liquid fixed point directly.

\section*{ACKNOWLEDGEMENT}

I. Jang was supported by Global Ph.D. Fellowship of the National Research Foundation of Korea (NRF-2015H1A2A1033126). K.-S. Kim was supported by the Ministry of Education, Science, and Technology (NRF-2021R1A2C1006453 and NRF-2021R1A4A3029839) of the National Research Foundation of Korea (NRF) and by TJ Park Science Fellowship of the POSCO TJ Park Foundation. I. Jang appreciates fruitful discussions with Kyong-Min Kim and Jae-Ho Han and encouragement from Jinho Yang for finishing this work.

\bibliographystyle{apsrev4-2}

\typeout{}

\bibliography{bibliographyOfDisorderedSDW}

\begin{widetext}
	
	\appendix

	\section{Setting for the Renormalization Group Theory} \label{Appendix:RGSetting}
	
	\subsection{Counterterms and renormalized effective field theory}
	
	We introduce a renormalized effective action, a counterterm action, and a bare effective action, respectively, as follows
	\begin{align}
		S_{eff,R}&=\sum_{a=1}^R\Bigg[\sum_{n=1}^4\sum_{\sigma=1}^{N_c}\sum_{i_f=1}^{N_f}\int dk \bar{\Psi}^a_{n,\sigma,i_f}[i\gamma_0k_0+i\mathbf{\Gamma}_\perp\cdot\mathbf{K}_\perp+i\gamma_{d-1}\epsilon_n(k;v)]\Psi^a_{n,\sigma,i_f}\nonumber\\
		&+\frac{1}{4}\int dq [q_0^2+c_\perp^2 |\mathbf{Q}_\perp|^2+c^2|q|^2]Tr [\Phi^a(-q)\Phi^a(q)]+i\frac{g\mu^{\epsilon/2}}{\sqrt{N_f}}\sum_{i_f=1}^{N_f}\sum_{\sigma,\sigma'=1}^{N_c}\int dk \int dq \bar{\Psi}^a_{\bar{n},\sigma,i_f}(k+q)\nonumber\\
		&\times \Phi^a_{\sigma,\sigma'}(q)\gamma_{d-1}\Psi^a_{n,\sigma,i_f}+\frac{u_1\mu^\epsilon}{4}\int dk_1dk_2dq {Tr[\Phi^a(k_1+q)\Phi^a(k_2-q)]Tr[\Phi^a(k_1)\Phi^a(k_2)]}\nonumber\\
		&+\frac{u_2\mu^\epsilon}{4}\int dk_1 dk_2 dqTr[\Phi^a(k_1+q)\Phi^a(k_2-q)\Phi^a(k_1)\Phi^a(k_2)]\Bigg]-\sum_{a,b=1}^R\int \frac{d\omega}{2\pi}\int \frac{d\omega'}{2\pi}\prod_{i=1}^4\int \frac{d^{d}\mathbf{k}_i}{(2\pi)^d}(2\pi)^d \nonumber\\
		&\times \Bigg[\sum_{i=1}^{27}\sum_{i_f,j_f=1}^{N_f}\sum_{\sigma,\sigma'=1}^{N_c}\frac{\Gamma_i\mu^{-1+\epsilon}}{2N_f}\delta(\mathbf{k}_1+\mathbf{k}_3-\mathbf{k}_2-\mathbf{k}_4) \Big([\bar{\Psi}^a_{n,\sigma,i_f}(\omega,\mathbf{k}_1) \mathcal{M}^i_{nm}\Psi^a_{m,\sigma,i_f}(\omega,\mathbf{k}_2)]\nonumber\\
		&\times [\bar{\Psi}^b_{k,\sigma',j_f}(\omega',\mathbf{k}_3) \tilde{\mathcal{M}}^i_{kl}\Psi^b_{l,\sigma',j_f}(\omega',\mathbf{k}_4)]+\cdots\Big)+\frac{\Gamma_M\mu^{\epsilon+\bar{\epsilon}}\Big(|\mathbf{K}_{1,\perp}+\mathbf{K}_{2,\perp}|^\alpha+\kappa |\vec{k}_1+\vec{k}_2|^\alpha \Big)}{8}\nonumber\\
		&\times Tr[\Phi^a(\omega,\vec{k}_1)\cdot\Phi^a(\omega,\vec{k}_2)]Tr[\Phi^b(\omega',\vec{k}_3)\Phi^b(\omega',\vec{k}_4)] \delta(\mathbf{k}_1+\mathbf{k}_2+\mathbf{k}_3+\mathbf{k}_4)\Bigg] ,
		\end{align}
	\begin{align}
		S_{eff,C}&=\sum_{a=1}^R\Bigg[\sum_{n=1}^4\sum_{\sigma=1}^{N_c}\sum_{i_f=1}^{N_f}\int dk \bar{\Psi}^a_{n,\sigma,i_f}\Big[iA_0\gamma_0k_0+iA_1\mathbf{\Gamma}_\perp\cdot\mathbf{K}_\perp+iA_3\gamma_{d-1}\epsilon_n\Big(k;\frac{A_2}{A_3}v\Big)\Big]\Psi^a_{n,\sigma,i_f}\nonumber\\
		&+\frac{1}{4}\int dq [A_4q_0^2+A_5c_\perp^2|\mathbf{Q}_\perp|^2+A_6c^2|q|^2]Tr [\Phi^a(-q)\Phi^a(q)]+iA_7\frac{g\mu^{\epsilon/2}}{\sqrt{N_f}}\sum_{i_f=1}^{N_f}\sum_{\sigma,\sigma'=1}^{N_c}\int dk \int dq \bar{\Psi}^a_{\bar{n},\sigma,i_f}(k+q)\nonumber\\
		&\times \Phi^a_{\sigma,\sigma'}(q)\gamma_{d-1}\Psi^a_{n,\sigma,i_f}+\frac{A_8u_1\mu^\epsilon}{4}\int dk_1dk_2dq {Tr[\Phi^a(k_1+q)\Phi^a(k_2-q)]Tr[\Phi^a(k_1)\Phi^a(k_2)]}\nonumber\\
		&+\frac{A_9u_2\mu^\epsilon}{4}\int dk_1 dk_2 dqTr[\Phi^a(k_1+q)\Phi^a(k_2-q)\Phi^a(k_1)\Phi^a(k_2)]\Bigg]-\sum_{a,b=1}^R\int \frac{d\omega}{2\pi}\int \frac{d\omega'}{2\pi}\prod_{i=1}^4\int \frac{d^{d}\mathbf{k}_i}{(2\pi)^d}(2\pi)^d\nonumber\\
		&\times \Bigg[\sum_{i_f,j_f=1}^{N_f}\sum_{i=1}^{27}\sum_{\sigma,\sigma'=1}^{N_c}\frac{A_{\Gamma_i}\Gamma_i\mu^{-1+\epsilon}}{2N_f}\delta(\mathbf{k}_1+\mathbf{k}_3-\mathbf{k}_2-\mathbf{k}_4) \Big([\bar{\Psi}^a_{n,\sigma,i_f}(\omega,\mathbf{k}_1) \mathcal{M}^i_{nm}\Psi^a_{m,\sigma,i_f}(\omega,\mathbf{k}_2)]\nonumber\\
		&\times [\bar{\Psi}^b_{k,\sigma',j_f}(\omega',\mathbf{k}_3) \tilde{\mathcal{M}}^i_{kl}\Psi^b_{l,\sigma',j_f}(\omega',\mathbf{k}_4)]+\cdots\Big)+\frac{A_{\Gamma_M}\Gamma_M\mu^{\epsilon+\bar{\epsilon}}\Big(|\mathbf{K}_{1,\perp}+\mathbf{K}_{2,\perp}|^\alpha+\kappa |\vec{k}_1+\vec{k}_2|^\alpha\Big)}{8}\nonumber\\
		&\times Tr[\Phi^a(\omega,\vec{k}_1)\cdot\Phi^a(\omega,\vec{k}_2)]Tr[\Phi^b(\omega',\vec{k}_3)\Phi^b(\omega',\vec{k}_4)] \delta(\mathbf{k}_1+\mathbf{k}_2+\mathbf{k}_3+\mathbf{k}_4)\Bigg] ,\\
		S_{eff,B}&=S_{eff,R}[\Psi\rightarrow \Psi_B, \; k\rightarrow k_B, \; g\rightarrow g_B,\; \cdots]
	\end{align}

	Here, $\mu$ is an energy scale introduced to make $g$, $u_1$, $u_2$, $\Gamma_i$, and $\Gamma_M$ dimensionless parameters. $\Psi_{n,\sigma,i_f}^a/\Phi_{\sigma,\sigma'}^a$ and $\Psi_{B,n,\sigma,i_f}^a/\Phi_{B,\sigma,\sigma'}$ are renormalized and bare fermion/boson fields, respectively.
	
	Considering $S_{eff,B}=S_{eff,R}+S_{eff,C}$ with introduction of renormalized constants as follows
	$	k_{0,B}=k_{0}Z_\tau,\; \mathbf{K}_{\perp,B}=\mathbf{K}_\perp Z_\perp,\; \vec{k}_{B}=\vec{k},\; \Psi_{B}=Z_{\psi}^{1/2}\Psi,\;\Phi_{B}=Z_{\phi}^{1/2}\Phi,\; Z_{i}=1+A_i $, we obtain the following renormalization conditions
	\begin{subequations}
		\begin{gather}
			(Z_\psi Z_\perp^{d-2}Z_\tau)Z_\tau=Z_0,\;\; (Z_\psi Z_\perp^{d-2}Z_\tau)Z_\perp=Z_1,\;\; Z_\psi Z_\perp^{d-2}Z_\tau=Z_3,\;\; (Z_\psi Z_\perp^{d-2}Z_\tau)v_B=Z_2v,\label{eq:RGCondition1}\\
			(Z_\phi Z_\perp^{d-2}Z_\tau)Z_\tau^2=Z_4,\;\; (Z_\phi Z_\perp^{d-2}Z_\tau)Z_{\perp}^2c_{\perp,B}^2=Z_5c_\perp^2,\;\; (Z_\phi Z_\perp^{d-2}Z_\tau)c_B^2=Z_6 c^2,\\
			(Z_\perp^{d-2}Z_\tau)^2 Z_\psi Z_\phi^{1/2}g_B=Z_7 g \mu^{\epsilon/2},\;\; (Z_\perp^{d-2}Z_\tau)^3Z_\phi^2 u_{1,B}=Z_8 u_1\mu^{\epsilon},\\
			(Z_\perp^{d-2}Z_\tau)^3Z_\phi^2 u_{2,B}=Z_9 u_2 \mu^{\epsilon},
			\;\; (Z_\perp^{d-2})^3Z_\tau^2Z_\psi^2\bar{\Gamma}_{j,B}=\mu^\epsilon Z_{\bar{\Gamma}_j}\bar{\Gamma}_j,\;\;(Z_\perp^{d-2})^3Z_\tau^2Z_\phi^2\Gamma_{M,B}=\mu^{\epsilon+\bar{\epsilon}} Z_{\Gamma_M}\Gamma_M . \label{eq:RGCondition2}
		\end{gather}
	\end{subequations}
	Here, we used $\bar{\Gamma}_i=\Gamma_i\Lambda_{FS}$, $Z_{\Gamma_i}=Z_{\bar{\Gamma}_i}$, and $\Lambda_{FS}=\mu\Lambda_{FS,B}$.
	
	\subsection{Feynman rules}

	\begin{figure}[h]
	\begin{subfigure}{0.49\textwidth}
	\includegraphics[scale=0.27]{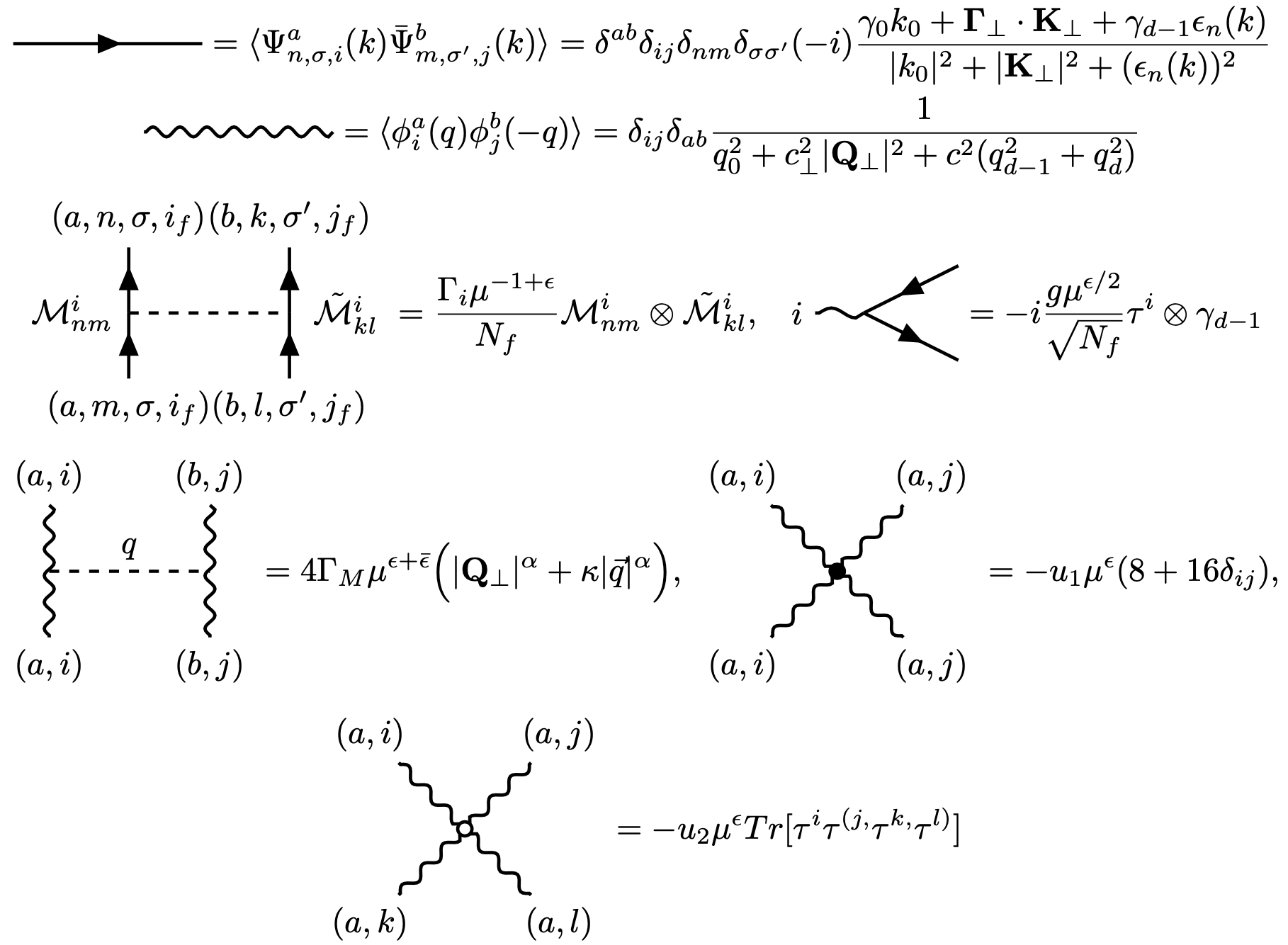}
	\end{subfigure}
	~
	\begin{subfigure}{0.49\textwidth}
	\includegraphics[scale=0.27]{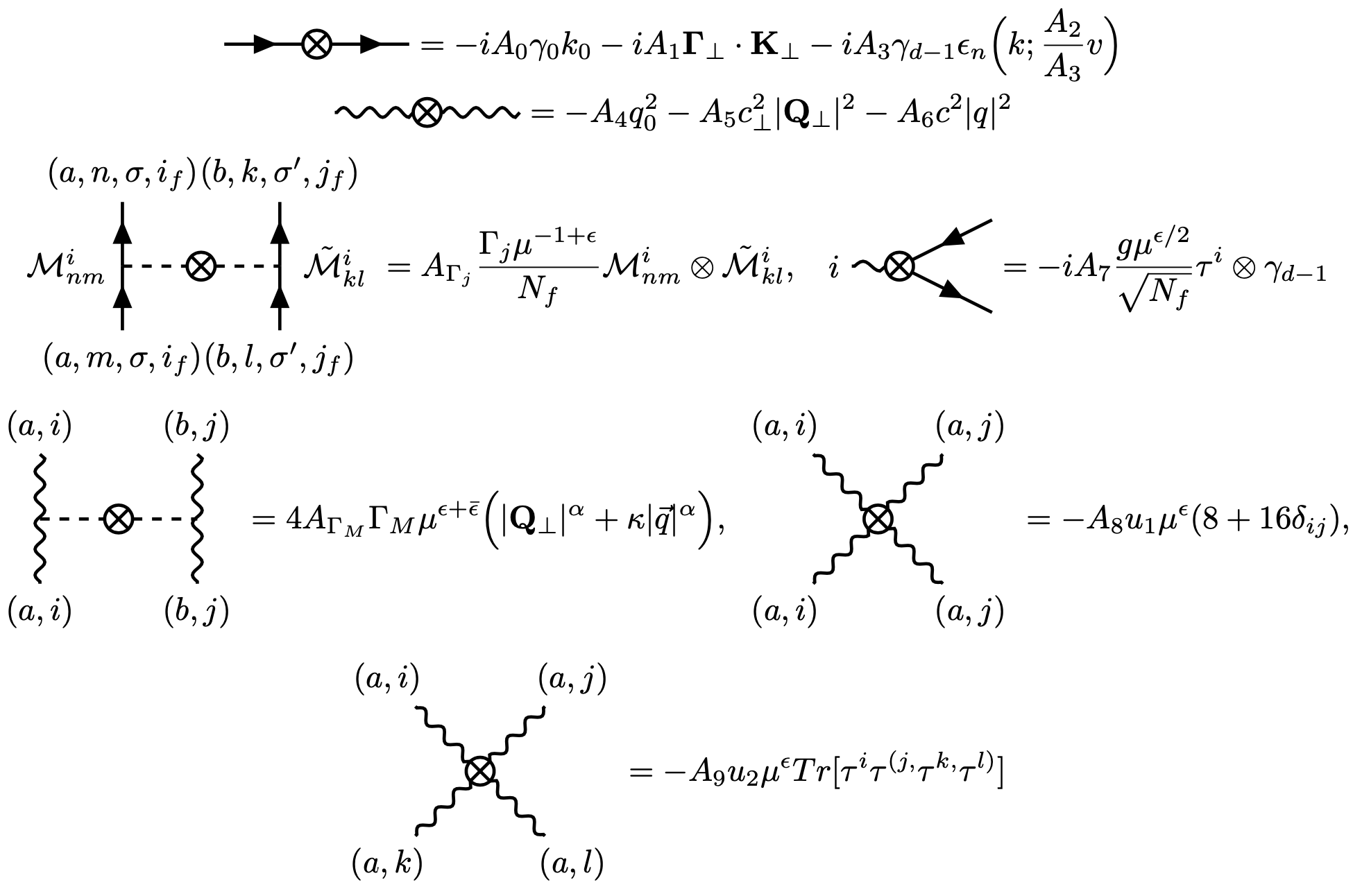}
	\end{subfigure}
	\end{figure}

	\subsection{Renormalization Group Equations}
	
	\subsubsection{Anomalous dimensions and renormalization group equations for $v$, $c$, $c_\perp$, $g$, $u_1$, $u_2$, $\{\Gamma_i\}$, and $\Gamma_M$}
	
	Anomalous scaling dimensions for fermions ($\eta_{\psi}$), bosons ($\eta_{\phi}$), $k_0$ ($z_\tau$), $\mathbf{K}_\perp$ ($z_{\perp}$), and RG beta functions of $v$, $c$, $c_\perp$, $g$, $u_1$, $u_2$, $\{\bar{\Gamma}_i\}$, and $\Gamma_M$ are defined as follows:
	\begin{subequations}
		\begin{gather}
			\eta_\psi=\frac{1}{2}\frac{\partial \ln Z_\psi}{\partial\ln \mu}, \; \;  \eta_\phi=\frac{1}{2}\frac{\partial \ln Z_\phi}{\partial \ln \mu},\;\; z_{\tau}=1+\frac{d\ln Z_{0}}{d\ln \mu},\;\; z_{\perp}=1+\frac{d\ln Z_\perp}{d\ln \mu}\label{eq:DefinitionAnomalousDimension}\\
			\beta_v \equiv \frac{d v}{d\ln \mu} , \;  \beta_c \equiv \frac{d c}{d\ln \mu},\; \beta_{c_\perp}=\frac{d c_\perp}{d\ln \mu},\; \beta_g \equiv \frac{d g}{d\ln \mu} , \;\beta_{u_1} \equiv \frac{d u_{1}}{d\ln \mu} , \;  \beta_{u_{2}} \equiv \frac{d u_{2}}{d\ln \mu},\;\beta_{\bar{\Gamma}_i}\equiv \frac{d\bar{\Gamma}_i}{d\ln \mu},\;\beta_{\Gamma_M}\equiv \frac{d\Gamma_{M}}{d\ln\mu}\label{eq:DefinitionBetaFunctions}
		\end{gather}
	\end{subequations}
	
	Considering $\frac{d\mathcal{O}_B}{d\ln\mu}=0$ with these equations, where $\mathcal{O}_B$ represents any bare or unrenormalized quantities such as frequency and momentum, fermion and boson fields, and all the coupling constants, we obtain
	\begin{gather*}
		-(\ln Z_0)'+2\eta_{\psi}+(d-2)(z_\perp-1)+2(z_\tau-1)=0 ,\;\; -(\ln Z_1)'+2\eta_\psi+(d-1)(z_\perp-1)+(z_\tau-1)=0, \\
		-(\ln Z_3)'+2\eta_\psi+(d-2)(z_\perp-1)+(z_\tau-1)=0,\; \; -(\ln Z_4)'+2\eta_\phi+(d-2)(z_\perp-1)+3(z_\tau-1)=0,\\
		\beta_v=[(\ln Z_3)'-(\ln Z_2)'] v_r,\;\; \beta_c=\frac{1}{2}[2\eta_\phi+(d-2)(z_\perp-1)+(z_\tau-1) -(\ln Z_6)']c_r\\
		\beta_{c_\perp}=\frac{1}{2}[2\eta_\phi +d(z_\perp-1)+(z_\tau-1)-(\ln Z_5)']c_{\perp,r},\;\; \beta_g=[-\frac{\epsilon}{2}+\eta_\phi+2\eta_\psi+2(d-2)(z_\perp-1)+2(z_\tau-1)-(\ln Z_7)']g\\
		\beta_{u_1}=[-\epsilon+4\eta_\phi+3(d-2)(z_\perp-1)+3(z_\tau-1)-(\ln Z_8)']u_{1},\;\; \beta_{u_2}=[-\epsilon+4\eta_\phi+3(d-2)(z_\perp-1) +3(z_\tau-1)-(\ln Z_9)']u_{2}\\
		\beta_{\bar{\Gamma}_i}=[-\epsilon+4\eta_{\psi}+3(d-2)(z_\perp-1)+2(z_\tau-1)-(\ln Z_{\bar{\Gamma}_i})']\bar{\Gamma}_{i}\\
		\beta_{\Gamma_M}=[-(\epsilon+\bar{\epsilon})+4\eta_\phi+3(d-2)(z_\perp-1)+2(z_\tau-1)-(\ln Z_{\Gamma_M})']\Gamma_M .
		\end{gather*}
	where short hand notation of $(\ln Z_i)' \equiv \frac{d\ln Z_i}{ d\ln \mu}$ is used. 
	
	Solving these coupled equations, we find renormalization group equations for $z_\tau$, $z_\perp$, $\eta_\psi$, $\eta_{\phi}$, $\beta_{v}$, $\beta_c$, $\beta_{c_\perp}$, $\beta_g$, $\beta_{u_1}$, $\beta_{u_{2}}$, $\beta_{\bar{\Gamma}_i}$, and $\beta_{\Gamma_M}$ in terms of all the coupling constants and $Z_i$. $z_\tau$, $z_\perp$, $\eta_\psi$ and $\eta_{\phi}$ are given as follows
	\begin{align}
		z_{\perp}&=\Big[1-\frac{\bar{\epsilon}}{\epsilon+\bar{\epsilon}}\Gamma_M(\bar{F}_{\Gamma_M,1}^{(1)}-\bar{F}_{\Gamma_M,3}^{(1)})\Big]\Big[1+\frac{1}{2}g\Big(F_{g,1}^{(1)}-F_{g,3}^{(1)}+\frac{\epsilon}{\epsilon+\bar{\epsilon}}(\bar{F}_{g,1}^{(1)}-\bar{F}_{g,3}^{(1)})\Big)+u_1\Big(F_{1,u_1}^{(1)}-F_{3,u_1}^{(1)}\nonumber\\
		&+\frac{\epsilon}{\epsilon+\bar{\epsilon}}(\bar{F}_{1,u_1}^{(1)}-\bar{F}_{3,u_1}^{(1)})\Big)+u_2\Big(F_{u_2,1}^{(1)}-F_{u_2,3}^{(1)}+\frac{\epsilon}{\epsilon+\bar{\epsilon}}(\bar{F}_{u_2,1}^{(1)}-\bar{F}_{u_2,3}^{(1)})\Big)+\sum_{\bar{\Gamma}_i}\bar{\Gamma}_i\Big(F_{\bar{\Gamma}_i,1}^{(1)}-F_{\bar{\Gamma}_i,3}^{(1)}+\frac{\epsilon}{\epsilon+\bar{\epsilon}}(\bar{F}_{\bar{\Gamma}_i,1}^{(1)}-\bar{F}_{\bar{\Gamma}_i,3}^{(1)})\Big)\nonumber\\
		&+\frac{\epsilon}{\epsilon+\bar{\epsilon}}\Gamma_M(\bar{F}_{\Gamma_M,1}^{(1)}-\bar{F}_{\Gamma_M,3}^{(1)})\Big]^{-1}\label{eq:BetaFunctionFirst}\\
		z_{\tau}&=-\frac{\bar{\epsilon}}{\epsilon+\bar{\epsilon}}\Gamma_M(\bar{F}_{\Gamma_M,0}^{(1)}-\bar{F}_{\Gamma_M,1}^{(1)})+z_{\perp}\Big[1+\frac{1}{2}g\Big(F_{g,1}^{(1)}-F_{g,0}^{(1)}+\frac{\epsilon}{\epsilon+\bar{\epsilon}}(\bar{F}_{g,1}^{(1)}-\bar{F}_{g,0}^{(1)})\Big)+u_1\Big(F_{u_1,1}^{(1)}-F_{u_1,0}^{(1)}\nonumber\\
		&+\frac{\epsilon}{\epsilon+\bar{\epsilon}}(\bar{F}_{u_1,1}^{(1)}-\bar{F}_{u_1,0}^{(1)})\Big)+u_2\Big(F_{u_2,1}^{(1)}-F_{u_2,0}^{(1)}+\frac{\epsilon}{\epsilon+\bar{\epsilon}}(\bar{F}_{u_2,1}^{(1)}-\bar{F}_{u_2,0}^{(1)})\Big)+\sum_{\bar{\Gamma}_i}\bar{\Gamma}_i\Big(F_{\bar{\Gamma}_i,1}^{(1)}-F_{\bar{\Gamma}_i,0}^{(1)}+\frac{\epsilon}{\epsilon+\bar{\epsilon}}(\bar{F}_{\bar{\Gamma}_i,1}^{(1)}-\bar{F}_{\bar{\Gamma}_i,0}^{(1)})\Big)\nonumber\\
		&+\Gamma_M\frac{\epsilon}{\epsilon+\bar{\epsilon}}(\bar{F}_{\Gamma_M,1}^{(1)}-\bar{F}_{\Gamma_M,0}^{(1)})\Big],\\
		\eta_{\psi}&=-\frac{1}{2}\Bigg[z_\perp\Big[1+\frac{1}{2}g\Big(F_{g,0}^{(1)}+\frac{\epsilon}{\epsilon+\bar{\epsilon}}\bar{F}_{g,0}^{(1)}\Big)+u_1\Big(F_{u_1,0}^{(1)}+\frac{\epsilon}{\epsilon+\bar{\epsilon}}\bar{F}_{u_1,0}^{(1)}\Big)+u_2\Big(F_{u_2,0}^{(1)}+\frac{\epsilon}{\epsilon+\bar{\epsilon}}\bar{F}_{u_2,0}^{(1)}\Big)\nonumber\\
		&+\sum_{\bar{\Gamma}_i}\bar{\Gamma}_i \Big(F_{\bar{\Gamma}_i,0}^{(1)}+\frac{\epsilon}{\epsilon+\bar{\epsilon}}\bar{F}_{\bar{\Gamma}_i,0}^{(1)}\Big)+\Gamma_M\frac{\epsilon}{\epsilon+\bar{\epsilon}}\bar{F}_{\Gamma_M,0}^{(1)}\Big]+\frac{\bar{\epsilon}}{\epsilon+\bar{\epsilon}}\Gamma_M\bar{F}_{\Gamma_M,0}^{(1)}+2z_\tau-3\Bigg]+\frac{z_\perp-1}{2}\epsilon\\
		\eta_{\phi}&=-\frac{1}{2}\Bigg[z_\perp\Big[1+\frac{1}{2}g\Big(F_{g,4}^{(1)}+\frac{\epsilon}{\epsilon+\bar{\epsilon}}\bar{F}_{g,4}^{(1)}\Big)+u_1\Big(F_{u_1,4}^{(1)}+\frac{\epsilon}{\epsilon+\bar{\epsilon}}\bar{F}_{u_1,4}^{(1)}\Big)+u_2\Big(F_{u_2,4}^{(1)}+\frac{\epsilon}{\epsilon+\bar{\epsilon}}\bar{F}_{u_2,4}^{(1)}\Big)\nonumber\\
		&+\sum_{\bar{\Gamma}_i}\bar{\Gamma}_i \Big(F_{\bar{\Gamma}_i,4}^{(1)}+\frac{\epsilon}{\epsilon+\bar{\epsilon}}\bar{F}_{\bar{\Gamma}_i,4}^{(1)}\Big)+\Gamma_M\frac{\epsilon}{\epsilon+\bar{\epsilon}}\bar{F}_{\Gamma_M,4}^{(1)}\Big]+\frac{\bar{\epsilon}}{\epsilon+\bar{\epsilon}}\Gamma_M\bar{F}_{\Gamma_M,4}^{(1)}+3z_\tau-4\Bigg]+\frac{z_\perp-1}{2}\epsilon
		\end{align}
	and $\beta_{v}$, $\beta_c$, $\beta_{c_\perp}$, $\beta_g$, $\beta_{u_1}$, $\beta_{u_{2}}$, $\beta_{\bar{\Gamma}_i}$, and $\beta_{\Gamma_M}$ are given by 
	\begin{align}
		\beta_v&=v\Bigg[z_\perp\Big[\frac{1}{2}g\Big(F_{g,2}^{(1)}-F_{g,3}^{(1)}+\frac{\epsilon}{\epsilon+\bar{\epsilon}}(\bar{F}_{g,2}^{(1)}-\bar{F}_{g,3}^{(1)})\Big)+u_1\Big(F_{u_1,2}^{(1)}-F_{u_1,3}^{(1)}+\frac{\epsilon}{\epsilon+\bar{\epsilon}}(\bar{F}_{u_1,2}^{(1)}-\bar{F}_{u_1,3}^{(1)})\Big)\nonumber\\
		&+u_2\Big(F_{u_2,2}^{(1)}-F_{u_2,3}^{(1)}+\frac{\epsilon}{\epsilon+\bar{\epsilon}}(\bar{F}_{u_2,2}^{(1)}-\bar{F}_{u_2,3}^{(1)})\Big)+\sum_{\bar{\Gamma}_i}\bar{\Gamma}_i\Big(F_{\bar{\Gamma}_i,2}^{(1)}-F_{\bar{\Gamma}_i,3}^{(1)}+\frac{\epsilon}{\epsilon+\bar{\epsilon}}(\bar{F}_{\bar{\Gamma}_i,2}^{(1)}-\bar{F}_{\bar{\Gamma}_i,3}^{(1)})\Big)\nonumber\\
		&+\Gamma_M\frac{\epsilon}{\epsilon+\bar{\epsilon}}(\bar{F}_{\Gamma_M,2}^{(1)}-\bar{F}_{\Gamma_M,3}^{(1)})\Big]+\frac{\bar{\epsilon}}{\epsilon+\bar{\epsilon}}\Gamma_M(\bar{F}_{\Gamma_M,2}^{(1)}-\bar{F}_{\Gamma_M,3}^{(1)})\Bigg]\\
		\beta_c&=\frac{c}{2}\Bigg[z_\perp\Big[\frac{1}{2}g\Big(F_{g,6}^{(1)}-F_{g,4}^{(1)}+\frac{\epsilon}{\epsilon+\bar{\epsilon}}(\bar{F}_{g,6}^{(1)}-\bar{F}_{g,4}^{(1)})\Big)+u_1\Big(F_{u_1,6}^{(1)}-F_{u_1,4}^{(1)}+\frac{\epsilon}{\epsilon+\bar{\epsilon}}(\bar{F}_{u_1,6}^{(1)}-\bar{F}_{u_1,4}^{(1)})\Big)\nonumber\\
		&+u_2\Big(F_{u_2,6}^{(1)}-F_{u_2,4}^{(1)}+\frac{\epsilon}{\epsilon+\bar{\epsilon}}(\bar{F}_{u_2,6}^{(1)}-\bar{F}_{u_2,4}^{(1)})\Big)+\sum_{\bar{\Gamma}_i}\bar{\Gamma}_i\Big(F_{\bar{\Gamma}_i,6}^{(1)}-F_{\bar{\Gamma}_i,4}^{(1)}+\frac{\epsilon}{\epsilon+\bar{\epsilon}}(\bar{F}_{\bar{\Gamma}_i,6}^{(1)}-\bar{F}_{\bar{\Gamma}_i,4}^{(1)})\Big)\nonumber\\
		&+\Gamma_M\frac{\epsilon}{\epsilon+\bar{\epsilon}}(\bar{F}_{\Gamma_M,6}^{(1)}-\bar{F}_{\Gamma_M,4}^{(1)})\Big]+\frac{\bar{\epsilon}}{\epsilon+\bar{\epsilon}}\Gamma_M(\bar{F}_{\Gamma_M,6}^{(1)}-\bar{F}_{\Gamma_M,4}^{(1)})+2(1-z_\tau)\Bigg]
		\end{align}
	\begin{align}
			\beta_{c_\perp}&=\frac{c_\perp}{2}\Bigg[z_\perp\Big[\frac{1}{2}g\Big(F_{g,5}^{(1)}-F_{g,4}^{(1)}+\frac{\epsilon}{\epsilon+\bar{\epsilon}}(\bar{F}_{g,5}^{(1)}-\bar{F}_{g,4}^{(1)})\Big)+u_1\Big(F_{u_1,5}^{(1)}-F_{u_1,4}^{(1)}+\frac{\epsilon}{\epsilon+\bar{\epsilon}}(\bar{F}_{u_1,5}^{(1)}-\bar{F}_{u_1,4}^{(1)})\Big)\nonumber\\
		&+u_2\Big(F_{u_2,5}^{(1)}-F_{u_2,4}^{(1)}+\frac{\epsilon}{\epsilon+\bar{\epsilon}}(\bar{F}_{u_2,5}^{(1)}-\bar{F}_{u_2,4}^{(1)})\Big)+\sum_{\bar{\Gamma}_i}\bar{\Gamma}_i\Big(F_{\bar{\Gamma}_i,5}^{(1)}-F_{\bar{\Gamma}_i,4}^{(1)}+\frac{\epsilon}{\epsilon+\bar{\epsilon}}(\bar{F}_{\bar{\Gamma}_i,5}^{(1)}-\bar{F}_{\bar{\Gamma}_i,4}^{(1)})\Big)\nonumber\\
		&+\Gamma_M\frac{\epsilon}{\epsilon+\bar{\epsilon}}(\bar{F}_{\Gamma_M,5}^{(1)}-\bar{F}_{\Gamma_M,4}^{(1)})\Big]+\frac{\bar{\epsilon}}{\epsilon+\bar{\epsilon}}\Gamma_M(\bar{F}_{\Gamma_M,5}^{(1)}-\bar{F}_{\Gamma_M,4}^{(1)})+2(z_\perp-z_\tau)\Bigg],\\
		\beta_g&=-\frac{\epsilon}{2}z_\perp g+g\Bigg[1+\frac{1}{2}z_\perp -\frac{3}{2}z_\tau +z_\perp\Big[\frac{1}{2}g\Big(F_{g,7}^{(1)}-F_{g,0}^{(1)}-\frac{1}{2}F_{g,4}^{(1)}+\frac{\epsilon}{\epsilon+\bar{\epsilon}}(\bar{F}_{g,7}^{(1)}-\bar{F}_{g,0}^{(1)}-\frac{1}{2}\bar{F}_{g,4}^{(1)})\Big)\nonumber\\
		&+u_1\Big(F_{u_1,7}^{(1)}-F_{u_1,0}^{(1)}-\frac{1}{2}F_{u_1,4}^{(1)}+\frac{\epsilon}{\epsilon+\bar{\epsilon}}(\bar{F}_{u_1,7}^{(1)}-\bar{F}_{u_1,0}^{(1)}-\frac{1}{2}\bar{F}_{u_1,4}^{(1)})\Big)+u_2\Big(F_{u_2,7}^{(1)}-F_{u_2,0}^{(1)}-\frac{1}{2}F_{u_2,4}^{(1)}\nonumber\\
		&+\frac{\epsilon}{\epsilon+\bar{\epsilon}}(\bar{F}_{u_2,7}^{(1)}-\bar{F}_{u_2,0}^{(1)}-\frac{1}{2}\bar{F}_{u_2,4}^{(1)})\Big)
		+\sum_{\bar{\Gamma}_i}\bar{\Gamma}_i\Big(F_{\bar{\Gamma}_i,7}^{(1)}-F_{\bar{\Gamma}_i,0}^{(1)}-\frac{1}{2}F_{\bar{\Gamma}_i,4}^{(1)}+\frac{\epsilon}{\epsilon+\bar{\epsilon}}(\bar{F}_{\bar{\Gamma}_i,7}^{(1)}-\bar{F}_{\bar{\Gamma}_i,0}^{(1)}-\frac{1}{2}\bar{F}_{\bar{\Gamma}_i,4}^{(1)})\Big)\nonumber\\
		&+\Gamma_M\frac{\epsilon}{\epsilon+\bar{\epsilon}}(\bar{F}_{\Gamma_M,7}^{(1)}-\bar{F}_{\Gamma_M,0}^{(1)}-\frac{1}{2}\bar{F}_{\Gamma_M,4}^{(1)})\Big]+\frac{\bar{\epsilon}}{\epsilon+\bar{\epsilon}}\Gamma_M\Big(\bar{F}_{\Gamma_M,7}^{(1)}-\bar{F}_{\Gamma_M,0}^{(1)}-\frac{1}{2}\bar{F}_{\Gamma_M,4}^{(1)}\Big)\Bigg]\\
		\beta_{u_1}&=-\epsilon z_\perp u_1+u_1\Bigg[2+z_\perp-3z_\tau+z_\perp\Big[\frac{1}{2}g\Big(F_{g,8}^{(1)}-2F_{g,4}^{(1)}+\frac{\epsilon}{\epsilon+\bar{\epsilon}}(\bar{F}_{g,8}^{(1)}-2\bar{F}_{g,4}^{(1)})\Big)+u_1\Big(F_{u_1,8}^{(1)}-2F_{u_1,4}^{(1)}\nonumber\\
		&+\frac{\epsilon}{\epsilon+\bar{\epsilon}}(\bar{F}_{u_1,8}^{(1)}-2\bar{F}_{u_1,4}^{(1)})\Big)+u_2\Big(F_{u_2,8}^{(1)}-2F_{u_2,4}^{(1)}+\frac{\epsilon}{\epsilon+\bar{\epsilon}}(\bar{F}_{u_2,8}^{(1)}-2\bar{F}_{u_2,4}^{(1)})\Big)+\sum_{\bar{\Gamma}_i}\bar{\Gamma}_i\Big(F_{\bar{\Gamma}_i,8}^{(1)}-2F_{\bar{\Gamma}_i,4}^{(1)}\nonumber\\
		&+\frac{\epsilon}{\epsilon+\bar{\epsilon}}(\bar{F}_{\bar{\Gamma}_i,8}^{(1)}-2\bar{F}_{\bar{\Gamma}_i,4}^{(1)})\Big)+\Gamma_M\frac{\epsilon}{\epsilon+\bar{\epsilon}}(\bar{F}_{\Gamma_M,8}-2\bar{F}_{\Gamma_M,4}^{(1)})\Big]+\Gamma_M\frac{\bar{\epsilon}}{\epsilon+\bar{\epsilon}}(\bar{F}_{\Gamma_M,8}-2\bar{F}_{\Gamma_M,4}^{(1)})\Bigg]\\
		\beta_{u_2}&=-\epsilon z_\perp u_2+u_2\Bigg[2+z_\perp-3z_\tau+z_\perp\Big[\frac{1}{2}g\Big(F_{g,9}^{(1)}-2F_{g,4}^{(1)}+\frac{\epsilon}{\epsilon+\bar{\epsilon}}(\bar{F}_{g,9}^{(1)}-2\bar{F}_{g,4}^{(1)})\Big)+u_1\Big(F_{u_1,9}^{(1)}-2F_{u_1,4}^{(1)}\nonumber\\
		&+\frac{\epsilon}{\epsilon+\bar{\epsilon}}(\bar{F}_{u_1,9}^{(1)}-2\bar{F}_{u_1,4}^{(1)})\Big)+u_2\Big(F_{u_2,9}^{(1)}-2F_{u_2,4}^{(1)}+\frac{\epsilon}{\epsilon+\bar{\epsilon}}(\bar{F}_{u_2,9}^{(1)}-2\bar{F}_{u_2,4}^{(1)})\Big)+\sum_{\bar{\Gamma}_i}\bar{\Gamma}_i\Big(F_{\bar{\Gamma}_i,9}^{(1)}-2F_{\bar{\Gamma}_i,4}^{(1)}\nonumber\\
		&+\frac{\epsilon}{\epsilon+\bar{\epsilon}}(\bar{F}_{\bar{\Gamma}_i,9}^{(1)}-2\bar{F}_{\bar{\Gamma}_i,4}^{(1)})\Big)+\Gamma_M\frac{\epsilon}{\epsilon+\bar{\epsilon}}(\bar{F}_{\Gamma_M,9}-2\bar{F}_{\Gamma_M,4}^{(1)})\Big]+\Gamma_M\frac{\bar{\epsilon}}{\epsilon+\bar{\epsilon}}(\bar{F}_{\Gamma_M,9}-2\bar{F}_{\Gamma_M,4}^{(1)})\Bigg],\\
		\beta_{\bar{\Gamma}_i}&=-\epsilon z_\perp\bar{\Gamma}_i+\bar{\Gamma}_i\Bigg[1+z_\perp-2z_\tau+z_\perp\Big[\frac{1}{2}g\Big(F_{g,\bar{\Gamma}_i}^{(1)}-2F_{g,0}^{(1)}+\frac{\epsilon}{\epsilon+\bar{\epsilon}}(\bar{F}_{g,\bar{\Gamma}_i}^{(1)}-2\bar{F}_{g,0}^{(1)})\Big)+u_1\Big(F_{u_1,\bar{\Gamma}_i}^{(1)}-2F_{u_1,0}^{(1)}\nonumber\\
		&+\frac{\epsilon}{\epsilon+\bar{\epsilon}}(\bar{F}_{u_1,\bar{\Gamma}_i}^{(1)}-2\bar{F}_{u_1,0}^{(1)})\Big)+u_2\Big(F_{u_2,\bar{\Gamma}_i}^{(1)}-2F_{u_2,0}^{(1)}+\frac{\epsilon}{\epsilon+\bar{\epsilon}}(\bar{F}_{u_2,\bar{\Gamma}_i}^{(1)}-2\bar{F}_{u_2,0}^{(1)})\Big)+\sum_{\bar{\Gamma}_j}\bar{\Gamma}_j\Big(F_{\bar{\Gamma}_j,\bar{\Gamma}_i}^{(1)}-2F_{\bar{\Gamma}_j,0}^{(1)}\nonumber\\
		&+\frac{\epsilon}{\epsilon+\bar{\epsilon}}(\bar{F}_{\bar{\Gamma}_j,\bar{\Gamma}_i}^{(1)}-2\bar{F}_{\bar{\Gamma}_j,0}^{(1)})\Big)+\Gamma_M\frac{\epsilon}{\epsilon+\bar{\epsilon}}(\bar{F}_{\Gamma_M,\bar{\Gamma}_i}^{(1)}-2\bar{F}_{\Gamma_M,0}^{(1)})\Big]+\frac{\bar{\epsilon}}{\epsilon+\bar{\epsilon}}\Gamma_M(\bar{F}_{\Gamma_M,\bar{\Gamma}_i}^{(1)}-2\bar{F}_{\Gamma_M,0}^{(1)})\Bigg]\\
		\beta_{\Gamma_M}&=-(z_\perp \epsilon+\bar{\epsilon})\Gamma_M+\Gamma_M\Bigg[3+z_\perp-4z_\tau+z_\perp\Big[\frac{1}{2}g\Big(F_{g,\Gamma_M}^{(1)}-2F_{g,4}^{(1)}+\frac{\epsilon}{\epsilon+\bar{\epsilon}}(\bar{F}_{g,\Gamma_M}^{(1)}-2\bar{F}_{g,4}^{(1)})\Big)+u_1\Big(F_{u_1,\Gamma_M}^{(1)}-2F_{u_1,4}^{(1)}\nonumber\\
		&+\frac{\epsilon}{\epsilon+\bar{\epsilon}}(\bar{F}_{u_1,\Gamma_M}^{(1)}-2\bar{F}_{u_1,4}^{(1)})\Big)+u_2\Big(F_{u_2,\Gamma_M}^{(1)}-2F_{u_2,4}^{(1)}+\frac{\epsilon}{\epsilon+\bar{\epsilon}}(\bar{F}_{u_2,\Gamma_M}^{(1)}-2\bar{F}_{u_2,4}^{(1)})\Big)+\sum_{\bar{\Gamma}_i}\bar{\Gamma}_i\Big(F_{\bar{\Gamma}_i,\Gamma_M}^{(1)}-2F_{\bar{\Gamma}_i,4}^{(1)}\nonumber\\
		&+\frac{\epsilon}{\epsilon+\bar{\epsilon}}(\bar{F}_{\bar{\Gamma}_i,\Gamma_M}^{(1)}-2\bar{F}_{\bar{\Gamma}_i,4}^{(1)})\Big)+\Gamma_M\frac{\epsilon}{\epsilon+\bar{\epsilon}}(\bar{F}_{\Gamma_M,\Gamma_M}^{(1)}-2\bar{F}_{\Gamma_M,4}^{(1)})\Big]+\frac{\bar{\epsilon}}{\epsilon+\bar{\epsilon}}\Gamma_M(\bar{F}_{\Gamma_M,\Gamma_M}^{(1)}-2\bar{F}_{\Gamma_M,4}^{(1)})\Bigg] . \label{eq:BetaFunctionFinal}
	\end{align}
	Here, we introduced $F_{\mathcal{O},i}^{(1)}=\partial_{\mathcal{O}}A_i^{(1)}$, $\bar{F}_{\mathcal{O},i}^{(1)}=\partial_{\mathcal{O}}\bar{A}_i^{(1)}$ where $A_i^{(1)}$ and $\bar{A}^{(1)}_i$ are coefficients of a term with the $\frac{1}{\epsilon}$-pole and a term with the $\frac{1}{\epsilon+\bar{\epsilon}}$-pole. respectively, in the counterterms. Since the $\frac{1}{\epsilon+\bar{\epsilon}}$-pole comes out only when random boson mass vertices are involved in Feynman diagrams, $\bar{A}^{(1)}_i$ always appears with the coupling parameter $\Gamma_M$ while $A^{(1)}_i$ dose not.

	\subsubsection{Callan-Symmanzik equation} \label{Appendix:CSEqs}
	
	Correlation functions in terms of bare and renormalized fermion and boson fields are defined by
	\begin{align}
		&\langle \Psi_B(k_{B,1})\cdots \Psi_B(k_{B,n_f})\bar{\Psi}_{B}(k_{B,n_f+1})\cdots \bar{\Psi}_{B}(k_{B,2n_f})\Phi_B(q_{B,1})\cdots \Phi_B(q_{B,n_b})\rangle\nonumber \\
		&=G_B^{(2n_f,n_b)}(k_{B,i},q_{B,i};v_B,c_B,c_{\perp,B},g_B,u_{1,B},u_{2,B},\{\bar{\Gamma}_{i,B}\},\Gamma_{M,B})\delta^{(d+1)}\Big(\sum_{i=1}^{n_f}(k_{B,i}-k_{B,i+n_f})+\sum_{j=1}^{n_b}q_{B,j}\Big),\\
		&\langle \Psi(k_{1})\cdots \Psi(k_{n_f})\bar{\Psi}(k_{n_f+1})\cdots \bar{\Psi}(k_{2n_f})\Phi(q_{1})\cdots \Phi(q_{n_b})\rangle\nonumber \\
		&=G^{(2n_f,n_b)}(k_{i},q_{i};v,c,c_\perp,g,u_{1},u_{2},\{\bar{\Gamma}_i\},\Gamma_M)\delta^{(d+1)}\Big(\sum_{i=1}^{n_f}(k_{i}-k_{i+f})+\sum_{j=1}^{n_b}q_{j}\Big) ,
	\end{align}
	where $2n_f$, $n_b$ are numbers of fermion fields and boson fields, respectively.
	
	Both Green's functions ($G_{B}^{(2n_f,n_b)}$ and $G^{(2n_f,n_b)}$) are related as follows
	\begin{gather}
		G_B^{(2n_f,n_b)}(k_{B,i},q_{B,i};v_B,c_B,c_{\perp,B},g_B,u_{1,B},u_{2,B},\{\bar{\Gamma}_{i,B}\},\Gamma_{M,B}) \nonumber\\
		= Z_0 Z_{\perp}^{d-2}Z_{\psi}^{n_f}Z_\phi^{\frac{n_b}{2}}G^{(2n_f,n_b)}(k_{i},q_{i};v,c,c_\perp,g,u_1,u_2,\{\bar{\Gamma}_i\},\Gamma_M;\mu)\label{eq:GreensFunctionRelation1}
	\end{gather}
	where $\delta(f(x))=\frac{\delta(x-x_0)}{f'(x)}$ has been used.
	
	Taking into account the classical scaling (engineering dimension) explicitly as follows: $\mathbf{K}=\mu\tilde{\mathbf{K}},\; k_{d-1}=\mu\tilde{k}_{d-1},\; k_{d}=\mu\tilde{k}_d,\; \Psi=\mu^{-\frac{d+2}{2}}\tilde{\Psi},\; \Phi=\mu^{-\frac{d+3}{2}}\tilde{\Phi}$,
	where $\mu$ is an energy scale for the RG transformation, we obtain
	\begin{align}
		&G_B^{(2n_f,n_b)}(k_{B,i},q_{B,i};v_B,c_B,c_{\perp,B},g_B,u_{1,B},u_{2,B},\{\bar{\Gamma}_{i,B}\},\Gamma_{M,B})\nonumber\\
		&=Z_0 Z_{\perp}^{d-2}Z_{\psi}^{n_f}Z_\phi^{\frac{n_b}{2}}\mu^{-n_f(d+2)-n_b\frac{d+3}{2}+d+1} \tilde{G}^{(2n_f,n_b)}(\tilde{k}_{i},\tilde{q}_{i};v,c,c_\perp,g,u_1,u_2,\{\bar{\Gamma}_i\},\Gamma_M;\mu) . \label{eq:GreensFunctionRelationFinal}
	\end{align}
	Here, we used
	\begin{gather*}
		G^{(2n_f,n_b)}(k_{i},q_{i};v,c,c_\perp,g,u_{1},u_{2},\{\Gamma_i\},\Gamma_M) = \mu^{-n_f(d+2)-n_b\frac{d+3}{2}+d+1}\tilde{G}^{(2n_f,n_b)}(\tilde{k}_{i},\tilde{q}_{i};v,c,c_\perp,g,u_1,u_2,\{\bar{\Gamma}_i\},\Gamma_M;\mu) .
	\end{gather*}
	
	Resorting to Eq. \eqref{eq:GreensFunctionRelationFinal} and considering that the bare Green's function is independent from the energy scale $\mu$; $\frac{d G_B^{(2n_f,n_b)}}{d\ln \mu}=0$, we obtain the Callan-Symanzik equation of a Green's function as follows
	\begin{align}\label{eq:CZequation}
		&\Big[\sum_{i=1}^{n_f}\Big(z_\tau \tilde{k}_{0}\partial_{\tilde{k}_0}+z_\perp\tilde{\mathbf{K}}_{\perp,i}\cdot\nabla_{\tilde{\mathbf{K}}_{\perp,i}}+\tilde{k}_{d-1}\partial_{\tilde{k}_{d-1}}+\tilde{k}_{d}\partial_{\tilde{k}_d}\Big)+\sum_{i=1}^{n_b}\Big(z_\tau \tilde{q}_{0}\partial_{\tilde{q}_0}+z_\perp\tilde{\mathbf{Q}}_{\perp,i}\cdot\nabla_{\tilde{\mathbf{Q}}_{\perp,i}}+\tilde{q}_{d-1}\partial_{\tilde{q}_{d-1}}+\tilde{q}_{d}\partial_{\tilde{q}_d}\Big)\nonumber\\
		&-\beta_{v}\partial_v-\beta_c\partial_c-\beta_{c_\perp}\partial_{c_\perp}-\beta_g\partial_g-\beta_{u_{1}}\partial_{u_{1}}-\beta_{u_{2}}\partial_{u_{2}}-\sum_{\tilde{\Gamma}_i}\beta_{\bar{\Gamma}_{i}}\partial_{\bar{\Gamma}_i}-\beta_{\Gamma_M}\partial_{\Gamma_M}+2n_f\Big(\frac{d+2}{2}-\eta_\psi\Big)+n_b\Big(\frac{d+3}{2}-\eta_\phi\Big)\nonumber \\
		&-(z_\tau+z_\perp(d-2)+2)\Big] \tilde{G}_r^{(2n_f,n_b)}=0 ,
	\end{align}
	where we considered Eqs. \eqref{eq:DefinitionAnomalousDimension}, \eqref{eq:DefinitionBetaFunctions}, and following equations
	\begin{gather*}
		\frac{dk_{b,0}}{d\ln\mu}=0 \rightarrow \frac{d\tilde{k}_{0}}{d\ln\mu}=-\Big(1+\frac{d\ln Z_0}{d\ln\mu}\Big)\equiv -z_{\tau}\tilde{k}_{0}\label{eq:betafunctionsfork1},\;\; \frac{d\mathbf{K}_{b,\perp}}{d\ln\mu}=0 \rightarrow \frac{d\tilde{\mathbf{K}}_{\perp}}{d\ln \mu}=-\Big(1+\frac{d\ln Z_\perp}{d\ln \mu}\Big)\equiv -z_\perp\tilde{\mathbf{K}}_\perp\\
		\frac{dk_{b,d-1}}{d\ln \mu}=0 \rightarrow \frac{d\tilde{k}_{d-1}}{d\ln \mu}=-\tilde{k}_{d-1},\;\; \frac{dk_{b,d}}{d\ln \mu}=0 \rightarrow \frac{d\tilde{k}_{d}}{d\ln \mu}=-\tilde{k}_{d}\label{eq:betafunctionsfork2}.
	\end{gather*}
	
	\section{Proof of a new expansion parameter ($\bar{\Gamma}_i=\Gamma_i\Lambda_{FS}$)} \label{Appendix:ProofOfNewExpansionParameter}
	
	We prove that $\bar{\Gamma}_i(=\Gamma_i\Lambda_{FS})$ is an expansion parameter for a random charge potential vertex rather than $\Gamma_i$ in all loops giving a log-divergence or the $\frac{1}{\epsilon} \& \frac{1}{\epsilon+\bar{\epsilon}}$-pole. Suppose an arbitrary Feynman diagram. Then, we obtain the following identities from Euler's formula:
	
	\begin{gather}
		V-E+L=1,\label{eq:proof1}\\
		V=V_g+V_{\Gamma}+V_{u}+\tilde{V}_{\Gamma_M}(2V_{\Gamma_{M}}=\tilde{V}_{\Gamma_M}),\;\; E=E_F+E_B+E_M,\;\; L=L_{F}+L_{BF}+L_{B}+L_{M} .
	\end{gather}
	Here, $V$, $E$, and $L$ are the total number of vertices, propagators, and loops, respectively. $V_g$, $V_\Gamma$, $V_{u}$, and $\tilde{V}_{\Gamma_M}$ are the number of the Yukawa vertices, random charge potential vertices, boson self-interaction vertices, and boson-mass disorder potential vertices, respectively, before disorder averaging. Note that there is always an even number of $\tilde{V}_{\Gamma_M}$ after disorder averaging. $E_F$, $E_B$, and $E_M$ are the number of fermion propagators, boson propagators, and mass disorder potentials (dotted lines in the Feynman diagram representation), respectively. $L_F$, $L_B$, $L_{BF}$, and $L_{M}$ are the number of loops involving only fermion propagators, number of loops involving only boson propagators, number of loops involving both fermion and boson propagators, and number of loops involving mass-disorder potential dotted lines.
	
	We can derive additional equations relating numbers of external lines, vertices, loops, and propagators as follows:
	\begin{gather}
		2V_g+4V_\Gamma=2E_F+N_F,\;\; V_g+4V_{u}+2\tilde{V}_{\Gamma_M}=2E_B+N_B,\;\; \tilde{V}_{\Gamma_M}=2E_{M}=2V_{\Gamma_M} ,
	\end{gather}
	where $N_F$ and $N_B$ are numbers of fermion and boson external lines, respectively. For Feynman diagrams to give the log-divergence, an order of momentum variables of a denominator and a numerator should be the same. This results in the following additional identity
	\begin{gather}
		(d-1)L_F+(d+1)(L_{BF}+L_{B})+dL_{M}-E_F-2E_B+\alpha E_{M}=P_{ex} , \label{eq:proof4}
	\end{gather}
	where $d$ is a spatial dimension, and $P_{ex}$ is an order of an external momentum. In this identity, we used the fact that loops involving only fermion propagators are related to random charge potential vertices, and only momentums perpendicular to the Fermi surface (line) are integral variables giving the log-divergence. On the other hand, the integration of the momentum parallel to the Fermi surface gives a $\Lambda_{FS}$ factor as discussed in the main text. Also, loops involving mass-disorder lines do not contain any frequency integrals due to the nature of the quenched disorder. However, all momentums and the frequency should be considered for all loops involving both fermion and boson propagators except for the $L_M$ cases. As a result, the coefficient of $L_F$ and $L_M$ is given by $(d-1)$ and $d$, respectively, while coefficients of $L_{BF}$ and $L_{B}$ are given by $(d+1)$.
	%
	%
	For the Yukawa vertex and the random charge potential vertices, $P_{ex}$ is given by zero. It is given by one, two, and $\alpha$ for the fermion self-energy, the boson self-energy, and the mass-random charge potential vertices, respectively.
	
	To prove that $\bar{\Gamma}_i$ is an expansion parameter for all loops, we need to show a relation between the number of $\Lambda_{FS}$ ($= L_{F}$) and the number of $\Gamma_i$ ($= V_{\Gamma}$), based on the above identities Eqs. \eqref{eq:proof1} $\sim$ \eqref{eq:proof4}. Dimension $d$ is set to be 3 and $\alpha$ is set to be 1. From Eqs. \eqref{eq:proof1} $\sim$ \eqref{eq:proof4}, we obtain
	\begin{gather}
		4(L_F-V_\Gamma)+2(L_M-V_{\Gamma_M})+3N_F+2N_B+2P_{ex}-8=0 .
	\end{gather}
	
	Using this identity, we obtain relations between $L_F$ and $V_{\Gamma}$ for the fermion self-energy (FS), the boson self-energy (BS), the Yukawa interaction vertex (YIV), the boson self-interaction vertex (BSIV), the random charge potential vertex (RCPV), and the random boson mass-random charge potential vertex (RBMV), given in Table \ref{Table:SumUp}:
	\begin{table}[h]
		\begin{tabular}{|c|c|c|c||c|c|c|}
			\hline
			& $N_F$ & $N_B$ & $P_{ex}$ & Relation & $L_F$ & $L_M$\\
			\hhline{|=|=|=|=||=|=l=l}
			FS & 2& 0& 1& $2(L_F-V_\Gamma)+L_M-V_{\Gamma_M}=0$ & $V_\Gamma$ & $V_{\Gamma_M}$ \\
			\hline
			BS & 0&2&2&$2(L_F-V_\Gamma)+L_M-V_{\Gamma_M}=0$ & $V_{\Gamma}$ & $V_{\Gamma_M}$ \\
			\hline
			YIV & 2 & 1 & 0 & $2(L_F-V_\Gamma)+L_M-V_{\Gamma_M}=0$ & $V_{\Gamma}$ & $V_{\Gamma_M}$ \\
			\hline
			BSIV & 0 & 4 & 0 & $2(L_F-V_\Gamma)+L_M-V_{\Gamma_M}=0$ & $V_{\Gamma}$ & $V_{\Gamma_M}$\\
			\hline
			RCPV& 4 & 0 & 0 & $2(L_F-V_\Gamma+1)+L_M-V_{\Gamma_M}=0$ & $V_{\Gamma}-1$ & $V_{\Gamma_M}$\\
			\hline
			RBMV & 0 & 4 & 1 & $2(L_F-V_\Gamma)+L_M-V_{\Gamma_M}+1=0$ & $V_{\Gamma}$ & $V_{\Gamma_M}-1$\\
			\hline
		\end{tabular}
		\caption{Summary of relations between number of loops ($L_F$ and $L_M$) and number of vertices ($V_\Gamma$ and $V_{\Gamma_M}$).} \label{Table:SumUp}
	\end{table}
	
	Since the fermion part and the boson part do not mix, we can consider them separately to satisfy these equations. Except for the random charge potential vertices (RCPV), the number of loops involving only fermion propagators ($L_F$) and the number of the random charge potential vertices is the same. This means that $\bar{\Gamma}(=\Gamma\Lambda_{FS})$ is an expansion parameter. In the case of the random charge potential vertices (RCPV), $L_F$ is less than $V_g$. This is because the random charge potential vertex in the effective action is written with $\Gamma$ rather than $\bar{\Gamma}$. Therefore, the renormalized random charge potential vertex parameter is given by $
		\Gamma_{r}\sim\Gamma_b+\Gamma_b(\Gamma_b\Lambda_{FS})^{V_\Gamma-1}+\cdots .
	$
	Multiplying $\Lambda_{FS}$ to both left and right sides of this equation gives
	$
		\bar{\Gamma}_r\sim\bar{\Gamma}_b+(\bar{\Gamma}_b)^{V_\Gamma}+\cdots .
$
	As a result, the expansion parameter for all Feynman diagrams giving the log-divergence is given by $\bar{\Gamma}$ $(=\Gamma\Lambda_{FS})$ rather than $\Gamma$.

	\section{Calculations of one-loop Feynman diagrams} \label{Appendix:CalOfOneLoopFeynmanDiagrams}
	
	\subsection{Useful identities}
	
	We change variables $k_{d-1}$ and $k_d$ to $\epsilon_{n}$ and $\epsilon_{m}$, where $\epsilon_{1}(k)$, $\epsilon_{2}(k)$, $\epsilon_{3}(k)$, and $\epsilon_{4}(k)$ are given in the main text, and $\epsilon(k)=vk_{d-1}+k_d$ and $\epsilon_{||}(k)=k_{d-1}-vk_d$ are used for the case of $n=m$. Measure factors are given by
	\begin{gather*}
		(n=m):\; dk_{d-1}dk_d=\frac{1}{1+v^2}d\epsilon d\epsilon_{||},\;\; 
		(n\neq m):\; d_{d-1}dk_d=\frac{1}{f_{nm}(v)}d\epsilon_n d\epsilon_m ,
	\end{gather*}
		where $f_{12}(v)=f_{21}(v)=f_{34}(v)=f_{43}(v)=1+v^2$, $f_{13}(v)=f_{31}(v)=f_{24}(v)=f_{42}(v)=2v$ and $f_{14}(v)=f_{41}(v)=f_{23}(v)=f_{32}(v)=1-v^2$.

	For calculations of one loops involving the Yukawa interaction vertex, we use the following identities
	\begin{gather}
		\sum_{i=1}^{N_c^2-1} \tau_{\alpha\beta}^i\tau_{\gamma\eta}^i=2\Big(\delta_{\alpha\eta}\delta_{\beta\gamma}-\frac{1}{N_c}\delta_{\alpha\beta}\delta_{\gamma\eta}\Big),\;\; \sum_{i=1}^{N_c^2-1}Tr[\tau^i\tau^i\tau^j\tau^j]=4\frac{N_c^2-1}{N_c},\; Tr[\tau^i\tau^i\tau^i\tau^i]=\frac{4}{N_c}.\label{eq:SpinorMatrixIdentity}
	\end{gather}
	
	In the one-loop calculation, we consider the following identity with the Feynman parametrization
	\begin{gather}
		\int \frac{d^{d-1}\mathbf{Q}}{(2\pi)^{d-1}}\int \frac{d\epsilon_1d\epsilon_2}{(2\pi)^2}\frac{a|\mathbf{Q}|^2\mathcal{M}_1+b\epsilon_1\epsilon_2 \mathcal{M}_2}{[\alpha |\mathbf{Q}|^2+\beta \epsilon_1^2+\gamma \epsilon_2^2+\eta \epsilon_1\epsilon_2]^3}=\frac{1}{(4\pi)^2}\alpha^{-\frac{d-1}{2}}[4\beta \gamma-\eta^2]^{-1/2}\frac{1}{\epsilon}\Big[\frac{2a}{\alpha}\mathcal{M}_1-\frac{2b\eta}{4\beta\gamma-\eta^2}\mathcal{M}_2\Big],\\
		\frac{1}{A_1^{\alpha_1}\cdots A_n^{\alpha_n}}=\frac{\Gamma(\alpha_1+\cdots+\alpha_n)}{\Gamma(\alpha_1)\cdots\Gamma(\alpha_n)}\int_0^1 du_1\cdots \int_0^1 du_n \frac{\delta (1-\sum_{k=1}^n u_k)u_1^{\alpha_1-1}\cdots u_n^{\alpha_n-1}}{(\sum_{k=1}^n u_kA_k)^{\sum_{k=1}^n \alpha_k}}.
	\end{gather}

	\subsection{One-loop Fermion self-energy corrections}
	
	\begin{figure}[h]
		\begin{subfigure}{0.35\textwidth}
			\begin{tikzpicture}[baseline=-0.1cm,scale=0.9]
				\begin{feynhand}
					\vertex (a) at (0,0) {$n,i,\sigma$}; \vertex (b) at (1,0); \vertex (c) at (3,0); \vertex (d) at (4,0) {$n,i,\sigma$};
					\propagator[fer] (a) to (b); \propagator[fer] (b) to (c); \propagator[fer] (c) to (d);
					\propagator[boson] (b) to [half left, looseness=1.5](c);
					\node at (1,-0.3) {$\mathcal{M}$}; \node at (3,-0.3) {$\mathcal{M}'$};
				\end{feynhand}
			\end{tikzpicture}
			\caption{1-loop Fermion self-energy from the Yukawa vertex}\label{selfEnergyYukawa}
		\end{subfigure}
		\begin{subfigure}{0.35\textwidth}
			\begin{tikzpicture}[baseline=-0.1cm,scale=0.9]
				\begin{feynhand}
					\vertex (a) at (0,0) {$n,i,\sigma$}; \vertex (b) at (1,0); \vertex (c) at (3,0); \vertex (d) at (4,0) {$n,i,\sigma$};
					\propagator[fer] (a) to (b); \propagator[fer] (b) to [edge label=$k$] (c); \propagator[fer] (c) to (d);
					\propagator[sca] (b) to [half left, looseness=1.5](c);
					\node at (1,-0.3) {$\mathcal{M}$}; \node at (3,-0.3) {$\mathcal{M}'$};
				\end{feynhand}
			\end{tikzpicture}
			\caption{1-loop Fermion self-energy from the random charge potential vertex}\label{selfEnergyDisorder}
		\end{subfigure}
		\caption{Two 1-loop Fermion self-energy Feynman diagrams; $n$, $i$, and $\sigma$ denote the replica index, the hot spot index, and the spin, respectively. $\mathcal{M}$ and $\mathcal{M}'$ are 2$\times$2-matrices given in the Supplementary Material\cite{Note1}.} \label{FSselfEnergy}
	\end{figure}
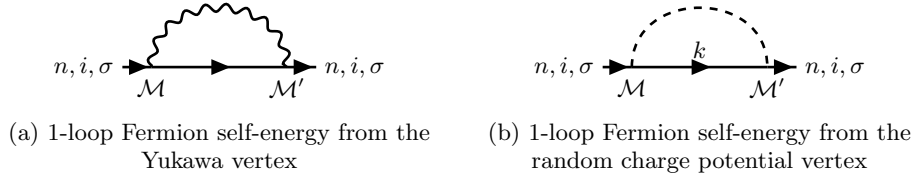
	
	Feynman diagrams of one loop fermion self-energy corrections are given in Fig. \ref{FSselfEnergy}. Based on the Feynman rules presented in Appendix \ref{Appendix:CalOfOneLoopFeynmanDiagrams}, we calculate the self-energy diagrams \ref{FSselfEnergy}. First, we calculate the diagram \ref{selfEnergyYukawa} as follows
	\begin{align}
		\Sigma_{f;n,i,\sigma}^{Yukawa}(p)&=-\frac{g^2}{N_f}\sum_{a=1}^{N_c^2-1}\sum_{\sigma'}(\tau^a_{\sigma,\sigma'}\tau^a_{\sigma',\sigma})\int\frac{d\Omega}{2\pi}\int\frac{d^{d-2}\mathbf{Q}_\perp}{(2\pi)^{d-2}}\int\frac{d^2\vec{q}}{(2\pi)^2}\gamma_{d-1}G_{f,\bar{n}}(\omega+\Omega,\mathbf{P}_\perp+\mathbf{Q}_\perp,\vec{p}+\vec{q})\gamma_{d-1} G_b(\Omega,\mathbf{Q}_\perp,\vec{q})\nonumber\\
		&=i\frac{N_c^2-1}{4\pi^2N_cN_f}\frac{g^2}{c}\frac{1}{\epsilon}\Big[-\omega \gamma_0 h_1(c,c_\perp,v)-\mathbf{P}_\perp\cdot\mathbf{\Gamma}_\perp h_2(c,c_\perp,v)+\epsilon_{\bar{n}}(p)\gamma_{d-1}h_3(c,c_\perp,v)\Big] ,
	\end{align}
	where
	\begin{gather}
		h_1(c,c_\perp,v)=\int_0^1dx\sqrt{\frac{x}{\Big(1-x+xc_\perp^2\Big)\Big((1+v^2)(1-x)+xc^2\Big)}}\\
		h_2(c,c_\perp,v)=c_\perp^2 \int_0^1 dx \sqrt{\frac{x}{\Big(1-x+xc_\perp^2\Big)^3\Big((1+v^2)(1-x)+xc^2\Big)}}\\
		h_3(c,c_\perp,v)=c^2 \int_0^1 dx \sqrt{\frac{x}{\Big(1-x+xc_\perp^2\Big)\Big((1+v^2)(1-x)+xc^2\Big)^{3}}} .
	\end{gather}
	
	Next, we consider the diagram \ref{selfEnergyDisorder} 
	\begin{align*}
		\Sigma^{dis}_{f;n,i,\sigma}(\omega)&=\frac{\Gamma_{i}\mu^{-1+\epsilon}}{N_f} \int \frac{d^{d-2}\mathbf{K}_\perp}{(2\pi)^{d-2}}\int \frac{d^2k}{(2\pi)^2} \mathcal{M}G_n(\omega,k)\mathcal{M}'=-i\frac{\Gamma_{i}\mu^{-1+\epsilon}}{N_f(1+v^2)} \int \frac{d^{d-2}\mathbf{K}_\perp}{(2\pi)^{d-2}}\int \frac{d\epsilon d\epsilon_{||}}{(2\pi)^2} \mathcal{M}\frac{\gamma_0\omega+\mathbf{\Gamma}_{\perp}\cdot\mathbf{K}_\perp+\gamma_{d-1}\epsilon}{\omega^2+|\mathbf{K}_\perp|^2+\epsilon}\mathcal{M}'\\
		&=-i\frac{\Gamma_i\mu^{-1+\epsilon}}{4\pi^2N_f(1+v^2)}\omega\frac{1}{\epsilon} \mathcal{M}\gamma_0\mathcal{M}' \int_{-\Lambda_{FS}\mu}^{\Lambda_{FS}\mu}d\epsilon_{||} =-i\frac{\bar{\Gamma}_i}{2\pi^2(1+v^2)}(\mathcal{M}\gamma_0\mathcal{M}')\omega\frac{1}{\epsilon}
	\end{align*}
	In this calculation the cut off $\Lambda_{FS}$ depicted in Fig. \ref{fig:SizeOfHotSpot} was introduced.
	
	Based on our classification scheme, we can identity that random charge potential vertices $\Gamma_0$, $\Gamma_{\theta_1}^e$, $\Gamma_{\theta_2}^e$, $\Gamma_{\pi/2}^e$, $\Gamma_{\pi-\theta_1}^e$, $\Gamma_{\pi-\theta_2}^e$, and $\Delta_\pi$ contribute to the self-energy. The final result is given as follows
	\begin{align*}
		\Sigma^{dis}_{f}(\omega)=-i\frac{\omega\gamma_0}{2\pi^2N_f(1+v^2)}\frac{1}{\epsilon}\Big[\bar{\Gamma}_0+\bar{\Gamma}_{\theta_1}^e +\bar{\Gamma}_{\theta_2}^e+2\bar{\Gamma}_{\pi/2}^e+\bar{\Gamma}_{\pi-\theta_2}^e+\bar{\Gamma}_{\pi-\theta_1}^e+\bar{\Delta}_\pi\Big] .
	\end{align*}
	
	\subsection{One-loop random charge potential vertex corrections}
	
	There are total six different one-loop diagrams shown in Fig. \ref{fig:SDWDisorderOneLoopDiagrams} for the random charge potential vertex. First four diagrams consist of only random charge potential vertices while the remaining two diagrams consist of both the random charge potential vertex and the Yukawa interaction vertex.
	
	\subsubsection{One-loop random charge potential vertex corrections involving only random charge potential vertices} \label{sec:OneLoopDisorderOnly}

	\begin{figure}[h]
		\begin{subfigure}[b]{0.2\textwidth}
			\begin{tikzpicture}[baseline=-0.1cm,scale=0.7]
				\begin{feynhand}
					\vertex (a1) at (0,0) {$a$}; \vertex (a2) at (0,1) ; \vertex (a3) at (0,2); \vertex (a4) at (0,3) {$a$};
					\vertex (b1) at (1.5,0) {$b$}; \vertex (b2) at (1.5,1); \vertex (b3) at (1.5,2); \vertex (b4) at (1.5,3) {$b$};
					\propag[fer] (a1) to (a2); \propagator[fer] (a2) to [edge label=$n$] (a3); \propag[fer] (a3) to (a4);
					\propag[fer] (b1) to (b2); \propagator[fer] (b2) to  [edge label=$m$] (b3); \propag[fer] (b3) to (b4);
					\propag[sca] (a2) to (b2); \propag[sca] (a3) to (b3);
					\node (a2l) at (-0.5,1) {$\mathcal{M}^j$}; \node (a3l) at (-0.5,2) {$\mathcal{M}^i$};
					\node (b2l) at (2,1)  {$\mathcal{\tilde{M}}^j$}; \node (b3l) at (2,2) {$\mathcal{\tilde{M}}^i$};
				\end{feynhand}
			\end{tikzpicture}
			\caption{particle-particle diagram}\label{ppdiagram}
		\end{subfigure}
		~
		\begin{subfigure}[b]{0.2\textwidth}
			\begin{tikzpicture}[baseline=-0.1cm,scale=0.7]
				\begin{feynhand}
					\vertex (a1) at (0,0) {$a$}; \vertex (a2) at (0,1); \vertex (a3) at (0,2); \vertex (a4) at (0,3) {$a$};
					\vertex (b1) at (1.5,0) {$b$}; \vertex (b2) at (1.5,1); \vertex (b3) at (1.5,2); \vertex (b4) at (1.5,3) {$b$};
					\propag[fer] (a1) to (a2); \propagator[fer] (a2) to [edge label=$n$](a3); \propag[fer] (a3) to (a4);
					\propag[fer] (b1) to (b2); \propagator[fer] (b2) to [edge label=$m$](b3); \propag[fer] (b3) to (b4);
					\propag[sca] (a2) to (b3); \propag[sca] (a3) to (b2);
					\node (a2l) at (-0.5,1) {$\mathcal{M}^j$}; \node (a3l) at (-0.5,2) {$\mathcal{M}^i$};
					\node (b2l) at (2,1)  {$\mathcal{\tilde{M}}^i$}; \node (b3l) at (2,2) {$\mathcal{\tilde{M}}^j$};
					
				\end{feynhand}
			\end{tikzpicture}
			\caption{particle-hole diagram}
		\end{subfigure}
		~
		\begin{subfigure}[b]{0.2\textwidth}
			\begin{tikzpicture}[baseline=-0.1cm,scale=0.7]
				\begin{feynhand}
					\vertex (a1) at (0,0) {$a$}; \vertex (a2) at (0,3/4); \vertex (a3) at (0,6/4); \vertex (a4) at (0, 9/4); \vertex(a5) at (0, 3) {$a$};
					\vertex (b1) at (1.5,0) {$b$}; \vertex (b2) at (1.5,3/4); \vertex (b3) at (1.5,6/4); \vertex (b4) at (1.5,9/4); \vertex(b5) at (1.5, 3) {$b$};
					\propag[fer] (a1) to (a2); \propagator[fer] (a2) to [edge label=$n$](a3); \propagator[fer] (a3) to [edge label=$m$] (a4); \propag[fer] (a4) to (a5);
					\propag[fer] (b1) to (b2); \propag[fer] (b2) to (b3); \propag[fer] (b3) to (b4); \propag[fer] (b4) to (b5);
					\propag[sca] (a3) to (b3); \propag[sca, looseness=1.5] (a2) to [out=180, in=180](a4);
					\node (a2l) at (0.3,3/4) {$\mathcal{\tilde{M}}^i$}; \node (a4l) at (0.3,9/4) {$\mathcal{M}^i$};
					\node (a3l) at (-0.3,6/4) {$\mathcal{M}^j$}; \node (b3l) at (1.8,6/4) {$\mathcal{\tilde{M}}^j$};
				\end{feynhand}
			\end{tikzpicture}
			\caption{left-loop diagram}
		\end{subfigure}
		~
		\begin{subfigure}[b]{0.2\textwidth}
			\begin{tikzpicture}[baseline=-0.1cm,scale=0.7]
				\begin{feynhand}
					\vertex (a1) at (0,0) {$a$}; \vertex (a2) at (0,3/4); \vertex (a3) at (0,6/4); \vertex (a4) at (0, 9/4); \vertex(a5) at (0, 3) {$a$};
					\vertex (b1) at (1.5,0) {$b$}; \vertex (b2) at (1.5,3/4); \vertex (b3) at (1.5,6/4); \vertex (b4) at (1.5,9/4); \vertex(b5) at (1.5, 3) {$b$};
					\propag[fer] (a1) to (a2); \propag[fer] (a2) to (a3); \propag[fer] (a3) to (a4); \propag[fer] (a4) to (a5);
					\propag[fer] (b1) to (b2); \propagator[fer] (b2) to [edge label=$n$](b3); \propagator[fer] (b3) to [edge label=$m$](b4); \propag[fer] (b4) to (b5);
					\propag[sca] (a3) to (b3); \propag[sca, looseness=1.5] (b2) to [out=0, in=0](b4);
					\node (a2l) at (1.5-0.3,3/4) {$\mathcal{\tilde{M}}^i$}; \node (a4l) at (1.5-0.3,9/4) {$\mathcal{M}^i$};
					\node (a3l) at (1.5+0.3,6/4) {$\mathcal{M}^j$}; \node (b3l) at (-0.3,6/4) {$\mathcal{\tilde{M}}^j$};
				\end{feynhand}
			\end{tikzpicture}
			\caption{right-loop diagram}\
		\end{subfigure}
		\caption{Four one-loop diagrams involving only random charge potential vertices for the random charge potential vertexes} \label{fig:DisorderInteractionFeynman}
	\end{figure}
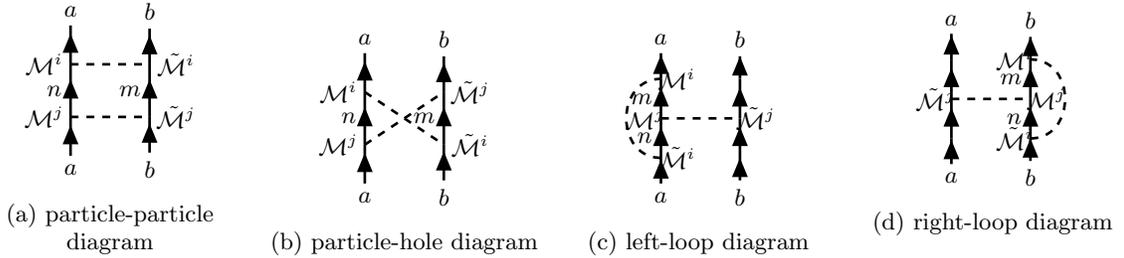
	
	First, let us consider the four one-loop Feynman diagrams involving only random charge potential vertices given in Fig. \ref{fig:DisorderInteractionFeynman}. Since calculations of the following one loop diagrams are almost same, we show details only for the particle-particle diagram (Fig. \ref{ppdiagram}), given by
	\begin{align}
		(pp)&=\frac{\Gamma_{i}\Gamma_{j}\mu^{2(-1+\epsilon)}}{N_f^2}\int \frac{d^{d-2}\mathbf{K}_\perp}{(2\pi)^{d-2}}\int\frac{d^2k}{(2\pi)^2}[\mathcal{M}^iG_n^a(0,k)\mathcal{M}^j][\tilde{\mathcal{M}}^iG_m^b(0,-k)\tilde{\mathcal{M}}^j]\nonumber\\
		&=\frac{\Gamma_{i}\Gamma_{j}\mu^{2(-1+\epsilon)}}{N_f^2}\int \frac{d^{d-2}\mathbf{K}_\perp}{(2\pi)^{d-2}}\frac{1}{f_{nm}(v)}\int \frac{d\epsilon_nd\epsilon_m}{(2\pi)^2}\frac{[\mathcal{M}^i(\mathbf{\Gamma}_\perp\cdot\mathbf{K}_\perp+\gamma_{d-1}\epsilon_n)\mathcal{M}^j][\tilde{\mathcal{M}}^i(\mathbf{\Gamma}_\perp\cdot\mathbf{K}_\perp+\gamma_{d-1}\epsilon_m)\tilde{\mathcal{M}}^j]}{[|\mathbf{K}_\perp|^2+\epsilon^2_n][|\mathbf{K}_\perp|^2+\epsilon^2_m]}\nonumber\\
		&=\Big\{\begin{array}{cc}
			\text{when }n=m:& \frac{\Gamma_i\Gamma_j\Lambda_{FS}\mu^{-1+2\epsilon}}{4\pi^2N_f^2(1+v^2)}\frac{1}{\epsilon}\Big[[\mathcal{M}^i\Gamma_{\perp,\mu}\mathcal{M}^j][\tilde{\mathcal{M}}^i\Gamma_{\perp,\mu}\tilde{\mathcal{M}}^j]+[\mathcal{M}^i\gamma_{d-1}\mathcal{M}^j][\tilde{\mathcal{M}}^i\gamma_{d-1}\tilde{\mathcal{M}}^j]\Big]\nonumber\\
			\text{when }n\neq m:& \frac{3}{4(4\pi)^{3/2}N_f^2}\frac{\Gamma_i\Gamma_j}{f_{nm}(v)}\Gamma\Big(\frac{\epsilon-1}{2}\Big)[\mathcal{M}^i\Gamma_{\perp,\mu}\mathcal{M}^j][\tilde{\mathcal{M}}^i\Gamma_{\perp,\mu}\tilde{\mathcal{M}}^j]\rightarrow \text{ no $\epsilon$ pole \& no physical term}
		\end{array}
		\\
		&=\delta_{nm}\frac{\mu^{-1+2\epsilon}\Gamma_i\Gamma_j\Lambda_{FS}}{4\pi^2N_f^2(1+v^2)}\frac{1}{\epsilon}\Big[[\mathcal{M}^i\Gamma_{\perp,\mu}\mathcal{M}^j][\tilde{\mathcal{M}}^i\Gamma_{\perp,\mu}\tilde{\mathcal{M}}^j]+[\mathcal{M}^i\gamma_{d-1}\mathcal{M}^j][\tilde{\mathcal{M}}^i\gamma_{d-1}\tilde{\mathcal{M}}^j]\Big]\label{eq:oneLoopCalDV}
	\end{align}
	This shows that there is a one-loop correction, regularized as an epsilon pole ($\frac{1}{\epsilon}$) only when $n=m$. However, this is an artifact of the co-dimensional regularization method. If we consider $(n,m)=(1,3),(3,1),(2,4),(4,2)$ cases, $f_{nm}(v)$ is given by $2v$. This means that the one-loop calculation goes to infinity as $v$ goes to zero even though there is no epsilon pole. This is closely related to how the UV infinity is regularized.
	
	To clarify this point, we calculate the same one-loop correction with a different approach in the following way
	\begin{align}
		(pp')&=\frac{\Gamma_i\Gamma_j}{N_f^2}\int \frac{d^{d-2}\mathbf{K}_\perp}{(2\pi)^{d-2}}\int\frac{d^2k}{(2\pi)^2}[\mathcal{M}^iG_1^a(0,k)\mathcal{M}^j][\tilde{\mathcal{M}}^iG_3^b(0,-k)\tilde{\mathcal{M}}^j]\nonumber\\
		&=\frac{\Gamma_i\Gamma_j}{4\pi^2N_f^2}\frac{1}{(4\pi)^{(d-2)/2}}\frac{1}{\Gamma(2)}\int_0^1 dz\int dX\int_{-\Lambda_{FS}}^{\Lambda_{FS}} dY \Big[[\mathcal{M}^i\Gamma_{\perp,\mu}\mathcal{M}^j][\tilde{\mathcal{M}}^i\Gamma_{\perp,\mu}\tilde{\mathcal{M}}^j]F_1(v,X,Y,z)\nonumber\\
		&+[\mathcal{M}^i\gamma_{d-1}\mathcal{M}^j][\tilde{\mathcal{M}}^i\gamma_{d-1}\tilde{\mathcal{M}}^j]F_2(v,X,Y,z)\Big] ,
	\end{align}
	where $X=\epsilon_1(k)=vk_{d-1}+k_d$, $Y=\epsilon_{1,||}(k)=k_{d-1}-vk_d$, 
	$F_1(v,X,Y,z)=\frac{1}{1+v^2}\frac{\Gamma(1-\frac{d-2}{2})}{2}\Big(\frac{1}{\Delta(X,Y,z)}\Big)^{1-\frac{d-2}{2}}$, $ F_2(v,X,Y,z)=\frac{X[2vY-(1-v^2)X]}{(1+v^2)^2}\Gamma(2-\frac{d-2}{2})\Big(\frac{1}{\Delta(X,Y,z)}\Big)^{2-\frac{d-2}{2}}$ and $\Delta(X,Y,z)=X^2(1-z)+\frac{z}{(1+v^2)^2}(2vY-(1-v^2)X)^2$. 
	Here, we calculated the one-loop diagram for the $(n,m)=(1,3)$ case. This can be easily generalized to other cases ($(n,m)=(3,1),(2,4),(4,2)$). In the previous calculations, we changed the variables $(k_{d-1},k_d)$ to $(\epsilon_n,\epsilon_m)$ while we used a new set $(\epsilon_n,\epsilon_{n,||})$ in the above calculation. In principle, there should be no difference between these two calculations. However, we find that there can appear a difference, depending on the UV regularization.
	
	To see this point clearly, we approximate the above result using the Taylor expansion near $v=0$. We find that the integration of $F_1$ and $F_2$ with respect to $z$, $X$, and $Y$ after the Taylor expansion are given as follows
	\begin{align*}
		\int_0^1dz\int dX\int dYF_{1}(v,X,Y,z)=\Lambda_{FS} \sqrt{\pi}\Gamma(\frac{3-d}{2})+(\cdots)\\
		\int_0^1dx\int dX\int dYF_{2}(v,X,Y,z)=-\Lambda_{FS} \sqrt{\pi}\Gamma(\frac{3-d}{2})+(\cdots) .
	\end{align*}
	Higher order terms in $v$ denoted by $(\cdots)$ do not have the $\frac{1}{\epsilon}$ pole. As a result, there is a one loop correction for the $(n,m)=(1,3),(3,1),(2,4),(4,2)$ cases in the $v\rightarrow 0$ limit, given by
	\begin{gather}
		(pp')=\frac{\Gamma_i\Gamma_j\Lambda_{FS}}{4\pi^{2}N_f^2}\frac{1}{\epsilon} \Big[[\mathcal{M}^i\Gamma_{\perp,\mu}\mathcal{M}^j][\tilde{\mathcal{M}}^i\Gamma_{\perp,\mu}\tilde{\mathcal{M}}^j] - [\mathcal{M}^i\gamma_{d-1}\mathcal{M}^j][\tilde{\mathcal{M}}^i\gamma_{d-1}\tilde{\mathcal{M}}^j]\Big] . \label{eq:oneLoopCalDVzerovlimit}
	\end{gather}
	
	Now, we encounter two different results Eq. \eqref{eq:oneLoopCalDV} and Eq. \eqref{eq:oneLoopCalDVzerovlimit} from the same one loop diagram. This can be understood in the Wilsonian Renormalization Group (RG) scheme. In the Wilsonian scheme, fields with high energies are integrated out to give renormalization effects. In the one loop calculation (Eq. \eqref{eq:oneLoopCalDV}), there are two Green's functions involved and they have two different dispersions given by $\epsilon_n(k)$ and $\epsilon_m(-k)$. When evaluating the one-loop diagram in the Wilsonian scheme, only phase spaces of momentum $k$ which satisfy $\Lambda-d\Lambda<|\epsilon_n(k)|,|\epsilon_m(-k)|<\Lambda$ need to be integrated out. Here, $\Lambda$ is a cut-off of energy. Do not confuse this cutoff scale with $\Lambda_{FS}$. In Fig. \ref{fig:PhaseSpaces}, it shows phase spaces of two different dispersions. Only overlapped regions of the two high energy parts (grey colored regimes) of the phase spaces contribute to the one-loop calculation.
	
	Let us consider the $n=m$ case first. High energy parts of two-phase spaces are fully overlapped since two dispersions are identical and the one-loop calculation gives the maximum value in this case. However, there is not much overlap in the case of $n \neq m$ as shown in Fig. \ref{fig:PhaseSpaces}. In this case, the phase-space overlap is strongly governed by an angle between two Fermi lines and also a ratio between $\Lambda$ and $\Lambda_{FS}$. Although we have finite overlapping in the case of $n \neq m$, considering the Wilsonian RG-analysis scheme, the story becomes a little bit tricky in the co-dimensional regularization method. Here, we set the energy cut-off $\Lambda$ to be infinity, where log-divergences are represented by $\frac{1}{\epsilon}$ poles. In the case of $n=m$, the overlapped phase space does not change even in the $\Lambda\rightarrow \infty$ limit. As a result, one-loop results are the same in two different regularization schemes: the finite cut-off regularization of the Wilsonian scheme and the co-dimensional regularization of the high-energy scheme. On the other hand, the overlapped phase space in $(n,m)=(1,3),(3,1),(2,4),(4,2)$ vanishes in the $\Lambda\rightarrow \infty$ limit as illustrated in Fig. \ref{fig:VanishingPhaseSpace}. Therefore, the one-loop calculation based on the co-dimensional regularization does not give a $\frac{1}{\epsilon}$ pole although there is a finite overlapped phase space before taking the $\Lambda\rightarrow \infty$ limit. If we take the $v\rightarrow 0$ limit for $(n,m)=(1,3),(3,1),(2,4),(4,2)$ first before considering the $\Lambda\rightarrow \infty$ limit, there is always a finite overlapped region of the phase space even in the $\Lambda\rightarrow \infty$ limit as illustrated in Fig. \ref{fig:zeroVPhaseSpace}.
	
	\begin{figure}
		\begin{subfigure}{0.3\textwidth}
			\includegraphics[scale=0.08]{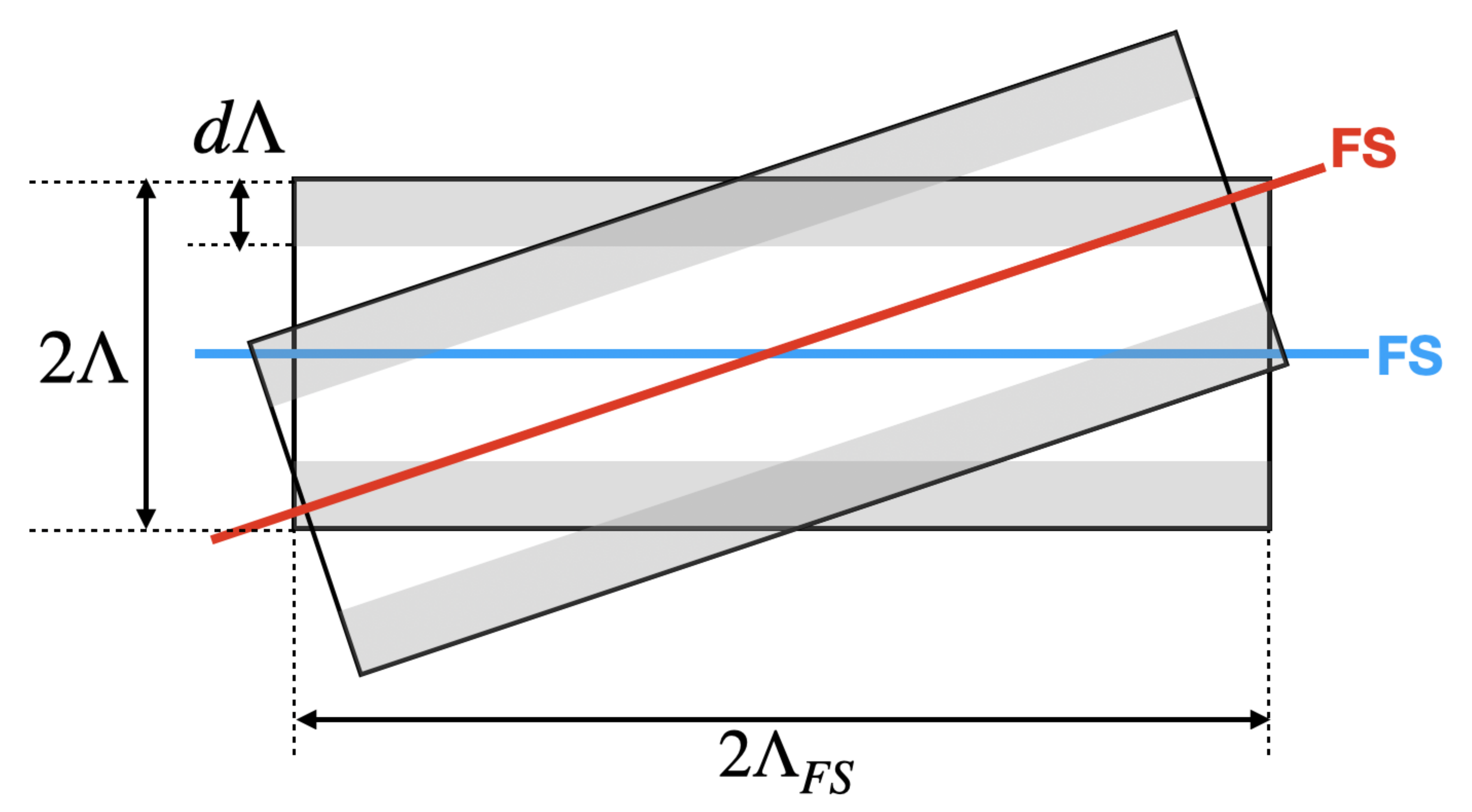}
			\caption{Two phase spaces of two different dispersions. Here, red and blue lines show two Fermi lines. $\Lambda$ and $\Lambda_{FS}$ denote an energy cut-off and a size of hot spots, respectively. Grey colored regions are high energy parts of the phase space.} \label{fig:PhaseSpaces}
		\end{subfigure}
	~
		\begin{subfigure}{0.3\textwidth}
			\includegraphics[scale=0.08]{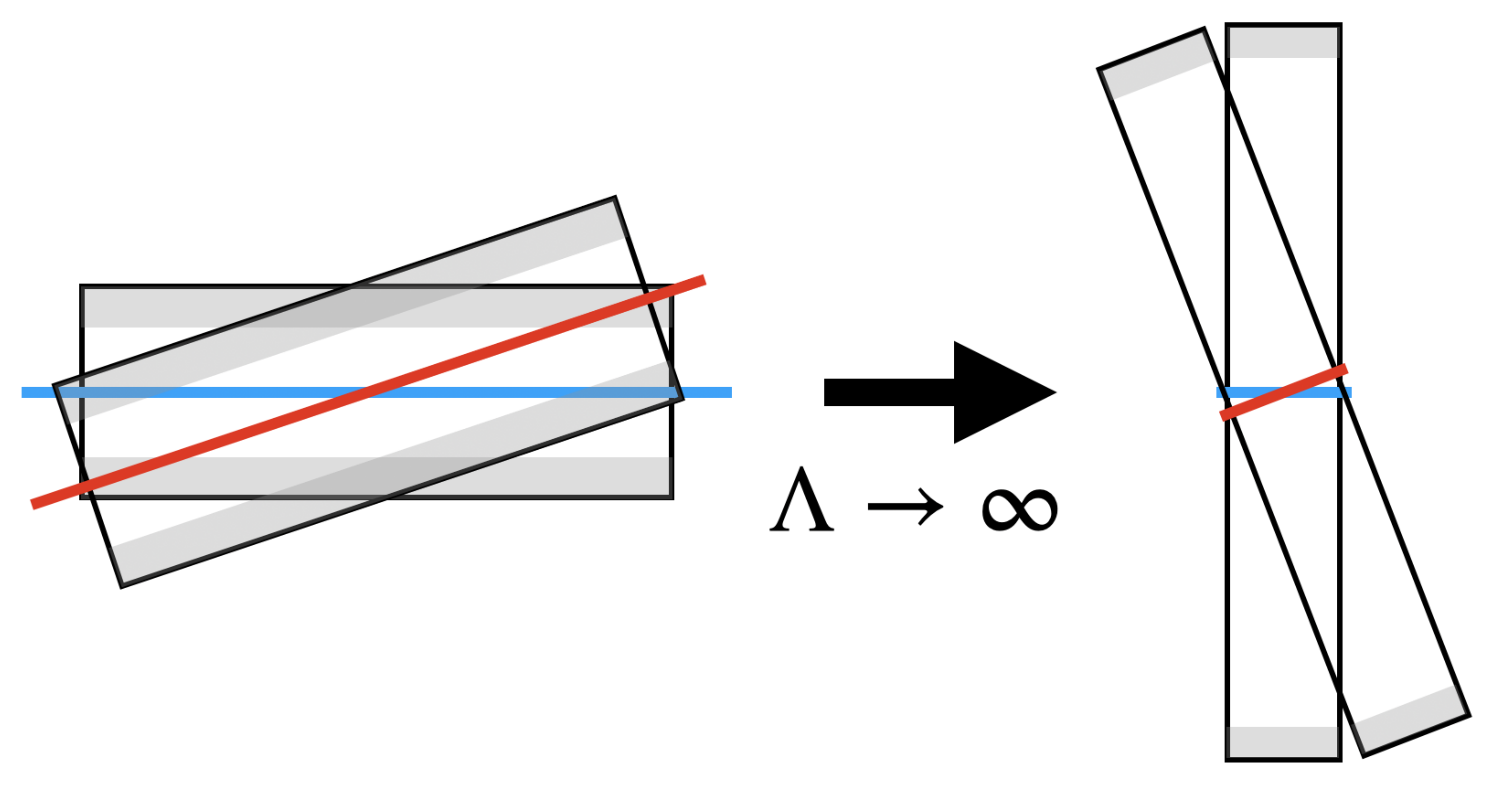}
			\caption{Vanishing overlapped phase spaces for $(n,m)=(1,3),(3,1),(2,4),(4,2)$ cases in the limit $\Lambda\rightarrow \infty$.} \label{fig:VanishingPhaseSpace}
		\end{subfigure}
		~
		\begin{subfigure}{0.3\textwidth}
			\includegraphics[scale=0.08]{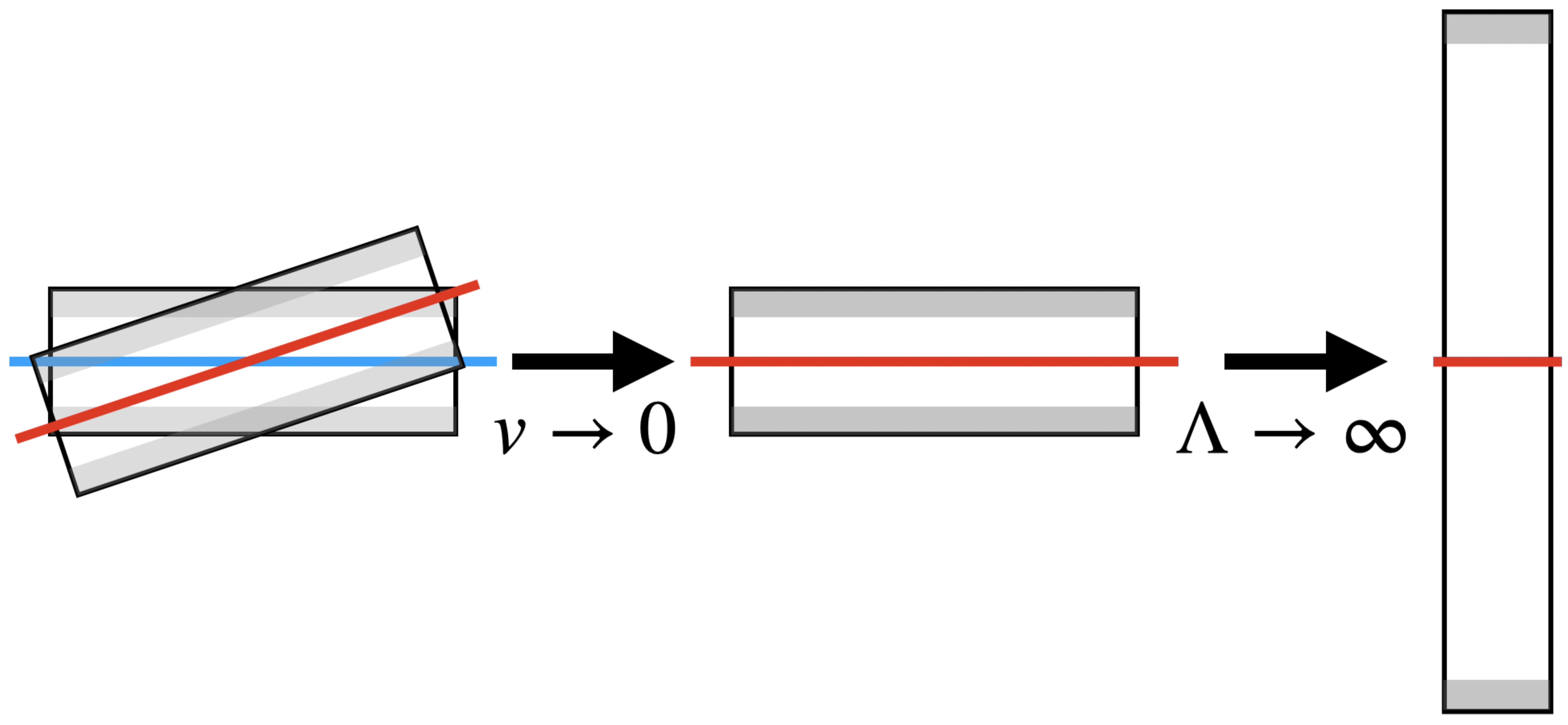}
			\caption{Finite overlapped phase space for $(n,m)=(1,3),(3,1),(2,4),(4,2)$ cases when the $v\rightarrow 0$ limit is taken first before the $\Lambda \rightarrow \infty$ limit.} \label{fig:zeroVPhaseSpace}
		\end{subfigure}
		\caption{Schematic figures for overlapped phase space.}
	\end{figure}
	
	These figures explain why the discrepancy happens between two integration results (Eq. \eqref{eq:oneLoopCalDV} and Eq. \eqref{eq:oneLoopCalDVzerovlimit}) from the same one-loop diagram. According to the previous RG results \cite{SurLee}, the overlapped phase space for $(n,m)=(1,3),(3,1),(2,4),(4,2)$ cases are almost same as that of the $n=m$ case since the Fermi velocity $v$ goes to zero as approaching to the low energy regime. Here, we choose our one-loop result based on the second approach Eq. \eqref{eq:oneLoopCalDVzerovlimit} with a suppression factor $e^{-v^2/v_c^2}$, where $v_c$ is an arbitrary value, introduced by hand. However, we confirm that $v$ goes to zero regardless of any choice of $v_c$, implying self-consistency in our RG analysis.
	
	We summarize our results for all the one-loop random charge potential vertex diagrams (Fig. \ref{fig:DisorderInteractionFeynman}):
	\begin{align}
		(pp)&=\Bigg\{\begin{array}{ll}
			\frac{\mu^{-1+2\epsilon}\Gamma_i\Gamma_j\Lambda_{FS}}{4\pi^2N_f^2(1+v^2)}\frac{1}{\epsilon}\Big[[\mathcal{M}^i\Gamma_{\perp,\mu}\mathcal{M}^j][\tilde{\mathcal{M}}^i\Gamma_{\perp,\mu}\tilde{\mathcal{M}}^j]+[\mathcal{M}^i\gamma_{d-1}\mathcal{M}^j][\tilde{\mathcal{M}}^i\gamma_{d-1}\tilde{\mathcal{M}}^j]\Big] &n=m\text{ case} \\
			e^{-v^2/v_c^2}\frac{\Gamma_i\Gamma_j\Lambda_{FS}}{4\pi^{2}N_f^2}\frac{1}{\epsilon}\Big[[\mathcal{M}^i\Gamma_{\perp,\mu}\mathcal{M}^j][\tilde{\mathcal{M}}^i\Gamma_{\perp,\mu}\tilde{\mathcal{M}}^j]-[\mathcal{M}^i\gamma_{d-1}\mathcal{M}^j][\tilde{\mathcal{M}}^i\gamma_{d-1}\tilde{\mathcal{M}}^j]\Big] &(n,m)=(1,3),(3,1),(2,4),(4,2) \\
			0 & \text{otherwise}
		\end{array}
	\\
		(ph)&=\Bigg\{\begin{array}{ll}
			-\frac{\mu^{-1+2\epsilon}\Gamma_i\Gamma_j\Lambda_{FS}}{4\pi^2N_f^2(1+v^2)}\frac{1}{\epsilon}\Big[[\mathcal{M}^i\Gamma_{\perp,\mu}\mathcal{M}^j][\tilde{\mathcal{M}}^j\Gamma_{\perp,\mu}\tilde{\mathcal{M}}^i]+[\mathcal{M}^i\gamma_{d-1}\mathcal{M}^j][\tilde{\mathcal{M}}^j\gamma_{d-1}\tilde{\mathcal{M}}^i]\Big] & n=m\text{ case} \\
			-e^{-v^2/v_c^2}\frac{\Gamma_i\Gamma_j\Lambda_{FS}}{4\pi^{2}N_f^2}\frac{1}{\epsilon}\Big[[\mathcal{M}^i\Gamma_{\perp,\mu}\mathcal{M}^j][\tilde{\mathcal{M}}^j\Gamma_{\perp,\mu}\tilde{\mathcal{M}}^i]-[\mathcal{M}^i\gamma_{d-1}\mathcal{M}^j][\tilde{\mathcal{M}}^j\gamma_{d-1}\tilde{\mathcal{M}}^i]\Big] &(n,m)=(1,3),(3,1),(2,4),(4,2) \\
			0 & \text{otherwise}
		\end{array}
\\
		(ll)&=\Bigg\{\begin{array}{ll}
			-\frac{\mu^{-1+2\epsilon}\Gamma_i\Gamma_j\Lambda_{FS}}{4\pi^2N_f^2(1+v^2)}\frac{1}{\epsilon}\Big[[\mathcal{M}^i\Gamma_{\perp,\mu}\mathcal{M}^j\Gamma_{\perp,\mu}\tilde{\mathcal{M}}^i][\mathcal{\tilde{M}}^{j}]+[\mathcal{M}^i\gamma_{d-1}\mathcal{M}^j\gamma_{d-1}\tilde{\mathcal{M}}^i][\mathcal{\tilde{M}}^{j}]\Big] & n=m\text{ case} \\
			-e^{-v^2/v_c^2}\frac{\Gamma_i\Gamma_j\Lambda_{FS}}{4\pi^{2}N_f^2}\frac{1}{\epsilon}\Big[[\mathcal{M}^j\Gamma_{\perp,\mu}\mathcal{M}^i\Gamma_{\perp,\mu}\mathcal{\tilde{M}}^j][\tilde{\mathcal{M}}^i]-[\mathcal{M}^j\gamma_{d-1}\mathcal{M}^i\gamma_{d-1}\mathcal{\tilde{M}}^j][\tilde{\mathcal{M}}^i]\Big] &(n,m)=(1,3),(3,1),(2,4),(4,2) \\
			0 & \text{otherwise}
		\end{array}
		\\
		(rl)&=\Bigg\{\begin{array}{ll}
			-\frac{\mu^{-1+2\epsilon}\Gamma_i\Gamma_j\Lambda_{FS}}{4\pi^2N_f^2(1+v^2)}\frac{1}{\epsilon}\Big[[\mathcal{\tilde{M}}^{j}][\mathcal{M}^i\Gamma_{\perp,\mu}\mathcal{M}^j\Gamma_{\perp,\mu}\tilde{\mathcal{M}}^i]+[\mathcal{\tilde{M}}^{j}][\mathcal{M}^i\gamma_{d-1}\mathcal{M}^j\gamma_{d-1}\tilde{\mathcal{M}}^i]\Big]& n=m\text{ case} \\
			-e^{-v^2/v_c^2}\frac{\Gamma_i\Gamma_j\Lambda_{FS}}{4\pi^{2}N_f^2}\frac{1}{\epsilon}\Big[[\mathcal{M}^i][\mathcal{M}^j\Gamma_{\perp,\mu}\mathcal{\tilde{M}}^i\Gamma_{\perp,\mu}\mathcal{\tilde{M}}^j]-[\mathcal{M}^i][\mathcal{M}^j\gamma_{d-1}\mathcal{\tilde{M}}^i\gamma_{d-1}\mathcal{\tilde{M}}^j]\Big] &(n,m)=(1,3),(3,1),(2,4),(4,2) \\
			0 & \text{otherwise}
		\end{array}
	\end{align}

	\subsubsection{One-loop random charge potential vertex corrections involving both the random charge potential vertex and the Yukawa interaction vertex}
	
	There are two kinds of one-loop Feynman diagrams for the random charge potential vertex, involving both the random charge potential vertex and the Yukawa interaction vertex, shown in Fig. \ref{fig:DisorderVertexFromBfD}. Here, the right-loop Feynman diagram is essentially the same as the left-loop one.
	
	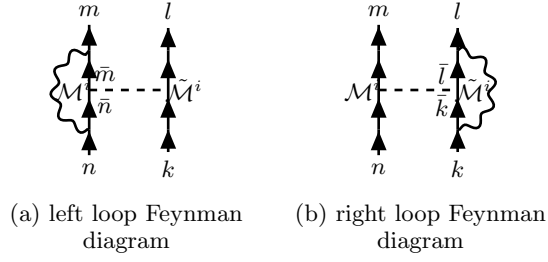
\begin{figure}[h]
		\begin{subfigure}[b]{0.2\textwidth}
			\begin{tikzpicture} [baseline=-0.1cm,scale=0.7]
				\begin{feynhand}
					\vertex (a1) at (0,0) {$n$}; \vertex (a2) at (0,3/4); \vertex (a3) at (0,6/4); \vertex (a4) at (0, 9/4); \vertex(a5) at (0, 3) {$m$};
					\vertex (b1) at (1.5,0) {$k$}; \vertex (b2) at (1.5,3/4); \vertex (b3) at (1.5,6/4); \vertex (b4) at (1.5,9/4); \vertex(b5) at (1.5, 3) {$l$};
					\propag[fer] (a1) to (a2); \propag[fer] (a2) to (a3); \propag[fer] (a3) to (a4); \propag[fer] (a4) to (a5);
					\propag[fer] (b1) to (b2); \propag[fer] (b2) to (b3); \propag[fer] (b3) to (b4); \propag[fer] (b4) to (b5);
					\propag[sca] (a3) to (b3); \propag[boson] (a2) to [out=180, in=180, looseness=1.5](a4);
					\node at (0.3,1.8) {$\bar{m}$}; \node at (0.3,1.2) {$\bar{n}$};
					\node at (-0.3,6/4) {$\mathcal{M}^i$};
					\node at (1.8,6/4) {$\tilde{\mathcal{M}}^i$};
				\end{feynhand}
			\end{tikzpicture}
			\caption{left loop Feynman diagram}\label{BfDa}
		\end{subfigure}
		~
		\begin{subfigure}[b]{0.2\textwidth}
			\begin{tikzpicture}[baseline=-0.1cm,scale=0.7]
				\begin{feynhand}
					\vertex (a1) at (0,0) {$n$}; \vertex (a2) at (0,3/4); \vertex (a3) at (0,6/4); \vertex (a4) at (0, 9/4); \vertex(a5) at (0, 3) {$m$};
					\vertex (b1) at (1.5,0) {$k$}; \vertex (b2) at (1.5,3/4); \vertex (b3) at (1.5,6/4); \vertex (b4) at (1.5,9/4); \vertex(b5) at (1.5, 3) {$l$};
					\propag[fer] (a1) to (a2); \propag[fer] (a2) to (a3); \propag[fer] (a3) to (a4); \propag[fer] (a4) to (a5);
					\propag[fer] (b1) to (b2); \propag[fer] (b2) to (b3); \propag[fer] (b3) to (b4); \propag[fer] (b4) to (b5);
					\propag[sca] (a3) to (b3); \propag[boson] (b2) to [out=0, in=0,looseness=1.5](b4);
					\node at (1.2,1.8) {$\bar{l}$}; \node at (1.2,1.2) {$\bar{k}$};
					\node at (-0.3,6/4) {$\mathcal{M}^i$};
					\node at (1.8,6/4) {$\tilde{\mathcal{M}}^i$};
				\end{feynhand}
			\end{tikzpicture}
			\caption{right loop Feynman diagram}\label{BfDb}
		\end{subfigure}
		\caption{Two kinds of one-loop Feynman diagrams involving both the random charge potential vertex and the Yukawa vertex} \label{fig:DisorderVertexFromBfD}
	\end{figure}
	
	One-loop RG results depend on $(\bar{n},\bar{m})$. There are four different cases;
	\begin{align*}
		&\text{(i)}\; (\bar{n},\bar{m})=(1,1),(2,2),(3,3),(4,4) ,\;\; \text{(ii)}\; (\bar{n},\bar{m})=(1,3),(3,1),(2,4),(4,2), \\
		&\text{(iii)}\; (\bar{n},\bar{m})=(1,2),(2,1),(3,4),(4,3) ,\;\; \text{(iv)}\; (\bar{n},\bar{m})=(1,4),(4,1),(2,3),(3,2) .
	\end{align*}
	Here, we present our detailed calculations for each case:
	\begin{itemize}
		\item[(i)] \underline{$(\bar{n},\bar{m})=(1,1),(2,2),(3,3),(4,4)$}
		
		In this case, results are all the same, given by
		\begin{align}
			(ll)&=\Big(-i\frac{g}{\sqrt{N_f}}\Big)^2\frac{\Gamma_i}{N_f}(\sum_{\sigma',i}\tau^i_{\sigma\sigma'}\tau^i_{\sigma'\sigma})\int \frac{d^{d-1}\mathbf{Q}}{(2\pi)^{d-1}}\int\frac{d^2q}{(2\pi)^2}\frac{1}{q_0^2+c_\perp^2|\mathbf{Q}_\perp|^2+c^2(q_{d-1}^2+q_d^2)}\nonumber\\
			&\times [\gamma_{d-1}G_{1,f}(q)\mathcal{M}^iG_{1,f}(q)\gamma_{d-1}]\otimes \tilde{\mathcal{M}}^i\nonumber\\
			&=\frac{g^2\Gamma_i (N_c^2-1)}{4\pi^2 N_f^2N_c c}\frac{1}{\epsilon}\Big(f_1(c,c_\perp,v)\gamma_{d-1}\gamma_0\mathcal{M}^i\gamma_0\gamma_{d-1}+f_2(c,c_\perp,v)\gamma_{d-1}\Gamma_{\perp,\mu}\mathcal{M}^i\Gamma_{\perp,\mu}\gamma_{d-1}+f_3(c,c_\perp,v)\mathcal{M}^i\Big)\otimes \tilde{\mathcal{M}}^i .
		\end{align}
		Here, we used $\epsilon(k)=vk_{d-1}+k_d$ and $\epsilon_{||}(k)=k_{d-1}-vk_d$. Functions of velocities are
		\begin{gather}
			f_1(c,c_\perp,v)=\frac{1}{2}\int_0^1 dx(1-x) x^{-1/2}\Big(xc^2+(1-x)(1+v^2)\Big)^{-1/2}(xc_\perp^2+1-x)^{-1/2},\\
			f_2(c,c_\perp,v)=\frac{1}{2}\int_0^1dx (1-x)x^{-1/2}\Big(xc^2+(1-x)(1+v^2)\Big)^{-1/2}(xc_\perp^2+1-x)^{-3/2},\\
			f_3(c,c_\perp,v)=\frac{1+v^2}{2}\int_0^1 dx(1-x)x^{-1/2} \Big(xc^2+(1-x)(1+v^2)\Big)^{-3/2} (xc_\perp^2+1-x)^{-1/2}.
		\end{gather}
		We explicitly check out that these functions satisfy
		\begin{gather}
			-f_1(c,c_\perp,v)+f_2(c,c_\perp,v)+f_3(c,c_\perp,v)=h_1(c,c_\perp,v),  \\
			f_1(c,c_\perp,v)-f_2(c,c_\perp,v)+f_3(c,c_\perp,v)=h_2(c,c_\perp,v), \\
			-f_1(c,c_\perp,v)-f_2(c,c_\perp,v)+f_3(c,c_\perp,v)=-h_3(c,c_\perp,v),
		\end{gather}
		originating from the Ward identity.
		
		\item[(ii)] \underline{$(\bar{n},\bar{m})=(1,3),(3,1),(2,4),(4,2)$}
		\begin{align}
			(ll)_{(1,3),(3,1)}&=\frac{2g^2\Gamma_i(N_c^2-1)}{N_f^2N_c}\int \frac{d^{d-1}Q}{(2\pi)^{d-1}}\int\frac{d^2q}{(2\pi)^2}\frac{1}{q_0^2+c_\perp^2|\mathbf{Q}_\perp|^2+c^2(q_{d-1}^2+q_d^2)}\nonumber\\
			&\times \frac{\gamma_{d-1}\Gamma_\mu \mathcal{M}^i\Gamma_\mu\gamma_{d-1}\frac{|\mathbf{Q}|^2}{d-1}+\epsilon_1(q)\epsilon_3(q)\mathcal{M}^i}{[|\mathbf{Q}|^2+\epsilon_1^2(q)][|\mathbf{Q}|^2+\epsilon_3^2(q)]}\otimes \tilde{\mathcal{M}}^i\nonumber \\
			&=\frac{(N_c^2-1)\Gamma_i}{8\pi^2N_f^2N_c}\frac{g^2}{c}\frac{1}{\epsilon}\Big[f_4(c,c_\perp,v)\gamma_{d-1}\gamma_0\mathcal{M}^i\gamma_0\gamma_{d-1}+f_5(c,c_\perp,v)\gamma_{d-1}\Gamma_{\perp,\mu}\mathcal{M}^i\Gamma_{\perp,\mu}\gamma_{d-1}-f_6(c,c_\perp,v)\mathcal{M}^i\Big]\otimes \tilde{\mathcal{M}}^i ,
		\end{align}
		Here, we used $\epsilon_1(q)=vq_{d-1}+q_d$ and $\epsilon_3(q)=vq_{d-1}-q_d$. Functions of the velocities are
		\begin{gather}
			f_4(c,c_\perp,v)=\int_0^1dx \int_0^{1-x} dy(xc_\perp^2+1-x)^{-1/2} \Big[c^2x^2+4\frac{v^2}{c^2}y(1-x-y)+x(1-x)(1+v^2)\Big]^{-1/2} , \\
			f_5(c,c_\perp,v)=\int_0^1dx \int_0^{1-x} dy(xc_\perp^2+1-x)^{-3/2} \Big[c^2x^2+4\frac{v^2}{c^2}y(1-x-y)+x(1-x)(1+v^2)\Big]^{-1/2} , \\
			f_6(c,c_\perp,v)=\frac{(xc_\perp^2+1-x)^{-1/2}x(1-v^2)}{\Big[c^2x^2+4\frac{v^2}{c^2}y(1-x-y)+x(1-x)(1+v^2)\Big]^{3/2}} .
		\end{gather}
		They satisfy
		\begin{gather}
			f_4(c,c_\perp,v)+f_5(c,c_\perp,v)+f_6(c,c_\perp,v)=\frac{h_4(c,c_\perp,v)}{\pi} .
		\end{gather}
		due to the Ward identity. The result for $(\bar{n},\bar{m})=(2,4),(4,2)$ is the same as that of $(\bar{n},\bar{m})=(1,3),(3,1)$.
		
		\item[(iii)] \underline{$(\bar{n},\bar{m})=(1,2),(2,1),(3,4),(4,1)$}
		\begin{align}
			(ll)_{(1,2),(2,1)}&=\frac{2g^2\Gamma_i(N_c^2-1)}{N_f^2N_c}\int \frac{d^{d-1}Q}{(2\pi)^{d-1}}\int\frac{d^2q}{(2\pi)^2}\frac{1}{q_0^2+c_\perp^2|\mathbf{Q}_\perp|^2+c^2(q_{d-1}^2+q_d^2)}\nonumber\\
			&\times \frac{\gamma_{d-1}\gamma_0\mathcal{M}^i\gamma_0\gamma_{d-1}q_0^2+\gamma_{d-1}\Gamma_{\perp,\mu} \mathcal{M}^i\Gamma_{\perp,\mu}\gamma_{d-1}\frac{|\mathbf{Q}_\perp|^2}{d-2}+\epsilon_1(q)\epsilon_2(q)\mathcal{M}^i}{[|\mathbf{Q}|^2+\epsilon_1^2(q)][|\mathbf{Q}|^2+\epsilon_2^2(q)]}\otimes \tilde{\mathcal{M}}^i\nonumber \\
			&=\frac{(N_c^2-1)g^2\Gamma_i}{8\pi^2N_f^2N_c}\frac{1}{\epsilon}\Big[f_7(c,c_\perp,v)\gamma_{d-1}\gamma_{0}\mathcal{M}^i\gamma_{0}\gamma_{d-1}\otimes \tilde{\mathcal{M}}^i+f_8(c,c_\perp,v)\gamma_{d-1}\Gamma_{\perp,\mu}\mathcal{M}^i\Gamma_{\perp,\mu}\gamma_{d-1}\otimes \tilde{\mathcal{M}}^i\Big]
		\end{align}
		Here, we used $\epsilon_1(q)=vq_{d-1}+q_d$ and $\epsilon_2(q)=-q_{d-1}+vq_d=-\epsilon_{1,||}$. Functions of the velocities are
		\begin{gather}
			f_7(c,c_\perp,v)=\int_0^1dx\int_0^{1-x}dy [xc^2+y(1+v^2)]^{-1/2}[xc^2+(1+v^2)(1-x-y)]^{-1/2}[xc_\perp^2+1-x]^{-1/2},\\
			f_8(c,c_\perp,v)=\int_0^1dx\int_0^{1-x}dy [xc^2+y(1+v^2)]^{-1/2}[xc^2+(1+v^2)(1-x-y)]^{-1/2}[xc_\perp^2+1-x]^{-3/2}.
		\end{gather}
		The result for $(\bar{n},\bar{m})=(3,4),(4,3)$ is the same as that of $(\bar{n},\bar{m})=(1,2),(2,1)$.
		
		\item[(iv)] \underline{$(\bar{n},\bar{m})=(1,4),(4,1),(2,3),(3,2)$}
		\begin{align}
			(ll)_{(1,4),(4,1)}&=\frac{2g^2\Gamma_i(N_c^2-1)}{N_f^2N_c}\int \frac{d^{d-1}Q}{(2\pi)^{d-1}}\int\frac{d^2q}{(2\pi)^2}\frac{1}{q_0^2+c_\perp^2|\mathbf{Q}_\perp|^2+c^2(q_{d-1}^2+q_d^2)}\nonumber\\
			&\times \frac{\gamma_{d-1}\Gamma_\mu \mathcal{M}^i\Gamma_\mu\gamma_{d-1}\frac{|\mathbf{Q}|^2}{d-1}+\epsilon_1(q)\epsilon_4(q)\mathcal{M}^i}{[|\mathbf{Q}|^2+\epsilon_1^2(q)][|\mathbf{Q}|^2+\epsilon_4^2(q)]}\otimes \tilde{\mathcal{M}}^i\nonumber \\
			&=\frac{(N_c^2-1)\Gamma_i}{8\pi^2N_f^2N_c }\frac{g^2}{c}\frac{1}{\epsilon}\Big[f_9(c,c_\perp,v)\gamma_{d-1}\gamma_0\mathcal{M}^i\gamma_0\gamma_{d-1}+f_{10}(c,c_\perp,v)\gamma_{d-1}\Gamma_{\perp,\mu}\mathcal{M}^i\Gamma_{\perp,\mu}\gamma_{d-1}\nonumber\\
			&+f_{11}(c,c_\perp,v)\mathcal{M}^i\Big]\otimes \tilde{\mathcal{M}}^i ,
		\end{align}
		where $\epsilon_1(q)=vq_{d-1}+q_d$ and $\epsilon_4(q)=q_{d-1}+vq_d$ have been used and
		\begin{gather}
			f_9(c,c_\perp,v)=\int_0^1dx \int_{0}^{1-x}dy [xc_\perp^2+1-x]^{-1/2}\Big[c^2x^2+(1+v^2)x(1-x)+\frac{(1-v^2)^2}{c^2}y(1-x-y)\Big]^{-1/2}\\
			f_{10}(c,c_\perp,v)=\int_0^1dx \int_{0}^{1-x}dy [xc_\perp^2+1-x]^{-3/2}\Big[c^2x^2+(1+v^2)x(1-x)+\frac{(1-v^2)^2}{c^2}y(1-x-y)\Big]^{-1/2}\\
			f_{11}(c,c_\perp,v)=2v\int_0^1dx \int_{0}^{1-x}dy [xc_\perp^2+1-x]^{-1/2}x\Big[c^2x^2+(1+v^2)x(1-x)+\frac{(1-v^2)^2}{c^2}y(1-x-y)\Big]^{-3/2} .
		\end{gather}
		
		\begin{align}
			(ll)_{(2,3),(3,2)}&=\frac{2g^2\Gamma_i(N_c^2-1)}{N_f^2N_c}\int \frac{d^{d-1}Q}{(2\pi)^{d-1}}\int\frac{d^2q}{(2\pi)^2}\frac{1}{q_0^2+c_\perp^2|\mathbf{Q}_\perp|^2+c^2(q_{d-1}^2+q_d^2)}\nonumber\\
			&\times \frac{\gamma_{d-1}\gamma_0\mathcal{M}^i\gamma_0\gamma_{d-1}q_0^2+\gamma_{d-1}\Gamma_{\perp,\mu} \mathcal{M}^i\Gamma_{\perp,\mu}\gamma_{d-1}\frac{|\mathbf{Q}_\perp|^2}{d-2}+\epsilon_2(q)\epsilon_3(q)\mathcal{M}^i}{[|\mathbf{Q}|^2+\epsilon_2^2(q)][|\mathbf{Q}|^2+\epsilon_3^2(q)]}\otimes \tilde{\mathcal{M}}^i\nonumber\\
			&=\frac{(N_c^2-1)\Gamma_i}{8\pi^2N_f^2N_c }\frac{g^2}{c}\frac{1}{\epsilon}\Big[f_9(c,c_\perp,v)\gamma_{d-1}\gamma_0\mathcal{M}^i\gamma_0\gamma_{d-1}+f_{10}(c,c_\perp,v)\gamma_{d-1}\Gamma_{\perp,\mu}\mathcal{M}^i\Gamma_{\perp,\mu}\gamma_{d-1}\nonumber\\
			&-f_{11}(c,c_\perp,v)\mathcal{M}^i\Big]\otimes \tilde{\mathcal{M}}^i ,
		\end{align}
		where $\epsilon_2(q)=\epsilon_1(R_{\pi/2}^{-1}q)=vq_d-q_{d-1}$ and $\epsilon_3(q)=-\epsilon_4(R_{\pi/2}^{-1}q)=-q_d+vq_{d-1}$ have been used. Note that The result of $(\bar{n},\bar{m})=(1,4),(4,1)$ is different from that of $(\bar{n},\bar{m})=(2,3),(3,2)$.
	\end{itemize}

	\subsection{One-loop Yukawa interaction vertex corrections}
	
	There are two types of one-loop Feynman diagrams, shown in Fig. \ref{fig:BFCorrections}, where either the Yukawa vertex or the random charge potential vertex is involved.
	
	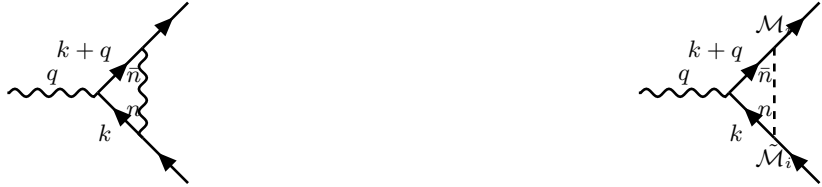
\begin{figure}[h]
		\begin{subfigure}{0.45\textwidth}
			\begin{tikzpicture}[baseline=-0.1cm, scale=1.2]
				\begin{feynhand}
					\vertex (a) at (0,0); \vertex (b) at (1,0);
					\vertex (c1) at (1.5,0.5); \vertex (d1) at (2,1);
					\vertex (c2) at (1.5,-0.5); \vertex (d2) at (2,-1);
					
					\propag[boson] (a) to [edge label=$q$] (b);
					\propag[fermion] (b) to [edge label=$k+q$](c1); \propag[fermion] (c1) to (d1);
					\propag[fermion] (c2) to [edge label=$k$ ](b); \propag[fermion] (d2) to (c2);
					\propag[boson] (c1) to (c2);
					
					\node at (1.4,0.2) {$\bar{n}$};
					\node at (1.4,-0.2) {$n$};
				\end{feynhand}
			\end{tikzpicture}
			\caption{1-loop Yukawa vertex correction from the Yukawa vertex}
			\label{fig:BFYukawaCorrection}
		\end{subfigure}
		~
		\begin{subfigure}{0.45\textwidth}
			\begin{tikzpicture}[baseline=-0.1cm, scale=1.2]
				\begin{feynhand}
					\vertex (a) at (0,0); \vertex (b) at (1,0);
					\vertex (c1) at (1.5,0.5); \vertex (d1) at (2,1);
					\vertex (c2) at (1.5,-0.5); \vertex (d2) at (2,-1);
					
					\propag[boson] (a) to [edge label=$q$] (b);
					\propag[fermion] (b) to [edge label=$k+q$](c1); \propag[fermion] (c1) to (d1);
					\propag[fermion] (c2) to [edge label=$k$ ](b); \propag[fermion] (d2) to (c2);
					\propag[scalar] (c1) to (c2);
					
					\node at (1.4,0.2) {$\bar{n}$};
					\node at (1.4,-0.2) {$n$};
					\node at (1.5,0.75) {$\mathcal{M}_i$};
					\node at (1.5,-0.7) {$\tilde{\mathcal{M}}_i$};
				\end{feynhand}
			\end{tikzpicture}
			\caption{1-loop Yukawa vertex correction from the random charge potential vertex}
			\label{fig:BFdisorderCorrection}
		\end{subfigure}
		\caption{Two types of 1-loop Yukawa vertex corrections} \label{fig:BFCorrections}
	\end{figure}

	First, we consider the Feynman diagram Fig. \ref{fig:BFYukawaCorrection} as follows:
	\begin{align}
		\Gamma_{bf}^{Yukawa}&=i\frac{g^3}{N_f^{3/2}}\Big(\sum_{b=1}^{N_c^2-1}\tau^b\tau^a\tau^b\Big)\int\frac{d\Omega}{2\pi}\int\frac{d^{d-2}\mathbf{Q}_\perp}{(2\pi)^{d-2}}\int\frac{d^2\mathbf{q}}{(2\pi)^2}\gamma_{d-1}G_{f,\bar{n}}(\Omega,\mathbf{Q}_\perp,\vec{q})\gamma_{d-1}G_{f,n}(\Omega,\mathbf{Q}_\perp,\vec{q})\gamma_{d-1}G_b(\Omega,\mathbf{Q}_\perp,\vec{q})\nonumber\\
		&= -i\frac{g^3}{8\pi^3c N_cN_f^{3/2}}\frac{1}{\epsilon}\gamma_{d-1}\tau^a \int_0^1dx\int_0^{1-x}dy h_4(c,c_\perp,v) ,
	\end{align}
	where
	\begin{align}
		h_4(c,c_\perp,v)&=\pi c\Big[\Big(1+g_3(c,c_\perp,v)\Big)\Big(g_1(c,c_\perp,v)g_2(c,c_\perp,v)-v^2(x-y)^2\Big)+g_3(c,c_\perp,v)\Big(g_2(c,c_\perp,v)-v^2g_1(c,c_\perp,v)\Big)\Big]\nonumber\\
		&\times \Big[g_3(c,c_\perp,v)\Big(g_1(c,c_\perp,v)g_2(c,c_\perp,v)-v^2(x-y)^2\Big)\Big]^{-3/2},\;
		\Bigg\{
		\begin{array}{l}
			g_1(c,c_\perp,v)=x+y+c^2(1-x-y),\\
			g_2(c,c_\perp,v)=(x+y)v^2+c^2(1-x-y),\\
			g_3(c,c_\perp,v)=x+y+c_\perp^2(1-x-y)
		\end{array}.
	\end{align}

	Next, we consider the Feynman diagram Fig. \ref{fig:BFdisorderCorrection}, which involves only a random charge potential vertex. Since we have $(n,\bar{n})=(1,3),(3,1),(2,4),(4,2)$, there is the same issue that we met in 1-loop corrections for the random charge potential vertex involving only random charge potential vertices. Here, we also set the $v \rightarrow 0$ limit first and get an $\epsilon$-pole with the introduction of the $e^{-v^2/v_c^2}$ factor as follows
	\begin{align*}
		\Gamma^{dis}_{bf}&=-i\frac{g\mu^{\epsilon/2}}{\sqrt{N_f}}\frac{\Gamma_i \mu^{-1+\epsilon}}{N_f}\int \frac{d^{d-2}\mathbf{K}_\perp}{(2\pi)^{d-2}}\int \frac{d^2k}{(2\pi)^2}\mathcal{M}_iG_{n,\sigma}(\mathbf{k}+\mathbf{q},\omega+\Omega)\gamma_{d-1}G_{\bar{n},\sigma'}(\mathbf{k},\omega)\tilde{\mathcal{M}}^i\\
		&=i\frac{g\bar{\Gamma}_{i}\mu^{3\epsilon/2}}{4\pi^2N_f^{3/2}}\frac{1}{\epsilon}\Big[\mathcal{M}_i\Gamma_{\perp,\mu}\gamma_{d-1}\Gamma_{\perp,\mu}\tilde{\mathcal{M}}_i-\mathcal{M}_i\gamma_{d-1}\tilde{\mathcal{M}}_i\Big]\Rightarrow e^{-v^2/v_c^2}i\frac{g\bar{\Gamma}_{i}\mu^{3\epsilon/2}}{4\pi^2N_f^{3/2}}\frac{1}{\epsilon} \Big[\mathcal{M}_i\Gamma_{\perp,\mu}\gamma_{d-1}\Gamma_{\perp,\mu}\tilde{\mathcal{M}}_i-\mathcal{M}_i\gamma_{d-1}\tilde{\mathcal{M}}_i\Big] .
	\end{align*}

	\subsection{One-loop boson self-energy corrections}
	
	There are two 1-loop boson self-energy Feynman diagrams shown in Fig. \ref{fig:BosonSelfEnergy}. Since the calculation of the Feynman diagram Fig. \ref{fig:BosonSelfEnergyYukawa} is the same as that of the clean case, we do not present details of the calculation here. We refer the details to Ref. \cite{SurLee}. Here. we only present the details for Fig. \ref{fig:BosonSelfEnergyRandomMass}.
	
	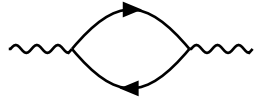
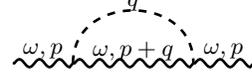
\begin{figure}[h]
		\begin{subfigure}{0.45\textwidth}
			\begin{tikzpicture}[baseline=-0.1cm,scale=1.2]
				\begin{feynhand}
					\vertex (a) at (0,0); \vertex (b) at (0.7,0); \vertex (c) at (2,0); \vertex (d) at (2.7,0);
					\propag[boson] (a) to (b); \propag[fermion] (b) to [out=50,in=130,looseness=1.5](c); \propag[fermion] (c) to [out=230, in=-50,looseness=1.5](b); \propag[boson] (c) to (d);
				\end{feynhand}
			\end{tikzpicture}
			\caption{1-loop boson self-energy from the Yukawa vertex.}\label{fig:BosonSelfEnergyYukawa}
		\end{subfigure}
		~
		\begin{subfigure}{0.45\textwidth}
			\begin{tikzpicture}[baseline=-0.1cm,scale=0.8]
				\begin{feynhand}
					\vertex (a) at (-2,0); \vertex (b) at (-1,0); \vertex (c) at (1,0); \vertex (d) at (2,0);
					\propag[boson] (a) to (b); \propag[boson] (b) to (c); \propag[boson] (c) to (d); \propag[sca] (b) to [out=90, in=90, looseness=1.5](c);
					\node at (-1.5,0.2) {$\omega,p$}; \node at (0,0.2) {$\omega,p+q$}; \node at (1.5,0.2) {$\omega,p$}; \node at (0,1) {$q$};
				\end{feynhand}
			\end{tikzpicture}
			\caption{1-loop boson self-energy from the random mass vertex $\Gamma_M$.}\label{fig:BosonSelfEnergyRandomMass}
		\end{subfigure}
		\caption{Two 1-loop boson self-energy Feynman diagrams.} \label{fig:BosonSelfEnergy}
	\end{figure}
	
	Our calculation for the Feynman diagram Fig. \ref{fig:BosonSelfEnergyRandomMass} is given by
	\begin{align}
		\Pi^{\Gamma_M}(p)&=4\Gamma_M \int\frac{d^{d-2}\mathbf{Q}_\perp}{(2\pi)^{d-2}}\int\frac{d^2 \vec{q}}{(2\pi)^2}\frac{|\mathbf{Q}_\perp|^{\alpha}+\kappa |\vec{q}|^{\alpha}}{\omega^2+c_\perp^2|\mathbf{Q}_\perp+\mathbf{P}_\perp|^2+c^2|\vec{q}+\vec{p}|^2}\nonumber\\
		&=-\frac{\Gamma_M}{2\pi^2 c^2c_\perp^2}\frac{2}{\epsilon+\bar{\epsilon}}\Big[\Big(1+\frac{\pi}{2}\frac{c_\perp}{c}\kappa\Big)\omega^2-c_\perp^2|\mathbf{P}_\perp|^2-\frac{\pi}{4}\kappa c c_\perp |\vec{p}|^2\Big] ,
	\end{align}

	Since the non-local term $(|\mathbf{Q}_\perp|^\alpha+\kappa |\vec{q}|^{\alpha})$ from the random mass vertex is involved in this calculation, which is not typical in the evaluation of Feynman diagrams, we check out this calculation, introducing a cut-off $\Lambda$ regularization in $d=3$ and $\alpha=1$ as follows:
	
	\begin{align}
		\Pi^{\Gamma_M}_{\Lambda}(p)&=4\Gamma_M\int\frac{dQ_\perp}{2\pi}\int\frac{d^2\vec{q}}{(2\pi)^2}\frac{|Q_\perp|+\kappa|\vec{q}|}{\omega^2+c_\perp^2|Q_\perp+P_\perp|+c^2|\vec{q}+\vec{p}|^2}\nonumber\\
		&=-\frac{\Gamma_M}{2\pi^2 c^2c_\perp^2}\Big[\omega^2\Big(1+\frac{\pi}{2}\frac{c_\perp}{c}\kappa\Big)-c_\perp^2 |P_\perp|^2-\frac{\pi}{4}\kappa c c_\perp |\vec{p}|^2\Big]\ln\Big(\frac{\Lambda^2}{\omega^2}\Big).
	\end{align}
	Identifying $\ln\Big(\frac{\Lambda^2}{\omega^2}\Big)$ with $\frac{2}{\epsilon+\bar{\epsilon}}$, we obtain the same result.

	\subsection{One-loop self-interaction corrections to $u_1$ and $u_2$}
	
	Since calculations of one-loop corrections from the $u_1$ and $u_2$ vertices are the same to that of the clean case except for a factor $\frac{1}{c_\perp}$ multiplied to the value of the clean case, we do not present detailed calculations of the one-loop Feynman diagrams composed of the $u_1$ and $u_2$ vertices only.
	
	\subsubsection{Corrections to $u_1$ from $\Gamma_M$}

	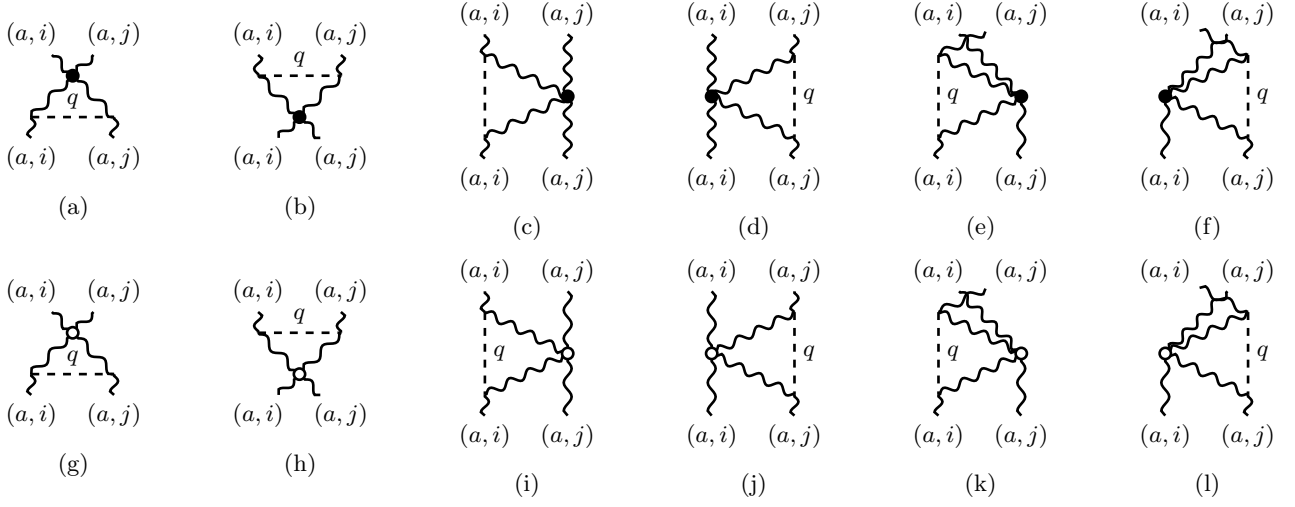
\begin{figure}[h]
		\begin{subfigure}{0.15\textwidth}
			\begin{tikzpicture}[scale=0.55]
				\begin{feynhand}
					\vertex (a) at (-1,1) {$(a,i)$}; \vertex (b) at (1,1) {$(a,j)$}; \vertex[dot] (c) at (0,0){}; \vertex (d) at (-1,-1); \vertex (e) at (1,-1); \vertex (f) at (-1,-2) {$(a,i)$}; \vertex (g) at (1,-2) {$(a,j)$};
					\propag[boson] (a) to (c); \propag[boson] (b) to (c); \propag[boson] (c) to (d); \propag[boson] (c) to (e);
					\propag[boson] (d) to (f); \propag[boson] (e) to (g);
					\propag[sca] (d) to[edge label=$q$] (e);
				\end{feynhand}
			\end{tikzpicture}
			\caption{}
		\end{subfigure}
		~
		\begin{subfigure}{0.15\textwidth}
			\begin{tikzpicture}[scale=0.55]
				\begin{feynhand}
					\vertex (a) at (-1,1); \vertex (b) at (1,1); \vertex [dot] (c) at (0,0) {}; \vertex (d) at (-1,-1) {$(a,i)$}; \vertex (e) at (1,-1){$(a,j)$}; \vertex (f) at (-1,2) {$(a,i)$}; \vertex (g) at (1,2) {$(a,j)$};
					\propag[boson] (a) to (c); \propag[boson] (b) to (c); \propag[boson] (c) to (d); \propag[boson] (c) to (e);
					\propag[boson] (a) to (f); \propag[boson] (b) to (g);
					\propag[sca] (a) to[edge label=$q$] (b);
				\end{feynhand}
			\end{tikzpicture}
			\caption{}
		\end{subfigure}
		~
		\begin{subfigure}{0.15\textwidth}
			\begin{tikzpicture}[scale=0.55]
				\begin{feynhand}
					\vertex (a) at (-1,2) {$(a,i)$}; \vertex (b) at (1,2) {$(a,j)$}; \vertex (c) at (-1,1); \vertex (d) at (-1,-1);  \vertex (e) at (1, 0);
					\vertex (f) at (-1,-2) {$(a,i)$}; \vertex (g) at (1,-2) {$(a,j)$}; \vertex [dot] (h) at (1,0) {};
					\propag[boson] (a) to (c); \propag[scalar] (c) to (d); \propag[boson] (d) to (f);
					\propag[boson] (c) to (e); \propag[boson] (d) to (e);   \propag[boson] (e) to (b); \propag[boson] (e) to (g);
				\end{feynhand}
			\end{tikzpicture}
			\caption{}
		\end{subfigure}
		~
		\begin{subfigure}{0.15\textwidth}
			\begin{tikzpicture}[scale=0.55]
				\begin{feynhand}
					\vertex (a) at (1,2) {$(a,j)$}; \vertex (b) at (-1,2) {$(a,i)$}; \vertex (c) at (1,1); \vertex (d) at (1,-1);  \vertex (e) at (-1, 0);
					\vertex (f) at (1,-2) {$(a,j)$}; \vertex (g) at (-1,-2) {$(a,i)$}; \vertex [dot] (h) at (-1,0) {};
					\propag[boson] (a) to (c); \propag[sca] (c) to [edge label=$q$](d); \propag[boson] (d) to (f);
					\propag[boson] (c) to (e); \propag[boson] (d) to (e);   \propag[boson] (e) to (b); \propag[boson] (e) to (g);
				\end{feynhand}
			\end{tikzpicture}
			\caption{}
		\end{subfigure}
		~
		\begin{subfigure}{0.15\textwidth}
			\begin{tikzpicture}[scale=0.55]
				\begin{feynhand}
					\vertex (a) at (-1,2) {$(a,i)$}; \vertex (b) at (1,2) {$(a,j)$}; \vertex (c) at (-1,1); \vertex (d) at (-1,-1);  \vertex[dot] (e) at (1, 0) {};
					\vertex (f) at (-1,-2) {$(a,i)$}; \vertex (g) at (1,-2) {$(a,j)$};
					\propag[boson] (a) to (e); \propag[sca] (c) to [edge label=$q$](d); \propag[boson] (d) to (f);
					\propag[boson] (c) to (e); \propag[boson] (d) to (e);   \propag[boson] (c) to (b); \propag[boson] (e) to (g);
				\end{feynhand}
			\end{tikzpicture}
			\caption{}
		\end{subfigure}
		~
		\begin{subfigure}{0.15\textwidth}
			\begin{tikzpicture}[scale=0.55]
				\begin{feynhand}
					\vertex (a) at (1,2) {$(a,j)$}; \vertex (b) at (-1,2) {$(a,i)$}; \vertex (c) at (1,1); \vertex (d) at (1,-1);  \vertex[dot] (e) at (-1, 0){};
					\vertex (f) at (1,-2) {$(a,j)$}; \vertex (g) at (-1,-2) {$(a,i)$};
					\propag[boson] (a) to (e); \propag[sca] (c) to [edge label=$q$](d); \propag[boson] (d) to (f);
					\propag[boson] (c) to (e); \propag[boson] (d) to (e);   \propag[boson] (c) to (b); \propag[boson] (e) to (g);
				\end{feynhand}
			\end{tikzpicture}
			\caption{}
		\end{subfigure}
		~
		\begin{subfigure}{0.15\textwidth}
			\begin{tikzpicture}[scale=0.55]
				\begin{feynhand}
					\vertex (a) at (-1,1) {$(a,i)$}; \vertex (b) at (1,1) {$(a,j)$}; \vertex[ringdot] (c) at (0,0){}; \vertex (d) at (-1,-1); \vertex (e) at (1,-1); \vertex (f) at (-1,-2) {$(a,i)$}; \vertex (g) at (1,-2) {$(a,j)$};
					\propag[boson] (a) to (c); \propag[boson] (b) to (c); \propag[boson] (c) to (d); \propag[boson] (c) to (e);
					\propag[boson] (d) to (f); \propag[boson] (e) to (g);
					\propag[sca] (d) to[edge label=$q$] (e);
				\end{feynhand}
			\end{tikzpicture}
			\caption{}
		\end{subfigure}
		~
		\begin{subfigure}{0.15\textwidth}
			\begin{tikzpicture}[scale=0.55]
				\begin{feynhand}
					\vertex (a) at (-1,1); \vertex (b) at (1,1); \vertex [ringdot] (c) at (0,0) {}; \vertex (d) at (-1,-1) {$(a,i)$}; \vertex (e) at (1,-1){$(a,j)$}; \vertex (f) at (-1,2) {$(a,i)$}; \vertex (g) at (1,2) {$(a,j)$};
					\propag[boson] (a) to (c); \propag[boson] (b) to (c); \propag[boson] (c) to (d); \propag[boson] (c) to (e);
					\propag[boson] (a) to (f); \propag[boson] (b) to (g);
					\propag[sca] (a) to[edge label=$q$] (b);
				\end{feynhand}
			\end{tikzpicture}
			\caption{}
		\end{subfigure}
		~
		\begin{subfigure}{0.15\textwidth}
			\begin{tikzpicture}[scale=0.55]
				\begin{feynhand}
					\vertex (a) at (-1,2) {$(a,i)$}; \vertex (b) at (1,2) {$(a,j)$}; \vertex (c) at (-1,1); \vertex (d) at (-1,-1);  \vertex[ringdot] (e) at (1, 0){};
					\vertex (f) at (-1,-2) {$(a,i)$}; \vertex (g) at (1,-2) {$(a,j)$};
					\propag[boson] (a) to (c); \propag[sca] (c) to [edge label=$q$](d); \propag[boson] (d) to (f);
					\propag[boson] (c) to (e); \propag[boson] (d) to (e);   \propag[boson] (e) to (b); \propag[boson] (e) to (g);
				\end{feynhand}
			\end{tikzpicture}
			\caption{}
		\end{subfigure}
		~
		\begin{subfigure}{0.15\textwidth}
			\begin{tikzpicture}[scale=0.55]
				\begin{feynhand}
					\vertex (a) at (1,2) {$(a,j)$}; \vertex (b) at (-1,2) {$(a,i)$}; \vertex (c) at (1,1); \vertex (d) at (1,-1);  \vertex[ringdot] (e) at (-1, 0){};
					\vertex (f) at (1,-2) {$(a,j)$}; \vertex (g) at (-1,-2) {$(a,i)$};
					\propag[boson] (a) to (c); \propag[sca] (c) to [edge label=$q$](d); \propag[boson] (d) to (f);
					\propag[boson] (c) to (e); \propag[boson] (d) to (e);   \propag[boson] (e) to (b); \propag[boson] (e) to (g);
				\end{feynhand}
			\end{tikzpicture}
			\caption{}
		\end{subfigure}
		~
		\begin{subfigure}{0.15\textwidth}
			\begin{tikzpicture}[scale=0.55]
				\begin{feynhand}
					\vertex (a) at (-1,2) {$(a,i)$}; \vertex (b) at (1,2) {$(a,j)$}; \vertex (c) at (-1,1); \vertex (d) at (-1,-1);  \vertex[ringdot] (e) at (1, 0) {};
					\vertex (f) at (-1,-2) {$(a,i)$}; \vertex (g) at (1,-2) {$(a,j)$};
					\propag[boson] (a) to (e); \propag[sca] (c) to [edge label=$q$](d); \propag[boson] (d) to (f);
					\propag[boson] (c) to (e); \propag[boson] (d) to (e);   \propag[boson] (c) to (b); \propag[boson] (e) to (g);
				\end{feynhand}
			\end{tikzpicture}
			\caption{}
		\end{subfigure}
		~
		\begin{subfigure}{0.15\textwidth}
			\begin{tikzpicture}[scale=0.55]
				\begin{feynhand}
					\vertex (a) at (1,2) {$(a,j)$}; \vertex (b) at (-1,2) {$(a,i)$}; \vertex (c) at (1,1); \vertex (d) at (1,-1);  \vertex[ringdot] (e) at (-1, 0){};
					\vertex (f) at (1,-2) {$(a,j)$}; \vertex (g) at (-1,-2) {$(a,i)$};
					\propag[boson] (a) to (e); \propag[sca] (c) to [edge label=$q$](d); \propag[boson] (d) to (f);
					\propag[boson] (c) to (e); \propag[boson] (d) to (e);   \propag[boson] (c) to (b); \propag[boson] (e) to (g);
				\end{feynhand}
			\end{tikzpicture}
			\caption{}
		\end{subfigure}
		\caption{One-loop corrections to $u_1$ from the $\Gamma_M$ vertex} \label{fig:FyenmanDiagramu1Corrections}
	\end{figure}
	
	Feynman diagrams of one-loop corrections to the $u_1$ vertex from the random mass vertex are given in Fig. \ref{fig:FyenmanDiagramu1Corrections}. Calculations of these diagrams are straightforward. Here, we present our results only.
	
	\begin{align}
		(a)_{u_1}&=(b)_{u_1}=(c)_{u_1}=(d)_{u_1}=(e)_{u_1}=(f)_{u_1}\nonumber\\
		&=-32\Gamma_M u_1\int\frac{d^{d-2}\mathbf{Q}_\perp}{(2\pi)^{d-2}}\int\frac{d^2\vec{q}}{(2\pi)^2}\frac{|\mathbf{Q}_\perp|^{\alpha}+\kappa |\vec{q}|^{\alpha}}{[c_\perp^2|\mathbf{Q}_\perp|^2+c^2|\vec{q}|^2]^2}=-\frac{4\Gamma_Mu_1}{\pi^2 c^2c_\perp^2}\Big(1+\frac{\pi}{2}\frac{c_\perp}{c}\kappa\Big)\frac{2}{\epsilon+\bar{\epsilon}}\\
		(g)_{u_1}&=(h)_{u_1}=(i)_{u_1}=(j)_{u_1}=(k)_{u_1}=(l)_{u_1}\nonumber\\
		&=-4\Gamma_Mu_2\Big(4Tr[\tau^i\tau^i\tau^j\tau^j]+2Tr[\tau^i\tau^j\tau^i\tau^j]\Big) \int\frac{d^{d-2}\mathbf{Q}_\perp}{(2\pi)^{d-2}}\int\frac{d^2\vec{q}}{(2\pi)^2}\frac{|\mathbf{Q}_\perp|^{\alpha}+\kappa |\vec{q}|^{\alpha}}{[c_\perp^2|\mathbf{Q}_\perp|^2+c^2|\vec{q}|^2]^2}\nonumber\\
		&=-\frac{4\Gamma_Mu_2}{\pi^2 c^2c_\perp^2 N_c}\Big(1+\frac{\pi}{2}\frac{c_\perp}{c}\kappa\Big)\frac{2}{\epsilon+\bar{\epsilon}} .
	\end{align}
	Here, we used $Tr[\tau^i\tau^i\tau^j\tau^j]=\frac{4}{N_c}(i\neq j)$ from Eq. \eqref{eq:SpinorMatrixIdentity}.

	\subsubsection{Corrections to $u_2$ from the $\Gamma_M$ vertex}
	
	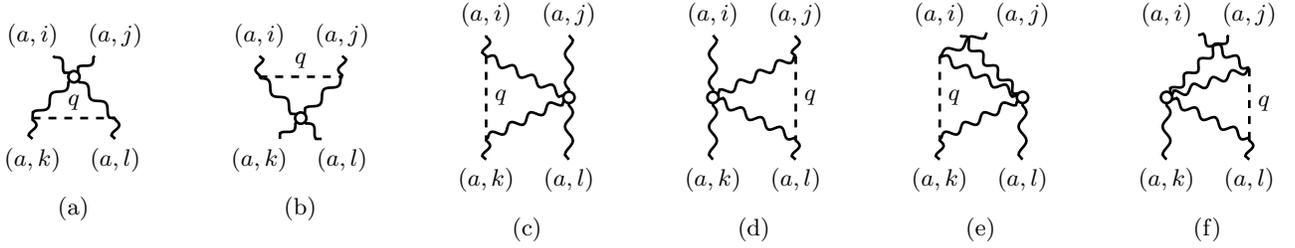
\begin{figure}
		\begin{subfigure}{0.15\textwidth}
			\begin{tikzpicture}[scale=0.55]
				\begin{feynhand}
					\vertex (a) at (-1,1) {$(a,i)$}; \vertex (b) at (1,1) {$(a,j)$}; \vertex[ringdot] (c) at (0,0){}; \vertex (d) at (-1,-1); \vertex (e) at (1,-1); \vertex (f) at (-1,-2) {$(a,k)$}; \vertex (g) at (1,-2) {$(a,l)$};
					\propag[boson] (a) to (c); \propag[boson] (b) to (c); \propag[boson] (c) to (d); \propag[boson] (c) to (e);
					\propag[boson] (d) to (f); \propag[boson] (e) to (g);
					\propag[sca] (d) to[edge label=$q$] (e);
				\end{feynhand}
			\end{tikzpicture}
			\caption{}
		\end{subfigure}
		~
		\begin{subfigure}{0.15\textwidth}
			\begin{tikzpicture}[scale=0.55]
				\begin{feynhand}
					\vertex (a) at (-1,1); \vertex (b) at (1,1); \vertex [ringdot] (c) at (0,0) {}; \vertex (d) at (-1,-1) {$(a,k)$}; \vertex (e) at (1,-1){$(a,l)$}; \vertex (f) at (-1,2) {$(a,i)$}; \vertex (g) at (1,2) {$(a,j)$};
					\propag[boson] (a) to (c); \propag[boson] (b) to (c); \propag[boson] (c) to (d); \propag[boson] (c) to (e);
					\propag[boson] (a) to (f); \propag[boson] (b) to (g);
					\propag[sca] (a) to[edge label=$q$] (b);
				\end{feynhand}
			\end{tikzpicture}
			\caption{}
		\end{subfigure}
		~
		\begin{subfigure}{0.15\textwidth}
			\begin{tikzpicture}[scale=0.55]
				\begin{feynhand}
					\vertex (a) at (-1,2) {$(a,i)$}; \vertex (b) at (1,2) {$(a,j)$}; \vertex (c) at (-1,1); \vertex (d) at (-1,-1);  \vertex[ringdot] (e) at (1, 0){};
					\vertex (f) at (-1,-2) {$(a,k)$}; \vertex (g) at (1,-2) {$(a,l)$};
					\propag[boson] (a) to (c); \propag[sca] (c) to [edge label=$q$](d); \propag[boson] (d) to (f);
					\propag[boson] (c) to (e); \propag[boson] (d) to (e);   \propag[boson] (e) to (b); \propag[boson] (e) to (g);
				\end{feynhand}
			\end{tikzpicture}
			\caption{}
		\end{subfigure}
		~
		\begin{subfigure}{0.15\textwidth}
			\begin{tikzpicture}[scale=0.55]
				\begin{feynhand}
					\vertex (a) at (1,2) {$(a,j)$}; \vertex (b) at (-1,2) {$(a,i)$}; \vertex (c) at (1,1); \vertex (d) at (1,-1);  \vertex[ringdot] (e) at (-1, 0){};
					\vertex (f) at (1,-2) {$(a,l)$}; \vertex (g) at (-1,-2) {$(a,k)$};
					\propag[boson] (a) to (c); \propag[sca] (c) to [edge label=$q$](d); \propag[boson] (d) to (f);
					\propag[boson] (c) to (e); \propag[boson] (d) to (e);   \propag[boson] (e) to (b); \propag[boson] (e) to (g);
				\end{feynhand}
			\end{tikzpicture}
			\caption{}
		\end{subfigure}
		~
		\begin{subfigure}{0.15\textwidth}
			\begin{tikzpicture}[scale=0.55]
				\begin{feynhand}
					\vertex (a) at (-1,2) {$(a,i)$}; \vertex (b) at (1,2) {$(a,j)$}; \vertex (c) at (-1,1); \vertex (d) at (-1,-1);  \vertex[ringdot] (e) at (1, 0) {};
					\vertex (f) at (-1,-2) {$(a,k)$}; \vertex (g) at (1,-2) {$(a,l)$};
					\propag[boson] (a) to (e); \propag[sca] (c) to [edge label=$q$](d); \propag[boson] (d) to (f);
					\propag[boson] (c) to (e); \propag[boson] (d) to (e);   \propag[boson] (c) to (b); \propag[boson] (e) to (g);
				\end{feynhand}
			\end{tikzpicture}
			\caption{}
		\end{subfigure}
		~
		\begin{subfigure}{0.15\textwidth}
			\begin{tikzpicture}[scale=0.55]
				\begin{feynhand}
					\vertex (a) at (1,2) {$(a,j)$}; \vertex (b) at (-1,2) {$(a,i)$}; \vertex (c) at (1,0.7); \vertex (d) at (1,-1);  \vertex[ringdot] (e) at (-1, 0){};
					\vertex (f) at (1,-2) {$(a,l)$}; \vertex (g) at (-1,-2) {$(a,k)$};
					\propag[boson] (a) to (e); \propag[sca] (c) to [edge label=$q$](d); \propag[boson] (d) to (f);
					\propag[boson] (c) to (e); \propag[boson] (d) to (e);   \propag[boson] (c) to (b); \propag[boson] (e) to (g);
				\end{feynhand}
			\end{tikzpicture}
			\caption{}
		\end{subfigure}
		\caption{One-loop corrections to $u_2$}
	\end{figure}
	
	\begin{align}
		(a)_{u_2}&=(b)_{u_2}=(c)_{u_2}=(d)_{u_2}=(e)_{u_2}=(f)_{u_2}\nonumber\\
		&=-4\Gamma_M u_2\Big[Tr[\tau^i\tau^j\tau^k\tau^l]+Tr[\tau^i\tau^j\tau^l\tau^k]+Tr[\tau^i\tau^k\tau^j\tau^l]+Tr[\tau^i\tau^k\tau^l\tau^j]\nonumber \\
		&+Tr[\tau^i\tau^l\tau^j\tau^k]+Tr[\tau^i\tau^l\tau^k\tau^j]\Big]\int\frac{d^{d-2}\mathbf{Q}_\perp}{(2\pi)^{d-2}}\int \frac{d^2q}{(2\pi)^2}\frac{c_\perp^2|\mathbf{Q}_\perp|^\alpha+\kappa |\vec{q}|^{\alpha}}{[c_\perp^2 |\mathbf{Q}_\perp|^2+c^2|\vec{q}|^2]^2}\nonumber\\
		&=-\frac{\Gamma_M u_2}{2\pi^2 c^2c_\perp^2}\Big(1+\frac{\pi}{2}\frac{c_\perp}{c}\kappa\Big)\frac{2}{\epsilon+\bar{\epsilon}}\Big[Tr[\tau^i\tau^j\tau^k\tau^l]+Tr[\tau^i\tau^j\tau^l\tau^k]+Tr[\tau^i\tau^k\tau^j\tau^l]+Tr[\tau^i\tau^k\tau^l\tau^j]\nonumber \\
		&+Tr[\tau^i\tau^l\tau^j\tau^k]+Tr[\tau^i\tau^l\tau^k\tau^j]\Big]
	\end{align}
	
	The above one-loop results for both the $u_1$ and $u_2$ vertices can be easily reproduced based on the cut-off regularization scheme.

	\subsection{One-loop corrections to the $\Gamma_M$ vertex} \label{Appendix:OneLoopRBMVertex}

		\begin{figure}[h]
		\begin{subfigure}[b]{0.15\textwidth}
			\begin{tikzpicture}[scale=0.7]
				\begin{feynhand}
					\vertex (a1) at (0,0) {$(a,i)$}; \vertex (a2) at (0,1) ; \vertex (a3) at (0,2); \vertex (a4) at (0,3) {$(a,i)$};
					\vertex (b1) at (1.5,0) {$(b,j)$}; \vertex (b2) at (1.5,1); \vertex (b3) at (1.5,2); \vertex (b4) at (1.5,3) {$(b,j)$};
					\propag[boson] (a1) to (a2); \propag[boson] (a2) to [edge label=$-q$](a3); \propag[boson] (a3) to (a4);
					\propag[boson] (b1) to (b2); \propag[boson] (b2) to [edge label=$q$](b3); \propag[boson] (b3) to (b4);
					\propag[sca] (a2) to [edge label=$q$](b2); \propag[sca] (a3) to [edge label=$-q-p$](b3);
				\end{feynhand}
			\end{tikzpicture}
			\caption{}
		\end{subfigure}
		~
		\begin{subfigure}[b]{0.15\textwidth}
			\begin{tikzpicture}[scale=0.7]
				\begin{feynhand}
					\vertex (a1) at (0,0) {$(a,i)$}; \vertex (a2) at (0,1); \vertex (a3) at (0,2); \vertex (a4) at (0,3) {$(a,i)$};
					\vertex (b1) at (1.5,0) {$(b,j)$}; \vertex (b2) at (1.5,1); \vertex (b3) at (1.5,2); \vertex (b4) at (1.5,3) {$(b,j)$};
					\propag[boson] (a1) to (a2); \propag[boson] (a2) to [edge label=$-q$](a3); \propag[boson] (a3) to (a4);
					\propag[boson] (b1) to (b2); \propag[boson] (b2) to (b3); \propag[boson] (b3) to (b4);
					\propag[sca] (a2) to (b3); \propag[sca] (a3) to (b2);
					\node at (2.1,1.5) {$-q-p$}; \node at (0.6, 1) {$q$}; \node at (0.9,2.1) {$-q-p$};
				\end{feynhand}
			\end{tikzpicture}
			\caption{}
		\end{subfigure}
		~
		\begin{subfigure}[b]{0.15\textwidth}
			\begin{tikzpicture}[scale=0.55]
				\begin{feynhand}
					\vertex (a1) at (0,0) {$(a,i)$}; \vertex (a2) at (0,1); \vertex (a3) at (0,2); \vertex (a4) at (0, 3); \vertex(a5) at (0, 4) {$(a,i)$};
					\vertex (b1) at (2,0) {$(b,j)$}; \vertex (b2) at (2,1); \vertex (b3) at (2,2); \vertex (b4) at (2,3); \vertex(b5) at (2, 4) {$(b,j)$};
					\propag[boson] (a1) to (a2); \propag[boson] (a2) to (a3); \propag[boson] (a3) to (a4); \propag[boson] (a4) to (a5);
					\propag[boson] (b1) to (b2); \propag[boson] (b2) to (b3); \propag[boson] (b3) to (b4); \propag[boson] (b4) to (b5);
					\propag[sca] (a3) to (b3); \propag[sca, looseness=1.5] (a2) to [out=180, in=180](a4);
				\end{feynhand}
			\end{tikzpicture}
			\caption{}
		\end{subfigure}
		~
		\begin{subfigure}[b]{0.15\textwidth}
			\begin{tikzpicture}[scale=0.55]
				\begin{feynhand}
					\vertex (a1) at (0,0) {$(a,i)$}; \vertex (a2) at (0,1); \vertex (a3) at (0,2); \vertex (a4) at (0, 3); \vertex(a5) at (0, 4) {$(a,i)$};
					\vertex (b1) at (2,0) {$(b,j)$}; \vertex (b2) at (2,1); \vertex (b3) at (2,2); \vertex (b4) at (2,3); \vertex(b5) at (2, 4) {$(b,j)$};
					\propag[boson] (a1) to (a2); \propag[boson] (a2) to (a3); \propag[boson] (a3) to (a4); \propag[boson] (a4) to (a5);
					\propag[boson] (b1) to (b2); \propag[boson] (b2) to (b3); \propag[boson] (b3) to (b4); \propag[boson] (b4) to (b5);
					\propag[sca] (a3) to (b3); \propag[sca, looseness=1.5] (b2) to [out=0, in=0](b4);
				\end{feynhand}
			\end{tikzpicture}
			\caption{}
		\end{subfigure}
		~
		\begin{subfigure}[b]{0.15\textwidth}
			\begin{tikzpicture}[scale=0.55]
				\begin{feynhand}
					\vertex (a1) at (0,0) {$(a,i)$}; \vertex (a2) at (0,1); \vertex (a3) at (0,2); \vertex (a4) at (0, 3); \vertex(a5) at (0, 4) {$(a,i)$};
					\vertex (b1) at (2,0) {$(b,j)$}; \vertex (b2) at (2,1); \vertex[dot] (b3) at (2,2){}; \vertex (c) at (1,2); \vertex (b4) at (2,3); \vertex(b5) at (2, 4) {$(b,j)$};
					\propag[boson] (a1) to (a2); \propag[boson] (a2) to (a3); \propag[boson] (a3) to (a4); \propag[boson] (a4) to (a5);
					\propag[boson] (b1) to (b2); \propag[boson] (b2) to(b3); \propag[boson] (b3) to (b4); \propag[boson] (b4) to (b5);
					\propag[sca] (a3) to (c); \propag[boson, looseness=1.5] (c) to [out=60, in=120](b3); \propag[boson, looseness=1.5] (c) to [out=-60, in=-120](b3);
				\end{feynhand}
			\end{tikzpicture}
			\caption{}
		\end{subfigure}
		~
		\begin{subfigure}[b]{0.15\textwidth}
			\begin{tikzpicture}[scale=0.55]
				\begin{feynhand}
					\vertex (a1) at (0,0) {$(a,i)$}; \vertex (a2) at (0,1); \vertex (a3) at (0,2); \vertex (a4) at (0, 3); \vertex(a5) at (0, 4) {$(a,i)$};
					\vertex (b1) at (2,0) {$(b,j)$}; \vertex (b2) at (2,1); \vertex[dot] (b3) at (2,2){}; \vertex (c) at (1,2); \vertex (b4) at (2,3); \vertex(b5) at (2, 4) {$(b,j)$};
					\propag[boson] (a1) to (a2); \propag[boson] (a2) to (a3); \propag[boson] (a3) to (a4); \propag[boson] (a4) to (a5);
					\propag[boson] (b1) to (b2); \propag[boson] (b2) to(b3); \propag[boson] (b3) to (b4); \propag[boson] (b4) to (b5);
					\propag[sca] (b3) to (c); \propag[boson, looseness=1.5] (c) to [out=120, in=60](a3); \propag[boson, looseness=1.5] (c) to [out=-120, in=-60](a3);
				\end{feynhand}
			\end{tikzpicture}
			\caption{}
		\end{subfigure}
		~
		\begin{subfigure}[b]{0.15\textwidth}
			\begin{tikzpicture}[scale=0.55]
				\begin{feynhand}
					\vertex (a1) at (0,0) {$(a,i)$}; \vertex (a2) at (0,1); \vertex (a3) at (0,2); \vertex (a4) at (0, 3); \vertex(a5) at (0, 4) {$(a,i)$};
					\vertex (b1) at (2,0) {$(b,j)$}; \vertex (b2) at (2,1); \vertex[ringdot] (b3) at (2,2){}; \vertex (c) at (1,2); \vertex (b4) at (2,3); \vertex(b5) at (2, 4) {$(b,j)$};
					\propag[boson] (a1) to (a2); \propag[boson] (a2) to (a3); \propag[boson] (a3) to (a4); \propag[boson] (a4) to (a5);
					\propag[boson] (b1) to (b2); \propag[boson] (b2) to(b3); \propag[boson] (b3) to (b4); \propag[boson] (b4) to (b5);
					\propag[sca] (a3) to (c); \propag[boson, looseness=1.5] (c) to [out=60, in=120](b3); \propag[boson, looseness=1.5] (c) to [out=-60, in=-120](b3);
				\end{feynhand}
			\end{tikzpicture}
			\caption{}
		\end{subfigure}
		~
		\begin{subfigure}[b]{0.15\textwidth}
			\begin{tikzpicture}[scale=0.55]
				\begin{feynhand}
					\vertex (a1) at (0,0) {$(a,i)$}; \vertex (a2) at (0,1); \vertex (a3) at (0,2); \vertex (a4) at (0, 3); \vertex(a5) at (0, 4) {$(a,i)$};
					\vertex (b1) at (2,0) {$(b,j)$}; \vertex (b2) at (2,1); \vertex[ringdot] (b3) at (2,2){}; \vertex (c) at (1,2); \vertex (b4) at (2,3); \vertex(b5) at (2, 4) {$(b,j)$};
					\propag[boson] (a1) to (a2); \propag[boson] (a2) to (a3); \propag[boson] (a3) to (a4); \propag[boson] (a4) to (a5);
					\propag[boson] (b1) to (b2); \propag[boson] (b2) to(b3); \propag[boson] (b3) to (b4); \propag[boson] (b4) to (b5);
					\propag[sca] (b3) to (c); \propag[boson, looseness=1.5] (c) to [out=120, in=60](a3); \propag[boson, looseness=1.5] (c) to [out=-120, in=-60](a3);
				\end{feynhand}
			\end{tikzpicture}
			\caption{}
		\end{subfigure}
		\caption{One-loop corrections to $\Gamma_M$}\label{fig:GammaMOneLoopDiagrams}
	\end{figure}
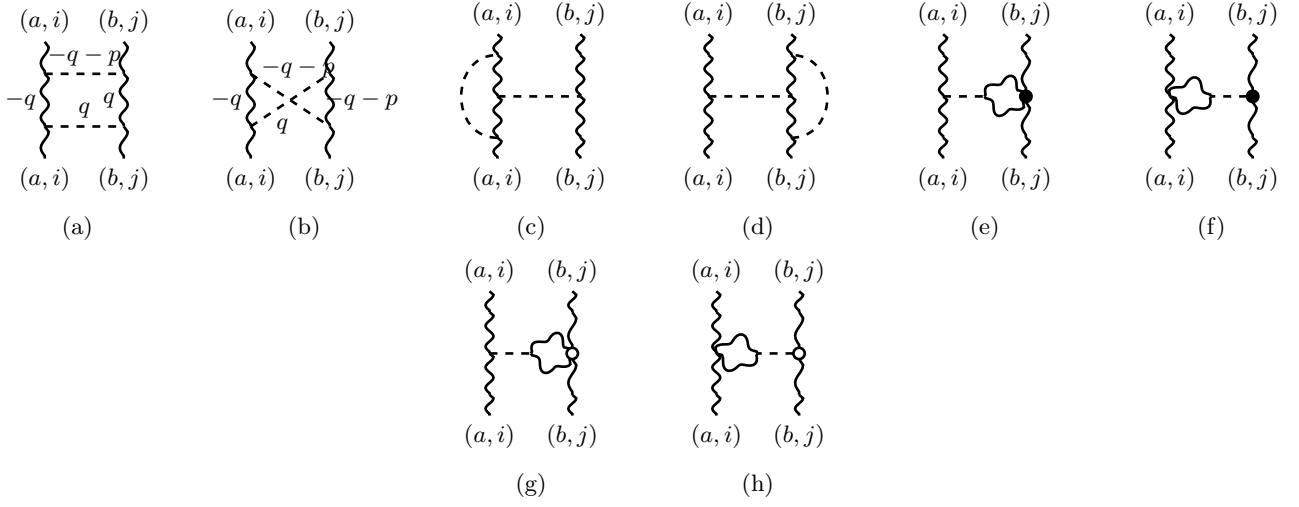

	There are four types of one-loop corrections to the $\Gamma_M$ vertex, shown in Fig. \ref{fig:GammaMOneLoopDiagrams}. It turns out that only Feynman diagrams (c) and (d) give corrections while Feynman diagrams (a) and (b) do not because of the non-local form of the $\Gamma_M$ vertex. On the other hand, it seems that there is a $\frac{1}{\epsilon+\bar{\epsilon}}$-pole in the calculation of the Feynman diagram (a), using the co-dimensional regularization. We suspect that this is an artifact of the naive calculation. We discuss the reason why it is not a true UV-correction in the cut-off regularization method. We present our calculations of diagram (a) and diagram (c), here. Calculations of (b) and (d) are similar to those of (a) and (b).
	
	First, we consider the Feynman diagram (a). The integral can be divided into four parts; $(a)_{\Gamma}^{(1)}$, $(a)_{\Gamma}^{(2)}$, $(a)_{\Gamma}^{(3)}$, and $(a)_{\Gamma}^{(4)}$ as follows
	\begin{align}
		(a)_{\Gamma_M}&=16\Gamma_M^2 \int\frac{d^{d-2}\mathbf{Q}_\perp}{(2\pi)^{d-2}}\int\frac{d^2q}{(2\pi)^2}\frac{\Big(|\mathbf{Q}_\perp|^\alpha+\kappa |\vec{q}|^\alpha \Big)\Big(|\mathbf{Q}_\perp+\mathbf{P}_\perp|^\alpha+\kappa |\vec{q}+\vec{p}|^\alpha \Big)}{[c_\perp^2|\mathbf{Q}_\perp|^2+c^2|\vec{q}|^2]^2}\nonumber\\
		&= \frac{16\Gamma_M^2}{c^2c_\perp^2}\Big[(a)_{\Gamma_M}^{(1)}+\kappa c^{-\alpha}c_{\perp}^\alpha(a)_{\Gamma_M}^{(2)}+\kappa c^{-\alpha}c_{\perp}^\alpha(a)_{\Gamma_M}^{(3)}+\kappa^2c^{-2\alpha}c_{\perp}^{2\alpha}(a)_{\Gamma_M}^{(4)}\Big]\nonumber\\
		&=-\frac{2\Gamma_M^2}{\pi^2 c^2c_\perp^2}\Big(1+\frac{\pi}{2}\frac{c_\perp}{c}\kappa\Big)\frac{2}{\epsilon+\bar{\epsilon}}\Big(|\mathbf{P}_\perp|+\kappa |\vec{p}|\Big) . \label{eq:(a)GammaM}
	\end{align}
	Here, $(a)_{\Gamma_M}^{(i)}$ are given by
	\begin{align}
		(a)_{\Gamma_M}^{(1)}&=\int\frac{d^{d-2}\mathbf{Q}_\perp}{(2\pi)^{d-2}}\int\frac{d^2q}{(2\pi)^2}\frac{1}{\Big[|\mathbf{Q}_\perp|^2+|\vec{q}'|^2\Big]^{2}\Big[|\mathbf{Q}_\perp|^2\Big]^{-\alpha/2}\Big[|\mathbf{Q}_\perp+\mathbf{P}_\perp|^2 \Big]^{-\alpha/2}} \nonumber\\
		&=\frac{|\mathbf{P}_\perp|^{-4+2\alpha+d}}{(4\pi)^{d/2}}\frac{\Gamma(2-\alpha-d/2)}{\Gamma(1-\alpha/2)\Gamma(-\alpha/2)}\frac{\Gamma(-2+(d+\alpha)/2)\Gamma(-1+(d+\alpha)/2)}{\Gamma(-3+d+\alpha)}=-\frac{|\mathbf{P}_\perp|}{8\pi^2}\frac{2}{\epsilon+\bar{\epsilon}}\\
		(a)_{\Gamma_M}^{(2)}&=\int\frac{d^{d-2}\mathbf{Q}_\perp}{(2\pi)^{d-2}}\int\frac{d^2q}{(2\pi)^2}\frac{1}{\Big[|\mathbf{Q}_\perp|^2+|\vec{q}'|^2\Big]^{2}\Big[|\mathbf{Q}_\perp|^2\Big]^{-\alpha/2}\Big[|\vec{q}+\vec{p}'|^2 \Big]^{-\alpha/2}} \nonumber\\
		&=\frac{|\vec{p}'|^{-4+2\alpha+d}}{(4\pi)^{d/2}}\frac{\Gamma((d+\alpha)/2-1)\Gamma(2-\alpha-d/2)}{\Gamma(d/2-1)\Gamma(-\alpha/2)} \frac{\Gamma(-2+(\alpha+d)/2)\Gamma(\alpha/2+1)}{\Gamma(-1+\alpha+d/2)}=-\frac{c|\vec{p}|}{8\pi^2}\frac{2}{\epsilon+\bar{\epsilon}}\\
		(a)_{\Gamma_M}^{(3)}&=\int\frac{d^{d-2}\mathbf{Q}_\perp}{(2\pi)^{d-2}}\int\frac{d^2q}{(2\pi)^2}\frac{1}{\Big[|\mathbf{Q}_\perp|^2+|\vec{q}|^2\Big]^{2}\Big[|\mathbf{Q}_\perp+\mathbf{P}_\perp|^2\Big]^{-\alpha/2}\Big[|\vec{q}|^2 \Big]^{-\alpha/2}}\nonumber\\
		&=\frac{|\mathbf{P}_\perp|^{-4+2\alpha+d}}{(4\pi)^{d/2}}\frac{\Gamma(\alpha/2+1)\Gamma(2-\alpha-d/2)}{\Gamma(-\alpha/2)} \frac{\Gamma(-2+(\alpha+d)/2)\Gamma(-1+(d+\alpha)/2)}{\Gamma(-3+\alpha+d)}=-\frac{|\mathbf{P}_\perp|}{16\pi }\frac{2}{\epsilon+\bar{\epsilon}}\\
		(a)_{\Gamma_M}^{(4)}&=\int\frac{d^{d-2}\mathbf{Q}_\perp}{(2\pi)^{d-2}}\int\frac{d^2q}{(2\pi)^2}\frac{1}{\Big[|\mathbf{Q}_\perp|^2+|\vec{q}|^2\Big]^{2}\Big[|\vec{q}+\vec{p}'|^2\Big]^{-\alpha/2}\Big[|\vec{q}|^2 \Big]^{-\alpha/2}}\nonumber\\
		&=\frac{c^{-4+2\alpha+d}|\vec{p}|^{-4+2\alpha+d}}{(4\pi)^{d/2}} \frac{\Gamma(3-d/2)\Gamma(2-\alpha-d/2)}{\Gamma(-\alpha/2)\Gamma(3-(\alpha+d)/2)} \frac{\Gamma(-2+(\alpha+d)/2)\Gamma(\alpha/2+1)}{\Gamma(-1+\alpha+d/2)}=-\frac{c|\vec{p}|}{16\pi}\frac{2}{\epsilon+\bar{\epsilon}} .
	\end{align}
	
	According to Eq. \ref{eq:(a)GammaM}, the diagram (a) gives a UV correction. However, it is an artifact from the dimensional regularization. We point out that $\frac{1}{\epsilon+\bar{\epsilon}}$-poles are coming from the $\Gamma(-2+(\alpha+d)/2)$ in $(a)_{\Gamma_M}^{(i)}$. It seems that there are UV corrections regularized as a $\frac{1}{\epsilon+\bar{\epsilon}}$-pole. However, $\epsilon+\bar{\epsilon}$ should be a negative value for $\Gamma(-2+(\alpha+d)/2)=\Gamma(-\frac{\epsilon+\bar{\epsilon}}{2})$ to be well defined as a finite value. This is not consistent with the fact that $\epsilon$ and $\bar{\epsilon}$ are positive values. More explicitly, we see that the diagram (a) does not give any UV correction, resorting to the cut-off regularization $\Lambda$ in $d=3$ and $\alpha=1$ as follows:
	\begin{align}
		(a)_{\Gamma_M,\Lambda}^{(1)}&=\int \frac{dQ_\perp}{2\pi}\int\frac{d^2q}{(2\pi)^2}\frac{|Q_\perp||Q_\perp+P_\perp|}{\Big[|Q_\perp|^2+|\vec{q}|^2\Big]^2}=\frac{1}{8\pi^2}\int_{-\Lambda}^{\Lambda}dQ_\perp\frac{|Q_\perp+P_\perp|}{|Q_\perp|}\nonumber\\
		&=\frac{1}{8\pi^2}\Big[2(\Lambda-|P_\perp|)+2|P_\perp|\ln|P_\perp|\Big]\rightarrow \text{No } |P_\perp| \ln\Lambda\\
		(a)_{\Gamma_M,\Lambda}^{(2)}&=\int \frac{dQ_\perp}{2\pi}\int\frac{d^2q}{(2\pi)^2}\frac{|Q_\perp||\vec{q}+\vec{p}'|}{\Big[|Q_\perp|^2+|\vec{q}|^2\Big]^2}=\frac{1}{2\pi}\int\frac{d^2q}{(2\pi)^2}\frac{|\vec{q}+\vec{p}'|}{|q|^2}\nonumber\\
		&=\frac{1}{(2\pi)^3}\int_0^{2\pi} d\theta \int_0^{\Lambda} dq\frac{\sqrt{q^2+|\vec{p}'|^2+2q|\vec{p}'|\cos\theta}}{q}\nonumber\\
		&\approx \frac{1}{(2\pi)^3}\int_0^{2\pi}d\theta \int_0^\Lambda dq \Big(1+\frac{|\vec{p}'|^2}{2q^2}(1-\cos^2\theta)\Big)\rightarrow \text{No } |\vec{p}'|\ln\Lambda\\
		(a)_{\Gamma_M,\Lambda}^{(3)}&=a_{\Gamma_M,\Lambda}^{(1)},\;\;	(a)_{\Gamma_M,\Lambda}^{(4)}=(a)_{\Gamma_M,\Lambda}^{(2)}.
	\end{align}
	Based on the same argument, the diagram (b) does not give any UV-correction, neither.
	
	Now let us calculate the diagram (c) in the following way
	
	\begin{align}
		(c)_{\Gamma_M}&=16\Gamma_M^2\Big[|\mathbf{P}_\perp|^\alpha+\kappa |\vec{p}|^\alpha \Big]\int\frac{d^{d-2}\mathbf{Q}_\perp}{(2\pi)^{d-2}}\int\frac{d^2q}{(2\pi)^2}\frac{\Big[|\mathbf{Q}_\perp|^\alpha +\kappa |\vec{q}|^{\alpha}\Big]}{\Big[c_\perp^2|\mathbf{Q}_\perp|^2+c^2|\vec{q}|^2\Big]\Big[c_\perp^2|\mathbf{Q}_\perp+\mathbf{P}_\perp|^2+c^2|\vec{q}+\vec{p}|^2\Big]}(\vec{p}'=c_\perp^{-1}c\vec{p})\nonumber\\
		&=\frac{16\Gamma_M^2}{c^2c_\perp^2}\Big[|\mathbf{P}_\perp|^\alpha+\kappa |\vec{p}|^{\alpha}\Big]\Big((c)_{\Gamma_M}^{(1)}+\kappa c^{-\alpha}c_\perp^\alpha(c)_{\Gamma_M}^{(2)}\Big)=\frac{2\Gamma_M^2}{\pi^2 c^2c_\perp^2}\Big[|\mathbf{P}_\perp|^\alpha+\kappa |\vec{p}|^{\alpha}\Big]\Big(1+\frac{\pi}{2}\frac{c_\perp}{c}\kappa\Big)\frac{2}{\epsilon+\bar{\epsilon}} ,
	\end{align}
	where
	\begin{align}
		(c)_{\Gamma_M}^{(1)}&=\int\frac{d^{d-2}\mathbf{Q}_\perp}{(2\pi)^{d-2}}\int\frac{d^2q}{(2\pi)^2}\frac{1}{\Big[|\mathbf{Q}|^2\Big]^{-\alpha/2}\Big[|\mathbf{Q}_\perp|^2+|\vec{q}|^2\Big]\Big[|\mathbf{Q}_\perp+\mathbf{P}_\perp|^2+|\vec{q}+\vec{p}'|^2\Big]}\approx \frac{1}{8\pi^2}\frac{2}{\epsilon+\bar{\epsilon}}\\
		(c)_{\Gamma_M}^{(2)}&=\int\frac{d^{d-2}\mathbf{Q}_\perp}{(2\pi)^{d-2}}\int\frac{d^2q}{(2\pi)^2}\frac{1}{\Big[|\vec{q}|^2\Big]^{-\alpha/2}\Big[|\mathbf{Q}_\perp|^2+|\vec{q}|^2\Big]\Big[|\mathbf{Q}_\perp+\mathbf{P}_\perp|^2+|\vec{q}+\vec{p}|^2\Big]}\approx \frac{1}{16\pi}\frac{2}{\epsilon+\bar{\epsilon}}
	\end{align}
	have been used. Using the cut-off regularization in $d=3$ and $\alpha=1$, the same result is reproduced as
	\begin{align}
		(c)_{\Gamma_M,\Lambda}=16\Gamma_M^2[|\mathbf{P}_\perp|+\kappa|\vec{p}|]\int\frac{d^3q}{(2\pi)^3}\frac{|Q_\perp|+\kappa|\vec{q}|}{[c_\perp^2|Q_\perp|^2+c^2|\vec{q}|^2]^2}=\frac{2\Gamma_M^2}{\pi^2 c^2c_\perp^2}[|\mathbf{P}_\perp|+\kappa|\vec{p}|]\Big(1+\frac{\pi}{2}\frac{c_\perp}{c}\kappa \Big)\ln\Lambda^2 .
	\end{align}
	The result of the diagram $(d)_{\Gamma_M}$ is the same as that of $(c)_{\Gamma_M}$.
	
	Finally, we consider the remaining Feynman diagrams ((e), (f), (g) and (h)). It is straightforward to perform as follows
	\begin{align}
		(e)_{\Gamma_M}=(f)_{\Gamma_M}&=-16u_1\Gamma_M(N_c^2+1)[|\mathbf{P}_\perp|^\alpha+\kappa |\vec{p}|^\alpha]\int\frac{d^d\mathbf{Q}}{(2\pi)^d}\int\frac{d^2q}{(2\pi)^2}\frac{1}{\Big[|Q_0|^2+c_\perp^2|\mathbf{Q}_\perp|^2+c^2|\vec{q}|^2\Big]^2}\nonumber\\
		&=-\frac{2(N_c^2+1)}{\pi^2c_\perp c^2}u_1\Gamma_M\frac{1}{\epsilon}\Big(|\mathbf{P}_\perp|^\alpha+\kappa|\vec{p}|^\alpha\Big)\\
		(g)_{\Gamma_M}=(h)_{\Gamma_M}&=-4u_2\Gamma_M\Big(2Tr[(\tau^j)^4]+\sum_{k=1}^{N_c^2-1}Tr[(\tau_k)^2(\tau_j)^2]\Big)\Big(|\mathbf{P}_\perp|^\alpha+\kappa|\vec{p}|^\alpha\Big)\int\frac{d^d\mathbf{Q}}{(2\pi)^d}\nonumber\\
		&\times \int\frac{d^2q}{(2\pi)^2}\frac{1}{\Big[|Q_0|^2+c_\perp^2|\mathbf{Q}_\perp|^2+c^2|\vec{q}|^2\Big]^2}=-\frac{2(N_c^2+1)}{N_c\pi^2c_\perp c^2}u_2\Gamma_M\frac{1}{\epsilon}\Big(|\mathbf{P}_\perp|^\alpha+\kappa|\vec{p}|^\alpha\Big) .
	\end{align}

	\section{Counter terms in one-loop calculations} \label{Appendix:OneLoopCounterTerms}
	
	Here, we present our results of one-loop counter terms. The fermion self-energy, boson self-energy, Yukawa vertex, and boson self-interaction corrections give the following counter terms from $A_0$ to $A_9$:
	\begin{align}
		A^{(1l)}_0&=-\frac{(N_c^2-1)}{4\pi^2N_cN_f}\frac{g^2}{c}h_1(c,c_\perp ,v)\frac{1}{\epsilon}-\frac{F_{dis}(\{\Gamma_i\},v)}{\epsilon}\\
		A^{(1l)}_{1}&=-\frac{(N_c^2-1)}{4\pi^2N_cN_f}\frac{g^2}{c}h_2(c,c_\perp,v)\frac{1}{\epsilon}\\
		A^{(1l)}_{2}&=\frac{(N_c^2-1)}{4\pi^2N_cN_f}\frac{g^2}{c}h_3(c,c_\perp,v)\frac{1}{\epsilon}\\
		A^{(1l)}_3&=-A^{1l}_2=-\frac{(N_c^2-1)}{4\pi^2N_cN_f}\frac{g^2}{c}h_3(c,c_\perp ,v)\frac{1}{\epsilon}\\
		A^{(1l)}_4&=-\frac{1}{4\pi}\frac{g^2}{v}\frac{1}{\epsilon}-\frac{\Gamma_M}{\pi^2c^2c_\perp^2}\Big(1+\frac{\pi}{2}\frac{c_\perp}{c}\kappa \Big)\frac{1}{\epsilon+\bar{\epsilon}}\\
		A^{(1l)}_5&=-\frac{1}{4\pi}\frac{g^2}{vc_\perp^2}\frac{1}{\epsilon}+\frac{\Gamma_M}{\pi^2 c^2c_\perp^2}\frac{1}{\epsilon+\bar{\epsilon}}\\
		A^{(1l)}_6&=\frac{\kappa \Gamma_M}{4\pi c^3c_\perp}\frac{1}{\epsilon+\bar{\epsilon}}\\
		A^{(1l)}_7&=-\frac{1}{8\pi^3N_cN_f}\frac{g^2}{c}h_4(c,c_\perp,v)\frac{1}{\epsilon}+\frac{G_{dis}(\{\Gamma_i\},v)}{\epsilon}\\
		A^{(1l)}_8&=\frac{1}{2\pi^2c^2c_\perp}\Big[(N_c^2+7)u_1+2\Big(2N_c-\frac{3}{N_c}\Big)u_2+3\Big(1+\frac{3}{N_c^2}\Big)\frac{u_2^2}{u_1}\Big]\frac{1}{\epsilon}-\frac{6\Gamma_M}{\pi^2 c^2c_\perp^2}\Big(1+\frac{\pi}{2}\frac{c_\perp}{c}\kappa\Big)\Big(1+\frac{u_2}{N_cu_1}\Big)\frac{1}{\epsilon+\bar{\epsilon}}\\
		A^{(1l)}_9&=\frac{1}{2\pi^2c^2c_\perp}\Big[12u_1+2\Big(N_c-\frac{9}{N_c}\Big)u_2\Big]\frac{1}{\epsilon}-\frac{6\Gamma_M}{\pi^2 c^2c_\perp^2}\Big(1+\frac{\pi}{2}\frac{c_\perp}{c}\kappa\Big)\frac{1}{\epsilon+\bar{\epsilon}}\\
		A^{(1l)}_{\Gamma_M}&=-\frac{2\Gamma_M}{\pi^2 c^2c_\perp^2}\Big(1+\frac{\pi}{2}\frac{c_\perp}{c}\kappa\Big)\frac{1}{\epsilon+\bar{\epsilon}}+\frac{N_c^2+1}{\pi^2c_\perp c^2}\Big(u_1+\frac{1}{N_c}u_2\Big)\frac{1}{\epsilon} ,
	\end{align}
	where
	\begin{align*}
		F_{dis}(\{\Gamma_{G}\},v)&=\frac{1}{2\pi^2N_f(1+v^2)}\Big(\Gamma_{G1}^d+\Gamma_{G5}^e+\Gamma_{G6}^e+\Gamma_{G7}^e+2\Gamma_{G8}^e+\Gamma_{G9}^e+\Gamma_{G10}^e\Big),\\
		G_{dis}(\{\Gamma_{G}\},v)&=\frac{e^{-v^2/v_c^2}}{2\pi^2N_f}\Big(\Gamma_{G2}^d+\Gamma_{G5}^e+\Gamma_{G11}^u+2\Gamma_{G13}^u+\Gamma_{G14}^u+2\Gamma_{G15}^u\Big),
		\end{align*}
	and
	\begin{align*}
		h_1(c,c_\perp,v)&=\int_0^1dx\sqrt{\frac{x}{\Big(1-x+xc_\perp^2\Big)\Big((1+v^2)(1-x)+xc^2\Big)}},\\
		h_2(c,c_\perp,v)&=c_\perp^2 \int_0^1 dx \sqrt{\frac{x}{\Big(1-x+xc_\perp^2\Big)^3\Big((1+v^2)(1-x)+xc^2\Big)}},\\
		h_3(c,c_\perp,v)&=c^2 \int_0^1 dx \sqrt{\frac{x}{\Big(1-x+xc_\perp^2\Big)\Big((1+v^2)(1-x)+xc^2\Big)^{3}}},\\
		h_4(c,c_\perp,v)&=\pi c\Big[\Big(1+g_3(c,c_\perp,v)\Big)\Big(g_1(c,c_\perp,v)g_2(c,c_\perp,v)-v^2(x-y)^2\Big)+g_3(c,c_\perp,v)\Big(g_2(c,c_\perp,v)-v^2g_1(c,c_\perp,v)\Big)\Big]\nonumber\\
		&\times \Big[g_3(c,c_\perp,v)\Big(g_1(c,c_\perp,v)g_2(c,c_\perp,v)-v^2(x-y)^2\Big)\Big]^{-3/2},\\
		&\Bigg\{
		\begin{array}{l}
			g_1(c,c_\perp,v)=x+y+c^2(1-x-y),\\
			g_2(c,c_\perp,v)=(x+y)v^2+c^2(1-x-y),\\
			g_3(c,c_\perp,v)=x+y+c_\perp^2(1-x-y)
		\end{array}.
	\end{align*}
	The factor of $e^{-v^2/v^2_c}$ in $G_{dis}(\{\Gamma_i,v\})$  is introduced to take into account the change of a phase space as the Fermi velocity $v$ is modified. More detailed explanations about this factor is given in Appendix \ref{Appendix:CalOfOneLoopFeynmanDiagrams}. 
	
	The counter terms of the random charge potential vertices are given by
	
	\begin{align}
		A_{\Gamma_{G1}^d}&=-\frac{1}{4 \pi^2 N_f  \Gamma_{G1}^d (1+v^2)}\frac{1}{\epsilon}\Big[(\Gamma_{G10}^e)^2+4 \Gamma_{G10}^e \Gamma_{G1}^d-(\Gamma_{G11}^u)^2+4 (\Gamma_{G1}^d)^2+4 \Gamma_{G2}^d \Gamma_{G5}^e+4 \Gamma_{G2}^d \Gamma_{G6}^e+4 \Gamma_{G3}^d \Gamma_{G7}^e\nonumber\\
		&+4 \Gamma_{G3}^d \Gamma_{G9}^e+8 \Gamma_{G4}^d \Gamma_{G8}^e+(\Gamma_{G5}^e)^2+(\Gamma_{G6}^e)^2+(\Gamma_{G7}^e)^2+2 (\Gamma_{G8}^e)^2+(\Gamma_{G9}^e)^2\Big]\nonumber\\
		&-\frac{ g^2  (N_c^2-1)}{2 \pi^2 c  N_c N_f}\frac{\Gamma_{G2}^d}{\Gamma_{G1}^d}\frac{-f_1(c,c_\perp,v)+f_2(c,c_\perp,v)+f_3(c,c_\perp,v)}{\epsilon}\\
		A_{\Gamma_{G2}^d}&=-\frac{1}{\pi^2N_f (1+v^2)\Gamma_{G2}^d}\frac{1}{\epsilon}\Big[\Gamma_{G10}^e \Gamma_{G2}^d+\Gamma_{G1}^d (\Gamma_{G2}^d+\Gamma_{G5}^e+\Gamma_{G6}^e)+2 \Gamma_{G3}^d \Gamma_{G8}^e+\Gamma_{G4}^d \Gamma_{G7}^e+\Gamma_{G4}^d \Gamma_{G9}^e\Big]\nonumber\\
		&+\frac{e^{-\frac{v^2}{v_c^2}}}{4 \pi^2 N_f \Gamma_{G2}^d}\frac{1}{\epsilon}\Big[2 (\Gamma_{G15}^u)^2+(\Gamma_{G11}^u)^2+2 (\Gamma_{G13}^u)^2+(\Gamma_{G14}^u)^2+(\Gamma_{G5}^e)^2-(\Gamma_{G6}^e)^2)\Big]\nonumber\\
		&-\frac{ g^2  (N_c^2-1)}{2 \pi^2 c N_c N_f}\frac{\Gamma_{G1}^d}{ \Gamma_{G2}^d}\frac{-f_1(c,c_\perp,v)+f_2(c,c_\perp,v)+f_3(c,c_\perp,v)}{\epsilon},\\
		A_{\Gamma_{G3}^d}&=-\frac{1}{\pi^2 N_f (1+v^2) \Gamma_{G3}^d }\frac{1}{\epsilon}\Big[\Gamma_{G10}^e \Gamma_{G3}^d+\Gamma_{G1}^d (\Gamma_{G3}^d+\Gamma_{G7}^e+\Gamma_{G9}^e)+2 \Gamma_{G2}^d \Gamma_{G8}^e+\Gamma_{G4}^d \Gamma_{G5}^e+\Gamma_{G4}^d \Gamma_{G6}^e\Big]\nonumber\\
		&-\frac{ g^2  (N_c^2-1)}{2 \pi^2 c  N_c N_f}\frac{\Gamma_{G4}^d}{ \Gamma_{G3}^d}\frac{-f_1(c,c_\perp,v)+f_2(c,c_\perp,v)+f_3(c,c_\perp,v)}{\epsilon}\\
		A_{\Gamma_{G4}^d}&=-\frac{1}{\pi^2 N_f(1+v^2) \Gamma_{G4}^d}\frac{1}{\epsilon}\Big[\Gamma_{G10}^e \Gamma_{G4}^d+\Gamma_{G1}^d (\Gamma_{G4}^d+2 \Gamma_{G8}^e)+\Gamma_{G2}^d \Gamma_{G7}^e+\Gamma_{G2}^d \Gamma_{G9}^e+\Gamma_{G3}^d \Gamma_{G5}^e+\Gamma_{G3}^d \Gamma_{G6}^e\Big]\nonumber\\
		&-\frac{ g^2  (N_c^2-1)}{2 \pi^2 c   N_c N_f}\frac{\Gamma_{G3}^d}{\Gamma_{G4}^d}\frac{-f_1(c,c_\perp,v)+f_2(c,c_\perp,v)+f_3(c,c_\perp,v)}{\epsilon}\\
		A_{\Gamma_{G5}^e}&=\frac{g^2 (N_c^2-1) }{4 \pi^2 c N_c N_f}\frac{f_4(c,c_\perp,v)-f_6(c,c_\perp,v)}{\epsilon}-\frac{e^{-\frac{v^2}{v_c^2}}}{2 \pi^2N_f \Gamma_{G5}^e\epsilon} \Big[-\Gamma_{G11}^u \Gamma_{G14}^u-2 \Gamma_{G13}^u \Gamma_{G15}^u+\Gamma_{G5}^e \Gamma_{G6}^e\Big]\nonumber\\
		&-\frac{1}{2\pi^2N_f (1+v^2)\Gamma_{G5}^e}\frac{1}{\epsilon}\Big[\Gamma_{G10}^e \Gamma_{G6}^e+\Gamma_{G1}^d \Gamma_{G5}^e+\Gamma_{G8}^e (\Gamma_{G7}^e+\Gamma_{G9}^e)\Big],
	\end{align}
\begin{align}
		A_{\Gamma_{G6}^e}&=\frac{g^2 (N_c^2-1)}{4 \pi^2 c  \Gamma_{G6}^e N_c N_f}\frac{\Gamma_{G11}^u\Big(f_4(c,c_\perp,v)+f_6(c,c_\perp,v)\Big)+\Gamma_{G6}^ef_5(c,c_\perp,v)}{\epsilon}-\frac{1}{2 \pi^2 N_f (1+v^2)\Gamma_{G6}^e}\frac{1}{\epsilon}\Big[\Gamma_{G10}^e \Gamma_{G5}^e+\Gamma_{G1}^d \Gamma_{G6}^e\nonumber\\
		&+\Gamma_{G8}^e (\Gamma_{G7}^e+\Gamma_{G9}^e)\Big]+\frac{e^{-\frac{v^2}{v_c^2}}}{2 \pi^2 N_f\Gamma_{G6}^e}\frac{1}{\epsilon}\Big[(\Gamma_{G11}^u)^2+\Gamma_{G11}^u (\Gamma_{G14}^u+\Gamma_{G2}^d+\Gamma_{G5}^e+\Gamma_{G6}^e)+4 \Gamma_{G12}^u (\Gamma_{G13}^u+\Gamma_{G15}^u)\nonumber\\
		&+\Gamma_{G6}^e (\Gamma_{G14}^u+\Gamma_{G5}^e)\Big],\\
		A_{\Gamma_{G7}^e}&=\frac{g^2 \Gamma_{G15}^u (N_c^2-1)}{4 \pi^2 c  \Gamma_{G7}^e   N_c N_f}\frac{f_9(c,c_\perp,v)-f_{11}(c,c_\perp,v)}{\epsilon}-\frac{1}{2 \pi^2N_f \Gamma_{G7}^e (1+v^2)}\frac{1}{\epsilon}\Big[\Gamma_{G10}^e \Gamma_{G9}^e+\Gamma_{G1}^d \Gamma_{G7}^e+\Gamma_{G8}^e (\Gamma_{G5}^e+\Gamma_{G6}^e)\Big]\\
		A_{\Gamma_{G8}^e}&=\frac{g^2 \Gamma_{G13}^u (N_c^2-1)}{4 \pi^2  \Gamma_{G8}^e N_c N_f}\frac{f_7(c,c_\perp,v)}{\epsilon}-\frac{1}{4 \pi^2N_f  \Gamma_{G8}^e (1+v^2)}\frac{1}{\epsilon}\Big[2 \Gamma_{G8}^e (\Gamma_{G10}^e+\Gamma_{G1}^d)+\Gamma_{G5}^e (\Gamma_{G7}^e+\Gamma_{G9}^e)\nonumber\\
		&+\Gamma_{G6}^e (\Gamma_{G7}^e+\Gamma_{G9}^e)\Big]\\
		A_{\Gamma_{G9}^e}&=\frac{g^2 \Gamma_{G15}^u (N_c^2-1)}{4 \pi^2 c  \Gamma_{G9}^e N_c N_f}\frac{f_9(c,c_\perp,v)+f_{11}(c,c_\perp,v)}{\epsilon}-\frac{1}{2 \pi^2N_f\Gamma_{G9}^e (1+v^2)}\frac{1}{\epsilon}\Big[\Gamma_{G10}^e \Gamma_{G7}^e+\Gamma_{G1}^d \Gamma_{G9}^e+\Gamma_{G8}^e (\Gamma_{G5}^e+\Gamma_{G6}^e)\Big]\\
		A_{\Gamma_{G10}^e}&=\frac{g^2 \Gamma_{G14}^u (N_c^2-1)}{2 \pi^2 c  \Gamma_{G10}^e N_c N_f}\frac{f_1(c,c_\perp,v)+f_3(c,c_\perp,v)}{\epsilon}-\frac{1}{2\pi^2N_f\Gamma_{G10}^e(1+v^2) }\frac{1}{\epsilon}\Big[\Gamma_{G10}^e \Gamma_{G1}^d-\Gamma_{G11}^u \Gamma_{G14}^u\nonumber\\
		&+2 \Big(\Gamma_{G5}^e \Gamma_{G6}^e+\Gamma_{G7}^e \Gamma_{G9}^e+(\Gamma_{G8}^e)^2\Big)\Big]\\
		A_{\Gamma_{G11}^u}&=\frac{e^{-\frac{v^2}{v_c^2}}}{2 \pi^2N_f \Gamma_{G11}^u }\frac{1}{\epsilon} \Big[\Gamma_{G11}^u (\Gamma_{G11}^u+\Gamma_{G14}^u+2 \Gamma_{G2}^d+\Gamma_{G6}^e)+\Gamma_{G5}^e (\Gamma_{G11}^u+\Gamma_{G14}^u+\Gamma_{G6}^e)+4 \Gamma_{G12}^u \Gamma_{G13}^u+4 \Gamma_{G12}^u \Gamma_{G15}^u\nonumber\\
		&+2 \Gamma_{G13}^u \Gamma_{G15}^u+\Gamma_{G14}^u \Gamma_{G6}^e+\Gamma_{G2}^d \Gamma_{G6}^e\Big]+\frac{1}{2 \pi^2N_f \epsilon (v^2+1)}\Gamma_{G1}^d\nonumber\\
		&+\frac{g^2 (N_c^2-1)}{4 \pi^2 c  \Gamma_{G11}^u N_c N_f}\frac{f_5(c,c_\perp,v)\Gamma_{G11}^u+\Gamma_{G6}^e\Big(f_4(c,c_\perp,v)+f_6(c,c_\perp,v)\Big)}{\epsilon}\\
		A_{\Gamma_{G12}^u}&=\frac{e^{-\frac{v^2}{v_c^2}}}{\pi^2 N_f  \Gamma_{G12}^u}\frac{1}{\epsilon}\Big[ \Gamma_{G11}^u (\Gamma_{G12}^u+\Gamma_{G13}^u+\Gamma_{G15}^u)+\Gamma_{G12}^u (\Gamma_{G14}^u+\Gamma_{G2}^d+\Gamma_{G5}^e)+\Gamma_{G6}^e (\Gamma_{G13}^u+\Gamma_{G15}^u)\Big]\nonumber\\
		&+\frac{g^2 (N_c^2-1)}{4 \pi^2 c  N_c N_f}\frac{ f_4(c,c_\perp,v)+f_5(c,c_\perp,v)+f_6(c,c_\perp,v)}{\epsilon}\\
		A_{\Gamma_{G13}^u}&=\frac{e^{-\frac{v^2}{v_c^2}}}{2 \pi^2 N_f \Gamma_{G13}^u} \frac{1}{\epsilon}\Big[\Gamma_{G15}^u (\Gamma_{G11}^u+\Gamma_{G5}^e)+\Gamma_{G13}^u (\Gamma_{G14}^u+\Gamma_{G2}^d)\Big]+\frac{ g^2 \Gamma_{G8}^e (N_c^2-1)}{4 \pi^2 \Gamma_{G13}^u N_c N_f}\frac{f_7(c,c_\perp,v)}{\epsilon},\\
		A_{\Gamma_{G14}^u}&=\frac{g^2 \Gamma_{G10}^e (N_c^2-1)}{2 \pi^2 c N_c N_f \Gamma_{G14}^u}\frac{f_1(c,c_\perp,v)+f_3(c,c_\perp,v)}{\epsilon}+\frac{e^{-\frac{v^2}{v_c^2}}}{\pi^2N_f \Gamma_{G14}^u}\frac{1}{\epsilon} \Big[\Gamma_{G11}^u \Gamma_{G5}^e+(\Gamma_{G13}^u)^2+\Gamma_{G14}^u \Gamma_{G2}^d+(\Gamma_{G15}^u)^2)\Big]\nonumber\\
		&+\frac{1}{2\pi^2N_f\Gamma_{G14}^u(1+v^2)}\frac{1}{\epsilon}\Big[\Gamma_{G10}^e \Gamma_{G11}^u+\Gamma_{G14}^u \Gamma_{G1}^d\Big]\\
		A_{\Gamma_{G15}^u}&=\frac{g^2 (N_c^2-1)}{8 \pi^2 c \epsilon \Gamma_{G15}^u N_c N_f}\frac{f_9(c,c_\perp,v) (\Gamma_{G7}^e+\Gamma_{G9}^e)+f_{11}(c,c_\perp,v) (\Gamma_{G7}^e-\Gamma_{G9}^e)}{\epsilon}+\frac{e^{-\frac{v^2}{v_c^2}}}{2 \pi^2 N_f \Gamma_{G15}^u}\frac{1}{\epsilon}\Big[ \Gamma_{G11}^u \Gamma_{G13}^u+\Gamma_{G13}^u \Gamma_{G5}^e\nonumber\\
		&+\Gamma_{G15}^u (\Gamma_{G14}^u+\Gamma_{G2}^d)\Big]
	\end{align}

	respectively, where
	\begin{gather*}
		f_1(c,c_\perp,v)=\frac{1}{2}\int_0^1 dx(1-x) x^{-1/2}\Big(xc^2+(1-x)(1+v^2)\Big)^{-1/2}(xc_\perp^2+1-x)^{-1/2},\\
		f_2(c,c_\perp,v)=\frac{1}{2}\int_0^1dx (1-x)x^{-1/2}\Big(xc^2+(1-x)(1+v^2)\Big)^{-1/2}(xc_\perp^2+1-x)^{-3/2},\\
		f_3(c,c_\perp,v)=\frac{1+v^2}{2}\int_0^1 dx(1-x)x^{-1/2} \Big(xc^2+(1-x)(1+v^2)\Big)^{-3/2} (xc_\perp^2+1-x)^{-1/2},\\
		f_4(c,c_\perp,v)=\int_0^1dx \int_0^{1-x} dy(xc_\perp^2+1-x)^{-1/2} \Big[c^2x^2+4\frac{v^2}{c^2}y(1-x-y)+x(1-x)(1+v^2)\Big]^{-1/2},\\
		f_5(c,c_\perp,v)=\int_0^1dx \int_0^{1-x} dy(xc_\perp^2+1-x)^{-3/2} \Big[c^2x^2+4\frac{v^2}{c^2}y(1-x-y)+x(1-x)(1+v^2)\Big]^{-1/2},\\
		f_6(c,c_\perp,v)=\frac{(xc_\perp^2+1-x)^{-1/2}x(1-v^2)}{\Big[c^2x^2+4\frac{v^2}{c^2}y(1-x-y)+x(1-x)(1+v^2)\Big]^{3/2}},\\
		f_7(c,c_\perp,v)=\int_0^1dx\int_0^{1-x}dy [xc^2+y(1+v^2)]^{-1/2}[xc^2+(1+v^2)(1-x-y)]^{-1/2}[xc_\perp^2+1-x]^{-1/2},\\
		f_8(c,c_\perp,v)=\int_0^1dx\int_0^{1-x}dy [xc^2+y(1+v^2)]^{-1/2}[xc^2+(1+v^2)(1-x-y)]^{-1/2}[xc_\perp^2+1-x]^{-3/2},\\
		f_9(c,c_\perp,v)=\int_0^1dx \int_{0}^{1-x}dy [xc_\perp^2+1-x]^{-1/2}\Big[c^2x^2+(1+v^2)x(1-x)+\frac{(1-v^2)^2}{c^2}y(1-x-y)\Big]^{-1/2},\\
		f_{10}(c,c_\perp,v)=\int_0^1dx \int_{0}^{1-x}dy [xc_\perp^2+1-x]^{-3/2}\Big[c^2x^2+(1+v^2)x(1-x)+\frac{(1-v^2)^2}{c^2}y(1-x-y)\Big]^{-1/2},\\
		f_{11}(c,c_\perp,v)=2v\int_0^1dx \int_{0}^{1-x}dy [xc_\perp^2+1-x]^{-1/2}x\Big[c^2x^2+(1+v^2)x(1-x)+\frac{(1-v^2)^2}{c^2}y(1-x-y)\Big]^{-3/2}.
	\end{gather*}

	\section{Detailed analysis on one loop beta functions} \label{Appendix:DetailedAnalysisOneLoopBetaFunctions}
	
	Here, we analyze our one-loop beta functions systematically. Since it is not easy to analyze the one-loop beta functions with all interaction, random charge potential, and random boson mass at the same time, we consider limiting cases first, where some of the vertices are ignored. Based on these limiting cases, we discuss the general case.
	
	\subsubsection{Clean case ($\lambda \neq 0$, $\Gamma_i=\gamma_M=0$)}
	
	\begin{figure}[h]
		\begin{subfigure}{0.23\textwidth}				 \includegraphics[scale=0.07]{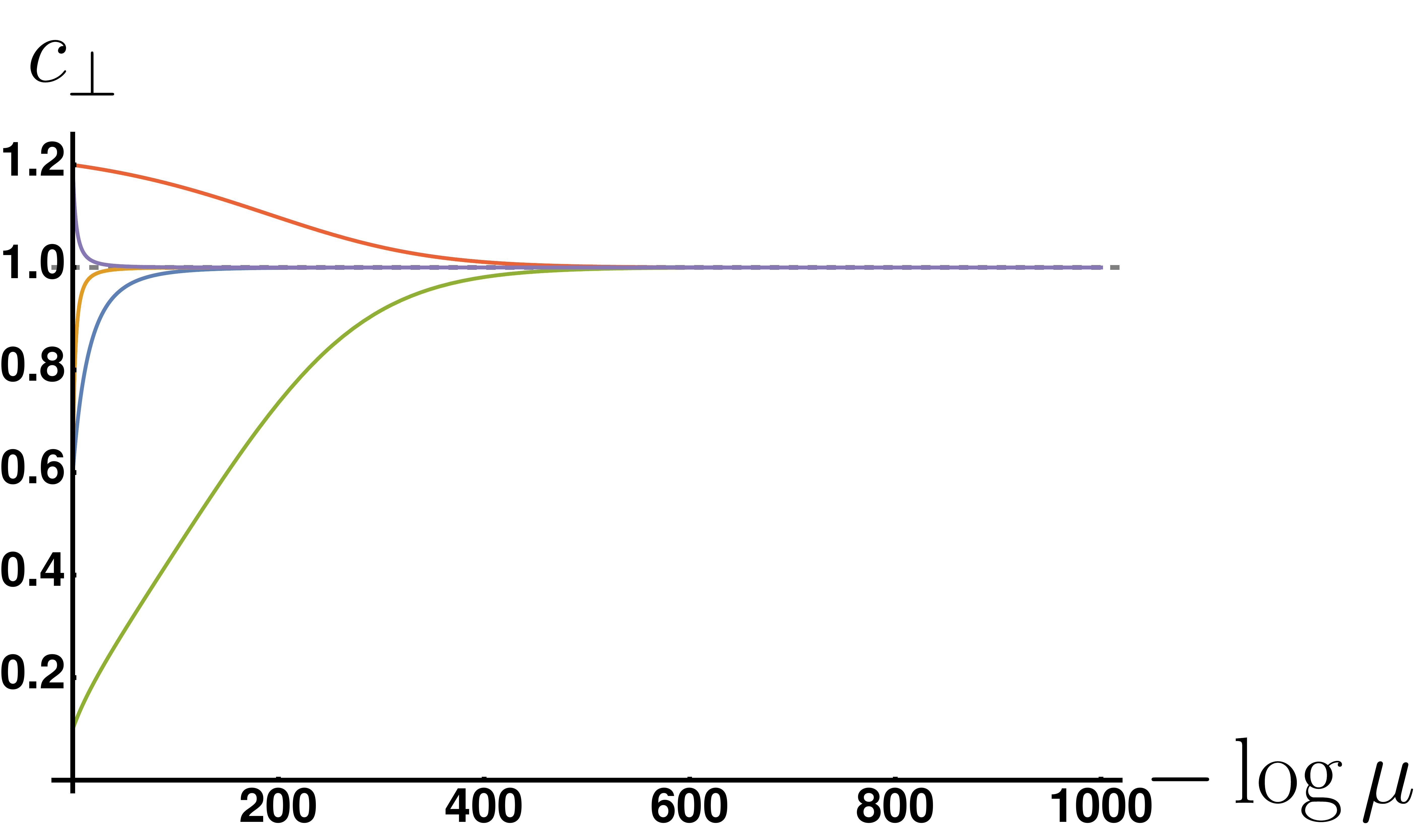}
			\caption{RG flow diagram of $c_\perp$}
		\end{subfigure}
		~
		\begin{subfigure}{0.23\textwidth}
			\includegraphics[scale=0.07]{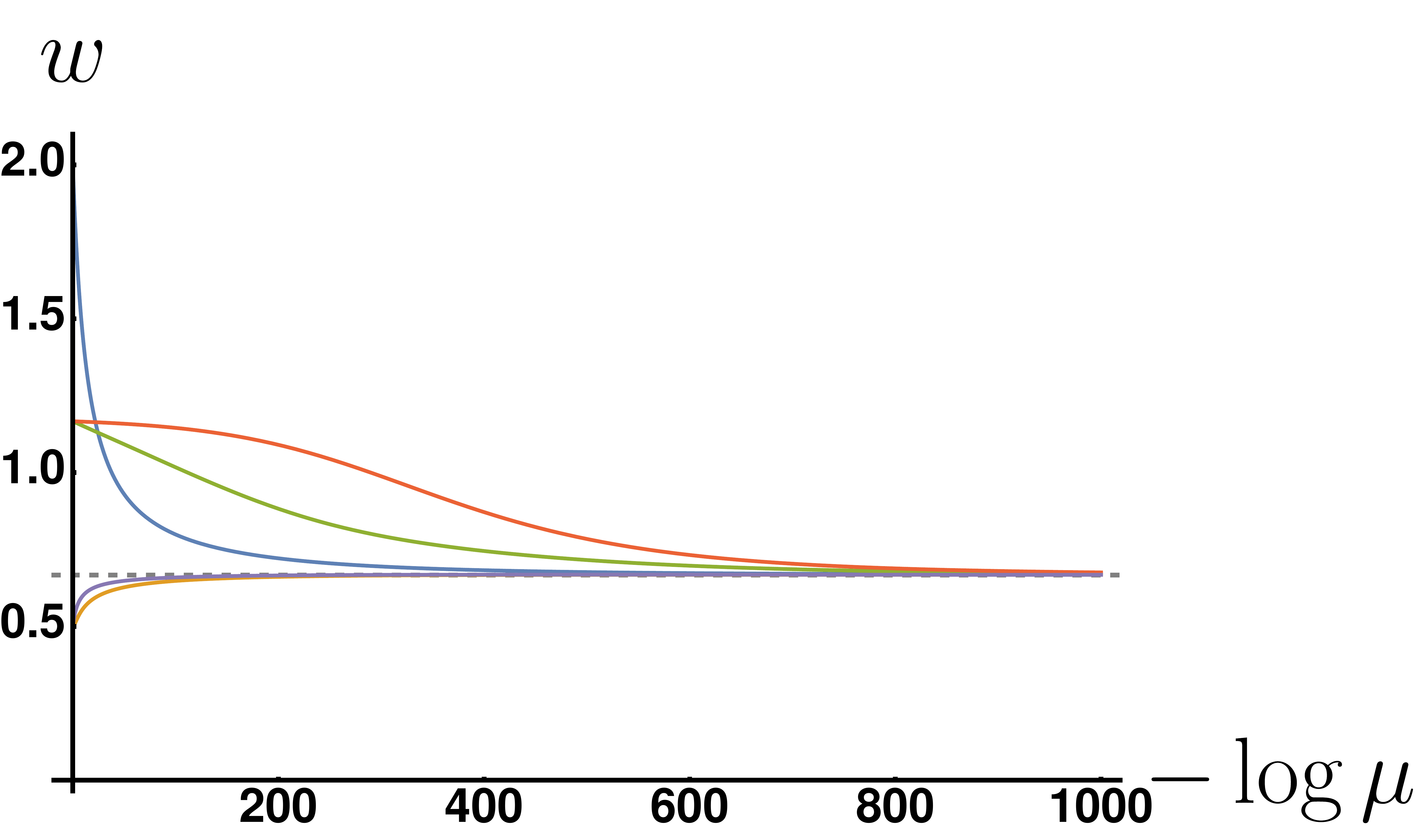}
			\caption{RG flow diagram of $w$}
		\end{subfigure}
		~
		\begin{subfigure}{0.23\textwidth}
			\includegraphics[scale=0.07]{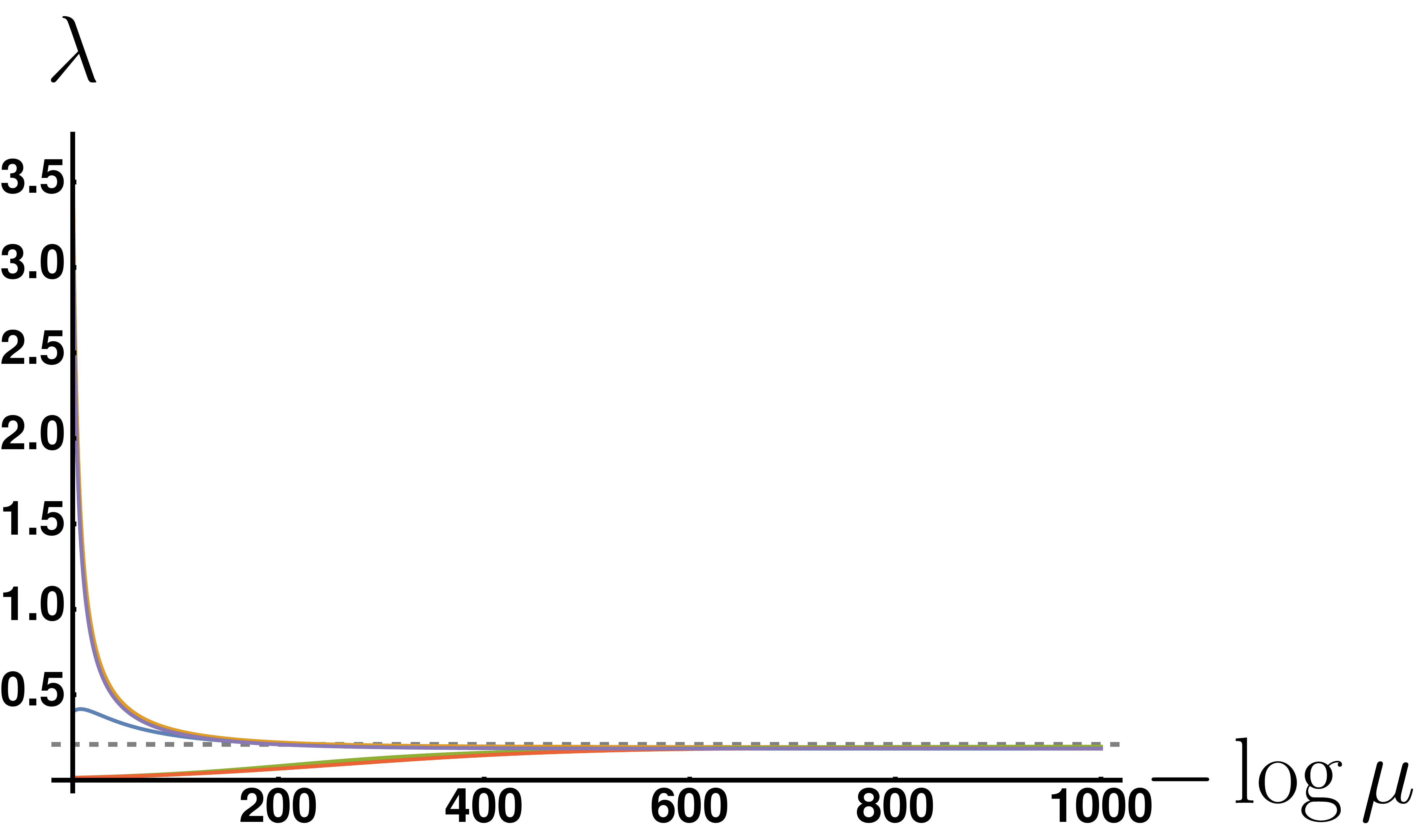}
			\caption{RG flow diagram of $\lambda$}
		\end{subfigure}
		~
		\begin{subfigure}{0.23\textwidth}
			\includegraphics[scale=0.07]{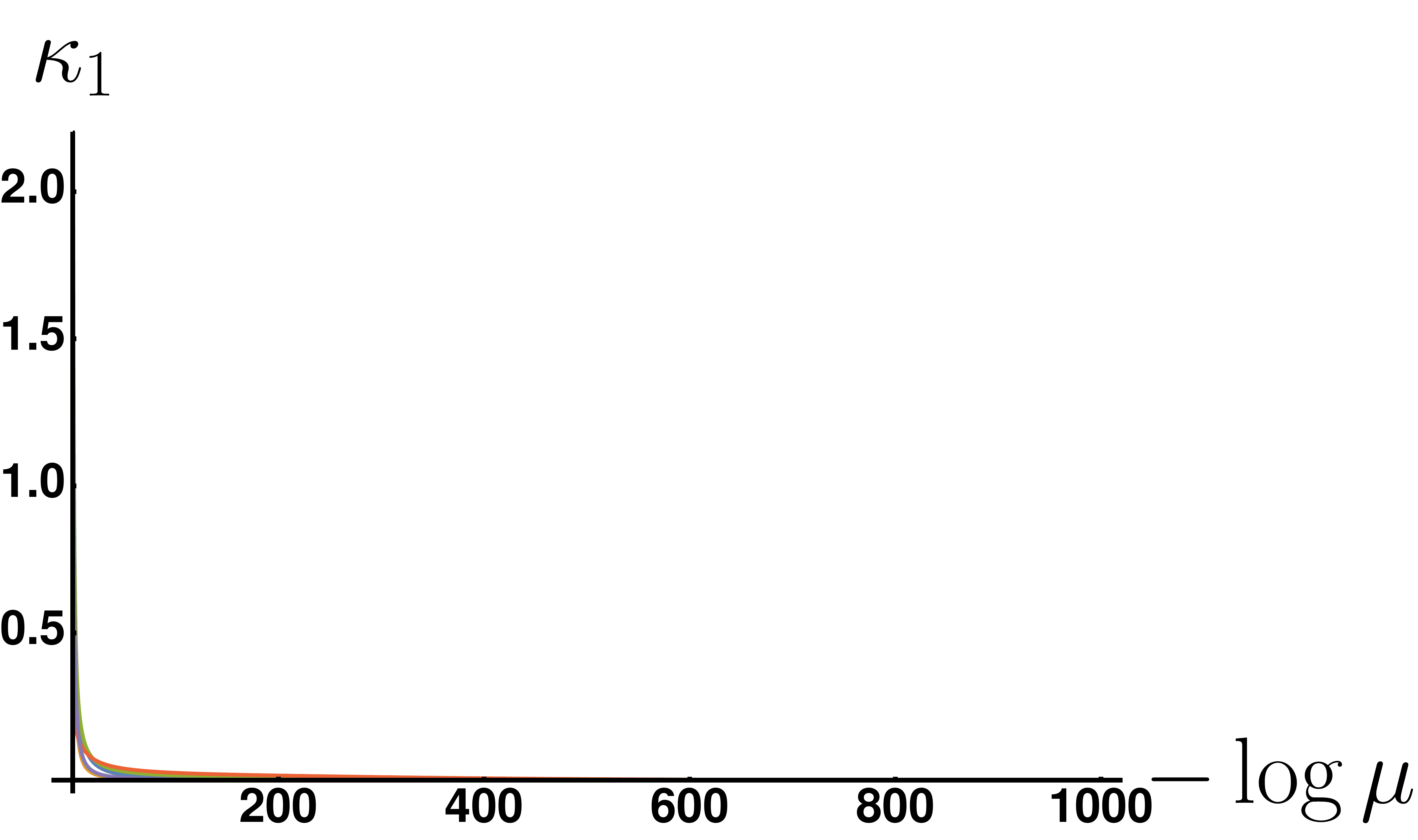}
			\caption{RG flow diagram of $\kappa_1$}
		\end{subfigure}
		\caption{RG flow diagrams in the clean case with various initial conditions. Here, dashed lines denote fixed point values obtained in Ref. \cite{SurLee}. We used $\epsilon=\bar{\epsilon}=0.01$, $N_f=1$, $N_c=2$, $v_c=0.05$, and $\kappa=1$.} \label{fig:2dBetaFunctionCleanCase}
	\end{figure}
	
	First, we consider the clean case without disorder effects. Ref. \cite{SurLee} discussed it with a fixed value of $c_\perp$ to 1. Here, we discuss the clean case with a non-fixed value of $c_\perp$, and show that it leads to the case obtained by Sur and Lee \cite{SurLee}. We introduce our beta function $\beta_{c_\perp}$ as
	\begin{align}
		\beta_{c_\perp}&=z_\perp\frac{c_\perp}{2}\Big[\frac{\lambda}{4\pi}\Big(1-\frac{1}{c_\perp^2}\Big)-\frac{N_c^2-1}{2\pi^2 N_cN_f}\lambda w[h_1(c,c_\perp,v)-h_2(c,c_\perp,v)]\Big]\nonumber\\
		&=z_\perp\frac{c_\perp}{2}(c_\perp^2-1)\Big[\frac{\lambda}{4\pi}\frac{1}{c_\perp^2}+\frac{N_c^2-1}{2\pi^2N_cN_f}\lambda w\int_0^1dx\sqrt{\frac{x(1-x)^2}{(1-x+xc_\perp^2)^3\Big((1+v^2)(1-x)+xc_\perp^2\Big)}}\Big] , \label{eq:Appendixbetacperp}
	\end{align}
	where $\lambda=\frac{g^2}{c}$ and $w=\frac{v}{c}$. This beta function shows that $c_\perp$ converges to 1 in the low energy limit. See Fig. \ref{fig:2dBetaFunctionCleanCase}.

	\subsubsection{No Yukawa interaction case ($\lambda=0$, $\Gamma_i\neq0$, $\gamma_M\neq0$)}\label{Appendix:DetailedAnalysisOneLoopBetaFunctions:NoYICase}
	
	Next, we consider the case without Yukawa interactions.
	%
	%
	From the beta functions with dimensionless variables in the main text, we obtain RG-flows of `No-Yukawa Interaction' case summarized in Table. \ref{table:NoInteractionCaseResult}.
	
	\begin{table}[h]
		\centering
		{\renewcommand{\arraystretch}{2}
			\begin{tabular}{|c|c||c|c|}
				\hline
				\multicolumn{2}{|c||}{When $\frac{\bar{\epsilon}}{\epsilon}<\frac{N_c^2-5}{N_c^2+7}$} &\multicolumn{2}{|c|}{When $\frac{\bar{\epsilon}}{\epsilon}>\frac{N_c^2-5}{N_c^2+7}$}   \\
				\hhline{====}
				$s<\frac{4}{\pi \kappa}\Rightarrow s\rightarrow 0$ & $s>\frac{4}{\pi \kappa}\Rightarrow s\nearrow$& $s<\frac{4}{\pi \kappa}\Rightarrow s\rightarrow 0$ &	$s>\frac{4}{\pi \kappa}\Rightarrow s\nearrow$   \\
				\hline
				\multicolumn{2}{|c||}{$\gamma_M\rightarrow 0$} & $\gamma_M\rightarrow \frac{\pi^2(N_c^2+7)}{9(N_c^2-1)}\Big(-\frac{N_c^2-5}{N_c^2+7}\epsilon+\bar{\epsilon}\Big)$ & $\gamma_M\rightarrow 0 $
				\\
				\multicolumn{2}{|c||}{$s\gamma_M\rightarrow 0$}&$s\gamma_M\rightarrow 0 $ & $s\gamma_M\rightarrow \frac{8\pi}{\kappa}\frac{N_c^2+7}{37N_c^2-29}\Big(-\frac{N_c^2-5}{N_c^2+7}\epsilon+\bar{\epsilon}\Big)$\\
				\multicolumn{2}{|c||}{$\kappa_1\rightarrow \frac{\pi^2(\epsilon+\bar{\epsilon})}{N_c^2+1}$}& $\kappa_1\rightarrow \frac{2\pi^2}{N_c^2+7}\Big[\frac{2}{3}\frac{N_c^2+7}{N_c^2-1}\Big(-\frac{N_c^2-5}{N_c^2+7}\epsilon+\bar{\epsilon}\Big)+\epsilon\Big]$ & $\kappa_1\rightarrow \frac{2\pi^2}{N_c^2+7}\Big[\frac{24(N_c^2+7)}{37N_c^2-29}\Big(-\frac{N_c^2-5}{N_c^2+7}\epsilon+\bar{\epsilon}\Big)+\epsilon\Big]$\\
				\hhline{----}
				\multicolumn{4}{|c|}{$w\rightarrow 0 $} \\
				\hline
			\end{tabular}
		}
		\caption{Summary of our beta-function analysis in the absence of Yukawa interactions} \label{table:NoInteractionCaseResult}
	\end{table}
	
	\begin{figure}
		\begin{subfigure}{0.45\textwidth}
			\includegraphics[scale=0.09]{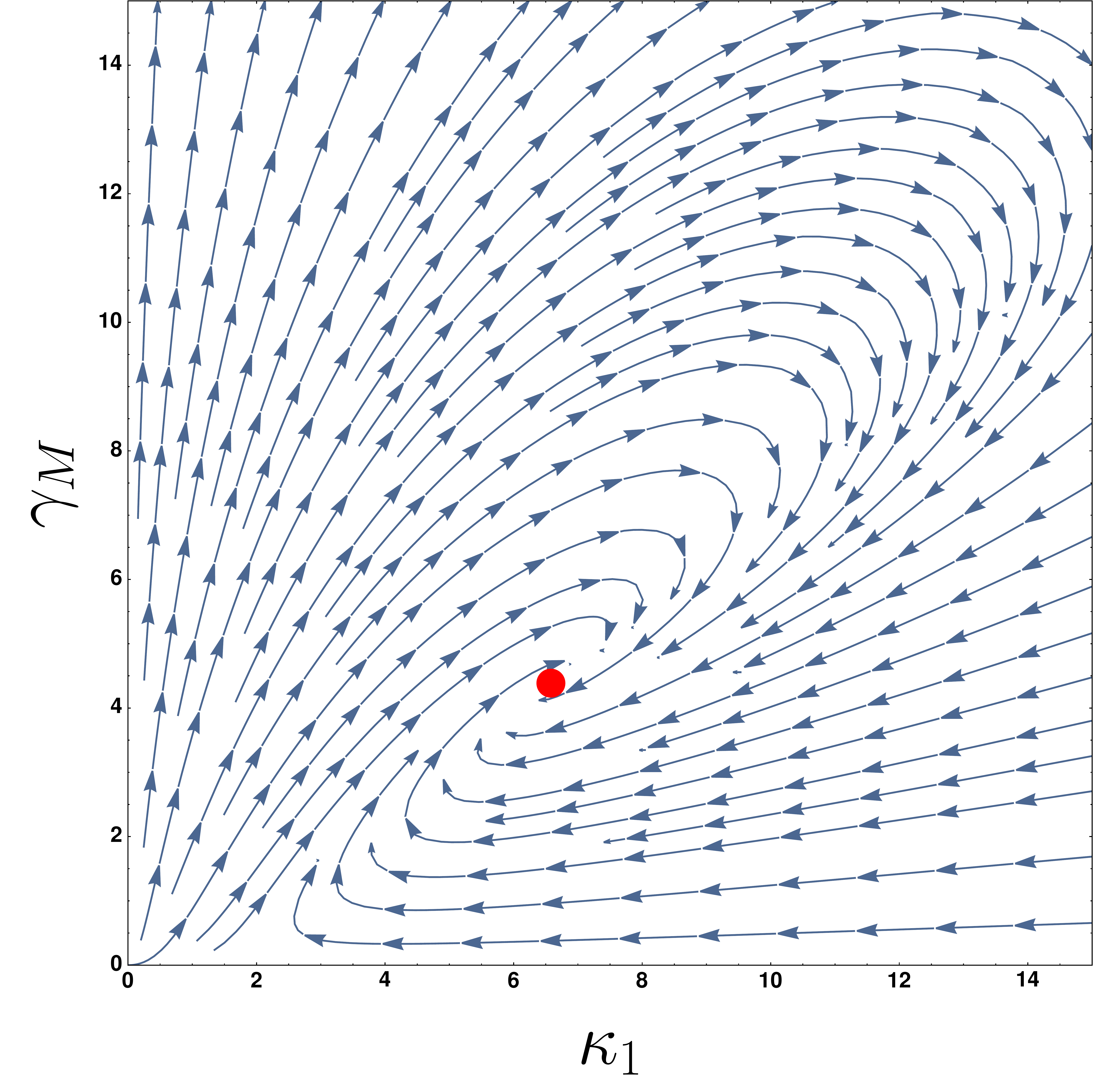}
			\caption{A two-dimensional RG flow diagram when $s<\frac{4}{\pi \kappa}$. A red dot denotes a fixed point with finite values of $(\kappa_1^*,\gamma_M^*)$ shown in Table \ref{table:NoInteractionCaseResult}. Here, $s\gamma_M$ is set to be zero and $\epsilon=\bar{\epsilon}=1$, $N_f=1$, $N_c=2$ have been used.}
		\end{subfigure}
		~
		\begin{subfigure}{0.45\textwidth}
			\includegraphics[scale=0.09]{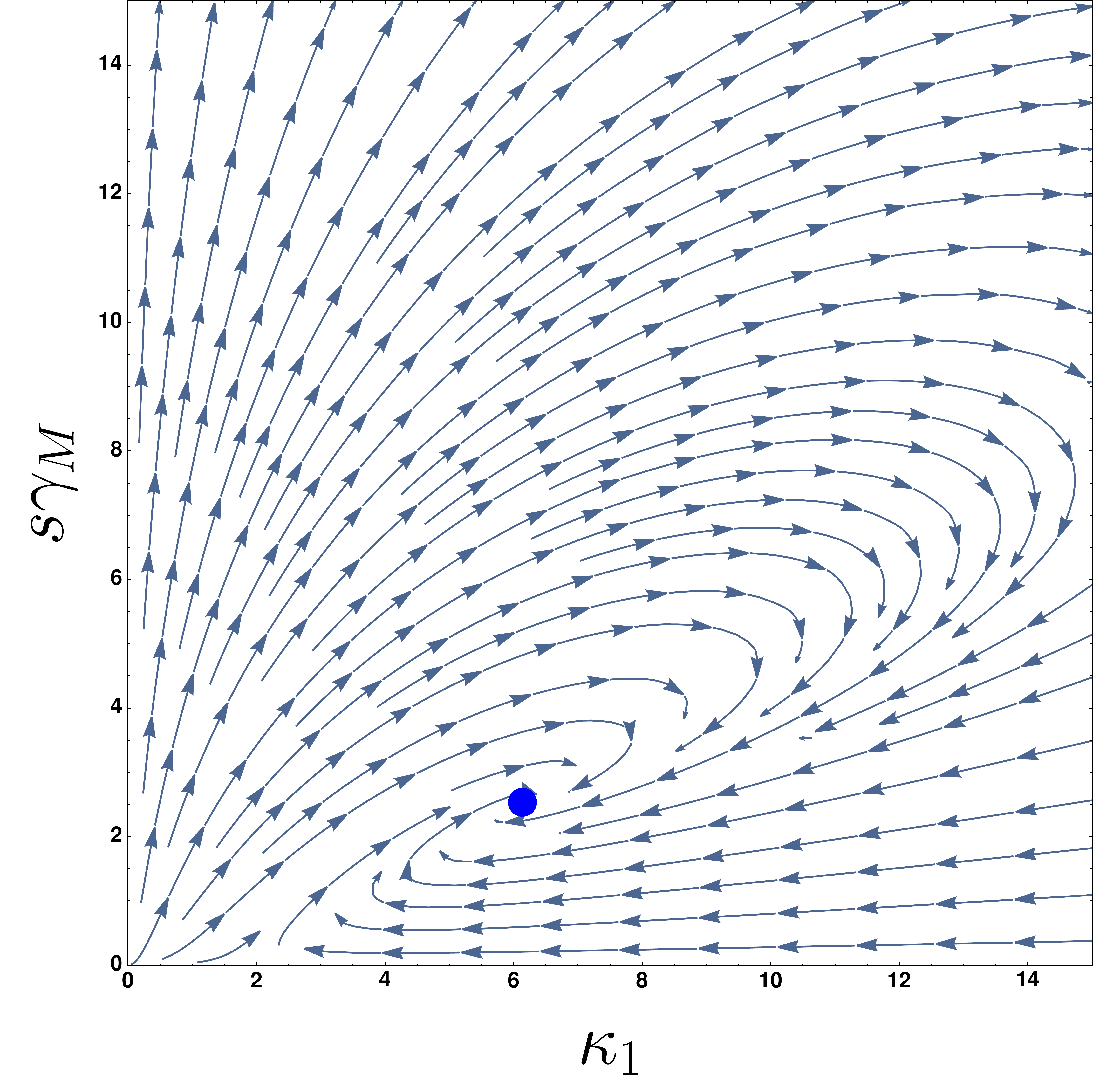}
			\caption{A two dimensional RG flow diagram when $s>\frac{4}{\pi \kappa}$. A blue dot denotes a fixed point with finite values of $(\kappa_1^*,(s\gamma_M)^*)$ shown in Table \ref{table:NoInteractionCaseResult}. Here, $\gamma_M$ is set to be zero and $\epsilon=\bar{\epsilon}=1$, $N_f=1$, $N_c=2$ have been used.}
		\end{subfigure}
		\caption{Two dimensional RG-flow diagrams without Yukawa interactions when $\frac{\bar{\epsilon}}{\epsilon}>\frac{N_c^2-5}{N_c^2+7}$.} \label{fig:2dFlowNoInterCase}
	\end{figure}
	
	Here, $\beta_{\Gamma_i}$ does not depend on other dimensionless variables. In other words, charge impurity potential vertices are decoupled from the other parameters. There appears an oscillating pattern of the RG flow as a result of the interplay between $\kappa_1$ and $\gamma_M$ (or $s\gamma_M$), shown in Fig. \ref{fig:2dFlowNoInterCase}. Here, eigenvalues of the linearized beta functions are given by complex numbers rather than real numbers. In Ref. \cite{KirkpatricBelitz}, there is a non-Gaussian fixed point, specified with finite fixed-point values of $\kappa_1$ and $\gamma_M$ (or $s\gamma_M$) and identified with a `Long-Range-Ordered' phase. Here, we use the same terminology for our analogous phase.

	\subsubsection{No random charge potential case ($\lambda\neq0$, $\Gamma_i=0$, $\gamma_M\neq 0$)}\label{Appendix:DetailedAnalysisOneLoopBetaFunctions:NorCPCase}
	\begin{figure}[h]
		\begin{subfigure}{0.23\textwidth}				 \includegraphics[scale=0.07]{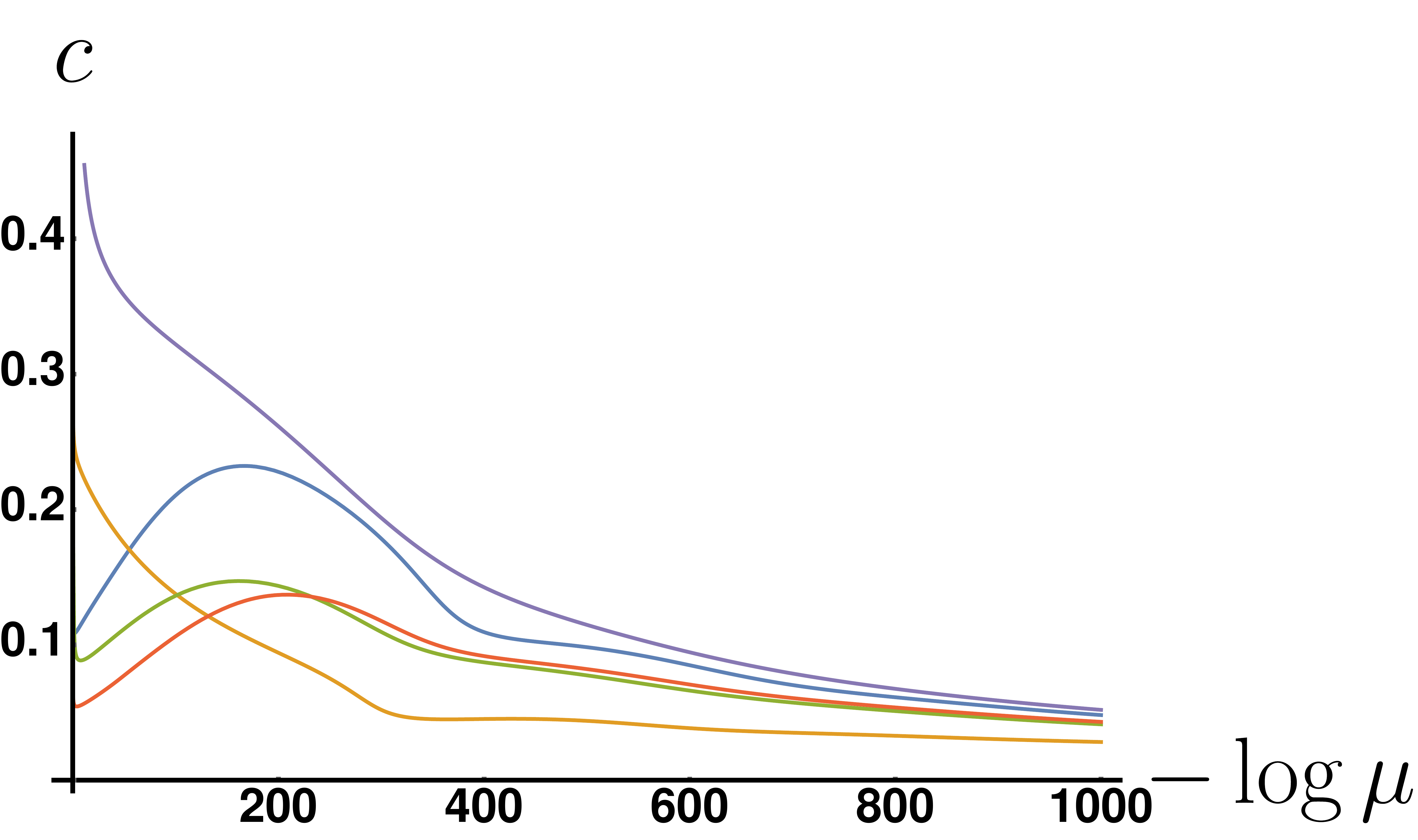}
			\caption{RG flow diagram of $c$}
		\end{subfigure}
		~
		\begin{subfigure}{0.23\textwidth}				 \includegraphics[scale=0.07]{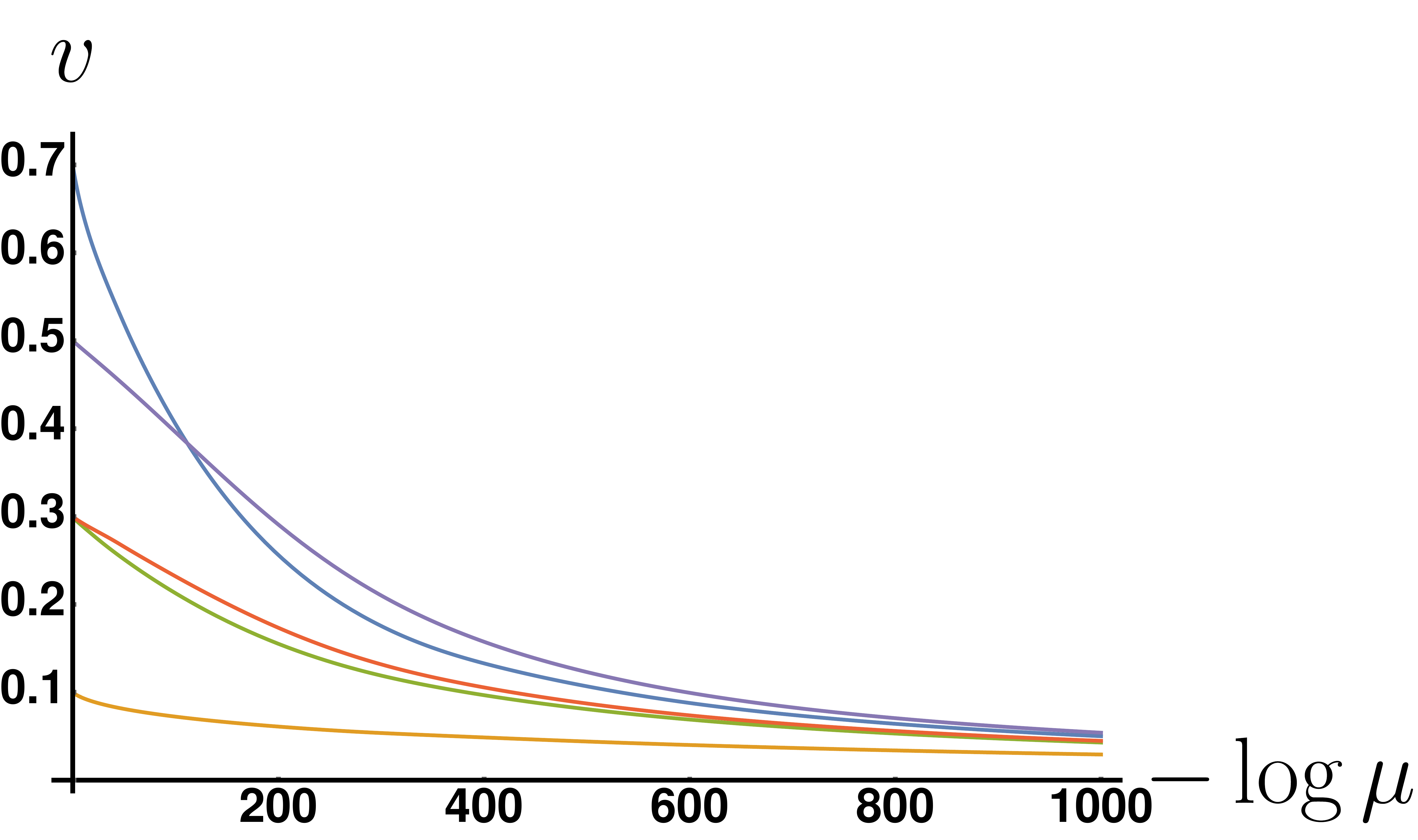}
			\caption{RG flow diagram of $vp$}
		\end{subfigure}
		~
		\begin{subfigure}{0.23\textwidth}				 \includegraphics[scale=0.07]{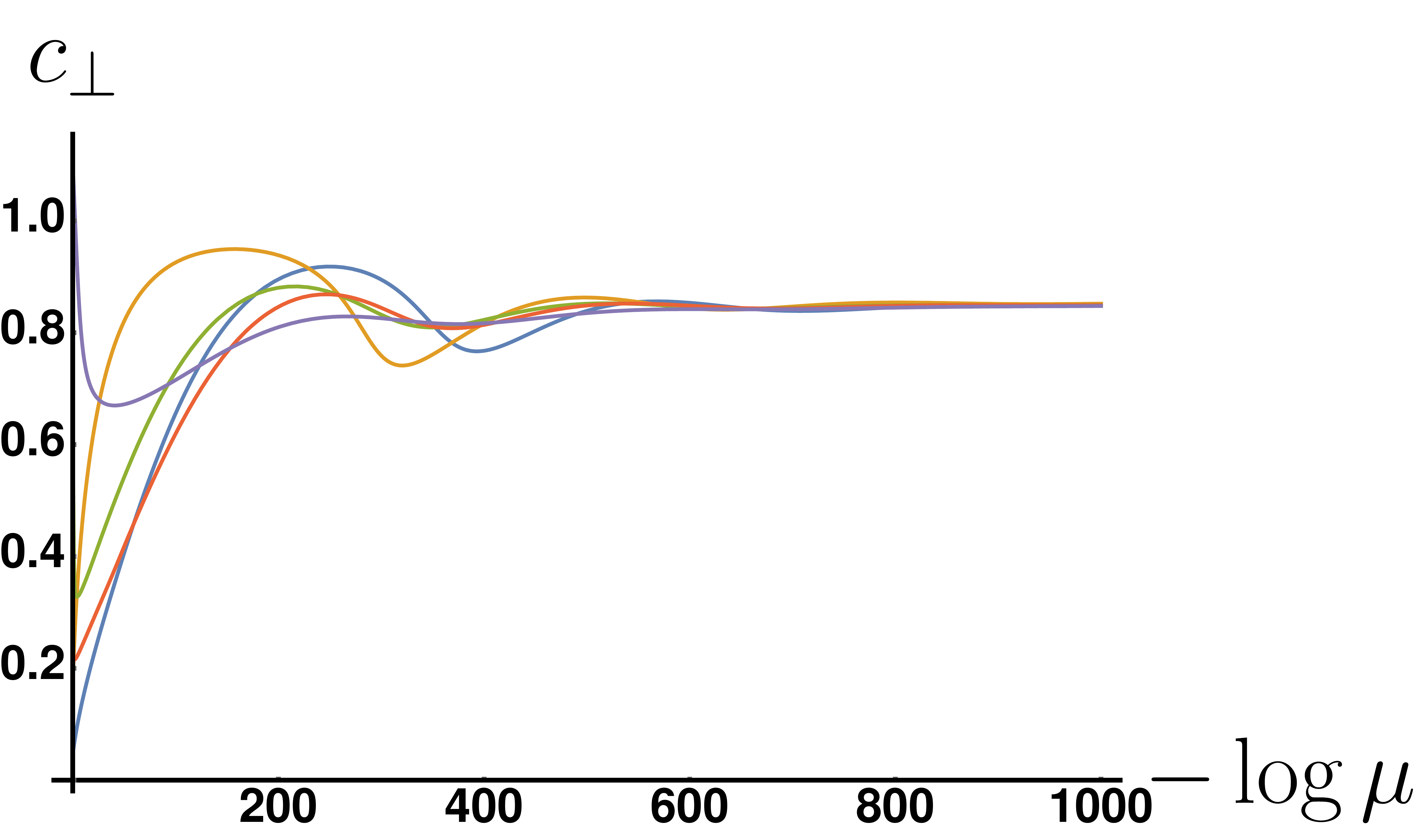}
			\caption{RG flow diagram of $c_\perp$}
		\end{subfigure}
		~
		\begin{subfigure}{0.23\textwidth}				 \includegraphics[scale=0.07]{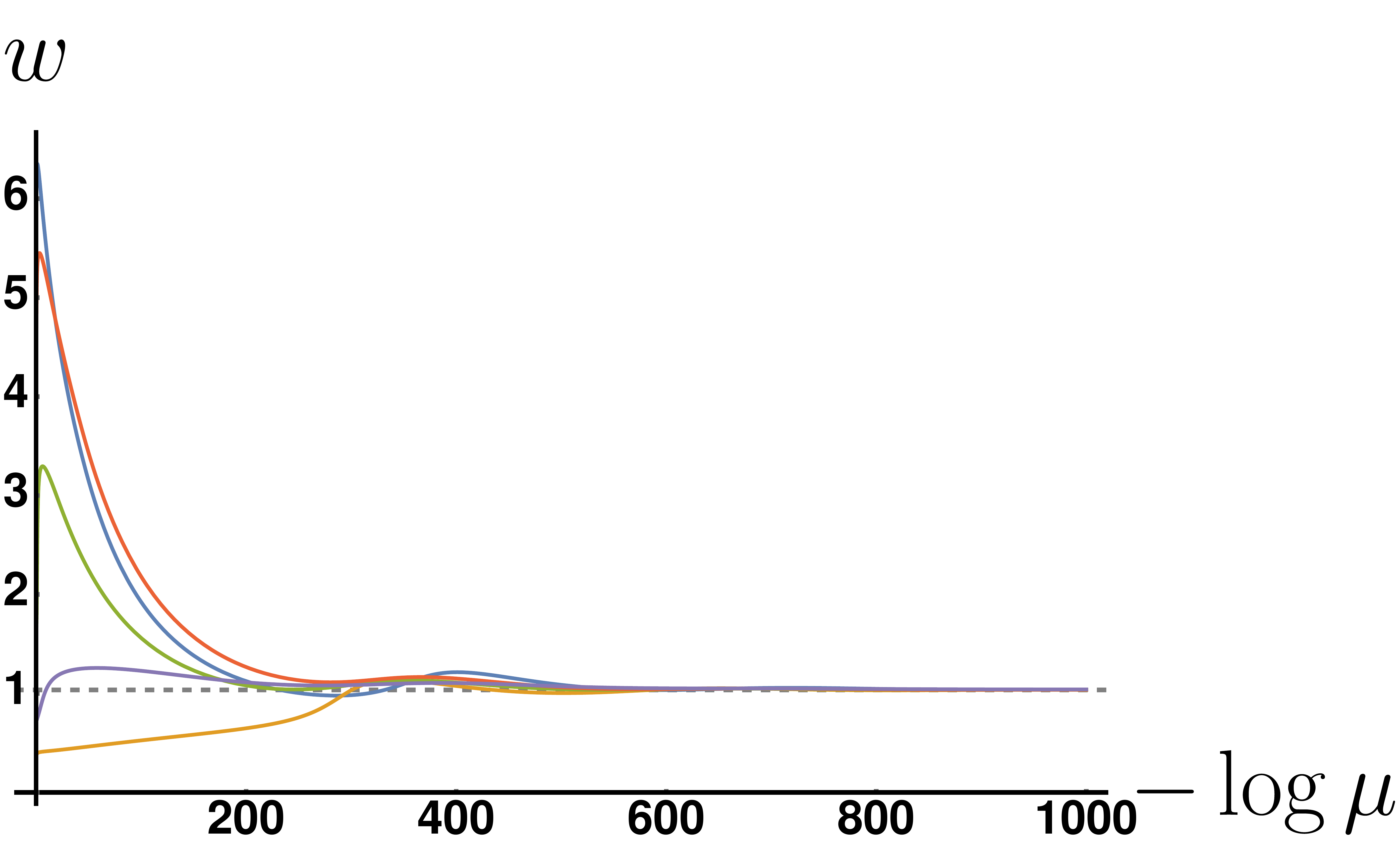}
			\caption{RG flow diagram of $w$}
		\end{subfigure}
		~
		\begin{subfigure}{0.23\textwidth}				 \includegraphics[scale=0.07]{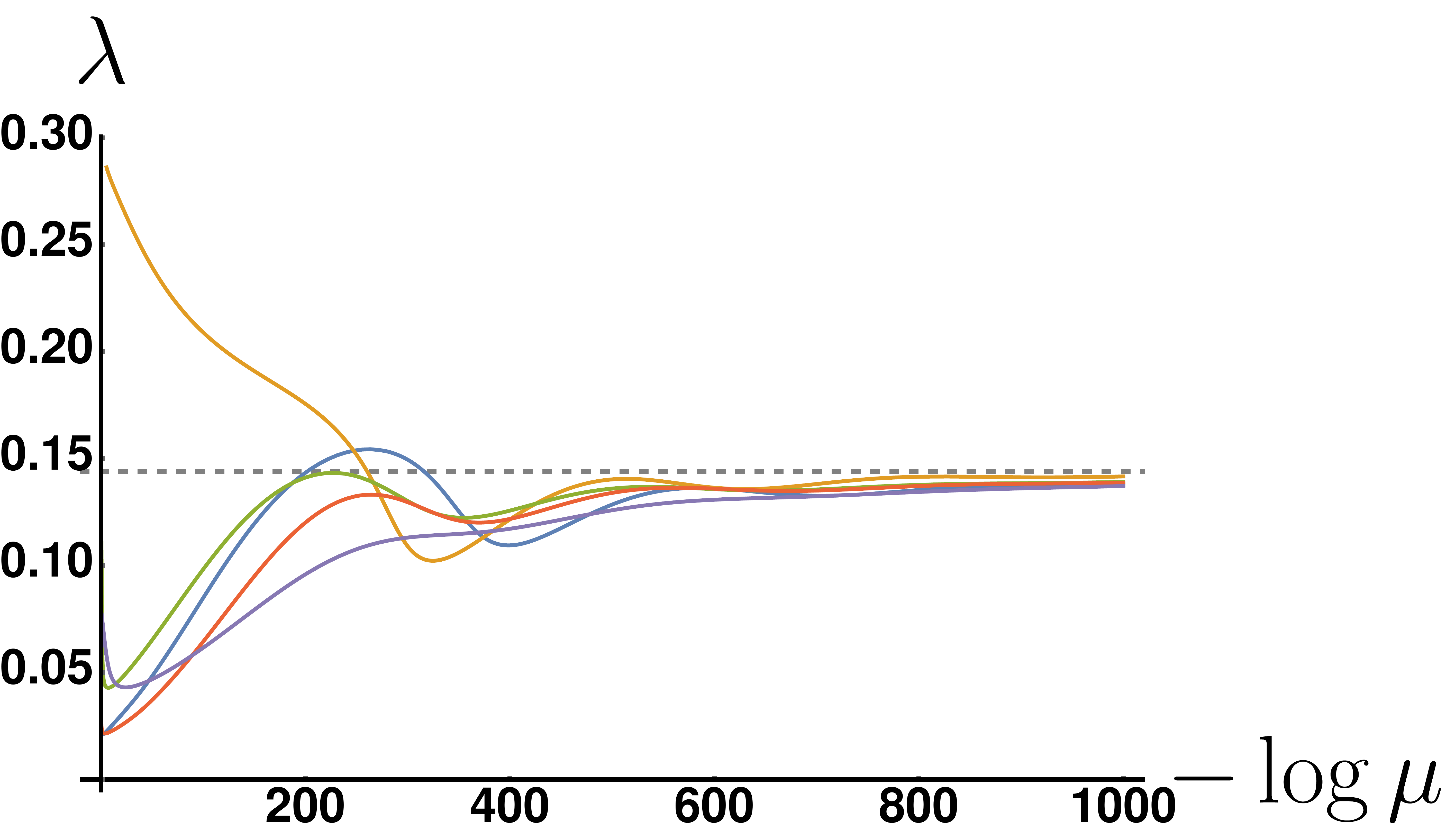}
			\caption{RG flow diagram of $\lambda$}
		\end{subfigure}
		~
		\begin{subfigure}{0.23\textwidth}				 \includegraphics[scale=0.07]{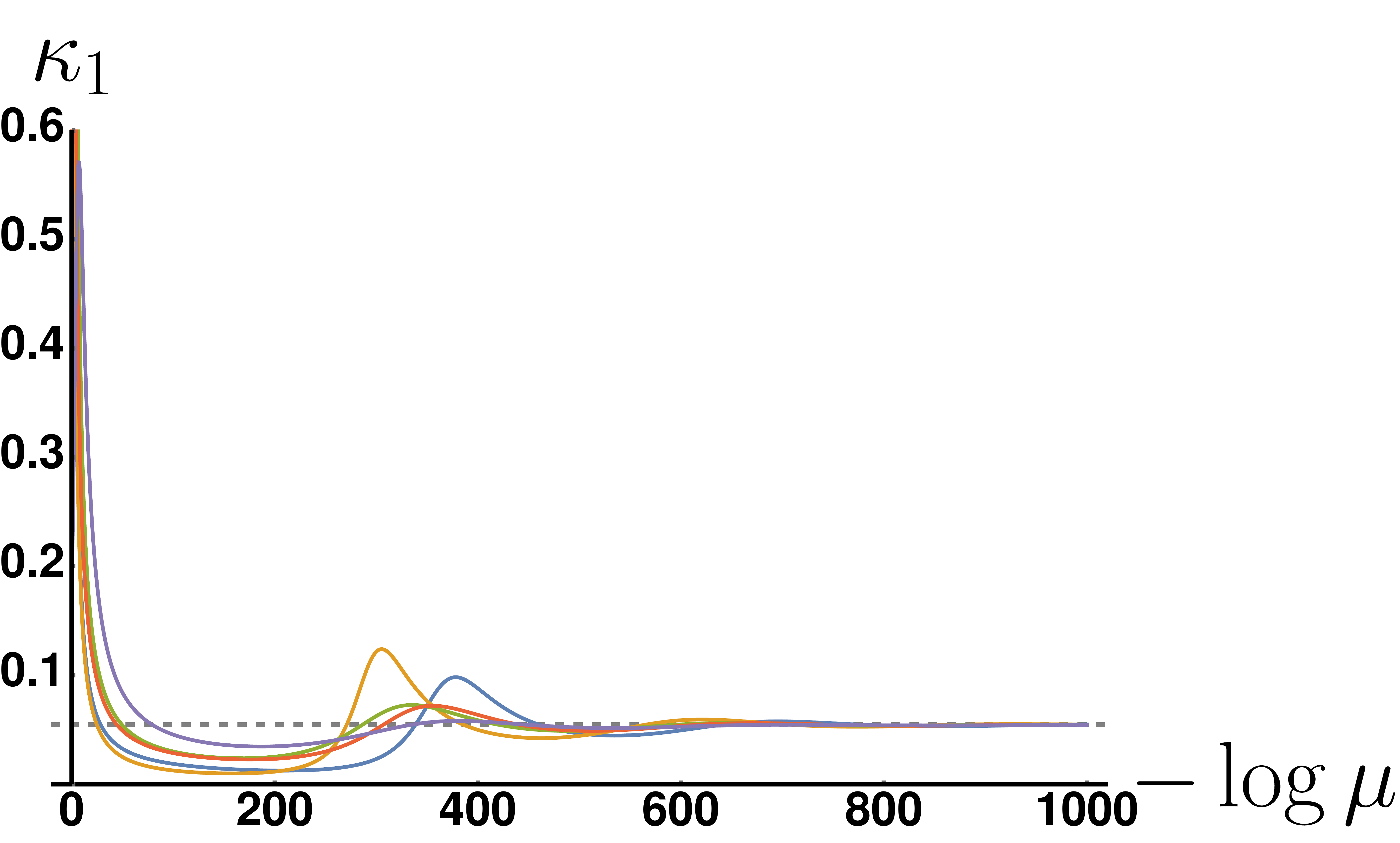}
			\caption{RG flow diagram of $\kappa_1$}
		\end{subfigure}
		~
		\begin{subfigure}{0.23\textwidth}				 \includegraphics[scale=0.07]{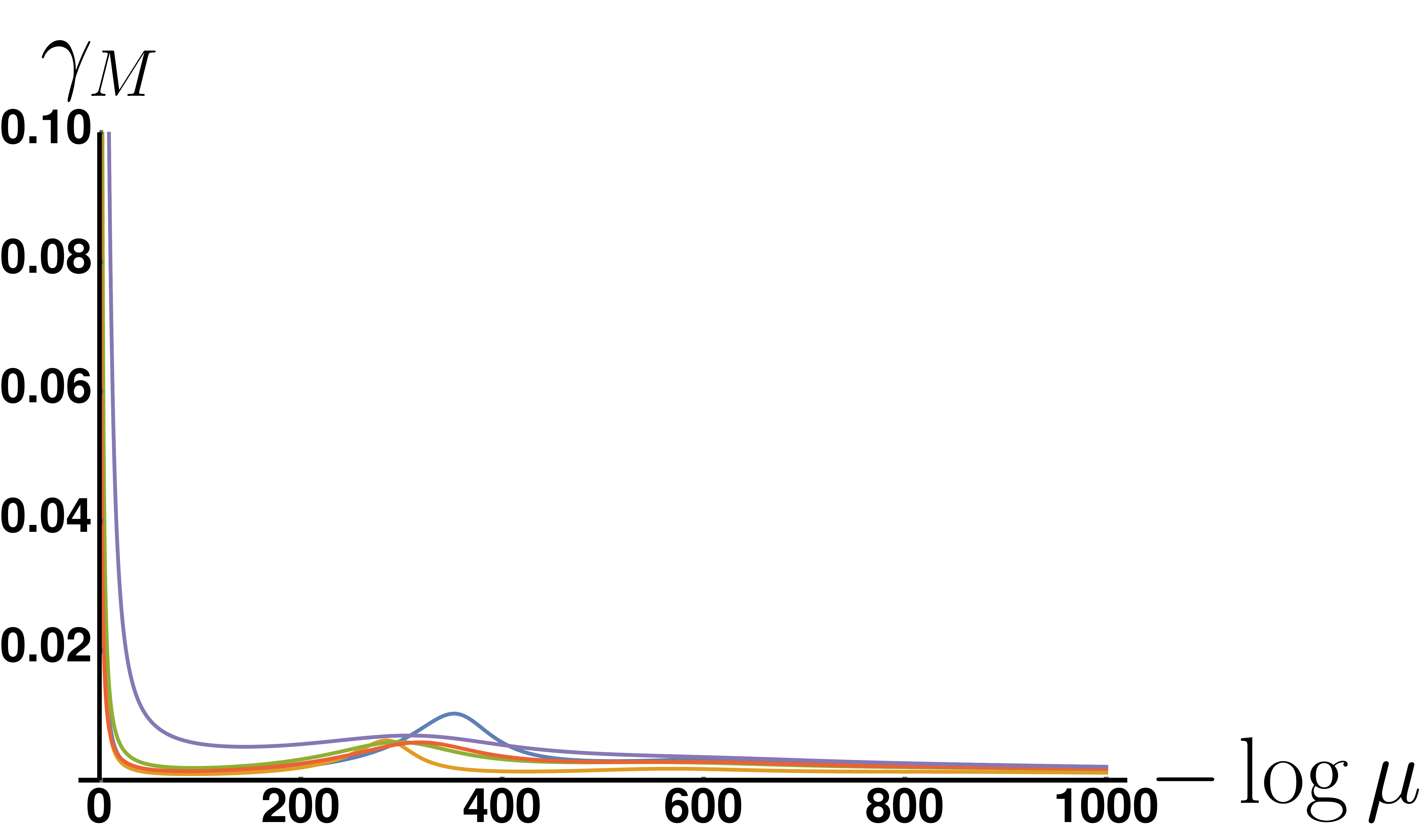}
			\caption{RG flow diagram of $\gamma_M$}
		\end{subfigure}
		~
		\begin{subfigure}{0.23\textwidth}				 \includegraphics[scale=0.07]{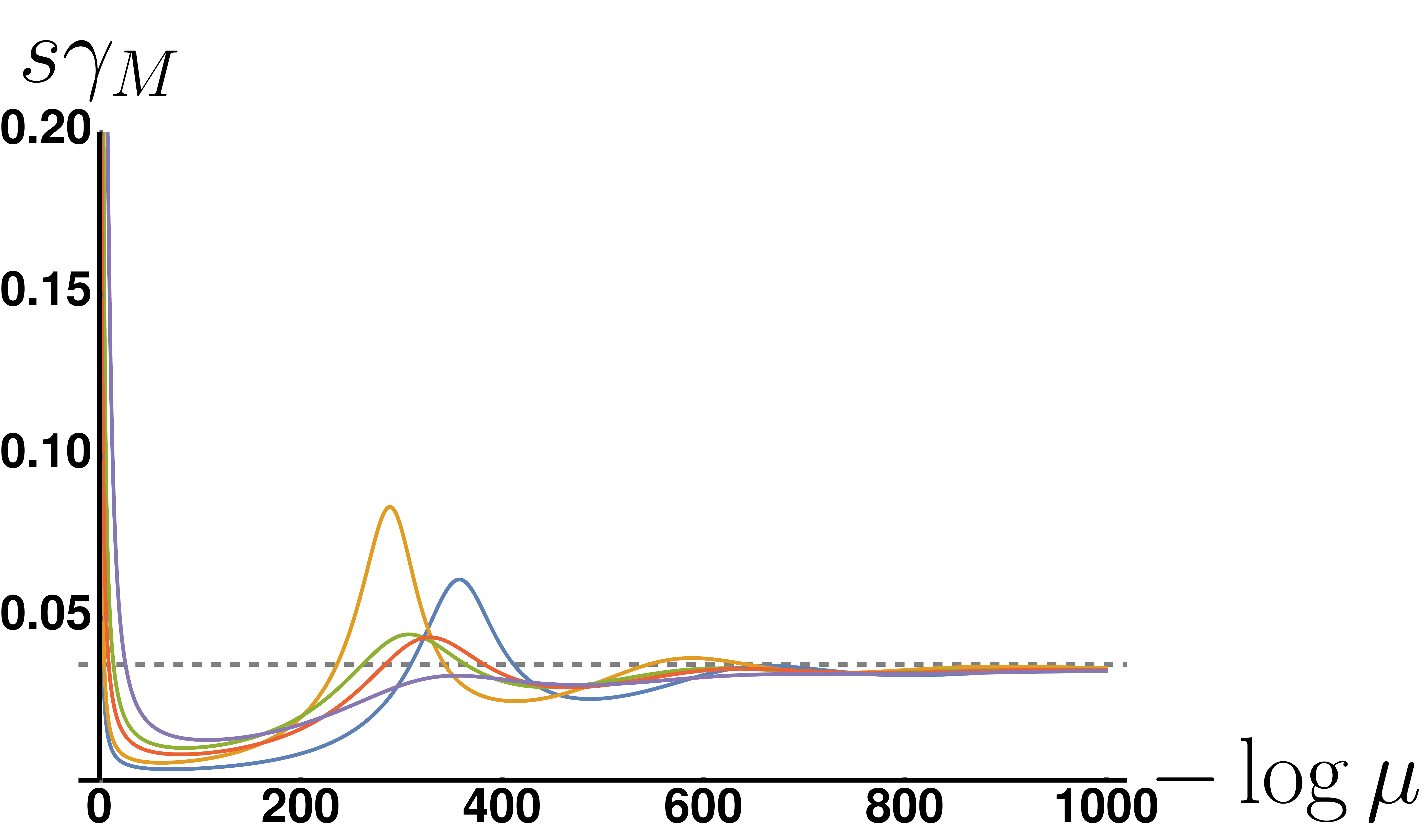}
			\caption{RG flow diagram of $s\gamma_M$}
		\end{subfigure}
		\caption{RG flow diagrams of the `No RCP' (No random charge potential) case with various initial conditions. Dashed lines of $c_\perp$, $w$, $\lambda$, $\kappa_1$, and $s\gamma_M$ denote fixed-point values (Eq. \eqref{eq:NoRCPFixedPoint}). Here, we used $\epsilon=\bar{\epsilon}=0.01$, $N_f=1$, $N_c=2$, $v_c=0.05$, and $\kappa=1$.} \label{fig:2dBetaFunctionNoRCPCase}
	\end{figure}
	
	 Since it is difficult to analyze the beta functions of this case in an analytic way, we first obtain numerical results given in Fig. \ref{fig:2dBetaFunctionNoRCPCase} and get some insight to investigate the analytical structure of these beta functions. We consider the case with Yukawa interactions and random boson mass fluctuations with a setting $N_f=1$, $N_c=2$, $\epsilon=\bar{\epsilon}=0.01$, and $\kappa=1$. RG flow diagrams in Fig. \ref{fig:2dBetaFunctionNoRCPCase} show that there is a stable fixed point with finite fixed point values of $c_\perp^*$, $w^*$, $\lambda^*$, $\kappa_1^*$, and $(s\gamma_M)^*$. Similar oscillating patterns observed in the `No-interacting' case are also found in this case. As explained in the `No-interacting' case, these oscillating patterns are due to the interplay between $\kappa_1$ and $\gamma_M$ (or $s\gamma_M$). Since this fixed point has a similar structure to the non-Gaussian fixed point discussed in the `No-Interacting' case and has a finite fixed-point value of the Yukawa interaction vertex $\lambda$, we call this fixed point as `Interacting Long-Range-Ordered' phase used in the main text.
	
	Now, let us determine fixed point values denoted with dashed lines in Fig. \ref{fig:2dBetaFunctionNoRCPCase}. Since RG beta functions involve functions $h_i(c,c_\perp,v)(i=1,2,3)$, it is necessary to simplify these functions. RG flows of $c$, $v$, and $\gamma_M$ in  Fig. \ref{fig:2dBetaFunctionNoRCPCase} show that $c$, $v$, and $\gamma_M$ are converging to zero in the low energy limit.
	%
	%
	With these fixed point values, $h_i(c,c_\perp,v)$ are approximated as follow:
	\begin{align}
		h_1(c,c_\perp,v)&\approx h_1(0,c_\perp,0)=\int_0^1dx \sqrt{\frac{x}{(1-x+xc_\perp^2)(1-x)}}=2\frac{E(1-c_\perp^2)-K(1-c_\perp^2)}{c_\perp^2-1},\\
		h_2(c,c_\perp,v)&\approx h_2(0,c_\perp,0)=c_\perp^2 \int_0^1dx \sqrt{\frac{x}{(1-x+xc_\perp^2)^3(1-x)}}=2\frac{E(1-c_\perp^2)-c_\perp^2K(1-c_\perp^2)}{1-c_\perp^2},\\
		h_3(c,c_\perp,v)&\approx h_3(0,c_\perp,0)=0,\\
		h_4(c,c_\perp,v)&\approx \lim_{c\rightarrow 0}h_4(c,c_\perp,cw)\nonumber\\
		&=\pi\int_0^1dx\int_0^{1-x}dy\Big[(1-x-y)\Big(x+y+c_\perp^2(1-x-y)\Big)\nonumber\\
		&+\Big(1+x+y+c_\perp^2(1-x-y)\Big) \Big(-w^2(x-y)^2+(x+y)\Big(1-x-y+w^2(x+y)\Big)\Big)\Big]\nonumber\\
		&\times \Big[\Big(x+y+c_\perp^2(1-x-y)\Big) \Big(-w^2(x-y)^2+(x+y)\Big(1-x-y+w^2(x+y)\Big)\Big)\Big]^{-3/2}\nonumber\\
		&\equiv h_4(c_\perp,w) .
	\end{align}
	Here, $E(x)$ and $K(x)$ are `$EllipticE(x)$' and `$EllipticK(x)$' functions defined in the Mathematica program.
	
	The resulting beta functions with $c=v=\gamma_M=0$ and $h_i(c,c_\perp,v)$ are given by
	\begin{subequations}
		\begin{align}
			z_\perp&=\Big(1-\frac{N_c^2-1}{4\pi^2 N_cN_f}w\lambda h_2(0,c_\perp,0)\Big)^{-1},\\
			\beta_{c_\perp}&=\frac{c_\perp}{2}\Big[z_\perp\Big\{\frac{\lambda}{4\pi}\Big(1-\frac{1}{c_\perp^2}\Big)-\frac{N_c^2-1}{2\pi^2N_cN_f}w\lambda[h_1(0,c_\perp,0)-h_2(0,c_\perp,0)]\Big\}+\frac{s\gamma_M}{2\pi}\frac{z_\perp\epsilon +\bar{\epsilon}}{\epsilon+\bar{\epsilon}}\Big],\\
			\beta_w&=w\Big[z_\perp\Big\{-\frac{\lambda}{8\pi}+\frac{N_c^2-1}{4\pi^2 N_cN_f}w\lambda h_1(0,c_\perp,0)\Big\}-\frac{3s\gamma_M}{8\pi}\frac{z_\perp \epsilon+\bar{\epsilon}}{\epsilon+\bar{\epsilon}}\Big],\\
			\beta_\lambda&=\lambda\Big[z_\perp\Big\{-\epsilon+\frac{\lambda}{4\pi}-\frac{N_c^2-1}{4\pi^2 N_cN_f}w\lambda [h_1(0,c_\perp,0)-h_2(0,c_\perp,0)]-\frac{w\lambda}{4\pi^3N_cN_f}h_4(c_\perp,w)\Big\}+\frac{s\gamma_M}{2\pi}\frac{z_\perp\epsilon+\bar{\epsilon}}{\epsilon+\bar{\epsilon}}\Big],\\
			\beta_{\kappa_1}&=\kappa_1\Big[z_\perp\Big\{-\epsilon+\frac{\lambda}{8\pi}\Big(1+\frac{1}{c_\perp^2}\Big)+\frac{N_c^2+7}{2\pi^2}\kappa_1\Big\}-\frac{3s\gamma_M}{\pi}\frac{z_\perp\epsilon+\bar{\epsilon}}{\epsilon+\bar{\epsilon}}\Big],\\
			\beta_{s\gamma_M}&=s\gamma_M\Big[-z_\perp \epsilon-\bar{\epsilon}+z_\perp\Big\{\frac{\lambda}{8\pi}\frac{1}{c_\perp^2}+\frac{N_c^2+1}{\pi^2}\kappa_1\Big\} -\frac{11s\gamma_M}{8\pi}\frac{z_\perp\epsilon+\bar{\epsilon}}{\epsilon+\bar{\epsilon}}\Big] ,
		\end{align}
		\label{eq:NoRCPCaseApproxBetaFunctions}
	\end{subequations}
	where $\kappa$ is set to be 1.
	
	Solving $\beta_{c_\perp}=\beta_w=\beta_\lambda=\beta_{\kappa_1}=\beta_{s\gamma_M}=0$, we obtain fixed point values of $c_\perp^*$, $w^*$, $\lambda^*$, $\kappa_1^*$, and $(s\gamma_M)^*$. Our numerical solution with $\epsilon=\bar{\epsilon}=0.01$, $N_f=1$, and $N_c=2$ is
	\begin{gather}
		c_\perp^*=0.856,\; w^*=1.035,\; \lambda^*=0.144,\; \kappa_1^*=0.055,\; (s\gamma_M)^*=0.036 . \label{eq:NoRCPFixedPoint}
	\end{gather}
	These fixed point values agree well with the limiting-case values of our numerical simulations as shown in Fig. \ref{fig:2dBetaFunctionNoRCPCase}. If we compare these fixed point values with those of the clean case with $\epsilon=\bar{\epsilon}=0.01$, $N_f=1$, and $N_c=2$, given by
	\begin{gather*}
		\text{Clean Case: } c_\perp^*=1,\; w^*=2/3\approx 0.67,\; \lambda^*=0.21,\; \kappa_1^*=0,
	\end{gather*}
	the velocities of boson fields $c$ and $c_\perp$ decrease due to random boson mass fluctuations, which leads to reduction of the effective Yukawa interaction. On the other hand, the effective boson-boson interaction parameter $\kappa_1$ increases. Physical interpretation of these low energy behaviors are given in the main text.

	\subsubsection{No random boson mass case ($\lambda \neq0$ ,$\Gamma_i\neq0$, $\gamma_M=0$)}\label{Appendix:DetailedAnalysisOneLoopBetaFunctions:NorBMCase}
	
	As the last limiting case before going to the general case, we consider the case with both Yukawa interaction and random charge potential effects. Here, only the $\Gamma_{G1}^d$ random charge potential channel is considered, mostly relevant as discussed in the main text. In the beta function $\beta_{\Gamma_{G1}^d}$, the second term proportional to the Yukawa interaction is summed up with the term of $A_{\Gamma_{G1}^d}^{(1)}$, coming from the Feynman diagram composed of Yukawa vertices and the random charge potential vertex.
	
	Assuming that all direct channels are the same for simplicity, we can re-write $\beta_{\Gamma_{G1}^d}$ as	
	
	\begin{align}
		\beta_{\Gamma_{G1}^d}&=z_\perp\Gamma_{G1}^d\Big[-\epsilon+\frac{N_c^2-1}{2\pi^2 N_cN_f}\lambda w[h_3(c,c_\perp,v)-h_2(c,c_\perp,v)]+\tilde{A}_{\Gamma_{G1}^d}^{(1)}\Big] ,
	\end{align}
	where
	\begin{align*}
		&A_{\Gamma_{G1}^d}^{(1)}\equiv \tilde{A}_{\Gamma_{G1}^d}^{(1)}-\frac{N_c^2-1}{2\pi^2}\frac{g^2}{c}[-f_1(c,c_\perp,v)+f_2(c,c_\perp,v)+f_3(c,c_\perp,v)],\\
		&h_2(c,c_\perp,v)+h_2(c,c_\perp,v)=2f_1(c,c_\perp,v),\\
		&f_1(c,c_\perp,v)-f_2(c,c_\perp,v)=h_1(c,c_\perp,v)-h_2(c,c_\perp,v),\\
		&f_1(c,c_\perp,v)-f_3(c,c_\perp,v)=-h_1(c,c_\perp,v)+h_3(c,c_\perp,v) .
	\end{align*}
	
	If we assume that the random charge potential is much bigger than other effects, $F_{dis}(\{\Gamma_i,v\})$ in these beta functions dominates over any other terms in the low energy limit. As a result, we can approximate the RG flows of the parameters as
	\begin{align*}
		\beta_c&\sim -z_\perp  c F_{dis}(\{\Gamma_i,v\})<0\Rightarrow c\nearrow,\;\; \beta_{c_\perp}\sim -z_\perp c_\perp F_{dis}(\{\Gamma_i,v\})<0\Rightarrow c_\perp\nearrow,\\
		\beta_{w}&\sim z_\perp w F_{dis}(\{\Gamma_i,v\})>0\Rightarrow w\searrow, \;\; \beta_{\lambda}\sim -z_\perp\lambda F_{dis}(\{\Gamma_i,v\})<0\Rightarrow \lambda \nearrow.
	\end{align*}
	
	We discuss this low energy physics in a more rigorous way. Let us start from the assumption that $c$ and $c_\perp$ are much larger than 1 and $v$, i.e, $c,c_\perp\gg1>v$, and show that this assumption is internally self-consistent. Physically, this assumption looks reasonable since the random charge potential enhances $c$ and $c_\perp$ rapidly in the low energy limit while $v$ gets screening by the Yukawa interaction. With this condition, we obtain the following approximate expressions of $h_1(c,c_\perp,v)$, $h_2(c,c_\perp,v)$, $h_3(c,c_\perp,v)$, and $h_4(c,c_\perp,v)$ as
	\begin{subequations}
		\label{eq:approximatedhi}
		\begin{align}
			h_1(c,c_\perp,v)&\approx \frac{1}{cc_\perp}\int_0^1dx \frac{1}{\sqrt{x}}=\frac{2}{cc_\perp},\\
			h_2(c,c_\perp,v)&\approx c_\perp^2\Big[\int_\alpha^1 dx \sqrt{\frac{1}{x^3c_\perp^6c^2}}\Big]=\frac{2}{c_\perp c}(\alpha^{-1/2}-1),\\
			h_3(c,c_\perp,v)&\approx c^2\int_{\alpha}^1\sqrt{\frac{1}{x^3 c_\perp^2 c^6}}=\frac{2}{c_\perp c}(\alpha^{-1/2}-1)\\
			h_4(c,c_\perp,v)&\approx \frac{\pi}{c^3c_\perp^3}\int_0^1dx\int_{0}^{\gamma(x)}dy \frac{c^2c_\perp^2(1-x-y)+c^2+c_\perp^2}{(1-x-y)^{5/2}}\Big(\gamma(x)=min[\frac{c^2}{1+c^2}-x,\frac{c^2}{v^2+c^2}-x,\frac{c_\perp^2}{1+c_\perp^2}-x]\Big)\nonumber\\
			&\approx 2\pi\frac{\beta^{-1/2}}{cc_\perp}\Big[1+\frac{1}{3}\beta^{-1}\Big(\frac{1}{c^2}+\frac{1}{c_\perp^2}\Big)\Big] ,
		\end{align}
	\end{subequations}
	where $\alpha=max[c_\perp^{-2},(1+v^2)c^{-2}]$ and $\beta=max[c^{-2},c^{-2}v^2,c_\perp^{-2}]$. In these approximate expressions, all $h_i(c,c_\perp,v)$ functions are proportional to $\frac{1}{c}$ or $\frac{1}{c_\perp}$ or $\frac{1}{c_\perp c}$ in the regime $c,c_\perp\gg1>v$.
	
	Now, we introduce $\beta_{w\lambda}$ from $\beta_w$ and $\beta_\lambda$ as follows
	\begin{align}
		\beta_{w\lambda}&=w\lambda\Big(\frac{\beta_w}{w}+\frac{\beta_\lambda}{\lambda}\Big)\nonumber\\
		&=w\lambda z_\perp\Big[-\epsilon+ \frac{\lambda}{8\pi}+\frac{\lambda w}{4\pi^3 N_cN_f}\Big(\pi(N_c^2-1)[h_2(c,c_\perp,v)+h_3(c,c_\perp,v)]-h_4(c,c_\perp,v)\Big)+2G_{dis}(\{\Gamma_i,v\})\Big]\nonumber\\
		&\approx w\lambda z_\perp\Big[-\epsilon+\frac{\lambda}{8\pi}+\frac{\lambda w}{4\pi^3 N_cN_f}\Big(4\pi(N_c^2-1)\frac{\alpha^{-1/2}}{c_\perp c}-2\pi\frac{\beta^{-1/2}}{cc_\perp}\Big)+2G_{dis}(\{\Gamma_i,v\})\Big]
	\end{align}
	In the regime of $c,c_\perp\gg 1>v$, $\beta_{w\lambda}$ is a positive valued function in the most region of the parameter space. As a result, $w\lambda$ decreases in the low energy limit. Then, the third term proportional to $\lambda w h_i(c,c_\perp,v)$ in $\beta_\lambda$ would be much smaller than other terms in the low energy limit. This leads to the fact that $\lambda\rightarrow 4\pi F_{dis}(\{\Gamma_i,v\})$ in the low energy limit.
	
	Using these results of $w\lambda\searrow$ and $\lambda\rightarrow 4\pi F_{dis}(\{\Gamma_i,v\})$ with our assumptions, we obtain the low energy behaviors of the remaining parameters as follows:
	\begin{subequations}
		\begin{align}
			\beta_{c}&\sim z_\perp\frac{c}{2}\Big[\frac{\lambda}{4\pi}-2F_{dis}(\{\Gamma_i,v\})\Big]\rightarrow z_\perp\frac{c}{2}\Big[-F_{dis}(\Gamma_i,v)\Big]\Rightarrow c\nearrow,\\
			\beta_{c_\perp}&\sim z_\perp\frac{c_\perp}{2}\Big[\frac{\lambda}{4\pi}-2F_{dis}(\{\Gamma_i,v\})\Big]\rightarrow z_\perp\frac{c_\perp}{2}\Big[-F_{dis}(\Gamma_i,v)\Big]\Rightarrow c_\perp\nearrow,\\
			\beta_s&\sim -z_\perp\frac{s}{2}\frac{\lambda}{4\pi c_\perp^2}\searrow\Rightarrow s\rightarrow s_{sat},\\
			\beta_{w}&\sim z_\perp w\Big[-\frac{\lambda}{8\pi}+F_{dis}(\{\Gamma_i,v\})\Big]\rightarrow z_\perp w\frac{F_{dis}(\{\Gamma_i,v\})}{2}\Rightarrow w\searrow,\\
			\beta_{\kappa_1}&\sim z_\perp \kappa_1\Big[-\epsilon+\frac{\lambda}{8\pi}+\frac{N_c^2+7}{2\pi^2}\kappa_1\Big]\rightarrow  z_\perp\kappa_1\Big[\frac{F_{dis}(\{\Gamma_i,v\})}{2}+\frac{1}{2\pi^2}(N_c^2+7)\kappa_1\Big]\Rightarrow \kappa_1\searrow ,\\
			\beta_{\Gamma_{G1}^d}&\sim z_\perp \Gamma_{G1}^d\Big[-\epsilon+\tilde{A}_{\Gamma_{G1}^d}^{(1)}\Big]\Rightarrow \Gamma_{G1}^d\nearrow.
		\end{align}
	\end{subequations}
	These low energy behaviors are consistent with our numerical results on various initial conditions, shown in Fig. \ref{fig:2dBetaFunctionNoRBMCase}. In addition, since $c$ and $c_\perp$ increase in the low energy limit, these results are consistent with the assumption we made in the beginning.
	
	\begin{figure}[h]
		\begin{subfigure}{0.23\textwidth}				 \includegraphics[scale=0.07]{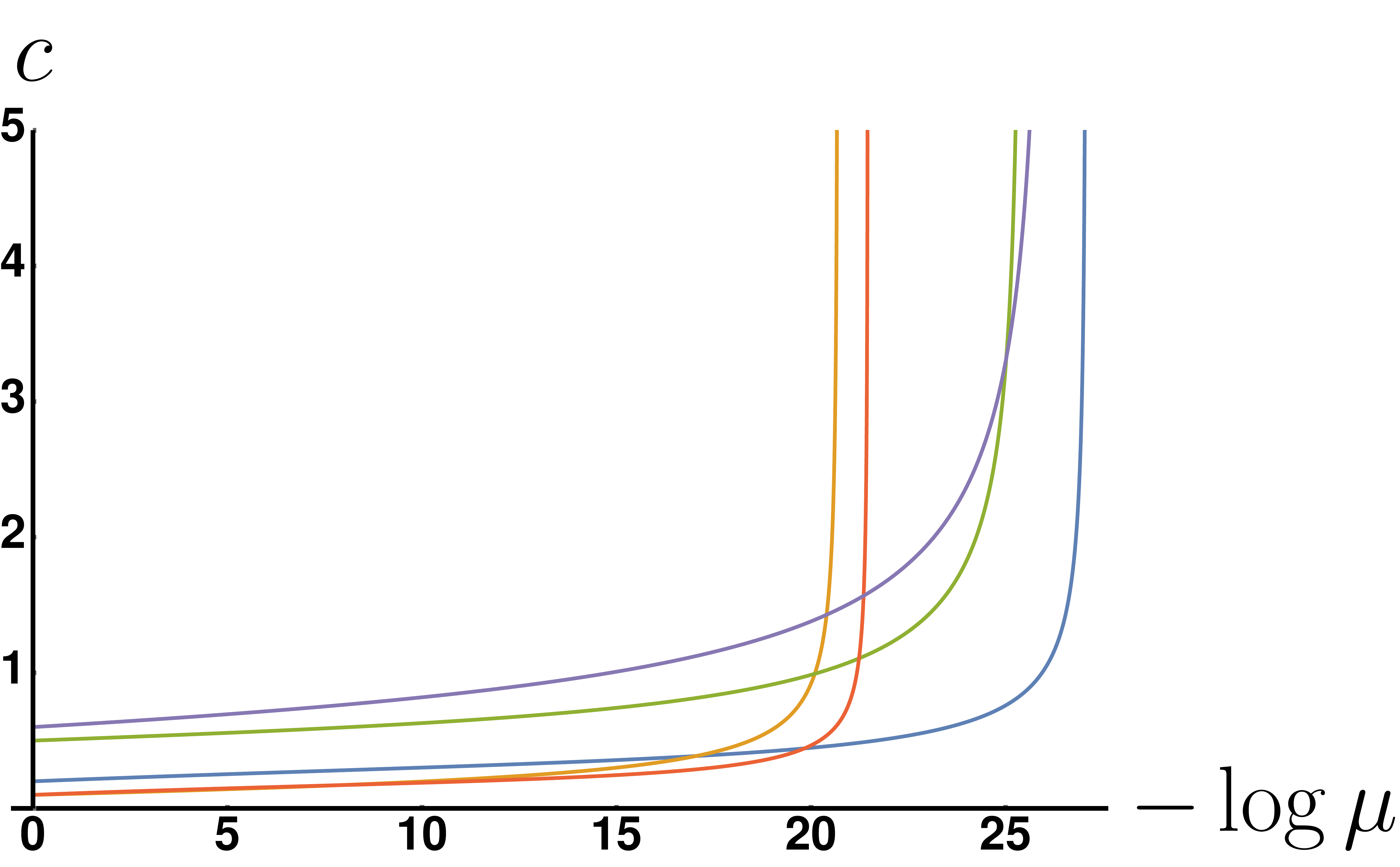}
			\caption{RG flow diagram of $c$}
		\end{subfigure}
		~
		\begin{subfigure}{0.23\textwidth}				 \includegraphics[scale=0.07]{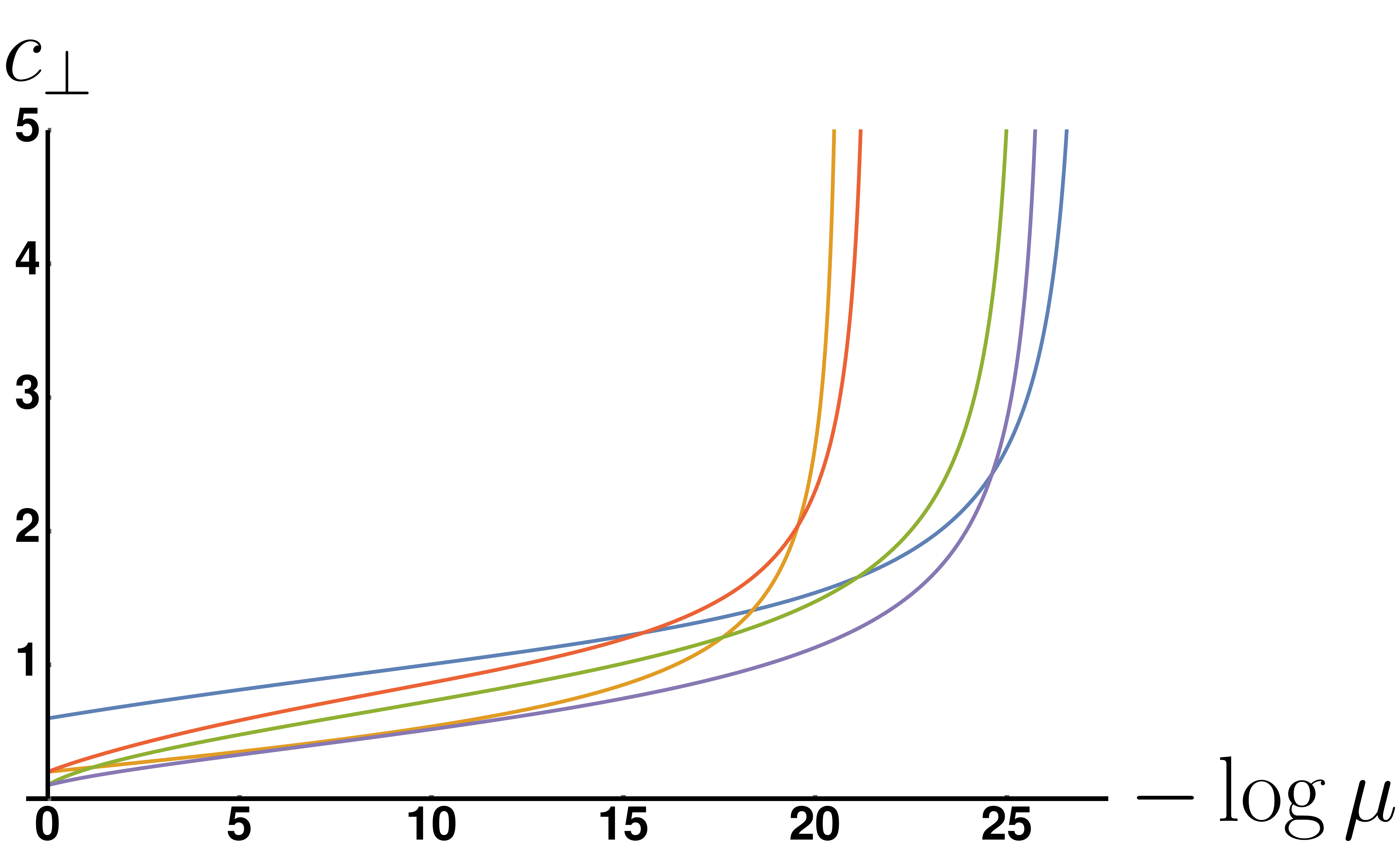}
			\caption{RG flow diagram of $c_\perp$}
		\end{subfigure}
		~
		\begin{subfigure}{0.23\textwidth}				 \includegraphics[scale=0.07]{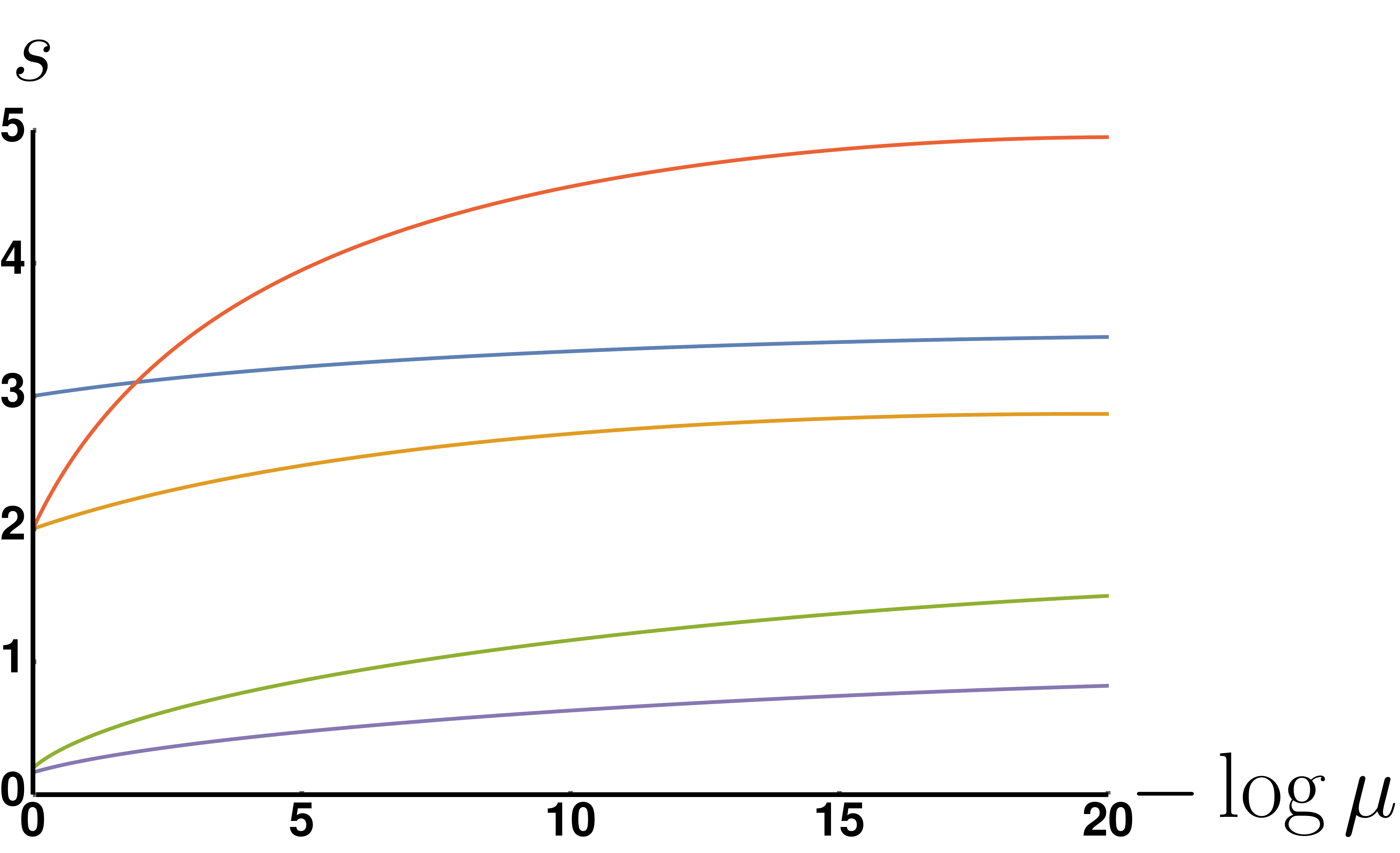}
			\caption{RG flow diagram of $s$}
		\end{subfigure}
		~
		\begin{subfigure}{0.23\textwidth}				 \includegraphics[scale=0.07]{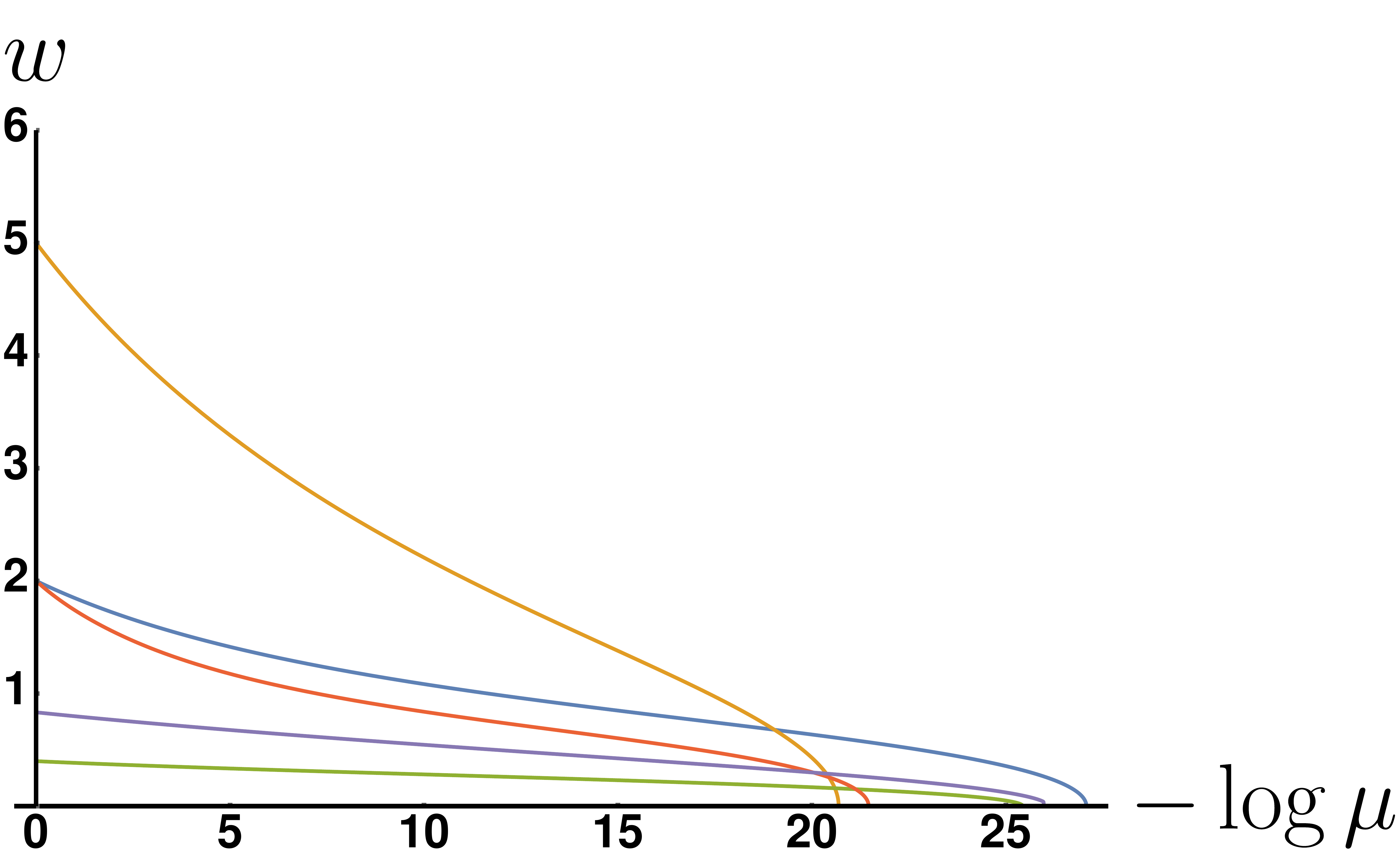}
			\caption{RG flow diagram of $w$}
		\end{subfigure}
		~
		\begin{subfigure}{0.23\textwidth}				 \includegraphics[scale=0.07]{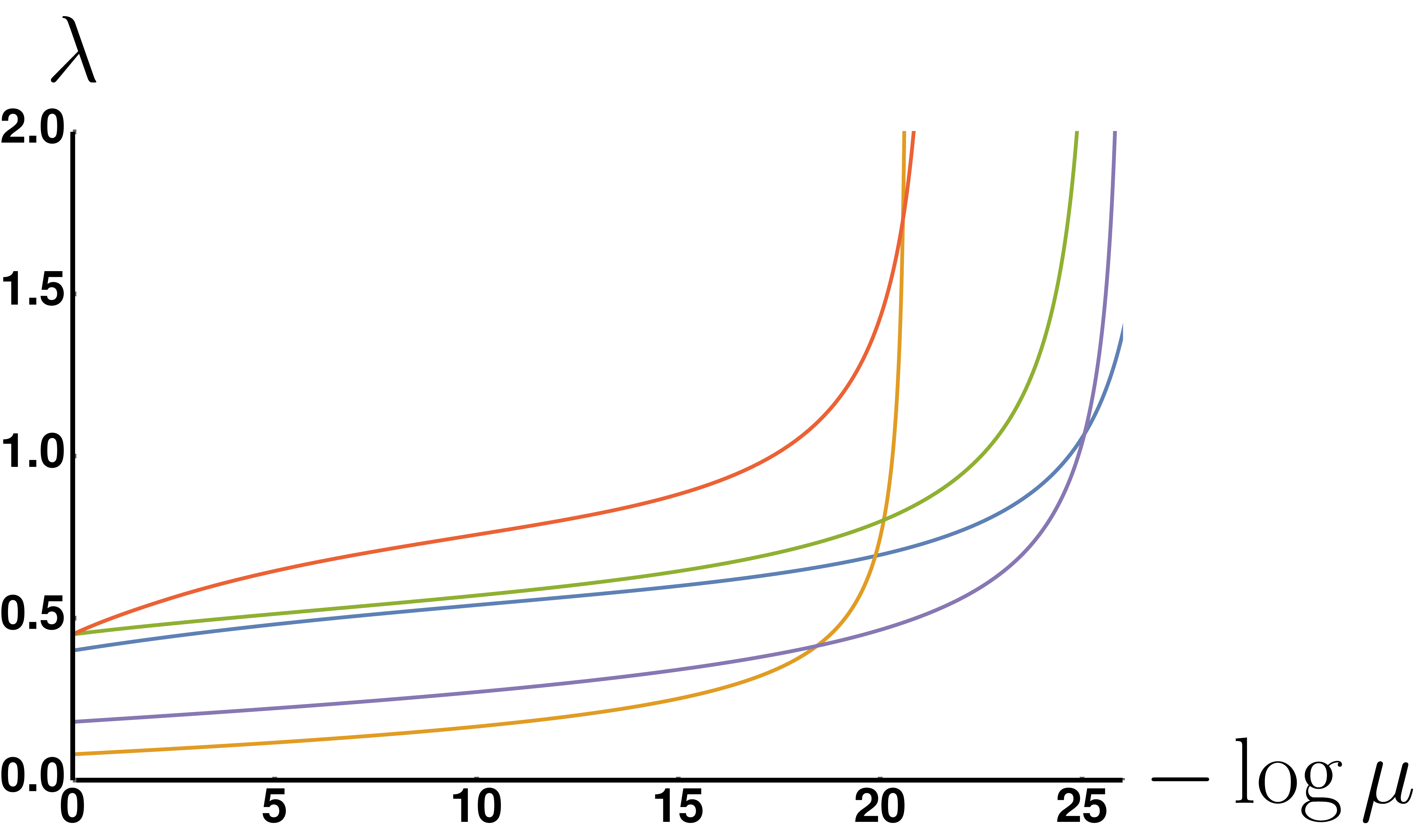}
			\caption{RG flow diagram of $\lambda$}
		\end{subfigure}
		~
		\begin{subfigure}{0.23\textwidth}				 \includegraphics[scale=0.07]{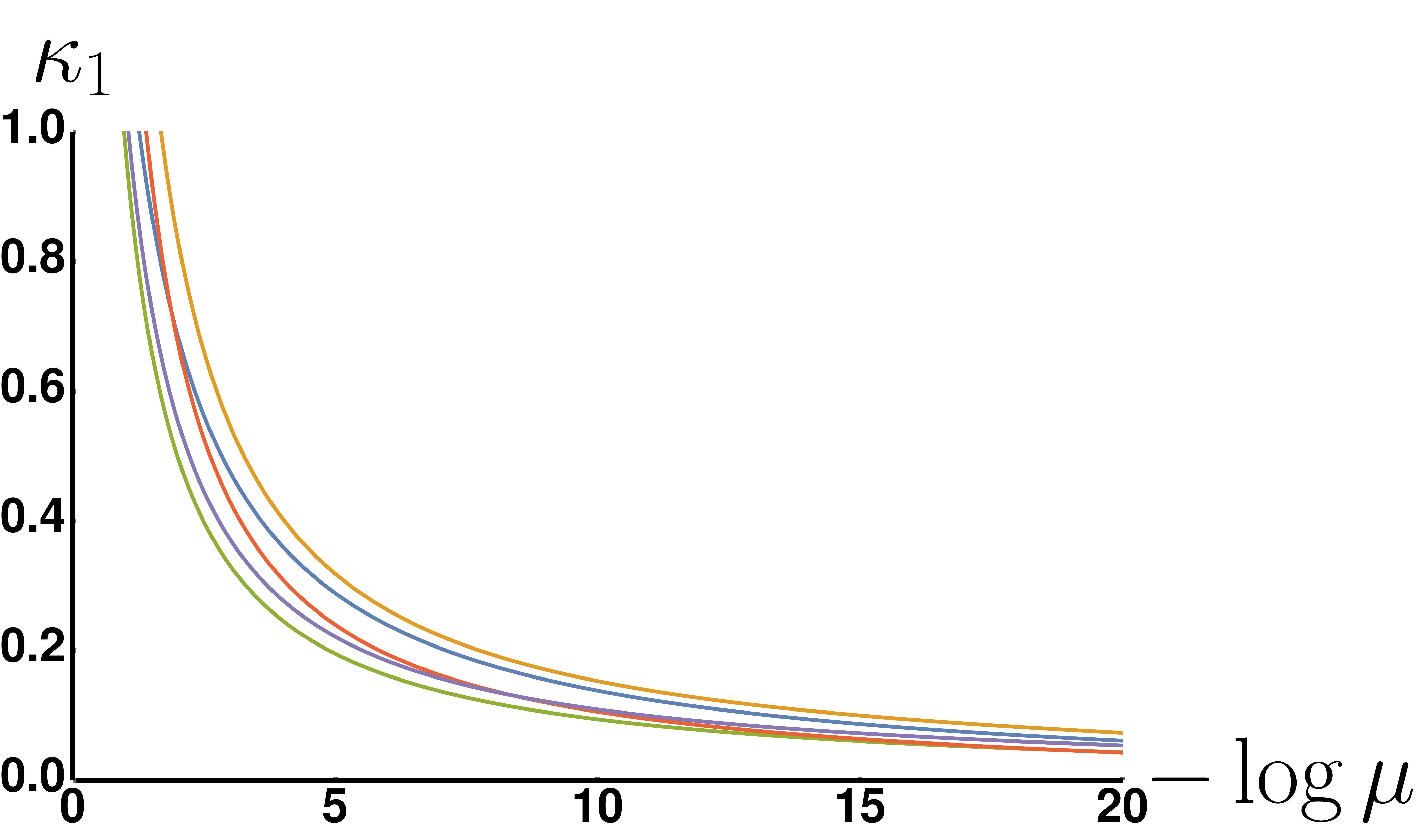}
			\caption{RG flow diagram of $\kappa_1$}
		\end{subfigure}
		~
		\begin{subfigure}{0.23\textwidth}				 \includegraphics[scale=0.07]{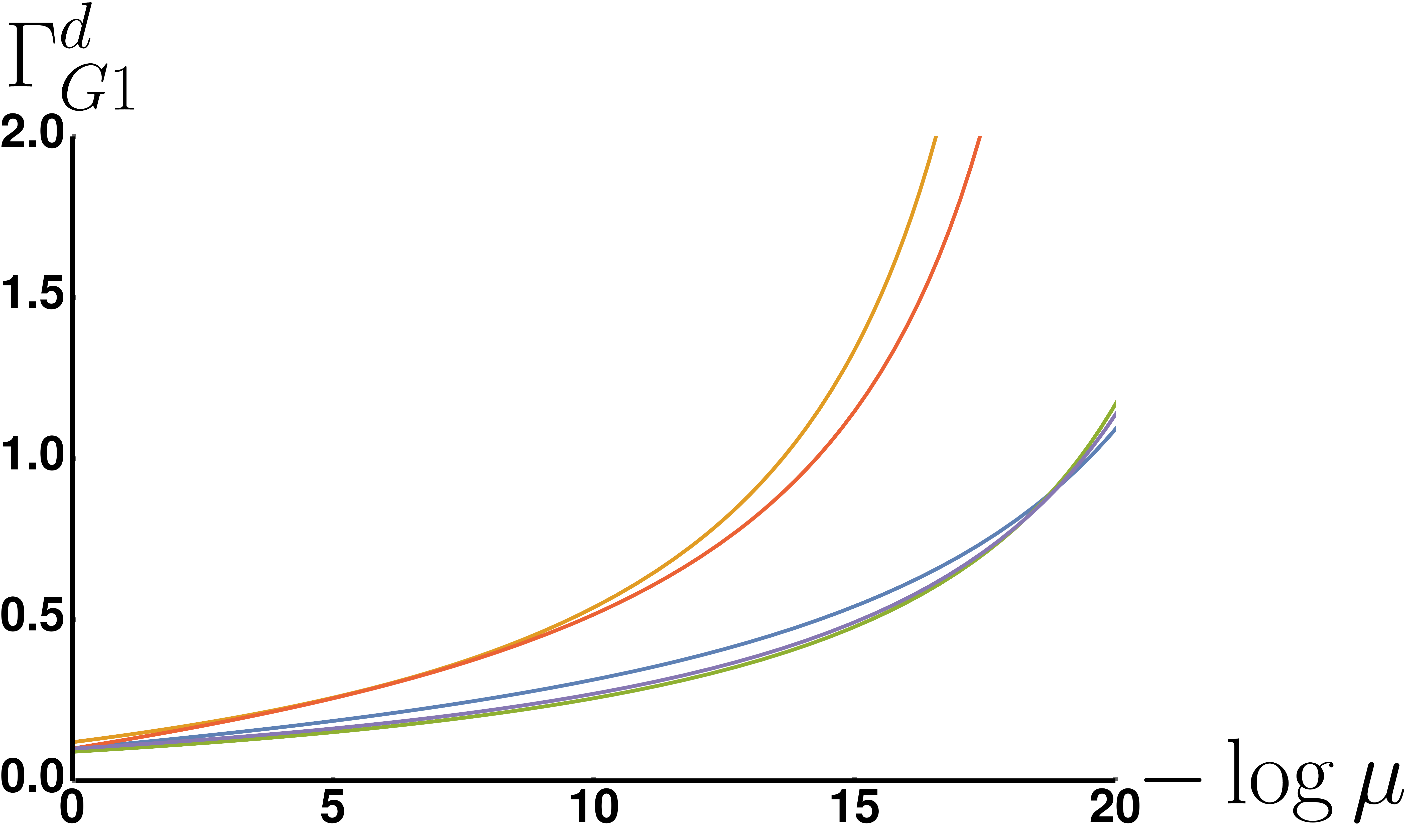}
			\caption{RG flow diagram of $\Gamma_{G1}^d$}
		\end{subfigure}
		\caption{RG flow diagrams of the `No RBM' (No random boson mass term) case with various initial conditions. We used $\epsilon=\bar{\epsilon}=0.01$, $N_f=1$, $N_c=2$, $v_c=0.05$, and $\kappa=1$.} \label{fig:2dBetaFunctionNoRBMCase}
	\end{figure}

	\subsubsection{General case ($g\neq0$ ,$\Gamma_i\neq 0$, $\Gamma_M\neq 0$)}\label{Appendix:DetailedAnalysisOneLoopBetaFunctions:GeneralCase}
	
	Finally, we consider the general case with all ingredients; Yukawa interaction, random charge potential, and random boson mass. We can obtain beta functions of $\frac{\lambda}{c_\perp^2}$, $\frac{\gamma_M}{\Gamma_i}$, and $w\lambda$ from the original beta functions. These ratio beta functions make our analysis easier.
	\begin{align}
		\beta_{\lambda/c_\perp^2}&=\frac{\lambda}{c_\perp^2}\Big[z_\perp\Big\{-\epsilon+\frac{\lambda}{4\pi}\frac{1}{c_\perp^2}+\frac{\lambda w}{4\pi^3 N_cN_f}\Big(\pi(N_c^2-1)[h_1(c,c_\perp,v)-h_2(c,c_\perp,v)]-h_4(c,c_\perp,v)\Big)\nonumber\\
		&+F_{dis}(\{\Gamma_i,v\})+2G_{dis}(\Gamma_i,v)\Big\}-\frac{\gamma_M}{\pi^2}\frac{z_\perp\epsilon+\bar{\epsilon}}{\epsilon+\bar{\epsilon}}\Big]\label{eq:generalCasebetalambdacperp},\\
		\beta_{\gamma_M/\Gamma_i}&=\frac{\gamma_M}{\Gamma_i}\Big[-\bar{\epsilon}+z_\perp\Big\{\frac{\lambda}{4\pi}\frac{1}{c_\perp^2}-\frac{N_c^2-1}{2\pi^2 N_cN_f}w\lambda h_2(c,c_\perp,v)+\frac{N_c^2+1}{\pi^2}\kappa_1 -A_{\Gamma_i}^{(1)}\Big\}-\frac{\gamma_M}{\pi^2}\Big(3+\frac{5\pi}{4}\kappa s \Big)\frac{z_\perp \epsilon+\bar{\epsilon}}{\epsilon+\bar{\epsilon}}\Big],\\
		\beta_{w\lambda}&=w\lambda\Big[z_\perp\Big\{-\epsilon+\frac{\lambda}{8\pi}+\frac{\lambda w}{4\pi^3 N_cN_f}\Big(-h_4(c,c_\perp,v)+\pi(N_c^2-1)[h_2(c,c_\perp,v)+h_3(c,c_\perp,v)]\Big)\nonumber\\
		&+2G_{dis}(\{\Gamma_{i},v\})\Big\}+\frac{\gamma_M}{2\pi^2}\Big(1+\frac{\pi}{4}\kappa s\Big)\frac{z_\perp \epsilon+\bar{\epsilon}}{\epsilon+\bar{\epsilon}}\Big] . \label{eq:generalCasebetawlambda}
	\end{align}
	
	\begin{figure}[h]
		\begin{subfigure}{0.23\textwidth}				 \includegraphics[scale=0.07]{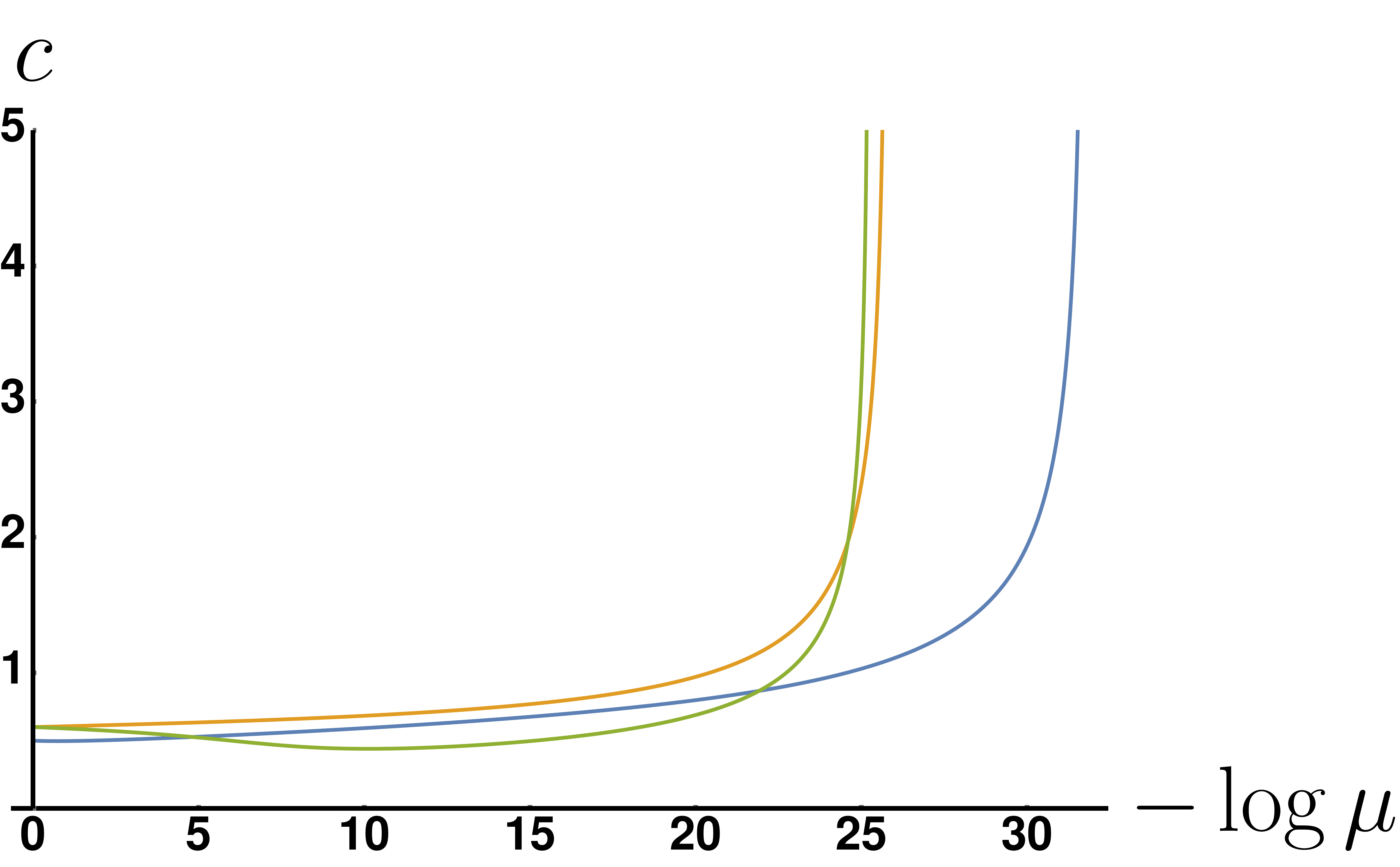}
			\caption{RG flow diagram of $c$}
		\end{subfigure}
		~
		\begin{subfigure}{0.23\textwidth}				 \includegraphics[scale=0.07]{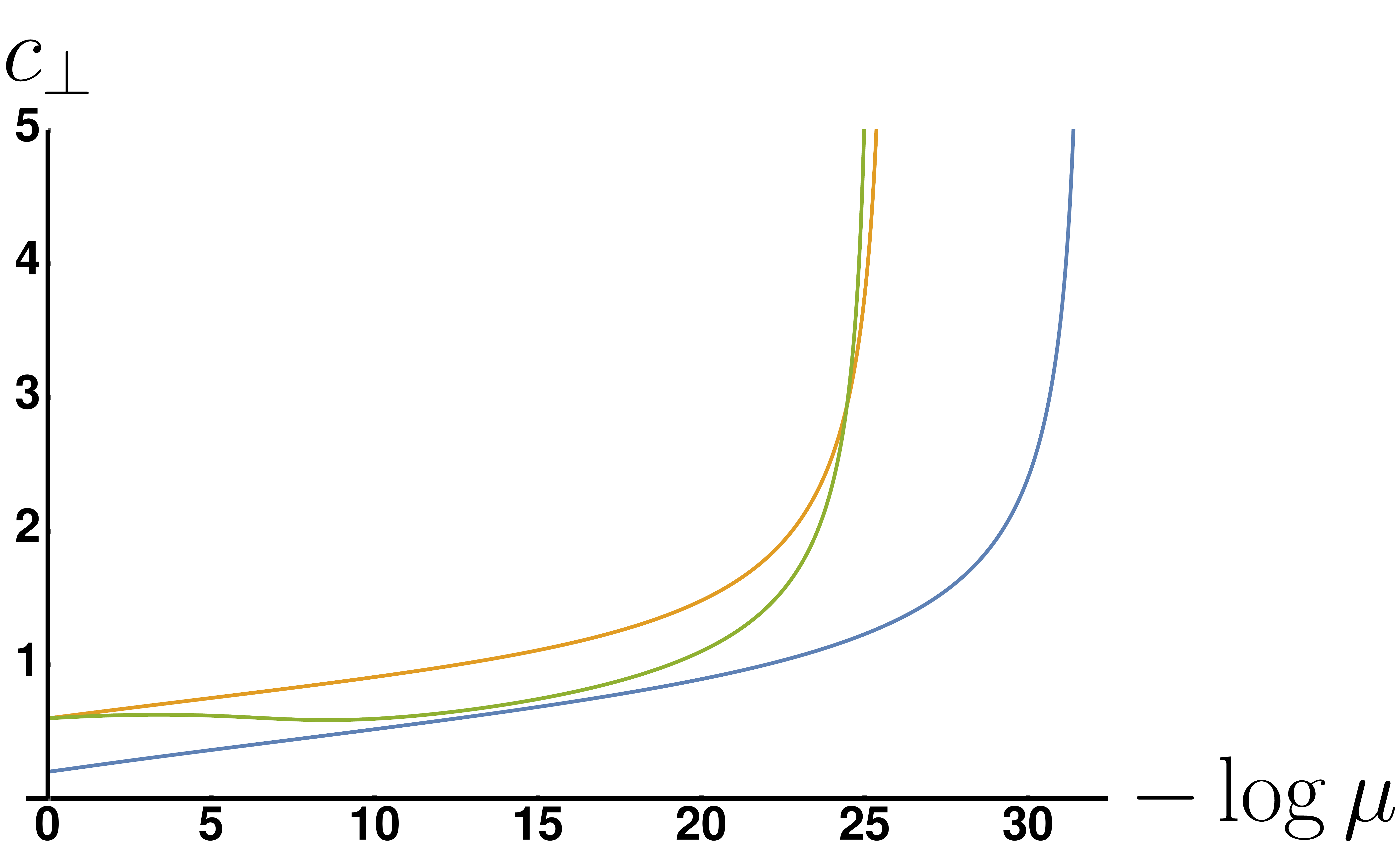}
			\caption{RG flow diagram of $c_\perp$}
		\end{subfigure}
		~
		\begin{subfigure}{0.23\textwidth}				 \includegraphics[scale=0.07]{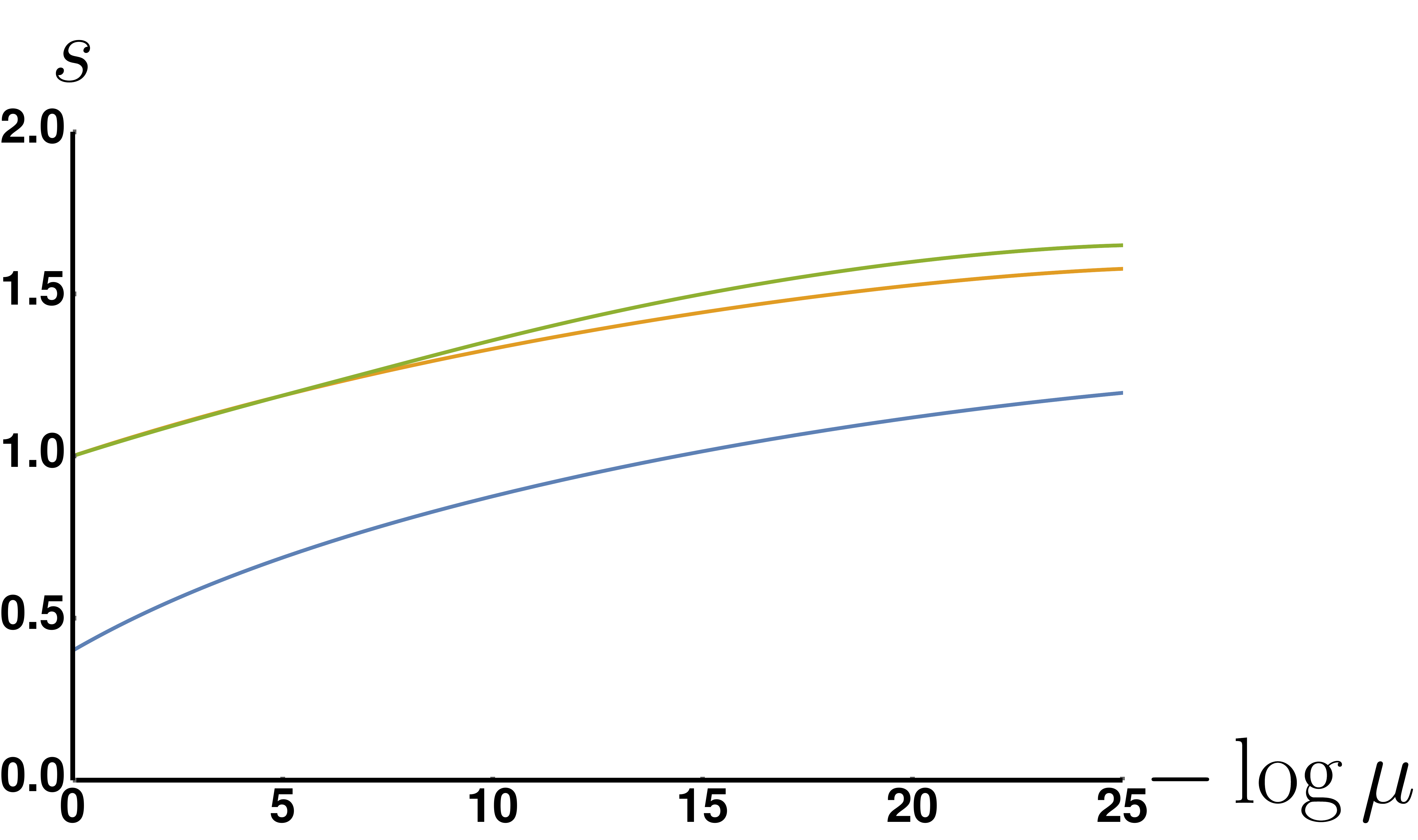}
			\caption{RG flow diagram of $s$}
		\end{subfigure}
		~
		\begin{subfigure}{0.23\textwidth}				 \includegraphics[scale=0.07]{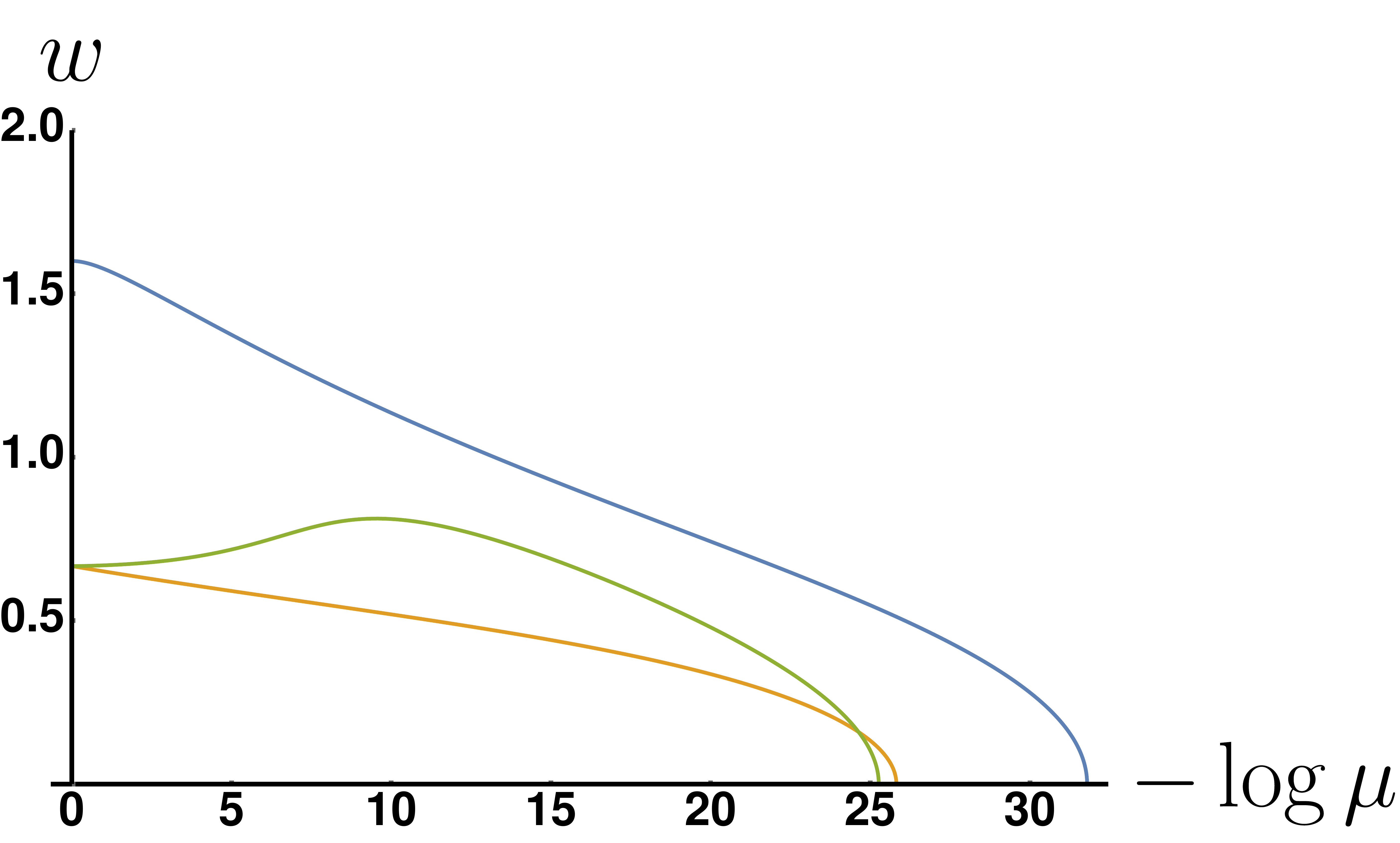}
			\caption{RG flow diagram of $w$}
		\end{subfigure}
		~
		\begin{subfigure}{0.23\textwidth}				 \includegraphics[scale=0.07]{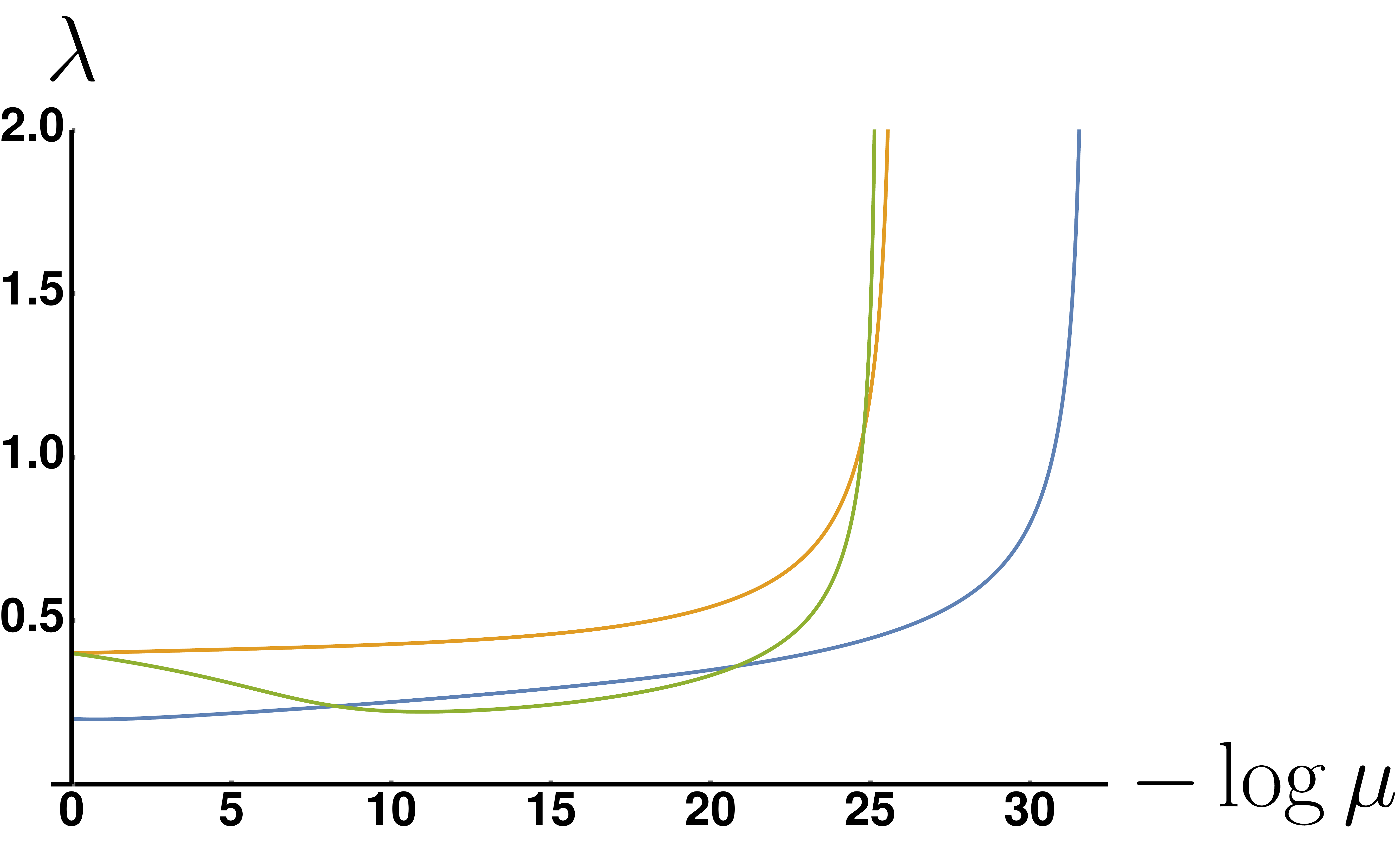}
			\caption{RG flow diagram of $\lambda$}
		\end{subfigure}
		~
		\begin{subfigure}{0.23\textwidth}				 \includegraphics[scale=0.07]{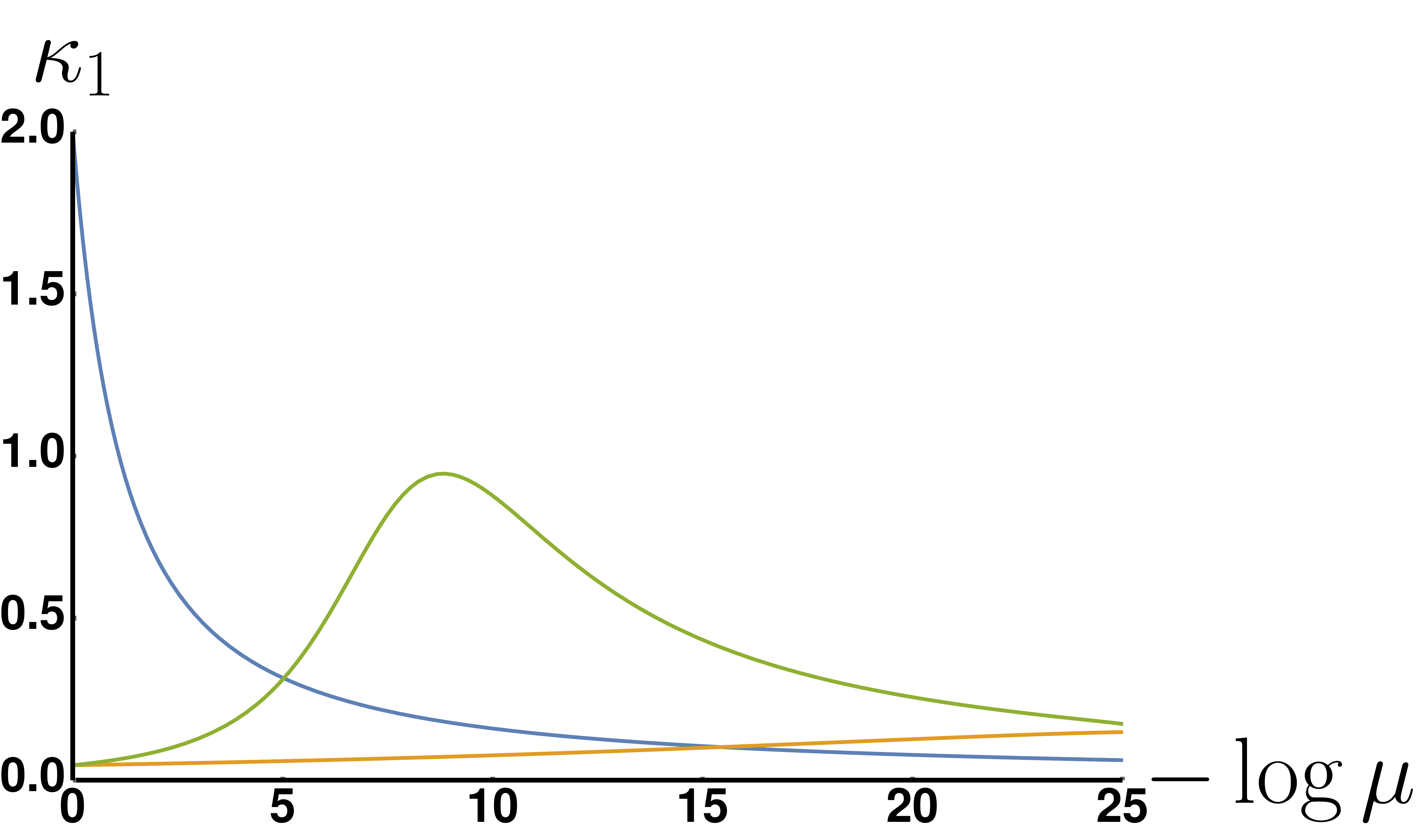}
			\caption{RG flow diagram of $\kappa_1$}
		\end{subfigure}
		~
		\begin{subfigure}{0.23\textwidth}				 \includegraphics[scale=0.07]{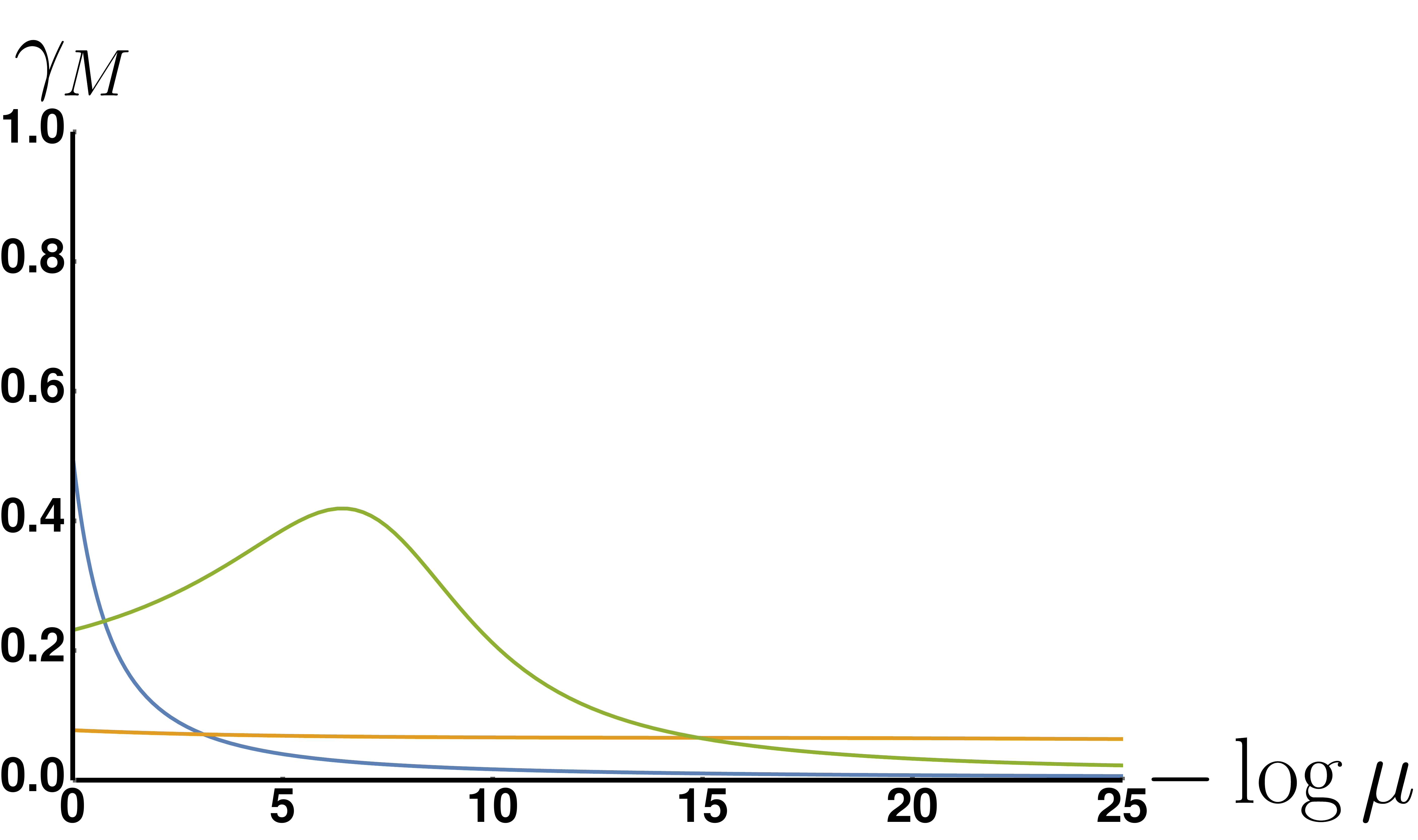}
			\caption{RG flow diagram of $\gamma_M$}
		\end{subfigure}
		~
		\begin{subfigure}{0.23\textwidth}				 \includegraphics[scale=0.07]{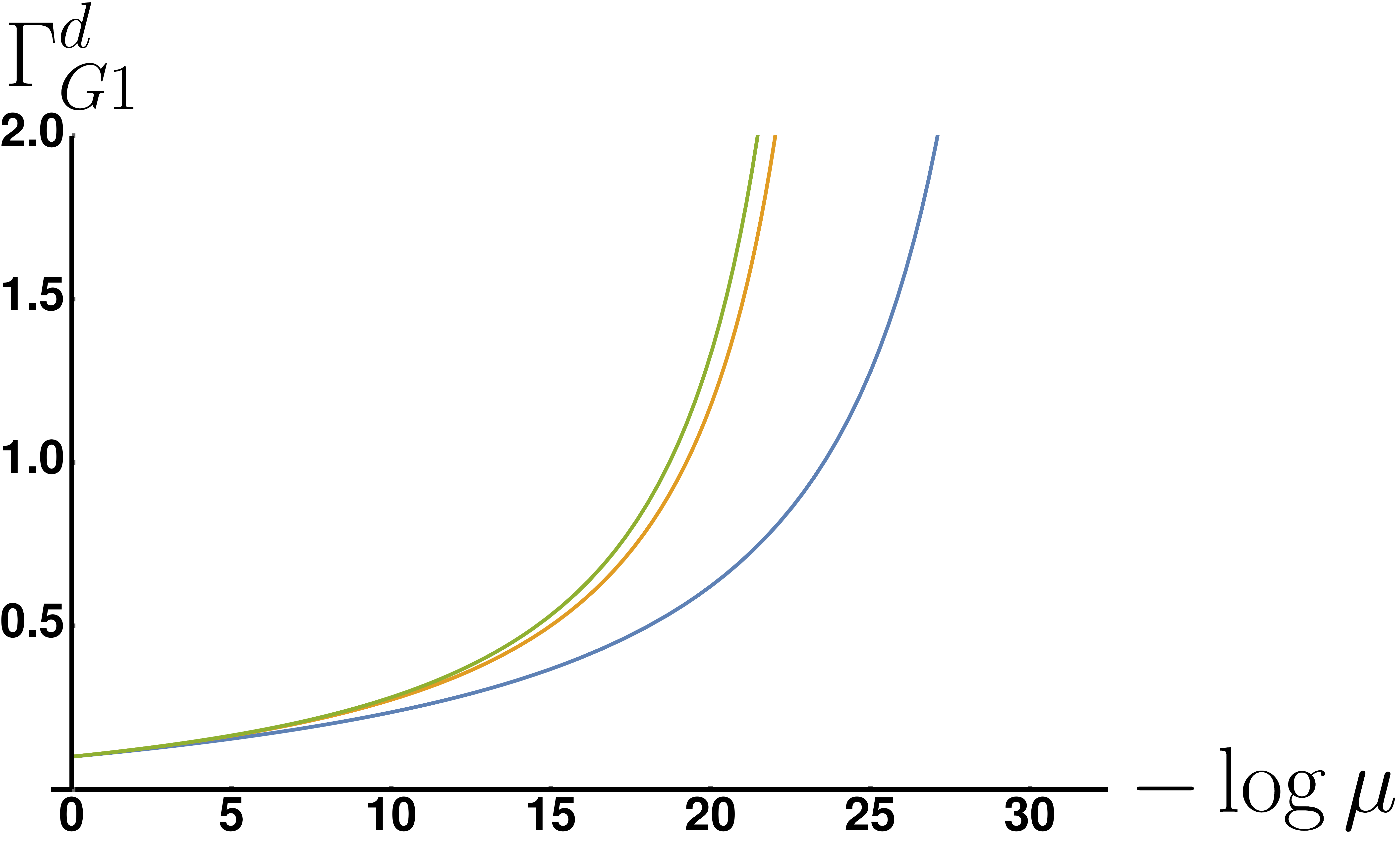}
			\caption{RG flow diagram of $\Gamma_{G1}^d$}
		\end{subfigure}
		\caption{RG flow diagrams of the general case with various initial conditions. Here, we used $\epsilon=\bar{\epsilon}=0.01$, $N_f=1$, $N_c=2$, $v_c=0.05$, and $\kappa=1$.} \label{fig:2dBetaFunctionGeneralCase}
	\end{figure}

	Two disorder effects, the random charge impurity potential and the random boson mass potential compete with each other as seen from the opposite sign of $F_{dis}(\{\Gamma_i,v\})$ and $\gamma_M$ in these beta functions.
	
	Our limiting-case studies of 'No random charge potential case' and `No random boson mass case' confirmed that effects of the random charge potential dominate over those of the random boson mass. The random boson mass gets screened by the boson-boson self-interaction ($\kappa_1$) while the random charge potential does not. This is indeed checked numerically as shown in Fig. \ref{fig:2dBetaFunctionGeneralCase}. In this respect, the low energy physics of the general case would be almost the same as that of the `No random boson mass' case, where the similarity between RG flow diagrams is shown in Fig. \ref{fig:2dBetaFunctionNoRBMCase} and Fig. \ref{fig:2dBetaFunctionGeneralCase}.
	
	To find analytical support for this observation, we use the same strategy as that of the `No Random Boson mass' case. Let us consider the parameter space, where the role of the random charge potential ($\sim F_{dis}(\{\Gamma_i,v\})$) is much stronger than that of the random boson mass ($\sim \gamma_M$). Based on this condition, we introduce an additional assumption: $c,\; c_\perp\gg 1>v$ as the `No random boson mass' case. Then, we apply the same approximation of the previous section to the function $h_i(c,c_\perp,v)$ here (Eq. \eqref{eq:approximatedhi}).
	
	First, let us consider $\beta_{w\lambda}$ (Eq. \eqref{eq:generalCasebetawlambda}) and $\beta_{\lambda/c_\perp^2}$ (Eq. \eqref{eq:generalCasebetalambdacperp}). Resulting approximate expressions of $\beta_{w\lambda}$ and $\beta_{\lambda/c_\perp^2}$ are given by
	
	\begin{align}
		\beta_{w\lambda}&\approx w\lambda\Big[z_\perp\Big\{-\epsilon+\frac{\lambda}{8\pi}+\frac{\lambda w}{4\pi^3 N_cN_f}\Big(4\pi(N_c^2-1)\frac{\alpha^{-1/2}}{c_\perp c}-2\pi\frac{\beta^{-1/2}}{cc_\perp}\Big)+2G_{dis}(\{\Gamma_i,v\})\Big\}\nonumber\\
		&+\frac{\gamma_M}{2\pi^2}\Big(1+\frac{\pi}{4}\kappa s\Big)\frac{z_\perp \epsilon+\bar{\epsilon}}{\epsilon+\bar{\epsilon}}\Big]>0\Rightarrow w\lambda \searrow,\\
		\beta_{\lambda/c_\perp^2}&\approx \frac{\lambda}{c_\perp^2} \Big[z_\perp \Big\{-\epsilon+\frac{\lambda }{4\pi}\frac{1}{c_\perp^2}+\frac{\lambda w}{4\pi^3 N_cN_f}\Big(2\pi(N_c^2-1)\frac{\alpha^{-1/2}}{cc_\perp}-2\pi \frac{\beta^{-1/2}}{cc_\perp}\Big)\nonumber\\
		&+F_{dis}(\{\Gamma_i,v\})+2G_{dis}(\{\Gamma_i,v\})\Big\}-\frac{\gamma_M}{\pi^2}\frac{z_\perp\epsilon+\bar{\epsilon}}{\epsilon+\bar{\epsilon}}\Big]>0\Rightarrow \frac{\lambda}{c_\perp^2}\searrow ,
	\end{align}
	where $\alpha=max[c_\perp^{-2},(1+v^2)c^{-2}]$ and $\beta=max[c^{-2},c^{-2}v^2,c_\perp^{-2}]$. In this derivation we resort to the assumption that $F_{dis}(\{\Gamma_i,v\}\}), G_{dis}(\{\Gamma_i,v\}) \gg \gamma_M, s\gamma_M$ and $c,c_\perp\gg 1>v$. Then, $\beta_{\lambda}$ can be approximated as
	\begin{align}
		\beta_{\lambda}\approx \lambda\Big[z_\perp\Big\{-\epsilon+\frac{\lambda}{4\pi}-F_{dis}(\{\Gamma_i,v\})+2G_{dis}(\{\Gamma_i,v\})\Big\}+\frac{\gamma_M}{\pi^2}\Big(1+\frac{\pi}{2}\kappa s \Big)\frac{z_\perp\epsilon+\bar{\epsilon}}{\epsilon+\bar{\epsilon}}\Big] .
	\end{align}
	Here, all the terms proportional to $w\lambda h_i(c,c_\perp,v)$ are ignored since they are much smaller than $\frac{\lambda}{4\pi}$ based on $c,c_\perp\gg 1>v$ and $h_i(c,c_\perp,v)\sim\frac{1}{c},\frac{1}{c_\perp}, \frac{1}{cc_\perp}$. $F_{dis}(\{\Gamma_i,v\})$ is larger than $G_{dis}(\{\Gamma_i,v\})$ since it involves more scattering channels in the `Direct' category. Additionally, the term proportional to $\gamma_M$ is much smaller than $F_{dis}(\{\Gamma_i,v\})$ based on the assumption. As a result, we find
	\begin{align}
		\frac{\lambda}{4\pi}\rightarrow \epsilon+F_{dis}(\{\Gamma_i,v\})-2G_{dis}(\{\Gamma_i,v\})-\frac{1}{z_\perp}\frac{\gamma_M}{\pi^2}\Big(1+\frac{\pi}{2}\kappa s \Big)\frac{z_\perp\epsilon+\bar{\epsilon}}{\epsilon+\bar{\epsilon}}\nearrow.
	\end{align}
	
	Using all these results on $w\lambda$, $\frac{\lambda}{c_\perp^2}$m and $\lambda$ with our assumptions, we find the low energy behaviors of all the remaining parameters as follows:
	\begin{subequations}
		\begin{align}
			\beta_{\Gamma_{G_i}^d}&\approx z_\perp \Gamma_{Gi}^d\Big[-\epsilon+A_{\Gamma_{Gi}^d}^{(1)}\Big]<0\Rightarrow \Gamma_{Gi}^d(\because A_{\Gamma_{Gi}^d}<0)\nearrow,\\
			\beta_c&\approx -z_\perp\frac{c_\perp}{2}F_{dis}(\{\Gamma_i,v\})<0\Rightarrow c\nearrow,\\
			\beta_{c_\perp}&\approx -z_\perp\frac{c_\perp}{2}F_{dis}(\{\Gamma_i,v\})<0\Rightarrow c_\perp\nearrow,\\
			\beta_w&\approx wz_\perp \frac{1}{2}F_{dis}(\{\Gamma_i,v\})>0\Rightarrow w\searrow,\\
			\beta_{\kappa_1}&\approx\kappa_1\Bigg[z_\perp\Big(-\frac{\epsilon}{2}+\frac{F_{dis}(\{\Gamma_i,v\})}{2}-G_{dis}(\{\Gamma_i,v\})+\frac{N_c^2+7}{2\pi^2}\kappa_1\Big)-\frac{\gamma_M}{\pi^2}\Big(7+\frac{7\pi}{2}\kappa s\Big)\frac{z_\perp\epsilon+\bar{\epsilon}}{\epsilon+\bar{\epsilon}}\Bigg]>0\Rightarrow \kappa_1\searrow,\\
			\beta_{\gamma_M/\Gamma_{Gi}^d}&\approx \frac{\gamma_M}{\Gamma_{Gi}^d}\Big[-z_\perp A_{\Gamma_{Gi}^d}^{(1)}-\frac{\gamma_M}{\pi^2}\Big(3+\frac{5\pi}{4}\kappa s\Big)\frac{z_\perp\epsilon+\bar{\epsilon}}{\epsilon+\bar{\epsilon}}\Big]>0(\because A_{\Gamma_{Gi}^d}^{(1)}<0,\; \Gamma_{Gi}^d\gg \gamma_M)\Rightarrow \frac{\gamma_M}{\Gamma_{Gi}^d}\searrow.
		\end{align}
	\end{subequations}
	
	In the case of $\gamma_M$ and $s\gamma_M$, it is not easy to obtain their low energy behaviors since there is no dominant term proportional to $\Gamma_{Gi}^d$. Oscillating behaviors of $\gamma_M$ due to $\kappa_1$, previously discussed in the `No random charge potential' case, makes our analysis more difficult since the low energy dynamics of $\gamma_M$ are is sensitive to its initial value. However, our numerical results in Fig. \ref{fig:2dBetaFunctionGeneralCase} shows that $\gamma_M$ and $s\gamma_M$ are presumed to be small in the low energy limit.
	
	We point out that the resulting low energy behaviors of all the parameters are consistent with our assumptions; $c,c_\perp\gg 1 >v (w\ll 1)$ and $\frac{\gamma_M}{\Gamma_{Gi}^d}\ll 1$. Therefore, we conclude that the low energy properties of the general case are essentially the same as those of the `No random boson mass' case. However, if the initial value of $\gamma_M$ is set to be rather large, compared to $\Gamma_{G1}^d$, one can find a parameter space, where $\gamma_M$ and $\kappa_1$ increase rapidly as energy is lowered. It is a vestige of the oscillating pattern of $\gamma_M$ and $\kappa_1$ discussed in the `No random charge impurity' case. This leads to the breakdown of the one-loop RG analysis and plays an important role in determining possible low-energy properties of the system discussed in the main text.

	\section{Breakdown of the one-loop results by the oscillating RG-flows}\label{Appendix:BreakdownOfOneLoopByOscillatingRG}
	
	\begin{figure}[h]
		\begin{subfigure}{0.4\textwidth}
			\includegraphics[scale=0.09]{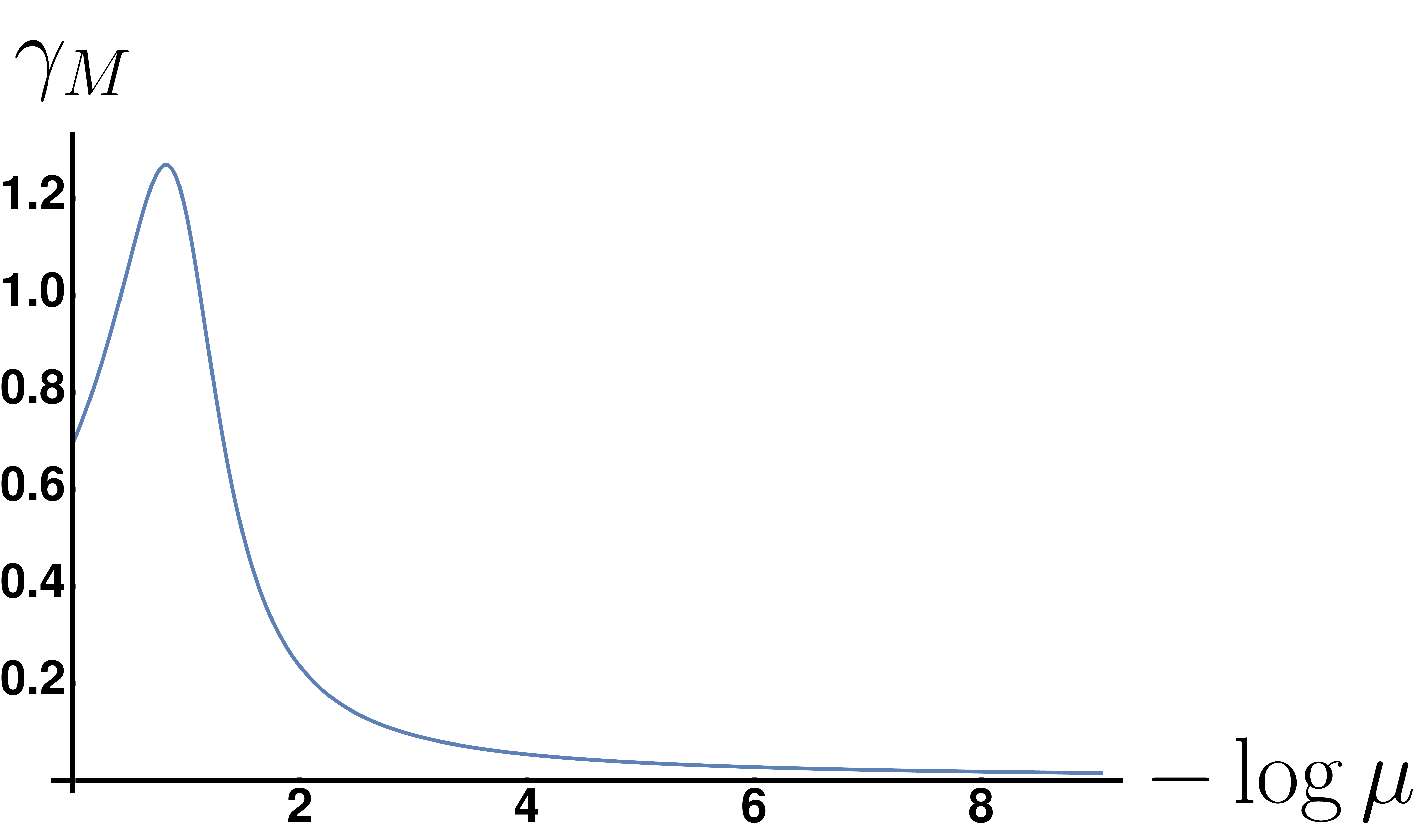}
			\caption{RG flow diagram of $\gamma_M$}
		\end{subfigure}
		~
		\begin{subfigure}{0.4\textwidth}
			\includegraphics[scale=0.09]{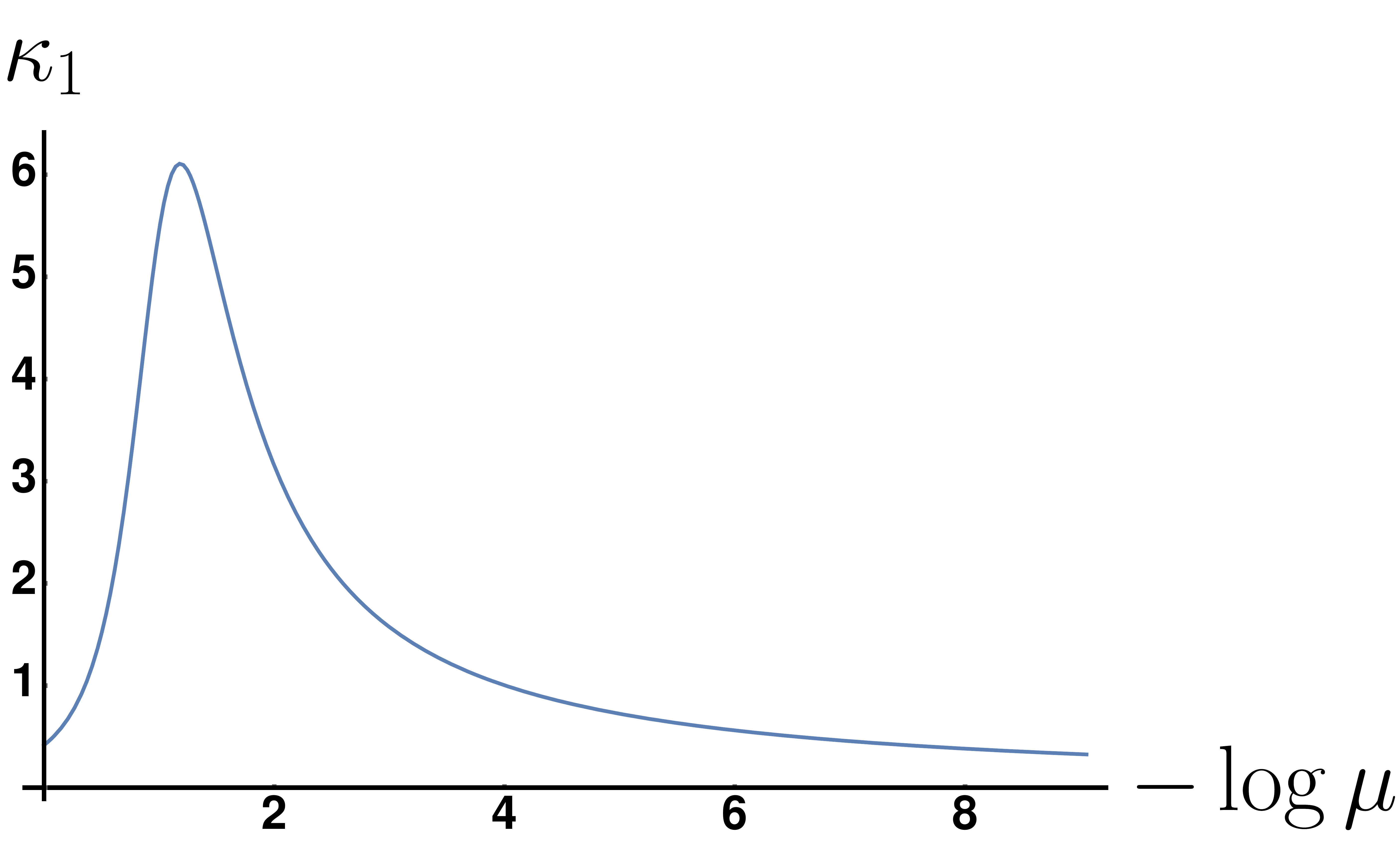}
			\caption{RG flow diagram of $\kappa_1$}
		\end{subfigure}
		\caption{RG flows of $\gamma_M$ and $\kappa_1$ in the general case. Here, an initial value of $\gamma_M$ sets to be larger than those of random charge potential vertices ($\Gamma_{i}$). Near the regime of $-\ln\mu\sim 1$, the one-loop RG analysis break down.} \label{fig:RGFlowsOneLoopBreakdown}
	\end{figure}

	Figure \ref{fig:RGFlowsOneLoopBreakdown} shows the RG-flows of $\gamma_M$ and $\kappa_1$ when an initial value of $\gamma_M$ is set to be larger than the value of random charge potential vertices. Even the RG-flows start from the small parameter regime where the one-loop analysis is valid, largely oscillatory RG flows allow large values of $\gamma_M$ and $\kappa_1$ outside the validity regime of the one-loop RG analysis at the early stage of the energy evolution, corresponding to the regime $-\log\mu\lesssim 1$ in Fig. \ref{fig:RGFlowsOneLoopBreakdown}. As a result, the one-loop RG analysis breaks down near the energy regime of $-\ln\mu\sim 1$.
	
	In the Ref. \cite{KirkpatricBelitz}, authors argued about the one-loop breakdown by the oscillating RG-flows in the interaction and disorder parameter space. In this study, we also observe such oscillating RG-flows of $\gamma_M$ and $\kappa_1$ as discussed in section \ref{sec:NoYIAndNorCPCase}.

	\section{Stability of the RBMD phase space} \label{Appendix:StabilityOfRBMDphase}
	
	Here, we discuss the stability RG flow in the RBMD phase space. Suppose that there is a stable phase space of $\gamma_M$ $ (s\gamma_M)$ and $\kappa_1$ with `large' values, which may correspond to the `random-singlet phase', discussed by Kirkpatrick and Belitz \cite{KirkpatricBelitz}. This is nothing but the condition that $\beta_{\gamma_M} < 0$ (or $\beta_{s\gamma_M}<0$) and $\beta_{\kappa_1}<0$. We will prove that there is a stable parameter space, where $w$ increases while $\Gamma_{G1}^d/\gamma_M$ or $\Gamma_{G1}^d/(s\gamma_M)$ and $\lambda$ decrease at least in the one-loop level, showing that $\beta_{w}<0$, $\beta_{\lambda}>0$, and $\beta_{\Gamma_{G1}^d/\gamma_M}>0$ or $\beta_{\Gamma_{G1}^d/(s\gamma_M)}>0$ in the parameter space.
	
	Following the strategy in section \ref{Appendix:DetailedAnalysisOneLoopBetaFunctions} to show the stability of a given phase, suppose that there is a parameter space where $max[s\gamma_M,\gamma_M]\gg \Gamma_{G1}^d$. Then, the effects of random boson mass fluctuations lead $c$ and $c_\perp$ to decrease in the low energy limit. As a result, it is reasonable to assume that $c,c_\perp\ll v<1 (w\gg 1)$ in the low energy limit, as discussed before. To simplify our arguments, let us consider two limiting cases depending on the hierarchy between $c$ and $c_\perp$: (i) $c\ll c_\perp$ and (ii) $c\gg c_\perp$.
	
	\subsection{$c\ll c_\perp$}
	
	First, we find approximate expressions of $h_1(c,c_\perp,v)$, $h_2(c,c_\perp,v)$, $h_3(c,c_\perp,v)$, and $h_4(c,c_\perp,v)$ as follows
	\begin{align}
		h_1(0,c_\perp,v)&=\frac{1}{\sqrt{1+v^2}}\int_0^1dx\sqrt{\frac{x}{\Big(1-x+xc_\perp^2\Big)(1-x)}}\approx \frac{1}{\sqrt{1+v^2}}\Big[\int_0^{\alpha}dx\frac{\sqrt{x}}{1-x}+\frac{1}{c_\perp}\int_{\alpha}^1 dx \frac{1}{\sqrt{1-x}}\Big]\nonumber\\
		&=\frac{2}{\sqrt{1+v^2}}\Big[-\sqrt{\alpha}+\frac{\sqrt{1-\alpha}}{c_\perp}+\tanh^{-1}\sqrt{\alpha}\Big]=\frac{2}{\sqrt{1+v^2}}\tanh^{-1}\Big(\frac{1}{\sqrt{1+c_\perp^2}}\Big)<\frac{2}{\sqrt{1+v^2}}\frac{1}{c_\perp},\\
		h_2(0,c_\perp,v)&=\frac{c_\perp^2}{\sqrt{1+v^2}}\int_0^1dx\sqrt{\frac{x}{\Big(1-x+xc_\perp^2\Big)^3(1-x)}}\approx \frac{c_\perp^2}{\sqrt{1+v^2}}\Big[\int_0^{\alpha}dx \frac{\sqrt{x}}{(1-x)^2}+\frac{1}{c_\perp^3}\int_\alpha^1 dx \frac{1}{x}\frac{1}{\sqrt{1-x}}\Big]\nonumber\\
		&=\frac{c_\perp^2}{\sqrt{1+v^2}}\Big[\frac{\sqrt{\alpha}}{1-\alpha}-\tanh^{-1}\sqrt{\alpha}+\frac{1}{c_\perp^3}\Big(2\ln(1+\sqrt{1-\alpha})-\ln \alpha \Big)\Big]\nonumber\\
		&=\frac{\sqrt{1+c_\perp^2}}{\sqrt{1+v^2}}-\frac{c_\perp^2}{\sqrt{1+v^2}}\tanh^{-1}\Big(\frac{1}{\sqrt{1+c_\perp^2}}\Big)+\frac{1}{\sqrt{1+v^2}}\frac{2}{c_\perp}\ln(c_\perp+\sqrt{1+c_\perp^2})\rightarrow \frac{3}{\sqrt{1+v^2}}(c_\perp\rightarrow 0)\\
		h_3(0,c_\perp,v)&=h_4(0,c_\perp,v)=0 ,
	\end{align}
	where $\alpha=\frac{1}{1+c_\perp^2}$. Here,  $c$ is set to be zero for simplicity. Note that $h_1(0,c_\perp,v)$ is a diverging function in the $c_\perp\rightarrow 0$ limit while $h_2(0,c_\perp,v)$ is not. As a result, $h_1(0,c_\perp,v)$ is usually larger than $h_2(0,c_\perp,v)$ in the low energy regime.
	
	With $h_i(c,c_\perp,v)$ and $c\ll c_\perp<v<1$, we find an approximate expression of the beta function $\beta_w$ as
	\begin{align}
		\beta_w&\approx w\Big[z_\perp\Big\{-\frac{\lambda}{8\pi}+\frac{N_c^2-1}{4\pi^2 N_cN_f}\lambda w h_1(0,c_\perp,v)+F_{dis}(\{\Gamma_i,v\})\Big\}-\frac{\gamma_M}{2\pi^2}\Big(1+\frac{3\pi}{4}\kappa s\Big)\frac{z_\perp \epsilon+\bar{\epsilon}}{\epsilon+\bar{\epsilon}}\Big] .
	\end{align}
	This expression gives $wh_1(0,c_\perp,v)\rightarrow \frac{4\pi^2 N_cN_f}{N_c^2-1}\frac{1}{\lambda}\Big(\frac{\lambda}{8\pi}-F_{dis}(\{\Gamma_i,v\})+\frac{1}{z_\perp}\frac{\gamma_M}{2\pi^2}\Big(1+\frac{3\pi}{4}\kappa s\Big)\frac{z_\perp \epsilon+\bar{\epsilon}}{\epsilon+\bar{\epsilon}}\Big)$ in the low energy limit.
	
	Accordingly, other beta functions and their RG flows are given by
	\begin{align}
		\beta_{\lambda/c_\perp^2}&\approx \frac{\lambda}{c_\perp^2}\Big[z_\perp\Big\{-\epsilon+\frac{\lambda}{4\pi}\frac{1}{c_\perp^2}+\frac{\lambda}{8\pi}+2G_{dis}(\{\Gamma_i,v\})\Big\}+\frac{\gamma_M}{2\pi^2}\Big(-1+\frac{3\pi}{4}\kappa s\Big)\frac{z_\perp\epsilon+\bar{\epsilon}}{\epsilon+\bar{\epsilon}}\Big]>0(\because s\gg 1)\Rightarrow \frac{\lambda}{c_\perp^2}\searrow,\\
		\beta_\lambda&\approx \lambda\Big[z_\perp\Big\{-\epsilon+\frac{\lambda}{8\pi}+2G_{dis}(\{\Gamma_i,v\})\Big\}+\frac{\gamma_M}{2\pi^2}\Big(1+\frac{\pi}{4}\kappa s\Big)\frac{z_\perp\epsilon+\bar{\epsilon}}{\epsilon+\bar{\epsilon}}\Big]>0\Rightarrow \lambda\searrow,\\
		\beta_{w\lambda}&\approx w\lambda\Big[z_\perp\Big\{-\epsilon+\frac{\lambda}{8\pi}+\frac{\lambda w(N_c^2-1)}{4\pi^2 N_cN_f}[h_2(0,c_\perp,v)+h_3(0,c_\perp,v)]+2G_{dis}(\{\Gamma_i,v\})\Big\}\nonumber\\
		&+\frac{\gamma_M}{2\pi^2}\Big(1+\frac{\pi}{4}\kappa s \Big)\frac{z_\perp \epsilon+\bar{\epsilon}}{\epsilon+\bar{\epsilon}}\Big]>0\Rightarrow w\lambda \searrow,\\
		\beta_s&\approx\frac{s}{2}\frac{\gamma_M}{\pi^2}\Big(1-\frac{\pi}{4}\kappa s\Big)\frac{z_\perp\epsilon+\bar{\epsilon}}{\epsilon+\bar{\epsilon}}<0(\because s\gg1 )\Rightarrow s\nearrow\\
		\beta_{\frac{\Gamma_{G1}^d}{\gamma_M}}&
		\approx \frac{\Gamma_{G1}^d}{\gamma_M}\frac{\gamma_M}{\pi^2}\Big(3+\frac{5\pi}{4}\kappa s\Big)\frac{z_\perp \epsilon+\bar{\epsilon}}{\epsilon+\bar{\epsilon}}>0\Rightarrow \frac{\Gamma_{G1}^d}{\gamma_M}\searrow,\\
		\beta_{\frac{\Gamma_{G1}^d}{s\gamma_M}}&\approx \frac{\Gamma_{G1}^d}{s\gamma_M}\frac{\gamma_M}{2\pi^2}\Big(5+\frac{11\pi}{4}\kappa s\Big)\frac{z_\perp\epsilon+\bar{\epsilon}}{\epsilon+\bar{\epsilon }}>0\Rightarrow \frac{\Gamma_{G1}^d}{(s\gamma_M)}\searrow .
	\end{align}
	
	$\beta_s$ gives a more specific condition: $s>\frac{4}{\pi \kappa}$ for our assumption to be consistent. If this assumption is specified as $\frac{4}{\pi\kappa}c\ll c_\perp\ll v<1$ ($s\gg \frac{4}{\pi \kappa}, w\gg 1$), we find that the low energy behaviors of the beta functions and the RG-flows of the parameters are consistent with this assumption. On the other hand, if $s$ is smaller than $\frac{4}{\pi \kappa }$, the consistency of the above analysis breaks down, where $s$ decreases in the low energy limit. We have to check out the case with $c_\perp\ll c\ll v<1$ or $s\ll 1, w\gg 1$.
	
	\subsection{$c\gg c_\perp$}
	
	As we discussed before, let us find approximate expressions of $h_i(c,c_\perp,v)$ first. For simplicity, we set $c_\perp=0$. Then, we obtain
	\begin{align}
		h_1(c,0,v)&=\int_0^1dx \sqrt{\frac{x}{(1-x)\Big((1+v^2)(1-x)+xc^2\Big)}}\approx \frac{2}{\sqrt{1+v^2}}\tanh^{-1}\Big(\sqrt{\frac{1+v^2}{1+v^2+c^2}}\Big),\\
		h_2(c,0,v)&=0,\\
		h_3(c,0,v)&=c^2\int_0^1dx\sqrt{\frac{x}{(1-x)\Big((1+v^2)(1-x)+xc^2\Big)^3}}\approx \frac{\sqrt{1+v^2+c^2}}{1+v^2}-\frac{c^2}{(1+v^2)^{3/2}}\tanh^{-1}\sqrt{\frac{1+v^2}{1+v^2+c^2}}\nonumber\\
		&+\frac{1}{c}\Big(2\ln\Big(1+\frac{c}{\sqrt{1+v^2+c^2}}\Big)-\frac{1}{2}\ln\Big(\frac{1+v^2}{1+v^2+c^2}\Big)\Big)\rightarrow const.(c\rightarrow0),\\
		h_4(c,0,v)&\rightarrow const.(c\rightarrow 0).
	\end{align}
	
	Like in the previous condition ($c\ll c_\perp<v<1$), only $h_1(c,0,v)$ is diverging in the $c\rightarrow 0$ limit. As a result, the same approximation is applicable in this case. Then, we obtain $wh_1(c,0,v) \rightarrow \frac{4\pi^2 N_cN_f}{N_c^2-1} \frac{1}{\lambda}\Big(\frac{\lambda}{8\pi}+F_{dis}(\{\Gamma_i,v\})+\frac{1}{z_\perp}\frac{\gamma_M}{2\pi^2}\Big(1+\frac{3\pi}{4}\kappa s \Big)\frac{z_\perp \epsilon+\bar{\epsilon}}{\epsilon+\bar{\epsilon}}\Big)$. Resorting to this fixed point equation, we find the low energy behaviours of the other beta functions as follows:
	\begin{align}
		\beta_\lambda&\approx \lambda\Big[z_\perp\Big\{-\epsilon+\frac{\lambda}{8\pi}+2G_{dis}(\{\Gamma_i,v\})\Big\}+\frac{\gamma_M}{2\pi^2}\Big(1+\frac{\pi}{4}\kappa s\Big)\frac{z_\perp\epsilon+\bar{\epsilon}}{\epsilon+\bar{\epsilon}}\Big]>0\Rightarrow \lambda\searrow,\\
		\beta_{w\lambda}&\approx w\lambda\Big[z_\perp\Big\{-\epsilon+\frac{\lambda}{8\pi}+\frac{\lambda w(N_c^2-1)}{4\pi^2 N_cN_f}[h_2(0,c_\perp,v)+h_3(0,c_\perp,v)]+2G_{dis}(\{\Gamma_i,v\})\Big\}\nonumber\\
		&+\frac{\gamma_M}{2\pi^2}\Big(1+\frac{\pi}{4}\kappa s \Big)\frac{z_\perp \epsilon+\bar{\epsilon}}{\epsilon+\bar{\epsilon}}\Big]>0\Rightarrow w\lambda \searrow,\\
		\beta_{\lambda/c_\perp^2}&\approx \frac{\lambda}{c_\perp^2}\Big[z_\perp\Big\{-\epsilon+\frac{\lambda}{4\pi}\frac{1}{c_\perp^2}+\frac{\lambda}{8\pi}+2G_{dis}(\{\Gamma_i,v\})\Big\}+\frac{\gamma_M}{2\pi^2}\Big(-1+\frac{3\pi}{4}\kappa s\Big)\frac{z_\perp\epsilon+\bar{\epsilon}}{\epsilon+\bar{\epsilon}}\Big]\nonumber\\
		&\Rightarrow \frac{\lambda}{4\pi}\frac{1}{c_\perp^2}\rightarrow \epsilon+\frac{1}{z_\perp}\frac{\gamma_M}{2\pi^2}\Big(1-\frac{3\pi}{4}\kappa s\Big)\frac{z_\perp\epsilon+\bar{\epsilon}}{\epsilon+\bar{\epsilon}}(\because s\ll1)\nearrow,\\
		\beta_s&\approx\frac{s}{2}\Big[-z_\perp\epsilon+\frac{\gamma_M}{2\pi^2}\Big(1+\frac{\pi}{4}\kappa s\Big)\frac{z_\perp\epsilon+\bar{\epsilon}}{\epsilon+\bar{\epsilon}}\Big]>0(\because \gamma_M\nearrow )\Rightarrow s\searrow,\\
		\beta_{\frac{\Gamma_{G1}^d}{\gamma_M}}&
		\approx \frac{\Gamma_{G1}^d}{\gamma_M}\frac{\gamma_M}{\pi^2}\Big(3+\frac{5\pi}{4}\kappa s\Big)\frac{z_\perp \epsilon+\bar{\epsilon}}{\epsilon+\bar{\epsilon}}>0\Rightarrow \frac{\Gamma_{G1}^d}{\gamma_M}\searrow,\\
		\beta_{\frac{\Gamma_{G1}^d}{s\gamma_M}}&\approx \frac{\Gamma_{G1}^d}{s\gamma_M}\frac{\gamma_M}{2\pi^2}\Big(5+\frac{11\pi}{4}\kappa s\Big)\frac{z_\perp\epsilon+\bar{\epsilon}}{\epsilon+\bar{\epsilon }}>0\Rightarrow \frac{\Gamma_{G1}^d}{(s\gamma_M)}\searrow .
	\end{align}
	
	Note that the low energy behaviors of all the parameters are consistent with each other, which means that this phase space is in the attractive regime in the low energy limit. Compared to the previous case, the low energy behaviors of $s$ and $\frac{\lambda}{c_\perp^2}$ are different while other parameters show the same tendencies.
	
	To sum up, there are two types of low energy stable phases when the random boson mass is dominant. Such two low energy phases are identified with different hierarchies of the velocities; $\frac{4}{\pi\kappa}c\ll c_\perp<v<1$ and $c_\perp\ll \frac{4}{\pi\kappa}c <v<1$, respectively. They show distinguishable low energy behaviors for the parameters of $s$ and $\frac{\lambda}{c_\perp^2}$, but the same low-energy physics for other remaining parameters; $\gamma_M\nearrow$, $\lambda\searrow$, $\kappa_1\nearrow$, and $w\nearrow$. This RBMD phase is stable at least in the one-loop level.

	\section{Superconducting Instability: Details of calculations} \label{Appendix:SCinstabilityCal}
	
	Superconducting instability channels we consider can be classified into two groups: Zero-momentum channels ($g$ and $d_{x^2-y^2}$) and $2k_F$-momentum channels ($s$ and $d_{xy}$). Here, we give the corresponding action representation of these channels and show details of the one-loop RG analysis.
	
	\subsection{Action representation of superconducting channels}

	Effective actions for superconducting instability channels are given as follows
	\begin{align}
		S^{(\pm)}_{A,\hat{\Omega}}&=\mu \Delta \sum_{i_f=1}^{N_f} \int\frac{d^{d+1}k}{(2\pi)^{d+1}}\Bigg[\Big[\Psi_{1,\sigma,i_f}^T(k)\hat{\Omega}\otimes A_{\sigma\sigma'}\Psi_{1,\sigma',i_f}(-k)+\Psi_{3,\sigma,i_f}^T(k)\hat{\Omega}\otimes A_{\sigma\sigma'}\Psi_{3,\sigma',i_f}(-k)\Big]\nonumber\\
		&\pm\Big[\Psi_{2,\sigma,i_f}^T(k)\hat{\Omega}\otimes A_{\sigma\sigma'}\Psi_{2,\sigma',i_f}(-k)+\Psi_{4,\sigma,i_f}^T(k)\hat{\Omega}\otimes A_{\sigma\sigma'}\Psi_{4,\sigma',i_f}(-k)\Big]\Bigg]\nonumber\\
		&+\mu\Delta^*\int\frac{d^{d+1}k}{(2\pi)^{d+1}}\Bigg[\Big[\bar{\Psi}_{1,\sigma,i_f}(k)(\gamma_0\hat{\Omega}^\dagger\gamma_0)\otimes A_{\sigma\sigma'}\bar{\Psi}^T_{1,\sigma,i_f}(-k)+\bar{\Psi}_{3,\sigma,i_f}(k)(\gamma_0\hat{\Omega}^\dagger\gamma_0)\otimes A_{\sigma\sigma'}\bar{\Psi}^T_{3,\sigma,i_f}(-k)\Big]\nonumber\\
		&\pm\Big[\bar{\Psi}_{2,\sigma,i_f}(k)(\gamma_0\hat{\Omega}^\dagger\gamma_0)\otimes A_{\sigma\sigma'}\bar{\Psi}^T_{2,\sigma,i_f}(-k)+\bar{\Psi}_{4,\sigma,i_f}(k)(\gamma_0\hat{\Omega}^\dagger\gamma_0)\otimes A_{\sigma\sigma'}\bar{\Psi}^T_{4,\sigma,i_f}(-k)\Big]\Bigg] ,
	\end{align}
	where $A_{\sigma\sigma'}=i\tau^y_{\sigma\sigma'}$ and $\hat{\Omega}=\{\hat{1},\gamma_{d-1}\}$, $S_{\Delta_{0,g}}=S_{A,\gamma_{d-1}}^{(+)}$, $S_{\Delta_{0,d_{x^2-y^2}}}=S_{A,\gamma_{d-1}}^{(-)}$, $S_{\Delta_{2k_F,s}}=S_{A,\hat{1}}^{(+)}$, and $S_{\Delta_{2k_F,d_{xy}}}=S_{A,\hat{1}}^{(-)}$.
	
	\subsection{Calculations of one-loop Feynman diagrams and RG beta functions}
	
	\subsubsection{Calculations of one-loop Feynman diagrams}
	
	Feynman rules for superconducting vertices are given by:
	\begin{align}
		&\begin{tikzpicture}[baseline=-0.1cm]
			\begin{feynhand}
				\vertex (a) at (0,0); \vertex [dot] (b) at (1,0) {}; \vertex (c) at (2,0);
				\propagator[fermion] (a) to [edge label=$k$] (b);
				\propagator[fermion] (c) to [edge label=$-k$] (b);
				\node at (-0.5,0) {$n,\sigma,i$};
				\node at (2.5,0) {$n,\sigma',i$};
			\end{feynhand}
		\end{tikzpicture}
		=-(-1)^{l_{\Delta}^n}\mu\Delta\Psi_{n,\sigma,i}^T(k)\hat{\Omega}\otimes A_{\sigma\sigma'}\Psi_{n,\sigma',i}(-k)\\
		&\begin{tikzpicture}[baseline=-0.1cm]
			\begin{feynhand}
				\vertex (a) at (0,0); \vertex [crossdot] (b) at (1,0) {}; \vertex (c) at (2,0);
				\propagator[fermion] (a) to [edge label=$k$] (b);
				\propagator[fermion] (c) to [edge label=$-k$] (b);
				\node at (-0.5,0) {$n,\sigma,i$};
				\node at (2.5,0) {$n,\sigma',i$};
			\end{feynhand}
		\end{tikzpicture}
		=-(-1)^{l_{\Delta}^n}\mu\Delta A_{SC} \Psi_{n,\sigma,i}^T(k)\hat{\Omega}\otimes A_{\sigma\sigma'}\Psi_{n,\sigma',i}(-k)
	\end{align}
	where
	$l_{\Delta}^n=\Big\{\begin{array}{ll}0 & \text{when }\Delta\in\{\Delta_{0,g},\Delta_{2k_F,s}\} \\n+1 & \text{when }\Delta\in\{\Delta_{0,d_{x^2-y^2}},\Delta_{2k_F,d_{xy}}\}\end{array}$ and $\hat{\Omega}=\{\hat{1},\gamma_{d-1}\}$.
	
	Using these Feynman rules, we find the one-loop correction to the counter term $A_{SC}$. The expectation value of $[\bar{\Psi}^{ex}_{n,\sigma,i_f}(k)]_\alpha [\Psi^{ex}_{n,\sigma',i_f}(-k)]_\beta$ is given by three contributions in the one-loop level:
	\begin{gather}
		\langle [\bar{\Psi}_{n,\sigma,i_f}^{ex}(k)]_\alpha [\bar{\Psi}_{n\sigma',i_f}^{ex}(-k)]_\beta \rangle=(a)+(b)+(c) 
	\end{gather}
	Here, $\alpha$ and $\beta$ are spinor-indices. (a), (b) (Fig. \ref{Fig:SConeLoop1}), and (c) (Fig. \ref{Fig:SConeLoop2}) are contributions from the counter term, the Yukawa interaction, and the random charge potential vertices, respectively. They are given as follows:
	\begin{align}
		(a)&=\langle [\bar{\Psi}_{n,\sigma,i_f}^{ex}(k)]_\alpha [\bar{\Psi}_{n,\sigma',i_f}(-k)]_{\beta}\int\frac{d^{d+1}k'}{(2\pi)^{d+1}}(-1)^{l_{\Delta}^m}(-\Delta\mu A_{SC})[\Psi_{m,\sigma_1,i_f}(k')]_a[\hat{\Omega}]_{ab}A_{\sigma_1\sigma_2}[\Psi_{m,\sigma_2,i_f}(-k')]_b\rangle_0\nonumber\\
		&=(-1)^{l_{\Delta}^n}2\mu\Delta A_{SC}A_{\sigma\sigma'}[G_{n,\sigma,i_f}^{ex,T}(k)\hat{\Omega}G_{n,\sigma',i_f}^{ex}(-k)]_{\alpha\beta}\\
		(b)&=\frac{1}{2}\Big\langle [\bar{\Psi}_{n\sigma,i_f}^{ex}(k)]_\alpha[\bar{\Psi}_{n\sigma',i_f}(-k)]_{\beta}i\frac{g}{\sqrt{N}_f}\int dk_1\int dq[\bar{\Psi}_{\bar{n}\sigma_1,i_f}(k_1-q)]_a\Phi_{\sigma_1\sigma_2}(-q)[\gamma_{d-1}]_{ab}[\Psi_{n\sigma_2,i_f}(k_1)]_b\nonumber\\
		&\times i\frac{g}{\sqrt{N}_f}\int dk_2\int dq'[\bar{\Psi}_{\bar{n}\sigma_1,i_f}(-k_2+q)]_c\Phi_{\sigma_3\sigma_4}(q')[\gamma_{d-1}]_{cd}[\Psi_{n\sigma_4,i_f}(-k_2)]_d\nonumber\\
		&\times (-1)^{l_{\Delta}^{\bar{n}}} (-\Delta\mu)\int dk_3[\Psi_{\bar{n},\sigma_5,i_f}(k_3)]_e[\hat{\Omega}]_{ef}A_{\sigma_5\sigma_6}[\Psi_{\bar{n}\sigma_6,i_f}(-k_3)]_f\Big\rangle_0\nonumber\\
		&=(-1)^{l_\Delta^{\bar{n}}}\frac{4(N_c-1)}{N_fN_c}\frac{g^2}{c}\Delta\mu A_{\sigma\sigma'}\frac{1}{16\pi^3}\frac{1}{\epsilon}\int_0^1dx \frac{\pi}{\sqrt{1+v^2}}x(1-x)^{-1/2}\Big(x+(1-x)c_\perp^2\Big)^{-1/2}\Big(x+\frac{c^2}{1+v^2}(1-x)\Big)^{-1/2}\nonumber\\
		&\times \Big[\frac{1}{x+\frac{c^2}{1+v^2}(1-x)}+Sign(\hat{\Omega})\Big(-1+\frac{1}{x+(1-x)c_\perp^2}\Big)\Big][G^{ex,T}_{n,\sigma,i_f}(k)\hat{\Omega}G^{ex}_{n,\sigma',i_f}(-k)],
		\end{align}
	\begin{align}
		(c)&=\Big\langle [\bar{\Psi}_{n,\sigma,i_f}^{ex}(k)]_\alpha [\bar{\Psi}_{n,\sigma',i_f}^{ex}(-k)]_\beta\frac{(-\Gamma)}{N_f}\int\frac{d\omega}{2\pi}\int\frac{d\omega'}{2\pi}\int d^dk_1d^dk_2d^dq[\bar{\Psi}_{m,\sigma_1,i_f}(\omega,k_1+q)]_a[\mathcal{M}]_{ab}[\Psi_{n,\sigma_1,i_f}(k_1)]_b\nonumber\\
		&\times [\bar{\Psi}_{m,\sigma_2,i_f}(\omega',k_2-q)]_{c}[\mathcal{\tilde{M}}]_{cd}[\Psi_{n,\sigma_2,i_f}(\omega',k_2)]_d(-1)^{l_{\Delta}^{m}}(-\Delta\mu)\int dk_3[\Psi_{m,\sigma_3,i_f}(k_3)]_e[\hat{\Omega}]_{ef}A_{\sigma_3\sigma_4}[\Psi_{m,\sigma_4,i_f}(-k_3)]_f\Big\rangle_0\nonumber\\
		&=-2(-1)^{l_{\Delta}^{m}}\Delta\mu(\Gamma\Lambda )A_{\sigma\sigma'}\frac{1}{4\pi}\frac{1}{1+v^2}\frac{1}{\epsilon}\Big(\frac{1}{\pi}\Big)[G_{n,\sigma,i_f}^{ex,T}(k)\Big(\mathcal{M}^T\Gamma_{\perp,i}\hat{\Omega}\Gamma_{\perp,i} \mathcal{\tilde{M}}+\mathcal{M}^T\gamma_{d-1}\hat{\Omega}\gamma_{d-1}\mathcal{\tilde{M}}\Big)G_{n,\sigma',i_f}^{ex}(-k)]_{\alpha\beta} ,
	\end{align}
	where $Sign(\hat{\Omega})=\Big\{\begin{array}{ll}1 &\text{ when }\hat{\Omega}=\hat{1}\\ -1&\text{ when }\hat{\Omega}=\gamma_{d-1} \end{array}$, $\sum_{i=1}^{N_c^2-1}[(\tau^i)^TA\tau^i]_{\sigma\sigma'}=-\frac{2(N_c+1)}{N_c}A_{\sigma\sigma'}$ and $\gamma_{0}^T=-\gamma_0,\; \gamma_i^{T}=\gamma_{i}(i>0)$ have been used.
	
	From the fact that all the $\frac{1}{\epsilon}$-poled should be absorbed to counter terms, we obtain the following result:
	\begin{align}
		A_{SC}\hat{\Omega}&=-\frac{2(N_c+1)}{N_cN_f}\frac{g^2}{c}\frac{1}{16\pi^3}\frac{1}{\epsilon}f^{(\hat{\Omega})}_{SC}(c,c_\perp,v)\hat{\Omega} +(-1)^{l_{\Delta}^n+l_{\Delta}^{m}}\frac{\Gamma}{4\pi^2N_f}\frac{1}{1+v^2}\frac{1}{\epsilon} \Big(\mathcal{M}^T\Gamma_{\perp,i}\hat{\Omega}\Gamma_{\perp,i}\mathcal{\tilde{M}}+ \mathcal{M}^T\gamma_{d-1}\hat{\Omega}\gamma_{d-1}\mathcal{\tilde{M}}\Big) ,
	\end{align}
	where
	\begin{align}
		f^{(\hat{\Omega})}_{SC}(c,c_\perp,v)&=\frac{\pi}{\sqrt{1+v^2}}\int_0^1dx\; x(1-x)^{-1/2}\Big(x+(1-x)c_\perp^2\Big)^{-1/2}\Big(x+\frac{c^2}{1+v^2}(1-x)\Big)^{-1/2}\nonumber\\
		&\times \Big[\frac{1}{x+\frac{c^2}{1+v^2}(1-x)}+Sign(\hat{\Omega})\Big(-1+\frac{1}{x+(1-x)c_\perp^2}\Big)\Big],\; \Big(Sign(\hat{\Omega})=\Big\{\begin{array}{ll}1 &\text{ when }\hat{\Omega}=\hat{1}\\ -1&\text{ when }\hat{\Omega}=\gamma_{d-1} \end{array}\Big) . \nonumber
	\end{align}
	
	After some classifications, we have all the counter terms as follows:
	\begin{align}
		&A_{SC,\Delta_{2k_F,s}}=A_{SC,\Delta_{2k_F,d_{xy}}}=-\frac{2(N_c+1)}{N_cN_f}\frac{g^2}{c}\frac{1}{16\pi^3}f^{(\hat{1})}_{SC}(c,c_\perp,v) -\frac{1}{2\pi^2N_f(1+v^2)}\frac{\Gamma_0+\Upsilon_{0}}{\epsilon} \\
		&A_{SC,\Delta_{0,g}}=A_{SC,\Delta_{0,d_{x^2-y^2}}}=-\frac{2(N_c+1)}{N_cN_f}\frac{g^2}{c}\frac{1}{16\pi^3}f^{(\gamma_{d-1})}_{SC}(c,c_\perp,v) .
	\end{align}

	\subsubsection{Beta functions of superconducting order parameters}
	
	We start from the renormalization conditions, $Z_{SC}=1+A_{SC}=Z_{\Delta}Z_3,\; \Delta_b=Z_{\Delta}\Delta_r\mu$, where $\Delta_b$ is a bare value of a superconducting gap and $\Delta_r$ is the renormalized one. We find the following beta function of the gap function:
	\begin{align}
		\beta_{\Delta}&=-\Big(1+\frac{d\ln Z_{SC}}{d\ln\mu}-\frac{d\ln Z_3}{d\ln\mu}\Big)\nonumber\\
		&=-\Big(1+\frac{1}{2}g(F_{g,3}^{(1)}-F_{g,SC}^{(1)})+u_1(F_{u_1,3}^{(1)}-F_{u_1,SC}^{(1)})+u_2(F_{u_2,3}^{(1)}-F_{u_2,SC}^{(1)})+\Gamma_i(F_{\Gamma_i,3}^{(1)}-F_{\Gamma_i,SC}^{(1)})\Big).
	\end{align}
	
	As a result, all the beta functions for each superconducting channel is given by
	\begin{align}
		\beta_{\Delta_{2k_F,s}}&=-\Big[1+\frac{1}{4\pi^2}\frac{N_c+1}{N_cN_f}\frac{g^2}{c}\Big(\frac{f^{(\hat{1})}_{SC}(c,c_\perp,v)}{2\pi}-(N_c-1)h_3(c,c_\perp,v)\Big)+\frac{1}{N_f}\frac{\Gamma_0+\Upsilon_{0}}{1+v^2}\Big]\Delta_{2k_F,s}\label{eq:betafunctionOfDelta1}\\
		\beta_{\Delta_{2k_F,d_{xy}}}&=-\Big[1+\frac{1}{4\pi^2}\frac{N_c+1}{N_cN_f}\frac{g^2}{c}\Big(\frac{f^{(\hat{1})}_{SC}(c,c_\perp,v)}{2\pi}-(N_c-1)h_3(c,c_\perp,v)\Big)+\frac{1}{N_f}\frac{\Gamma_0+\Upsilon_{0}}{1+v^2}\Big]\Delta_{2k_F,d_{xy}}\\
		\beta_{\Delta_{0,g}}&=-\Big[1+\frac{1}{4\pi^2}\frac{N_c+1}{N_cN_f}\frac{g^2}{c}\Big(\frac{f^{(\gamma_{d-1})}_{SC}(c,c_\perp,v)}{2\pi}-(N_c-1)h_3(c,c_\perp,v)\Big)\Big]\Delta_{0,g}\\
		\beta_{\Delta_{0,d_{x^2-y^2}}}&=-\Big[1+\frac{1}{4\pi^2}\frac{N_c+1}{N_cN_f}\frac{g^2}{c} \Big(\frac{f^{(\gamma_{d-1})}_{SC}(c,c_\perp,v)}{2\pi}-(N_c-1)h_3(c,c_\perp,v)\Big)\Big]\Delta_{0,d_{x^2-y^2}} . \label{eq:betafunctionOfDelta4}
	\end{align}
	Note that random charge potential vertexes, $\Gamma_0$ and $\Upsilon_{0}$, enhance only the finite-momentum superconducting instability channels: $\Delta_{2k_F,s}$ and $\Delta_{2k_F,d_{xy}}$.

\end{widetext}

\end{document}


\title{Supplementary Material for\\ ``Effects of general non-magnetic quenched disorder on a spin-density-wave quantum critical metallic system in two spatial dimension''}
		
		\author{{ Iksu Jang$^{1,2}$ and Ki-Seok Kim$^{1,3}$}\\
			{\small  $^1$Department of Physics, Pohang University of Science and Technology, Pohang, Gyeongbuk 790-784, Korea \\ 
		$^2$Department of Physics, National Tsing Hua University, Hsinchu 30013, Taiwan} \\
$^3$Asia Pacific Center for Theoretical Physics (APCTP), Pohang, Gyeongbuk 37673, Korea	
}
		
		\date{\today}
		\maketitle
	
	\section{Explicit translational symmetry breaking in the co-dimensional regularization} \label{Appendix:TranslationSymBreaking}

Acting the translation operator $\mathcal{T}_{\mathbf{a}}$ on a Bloch wave function, we obtain
\begin{gather*}
	\mathcal{T}_{\mathbf{a}}|\mathbf{r}\rangle=|\mathbf{r}+\mathbf{a}\rangle,\;\langle \mathbf{r}|\mathcal{T}_{\mathbf{a}}=\langle \mathbf{r}-\mathbf{a}|, \; \mathcal{T}_{\mathbf{a}}=e^{i\hat{p}\cdot \mathbf{a}},\\
	\text{Bloch function: }\langle \mathbf{r}|\psi_\mathbf{k}\rangle=\psi_{\mathbf{k}}(\mathbf{r})=e^{i\mathbf{k}\cdot\mathbf{r}}u_{\mathbf{k}}(\mathbf{r}),\\
	\Rightarrow \langle \mathbf{r}|\mathcal{T}_{\mathbf{a}}|\psi_{\mathbf{k}}\rangle=\langle \mathbf{r}-\mathbf{a}|\psi_{\mathbf{k}}\rangle=e^{-i\mathbf{k}\cdot\mathbf{a}}\psi_{\mathbf{k}}(\mathbf{r})=e^{-i\mathbf{k}\cdot\mathbf{a}}\langle \mathbf{r}|\psi_{\mathbf{k}}\rangle\\
	\therefore \mathcal{T}_{\mathbf{a}}|\psi_{\mathbf{k}}\rangle=e^{-i\mathbf{k}\cdot\mathbf{a}}|\psi_{\mathbf{k}} \rangle .
\end{gather*}
Here, $\hat{p}=\frac{1}{i}\nabla$ is the momentum operator, $\mathbf{a}$ is a lattice constant, and $|\psi_{\mathbf{k}}\rangle$ is a Bloch wave function.

Introducing a creation operator for the Bloch state, we obtain a 
\begin{gather}		 \mathcal{T}_{\mathbf{a}}|\psi_{\mathbf{k}}\rangle=\mathcal{T}_{\mathbf{a}}c_{\mathbf{k}}^\dagger|0\rangle=\mathcal{T}_{\mathbf{a}}c_{\mathbf{k}}^\dagger\mathcal{T}^{-1}\mathcal{T}|0\rangle=\mathcal{T}_{\mathbf{a}}c_{\mathbf{k}}^\dagger\mathcal{T}_{\mathbf{a}}^{-1}|0\rangle(\because \mathcal{T}_{\mathbf{a}}|0\rangle=|0\rangle)=e^{-i\mathbf{k}\cdot\mathbf{a}}c_k^\dagger|0\rangle\nonumber\\
	\therefore \mathcal{T}_{\mathbf{a}}c_{\mathbf{k}}^\dagger\mathcal{T}_{\mathbf{a}}^{-1} = e^{-i\mathbf{k}\cdot\mathbf{a}}c_{\mathbf{k}}^\dagger . \label{eq:TranslationOpRep1}
\end{gather}

Using the above result Eq. \eqref{eq:TranslationOpRep1}, we find a representation of the translation operator in terms of the fermion field with hot-spot indexes; $\psi_{a,n,\sigma}^{(m)}(\mathbf{k})$, where $a$ and $\sigma$ are replica and spin index, respectively, and $(n,m)$ denotes the hot spot-index. Considering the Fermi wave-vector $\mathbf{k}_F^{(i)}$, $\psi_{a,n,\sigma}^{(m)}(\mathbf{k})$ can be re-expressed as follows
\begin{gather}
	\psi^{(m)\dagger}_{a,n,\sigma}(\mathbf{k})=\psi^\dagger_{a,\sigma}(\mathbf{k}_F^{(i)}+\mathbf{k})\nonumber\\
	\Rightarrow \mathcal{T}_{\mathbf{a}}\psi_{a,n,\sigma}^{(m)\dagger}(\mathbf{k}) \mathcal{T}_{\mathbf{a}}^{-1}=\mathcal{T}_{\mathbf{a}}\psi_{a,\sigma}^{\dagger}(\mathbf{k}_F^{(i)}+\mathbf{k}) \mathcal{T}_{\mathbf{a}}^{-1}=e^{-i(\mathbf{k}_F^{(i)}+\mathbf{k}) \cdot \mathbf{r}}\psi_{a,\sigma}^\dagger(\mathbf{k}_F^{(i)}+\mathbf{k}) = e^{-i(\mathbf{k}_F^{(i)}+\mathbf{k})\cdot\mathbf{r}}\psi_{a,n,\sigma}^{(m)\dagger}(\mathbf{k})\nonumber\\
	\therefore \mathcal{T}_{\mathbf{a}}\psi_{a,n,\sigma}^{(m)\dagger}(\mathbf{k}) \mathcal{T}_{\mathbf{a}}^{-1}=e^{-i(\mathbf{k}_F^{(i)}+\mathbf{k}) \cdot \mathbf{r}} \psi_{a,n,\sigma}^{(m)\dagger}(\mathbf{k}) . \label{eq:TranslationOpRep2}
\end{gather}

Since the regularized effective action is re-written in terms of gamma matrices and spinors, a representation of the translation operator in a matrix form, using Eq. \eqref{eq:TranslationOpRep2}, is given by
\begin{align} \mathcal{T}_{\mathbf{a}}\Psi^a_{n,\sigma}(\mathbf{k})\mathcal{T}^{-1}_{\mathbf{a}}=e^{i\mathbf{k}\cdot\mathbf{a}}\left(\begin{array}{cc}e^{i\mathbf{k}_{F,n}\cdot\mathbf{a}} & 0 \\ 0 & e^{-i\mathbf{k}_{F,n}\cdot\mathbf{a}}\end{array}\right)\Psi_{n,\sigma}^{a}(\mathbf{k}) ,
\end{align}
where $\mathbf{k}_{F,1}=\mathbf{k}_F^{(1,+)}$, $\mathbf{k}_{F,2}=\mathbf{k}_F^{(2,+)}$, $\mathbf{k}_{F,3}=\mathbf{k}_F^{(1,-)}$, and $\mathbf{k}_{F,4}=\mathbf{k}_F^{(2,-)}$.

Generalizing the lattice constant $\mathbf{a}$ to $\mathbf{r}=\sum_{n_i}n_i\mathbf{a}_i(n\in Z)$, we obtain
\begin{align}
	\mathcal{T}_{\mathbf{r}}\Psi^a_{n,\sigma}(\mathbf{k})\mathcal{T}_{\mathbf{r}}^{-1}= e^{i\mathbf{k}\cdot\mathbf{r}}\left(\begin{array}{cc}e^{i\mathbf{k}_{F,n}\cdot\mathbf{r}} & 0 \\ 0 & e^{-i\mathbf{k}_{F,n}\cdot\mathbf{r}}\end{array}\right)\Psi^a_{n,\sigma}(\mathbf{k}) = e^{i\mathbf{k}\cdot\mathbf{r}}e^{i\sigma_z\mathbf{k}_{F,n}\cdot\mathbf{r}}\Psi_{n,\sigma}^{a}(\mathbf{k}) . \label{eq:TranslationOpRepFinal}
\end{align}
A representation of the translation operator for the spin density order parameter is given by
\begin{gather}
	\mathcal{T}_{\mathbf{a}} \vec{S}(\mathbf{r})\mathcal{T}_{\mathbf{a}}^{-1}=\vec{S}(\mathbf{r}-\mathbf{a})
	\rightarrow \mathcal{T}_{\mathbf{a}} \vec{\phi}(\mathbf{r})\mathcal{T}_{\mathbf{a}}^{-1}= e^{i\mathbf{Q}\cdot\mathbf{a}}\vec{\phi}(\mathbf{r}-\mathbf{a})\;(\because \vec{S}(\mathbf{r})=e^{-i\mathbf{Q}\cdot\mathbf{r}}\vec{\phi}(\mathbf{r})+h.c)\nonumber\\
	\rightarrow \mathcal{T}_{\mathbf{a}}\vec{\phi}(\mathbf{q})\mathcal{T}_{\mathbf{a}}^{-1} = e^{i\mathbf{Q}\cdot\mathbf{a}}e^{i\mathbf{q}\cdot\mathbf{a}}\vec{\phi}(\mathbf{q})\; (\because \vec{\phi}(\mathbf{q})=\int d^d\mathbf{r} e^{i\mathbf{q}\cdot\mathbf{r}}\vec{\phi}(\mathbf{r}))
	\Rightarrow \mathcal{T}_{\mathbf{r}} \vec{\phi}(\mathbf{q}) \mathcal{T}_{\mathbf{r}}^{-1} = e^{i\mathbf{Q}\cdot\mathbf{r}}e^{i\mathbf{q}\cdot\mathbf{r}}\vec{\phi}(\mathbf{q}) , \label{eq:TranslationOpRepBoson}
\end{gather}
where $\mathbf{Q}$ represent nesting vectors $(\mathbf{Q}=(\pi/a,\pi/a),(\pi/a,-\pi/a),(-\pi/a,\pi/a),(-\pi/a,-\pi/a))$.

Now, we are ready to check out whether the original effective action defined at $d=2$ is invariant under a translation operation. The result is given as follows:
\begin{align}
	\mathcal{T}_{\mathbf{r}}S_{eff}\mathcal{T}^{-1}_{\mathbf{r}}&=\sum_{a=1}^R\Bigg[\sum_{n=1}^4\sum_{\sigma}\int dk \Psi^{a\dagger}_{n,\sigma}(k)e^{-i\sigma_z\mathbf{k}_{F,n}\cdot\mathbf{r}}\gamma_0[i\gamma_0k_0+i\gamma_1\epsilon_n(
	\mathbf{k})]e^{i\mathbf{k}_{F,n}\cdot\mathbf{r}}\Psi^a_{n,\sigma}(k)\nonumber\\
	&+ig\sum_{n=1}^4\sum_{\sigma,\sigma'}\int dk\int dq \Psi^{a\dagger}_{\bar{n},\sigma}(k+q)e^{-i\sigma_z\mathbf{k}_{F,\bar{n}} \cdot \mathbf{r}} \gamma_0 e^{i\mathbf{Q} \cdot \mathbf{r}} \Phi_{\sigma,\sigma'}^a(q) \gamma_1 e^{i\sigma_z\mathbf{k}_{F,n}\cdot\mathbf{r}}\Psi^a_{n,\sigma'}(k) + \cdots =S_{eff} .
\end{align}
where $\gamma_0=\sigma_y$, $\gamma_1=\sigma_x$. Here we used a fact that $\{\sigma_z,\gamma_i\}=0$, $e^{i n\mathbf{Q}\cdot\mathbf{r}}=1$ $(n \in 2Z)$, $e^{i(\mathbf{k}_{F,\bar{n}}-\mathbf{k}_{F,n}\pm \mathbf{Q})\cdot\mathbf{r}}=1$. We also used explicit forms of random charge potential vertices in the \ref{Appendix:RegularizedDisorderVertices}. As a result, the original effective action constructed at $d=2$ is invariant under translation.

On the other hand, we find that the regularized effective action at $d=3$ is not invariant under the translation operation as follows
\begin{align}
	\mathcal{T}_{\mathbf{r}}S_{eff}\mathcal{T}^{-1}_{\mathbf{r}}&=\sum_{a=1}^R\Bigg[\sum_{i_f=1}^{N_f}\sum_{n=1}^4\sum_{\sigma=1}^{N_c}\int dk \Psi^{a\dagger}_{n,\sigma,i_f}(k)e^{-i\sigma_z\mathbf{k}_{F,n}\cdot\mathbf{r}}\gamma_0[i\gamma_0k_0+i\gamma_1k_1+i\gamma_2\epsilon_n(k)]e^{i\sigma_z\mathbf{k}_{F,n}\cdot\mathbf{r}}\Psi^a_{n,\sigma,i_f}(k)\nonumber\\
	&+ig\sum_{i_f=1}^{N_f}\sum_{n=1}^4\sum_{\sigma,\sigma'=1}^{N_c}\int dk\int dq \Psi^{a\dagger}_{\bar{n},\sigma,i_f}(k+q)e^{-i\sigma_z\mathbf{k}_{F,\bar{n}}\cdot\mathbf{r}}\gamma_0e^{i\mathbf{Q}\cdot\mathbf{r}}\Phi_{\sigma,\sigma'}^a(q)\gamma_{d-1}e^{-i\sigma_z\mathbf{k}_{F,n}\cdot\mathbf{r}}\Psi^a_{n,\sigma',i_f}(k)+\cdots\nonumber\\
	&\neq S_{eff} .
\end{align}
where $\gamma_0=\sigma_y,\; \gamma_1=\sigma_z,\; \gamma_2=\sigma_x$. The fermion kinetic energy is not invariant under the translation in $d=3$ for the regularized action since $\{\sigma_z,\gamma_1\} \neq 0$. This explicit translation symmetry breaking in the co-dimensional regularized action is a consequence of considering an additional dimension in a form of $p_z$-wave charge-density-wave ordering. Here, the nesting vector is $|\mathbf{Q}_{CDW}|=2|K_F|$, where $K_F$ is a magnitude of the Fermi wave vector.

	\section{Explicit form of random charge potential vertices} \label{Appendix:DisorderVertices}

\subsection{Random charge potential vertices with the representation of fermion fields $\psi_{n,\sigma}^{(m)}(k)$}

All possible random charge potential vertices classified in the main text are presented here with following short-hand notations:
\begin{gather*}
	\int dk=\int d\omega\int d\omega'\int\frac{d^2\mathbf{k}_1}{(2\pi)^2}\int\frac{d^2\mathbf{k}_2}{(2\pi)^2}\int\frac{d^2\mathbf{k}_3}{(2\pi)^2}\int\frac{d^2\mathbf{k}_4}{(2\pi)^2}\\
	\psi_{a,n_1,\sigma}^{m_1*}\psi_{a,n_2,\sigma}^{m_2}\psi_{b,n_3,\sigma'}^{m_3*}\psi_{b,n_4,\sigma'}^{m_4}=\psi_{a,\sigma}^*(\omega,\mathbf{k}_F^{(i_1)}+\mathbf{k}_1)\psi_{a,\sigma}(\omega,\mathbf{k}_F^{(i_2)}+\mathbf{k}_2)\psi_{b,\sigma'}^*(\omega',\mathbf{k}_F^{(i_3)}+\mathbf{k}_3)\psi_{b,\sigma'}(\omega',\mathbf{k}_F^{(i_4)}+\mathbf{k}_4).
\end{gather*}

\subsubsection{Normal process}\label{Appendix:NormalScatteringChannels}

We show all the normal processes as follows:
\begin{align}
	S_{nor,0}&=-\frac{\Gamma_0}{2}\sum_{a,b=1}^R\sum_{n=1}^4\sum_{\sigma,\sigma'=\uparrow,\downarrow}\sum_{m=\pm}\int dk  \psi_{a,n,\sigma}^{(m)*}\psi_{a,n,\sigma}^{(m)}\psi_{b,n,\sigma'}^{(m)*}\psi_{b,n,\sigma'}^{(m)}\delta(\mathbf{k}_1+\mathbf{k}_3-\mathbf{k}_2-\mathbf{k}_4)
	\end{align}
\begin{align}
	S^d_{nor,\theta_1}&=- \frac{\Gamma_{\theta_1}^{d}}{2}\sum_{a,b=1}^R\sum_{\sigma,\sigma'=\uparrow,\downarrow}\int dk\Big[\psi_{a,1,\sigma}^{(+)*}\psi_{a,1,\sigma}^{(+)}\psi_{b,3,\sigma'}^{(-)*}\psi_{b,3,\sigma'}^{(-)}+\psi_{a,4,\sigma}^{(+)*}\psi_{a,4,\sigma}^{(+)}\psi_{b,2,\sigma'}^{(-)*}\psi_{b,2,\sigma'}^{(-)}+\psi_{a,3,\sigma}^{(+)*}\psi_{a,3,\sigma}^{(+)}\psi_{b,1,\sigma'}^{(-)*}\psi_{b,1,\sigma'}^{(-)}\nonumber \\
	&+\psi_{a,2,\sigma}^{(+)*}\psi_{a,2,\sigma}^{(+)}\psi_{b,4,\sigma'}^{(-)*}\psi_{b,4,\sigma'}^{(-)}+(a\leftrightarrow b)\Big]\delta(\mathbf{k}_1+\mathbf{k}_3-\mathbf{k}_2-\mathbf{k}_4)\\
	S^{e}_{nor,\theta_1}&=-\frac{\Gamma_{\theta_1}^{e}}{2}\sum_{a,b=1}^R\sum_{a,b=1}^R\sum_{\sigma,\sigma'=\uparrow,\downarrow}\int dk\Big[\psi_{a,1,\sigma}^{(+)*}\psi_{a,3,\sigma}^{(-)}\psi_{b,3,\sigma'}^{(-)*}\psi_{b,1,\sigma'}^{(+)}+\psi_{a,4,\sigma}^{(+)*}\psi_{a,2,\sigma}^{(-)}\psi_{b,2,\sigma'}^{(-)*}\psi_{b,4,\sigma'}^{(+)}+\psi_{a,3,\sigma}^{(+)*}\psi_{a,1,\sigma}^{(-)}\psi_{b,1,\sigma'}^{(-)*}\psi_{b,3,\sigma'}^{(+)}\nonumber \\
	&+\psi_{a,2,\sigma}^{(+)*}\psi_{a,4,\sigma}^{(-)}\psi_{b,4,\sigma'}^{(-)*}\psi_{b,2,\sigma'}^{(+)}+(a\leftrightarrow b)\Big]\delta(\mathbf{k}_1+\mathbf{k}_3-\mathbf{k}_2-\mathbf{k}_4)\\
	S^{d}_{nor,\theta_2}&=-\frac{\Gamma_{\theta_2}^{d}}{2}\sum_{a,b=1}^R\sum_{\sigma,\sigma'=\uparrow,\downarrow}\int dk \Big[\psi_{a,1,\sigma}^{(+)*}\psi_{a,1,\sigma}^{(+)}\psi_{b,4,\sigma'}^{(-)*}\psi_{b,4,\sigma'}^{(-)}+\psi_{a,2,\sigma}^{(+)*}\psi_{a,2,\sigma}^{(+)}\psi_{b,1,\sigma'}^{(-)*}\psi_{b,1,\sigma'}^{(-)}+\psi_{a,3,\sigma}^{(+)*}\psi_{a,3,\sigma}^{(+)}\psi_{b,2,\sigma'}^{(-)*}\psi_{b,2,\sigma'}^{(-)}\nonumber \\
	&+\psi_{a,4,\sigma}^{(+)*}\psi_{a,4,\sigma}^{(+)}\psi_{b,3,\sigma'}^{(-)*}\psi_{b,3,\sigma'}^{(-)}+(a\leftrightarrow b)\Big]\delta(\mathbf{k}_1+\mathbf{k}_3-\mathbf{k}_2-\mathbf{k}_4)\\
	S^{e}_{nor,\theta_2}&=-\frac{\Gamma_{\theta_2}^{e}}{2}\sum_{a,b=1}^R\sum_{\sigma,\sigma'=\uparrow,\downarrow}\int dk \Big[\psi_{a,1,\sigma}^{(+)*}\psi_{a,4,\sigma}^{(-)}\psi_{b,4,\sigma'}^{(-)*}\psi_{b,1,\sigma'}^{(+)}+\psi_{a,2,\sigma}^{(+)*}\psi_{a,1,\sigma}^{(-)}\psi_{b,1,\sigma'}^{(-)*}\psi_{b,2,\sigma'}^{(+)}+\psi_{a,3,\sigma}^{(+)*}\psi_{a,2,\sigma}^{(-)}\psi_{b,2,\sigma'}^{(-)*}\psi_{b,3,\sigma'}^{(+)}\nonumber \\
	&+\psi_{a,4,\sigma}^{(+)*}\psi_{a,3,\sigma}^{(-)}\psi_{b,3,\sigma'}^{(-)*}\psi_{b,4,\sigma'}^{(+)}+(a\leftrightarrow b)\Big]\delta(\mathbf{k}_1+\mathbf{k}_3-\mathbf{k}_2-\mathbf{k}_4)\\
	S^{d}_{nor,\pi/2}&=-\frac{\Gamma_{\pi/2}^{d}}{2}\sum_{a,b=1}^R\sum_{m=\pm}\sum_{\sigma,\sigma'=\uparrow,\downarrow}\int dk \Big[\psi_{a,1,\sigma}^{(m)*}\psi_{a,1,\sigma}^{(m)}\psi_{b,2,\sigma'}^{(m)*}\psi_{b,2,\sigma'}^{(m)}+\psi_{a,2,\sigma}^{(m)*}\psi_{a,2,\sigma}^{(m)}\psi_{b,3,\sigma'}^{(m)*}\psi_{b,3,\sigma'}^{(m)}+\psi_{a,3,\sigma}^{(m)*}\psi_{a,3,\sigma}^{(m)}\psi_{b,4,\sigma'}^{(+)*}\psi_{b,4,\sigma'}^{(m)}\nonumber \\
	&+\psi_{a,4,\sigma}^{(m)*}\psi_{a,4,\sigma}^{(m)}\psi_{b,1,\sigma'}^{(m)*}\psi_{b,1,\sigma'}^{(m)}+(a\leftrightarrow b)\Big]\delta(\mathbf{k}_1+\mathbf{k}_3-\mathbf{k}_2-\mathbf{k}_4),\\
	S^{e}_{nor,\pi/2}&=-\frac{\Gamma_{\pi/2}^{e}}{2}\sum_{a,b=1}^R\sum_{m=\pm}\sum_{\sigma,\sigma'=\uparrow,\downarrow}\int dk\Big[\psi_{a,1,\sigma}^{(m)*}\psi_{a,2,\sigma}^{(m)}\psi_{b,2,\sigma'}^{(m)*}\psi_{b,1,\sigma'}^{(m)}
	+\psi_{a,2,\sigma}^{(m)*}\psi_{a,3,\sigma}^{(m)}\psi_{b,3,\sigma'}^{(m)*}\psi_{b,2,\sigma'}^{(m)}+\psi_{a,3,\sigma}^{(m)*}\psi_{a,4,\sigma}^{(m)}\psi_{b,4,\sigma'}^{(m)*}\psi_{b,3,\sigma'}^{(m)}\nonumber \\
	&+\psi_{a,4,\sigma}^{(m)*}\psi_{a,1,\sigma}^{(m)}\psi_{b,1,\sigma'}^{(m)*}\psi_{b,4,\sigma'}^{(m)}+(a\leftrightarrow b)\Big]\delta(\mathbf{k}_1+\mathbf{k}_3-\mathbf{k}_2-\mathbf{k}_4)\\
	S^{d}_{nor,\pi-\theta_1}&=-\sum_{a,b=1}^R\sum_{\sigma,\sigma'=\uparrow,\downarrow}\int dk \frac{\Gamma_{\pi-\theta_1}^{d}}{2}\Big[\psi_{a,1,\sigma}^{(+)*}\psi_{a,1,\sigma}^{(+)}\psi_{b,1,\sigma'}^{(-)*}\psi_{b,1,\sigma'}^{(-)}+\psi_{a,2,\sigma}^{(+)*}\psi_{a,2,\sigma}^{(+)}\psi_{b,2,\sigma'}^{(-)*}\psi_{b,2,\sigma'}^{(-)}+\psi_{a,3,\sigma}^{(+)*}\psi_{a,3,\sigma}^{(+)}\psi_{b,3,\sigma'}^{(-)*}\psi_{b,3,\sigma'}^{(-)}\nonumber \\
	&+\psi_{a,4,\sigma}^{(+)*}\psi_{a,4,\sigma}^{(+)}\psi_{b,4,\sigma'}^{(-)*}\psi_{b,4,\sigma'}^{(-)}+(a\leftrightarrow b)\Big]\delta(\mathbf{k}_1+\mathbf{k}_3-\mathbf{k}_2-\mathbf{k}_4)\\
	S^{e}_{nor,\pi-\theta_1}&=-\sum_{a,b=1}^R\sum_{\sigma,\sigma'=\uparrow,\downarrow}\int dk \frac{\Gamma_{\pi-\theta_1}^{e}}{2}\Big[\psi_{a,1,\sigma}^{(+)*}\psi_{a,1,\sigma}^{(-)}\psi_{b,1,\sigma'}^{(-)*}\psi_{b,1,\sigma'}^{(+)}+\psi_{a,2,\sigma}^{(+)*}\psi_{a,2,\sigma}^{(-)}\psi_{b,2,\sigma'}^{(-)*}\psi_{b,2,\sigma'}^{(+)}+\psi_{a,3,\sigma}^{(+)*}\psi_{a,3,\sigma}^{(-)}\psi_{b,3,\sigma'}^{(-)*}\psi_{b,3,\sigma'}^{(+)}\nonumber \\
	&+\psi_{a,4,\sigma}^{(+)*}\psi_{a,4,\sigma}^{(-)}\psi_{b,4,\sigma'}^{(-)*}\psi_{b,4,\sigma'}^{(+)}+(a\leftrightarrow b)\Big]\delta(\mathbf{k}_1+\mathbf{k}_3-\mathbf{k}_2-\mathbf{k}_4)\\
	S^{d}_{nor,\pi-\theta_2}&=-\sum_{a,b=1}^R\sum_{\sigma,\sigma'=\uparrow,\downarrow}\int dk \frac{\Gamma_{\pi-\theta_2}^{d}}{2}\Big[\psi_{a,3,\sigma}^{(-)*}\psi_{a,3,\sigma}^{(-)}\psi_{b,2,\sigma'}^{(+)*}\psi_{b,2,\sigma'}^{(+)}+\psi_{a,4,\sigma}^{(-)*}\psi_{a,4,\sigma}^{(-)}\psi_{b,3,\sigma'}^{(+)*}\psi_{b,3,\sigma'}^{(+)}+\psi_{a,1,\sigma}^{(-)*}\psi_{a,1,\sigma}^{(-)}\psi_{b,4,\sigma'}^{(+)*}\psi_{b,4,\sigma'}^{(+)}\nonumber \\
	&+\psi_{a,2,\sigma}^{(-)*}\psi_{a,2,\sigma}^{(-)}\psi_{b,1,\sigma'}^{(+)*}\psi_{b,1,\sigma'}^{(+)}+(a\leftrightarrow b)\Big]\delta(\mathbf{k}_1+\mathbf{k}_3-\mathbf{k}_2-\mathbf{k}_4)\\
	S^{e}_{nor,\pi-\theta_2}&=-\sum_{a,b=1}^R\sum_{\sigma,\sigma'=\uparrow,\downarrow}\int dk \frac{\Gamma_{\pi-\theta_2}^{e}}{2}\Big[\psi_{a,3,\sigma}^{(-)*}\psi_{a,2,\sigma}^{(+)}\psi_{b,2,\sigma'}^{(+)*}\psi_{b,3,\sigma'}^{(-)}+\psi_{a,4,\sigma}^{(-)*}\psi_{a,3,\sigma}^{(+)}\psi_{b,3,\sigma'}^{(+)*}\psi_{b,4,\sigma'}^{(-)}+\psi_{a,1,\sigma}^{(-)*}\psi_{a,4,\sigma}^{(+)}\psi_{b,4,\sigma'}^{(+)*}\psi_{b,1,\sigma'}^{(-)}\nonumber \\
	&+\psi_{a,2,\sigma}^{(-)*}\psi_{a,1,\sigma}^{(+)}\psi_{b,1,\sigma'}^{(+)*}\psi_{b,2,\sigma'}^{(-)}+(a\leftrightarrow b)\Big]\delta(\mathbf{k}_1+\mathbf{k}_3-\mathbf{k}_2-\mathbf{k}_4)\\
	S^c_{nor,0}&=-\frac{\Delta_{0}}{2}\sum_{a,b=1}^R\sum_{\sigma,\sigma'=\uparrow,\downarrow}\int dk \Big[\psi_{a,1,\sigma}^{(+)*}\psi_{a,1,\sigma}^{(+)}\psi_{b,3,\sigma'}^{(+)*}\psi_{b,3,\sigma'}^{(+)}+\psi_{a,4,\sigma}^{(-)*}\psi_{a,4,\sigma}^{(-)}\psi_{b,2,\sigma'}^{(-)*}\psi_{b,2,\sigma'}^{(-)}+\psi_{a,2,\sigma}^{(+)*}\psi_{a,2,\sigma}^{(+)}\psi_{b,4,\sigma'}^{(+)*}\psi_{b,4,\sigma'}^{(+)}\nonumber \\
	&+\psi_{a,1,\sigma}^{(-)*}\psi_{a,1,\sigma}^{(-)}\psi_{b,3,\sigma'}^{(-)*}\psi_{b,3,\sigma'}^{(-)}+(a\leftrightarrow b)\Big]\delta(\mathbf{k}_1+\mathbf{k}_3-\mathbf{k}_2-\mathbf{k}_4)
\end{align}
\begin{align}
	S^c_{nor,\pi}&=-\frac{\Delta_{\pi}}{2}\sum_{a,b=1}^R\sum_{\sigma,\sigma'=\uparrow,\downarrow}\int dk\Big[\psi_{a,1,\sigma}^{(+)*}\psi_{a,3,\sigma}^{(+)}\psi_{b,3,\sigma'}^{(+)*}\psi_{b,1,\sigma'}^{(+)}+\psi_{a,4,\sigma}^{(-)*}\psi_{a,2,\sigma}^{(-)}\psi_{b,2,\sigma'}^{(-)*}\psi_{b,4,\sigma'}^{(-)}+\psi_{a,2,\sigma}^{(+)*}\psi_{a,4,\sigma}^{(+)}\psi_{b,4,\sigma'}^{(+)*}\psi_{b,2,\sigma'}^{(+)}\nonumber \\
	&+\psi_{a,1,\sigma}^{(-)*}\psi_{a,3,\sigma}^{(-)}\psi_{b,3,\sigma'}^{(-)*}\psi_{b,1,\sigma'}^{(-)}+(a\leftrightarrow b)\Big]\delta(\mathbf{k}_1+\mathbf{k}_3-\mathbf{k}_2-\mathbf{k}_4)\\
	S^c_{nor,\theta_1}&=-\frac{\Delta_{\theta_1}}{2}\sum_{a,b=1}^R\sum_{\sigma,\sigma'=\uparrow,\downarrow}\int dk \Big[\psi_{a,3,\sigma}^{(-)*}\psi_{a,1,\sigma}^{(+)}\psi_{b,1,\sigma'}^{(-)*}\psi_{b,3,\sigma'}^{(+)}+\psi_{a,4,\sigma}^{(-)*}\psi_{a,2,\sigma}^{(+)}\psi_{b,2,\sigma'}^{(-)*}\psi_{b,4,\sigma'}^{(+)}+\psi_{a,1,\sigma}^{(+)*}\psi_{a,3,\sigma}^{(-)}\psi_{b,3,\sigma'}^{(+)*}\psi_{b,1,\sigma'}^{(-)}\nonumber \\
	&+\psi_{a,2,\sigma}^{(+)*}\psi_{a,4,\sigma}^{(-)}\psi_{b,4,\sigma'}^{(+)*}\psi_{b,2,\sigma'}^{(-)}+(a\leftrightarrow b)\Big]\delta(\mathbf{k}_1+\mathbf{k}_3-\mathbf{k}_2-\mathbf{k}_4)\\
	S^c_{nor,\pi-\theta_1}&=-\frac{\Delta_{\pi-\theta_1}}{2}\sum_{a,b=1}^R\sum_{\sigma,\sigma'=\uparrow,\downarrow}\int dk\Big[\psi_{a,3,\sigma}^{(-)*}\psi_{a,3,\sigma}^{(+)}\psi_{b,1,\sigma'}^{(-)*}\psi_{b,1,\sigma'}^{(+)}+\psi_{a,4,\sigma}^{(-)*}\psi_{a,4,\sigma}^{(+)}\psi_{b,2,\sigma'}^{(-)*}\psi_{b,2,\sigma'}^{(+)}+\psi_{a,3,\sigma}^{(+)*}\psi_{a,3,\sigma}^{(-)}\psi_{b,1,\sigma'}^{(+)*}\psi_{b,1,\sigma'}^{(-)}\nonumber \\
	&+\psi_{a,4,\sigma}^{(+)*}\psi_{a,4,\sigma}^{(-)}\psi_{b,2,\sigma'}^{(+)*}\psi_{b,2,\sigma'}^{(-)}+(a\leftrightarrow b)\Big]\delta(\mathbf{k}_1+\mathbf{k}_3-\mathbf{k}_2-\mathbf{k}_4)\\
	S^c_{nor,\theta_2}&=-\frac{\Delta_{\theta_2}}{2}\sum_{a,b=1}^R\sum_{\sigma,\sigma'=\uparrow,\downarrow}\int dk \Big[\psi_{a,1,\sigma}^{(+)*}\psi_{a,4,\sigma}^{(-)}\psi_{b,3,\sigma'}^{(+)*}\psi_{b,2,\sigma'}^{(-)}+\psi_{a,2,\sigma}^{(+)*}\psi_{a,1,\sigma}^{(-)}\psi_{b,4,\sigma'}^{(+)*}\psi_{b,3,\sigma'}^{(-)}+\psi_{a,4,\sigma}^{(-)*}\psi_{a,1,\sigma}^{(+)}\psi_{b,2,\sigma'}^{(-)*}\psi_{b,3,\sigma'}^{(+)}\nonumber \\
	&+\psi_{a,1,\sigma}^{(-)*}\psi_{a,2,\sigma}^{(+)}\psi_{b,3,\sigma'}^{(-)*}\psi_{b,4,\sigma'}^{(+)}+(a\leftrightarrow b)\Big]\delta(\mathbf{k}_1+\mathbf{k}_3-\mathbf{k}_2-\mathbf{k}_4)\\
	S^c_{nor,\pi-\theta_2}&=- \frac{\Delta_{\pi-\theta_2}}{2}\sum_{a,b=1}^R\sum_{\sigma,\sigma'=\uparrow,\downarrow}\int dk\Big[\psi_{a,1,\sigma}^{(+)*}\psi_{a,2,\sigma}^{(-)}\psi_{b,3,\sigma'}^{(+)*}\psi_{b,4,\sigma'}^{(-)}+\psi_{a,2,\sigma}^{(+)*}\psi_{a,3,\sigma}^{(-)}\psi_{b,4,\sigma'}^{(+)*}\psi_{b,1,\sigma'}^{(-)}+\psi_{a,2,\sigma}^{(-)*}\psi_{a,1,\sigma}^{(+)}\psi_{b,4,\sigma'}^{(-)*}\psi_{b,3,\sigma'}^{(+)}\nonumber \\
	&+\psi_{a,3,\sigma}^{(-)*}\psi_{a,2,\sigma}^{(+)}\psi_{b,1,\sigma'}^{(-)*}\psi_{b,4,\sigma'}^{(+)}+(a\leftrightarrow b)\Big]\delta(\mathbf{k}_1+\mathbf{k}_3-\mathbf{k}_2-\mathbf{k}_4)\\
	S^c_{nor,\pi/2}&=-\frac{\Delta_{\pi/2}}{2}\sum_{a,b=1}^R\sum_{\sigma,\sigma'=\uparrow,\downarrow}\int dk\Big[\psi_{a,3,\sigma}^{(-)*}\psi_{a,4,\sigma}^{(-)}\psi_{b,1,\sigma'}^{(-)*}\psi_{b,2,\sigma'}^{(-)}+\psi_{a,4,\sigma}^{(-)*}\psi_{a,1,\sigma}^{(-)}\psi_{b,2,\sigma'}^{(-)*}\psi_{b,3,\sigma'}^{(-)}+\psi_{a,4,\sigma}^{(-)*}\psi_{a,3,\sigma}^{(-)}\psi_{b,2,\sigma'}^{(-)*}\psi_{b,1,\sigma'}^{(-)}\nonumber\\
	&+\psi_{a,1,\sigma}^{(-)*}\psi_{a,4,\sigma}^{(-)}\psi_{b,3,\sigma'}^{(-)*}\psi_{b,2,\sigma'}^{(-)}+\psi_{a,2,\sigma}^{(+)*}\psi_{a,1,\sigma}^{(+)}\psi_{b,4,\sigma'}^{(+)*}\psi_{b,3,\sigma'}^{(+)}+\psi_{a,1,\sigma}^{(+)*}\psi_{a,4,\sigma}^{(+)}\psi_{b,3,\sigma'}^{(+)*}\psi_{b,2,\sigma'}^{(+)}+\psi_{a,1,\sigma}^{(+)*}\psi_{a,2,\sigma}^{(+)}\psi_{b,3,\sigma'}^{(+)*}\psi_{b,4,\sigma'}^{(+)}\nonumber\\
	&+\psi_{a,4,\sigma}^{(+)*}\psi_{a,1,\sigma}^{(+)}\psi_{b,2,\sigma'}^{(+)*}\psi_{b,3,\sigma'}^{(+)}+(a\leftrightarrow b)\Big]\delta(\mathbf{k}_1+\mathbf{k}_3-\mathbf{k}_2-\mathbf{k}_4)
\end{align}

\subsubsection{Umklapp process}\label{Appendix:UmklappScatteringChannels}

We show all the Umklapp processes as follows:
\begin{align}
	S_{umk1, 0}&=-\frac{\Upsilon_0}{2}\sum_{a,b=1}^R\sum_{\sigma,\sigma'}\int dk\Big[\psi_{a,1,\sigma}^{(+)*}\psi_{a,1,\sigma}^{(-)}\psi_{b,1,\sigma'}^{(+)*}\psi_{b,1,\sigma'}^{(-)}+\psi_{a,2,\sigma}^{(+)*}\psi_{a,2,\sigma}^{(-)}\psi_{b,2,\sigma'}^{(+)*}\psi_{b,2,\sigma'}^{(-)}+\psi_{a,3,\sigma}^{(+)*}\psi_{a,3,\sigma}^{(-)}\psi_{b,3,\sigma'}^{(+)*}\psi_{b,3,\sigma'}^{(-)}\nonumber\\
	&+\psi_{a,4,\sigma}^{(+)*}\psi_{a,4,\sigma}^{(-)}\psi_{b,4,\sigma'}^{(+)*}\psi_{b,4,\sigma'}^{(-)}+\psi_{a,1,\sigma}^{(-)*}\psi_{a,1,\sigma}^{(+)}\psi_{b,1,\sigma'}^{(-)*}\psi_{b,1,\sigma'}^{(+)}+\psi_{a,2,\sigma}^{(-)*}\psi_{a,2,\sigma}^{(+)}\psi_{b,2,\sigma'}^{(-)*}\psi_{b,2,\sigma'}^{(+)}+\psi_{a,3,\sigma}^{(-)*}\psi_{a,3,\sigma}^{(+)}\psi_{b,3,\sigma'}^{(-)*}\psi_{b,3,\sigma'}^{(+)}\nonumber \\
	&+\psi_{a,4,\sigma}^{(-)*}\psi_{a,4,\sigma}^{(+)}\psi_{b,4,\sigma'}^{(-)*}\psi_{b,4,\sigma'}^{(+)}\Big]\delta(\mathbf{k}_1+\mathbf{k}_3-\mathbf{k}_2-\mathbf{k}_4)\\
	S_{umk1, \theta_1}^d&=-\frac{\Upsilon_{\theta_1}^d}{2}\sum_{a,b=1}^R\sum_{\sigma,\sigma'}\int dk\Big[\psi_{a,1,\sigma}^{(+)*}\psi_{a,1,\sigma}^{(-)}\psi_{b,3,\sigma'}^{(-)*}\psi_{b,3,\sigma'}^{(+)}+\psi_{a,2,\sigma}^{(+)*}\psi_{a,2,\sigma}^{(-)}\psi_{b,4,\sigma'}^{(-)*}\psi_{b,4,\sigma'}^{(+)}+\psi_{a,3,\sigma}^{(+)*}\psi_{a,3,\sigma}^{(-)}\psi_{b,1,\sigma'}^{(-)*}\psi_{b,1,\sigma'}^{(+)}\nonumber\\
	&+\psi_{a,4,\sigma}^{(+)*}\psi_{a,4,\sigma}^{(-)}\psi_{b,2,\sigma'}^{(-)*}\psi_{b,2,\sigma'}^{(+)}+(a \leftrightarrow b)\Big]\delta(\mathbf{k}_1+\mathbf{k}_3-\mathbf{k}_2-\mathbf{k}_4)\\
	S_{umk1, \theta_1}^e&=-\frac{\Upsilon_{\theta_1}^e}{2}\sum_{a,b=1}^R\sum_{\sigma,\sigma'}\int dk\Big[\psi_{a,1,\sigma}^{(+)*}\psi_{a,3,\sigma}^{(+)}\psi_{b,3,\sigma'}^{(-)*}\psi_{b,1,\sigma'}^{(-)}+\psi_{a,2,\sigma}^{(+)*}\psi_{a,4,\sigma}^{(+)}\psi_{b,4,\sigma'}^{(-)*}\psi_{b,2,\sigma'}^{(-)}+\psi_{a,3,\sigma}^{(+)*}\psi_{a,1,\sigma}^{(+)}\psi_{b,1,\sigma'}^{(-)*}\psi_{b,3,\sigma'}^{(-)}\nonumber\\
	&+\psi_{a,4,\sigma}^{(+)*}\psi_{a,2,\sigma}^{(+)}\psi_{b,2,\sigma'}^{(-)*}\psi_{b,4,\sigma'}^{(-)}+(a \leftrightarrow b)\Big]\delta(\mathbf{k}_1+\mathbf{k}_3-\mathbf{k}_2-\mathbf{k}_4)
\end{align}
\begin{align}
	S_{umk2,\theta_1}^d&=-\frac{\Xi_{\theta_1}^d}{2}\sum_{a,b=1}^R\sum_{\sigma,\sigma'}\int dk\Big[\psi_{a,1,\sigma}^{(+)*}\psi_{a,2,\sigma}^{(+)}\psi_{b,3,\sigma'}^{(-)*}\psi_{b,4,\sigma'}^{(-)}+\psi_{a,2,\sigma}^{(+)*}\psi_{a,3,\sigma}^{(+)}\psi_{b,4,\sigma'}^{(-)*}\psi_{b,1,\sigma'}^{(-)}+\psi_{a,3,\sigma}^{(+)*}\psi_{a,4,\sigma}^{(+)}\psi_{b,1,\sigma'}^{(-)*}\psi_{b,2,\sigma'}^{(-)}\nonumber\\
	&+\psi_{a,4,\sigma}^{(+)*}\psi_{a,1,\sigma}^{(+)}\psi_{b,2,\sigma'}^{(-)*}\psi_{b,3,\sigma'}^{(-)}+\psi_{a,1,\sigma}^{(+)*}\psi_{a,4,\sigma}^{(+)}\psi_{b,3,\sigma'}^{(-)*}\psi_{b,2,\sigma'}^{(-)}+\psi_{a,2,\sigma}^{(+)*}\psi_{a,1,\sigma}^{(+)}\psi_{b,4,\sigma'}^{(-)*}\psi_{b,3,\sigma'}^{(-)}+\psi_{a,3,\sigma}^{(+)*}\psi_{a,2,\sigma}^{(+)}\psi_{b,1,\sigma'}^{(-)*}\psi_{b,4,\sigma'}^{(-)}\nonumber \\
	&+\psi_{a,4,\sigma}^{(+)*}\psi_{a,3,\sigma}^{(+)}\psi_{b,2,\sigma'}^{(-)*}\psi_{b,1,\sigma'}^{(-)}+(a\leftrightarrow b)\Big]\delta(\mathbf{k}_1+\mathbf{k}_3-\mathbf{k}_2-\mathbf{k}_4)\\
	S_{umk2,\theta_1}^e&=-\frac{\Xi_{\theta_1}^e}{2}\sum_{a,b=1}^R\sum_{\sigma,\sigma'}\int dk\Big[\psi_{a,1,\sigma}^{(+)*}\psi_{a,4,\sigma}^{(-)}\psi_{b,3,\sigma'}^{(-)*}\psi_{b,2,\sigma'}^{(+)}+\psi_{a,2,\sigma}^{(+)*}\psi_{a,1,\sigma}^{(-)}\psi_{b,4,\sigma'}^{(-)*}\psi_{b,3,\sigma'}^{(+)}+\psi_{a,3,\sigma}^{(+)*}\psi_{a,2,\sigma}^{(-)}\psi_{b,1,\sigma'}^{(-)*}\psi_{b,4,\sigma'}^{(+)}\nonumber\\
	&+\psi_{a,4,\sigma}^{(+)*}\psi_{a,3,\sigma}^{(-)}\psi_{b,2,\sigma'}^{(-)*}\psi_{b,1,\sigma'}^{(+)}+\psi_{a,1,\sigma}^{(+)*}\psi_{a,2,\sigma}^{(-)}\psi_{b,3,\sigma'}^{(-)*}\psi_{b,4,\sigma'}^{(+)}+\psi_{a,2,\sigma}^{(+)*}\psi_{a,3,\sigma}^{(-)}\psi_{b,4,\sigma'}^{(-)*}\psi_{b,1,\sigma'}^{(+)}+\psi_{a,3,\sigma}^{(+)*}\psi_{a,4,\sigma}^{(-)}\psi_{b,1,\sigma'}^{(-)*}\psi_{b,2,\sigma'}^{(+)}\nonumber \\
	&+\psi_{a,4,\sigma}^{(+)*}\psi_{a,1,\sigma}^{(-)}\psi_{b,2,\sigma'}^{(-)*}\psi_{b,3,\sigma'}^{(+)}+(a\leftrightarrow b)\Big]\delta(\mathbf{k}_1+\mathbf{k}_3-\mathbf{k}_2-\mathbf{k}_4),\\
	S_{umk2,\theta_2}^d&=-\frac{\Xi_{\theta_2}^d}{2}\sum_{a,b=1}^R\sum_{\sigma,\sigma'}\int dk\Big[\psi_{a,1,\sigma}^{(+)*}\psi_{a,1,\sigma}^{(-)}\psi_{b,2,\sigma'}^{(-)*}\psi_{b,2,\sigma'}^{(+)}+\psi_{a,4,\sigma}^{(+)*}\psi_{a,4,\sigma}^{(-)}\psi_{b,1,\sigma'}^{(-)*}\psi_{b,1,\sigma'}^{(+)}+\psi_{a,3,\sigma}^{(+)*}\psi_{a,3,\sigma}^{(-)}\psi_{b,4,\sigma'}^{(-)*}\psi_{b,4,\sigma'}^{(+)}\nonumber\\
	&+\psi_{a,2,\sigma}^{(+)*}\psi_{a,2,\sigma}^{(-)}\psi_{b,3,\sigma'}^{(-)*}\psi_{b,3,\sigma'}^{(+)}+\psi_{a,1,\sigma}^{(-)*}\psi_{a,1,\sigma}^{(+)}\psi_{b,2,\sigma'}^{(+)*}\psi_{b,2,\sigma'}^{(-)}+\psi_{a,4,\sigma}^{(-)*}\psi_{a,4,\sigma}^{(+)}\psi_{b,1,\sigma'}^{(+)*}\psi_{b,1,\sigma'}^{(-)}+\psi_{a,3,\sigma}^{(-)*}\psi_{a,3,\sigma}^{(+)}\psi_{b,4,\sigma'}^{(+)*}\psi_{b,4,\sigma'}^{(-)}\nonumber \\
	&+\psi_{a,2,\sigma}^{(-)*}\psi_{a,2,\sigma}^{(+)}\psi_{b,3,\sigma'}^{(+)*}\psi_{b,3,\sigma'}^{(-)}+(a\leftrightarrow b)\Big]\delta(\mathbf{k}_1+\mathbf{k}_3-\mathbf{k}_2-\mathbf{k}_4)\\
	S_{umk2,\theta_2}^e&=-\frac{\Xi_{\theta_2}^e}{2}\sum_{a,b=1}^R\sum_{\sigma,\sigma'}\int dk\Big[\psi_{a,1,\sigma}^{(+)*}\psi_{a,2,\sigma}^{(+)}\psi_{b,2,\sigma'}^{(-)*}\psi_{b,1,\sigma'}^{(-)}+\psi_{a,4,\sigma}^{(+)*}\psi_{a,1,\sigma}^{(+)}\psi_{b,1,\sigma'}^{(-)*}\psi_{b,4,\sigma'}^{(-)}+\psi_{a,3,\sigma}^{(+)*}\psi_{a,4,\sigma}^{(+)}\psi_{b,4,\sigma'}^{(-)*}\psi_{b,3,\sigma'}^{(-)}\nonumber\\
	&+\psi_{a,2,\sigma}^{(+)*}\psi_{a,3,\sigma}^{(+)}\psi_{b,3,\sigma'}^{(-)*}\psi_{b,2,\sigma'}^{(-)}+\psi_{a,1,\sigma}^{(-)*}\psi_{a,2,\sigma}^{(-)}\psi_{b,2,\sigma'}^{(+)*}\psi_{b,1,\sigma'}^{(+)}+\psi_{a,4,\sigma}^{(-)*}\psi_{a,1,\sigma}^{(-)}\psi_{b,1,\sigma'}^{(+)*}\psi_{b,4,\sigma'}^{(+)}+\psi_{a,3,\sigma}^{(-)*}\psi_{a,4,\sigma}^{(-)}\psi_{b,4,\sigma'}^{(+)*}\psi_{b,3,\sigma'}^{(+)}\nonumber \\
	&+\psi_{a,2,\sigma}^{(-)*}\psi_{a,3,\sigma}^{(-)}\psi_{b,3,\sigma'}^{(+)*}\psi_{b,2,\sigma'}^{(+)}+(a\leftrightarrow b)\Big]\delta(\mathbf{k}_1+\mathbf{k}_3-\mathbf{k}_2-\mathbf{k}_4)\\
	S_{umk2,\pi/2}^d&=-\frac{\Xi_{\pi/2}^d}{2}\sum_{a,b=1}^R\sum_{\sigma,\sigma'}\int dk\Big[\psi_{a,1,\sigma}^{(+)*}\psi_{a,1,\sigma}^{(-)}\psi_{b,4,\sigma'}^{(+)*}\psi_{b,4,\sigma'}^{(-)}+\psi_{a,4,\sigma}^{(+)*}\psi_{a,4,\sigma}^{(-)}\psi_{b,3,\sigma'}^{(+)*}\psi_{b,3,\sigma'}^{(-)}+\psi_{a,3,\sigma}^{(+)*}\psi_{a,3,\sigma}^{(-)}\psi_{b,2,\sigma'}^{(+)*}\psi_{b,2,\sigma'}^{(-)}\nonumber\\
	&+\psi_{a,2,\sigma}^{(+)*}\psi_{a,2,\sigma}^{(-)}\psi_{b,1,\sigma'}^{(+)*}\psi_{b,1,\sigma'}^{(-)}+\psi_{a,1,\sigma}^{(-)*}\psi_{a,1,\sigma}^{(+)}\psi_{b,4,\sigma'}^{(-)*}\psi_{b,4,\sigma'}^{(+)}+\psi_{a,4,\sigma}^{(-)*}\psi_{a,4,\sigma}^{(+)}\psi_{b,3,\sigma'}^{(-)*}\psi_{b,3,\sigma'}^{(+)}+\psi_{a,3,\sigma}^{(-)*}\psi_{a,3,\sigma}^{(+)}\psi_{b,2,\sigma'}^{(-)*}\psi_{b,2,\sigma'}^{(+)}\nonumber \\
	&+\psi_{a,2,\sigma}^{(-)*}\psi_{a,2,\sigma}^{(+)}\psi_{b,1,\sigma'}^{(-)*}\psi_{b,1,\sigma'}^{(+)}+(a\leftrightarrow b)\Big]\delta(\mathbf{k}_1+\mathbf{k}_3-\mathbf{k}_2-\mathbf{k}_4),\\
	S_{umk2,\pi/2}^e&=-\frac{\Xi_{\pi/2}^e}{2}\sum_{a,b=1}^R\sum_{\sigma,\sigma'}\int dk\Big[\psi_{a,1,\sigma}^{(+)*}\psi_{a,4,\sigma}^{(-)}\psi_{b,4,\sigma'}^{(+)*}\psi_{b,1,\sigma'}^{(-)}+\psi_{a,4,\sigma}^{(+)*}\psi_{a,3,\sigma}^{(-)}\psi_{b,3,\sigma'}^{(+)*}\psi_{b,4,\sigma'}^{(-)}+\psi_{a,3,\sigma}^{(+)*}\psi_{a,2,\sigma}^{(-)}\psi_{b,2,\sigma'}^{(+)*}\psi_{b,3,\sigma'}^{(-)}\nonumber\\
	&+\psi_{a,2,\sigma}^{(+)*}\psi_{a,1,\sigma}^{(-)}\psi_{b,1,\sigma'}^{(+)*}\psi_{b,2,\sigma'}^{(-)}+\psi_{a,1,\sigma}^{(-)*}\psi_{a,4,\sigma}^{(+)}\psi_{b,4,\sigma'}^{(-)*}\psi_{b,1,\sigma'}^{(+)}+\psi_{a,4,\sigma}^{(-)*}\psi_{a,3,\sigma}^{(+)}\psi_{b,3,\sigma'}^{(-)*}\psi_{b,4,\sigma'}^{(+)}+\psi_{a,3,\sigma}^{(-)*}\psi_{a,2,\sigma}^{(+)}\psi_{b,2,\sigma'}^{(-)*}\psi_{b,3,\sigma'}^{(+)}\nonumber \\
	&+\psi_{a,2,\sigma}^{(-)*}\psi_{a,1,\sigma}^{(+)}\psi_{b,1,\sigma'}^{(-)*}\psi_{b,2,\sigma'}^{(+)}+(a\leftrightarrow b)\Big]\delta(\mathbf{k}_1+\mathbf{k}_3-\mathbf{k}_2-\mathbf{k}_4)
\end{align}

\subsection{Regularized random charge potential vertices} \label{Appendix:RegularizedDisorderVertices}

Here, co-dimensional regularized random charge potential vertices are given with following short-hand notations:
\begin{gather*}
	A=\frac{\gamma_0+i\gamma_{d-1}}{2},\; B=\frac{\gamma_0-i\gamma_{d-1}}{2},\\
	\int d\tilde{k}=\int \frac{d\omega}{2\pi} \int\frac{d\omega'}{2\pi} \int\frac{d^d\mathbf{k}_1}{(2\pi)^d}\int\frac{d^d\mathbf{k}_2}{(2\pi)^d}\int\frac{d^d\mathbf{k}_3}{(2\pi)^d}\int\frac{d^d\mathbf{k}_4}{(2\pi)^d} \\
	\bar{\Psi}^a_{n,\sigma,i_f}\mathcal{M}_{nm}\Psi_{m,\sigma,i_f}\bar{\Psi}^b_{k,\sigma',i_f}\tilde{\mathcal{M}}_{kl}\Psi^b_{l,\sigma',i_f}=\bar{\Psi}^a_{n,\sigma,i_f}(\omega,\mathbf{k}_1)\mathcal{M}_{nm}\Psi_{m,\sigma,i_f}(\omega,\mathbf{k}_2)\bar{\Psi}^b_{k,\sigma',i_f}(\omega',\mathbf{k}_3)\tilde{\mathcal{M}}_{kl}\Psi^b_{l,\sigma',i_f}(\omega',\mathbf{k}_4).
\end{gather*}

\subsubsection{Normal process}

\begin{align*}
	S_{dis,0}&=-\frac{\Gamma_0}{2} \sum_{a,b=1}^R\sum_{i_f=1}^{N_f}\sum_{n=1}^4\sum_{\sigma,\sigma'=1}^{N_c}\int d\tilde{k}  \Big[\bar{\Psi}^a_{n,\sigma,i_f}A\Psi^a_{n,\sigma,i_f}\bar{\Psi}^b_{n,\sigma',i_f}A\Psi^b_{n,\sigma',i_f}+\bar{\Psi}^a_{n,\sigma,i_f}B\Psi^a_{n,\sigma,i_f}\bar{\Psi}^{b}_{n,\sigma',i_f}B\Psi^b_{n,\sigma',i_f}\Big]\nonumber\\
	&\times \delta(\mathbf{k}_1+\mathbf{k}_3-\mathbf{k}_2-\mathbf{k}_4)\\
	S_{dis,\theta_1}^d&=-\frac{\Gamma_{\theta_1}^d}{2}\sum_{a,b=1}^R\sum_{i_f=1}^{N_f}\sum_{\sigma,\sigma'=1}^{N_c}\int d\tilde{k}\Big[\bar{\Psi}^a_{1,\sigma,i_f}A\Psi^a_{1,\sigma,i_f}\bar{\Psi}^b_{3,\sigma',i_f}B\Psi^b_{3,\sigma',i_f}+\bar{\Psi}^a_{2,\sigma,i_f}B\Psi^a_{2,\sigma,i_f}\bar{\Psi}^b_{4,\sigma',i_f}A\Psi^b_{4,\sigma',i_f}\\
	&+\bar{\Psi}^a_{1,\sigma,i_f}B\Psi^a_{1,\sigma,i_f}\bar{\Psi}^b_{3,\sigma',i_f}A\Psi^b_{3,\sigma',i_f}+\bar{\Psi}^a_{2,\sigma,i_f}A\Psi^a_{2,\sigma,i_f}\bar{\Psi}^b_{4,\sigma',i_f}B\Psi^b_{4,\sigma',i_f}+(a \leftrightarrow b)\Big]\delta(\mathbf{k}_1+\mathbf{k}_3-\mathbf{k}_2-\mathbf{k}_4)\\
	S_{dis,\theta_1}^e&=-\frac{\Gamma_{\theta_1}^e}{2}\sum_{a,b=1}^R\sum_{i_f=1}^{N_f}\sum_{\sigma,\sigma'=1}^{N_c}\int d\tilde{k}\Big[\bar{\Psi}^a_{1,\sigma,i_f}AB\Psi^a_{3,\sigma,i_f}\bar{\Psi}^b_{3,\sigma',i_f}BA\Psi^b_{1,\sigma',i_f}+\bar{\Psi}^a_{2,\sigma,i_f}BA\Psi^a_{4,\sigma,i_f}\bar{\Psi}^b_{4,\sigma',i_f}AB\Psi^b_{2,\sigma',i_f}\\
	&+\bar{\Psi}^a_{1,\sigma,i_f}BA\Psi^a_{3,\sigma,i_f}\bar{\Psi}^b_{3,\sigma',i_f}AB\Psi^b_{1,\sigma',i_f}+\bar{\Psi}^a_{2,\sigma,i_f}AB\Psi^a_{4,\sigma,i_f}\bar{\Psi}^b_{4,\sigma',i_f}BA\Psi^b_{2,\sigma',i_f}+(a \leftrightarrow b)\Big]\delta(\mathbf{k}_1+\mathbf{k}_3-\mathbf{k}_2-\mathbf{k}_4)\\
	S_{dis,\theta_2}^d&=-\frac{\Gamma_{\theta_2}^d}{2}\sum_{a,b=1}^R\sum_{i_f=1}^{N_f}\sum_{\sigma,\sigma'=1}^{N_c}\int d\tilde{k}\Big[\bar{\Psi}^a_{1,\sigma,i_f}A\Psi^a_{1,\sigma,i_f}\bar{\Psi}^b_{4,\sigma',i_f}B\Psi^b_{4,\sigma',i_f}+\bar{\Psi}^a_{2,\sigma,i_f}A\Psi^a_{2,\sigma,i_f}\bar{\Psi}^b_{3,\sigma',i_f}A\Psi^b_{3,\sigma',i_f}\\
	&+\bar{\Psi}^a_{1,\sigma,i_f}B\Psi^a_{1,\sigma,i_f}\bar{\Psi}^b_{4,\sigma',i_f}A\Psi^b_{4,\sigma',i_f}+\bar{\Psi}^a_{2,\sigma,i_f}B\Psi^a_{2,\sigma,i_f}\bar{\Psi}^b_{3,\sigma',i_f}B\Psi^b_{3,\sigma',i_f}+(a \leftrightarrow b)\Big]\delta(\mathbf{k}_1+\mathbf{k}_3-\mathbf{k}_2-\mathbf{k}_4)\\
	S_{dis,\theta_2}^e&=-\frac{\Gamma_{\theta_2}^e}{2}\sum_{a,b=1}^R\sum_{i_f=1}^{N_f}\sum_{\sigma,\sigma'=1}^{N_c}\int d\tilde{k}\Big[\bar{\Psi}^a_{1,\sigma,i_f}AB\Psi^a_{4,\sigma,i_f}\bar{\Psi}^b_{4,\sigma',i_f}BA\Psi^b_{1,\sigma',i_f}+\bar{\Psi}^a_{2,\sigma,i_f}A\Psi^a_{3,\sigma,i_f}\bar{\Psi}^b_{3,\sigma',i_f}A\Psi^b_{2,\sigma',i_f}\\
	&+\bar{\Psi}^a_{1,\sigma,i_f}BA\Psi^a_{4,\sigma,i_f}\bar{\Psi}^b_{4,\sigma',i_f}AB\Psi^b_{1,\sigma',i_f}+\bar{\Psi}^a_{2,\sigma,i_f}B\Psi^a_{3,\sigma,i_f}\bar{\Psi}^b_{3,\sigma',i_f}B\Psi^b_{2,\sigma',i_f}+(a\leftrightarrow b)\Big]\delta(\mathbf{k}_1+\mathbf{k}_3-\mathbf{k}_2-\mathbf{k}_4)\\
	S_{dis,\pi/2}^d&=-\frac{\Gamma_{\pi/2}^d}{2}\sum_{a,b=1}^R\sum_{i_f=1}^{N_f}\sum_{\sigma,\sigma'=1}^{N_c}\int d\tilde{k}\Big[\bar{\Psi}^a_{1,\sigma,i_f}A\Psi^a_{1,\sigma,i_f}\bar{\Psi}^b_{2,\sigma}A\Psi^b_{2,\sigma',i_f}+\bar{\Psi}^a_{2,\sigma,i_f}A\Psi^a_{2,\sigma,i_f}\bar{\Psi}^b_{1,\sigma',i_f}B\Psi^b_{1,\sigma',i_f}\\
	&+\bar{\Psi}^a_{1,\sigma,i_f}B\Psi^a_{1,\sigma,i_f}\bar{\Psi}^b_{2,\sigma',i_f}B\Psi^b_{2,\sigma',i_f}+\bar{\Psi}^a_{2,\sigma,i_f}B\Psi^a_{2,\sigma,i_f}\bar{\Psi}^b_{1,\sigma',i_f}A\Psi^b_{1,\sigma',i_f}+\bar{\Psi}^a_{3,\sigma,i_f}B\Psi^a_{3,\sigma,i_f}\bar{\Psi}^b_{4,\sigma',i_f}B\Psi^b_{4,\sigma',i_f}\\
	&+\bar{\Psi}^a_{4,\sigma,i_f}B\Psi^a_{4,\sigma,i_f}\bar{\Psi}^b_{3,\sigma',i_f}A\Psi^b_{3,\sigma',i_f}+\bar{\Psi}^a_{3,\sigma,i_f}A\Psi^a_{3,\sigma,i_f}\bar{\Psi}^b_{4,\sigma',i_f}A\Psi^b_{4,\sigma',i_f}+\bar{\Psi}^a_{4,\sigma,i_f}A\Psi^a_{4,\sigma,i_f}\bar{\Psi}^b_{3,\sigma',i_f}B\Psi^b_{3,\sigma',i_f}\nonumber\\
	&+(a \leftrightarrow b)\Big]\delta(\mathbf{k}_1+\mathbf{k}_3-\mathbf{k}_2-\mathbf{k}_4)\\
	S_{dis,\pi/2}^e&=-\frac{\Gamma_{\pi/2}^e}{2}\sum_{a,b=1}^R\sum_{i_f=1}^{N_f}\sum_{\sigma,\sigma'=1}^{N_c}\int d\tilde{k}\Big[\bar{\Psi}^a_{1,\sigma,i_f}A\Psi^a_{2,\sigma,i_f}\bar{\Psi}^b_{2,\sigma',i_f}A\Psi^b_{1,\sigma',i_f}+\bar{\Psi}^a_{2,\sigma,i_f}AB\Psi^a_{1,\sigma,i_f}\bar{\Psi}^b_{1,\sigma',i_f}BA\Psi^b_{2,\sigma',i_f}\\
	&+\bar{\Psi}^a_{1,\sigma,i_f}B\Psi^a_{2,\sigma,i_f}\bar{\Psi}^b_{2,\sigma',i_f}B\Psi^b_{1,\sigma',i_f}+\bar{\Psi}^a_{2,\sigma,i_f}BA\Psi^a_{1,\sigma,i_f}\bar{\Psi}^b_{1,\sigma',i_f}AB\Psi^b_{2,\sigma',i_f}+\bar{\Psi}^a_{3,\sigma,i_f}B\Psi^a_{4,\sigma,i_f}\bar{\Psi}^b_{4,\sigma',i_f}B\Psi^b_{3,\sigma',i_f}\\
	&+\bar{\Psi}^a_{4,\sigma,i_f}BA\Psi^a_{3,\sigma,i_f}\bar{\Psi}^b_{3,\sigma',i_f}AB\Psi^b_{4,\sigma',i_f}+\bar{\Psi}^a_{3,\sigma,i_f}A\Psi^a_{4,\sigma,i_f}\bar{\Psi}^b_{4,\sigma',i_f}A\Psi^b_{3,\sigma',i_f}+\bar{\Psi}^a_{4,\sigma,i_f}AB\Psi^a_{3,\sigma,i_f}\bar{\Psi}^b_{3,\sigma',i_f}BA\Psi^b_{4,\sigma',i_f}\nonumber\\
	&+(a \leftrightarrow b)\Big]\delta(\mathbf{k}_1+\mathbf{k}_3-\mathbf{k}_2-\mathbf{k}_4)\\
	S_{dis,\pi-\theta_1}^d&=-\frac{\Gamma_{\pi-\theta_1}^d}{2}\sum_{a,b=1}^R\sum_{i_f=1}^{N_f}\sum_{\sigma,\sigma'=1}^{N_c}\int d\tilde{k}\Big[\bar{\Psi}^a_{1,\sigma,i_f}A\Psi^a_{1,\sigma,i_f}\bar{\Psi}^b_{3,\sigma',i_f}A\Psi^b_{3,\sigma',i_f}+\bar{\Psi}^a_{2,\sigma,i_f}A\Psi^a_{2,\sigma,i_f}\bar{\Psi}^b_{4,\sigma',i_f}A\Psi^b_{4,\sigma',i_f}\\
	&+\bar{\Psi}^a_{1,\sigma,i_f}B\Psi^a_{1,\sigma}\bar{\Psi}^b_{3,\sigma',i_f}B\Psi^b_{3,\sigma',i_f}+\bar{\Psi}^a_{2,\sigma,i_f}B\Psi^a_{2,\sigma,i_f}\bar{\Psi}^b_{4,\sigma',i_f}B\Psi^b_{4,\sigma',i_f}+(a \leftrightarrow b)\Big]\delta(\mathbf{k}_1+\mathbf{k}_3-\mathbf{k}_2-\mathbf{k}_4),\\
	S_{dis,\pi-\theta_1}^e&=-\frac{\Gamma_{\pi-\theta_1}^e}{2}\sum_{a,b=1}^R\sum_{i_f=1}^{N_f}\sum_{\sigma,\sigma'=1}^{N_c}\int d\tilde{k}\Big[\bar{\Psi}^a_{1,\sigma,i_f}A\Psi^a_{3,\sigma,i_f}\bar{\Psi}^b_{3,\sigma',i_f}A\Psi^b_{1,\sigma',i_f}+\bar{\Psi}^a_{2,\sigma,i_f}A\Psi^a_{4,\sigma,i_f}\bar{\Psi}^b_{4,\sigma',i_f}A\Psi^b_{2,\sigma',i_f}\\
	&+\bar{\Psi}^a_{1,\sigma,i_f}B\Psi^a_{3,\sigma,i_f}\bar{\Psi}^b_{3,\sigma',i_f}B\Psi^b_{1,\sigma',i_f}+\bar{\Psi}^a_{2,\sigma,i_f}B\Psi^a_{4,\sigma,i_f}\bar{\Psi}^b_{4,\sigma',i_f}B\Psi^b_{2,\sigma',i_f}+(a \leftrightarrow b)\Big]\delta(\mathbf{k}_1+\mathbf{k}_3-\mathbf{k}_2-\mathbf{k}_4),
\end{align*}
\begin{align*}
	S_{dis,\pi-\theta_2}^d&=-\frac{\Gamma_{\pi-\theta_2}^d}{2}\sum_{a,b=1}^R\sum_{i_f=1}^{N_f}\sum_{\sigma,\sigma'=1}^{N_c}\int d\tilde{k}\Big[\bar{\Psi}^a_{3,\sigma,i_f}B\Psi^a_{3,\sigma,i_f}\bar{\Psi}^b_{2,\sigma',i_f}A\Psi^b_{2,\sigma',i_f}+\bar{\Psi}^a_{4,\sigma,i_f}B\Psi^a_{4,\sigma,i_f}\bar{\Psi}^b_{1,\sigma',i_f}B\Psi^b_{1,\sigma',i_f}\\
	&+\bar{\Psi}^a_{3,\sigma,i_f}A\Psi^a_{3,\sigma,i_f}\bar{\Psi}^b_{2,\sigma',i_f}B\Psi^b_{2,\sigma',i_f}+\bar{\Psi}^a_{4,\sigma,i_f}A\Psi^a_{4,\sigma,i_f}\bar{\Psi}^b_{1,\sigma',i_f}A\Psi^b_{1,\sigma',i_f}+(a \leftrightarrow b)\Big]\delta(\mathbf{k}_1+\mathbf{k}_3-\mathbf{k}_2-\mathbf{k}_4)\\
	S_{dis,\pi-\theta_2}^e&=-\frac{\Gamma_{\pi-\theta_2}^e}{2}\sum_{a,b=1}^R\sum_{i_f=1}^{N_f}\sum_{\sigma,\sigma'=1}^{N_c}\int d\tilde{k}\Big[\bar{\Psi}^a_{3,\sigma,i_f}BA\Psi^a_{2,\sigma,i_f}\bar{\Psi}^b_{2,\sigma',i_f}AB\Psi^b_{3,\sigma',i_f}+\bar{\Psi}^a_{4,\sigma,i_f}B\Psi^a_{1,\sigma,i_f}\bar{\Psi}^b_{1,\sigma',i_f}B\Psi^b_{4,\sigma',i_f}\\
	&+\bar{\Psi}^a_{3,\sigma,i_f}AB\Psi^a_{2,\sigma,i_f}\bar{\Psi}^b_{2,\sigma',i_f}BA\Psi^b_{3,\sigma',i_f}+\bar{\Psi}^a_{4,\sigma,i_f}A\Psi^a_{1,\sigma,i_f}\bar{\Psi}^b_{1,\sigma',i_f}A\Psi^b_{4,\sigma',i_f}+(a \leftrightarrow b)\Big]\delta(\mathbf{k}_1+\mathbf{k}_3-\mathbf{k}_2-\mathbf{k}_4)\\
	S_{dis,0}&=-\frac{\Delta_{0}}{2}\sum_{a,b=1}^R\sum_{i_f=1}^{N_f}\sum_{\sigma,\sigma'=1}^{N_c}\int d\tilde{k}\Big[\bar{\Psi}^a_{1,\sigma,i_f}A\Psi^a_{1,\sigma,i_f}\bar{\Psi}^b_{1,\sigma',i_f}B\Psi^b_{1,\sigma',i_f}+\bar{\Psi}^a_{4,\sigma,i_f}B\Psi^a_{4,\sigma,i_f}\bar{\Psi}^b_{4,\sigma',i_f}A\Psi^b_{4,\sigma',i_f}\\
	&+\bar{\Psi}^a_{2,\sigma,i_f}A\Psi^a_{2,\sigma,i_f}\bar{\Psi}^b_{2,\sigma',i_f}B\Psi^b_{2,\sigma',i_f}+\bar{\Psi}^a_{3,\sigma,i_f}A\Psi^a_{3,\sigma,i_f}\bar{\Psi}^b_{3,\sigma',i_f}B\Psi^b_{3,\sigma',i_f}+(a \leftrightarrow b)\Big]\delta(\mathbf{k}_1+\mathbf{k}_3-\mathbf{k}_2-\mathbf{k}_4)\\
	S_{dis,\pi}&=-\frac{\Delta_{\pi}}{2}\sum_{a,b=1}^R\sum_{i_f=1}^{N_f}\sum_{\sigma,\sigma'=1}^{N_c}\int d\tilde{k}\Big[\bar{\Psi}^a_{1,\sigma,i_f}AB\Psi^a_{1,\sigma,i_f}\bar{\Psi}^b_{1,\sigma',i_f}BA\Psi^b_{1,\sigma',i_f}+\bar{\Psi}^a_{4,\sigma,i_f}AB\Psi^a_{4,\sigma,i_f}\bar{\Psi}^b_{4,\sigma',i_f}BA\Psi^b_{4,\sigma',i_f}\nonumber\\
	&+\bar{\Psi}^a_{2,\sigma,i_f}AB\Psi^a_{2,\sigma,i_f}\bar{\Psi}^b_{2,\sigma',i_f}BA\Psi^b_{2,\sigma',i_f}+\bar{\Psi}^a_{3,\sigma,i_f}AB\Psi^a_{3,\sigma,i_f}\bar{\Psi}^b_{3,\sigma',i_f}BA\Psi^b_{3,\sigma',i_f}+(a\leftrightarrow b)\Big]\delta(\mathbf{k}_1+\mathbf{k}_3-\mathbf{k}_2-\mathbf{k}_4)\\
	S_{dis,\theta_1}&=-\frac{\Delta_{\theta_1}}{2}\sum_{a,b=1}^R\sum_{i_f=1}^{N_f}\sum_{\sigma,\sigma'=1}^{N_c}\int d\tilde{k}\Big[\bar{\Psi}^a_{3,\sigma,i_f}iBA\Psi^a_{1,\sigma,i_f}\bar{\Psi}^b_{3,\sigma',i_f}iAB\Psi^b_{1,\sigma',i_f}+\bar{\Psi}^a_{4,\sigma,i_f}iBA\Psi^a_{2,\sigma,i_f}\bar{\Psi}^b_{4,\sigma',i_f}iAB\Psi^b_{2,\sigma',i_f}\nonumber\\
	&+\bar{\Psi}^a_{1,\sigma,i_f}iAB\Psi^a_{3,\sigma,i_f}\bar{\Psi}^b_{1,\sigma',i_f}iBA\Psi^b_{3,\sigma',i_f}+\bar{\Psi}^a_{2,\sigma,i_f}iAB\Psi^a_{4,\sigma,i_f}\bar{\Psi}^b_{2,\sigma',i_f}iBA\Psi^b_{4,\sigma',i_f}+(a \leftrightarrow b)\Big]\delta(\mathbf{k}_1+\mathbf{k}_3-\mathbf{k}_2-\mathbf{k}_4)
\\
	S_{dis,\pi-\theta_1}&=-\frac{\Delta_{\pi-\theta_1}}{2}\sum_{a,b=1}^R\sum_{i_f=1}^{N_f}\sum_{\sigma,\sigma'=1}^{N_c}\int d\tilde{k}\Big[-\bar{\Psi}^a_{3,\sigma,i_f}B\Psi^a_{1,\sigma,i_f}\bar{\Psi}^b_{3,\sigma',i_f}A\Psi^b_{1,\sigma',i_f}-\bar{\Psi}^a_{4,\sigma,i_f}B\Psi^a_{2,\sigma,i_f}\bar{\Psi}^b_{4,\sigma',i_f}A\Psi^b_{2,\sigma',i_f}\nonumber\\
	&-\bar{\Psi}^a_{1,\sigma,i_f}B\Psi^a_{3,\sigma,i_f}\bar{\Psi}^b_{1,\sigma',i_f}A\Psi^b_{3,\sigma',i_f}-\bar{\Psi}^a_{2,\sigma,i_f}B\Psi^a_{4,\sigma,i_f}\bar{\Psi}^b_{2,\sigma',i_f}A\Psi^b_{4,\sigma',i_f}+(a \leftrightarrow b)\Big]\delta(\mathbf{k}_1+\mathbf{k}_3-\mathbf{k}_2-\mathbf{k}_4)\\
	S_{dis,\theta_2}&=-\frac{\Delta_{\theta_2}}{2}\sum_{a,b=1}^R\sum_{i_f=1}^{N_f}\sum_{\sigma,\sigma'=1}^{N_c}\int d\tilde{k}\Big[\bar{\Psi}^a_{1,\sigma,i_f}iAB\Psi^a_{4,\sigma,i_f}\bar{\Psi}^b_{1,\sigma',i_f}iBA\Psi^b_{4,\sigma',i_f}-\bar{\Psi}^a_{2,\sigma,i_f}A\Psi^a_{3,\sigma,i_f}\bar{\Psi}^b_{2,\sigma',i_f}B\Psi^b_{3,\sigma',i_f}\nonumber\\
	&+\bar{\Psi}^a_{4,\sigma,i_f}iBA\Psi^a_{1,\sigma,i_f}\bar{\Psi}^b_{4,\sigma',i_f}iAB\Psi^b_{1,\sigma',i_f}-\bar{\Psi}^a_{3,\sigma,i_f}A\Psi^a_{2,\sigma,i_f}\bar{\Psi}^b_{3,\sigma',i_f}B\Psi^b_{2,\sigma',i_f}+(a \leftrightarrow b)\Big]\delta(\mathbf{k}_1+\mathbf{k}_3-\mathbf{k}_2-\mathbf{k}_4)\\
	S_{dis,\pi-\theta_2}&=-\frac{\Delta_{\pi-\theta_2}}{2}\sum_{a,b=1}^R\sum_{i_f=1}^{N_f}\sum_{\sigma,\sigma'=1}^{N_c}\int d\tilde{k}\Big[-\bar{\Psi}^a_{1,\sigma,i_f}A\Psi^a_{4,\sigma,i_f}\bar{\Psi}^b_{1,\sigma',i_f}B\Psi^b_{4,\sigma',i_f}+\bar{\Psi}^a_{2,\sigma,i_f}iAB\Psi^a_{3,\sigma,i_f}\bar{\Psi}^b_{2,\sigma',i_f}iBA\Psi^b_{3,\sigma',i_f}\nonumber\\
	&-\bar{\Psi}^a_{4,\sigma,i_f}A\Psi^a_{1,\sigma,i_f}\bar{\Psi}^b_{4,\sigma',i_f}B\Psi^b_{1,\sigma',i_f}+\bar{\Psi}^a_{3,\sigma,i_f}iBA\Psi^a_{2,\sigma,i_f}\bar{\Psi}^b_{3,\sigma',i_f}iAB\Psi^b_{2,\sigma',i_f}+(a \leftrightarrow b)\Big]\delta(\mathbf{k}_1+\mathbf{k}_3-\mathbf{k}_2-\mathbf{k}_4)\\
	S_{dis,\pi/2}&=-\frac{\Delta_{\pi/2}}{2}\sum_{a,b=1}^R\sum_{i_f=1}^{N_f}\sum_{\sigma,\sigma'=1}^{N_c}\int d\tilde{k}\Big[\bar{\Psi}^a_{3,\sigma,i_f}B\Psi^a_{4,\sigma,i_f}\bar{\Psi}^b_{3,\sigma',i_f}A\Psi^b_{4,\sigma',i_f}+\bar{\Psi}^a_{4,\sigma,i_f}BA\Psi^a_{3,\sigma,i_f}\bar{\Psi}^b_{4,\sigma',i_f}AB\Psi^b_{3,\sigma',i_f}\nonumber\\
	&+\bar{\Psi}^a_{4,\sigma,i_f}B\Psi^a_{3,\sigma,i_f}\bar{\Psi}^b_{4,\sigma',i_f}A\Psi^b_{3,\sigma',i_f}+\bar{\Psi}^a_{3,\sigma,i_f}AB\Psi^a_{4,\sigma,i_f}\bar{\Psi}^b_{3,\sigma',i_f}BA\Psi^b_{4,\sigma',i_f}\\
	&+\bar{\Psi}^a_{2,\sigma,i_f}A\Psi^a_{1,\sigma,i_f}\bar{\Psi}^b_{2,\sigma',i_f}B\Psi^b_{1,\sigma',i_f}+\bar{\Psi}^a_{1,\sigma,i_f}AB\Psi^a_{2,\sigma,i_f}\bar{\Psi}^b_{1,\sigma',i_f}BA\Psi^b_{2,\sigma',i_f}+\bar{\Psi}^a_{1,\sigma,i_f}A\Psi^a_{2,\sigma,i_f}\bar{\Psi}^b_{1,\sigma',i_f}B\Psi^b_{2,\sigma',i_f}\\
	&+\bar{\Psi}^a_{2,\sigma,i_f}BA\Psi^b_{1,\sigma',i_f}\bar{\Psi}^b_{2,\sigma',i_f}AB\Psi^b_{1,\sigma',i_f}+(a\leftrightarrow b)\Big]\delta(\mathbf{k}_1+\mathbf{k}_3-\mathbf{k}_2-\mathbf{k}_4)
\end{align*}

\subsubsection{Umklapp process}

\begin{align*}
	S_{umk1, 0}&=-\frac{\Upsilon_{0}}{2}\sum_{a,b=1}^R\sum_{\sigma,\sigma'=1}^{N_c}\int d\tilde{k}\Big[\bar{\Psi}^a_{1,\sigma,i_f}A\Psi^a_{3,\sigma,i_f}\bar{\Psi}^b_{1,\sigma',i_f}A\Psi^b_{3,\sigma',i_f}+\bar{\Psi}^a_{2,\sigma,i_f}A\Psi^a_{4,\sigma,i_f}\bar{\Psi}^b_{2,\sigma',i_f}A\Psi^b_{4,\sigma',i_f}\nonumber\\
	&+\bar{\Psi}^a_{1,\sigma,i_f}B\Psi^a_{3,\sigma,i_f}\bar{\Psi}^b_{1,\sigma',i_f}B\Psi^b_{3,\sigma',i_f}+\bar{\Psi}^a_{2,\sigma,i_f}B\Psi^a_{4,\sigma,i_f}\bar{\Psi}^b_{2,\sigma',i_f}B\Psi^b_{4,\sigma',i_f}+\bar{\Psi}^a_{3,\sigma,i_f}A\Psi^a_{1,\sigma,i_f}\bar{\Psi}^b_{3,\sigma',i_f}A\Psi^b_{1,\sigma',i_f}\nonumber\\
	&+\bar{\Psi}^a_{4,\sigma,i_f}A\Psi^a_{2,\sigma,i_f}\bar{\Psi}^b_{4,\sigma',i_f}A\Psi^b_{2,\sigma',i_f}+\bar{\Psi}^a_{3,\sigma,i_f}B\Psi^a_{1,\sigma,i_f}\bar{\Psi}^b_{3,\sigma',i_f}B\Psi^b_{1,\sigma',i_f}+\bar{\Psi}^a_{4,\sigma,i_f}B\Psi^a_{2,\sigma,i_f}\bar{\Psi}^b_{4,\sigma',i_f}B\Psi^b_{2,\sigma',i_f}
	\Big]\nonumber\\
	&\times \delta(\mathbf{k}_1+\mathbf{k}_3-\mathbf{k}_2-\mathbf{k}_4)\\
	S_{umk1, \theta_1}^d&=-\frac{\Upsilon_{\theta_1}^d}{2}\sum_{a,b=1}^R\sum_{\sigma,\sigma'=1}^{N_c}\int d\tilde{k}\Big[-\bar{\Psi}^a_{1,\sigma,i_f}A\Psi^a_{3,\sigma,i_f}\bar{\Psi}^b_{3,\sigma',i_f}B\Psi^b_{1,\sigma',i_f}-\bar{\Psi}^a_{2,\sigma,i_f}A\Psi^a_{4,\sigma,i_f}\bar{\Psi}^b_{4,\sigma',i_f}B\Psi^b_{2,\sigma',i_f}\\
	&-\bar{\Psi}^a_{1,\sigma,i_f}B\Psi^a_{3,\sigma,i_f}\bar{\Psi}^b_{3,\sigma',i_f}A\Psi^b_{1,\sigma',i_f}-\bar{\Psi}^a_{2,\sigma,i_f}B\Psi^a_{4,\sigma,i_f}\bar{\Psi}^b_{4,\sigma',i_f}A\Psi^b_{2,\sigma',i_f}+(a\leftrightarrow b)\Big]\nonumber\\
	&\times \delta(\mathbf{k}_1+\mathbf{k}_3-\mathbf{k}_2-\mathbf{k}_4)\\
	S_{umk1, \theta_1}^e&=-\frac{\Upsilon_{\theta_1}^e}{2}\sum_{a,b=1}^R\sum_{\sigma,\sigma'=1}^{N_c}\int d\tilde{k}\Big[\bar{\Psi}^a_{1,\sigma,i_f}iAB\Psi^a_{1,\sigma,i_f}\bar{\Psi}^b_{3,\sigma',i_f}iBA\Psi^b_{3,\sigma',i_f}+\bar{\Psi}^a_{2,\sigma,i_f}iAB\Psi^a_{2,\sigma,i_f}\bar{\Psi}^b_{4,\sigma',i_f}iBA\Psi^b_{4,\sigma',i_f}\\
	&+\bar{\Psi}^a_{1,\sigma,i_f}iBA\Psi^a_{1,\sigma,i_f}\bar{\Psi}^b_{3,\sigma',i_f}iAB\Psi^b_{3,\sigma',i_f}+\bar{\Psi}^a_{2,\sigma,i_f}iBA\Psi^a_{2,\sigma,i_f}\bar{\Psi}^b_{4,\sigma',i_f}iAB\Psi^b_{4,\sigma',i_f}+(a\leftrightarrow b)\Big]\nonumber\\
	&\times \delta(\mathbf{k}_1+\mathbf{k}_3-\mathbf{k}_2-\mathbf{k}_4)\\
	S_{umk2,\theta_1}^d&=-\frac{\Xi_{\theta_1}^d}{2}\sum_{a,b=1}^R\sum_{\sigma,\sigma'=1}^{N_c}\int d\tilde{k}\Big[\bar{\Psi}^a_{1,\sigma,i_f}A\Psi^a_{2,\sigma,i_f}\bar{\Psi}^b_{3,\sigma',i_f}B\Psi^b_{4,\sigma',i_f}+\bar{\Psi}^a_{1,\sigma,i_f}B\Psi^a_{2,\sigma,i_f}\bar{\Psi}^b_{3,\sigma',i_f}A\Psi^b_{4,\sigma',i_f}\\
	&+\bar{\Psi}^a_{2,\sigma,i_f}A\Psi^a_{1,\sigma,i_f}\bar{\Psi}^b_{4,\sigma',i_f}B\Psi^b_{3,\sigma',i_f}+\bar{\Psi}^a_{2,\sigma,i_f}B\Psi^a_{1,\sigma,i_f}\bar{\Psi}^b_{4,\sigma',i_f}A\Psi^b_{3,\sigma',i_f}+\bar{\Psi}^a_{2,\sigma,i_f}iAB\Psi^a_{1,\sigma,i_f}\bar{\Psi}^b_{4,\sigma',i_f}iBA\Psi^b_{3,\sigma',i_f}\\
	&+\bar{\Psi}^a_{2,\sigma,i_f}iBA\Psi^a_{1,\sigma,i_f}\bar{\Psi}^b_{4,\sigma',i_f}iAB\Psi^b_{3,\sigma',i_f}+\bar{\Psi}^a_{1,\sigma,i_f}iAB\Psi^a_{2,\sigma,i_f}\bar{\Psi}^b_{3,\sigma',i_f}iBA\Psi^b_{4,\sigma',i_f}\nonumber\\
	&+\bar{\Psi}^a_{1,\sigma,i_f}iBA\Psi^a_{2,\sigma,i_f}\bar{\Psi}^b_{3,\sigma',i_f}iAB\Psi^b_{4,\sigma',i_f}+(a\leftrightarrow b)\Big]\delta(\mathbf{k}_1+\mathbf{k}_3-\mathbf{k}_2-\mathbf{k}_4)\\
	S_{umk2,\theta_1}^e&=-\frac{\Xi_{\theta_1}^e}{2}\sum_{a,b=1}^R\sum_{\sigma,\sigma'=1}^{N_c}\int d\tilde{k}\Big[\bar{\Psi}^a_{1,\sigma,i_f}AB\Psi^a_{4,\sigma,i_f}\bar{\Psi}^b_{3,\sigma',i_f}BA\Psi^b_{2,\sigma',i_f}+\bar{\Psi}^a_{1,\sigma,i_f}BA\Psi^a_{4,\sigma,i_f}\bar{\Psi}^b_{3,\sigma',i_f}AB\Psi^b_{2,\sigma',i_f}\\
	&+\bar{\Psi}^a_{2,\sigma,i_f}AB\Psi^a_{3,\sigma,i_f}\bar{\Psi}^b_{4,\sigma',i_f}BA\Psi^b_{1,\sigma',i_f}+\bar{\Psi}^a_{2,\sigma,i_f}BA\Psi^a_{3,\sigma,i_f}\bar{\Psi}^b_{4,\sigma',i_f}AB\Psi^b_{1,\sigma',i_f}\nonumber\\
	&-\bar{\Psi}^a_{2,\sigma,i_f}A\Psi^a_{3,\sigma,i_f}\bar{\Psi}^b_{4,\sigma',i_f}B\Psi^b_{1,\sigma',i_f}-\bar{\Psi}^a_{2,\sigma,i_f}B\Psi^a_{3,\sigma,i_f}\bar{\Psi}^b_{4,\sigma',i_f}A\Psi^b_{1,\sigma',i_f}-\bar{\Psi}^a_{1,\sigma,i_f}A\Psi^a_{4,\sigma,i_f}\bar{\Psi}^b_{3,\sigma',i_f}B\Psi^b_{2,\sigma',i_f}\nonumber\\
	&-\bar{\Psi}^a_{1,\sigma,i_f}B\Psi^a_{4,\sigma,i_f}\bar{\Psi}^b_{3,\sigma',i_f}A\Psi^b_{2,\sigma',i_f}+(a\leftrightarrow b)\Big]\delta(\mathbf{k}_1+\mathbf{k}_3-\mathbf{k}_2-\mathbf{k}_4)\\
	S_{umk2,\theta_2}^d&=-\frac{\Xi_{\theta_2}^d}{2}\sum_{a,b=1}^R\sum_{\sigma,\sigma'=1}^{N_c}\int d\tilde{k}\Big[\bar{\Psi}^a_{1,\sigma,i_f}A\Psi^a_{3,\sigma,i_f}\bar{\Psi}^b_{4,\sigma',i_f}A\Psi^b_{2,\sigma',i_f}+\bar{\Psi}^a_{1,\sigma,i_f}B\Psi^a_{3,\sigma,i_f}\bar{\Psi}^b_{4,\sigma',i_f}B\Psi^b_{2,\sigma',i_f}\\
	&+\bar{\Psi}^a_{3,\sigma,i_f}A\Psi^a_{1,\sigma,i_f}\bar{\Psi}^b_{2,\sigma',i_f}A\Psi^b_{4,\sigma',i_f}+\bar{\Psi}^a_{3,\sigma,i_f}B\Psi^a_{1,\sigma,i_f}\bar{\Psi}^b_{2,\sigma',i_f}B\Psi^b_{4,\sigma',i_f}-\bar{\Psi}^a_{2,\sigma,i_f}B\Psi^a_{4,\sigma,i_f}\bar{\Psi}^b_{3,\sigma',i_f}A\Psi^b_{1,\sigma',i_f}\\
	&-\bar{\Psi}^a_{2,\sigma,i_f}A\Psi^a_{4,\sigma,i_f}\bar{\Psi}^b_{3,\sigma',i_f}B\Psi^b_{1,\sigma',i_f}-\bar{\Psi}^a_{4,\sigma,i_f}B\Psi^a_{2,\sigma,i_f}\bar{\Psi}^b_{1,\sigma',i_f}A\Psi^b_{3,\sigma',i_f}-\bar{\Psi}^a_{4,\sigma,i_f}A\Psi^a_{2,\sigma,i_f}\bar{\Psi}^b_{1,\sigma',i_f}B\Psi^b_{3,\sigma',i_f}\nonumber\\
	&+(a\leftrightarrow b)\Big]\delta(\mathbf{k}_1+\mathbf{k}_3-\mathbf{k}_2-\mathbf{k}_4)\\
	S_{umk2,\theta_2}^e&=-\frac{\Xi_{\theta_2}^e}{2}\sum_{a,b=1}^R\sum_{\sigma,\sigma'=1}^{N_c}\int d\tilde{k}\Big[\bar{\Psi}^a_{1,\sigma,i_f}A\Psi^a_{2,\sigma,i_f}\bar{\Psi}^b_{4,\sigma',i_f}A\Psi^b_{3,\sigma',i_f}+\bar{\Psi}^a_{1,\sigma,i_f}B\Psi^a_{2,\sigma,i_f}\bar{\Psi}^b_{4,\sigma',i_f}B\Psi^b_{3,\sigma',i_f}\\
	&+\bar{\Psi}^a_{3,\sigma,i_f}A\Psi^a_{4,\sigma,i_f}\bar{\Psi}^b_{2,\sigma',i_f}A\Psi^b_{1,\sigma',i_f}+\bar{\Psi}^a_{3,\sigma,i_f}B\Psi^a_{4,\sigma,i_f}\bar{\Psi}^b_{2,\sigma',i_f}B\Psi^b_{1,\sigma',i_f}+\bar{\Psi}^a_{2,\sigma,i_f}iBA\Psi^a_{1,\sigma,i_f}\bar{\Psi}^b_{3,\sigma',i_f}iAB\Psi^b_{4,\sigma',i_f}\\
	&+\bar{\Psi}^a_{2,\sigma,i_f}iAB\Psi^a_{1,\sigma,i_f}\bar{\Psi}^b_{3,\sigma',i_f}iBA\Psi^b_{4,\sigma',i_f}+\bar{\Psi}^a_{4,\sigma,i_f}iBA\Psi^a_{3,\sigma,i_f}\bar{\Psi}^b_{1,\sigma',i_f}iAB\Psi^b_{2,\sigma',i_f}\nonumber\\
	&+\bar{\Psi}^a_{4,\sigma,i_f}iAB\Psi^a_{3,\sigma,i_f}\bar{\Psi}^b_{1,\sigma',i_f}iBA\Psi^b_{2,\sigma',i_f}+(a\leftrightarrow b)\Big]\delta(\mathbf{k}_1+\mathbf{k}_3-\mathbf{k}_2-\mathbf{k}_4)
\end{align*}
\begin{align*}
	S_{umk2,\pi/2}^d&=-\frac{\Xi_{\pi/2}^d}{2}\sum_{a,b=1}^R\sum_{\sigma,\sigma'=1}^{N_c}\int d\tilde{k}\Big[-\bar{\Psi}^a_{1,\sigma,i_f}A\Psi^a_{3,\sigma,i_f}\bar{\Psi}^b_{2,\sigma',i_f}B\Psi^b_{4,\sigma',i_f}-\bar{\Psi}^a_{1,\sigma,i_f}B\Psi^a_{3,\sigma,i_f}\bar{\Psi}^b_{2,\sigma',i_f}A\Psi^b_{4,\sigma',i_f}\\
	&-\bar{\Psi}^a_{3,\sigma,i_f}A\Psi^a_{1,\sigma,i_f}\bar{\Psi}^b_{4,\sigma',i_f}B\Psi^b_{2,\sigma',i_f}-\bar{\Psi}^a_{3,\sigma,i_f}B\Psi^a_{1,\sigma,i_f}\bar{\Psi}^b_{4,\sigma',i_f}A\Psi^b_{2,\sigma',i_f}+\bar{\Psi}^a_{2,\sigma,i_f}B\Psi^a_{4,\sigma,i_f}\bar{\Psi}^b_{1,\sigma',i_f}B\Psi^b_{3,\sigma',i_f}\\
	&+\bar{\Psi}^a_{2,\sigma,i_f}A\Psi^a_{4,\sigma,i_f}\bar{\Psi}^b_{1,\sigma',i_f}A\Psi^b_{3,\sigma',i_f}+\bar{\Psi}^a_{4,\sigma,i_f}B\Psi^a_{2,\sigma,i_f}\bar{\Psi}^b_{3,\sigma',i_f}B\Psi^b_{1,\sigma',i_f}+\bar{\Psi}^a_{4,\sigma,i_f}A\Psi^a_{2,\sigma,i_f}\bar{\Psi}^b_{3,\sigma',i_f}A\Psi^b_{1,\sigma',i_f}\nonumber\\
	&+(a\leftrightarrow b)\Big]\delta(\mathbf{k}_1+\mathbf{k}_3-\mathbf{k}_2-\mathbf{k}_4)\\
	S_{umk2,\pi/2}^e&=-\frac{\Xi_{\pi/2}^e}{2}\sum_{a,b=1}^R\sum_{\sigma,\sigma'=1}^{N_c}\int d\tilde{k}\Big[\bar{\Psi}^a_{1,\sigma,i_f}iAB\Psi^a_{4,\sigma,i_f}\bar{\Psi}^b_{2,\sigma',i_f}iBA\Psi^b_{3,\sigma',i_f}+\bar{\Psi}^a_{1,\sigma,i_f}iBA\Psi^a_{4,\sigma,i_f}\bar{\Psi}^b_{2,\sigma',i_f}iAB\Psi^b_{3,\sigma',i_f}\\
	&+\bar{\Psi}^a_{3,\sigma,i_f}iAB\Psi^a_{2,\sigma,i_f}\bar{\Psi}^b_{4,\sigma',i_f}iBA\Psi^b_{1,\sigma',i_f}+\bar{\Psi}^a_{3,\sigma,i_f}iBA\Psi^a_{2,\sigma,i_f}\bar{\Psi}^b_{4,\sigma',i_f}iAB\Psi^b_{1,\sigma',i_f}\nonumber\\
	&+\bar{\Psi}^a_{2,\sigma,i_f}B\Psi^a_{3,\sigma,i_f}\bar{\Psi}^b_{1,\sigma',i_f}B\Psi^b_{4,\sigma',i_f}+\bar{\Psi}^a_{2,\sigma,i_f}A\Psi^a_{3,\sigma,i_f}\bar{\Psi}^b_{1,\sigma',i_f}A\Psi^b_{4,\sigma',i_f}+\bar{\Psi}^a_{4,\sigma,i_f}B\Psi^a_{1,\sigma,i_f}\bar{\Psi}^b_{3,\sigma',i_f}B\Psi^b_{2,\sigma',i_f}\nonumber\\
	&+\bar{\Psi}^a_{4,\sigma,i_f}A\Psi^a_{1,\sigma,i_f}\bar{\Psi}^b_{3,\sigma',i_f}A\Psi^b_{2,\sigma',i_f}+(a\leftrightarrow b)\Big]\delta(\mathbf{k}_1+\mathbf{k}_3-\mathbf{k}_2-\mathbf{k}_4)
\end{align*}

	\section{Calculations of two-loop Feynman diagrams (random charge potential vertices only)} \label{Appendix:CalOfTwoLoopFeynmanDiagrams}

In this paper, we consider two-loop corrections composed of only the random charge potential vertices, where disorder coupling constants are mostly fast growing, compared to $g$ and $\lambda$. In addition, it is technically too complicated to consider all two-loop Feynman diagrams. We recommend the textbook \cite{CriticalPhi4} for the calculations of two-loop Feynman diagrams.

\subsection{Useful techniques for calculations}

\subsubsection{$\alpha_{nm}$ and $\beta_{nm}$}

To calculate two-loop Feynman diagrams, we use variables $\epsilon_n(k)$ and $\epsilon_{n}^{||}(k)$ instead of $k_{d-1}$, $k_d$. Since there are two sets of $\epsilon$ variables in two-loop calculations, we introduce how $\epsilon_n(k)$ can be expressed in terms of $\epsilon_m(k)$ and $\epsilon_{m}^{||}(k)$ first. 

We consider the following dispersion relations:
\begin{gather*}
	\epsilon_{1}(k)=vk_{d-1}+k_d,\; \epsilon_{1}^{||}(k)=k_{d-1}-vk_d,\;\; \epsilon_{2}(k)=vk_{d}-k_{d-1},\; \epsilon_{2}^{||}(k)=k_d+vk_{d-1},\\
	\epsilon_{3}(k)=vk_{d-1}-k_d,\; \epsilon_{3}^{||}(k)=k_{d-1}+vk_d,\;\; \epsilon_4(k)=vk_d+k_{d-1},\; \epsilon_{4}^{||}(k)=k_d-vk_{d-1}.
\end{gather*}

Setting $\epsilon_{n}(k)=\alpha_{nm}\epsilon_{m}(k)+\beta_{nm}\epsilon_{m}^{||}(k)$, we obtain
\begin{gather*}
	\alpha_{nn}=1,\; \beta_{nn}=0,\; \alpha_{12}=\alpha_{21}=0,\; \beta_{12}=-\beta_{21}=1,\\
	\alpha_{13}=\alpha_{31}=-\frac{1-v^2}{1+v^2},\; \beta_{13}=\beta_{31}=\frac{2v}{1+v^2},\; \alpha_{14}=\alpha_{41}=\frac{2v}{1+v^2},\; \beta_{14}=\beta_{41}=\frac{1-v^2}{1+v^2},\\
	\alpha_{23}=\alpha_{32}=-\frac{2v}{1+v^2},\; \beta_{23}=\beta_{32}=-\frac{1-v^2}{1+v^2},\; \alpha_{24}=\alpha_{42}=-\frac{1-v^2}{1+v^2},\; \beta_{24}=\beta_{42}=\frac{2v}{1+v^2},\\
	\alpha_{34}=\alpha_{43}=0,\; \beta_{34}=-\beta_{43}=-1.
\end{gather*}

\subsubsection{Extraction of $\epsilon$-pole from integration of Feynman parameters} \label{sec:ExtractionOfEpsilonPole}

We need to deal with Feynman parameters to extract out the $\frac{1}{\epsilon}$ pole. Some $\frac{1}{\epsilon}$-poles are hidden in the integration of Feynman parameters, which can be trickier to deal with, compared to the one-loop analysis. Here, we introduce a systematic way to extract out the $\frac{1}{\epsilon}$-pole from the integral of Feynman parameters.

Consider two functions, $f(x,\epsilon)$ and $g(x,\epsilon)$, where $x$ is a Feynman parameter and $\epsilon$ is a tuning parameter of dimension. The function $f(x,\epsilon)$ has a simple or higher order pole at $x=x_0$ in the limit of $\epsilon \rightarrow 0$. On the other hand, the function $g(x,\epsilon)$ does not have any poles in the regime $0<x<1$ and in the limit of $\epsilon \rightarrow 0$. 

Suppose $f(x,\epsilon)$ is given by $f(x,\epsilon)\sim (x-x_0)^{-n+\epsilon}$ where $n\geq 1$. Suppose integration of the product of $f(x,\epsilon)$ and $g(x,\epsilon)$ with respect to $x$ as follows: $F(\epsilon)=\int_0^1dxf(x,\epsilon)g(x,\epsilon)$. To extract out $\epsilon$-poles in $F(\epsilon)$, we perform the series expansion of $g(x,\epsilon)$ at $x=x_0$. Then, we obtain
\begin{align}
	F(\epsilon)&=\int_0^1dx f(x,\epsilon)\sum_{l=0}^\infty\frac{g^{(l)}(x_0,\epsilon)}{l!}(x-x_0)^l=\sum_{l=0}^{n-1}g^{(l)}(x_0,\epsilon)\int_0^1dxf(x,\epsilon)(x-x_0)^l+\sum_{l=n}^{\infty}g^{(l)}(x_0,\epsilon) \int_0^1dxf(x,\epsilon)(x-x_0)^l\nonumber\\
	&\equiv F^{(1)}(\epsilon)+F^{(2)}(\epsilon).
\end{align}
The first term $F^{(1)}(\epsilon)$ contains the $\epsilon$-poles while the second term $F^{(2)}(\epsilon)$ does not have $\frac{1}{\epsilon}$-poles. Since the second term does not have any epsilon poles, it can be represented as follows $F^{(2)}(\epsilon)=\sum_{m=0}^{\infty}\frac{F^{(2,m)}(0)}{m!}\epsilon^m$. As a result, we find $F(\epsilon)=F^{(1)}(\epsilon)+F^{(2)}(0)+\mathcal{O}(\epsilon)$. Note that we keep terms up to the zeroth order in $\epsilon$ ($\mathcal{O}(\epsilon^0)$) since there can be another $\frac{1}{\epsilon}$-pole coming from the integration of momentum variables multiplied to the zeroth order term.

For the case of $n=1$, we obtain
\begin{align}
	F(\epsilon)=\int_0^1dxf(x,\epsilon)g(x_0,\epsilon)+\int_0^1dxf(x,0)[g(x,0)-g(x_0,0)]+\mathcal{O}(\epsilon) ,
\end{align}
where we used the fact that $g(x,0)\approx g(x_0,0)+g^{(1)}(x_0)(x-x_0)+\cdots\Rightarrow \sum_{l=1}^{\infty}g^{(l)}(x_0)(x-x_0)^l=g(x,0)-g(x_0,0)$.

\subsubsection{Useful integrations}

For two-loop Feynman diagrams composed of only random charge potential vertices, all integrations have the same form of the integration introduced below:
\begin{align}
	(*)=&\int\frac{d^{d-2}\mathbf{K}_\perp}{(2\pi)^{d-2}}\int\frac{d^{d-2}\mathbf{P}_\perp}{(2\pi)^{d-2}}\int\frac{d^2k}{(2\pi)^2}\int\frac{d^2p}{(2\pi)^2}G_{m_1,\sigma_1}^{f(1)}(\omega,k)G_{m_2,\sigma_2}^{f(2)}(\omega,p)G_{m_3,\sigma_3}^{f(3)}(\omega,ap)G_{m_4,\sigma}^{f(4)}(\omega,bk+cp)\nonumber\\
	&=\int\frac{d^{d-2}\mathbf{K}_\perp}{(2\pi)^{d-2}}\int\frac{d^{d-2}\mathbf{P}_\perp}{(2\pi)^{d-2}}\int\frac{d^2k}{(2\pi)^2}\int\frac{d^2p}{(2\pi)^2}\frac{\gamma_0^{(1)}\omega+\mathbf{\Gamma}^{(1)}_\perp\cdot\mathbf{K}_\perp+\gamma^{(1)}_{d-1}\epsilon_{m_1}(k)}{\omega^2+|\mathbf{K}_\perp|^2+\epsilon_{m_1}^2(k)}\frac{\gamma_0^{(2)}\omega+\mathbf{\Gamma}^{(2)}_\perp\cdot\mathbf{P}_\perp+\gamma^{(2)}_{d-1}\epsilon_{m_2}(p)}{\omega^2+|\mathbf{P}_\perp|^2+\epsilon^2_{m_2}(p)}\nonumber\\
	&\times\frac{\gamma_0^{(3)}\omega+a\mathbf{\Gamma}^{(3)}_\perp\cdot\mathbf{P}_\perp+a\gamma^{(3)}_{d-1}\epsilon_{m_3}(p)}{\omega^2+|\mathbf{P}_\perp|^2+\epsilon_{m_3}^2(p)} \frac{\gamma_0^{(4)}\omega+\mathbf{\Gamma}^{(4)}_\perp\cdot(b\mathbf{K}_\perp+c\mathbf{P}_\perp)+\gamma^{(4)}_{d-1}[b\epsilon_{m_4}(k)+c\epsilon_{m_4}(p)]}{\omega^2+|b\mathbf{K}_\perp+c\mathbf{P}_\perp|^2+[b\epsilon_{m_4}(k)+c\epsilon_{m_4}(p)]^2}\nonumber\\
	&=\frac{1}{(1+v^2)^2}\int\frac{d^{d-2}\mathbf{K}_\perp}{(2\pi)^{d-2}}\int\frac{d^{d-2}\mathbf{P}_\perp}{(2\pi)^{d-2}}\int\frac{d\epsilon_1}{2\pi}\int_{-\Lambda_{FS}}^{\Lambda_{FS}}\frac{d\epsilon_1^{||}}{2\pi}\int\frac{d\epsilon_2}{2\pi}\int_{-\Lambda_{FS}}^{\Lambda_{FS}}\frac{d\epsilon_2^{||}}{2\pi}\frac{\gamma_0^{(1)}\omega+\mathbf{\Gamma}^{(1)}_\perp\cdot\mathbf{K}_\perp+\gamma^{(1)}_{d-1}\epsilon_1}{\omega^2+|\mathbf{K}_\perp|^2+\epsilon_1^2}\nonumber\\
	&\times \frac{\gamma_0^{(2)}\omega+\mathbf{\Gamma}^{(2)}_\perp\cdot\mathbf{P}_\perp+\gamma^{(2)}_{d-1}\epsilon_2}{\omega^2+|\mathbf{P}_\perp|^2+\epsilon_2^2} \frac{\gamma_0^{(3)}\omega+a\mathbf{\Gamma}^{(3)}_\perp\cdot\mathbf{P}_\perp+a\gamma^{(3)}_{d-1}[\alpha_{m_3m_2}\epsilon_2+\beta_{m_3m_2}\epsilon_2^{||}]}{\omega^2+|\mathbf{P}_\perp|^2+[\alpha_{m_3m_2}\epsilon_2+\beta_{m_3m_2}\epsilon_2^{||}]^2}\nonumber\\
	&\times \frac{\gamma_0^{(4)}\omega+\mathbf{\Gamma}^{(4)}_\perp\cdot(b\mathbf{K}_\perp+c\mathbf{P}_\perp) +\gamma^{(4)}_{d-1}[b(\alpha_{m_4m_1}\epsilon_1+\beta_{m_4m_1}\epsilon_1^{||})+c(\alpha_{m_4m_2}\epsilon_2 +\beta_{m_4m_2}\epsilon_2^{||})]}{\omega^2+|b\mathbf{K}_\perp+c\mathbf{P}_\perp|^2+[b(\alpha_{m_4m_1}\epsilon_1 +\beta_{m_4m_1}\epsilon_1^{||})+c(\alpha_{m_4m_2}\epsilon_2+\beta_{m_4m_2}\epsilon_2^{||})]^2} ,
\end{align}
where $a^2=b^2=c^2=1$. Based on the dimensional counting or phase space argument or explicit calculations, we can check out that only when $\beta_{m_3m_2}=\beta_{m_4m_1}=\beta_{m_4m_2}=0$, we get a $\frac{1}{\epsilon}$-pole. This is similar to the case of the one-loop analysis . Here, we also consider $v=0$ ($\alpha_{m_3m_2}^2=\alpha_{m_4m_1}^2=\alpha_{m_4m_2}^2=1$) and introduce the $e^{-v^2/v_c^2}$ factor as we did in the one-loop calculations. 

As a result, the integration is reduced to as follows:
\begin{align}
	(*)&=\Big(\frac{\Lambda_{FS}}{\pi}\Big)^2\frac{ab \mathcal{C}_{m_1m_2m_3m_4}(v)}{(4\pi)^{d-1}}\frac{1}{4}\omega^{-2(3-d)}\Gamma(3-d)\int_0^1dx\int_0^1dy\frac{[x(1-x)]^{\frac{3-d}{2}}}{[y+(1-y)x(1-x)]^{3-d}}\Bigg[y^{\frac{1-d}{2}}(1-y)\Bigg(\Gamma_{\perp,i}^{(1)}\Gamma_{\perp,j}^{(2)}\Gamma_{\perp,j}^{(3)}\Gamma_{\perp,i}^{(4)}\nonumber\\
	&+\alpha_{m_3m_2}\Gamma_{\perp,i}^{(1)}\gamma_{d-1}^{(2)}\gamma_{d-1}^{(3)}\Gamma_{\perp,i}^{(4)}+\alpha_{m_4m_1}\gamma_{d-1}^{(1)}\Gamma_{\perp,j}^{(2)}\Gamma_{\perp,j}^{(3)}\gamma_{d-1}^{(4)}+\alpha_{m_3m_2}\alpha_{m_4m_1}\gamma_{d-1}^{(1)}\gamma_{d-1}^{(2)}\gamma_{d-1}^{(3)}\gamma_{d-1}^{(4)}\Bigg)\nonumber\\
	&-y^{\frac{3-d}{2}}(1-y)\Bigg(\Gamma_{\perp,i}^{(1)}\Gamma_{\perp,i}^{(2)}\Gamma_{\perp,j}^{(3)}\Gamma_{\perp,j}^{(4)}+\Gamma_{\perp,i}^{(1)}\Gamma_{\perp,j}^{(2)}\Gamma_{\perp,i}^{(3)}\Gamma_{\perp,j}^{(4)}+\Gamma_{\perp,i}^{(1)}\Gamma_{\perp,j}^{(2)}\Gamma_{\perp,j}^{(3)}\Gamma_{\perp,i}^{(4)}+3\alpha_{m_3m_2}\alpha_{m_4m_1}\gamma_{d-1}^{(1)}\gamma_{d-1}^{(2)}\gamma_{d-1}^{(3)}\gamma_{d-1}^{(4)}\nonumber\\
	&+\alpha_{m_4m_1}\alpha_{m_4m_2}\gamma_{d-1}^{(1)}\gamma_{d-1}^{(2)}\Gamma_{\perp,i}^{(3)}\Gamma_{\perp,i}^{(4)}+\alpha_{m_3m_2}\alpha_{m_4m_2}\Gamma_{\perp,i}^{(1)} \Gamma_{\perp,i}^{(2)} \gamma_{d-1}^{(3)} \gamma_{d-1}^{(4)}+\alpha_{m_4m_1}\alpha_{m_4m_2}\alpha_{m_3m_2}\gamma_{d-1}^{(1)} \Gamma_{\perp,i}^{(2)}\gamma_{d-1}^{(3)}\Gamma_{\perp,i}^{(4)}\nonumber\\
	&+\alpha_{m_3m_2}\Gamma_{\perp,i}^{(1)}\gamma_{d-1}^{(2)}\gamma_{d-1}^{(3)}\Gamma_{\perp,i}^{(4)}+\alpha_{m_4m_1}\gamma_{d-1}^{(1)}\Gamma_{\perp,i}^{(2)}\Gamma_{\perp,i}^{(3)}\gamma_{d-1}^{(4)}+\alpha_{m_4m_2}\Gamma_{\perp,i}^{(1)}\gamma_{d-1}^{(2)}\Gamma_{\perp,i}^{(3)}\gamma_{d-1}^{(4)}\Bigg)\Bigg]\nonumber\\
	&\approx \Big(\frac{\Lambda_{FS}}{\pi}\Big)^2\frac{ab\mathcal{C}_{m_1m_2m_3m_4}(v)}{(4\pi)^{2}}\frac{1}{4}\Bigg[\Bigg(\frac{2}{\epsilon^2}+\frac{1-2\gamma_E+2\ln\Big(\frac{4\pi}{\omega^2}\Big)}{\epsilon}\Bigg)\Big[\Gamma_{\perp,i}^{(1)}\Gamma_{\perp,j}^{(2)}\Gamma_{\perp,j}^{(3)}\Gamma_{\perp,i}^{(4)}+\alpha_{m_3m_2}\Gamma_{\perp,i}^{(1)}\gamma_{d-1}^{(2)}\gamma_{d-1}^{(3)}\Gamma_{\perp,i}^{(4)}\nonumber\\
	&+\alpha_{m_4m_1}\gamma_{d-1}^{(1)}\Gamma_{\perp,j}^{(2)}\Gamma_{\perp,j}^{(3)}\gamma_{d-1}^{(4)}+\alpha_{m_3m_2}\alpha_{m_4m_1}\gamma_{d-1}^{(1)}\gamma_{d-1}^{(2)}\gamma_{d-1}^{(3)}\gamma_{d-1}^{(4)}\Big]-\Bigg(\frac{1}{2\epsilon}\Bigg)\Big[\Gamma_{\perp,i}^{(1)}\Gamma_{\perp,i}^{(2)}\Gamma_{\perp,j}^{(3)}\Gamma_{\perp,j}^{(4)}+\Gamma_{\perp,i}^{(1)}\Gamma_{\perp,j}^{(2)}\Gamma_{\perp,i}^{(3)}\Gamma_{\perp,j}^{(4)}\nonumber\\
	&+\Gamma_{\perp,i}^{(1)}\Gamma_{\perp,j}^{(2)}\Gamma_{\perp,j}^{(3)}\Gamma_{\perp,i}^{(4)}+3\alpha_{m_3m_2}\alpha_{m_4m_1}\gamma_{d-1}^{(1)}\gamma_{d-1}^{(2)}\gamma_{d-1}^{(3)}\gamma_{d-1}^{(4)}+\alpha_{m_4m_1}\alpha_{m_4m_2}\gamma_{d-1}^{(1)}\gamma_{d-1}^{(2)}\Gamma_{\perp,i}^{(3)}\Gamma_{\perp,i}^{(4)}\nonumber\\
	&+\alpha_{m_3m_2}\alpha_{m_4m_2}\Gamma_{\perp,i}^{(1)} \Gamma_{\perp,i}^{(2)} \gamma_{d-1}^{(3)} \gamma_{d-1}^{(4)}+\alpha_{m_4m_1}\alpha_{m_4m_2}\alpha_{m_3m_2}\gamma_{d-1}^{(1)} \Gamma_{\perp,i}^{(2)}\gamma_{d-1}^{(3)}\Gamma_{\perp,i}^{(4)}+\alpha_{m_3m_2}\Gamma_{\perp,i}^{(1)}\gamma_{d-1}^{(2)}\gamma_{d-1}^{(3)}\Gamma_{\perp,i}^{(4)}\nonumber\\
	&+\alpha_{m_4m_1}\gamma_{d-1}^{(1)}\Gamma_{\perp,i}^{(2)}\Gamma_{\perp,i}^{(3)}\gamma_{d-1}^{(4)} +\alpha_{m_4m_2}\Gamma_{\perp,i}^{(1)}\gamma_{d-1}^{(2)}\Gamma_{\perp,i}^{(3)}\gamma_{d-1}^{(4)}\Big]\Bigg] +\mathcal{O}(\epsilon^0) , \label{eq:TwoLoopUsefulIntegration}
\end{align}
where $\gamma_{E}$ is the Euler constant and
\begin{align}
	\mathcal{C}_{m_1m_2m_3m_4}(v)=\Big\{\begin{array}{ll}\frac{1}{(1+v^2)^2} & \text{when }m_1=m_2=m_3=m_4 \\ e^{-v^2/v_c^2} &\text{when }(m_1=m_2=m_3=m_4)! \; \& \; \lim_{v\rightarrow 0} \beta_{m_im_j}=0\end{array}.
\end{align}

There are two Feynman integrals:
\begin{gather*}
	\int_0^1dx\int_0^1dy \frac{[x(1-x)]^{\frac{3-d}{2}}(1-y)y^{\frac{1-d}{2}}}{[y+(1-y)x(1-x)]^{3-d}} , \;\;\;\;\; \int_0^1dx\int_0^1dy \frac{[x(1-x)]^{\frac{3-d}{2}}(1-y)y^{\frac{3-d}{2}}}{[y+(1-y)x(1-x)]^{3-d}} .
\end{gather*}
Near $d=3$ $(\epsilon=0)$, only the first integral contains divergences because of the term $y^{\frac{1-d}{2}}$. On the other hand, the second integral gives a value $\frac{1}{2}$ at $d=3$. To extract our the hidden $\frac{1}{\epsilon}$-pole from the integrals of Feynman parameters in the first integral, we use the method introduced in section \ref{sec:ExtractionOfEpsilonPole} as follows:
\begin{align*}
	F(\epsilon)&=\int_0^1dx\int_0^1dy \frac{[x(1-x)]^{\frac{\epsilon }{2}}(1-y)y^{\frac{\epsilon}{2}-1}}{[y+(1-y)x(1-x)]^{\epsilon}}\nonumber\\
	&=\int_0^1dx\int_0^1dy f(y,\epsilon)g(x,y,\epsilon) \; \Big(f(y,\epsilon)=y^{\frac{\epsilon}{2}-1}(1-y),\; g(x,y,\epsilon)=\frac{[x(1-x)]^{\frac{\epsilon }{2}}}{[y+(1-y)x(1-x)]^{\epsilon}}\Big)\nonumber\\
	&=\int_0^1dx\int_0^1dy f(y,\epsilon)g(x,y=0,\epsilon)+\int_0^1dx\int_0^1dyf(y,0)[g(x,y,0)-g(x,y=0,0)]+\mathcal{O}(\epsilon)\nonumber\\
	&=\int_0^1dx[x(1-x)]^{-\epsilon/2}\int_0^1y^{\frac{\epsilon}{2}-1}(1-y)+\mathcal{O}(\epsilon)=\frac{[\Gamma(1-\epsilon/2)]^2}{\Gamma(2-\epsilon)}\frac{\Gamma(\epsilon/2)}{\Gamma(2+\epsilon/2)}+\mathcal{O}(\epsilon) ,
\end{align*}
where $\int_0^1 x^{\alpha-1}(1-x)^{\beta-1}=\frac{\Gamma(\alpha)\Gamma(\beta)}{\Gamma(\alpha+\beta)}$ has been used.

\subsection{Two-loop fermion self-energy corrections}

\begin{figure}[h]
	\begin{subfigure}{0.4\textwidth}
		\begin{tikzpicture}[baseline=-0.1cm]
			\begin{feynhand}
				\vertex (c1) at (0,0); \vertex (c2) at (4,0);
				\vertex (a1) at (0.7,0); \vertex (a2) at (3.3,0);
				\vertex (b1) at (1.4,0); \vertex (b2) at (2.6,0);
				\propag[fer] (c1) to (a1); \propag[fer] (a1) to [edge label=$n_1$](b1); \propag[fer] (b1) to [edge label=$n_2$](b2); \propag[fer] (b2) to [edge label=$n_3$](a2); \propag[fer] (a2) to (c2);
				\propag[sca] (a1) to [out=90, in=90](b2); \propag[sca] (b1) to [out=90, in=90](a2);
				\node at (0.7, -0.2) {\tiny $\mathcal{M}^i$}; \node at (3.3,-0.2) {\tiny $\tilde{\mathcal{M}}^j$};
				\node at (1.4, -0.2) {\tiny $\mathcal{M}^j$}; \node at (2.6,-0.2) {\tiny $\tilde{\mathcal{M}}^i$};
			\end{feynhand}
		\end{tikzpicture}
		\caption{}
	\end{subfigure}
	~
	\begin{subfigure}{0.4\textwidth}
		\begin{tikzpicture}[baseline=-0.1cm]
			\begin{feynhand}
				\vertex (c1) at (0,0); \vertex (c2) at (4,0);
				\vertex (a1) at (0.7,0); \vertex (a2) at (3.3,0);
				\vertex[crossdot] (b) at (2,1) {};
				\propag[fer] (c1) to (a1); \propag[fer] (a1) to (a2); \propag[fer] (a2) to (c2);
				\propag[sca] (a1) to [out=90, in=180](b); \propag[sca] (b) to [out=0, in=90](a2);
			\end{feynhand}
		\end{tikzpicture}
		\caption{}
	\end{subfigure}
	\caption{(a) Two-loop fermion self-energy corrections composed of only random charge potential vertices and (b) Counter one-loop diagram.}
\end{figure}
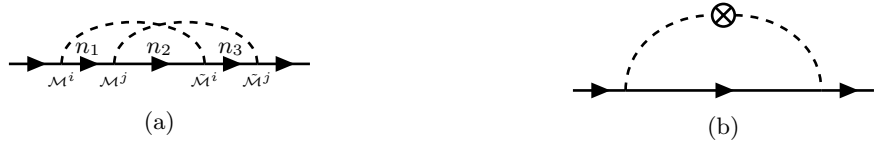

The two-loop fermion self-energy is given by
\begin{align}
	\Sigma_f^{(a)}(\omega)&=\frac{\Gamma_i\Gamma_j\mu^{2(-1+\epsilon)}}{N_f^3}\int\frac{d^2\mathbf{k}}{(2\pi)^2}\int\frac{d^2\mathbf{p}}{(2\pi)^2}\int\frac{d^{d-2}\mathbf{K}_\perp}{(2\pi)^{d-2}}\int\frac{d^{d-2}\mathbf{P}_\perp}{(2\pi)^{d-2}}\mathcal{M}^iG_{n_1}(\omega, k)\mathcal{M}^jG_{n_2}(\omega,p)\tilde{\mathcal{M}}^iG_{n_3}(\omega,p-k)\tilde{\mathcal{M}}^j\nonumber\\
	&=i\frac{\Gamma_i\Gamma_j\mu^{2(-1+\epsilon)}}{N_f^3}\frac{1}{(1+v^2)^2}\int\frac{d\epsilon_1 d\epsilon_{1}^{||}}{(2\pi)^2}\int\frac{d\epsilon_2 d\epsilon_2^{||}}{(2\pi)^2}\int\frac{d^{d-2}\mathbf{K}_\perp}{(2\pi)^{d-2}}\int\frac{d^{d-2}\mathbf{P}_\perp}{(2\pi)^{d-2}}\frac{\gamma_0^{(1)}\omega+\mathbf{\Gamma}_\perp^{(1)}\cdot\mathbf{K}_\perp+\gamma_{d-1}^{(1)}\epsilon_1} {\omega^2+|\mathbf{K}_\perp|^2+\epsilon_1^2}\nonumber\\
	&\times \frac{\gamma_0^{(2)}\omega+\mathbf{\Gamma}_\perp^{(2)}\cdot\mathbf{P}_\perp+\gamma_{d-1}^{(2)}\epsilon_2}{\omega^2+|\mathbf{P}_\perp|^2+\epsilon_2^2} \frac{\gamma_0^{(3)}\omega+\mathbf{\Gamma}_\perp^{(3)}\cdot(\mathbf{P}_\perp-\mathbf{K}_\perp)+\gamma_{d-1}^{(3)} [\alpha_{n_3n_2}\epsilon_2+\beta_{n_3n_2}\epsilon_2^{||}-\alpha_{n_3n_1}\epsilon_1-\beta-{n_3n_1}\epsilon_1^{||}]}{\omega^2+|\mathbf{P}_\perp -\mathbf{K}_\perp|^2+[\alpha_{n_3n_2}\epsilon_2+\beta_{n_3n_2}\epsilon_2^{||}-\alpha_{n_3n_1}\epsilon_1-\beta_{n_3n_1}\epsilon_1^{||}]^2} . \nonumber\\
\end{align}

According to the dimensional counting or phase space arguments, there is an epsilon pole only when $\beta_{n_3n_2}=\beta_{n_3n_1}=0$. As a result, there is finite two-loop corrections only for the $n_1=n_2=n_3$ case. However, if consider the $v=0$ limit, cases with $\lim_{v\rightarrow 0}\beta_{nm}=0$ also give finite corrections, which we sincerely discussed in our one-loop calculations. We consider both cases here and introduce the $e^{-v^2/v_c^2}$ factor in the second case as we did in the one-loop calculation. We find	
\begin{align}
	\Sigma_f^{(a)}(\omega)&=i\frac{\Gamma_i\Gamma_j\mu^{2\epsilon}}{N_f^2}\mathcal{C}_{n_1n_2n_3}(v)\Big(\frac{\Lambda_{FS}}{\pi}\Big)^2\int\frac{d\epsilon_1}{2\pi}\int\frac{d\epsilon_2}{2\pi}\int\frac{d^{d-2}\mathbf{K}_\perp}{(2\pi)^{d-2}}\int\frac{d^{d-2}\mathbf{P}_\perp}{(2\pi)^{d-2}}\mathcal{M}^i\frac{\gamma_0\omega+\mathbf{\Gamma}_\perp\cdot\mathbf{K}_\perp+\gamma_{d-1}\epsilon_1}{\omega^2+|\mathbf{K}_\perp|^2+\epsilon_1^2} \mathcal{M}^j\nonumber\\
	&\times\frac{\gamma_0\omega+\mathbf{\Gamma}_\perp\cdot\mathbf{P}_\perp+\gamma_{d-1}\epsilon_2}{\omega^2+|\mathbf{P}_\perp|^2+\epsilon_2^2}\tilde{\mathcal{M}}^i\frac{\gamma_0\omega+\mathbf{\Gamma}_\perp\cdot(\mathbf{P}_\perp-\mathbf{K}_\perp)+\gamma_{d-1}[\alpha_{n_3n_2}\epsilon_2-\alpha_{n_3n_1}\epsilon_1]}{\omega^2+|\mathbf{P}_\perp-\mathbf{K}_\perp|^2+[\alpha_{n_3n_2}\epsilon_2-\alpha_{n_3n_1}\epsilon_1]^2}\tilde{\mathcal{M}}^j\nonumber\\
	&=i\frac{\Gamma_i\Gamma_j}{N_f^2}\mathcal{C}_{n_1n_2n_3}(v)\Big(\frac{\Lambda_{FS}}{\pi}\Big)^2\frac{1}{(4\pi)^2}\frac{\omega}{2}\Gamma(\epsilon)\Big(\frac{4\pi\mu^2}{\omega^2}\Big)^\epsilon \int_0^1dx\int_0^1\Bigg[ dy [x(1-x)]^{-\frac{\epsilon}{2}}y^{-1+\frac{\epsilon}{2}}\frac{[x(1-x)]^{\epsilon}}{[y+(1-y)x(1-x)]^{\epsilon}}\nonumber\\
	&\times \Big(\mathcal{M}^i\gamma_0\mathcal{M}^j\Gamma_{\perp,i}\tilde{\mathcal{M}}^i\Gamma_{\perp,i}\tilde{\mathcal{M}}^j+\alpha_{n_3n_2}\mathcal{M}^i\gamma_0\mathcal{M}^j\gamma_{d-1}\tilde{\mathcal{M}}^i\gamma_{d-1}\tilde{\mathcal{M}}^j\Big)+[x(1-x)]^{-1-\frac{\epsilon}{2}}y^{\epsilon/2}\frac{[x(1-x)]^{\epsilon}}{[y+(1-y)x(1-x)]^\epsilon}\nonumber\\
	&\times \Big[-x(1-x)\Big(\mathcal{M}^i\gamma_0\mathcal{M}^j\Gamma_{\perp,i}\tilde{\mathcal{M}}^i\Gamma_{\perp,i}\tilde{\mathcal{M}}^j+\alpha_{n_3n_2}\mathcal{M}^i\gamma_0\mathcal{M}^j\gamma_{d-1}\tilde{\mathcal{M}}^i\gamma_{d-1}\tilde{\mathcal{M}}^j\Big)+x\Big(-\mathcal{M}^i\Gamma_{\perp,i}\mathcal{M}^j\gamma_0\tilde{\mathcal{M}}^i\Gamma_{\perp,i}\tilde{\mathcal{M}}^j\nonumber\\
	&-\alpha_{n_3n_1}\mathcal{M}^i\gamma_{d-1}\mathcal{M}^j\gamma_0\tilde{\mathcal{M}}^i\gamma_{d-1}\tilde{\mathcal{M}}^j+\mathcal{M}^i\Gamma_{\perp,i}\mathcal{M}^j\Gamma_{\perp,i}\tilde{\mathcal{M}}^i\gamma_0\tilde{\mathcal{M}}^j+\alpha_{n_3n_1}\alpha_{n_3n_2}\mathcal{M}^i\gamma_{d-1}\mathcal{M}^j\gamma_{d-1}\gamma_0\Big)\Big]\Bigg]\nonumber\\
	&= i\frac{\Gamma_i\Gamma_j}{N_f^2}\mathcal{C}_{n_1n_2n_3}(v) \Big(\frac{\Lambda_{FS}}{\pi}\Big)^2\frac{1}{(4\pi)^2}\frac{\omega}{2}\Big(\frac{2}{\epsilon^2}+\frac{1-2\gamma_E+2\ln\Big(\frac{4\pi\mu^2}{\omega^2}\Big)}{\epsilon}\Big)\Big[\mathcal{M}^i\gamma_0\mathcal{M}^j\Gamma_{\perp,i}\tilde{\mathcal{M}}^i\Gamma_{\perp,i}\tilde{\mathcal{M}}^j\nonumber\\
	&+\alpha_{n_3n_2}\mathcal{M}^i\gamma_0\mathcal{M}^j\gamma_{d-1}\tilde{\mathcal{M}}^i\gamma_{d-1}\tilde{\mathcal{M}}^j-\mathcal{M}^i\Gamma_{\perp,i}\mathcal{M}^j\gamma_0\tilde{\mathcal{M}}^i\Gamma_{\perp,i}\tilde{\mathcal{M}}^j-\alpha_{n_3n_1}\mathcal{M}^i\gamma_{d-1}\mathcal{M}^j\gamma_0\tilde{\mathcal{M}}^i\gamma_{d-1}\tilde{\mathcal{M}}^j\nonumber\\
	&+\mathcal{M}^i\Gamma_{\perp,i}\mathcal{M}^j\Gamma_{\perp,i}\gamma_0\tilde{\mathcal{M}}^j+\alpha_{n_3n_1}\alpha_{n_3n_2} \mathcal{M}^i\gamma_{d-1}\mathcal{M}^j\gamma_{d-1}\tilde{\mathcal{M}}^i\gamma_0\tilde{\mathcal{M}}^j\Big] ,
\end{align}
where $\mathcal{C}_{n_1n_2n_3}(v)=\Big\{\begin{array}{ll}\frac{1}{(1+v^2)^2} & \text{when }n_1=n_2=n_3 \\ e^{-v^2/v_c^2} &\text{when }(n_1=n_2=n_3)! \; \& \; \lim_{v\rightarrow 0} \beta_{n_in_j}=0\end{array}
$ and
\begin{align}
	&\Gamma(\epsilon)\int [x(1-x)]^{-\epsilon/2}y^{\epsilon/2}\frac{[x(1-x)]^\epsilon}{[y+(1-y)x(1-x)]^\epsilon}\Big(\frac{4\pi \mu^2}{\omega^2}\Big)^\epsilon\approx \frac{1}{\epsilon}+\mathcal{O}(\epsilon^0)\\
	&\Gamma(\epsilon)\int x[x(1-x)]^{-1-\epsilon/2}y^{\epsilon/2}\frac{[x(1-x)]^\epsilon}{[y+(1-y)x(1-x)]^\epsilon}\Big(\frac{4\pi \mu^2}{\omega^2}\Big)^\epsilon\approx \frac{2}{\epsilon^2}+\frac{1-2\gamma_E+2\ln\Big(\frac{4\pi\mu^2}{\omega^2}\Big)}{\epsilon}+\mathcal{O}(\epsilon^0)
\end{align}
have been used.

In the case of the one-loop counter diagram (b), the procedure is the same as that of the usual one-loop calculation, given by
\begin{align}
	\Sigma_{f,C}^{(b)}(\omega)&=-\frac{i\Gamma_i\Gamma_j}{(4\pi)^2N_f^2}\mathcal{C}_{n_1n_2n_3}(v)\Big(\frac{\Lambda_{FS}}{\pi}\Big)^2\frac{\omega }{2}\Big[\frac{2}{\epsilon^2}-\frac{\gamma_E}{\epsilon}+\frac{\ln\Big(\frac{4\pi}{\omega^2}\Big)}{\epsilon}\Big]\times\Big[\mathcal{M}^i\gamma_0\mathcal{M}^j\Gamma_{\perp,i}\tilde{\mathcal{M}}^i\Gamma_{\perp,i}\tilde{\mathcal{M}}^j\nonumber\\
	&+\alpha_{n_3n_2}\mathcal{M}^i\gamma_0\mathcal{M}^j\gamma_{d-1}\tilde{\mathcal{M}}^i\gamma_{d-1}\tilde{\mathcal{M}}^j-\mathcal{M}^i\Gamma_{\perp,i}\mathcal{M}^j\gamma_0\tilde{\mathcal{M}}^i\Gamma_{\perp,i}\tilde{\mathcal{M}}^j-\alpha_{n_3n_1}\mathcal{M}^i\gamma_{d-1}\mathcal{M}^j\gamma_0\tilde{\mathcal{M}}^i\gamma_{d-1}\tilde{\mathcal{M}}^j\nonumber\\
	&+\mathcal{M}^i\Gamma_{\perp,i}\mathcal{M}^j\Gamma_{\perp,i}\tilde{\mathcal{M}}^i\gamma_0\tilde{\mathcal{M}}^j+\alpha_{n_3n_1}\alpha_{n_3n_2}\mathcal{M}^i\gamma_{d-1}\mathcal{M}^j\gamma_{d-1}\tilde{\mathcal{M}}^i\gamma_0\tilde{\mathcal{M}}^j\Big]
\end{align}

Summing up results of (a) and (b) give the final answer:
\begin{align}
	\Sigma_f^{(a)}+\Sigma_{f,C}^{(b)}&=i\frac{\Gamma_i\Gamma_j}{N_f^2}\mathcal{C}_{n_1n_2n_3}(v)\Big(\frac{\Lambda_{FS}}{\pi}\Big)^2\frac{\omega}{(4\pi)^2}\Big[-\frac{1}{\epsilon^2}+\frac{1}{2\epsilon}\Big]\Big[\mathcal{M}^i\gamma_0\mathcal{M}^j\Gamma_{\perp,i}\tilde{\mathcal{M}}^i\Gamma_{\perp,i}\tilde{\mathcal{M}}^j+\alpha_{n_3n_2}\mathcal{M}^i\gamma_0\mathcal{M}^j\gamma_{d-1}\tilde{\mathcal{M}}^i\gamma_{d-1}\tilde{\mathcal{M}}^j\nonumber\\
	&-\mathcal{M}^i\Gamma_{\perp,i}\mathcal{M}^j\gamma_0\tilde{\mathcal{M}}^i\Gamma_{\perp,i}\tilde{\mathcal{M}}^j-\alpha_{n_3n_1}\mathcal{M}^i\gamma_{d-1}\mathcal{M}^j\gamma_0\tilde{\mathcal{M}}^i\gamma_{d-1}\tilde{\mathcal{M}}^j+\mathcal{M}^i\Gamma_{\perp,i}\mathcal{M}^j\Gamma_{\perp,i}\tilde{\mathcal{M}}^i\gamma_0\tilde{\mathcal{M}}^j\nonumber\\
	&+\alpha_{n_3n_1}\alpha_{n_3n_2}\mathcal{M}^i\gamma_{d-1}\mathcal{M}^j\gamma_{d-1}\tilde{\mathcal{M}}^i\gamma_0\tilde{\mathcal{M}}^j\Big] .
\end{align}

Note that the non-local divergence ($\frac{\ln\Big(\frac{4\pi\mu^2}{\omega^2}\Big)}{\epsilon}$) in the two-loop calculation disappears due to introduction of the one-loop counter diagram. Only local divergence terms remain.

\newpage
\subsection{Two-loop random charge potential vertex corrections}

\begin{figure}[H]
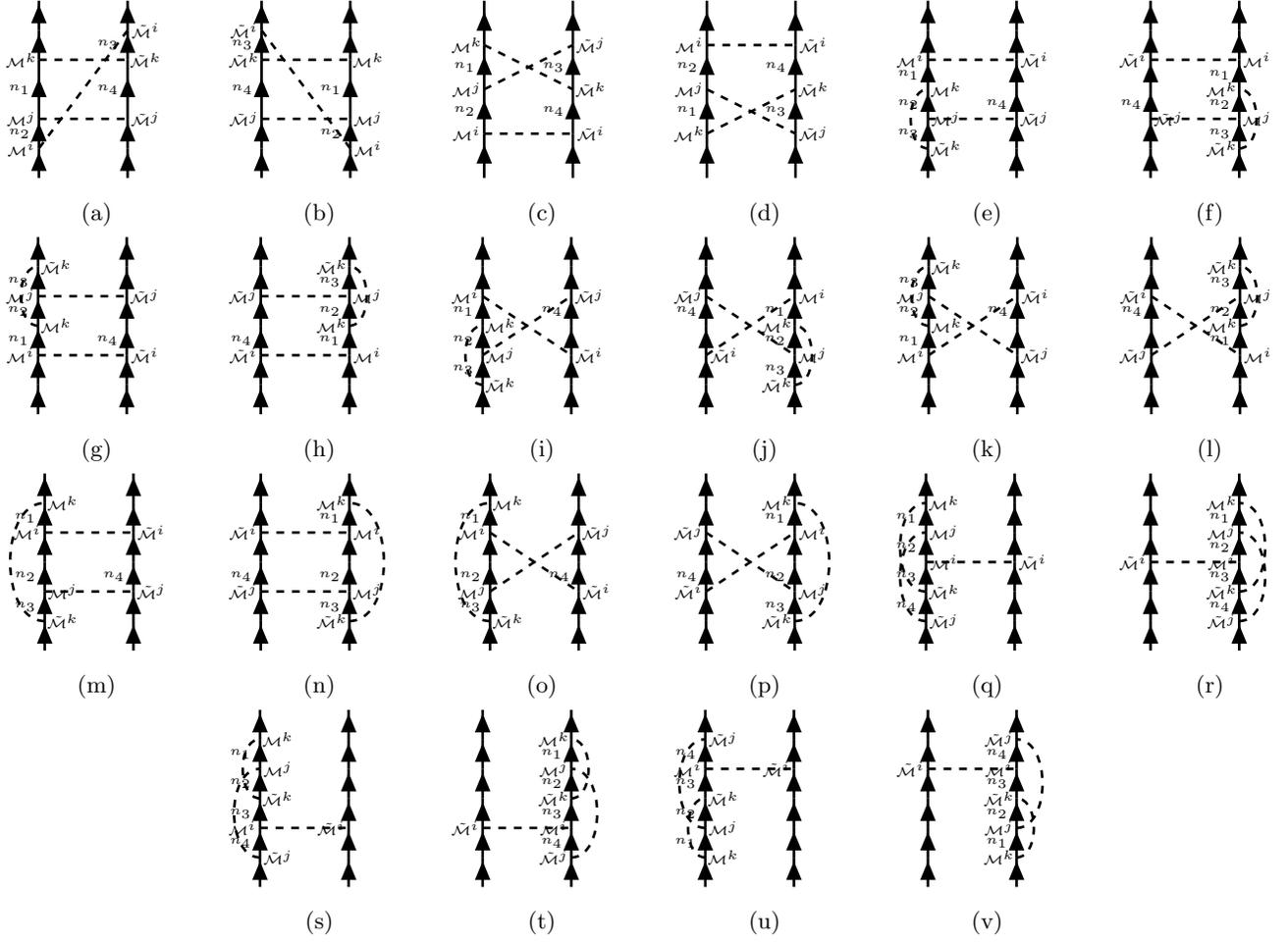

	\centering
	\begin{subfigure}{0.15\textwidth}

		\caption{}
	\end{subfigure}
	~
	\begin{subfigure}{0.15\textwidth}
		%
		\caption{}
	\end{subfigure}
	~
	\begin{subfigure}{0.15\textwidth}
		%
		\caption{}
	\end{subfigure}
	~
	\begin{subfigure}{0.15\textwidth}
		%
		\caption{}
	\end{subfigure}
	\caption{Two loop relevant disorder vertex diagrams}
\end{figure}

\begin{figure}[h]
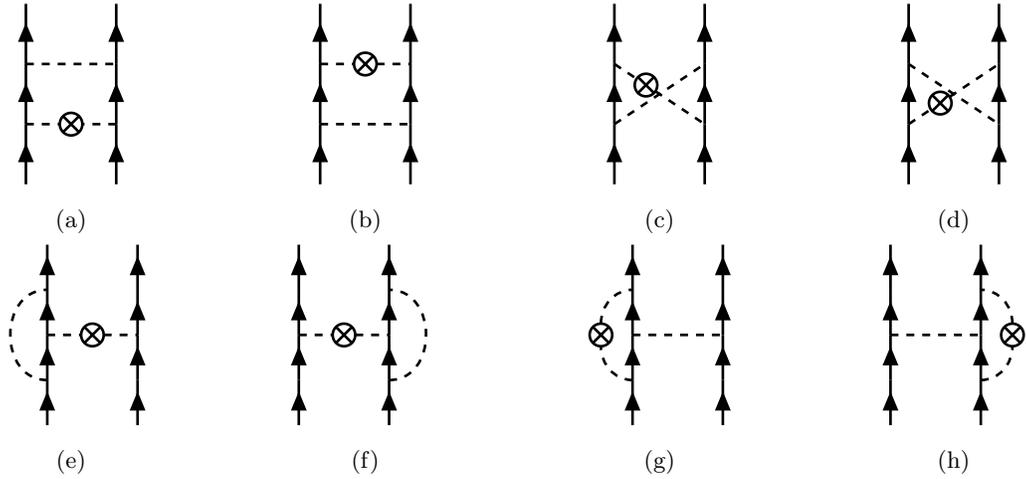

	\begin{subfigure}{0.2\textwidth}
		%
		\caption{}
	\end{subfigure}
	\caption{One loop counter terms for disorder interactions}
\end{figure}

Using the identity Eq. \eqref{eq:TwoLoopUsefulIntegration}, all two-loop random charge potential vertices can be easily calculated. Also, many Feynman diagrams are related to each other by changing the order of the vertex matrix $\mathcal{M}$ or the fermion index $n$. Here, we present only the result of the diagram $(a)$ and its one-loop counter diagram $(a)_{CT}$, given by

\begin{align}
	(a)&=\frac{\Gamma_i\Gamma_j\Gamma_k\mu^{3(-1+\epsilon)}}{N_f^3}\int\frac{d^dk}{(2\pi)^d}\int\frac{d^dp}{(2\pi)^d}[\mathcal{M}^kG_{n_1,\sigma}^f(\omega,k)\mathcal{M}^jG_{n_2,\sigma}^f(\omega,k+p)\mathcal{M}^i]\otimes[\tilde{\mathcal{M}}^iG_{n_3,\sigma'}^f(\omega,k+p)\tilde{\mathcal{M}}^kG_{n_4,\sigma'}^f(\omega,p)\tilde{\mathcal{M}}^j]\nonumber\\
	&= -\frac{\Gamma_i\Gamma_j\Gamma_k}{N_f^3}\Big(\frac{\Lambda_{FS}}{\pi}\Big)^2\frac{\mathcal{C}_{n_1n_2n_3n_4}(v)}{(4\pi)^{2}}\frac{1}{4}\Bigg[\Bigg(\frac{2}{\epsilon^2}+\frac{1-2\gamma_E+2\ln\Big(\frac{4\pi}{\omega^2}\Big)}{\epsilon}\Bigg)\Big[[\mathcal{M}^k\Gamma_{\perp,i}\mathcal{M}^j\Gamma_{\perp,j}\mathcal{M}^i]\otimes[\tilde{\mathcal{M}}^i\Gamma_{\perp,j}\tilde{\mathcal{M}}^k\Gamma_{\perp,i}\tilde{\mathcal{M}}^j]\nonumber\\
	&+\alpha_{n_3n_2}[\mathcal{M}^k\Gamma_{\perp,i}\mathcal{M}^j\gamma_{d-1}\mathcal{M}^i]\otimes[\tilde{\mathcal{M}}^i\gamma_{d-1}\tilde{\mathcal{M}}^k\Gamma_{\perp,i}\tilde{\mathcal{M}}^j]+\alpha_{n_4n_1}[\mathcal{M}^k\gamma_{d-1}\mathcal{M}^j\Gamma_{\perp,j}\mathcal{M}^i]\otimes[\tilde{\mathcal{M}}^i\Gamma_{\perp,j}\tilde{\mathcal{M}}^k\gamma_{d-1}\tilde{\mathcal{M}}^j]\nonumber\\
	&+\alpha_{n_3n_2}\alpha_{n_4n_1}[\mathcal{M}^k\gamma_{d-1}\mathcal{M}^j\gamma_{d-1}\mathcal{M}^i]\otimes[\tilde{\mathcal{M}}^i\gamma_{d-1}\tilde{\mathcal{M}}^k\gamma_{d-1}\tilde{\mathcal{M}}^j]\Big]-\Bigg(\frac{1}{2\epsilon}\Bigg)\Big[\mathcal{M}^k\Gamma_{\perp,i}\mathcal{M}^j\Gamma_{\perp,i}\mathcal{M}^i]\otimes[\tilde{\mathcal{M}}^i\Gamma_{\perp,j}\tilde{\mathcal{M}}^k\Gamma_{\perp,j}\tilde{\mathcal{M}}^j]\nonumber\\
	&+[\mathcal{M}^k\Gamma_{\perp,i}\mathcal{M}^j\Gamma_{\perp,j}\mathcal{M}^i]\otimes[\tilde{\mathcal{M}}^i\Gamma_{\perp,i}\tilde{\mathcal{M}}^k\Gamma_{\perp,j}\tilde{\mathcal{M}}^j]+[\mathcal{M}^k\Gamma_{\perp,i}\mathcal{M}^j\Gamma_{\perp,j}\mathcal{M}^i]\otimes[\tilde{\mathcal{M}}^i\Gamma_{\perp,j}\tilde{\mathcal{M}}^k\Gamma_{\perp,i}\tilde{\mathcal{M}}^j]\nonumber\\
	&+3\alpha_{n_3n_2}\alpha_{n_4n_1}[\mathcal{M}^k\gamma_{d-1}\mathcal{M}^j\gamma_{d-1}\mathcal{M}^i]\otimes[\tilde{\mathcal{M}}^i\gamma_{d-1}\tilde{\mathcal{M}}^k\gamma_{d-1}\tilde{\mathcal{M}}^j]\nonumber\\
	&+\alpha_{n_4n_1}\alpha_{n_4n_2}[\mathcal{M}^k\gamma_{d-1}\mathcal{M}^j\gamma_{d-1}\mathcal{M}^i]\otimes[\tilde{\mathcal{M}}^i\Gamma_{\perp,i}\tilde{\mathcal{M}}^k\Gamma_{\perp,i}\tilde{\mathcal{M}}^j]\nonumber\\
	&+\alpha_{n_3n_2}\alpha_{n_4n_2}[\mathcal{M}^k\Gamma_{\perp,i}\mathcal{M}^j \Gamma_{\perp,i}\mathcal{M}^i]\otimes[\tilde{\mathcal{M}}^i\gamma_{d-1}\tilde{\mathcal{M}}^k \gamma_{d-1}\tilde{\mathcal{M}}^j]\nonumber\\
	&+\alpha_{n_4n_1}\alpha_{n_4n_2}\alpha_{n_3n_2}[\mathcal{M}^k\gamma_{d-1} \mathcal{M}^j\Gamma_{\perp,i}\mathcal{M}^i]\otimes[\tilde{\mathcal{M}}^i\gamma_{d-1}\tilde{\mathcal{M}}^k\Gamma_{\perp,i}\tilde{\mathcal{M}}^j]\nonumber\\
	&+\alpha_{n_3n_2}[\mathcal{M}^k\Gamma_{\perp,i}\mathcal{M}^j\gamma_{d-1}\mathcal{M}^i]\otimes[\tilde{\mathcal{M}}^i\gamma_{d-1}\tilde{\mathcal{M}}^k\Gamma_{\perp,i}\tilde{\mathcal{M}}^j]\nonumber\\
	&+\alpha_{n_4n_1}[\mathcal{M}^k\gamma_{d-1}\mathcal{M}^j\Gamma_{\perp,i}\mathcal{M}^i]\otimes[\tilde{\mathcal{M}}^i\Gamma_{\perp,i}\tilde{\mathcal{M}}^k\gamma_{d-1}\tilde{\mathcal{M}}^j]+\alpha_{n_4n_2}[\Gamma_{\perp,i}\mathcal{M}^j\gamma_{d-1}\mathcal{M}^i]\otimes[\tilde{\mathcal{M}}^i\Gamma_{\perp,i}\tilde{\mathcal{M}}^k\gamma_{d-1}\tilde{\mathcal{M}}^j]\Big]\Bigg]
\end{align}

\begin{align}
	(a)_{CT}&=\frac{\Gamma_i\Gamma_j\Gamma_k}{N_f^3}\Big(\frac{\Lambda_{FS}}{\pi}\Big)^2\frac{\mathcal{C}_{n_1n_2n_3n_4}(v)}{(4\pi)^2}\frac{1}{2}\Big[\frac{2}{\epsilon^2}-\frac{\gamma_E}{\epsilon}+\frac{\ln\Big(\frac{4\pi}{\omega^2}\Big)}{\epsilon}\Big]\Bigg[[\mathcal{M}^k\Gamma_{\perp,i}\mathcal{M}^j\Gamma_{\perp,j}\mathcal{M}^i]\otimes[\tilde{\mathcal{M}}^i\Gamma_{\perp,i}\tilde{\mathcal{M}}^k\Gamma_{\perp,j}\tilde{\mathcal{M}}^j]\nonumber\\
	&+\alpha_{n_3n_2}[\mathcal{M}^k\Gamma_{\perp,i}\mathcal{M}^j\gamma_{d-1}\mathcal{M}^i]\otimes[\tilde{\mathcal{M}}^i\gamma_{d-1}\tilde{\mathcal{M}}^k\Gamma_{\perp,i}\tilde{\mathcal{M}}^j]+\alpha_{n_4n_1}[\mathcal{M}^k\gamma_{d-1}\mathcal{M}^j\Gamma_{\perp,i}\mathcal{M}^i]\otimes[\tilde{\mathcal{M}}^i\Gamma_{\perp,\mu}\tilde{\mathcal{M}}^k\gamma_{d-1}\tilde{\mathcal{M}}^j]\nonumber\\
	&+\alpha_{n_3n_2}\alpha_{n_4n_1}[\mathcal{M}^k\gamma_{d-1}\mathcal{M}^j\gamma_{d-1}\mathcal{M}^i]\otimes[\tilde{\mathcal{M}}^i\gamma_{d-1}\tilde{\mathcal{M}}^k\gamma_{d-1}\tilde{\mathcal{M}}^j]\Bigg]
\end{align}

Adding $(a)$ to $(a)_{CT}$ gives the following final result:
\begin{align}
	(a)+(a)_{CT}&=-\frac{\Gamma_i\Gamma_j\Gamma_k}{N_f^3} \Big(\frac{\Lambda_{FS}}{\pi}\Big)^2\frac{\mathcal{C}_{n_1n_2n_3n_4}(v)}{(4\pi)^{2}}\frac{1}{4}\Bigg[-\frac{2}{\epsilon^2}\Big[\alpha_{n_3n_2}\alpha_{n_4n_1}[\mathcal{M}^k\gamma_{d-1}\mathcal{M}^j\gamma_{d-1}\mathcal{M}^i]\otimes[\tilde{\mathcal{M}}^i\gamma_{d-1}\tilde{\mathcal{M}}^k\gamma_{d-1}\tilde{\mathcal{M}}^j]\Big]\nonumber\\
	&-\Bigg(\frac{1}{2\epsilon}\Bigg)\Big[\mathcal{M}^k\Gamma_{\perp,1}\mathcal{M}^j\Gamma_{\perp,1}\mathcal{M}^i]\otimes[\tilde{\mathcal{M}}^i\Gamma_{\perp,1}\tilde{\mathcal{M}}^k\Gamma_{\perp,1}\tilde{\mathcal{M}}^j]\nonumber\\
	&+\alpha_{n_3n_2}\alpha_{n_4n_1}[\mathcal{M}^k\gamma_{d-1}\mathcal{M}^j\gamma_{d-1}\mathcal{M}^i]\otimes[\tilde{\mathcal{M}}^i\gamma_{d-1}\tilde{\mathcal{M}}^k\gamma_{d-1}\tilde{\mathcal{M}}^j]\nonumber\\
	&+\alpha_{n_4n_1}\alpha_{n_4n_2}[\mathcal{M}^k\gamma_{d-1}\mathcal{M}^j\gamma_{d-1}\mathcal{M}^i]\otimes[\tilde{\mathcal{M}}^i\Gamma_{\perp,1}\tilde{\mathcal{M}}^k\Gamma_{\perp,1}\tilde{\mathcal{M}}^j]\nonumber\\
	&+\alpha_{n_3n_2}\alpha_{n_4n_2}[\mathcal{M}^k\Gamma_{\perp,1}\mathcal{M}^j \Gamma_{\perp,1}\mathcal{M}^i]\otimes[\tilde{\mathcal{M}}^i\gamma_{d-1}\tilde{\mathcal{M}}^k \gamma_{d-1}\tilde{\mathcal{M}}^j]\Big]\Bigg]+\mathcal{O}(\epsilon^0) .
\end{align}
Since only the $\frac{1}{\epsilon}$-pole contributes to beta functions \cite{CriticalPhi4}, we can ignore the $\frac{1}{\epsilon^2}$-pole safely.

\subsection{Two-loop Yukawa interaction vertex corrections}

\begin{figure}[H]
	\centering
	\begin{subfigure}{0.15\textwidth}
		\begin{tikzpicture}[baseline=-0.1cm, scale=0.5]
			\begin{feynhand}
				\vertex (a) at (-3,0); \vertex (b) at (0,0); \vertex (c1) at (2.5/3, 2.5/3); \vertex (c2) at (2.5/3, -2.5/3); \vertex (d1) at (5/3, 5/3); \vertex (d2) at (5/3,-5/3); \vertex (e1) at (2.5,2.5); \vertex (e2) at (2.5,-2.5);
				\propag[boson] (a) to (b); \propag[fer] (b) to [edge label={\tiny $n_2$}](c1); \propag[fer] (c1) to [edge label={\tiny $n_1$}](d1); \propag[fer] (d1) to (e1);
				\propag[fer] (c2) to [edge label={\tiny $n_3$}](b); \propag[fer] (d2) to [edge label={\tiny $n_4$}](c2); \propag[fer] (e2) to (d2);
				\propag[sca] (c1) to (d2); \propag[sca] (d1) to (c2);
				\node at (2.5/3,2.5/3+0.6) {\tiny $\mathcal{M}^i$}; \node at (2.5/3,-2.5/3-1) {\tiny $\tilde{\mathcal{M}}^j$};
				\node at (5/3, 5/3) {\tiny $\mathcal{M}^j$}; \node at (5/3, -5/3-1) {\tiny $\tilde{\mathcal{M}}^i$};
			\end{feynhand}
		\end{tikzpicture}
		\caption{}
	\end{subfigure}
	~
	\begin{subfigure}{0.15\textwidth}
		\begin{tikzpicture}[baseline=-0.1cm, scale=0.5]
			\begin{feynhand}
				\vertex (a) at (-3,0); \vertex (b) at (0,0); \vertex (c1) at (2.5/4, 2.5/4); \vertex (c2) at (2.5/4, -2.5/4); \vertex (d1) at (5/4, 5/4); \vertex (d2) at (5/4,-5/4); \vertex (e1) at (7.5/4,7.5/4); \vertex (e2) at (7.5/4,-7.5/4); \vertex (f1) at (2.5,2.5); \vertex (f2) at (2.5,-2.5);
				\propag[boson] (a) to (b); \propag[fer] (b) to [edge label={\tiny $n_3$}](c1); \propag[fer] (c1) to [edge label={\tiny $n_2$}](d1); \propag[fer] (d1) to [edge label={\tiny $n_1$}](e1); \propag[fer] (e1) to (f1);
				\propag[fer] (c2) to [edge label={\tiny $n_4$}](b); \propag[fer] (d2) to (c2); \propag[fer] (e2) to (d2); \propag[fer] (f2) to (e2);
				\propag[sca] (d1) to (d2); \propag[sca] (c1) to [out=135, in=135](e1);
				\node at (2.5/4,2.5/4-0.5) {\tiny $\tilde{\mathcal{M}}^j$}; \node at (7.5/4+0.5,7.5/4) {\tiny $\mathcal{M}^j$};
				\node at (5/4+0.5, 5/4) {\tiny $\mathcal{M}^i$}; \node at (5/4-0.3, -5/4-0.5) {\tiny $\tilde{\mathcal{M}}^i$};
			\end{feynhand}
		\end{tikzpicture}
		\caption{}
	\end{subfigure}
	~
	\begin{subfigure}{0.15\textwidth}
		\begin{tikzpicture}[baseline=-0.1cm, scale=0.5]
			\begin{feynhand}
				\vertex (a) at (-3,0); \vertex (b) at (0,0); \vertex (c1) at (2.5/4, 2.5/4); \vertex (c2) at (2.5/4, -2.5/4); \vertex (d1) at (5/4, 5/4); \vertex (d2) at (5/4,-5/4); \vertex (e1) at (7.5/4,7.5/4); \vertex (e2) at (7.5/4,-7.5/4); \vertex (f1) at (2.5,2.5); \vertex (f2) at (2.5,-2.5);
				\propag[boson] (a) to (b); \propag[fer] (b) to (c1); \propag[fer] (c1) to [edge label={\tiny $n_1$}](d1); \propag[fer] (d1) to (e1); \propag[fer] (e1) to (f1);
				\propag[fer] (c2) to [edge label={\tiny $n_2$}](b); \propag[fer] (d2) to [edge label={\tiny $n_3$}](c2); \propag[fer] (e2) to [edge label={\tiny $n_4$}](d2); \propag[fer] (f2) to (e2);
				\propag[sca] (d1) to (d2); \propag[sca] (c2) to [out=225, in=225](e2);
				\node at (2.5/4-0.6,-2.5/4) {\tiny $\tilde{\mathcal{M}}^j$}; \node at (7.5/4,-7.5/4-0.6) {\tiny $\mathcal{M}^j$};
				\node at (5/4+0.6, 5/4) {\tiny $\tilde{\mathcal{M}}^i$}; \node at (5/4+0.6, -5/4) {\tiny $\mathcal{M}^i$};
			\end{feynhand}
		\end{tikzpicture}
		\caption{}
	\end{subfigure}
	~
	\begin{subfigure}{0.15\textwidth}
		\begin{tikzpicture}[baseline=-0.1cm, scale=0.5]
			\begin{feynhand}
				\vertex (a) at (-3,0); \vertex (b) at (0,0); \vertex (c1) at (2.5/2, 2.5/2); \vertex (c2) at (2.5/2, -2.5/2); \vertex (d1) at (2.5,2.5); \vertex (d2) at (2.5,-2.5); \vertex[crossdot] (f) at (2.5/2,0) {};
				\propag[boson] (a) to (b); \propag[fer] (b) to (c1); \propag[fer] (c1) to (d1);
				\propag[fer] (c2) to (b); \propag[fer] (d2) to (c2);
				\propag[sca] (c1) to (f); \propag[sca] (f) to (c2);
			\end{feynhand}
		\end{tikzpicture}
		\caption{}
	\end{subfigure}
\end{figure}

We can use the same identity Eq. \eqref{eq:TwoLoopUsefulIntegration} to calculate two-loop Yukawa interaction vertices. Here, we only present our results as follows:

\begin{align}
	&(a)+(a)_{CT}=i\tau_{\sigma\sigma'}^a\frac{g\Gamma_i\Gamma_j}{N_f^{3/2}}\Big(\frac{\Lambda_{FS}}{\pi}\Big)^2\frac{\mathcal{C}_{n_1n_2n_3n_4}(v)}{(4\pi)^2}\frac{1}{2}\Bigg[-\frac{1}{\epsilon^2}\Big[\mathcal{M}^j\Gamma_{\perp,i}\mathcal{M}^i\Gamma_{\perp,j}\gamma_{d-1}\Gamma_{\perp,j}\tilde{\mathcal{M}}^j\Gamma_{\perp,i}\tilde{\mathcal{M}}^i\nonumber\\
	&-\mathcal{M}^j\Gamma_{\perp,i}\mathcal{M}^i\gamma_{d-1}\tilde{\mathcal{M}}^j\Gamma_{\perp,i}\tilde{\mathcal{M}}^i+\alpha_{n_4n_1}\mathcal{M}^j\gamma_{d-1}\mathcal{M}^i\Gamma_{\perp,j}\gamma_{d-1}\Gamma_{\perp,j}\tilde{\mathcal{M}}^j\gamma_{d-1}\tilde{\mathcal{M}}^i-\alpha_{n_4n_1}\mathcal{M}^j\gamma_{d-1}\mathcal{M}^i\gamma_{d-1}\tilde{\mathcal{M}}^j\gamma_{d-1}\tilde{\mathcal{M}}^i\Big]\nonumber\\
	&-\frac{1}{4\epsilon}\Big[\mathcal{M}^j\Gamma_{\perp,i}\mathcal{M}^i\Gamma_{\perp,i}\gamma_{d-1}\Gamma_{\perp,j}\tilde{\mathcal{M}}^j\Gamma_{\perp,j}\tilde{\mathcal{M}}^i+\mathcal{M}^j\Gamma_{\perp,i}\mathcal{M}^i\Gamma_{\perp,j}\gamma_{d-1}\Gamma_{\perp,i}\tilde{\mathcal{M}}^j\Gamma_{\perp,j}\tilde{\mathcal{M}}^i\nonumber\\
	&-\mathcal{M}^j\Gamma_{\perp,i}\mathcal{M}^i\Gamma_{\perp,j}\gamma_{d-1}\Gamma_{\perp,j}\tilde{\mathcal{M}}^j\Gamma_{\perp,i}\tilde{\mathcal{M}}^i-\alpha_{n_4n_1}\mathcal{M}^j\gamma_{d-1}\mathcal{M}^i\gamma_{d-1}\tilde{\mathcal{M}}^j\gamma_{d-1}\tilde{\mathcal{M}}^i+\alpha_{n_4n_1}\alpha_{n_4n_2}\mathcal{M}^j\gamma_{d-1}\mathcal{M}^i\Gamma_{\perp,i}\tilde{\mathcal{M}}^j\Gamma_{\perp,i}\tilde{\mathcal{M}}^i\nonumber\\
	&-\alpha_{n_4n_2}\mathcal{M}^j\Gamma_{\perp,i}\mathcal{M}^i \Gamma_{\perp,i}\tilde{\mathcal{M}}^j \gamma_{d-1}\tilde{\mathcal{M}}^i-\alpha_{n_4n_1}\alpha_{n_4n_2}\mathcal{M}^j\gamma_{d-1}\mathcal{M}^i \Gamma_{\perp,i}\tilde{\mathcal{M}}^j\Gamma_{\perp,i}\tilde{\mathcal{M}}^i+\mathcal{M}^j\Gamma_{\perp,i}\mathcal{M}^i\gamma_{d-1}\tilde{\mathcal{M}}^j\Gamma_{\perp,i}\tilde{\mathcal{M}}^i\nonumber\\
	&-\alpha_{n_4n_1}\mathcal{M}^j\gamma_{d-1}\mathcal{M}^i\Gamma_{\perp,i}\gamma_{d-1}\Gamma_{\perp,i}\tilde{\mathcal{M}}^j\gamma_{d-1}\tilde{\mathcal{M}}^i+\alpha_{n_4n_2}\mathcal{M}^j\Gamma_{\perp,i}\mathcal{M}^i\Gamma_{\perp,i}\tilde{\mathcal{M}}^j\gamma_{d-1}\tilde{\mathcal{M}}^i\Big]\Bigg]
\end{align}

\begin{align}
	&(b)+(b)_{CT}=-i\tau_{\sigma\sigma'}^a\frac{g\Gamma_i\Gamma_j}{N_f^{3/2}}\Big(\frac{\Lambda_{FS}}{\pi}\Big)^2\frac{\mathcal{C}_{n_1n_2n_3n_4}(v)}{(4\pi)^2}\frac{1}{2}\Bigg[-\frac{1}{\epsilon^2}\Big[\mathcal{M}^j\Gamma_{\perp,i}\mathcal{M}^i\Gamma_{\perp,i}\tilde{\mathcal{M}}^j\Gamma_{\perp,j}\gamma_{d-1}\Gamma_{\perp,j}\tilde{\mathcal{M}}^i\nonumber\\
	&-\mathcal{M}^j\Gamma_{\perp,i}\mathcal{M}^i\Gamma_{\perp,i}\tilde{\mathcal{M}}^j\gamma_{d-1}\tilde{\mathcal{M}}^i+\alpha_{n_2n_1}\mathcal{M}^j\gamma_{d-1}\mathcal{M}^i\gamma_{d-1}\tilde{\mathcal{M}}^j\Gamma_{\perp,j}\gamma_{d-1}\Gamma_{\perp,j}\tilde{\mathcal{M}}^i-\alpha_{n_2n_1}\mathcal{M}^j\gamma_{d-1}\mathcal{M}^i\gamma_{d-1}\tilde{\mathcal{M}}^j\gamma_{d-1}\tilde{\mathcal{M}}^i\Big]\nonumber\\
	&-\frac{1}{4\epsilon}\Big[\mathcal{M}^j\Gamma_{\perp,i}\mathcal{M}^i\Gamma_{\perp,j}\tilde{\mathcal{M}}^j\Gamma_{\perp,j}\gamma_{d-1}\Gamma_{\perp,i}\tilde{\mathcal{M}}^i+\mathcal{M}^j\Gamma_{\perp,i}\mathcal{M}^i\Gamma_{\perp,j}\tilde{\mathcal{M}}^j\Gamma_{\perp,i}\gamma_{d-1}\Gamma_{\perp,j}\tilde{\mathcal{M}}^i\nonumber\\
	&-\mathcal{M}^j\Gamma_{\perp,i}\mathcal{M}^i\Gamma_{\perp,i}\tilde{\mathcal{M}}^j\Gamma_{\perp,j}\gamma_{d-1}\Gamma_{\perp,j}\tilde{\mathcal{M}}^i-\alpha_{n_2n_1}\mathcal{M}^j\gamma_{d-1}\mathcal{M}^i\gamma_{d-1}\tilde{\mathcal{M}}^j\gamma_{d-1}\tilde{\mathcal{M}}^i+\alpha_{n_2n_1}\alpha_{n_2\bar{n}_3}\mathcal{M}^j\gamma_{d-1}\mathcal{M}^i\Gamma_{\perp,i}\tilde{\mathcal{M}}^j\Gamma_{\perp,i}\tilde{\mathcal{M}}^i\nonumber\\
	&-\alpha_{n_2\bar{n}_3}\mathcal{M}^j\Gamma_{\perp,i}\mathcal{M}^i \gamma_{d-1}\tilde{\mathcal{M}}^j  \Gamma_{\perp,i}\tilde{\mathcal{M}}^i-\alpha_{n_2n_1}\alpha_{n_2\bar{n}_3}\mathcal{M}^j\gamma_{d-1}\mathcal{M}^i\Gamma_{\perp,i}\tilde{\mathcal{M}}^j \Gamma_{\perp,i}\tilde{\mathcal{M}}^i+\mathcal{M}^j\Gamma_{\perp,i}\mathcal{M}^i\Gamma_{\perp,i}\tilde{\mathcal{M}}^j\gamma_{d-1}\tilde{\mathcal{M}}^i\nonumber\\
	&-\alpha_{n_2n_1}\mathcal{M}^j\gamma_{d-1}\mathcal{M}^i\gamma_{d-1}\tilde{\mathcal{M}}^j\Gamma_{\perp,i}\gamma_{d-1}\Gamma_{\perp,i}\tilde{\mathcal{M}}^i+\alpha_{n_2\bar{n}_3}\mathcal{M}^j\Gamma_{\perp,i}\mathcal{M}^i\gamma_{d-1}\tilde{\mathcal{M}}^j\Gamma_{\perp,i}\tilde{\mathcal{M}}^i\Big]\Bigg]
\end{align}

\begin{align}
	&(c)+(c)_{CT}=-i\tau_{\sigma\sigma'}^a\frac{g\Gamma_i\Gamma_j}{N_f^{3/2}}\Big(\frac{\Lambda_{FS}}{\pi}\Big)^2\frac{\mathcal{C}_{n_1n_2n_3n_4}(v)}{(4\pi)^2}\frac{1}{2}\Bigg[-\frac{1}{\epsilon^2}\Big[\tilde{\mathcal{M}}^i\Gamma_{\perp,j}\gamma_{d-1}\Gamma_{\perp,j}\tilde{\mathcal{M}}^j\Gamma_{\perp,i}\mathcal{M}^i\Gamma_{\perp,i}\mathcal{M}^j\nonumber\\
	&-\tilde{\mathcal{M}}^i\gamma_{d-1}\tilde{\mathcal{M}}^j\Gamma_{\perp,i}\mathcal{M}^i\Gamma_{\perp,i}\mathcal{M}^j+\alpha_{n_4n_3}\tilde{\mathcal{M}}^i\Gamma_{\perp,j}\gamma_{d-1}\Gamma_{\perp,j}\tilde{\mathcal{M}}^j\gamma_{d-1}\mathcal{M}^i\gamma_{d-1}\mathcal{M}^j-\alpha_{n_4n_3}\tilde{\mathcal{M}}^i\gamma_{d-1}\tilde{\mathcal{M}}^j\gamma_{d-1}\mathcal{M}^i\gamma_{d-1}\mathcal{M}^j\Big]\nonumber\\
	&-\Bigg(\frac{1}{4\epsilon}\Bigg)\Big[\tilde{\mathcal{M}}^i\Gamma_{\perp,j}\gamma_{d-1}\Gamma_{\perp,i}\tilde{\mathcal{M}}^j\Gamma_{\perp,i}\mathcal{M}^i\Gamma_{\perp,j}\mathcal{M}^j+\tilde{\mathcal{M}}^i\Gamma_{\perp,i}\gamma_{d-1}\Gamma_{\perp,j}\tilde{\mathcal{M}}^j\Gamma_{\perp,i}\mathcal{M}^i\Gamma_{\perp,j}\mathcal{M}^j\nonumber\\
	&-\tilde{\mathcal{M}}^i\Gamma_{\perp,j}\gamma_{d-1}\Gamma_{\perp,j}\tilde{\mathcal{M}}^j\Gamma_{\perp,i}\mathcal{M}^i\Gamma_{\perp,i}\mathcal{M}^j-\alpha_{n_4n_3}\tilde{\mathcal{M}}^i\gamma_{d-1}\tilde{\mathcal{M}}^j\gamma_{d-1}\mathcal{M}^i\gamma_{d-1}\mathcal{M}^j+\alpha_{n_4n_3}\alpha_{n_4\bar{n}_1}\tilde{\mathcal{M}}^i\Gamma_{\perp,i}\tilde{\mathcal{M}}^j\gamma_{d-1}\mathcal{M}^i\Gamma_{\perp,i}\mathcal{M}^j\nonumber\\
	&-\alpha_{n_4\bar{n}_1} \tilde{\mathcal{M}}^i \Gamma_{\perp,i}\tilde{\mathcal{M}}^j\Gamma_{\perp,i}\mathcal{M}^i \gamma_{d-1}\mathcal{M}^j-\alpha_{n_4n_3}\alpha_{n_4\bar{n}_1}\tilde{\mathcal{M}}^i\Gamma_{\perp,i}\tilde{\mathcal{M}}^j\gamma_{d-1}\mathcal{M}^i\Gamma_{\perp,i}\mathcal{M}^j+\tilde{\mathcal{M}}^i\gamma_{d-1}\tilde{\mathcal{M}}^j\Gamma_{\perp,i}\mathcal{M}^i\Gamma_{\perp,i}\mathcal{M}^j\nonumber\\
	&-\alpha_{n_4n_3}\tilde{\mathcal{M}}^i\Gamma_{\perp,i}\gamma_{d-1}\Gamma_{\perp,i}\tilde{\mathcal{M}}^j\gamma_{d-1}\mathcal{M}^i\gamma_{d-1}\mathcal{M}^j+\alpha_{n_4\bar{n}_1}\tilde{\mathcal{M}}^i\Gamma_{\perp,i}\tilde{\mathcal{M}}^j\Gamma_{\perp,i}\mathcal{M}^i\gamma_{d-1}\mathcal{M}^j\Big]\Bigg]
\end{align}

Surprisingly, it turns out that all two loop contributions to the Yukawa interaction vertex from only random charge potential vertices are zero.

\section{Counter terms in two-loop calculations} \label{Appendix:TwoLoopCounterTerms}

Using the Mathematica programming, we are able to get all two-loop counter terms as follows:
\begin{align}
	A_0^{(2l)}&=\frac{1}{32 \pi^4N_f^2(1+v^2)^2}\frac{1}{\epsilon}\Big[(\Delta_{0}-\Delta_{\pi})^2-2 (\Delta_{0}+\Delta_{\pi}) \Gamma_{0}-(\Gamma_{0})^2\Big]+\frac{e^{-\frac{v^2}{v_c^2}}}{32\pi^4N_f^2}\frac{1}{\epsilon} \Big[(\Delta_{\pi-\theta_1})^2-2 \Delta_{\pi-\theta_1} \Delta_{\theta_1}+(\Delta_{\theta_1})^2\nonumber\\
	&-2 \Gamma_{\pi-\theta_1}^d \Gamma_{\pi-\theta_1}^e-2 \Gamma_{\pi-\theta_1}^e \Gamma_{\theta_1}^d-2 \Gamma_{\pi-\theta_1}^d \Gamma_{\theta_1}^e-2 \Gamma_{\theta_1}^d \Gamma_{\theta_1}^e+2 \Delta_{\pi-\theta_1} \Upsilon_{0}+2 \Delta_{\theta_1} \Upsilon_{0}+3 (\Upsilon_{0})^2+2 \Gamma_{\pi-\theta_1}^d \Upsilon_{\theta_1}^d\nonumber\\
	&-2 \Gamma_{\theta_1}^d \Upsilon_{\theta_1}^d+2 \Gamma_{\pi-\theta_1}^e \Upsilon_{\theta_1}^e-2 \Gamma_{\theta_1}^e \Upsilon_{\theta_1}^e+6 \Upsilon_{\theta_1}^d \Upsilon_{\theta_1}^e\Big]\\
	A^{(2l)}_{\Gamma_0}&=\frac{e^{-\frac{v^2}{v_c^2}}}{64\Gamma_{0}\pi^4N_f^2 } \frac{1}{\epsilon} \Big[\Delta_{\pi} (\Delta_{\pi-\theta_1})^2+2 \Delta_{\pi} \Delta_{\pi-\theta_1} \Delta_{\theta_1}+\Delta_{\pi} (\Delta_{\theta_1})^2-2 \Delta_{\pi-\theta_1} \Delta_{\theta_1} \Gamma_{\pi-\theta_1}^d+2 (\Delta_{\theta_1})^2 \Gamma_{\pi-\theta_1}^d\nonumber\\
	&+(\Delta_{\pi-\theta_1})^2 \Gamma_{\pi-\theta_1}^e-2 \Delta_{\pi-\theta_1} \Delta_{\theta_1} \Gamma_{\pi-\theta_1}^e-(\Delta_{\theta_1})^2 \Gamma_{\pi-\theta_1}^e+2 (\Gamma_{\pi-\theta_1}^d)^2 \Gamma_{\pi-\theta_1}^e-\Delta_{\pi} (\Gamma_{\pi-\theta_1}^e)^2+(\Gamma_{\pi-\theta_1}^e)^3\nonumber\\
	&-2 (\Delta_{\pi-\theta_1})^2 \Gamma_{\theta_1}^d+6 \Delta_{\pi-\theta_1} \Delta_{\theta_1} \Gamma_{\theta_1}^d+2 \Gamma_{\pi-\theta_1}^d \Gamma_{\pi-\theta_1}^e \Gamma_{\theta_1}^d+\Gamma_{\pi-\theta_1}^e (\Gamma_{\theta_1}^d)^2+(\Delta_{\pi-\theta_1})^2 \Gamma_{\theta_1}^e+2 \Delta_{\pi-\theta_1} \Delta_{\theta_1} \Gamma_{\theta_1}^e\nonumber\\
	&-(\Delta_{\theta_1})^2 \Gamma_{\theta_1}^e-(\Gamma_{\pi-\theta_1}^d)^2 \Gamma_{\theta_1}^e-(\Gamma_{\pi-\theta_1}^e)^2 \Gamma_{\theta_1}^e+2 (\Gamma_{\theta_1}^d)^2 \Gamma_{\theta_1}^e-\Delta_{\pi} (\Gamma_{\theta_1}^e)^2+\Gamma_{\pi-\theta_1}^e (\Gamma_{\theta_1}^e)^2-(\Gamma_{\theta_1}^e)^3+2 \Delta_{\pi} \Delta_{\pi-\theta_1} \Upsilon_{0}\nonumber\\
	&+2 \Delta_{\pi} \Delta_{\theta_1} \Upsilon_{0}-2 \Delta_{\theta_1} \Gamma_{\pi-\theta_1}^d \Upsilon_{0}+2 \Delta_{\theta_1} \Gamma_{\theta_1}^d \Upsilon_{0}+2 \Delta_{\pi-\theta_1} \Gamma_{\theta_1}^e \Upsilon_{0}+2 \Delta_{\theta_1} \Gamma_{\theta_1}^e \Upsilon_{0}+\Delta_{\pi} (\Upsilon_{0})^2+4 \Gamma_{\pi-\theta_1}^d (\Upsilon_{0})^2\nonumber\\
	&-\Gamma_{\pi-\theta_1}^e (\Upsilon_{0})^2+6 \Gamma_{\theta_1}^d (\Upsilon_{0})^2+\Gamma_{\theta_1}^e (\Upsilon_{0})^2+2 (\Gamma_{\pi-\theta_1}^d)^2 \Upsilon_{\theta_1}^d-2 \Delta_{\pi} \Gamma_{\pi-\theta_1}^e \Upsilon_{\theta_1}^d+4 \Gamma_{\pi-\theta_1}^d \Gamma_{\pi-\theta_1}^e \Upsilon_{\theta_1}^d+2 (\Gamma_{\pi-\theta_1}^e)^2 \Upsilon_{\theta_1}^d\nonumber\\
	&-\Gamma_{\pi-\theta_1}^d \Gamma_{\theta_1}^d \Upsilon_{\theta_1}^d+4 \Gamma_{\pi-\theta_1}^e \Gamma_{\theta_1}^d \Upsilon_{\theta_1}^d-2 \Gamma_{\pi-\theta_1}^e \Gamma_{\theta_1}^e \Upsilon_{\theta_1}^d-2 \Delta_{\pi-\theta_1} \Upsilon_{0} \Upsilon_{\theta_1}^d+2 \Delta_{\theta_1} \Upsilon_{0} \Upsilon_{\theta_1}^d-2 (\Upsilon_{0})^2 \Upsilon_{\theta_1}^d-\Delta_{\pi} (\Upsilon_{\theta_1}^d)^2\nonumber\\
	&+\Gamma_{\pi-\theta_1}^e (\Upsilon_{\theta_1}^d)^2-\Gamma_{\theta_1}^e (\Upsilon_{\theta_1}^d)^2+\Delta_{0}\Big [-(\Delta_{\pi-\theta_1})^2+(\Delta_{\theta_1})^2+\Delta_{\pi-\theta_1} \Upsilon_{0}+2 \Delta_{\theta_1} (\Delta_{\pi-\theta_1}+\Upsilon_{0})+2 [(\Upsilon_{0})^2\nonumber\\
	&+(-\Gamma_{\pi-\theta_1}^d+\Gamma_{\theta_1}^d) (\Gamma_{\pi-\theta_1}^e-\Gamma_{\theta_1}^e+\Upsilon_{\theta_1}^d)]\Big]-2 \Delta_{\pi} \Gamma_{\pi-\theta_1}^e \Upsilon_{\theta_1}^e+\Gamma_{\pi-\theta_1}^d \Gamma_{\pi-\theta_1}^e \Upsilon_{\theta_1}^e+2 (\Gamma_{\pi-\theta_1}^e)^2 \Upsilon_{\theta_1}^e-4 \Gamma_{\pi-\theta_1}^e \Gamma_{\theta_1}^d \Upsilon_{\theta_1}^e\nonumber\\
	&+2 \Delta_{\pi} \Gamma_{\theta_1}^e \Upsilon_{\theta_1}^e+4 \Gamma_{\pi-\theta_1}^d \Gamma_{\theta_1}^e \Upsilon_{\theta_1}^e-2 \Gamma_{\theta_1}^d \Gamma_{\theta_1}^e \Upsilon_{\theta_1}^e-2 (\Gamma_{\theta_1}^e)^2 \Upsilon_{\theta_1}^e+2 \Delta_{\pi-\theta_1} \Upsilon_{0} \Upsilon_{\theta_1}^e+2 \Delta_{\theta_1} \Upsilon_{0} \Upsilon_{\theta_1}^e-2 (\Upsilon_{0})^2 \Upsilon_{\theta_1}^e\nonumber\\
	&-2 \Delta_{\pi} \Upsilon_{\theta_1}^d \Upsilon_{\theta_1}^e+4 \Gamma_{\pi-\theta_1}^d \Upsilon_{\theta_1}^d \Upsilon_{\theta_1}^e-2 \Gamma_{\pi-\theta_1}^e \Upsilon_{\theta_1}^d \Upsilon_{\theta_1}^e-2 \Gamma_{\theta_1}^e \Upsilon_{\theta_1}^d \Upsilon_{\theta_1}^e-\Gamma_{0} \Big(3 (\Delta_{\pi-\theta_1})^2+3 (\Delta_{\theta_1})^2-4 \Gamma_{\pi-\theta_1}^d \Gamma_{\pi-\theta_1}^e\nonumber\\
	&-4 \Gamma_{\theta_1}^d \Gamma_{\theta_1}^e+6 \Delta_{\pi-\theta_1} \Upsilon_{0}+3 (\Upsilon_{0})^2+\Delta_{\theta_1} (2 \Delta_{\pi-\theta_1}+\Upsilon_{0})-4 \Gamma_{\pi-\theta_1}^d \Upsilon_{\theta_1}^d-2 \Gamma_{\pi-\theta_1}^e \Upsilon_{\theta_1}^e+2 \Gamma_{\theta_1}^e \Upsilon_{\theta_1}^e-2 \Upsilon_{\theta_1}^d \Upsilon_{\theta_1}^e\Big)\Big]\nonumber\\
	&+\frac{1}{64\Gamma_0\pi^4 N_f^2(1+v^2)^2}\frac{1}{\epsilon}\Big[(\Delta_{0})^3+\Delta_{0} (\Delta_{\pi})^2+2 (\Delta_{\pi})^3+[-2 (\Delta_{0})^2+8 \Delta_{0} \Delta_{\pi}-5 (\Delta_{\pi})^2] \Gamma_{0}+(-4 \Delta_{0}+\Delta_{\pi}) (\Gamma_{0})^2\nonumber\\
	&+(\Gamma_{0})^3) \Big]
\end{align}

\begin{align}
	A^{(2l)}_{\Gamma_{\theta_1}^d}&=-\frac{e^{-\frac{v^2}{v_c^2}}}{64\Gamma_{\theta_1}^d  \pi^4N_f^2}\frac{1}{\epsilon} \Big[-2 \Delta_{0} (\Delta_{\pi-\theta_1})^2+\Delta_{\pi} (\Delta_{\pi-\theta_1})^2+2 \Delta_{0} \Delta_{\pi-\theta_1} \Delta_{\theta_1}-2 \Delta_{\pi} \Delta_{\pi-\theta_1} \Delta_{\theta_1}-2 \Delta_{0} (\Delta_{\theta_1})^2+\Delta_{\pi} (\Delta_{\theta_1})^2\nonumber\\
	&+3 (\Delta_{\pi-\theta_1})^2 \Gamma_{\pi-\theta_1}^d-2 \Delta_{\pi-\theta_1} \Delta_{\theta_1} \Gamma_{\pi-\theta_1}^d-(\Delta_{\theta_1})^2 \Gamma_{\pi-\theta_1}^d-3 (\Delta_{\pi-\theta_1})^2 \Gamma_{\pi-\theta_1}^e+2 \Delta_{\pi-\theta_1} \Delta_{\theta_1} \Gamma_{\pi-\theta_1}^e-(\Delta_{\theta_1})^2 \Gamma_{\pi-\theta_1}^e\nonumber\\
	&-2 \Delta_{0} \Gamma_{\pi-\theta_1}^d \Gamma_{\pi-\theta_1}^e+2 (\Gamma_{\pi-\theta_1}^d)^2 \Gamma_{\pi-\theta_1}^e-\Delta_{\pi} (\Gamma_{\pi-\theta_1}^e)^2+(\Gamma_{\pi-\theta_1}^e)^3+2 (\Gamma_{\theta_1}^d)^3+(\Delta_{\pi-\theta_1})^2 \Gamma_{\theta_1}^e+2 \Delta_{\pi-\theta_1} \Delta_{\theta_1} \Gamma_{\theta_1}^e\nonumber\\
	&-(\Delta_{\theta_1})^2 \Gamma_{\theta_1}^e+\Delta_{0} \Gamma_{\pi-\theta_1}^d \Gamma_{\theta_1}^e-2 (\Gamma_{\pi-\theta_1}^d)^2 \Gamma_{\theta_1}^e+2 \Delta_{\pi} \Gamma_{\pi-\theta_1}^e \Gamma_{\theta_1}^e-(\Gamma_{\pi-\theta_1}^e)^2 \Gamma_{\theta_1}^e+2 \Delta_{0} (\Gamma_{\theta_1}^e)^2-\Delta_{\pi} (\Gamma_{\theta_1}^e)^2\nonumber\\
	&+2 \Gamma_{\pi-\theta_1}^d (\Gamma_{\theta_1}^e)^2+\Gamma_{\pi-\theta_1}^e (\Gamma_{\theta_1}^e)^2-(\Gamma_{\theta_1}^e)^3-4 (\Gamma_{\theta_1}^d)^2 (-\Gamma_{\pi-\theta_1}^d+\Gamma_{\theta_1}^e)-2 \Delta_{0} \Delta_{\pi-\theta_1} \Upsilon_{0}+2 \Delta_{\pi} \Delta_{\pi-\theta_1} \Upsilon_{0}\nonumber\\
	&+2 \Delta_{0} \Delta_{\theta_1} \Upsilon_{0}-3 \Delta_{\pi-\theta_1} \Gamma_{\pi-\theta_1}^d \Upsilon_{0}-2 \Delta_{\theta_1} \Gamma_{\pi-\theta_1}^d \Upsilon_{0}-4 \Delta_{\pi-\theta_1} \Gamma_{\pi-\theta_1}^e \Upsilon_{0}+2 \Delta_{\pi-\theta_1} \Gamma_{\theta_1}^e \Upsilon_{0}+2 \Delta_{\theta_1} \Gamma_{\theta_1}^e \Upsilon_{0}\nonumber\\
	&+\Delta_{\pi} (\Upsilon_{0})^2-2 \Gamma_{\pi-\theta_1}^d (\Upsilon_{0})^2-\Gamma_{\pi-\theta_1}^e (\Upsilon_{0})^2+\Gamma_{\theta_1}^e (\Upsilon_{0})^2-2 \Delta_{0} \Gamma_{\pi-\theta_1}^d \Upsilon_{\theta_1}^d+2 (\Gamma_{\pi-\theta_1}^d)^2 \Upsilon_{\theta_1}^d-2 \Delta_{0} \Gamma_{\pi-\theta_1}^e \Upsilon_{\theta_1}^d\nonumber\\
	&-2 \Delta_{\pi} \Gamma_{\pi-\theta_1}^e \Upsilon_{\theta_1}^d+2 \Gamma_{\pi-\theta_1}^d \Gamma_{\pi-\theta_1}^e \Upsilon_{\theta_1}^d+2 (\Gamma_{\pi-\theta_1}^e)^2 \Upsilon_{\theta_1}^d-2 \Gamma_{\pi-\theta_1}^e \Gamma_{\theta_1}^e \Upsilon_{\theta_1}^d+2 \Delta_{\pi-\theta_1} \Upsilon_{0} \Upsilon_{\theta_1}^d-2 \Delta_{\theta_1} \Upsilon_{0} \Upsilon_{\theta_1}^d+2 (\Upsilon_{0})^2 \Upsilon_{\theta_1}^d\nonumber\\
	&-2 \Delta_{0} (\Upsilon_{\theta_1}^d)^2-\Delta_{\pi} (\Upsilon_{\theta_1}^d)^2-2 \Gamma_{\pi-\theta_1}^d (\Upsilon_{\theta_1}^d)^2+\Gamma_{\pi-\theta_1}^e (\Upsilon_{\theta_1}^d)^2-\Gamma_{\theta_1}^e (\Upsilon_{\theta_1}^d)^2-2 (\Delta_{\pi-\theta_1})^2 \Upsilon_{\theta_1}^e-2 \Delta_{\pi-\theta_1} \Delta_{\theta_1} \Upsilon_{\theta_1}^e\nonumber\\
	&-\Delta_{0} \Gamma_{\pi-\theta_1}^e \Upsilon_{\theta_1}^e-2 \Delta_{\pi} \Gamma_{\pi-\theta_1}^e \Upsilon_{\theta_1}^e-4 \Delta_{0} \Gamma_{\theta_1}^e \Upsilon_{\theta_1}^e+2 \Delta_{\pi} \Gamma_{\theta_1}^e \Upsilon_{\theta_1}^e-2 (\Gamma_{\theta_1}^e)^2 \Upsilon_{\theta_1}^e+2 \Delta_{\pi-\theta_1} \Upsilon_{0} \Upsilon_{\theta_1}^e-2 \Delta_{\theta_1} \Upsilon_{0} \Upsilon_{\theta_1}^e\nonumber\\
	&-4 \Delta_{0} \Upsilon_{\theta_1}^d \Upsilon_{\theta_1}^e-2 \Delta_{\pi} \Upsilon_{\theta_1}^d \Upsilon_{\theta_1}^e-2 \Gamma_{\pi-\theta_1}^e \Upsilon_{\theta_1}^d \Upsilon_{\theta_1}^e+2 (\Upsilon_{\theta_1}^d)^2 \Upsilon_{\theta_1}^e-2 \Delta_{0} (\Upsilon_{\theta_1}^e)^2+2 \Delta_{\pi} (\Upsilon_{\theta_1}^e)^2-2 \Gamma_{\pi-\theta_1}^d (\Upsilon_{\theta_1}^e)^2\nonumber\\
	&+\Gamma_{\theta_1}^d (4 (\Delta_{\pi-\theta_1})^2+2 (\Delta_{\theta_1})^2+2 (\Gamma_{\pi-\theta_1}^d)^2-2 \Delta_{0} \Gamma_{\pi-\theta_1}^e-\Gamma_{0} \Gamma_{\pi-\theta_1}^e-6 \Gamma_{\pi-\theta_1}^d \Gamma_{\pi-\theta_1}^e+(\Gamma_{\pi-\theta_1}^e)^2+2 \Delta_{0} \Gamma_{\theta_1}^e-2 \Gamma_{0} \Gamma_{\theta_1}^e\nonumber\\
	&+2 \Gamma_{\pi-\theta_1}^d \Gamma_{\theta_1}^e+3 (\Gamma_{\theta_1}^e)^2+2 \Delta_{\pi-\theta_1} \Upsilon_{0}+2 (\Upsilon_{0})^2+\Delta_{\theta_1} (2 \Delta_{\pi-\theta_1}+\Upsilon_{0})-2 \Delta_{0} \Upsilon_{\theta_1}^d-6 \Gamma_{\pi-\theta_1}^d \Upsilon_{\theta_1}^d+4 \Gamma_{\pi-\theta_1}^e \Upsilon_{\theta_1}^d\nonumber\\
	&-(\Upsilon_{\theta_1}^d)^2-2 \Gamma_{\pi-\theta_1}^e \Upsilon_{\theta_1}^e+2 \Gamma_{\theta_1}^e \Upsilon_{\theta_1}^e-2 \Upsilon_{\theta_1}^d \Upsilon_{\theta_1}^e)-\Gamma_{0}\Big((\Delta_{\pi-\theta_1})^2+6 \Delta_{\pi-\theta_1} \Delta_{\theta_1}+(\Delta_{\theta_1})^2-(\Gamma_{\pi-\theta_1}^e)^2+2 \Gamma_{\pi-\theta_1}^d \Gamma_{\theta_1}^e\nonumber\\
	&-3 (\Gamma_{\theta_1}^e)^2+4 \Delta_{\pi-\theta_1} \Upsilon_{0}+2 \Delta_{\theta_1} \Upsilon_{0}+3 (\Upsilon_{0})^2-3 \Gamma_{\pi-\theta_1}^d \Upsilon_{\theta_1}^d+(\Upsilon_{\theta_1}^d)^2-4 \Gamma_{\pi-\theta_1}^e \Upsilon_{\theta_1}^e-2 \Gamma_{\theta_1}^e \Upsilon_{\theta_1}^e+2 (\Upsilon_{\theta_1}^e)^2\Big)\Big] \nonumber\\
	&+\frac{1}{64 \Gamma_{\theta_1}^d \pi^4 N_f^2(1+v^2)^2}\frac{1}{\epsilon}\Big[\Big((\Delta_{0})^2+(\Delta_{\pi})^2\Big) \Gamma_{\pi-\theta_1}^d+(\Delta_{0}+4 \Delta_{\pi}) \Gamma_{0} \Gamma_{\pi-\theta_1}^d-\Big((\Delta_{0})^2-4 \Delta_{0} \Delta_{\pi}+3 (\Delta_{\pi})^2\Big) \Gamma_{\theta_1}^d\nonumber\\
	&+(2 \Delta_{0}+\Delta_{\pi}) \Gamma_{0} \Gamma_{\theta_1}^d+(\Gamma_{0})^2 (-2 \Gamma_{\pi-\theta_1}^d+3 \Gamma_{\theta_1}^d)\Big]\\
	A^{(2l)}_{\Gamma_{\theta_1}^e}&=-\frac{e^{-\frac{v^2}{v_c^2}}}{64\Gamma_{\theta_1}^e \pi^4N_f^2}\frac{1}{\epsilon}\Big[-4 \Delta_{\pi} (\Delta_{\pi-\theta_1})^2-4 \Delta_{0} \Delta_{\pi-\theta_1} \Delta_{\theta_1}+2 \Delta_{\pi} \Delta_{\pi-\theta_1} \Delta_{\theta_1}-2 \Delta_{\pi} (\Delta_{\theta_1})^2+2 \Delta_{\pi-\theta_1} \Delta_{\theta_1} \Gamma_{\pi-\theta_1}^d-2 (\Delta_{\theta_1})^2 \Gamma_{\pi-\theta_1}^d\nonumber\\
	&-2 \Delta_{\pi-\theta_1} \Delta_{\theta_1} \Gamma_{\pi-\theta_1}^e+2 (\Delta_{\pi-\theta_1})^2 \Gamma_{\theta_1}^e-2 \Delta_{\pi-\theta_1} \Delta_{\theta_1} \Gamma_{\theta_1}^e+4 (\Delta_{\theta_1})^2 \Gamma_{\theta_1}^e-2 \Delta_{0} \Gamma_{\pi-\theta_1}^d \Gamma_{\theta_1}^e-2 \Delta_{\pi} \Gamma_{\pi-\theta_1}^d \Gamma_{\theta_1}^e+2 (\Gamma_{\pi-\theta_1}^d)^2 \Gamma_{\theta_1}^e\nonumber\\
	&-2 \Delta_{0} \Gamma_{\pi-\theta_1}^e \Gamma_{\theta_1}^e+2 \Delta_{\pi} \Gamma_{\pi-\theta_1}^e \Gamma_{\theta_1}^e-2 \Gamma_{\pi-\theta_1}^d \Gamma_{\pi-\theta_1}^e \Gamma_{\theta_1}^e+(\Gamma_{\pi-\theta_1}^e)^2 \Gamma_{\theta_1}^e+6 (\Gamma_{\theta_1}^d)^2 \Gamma_{\theta_1}^e-2 \Delta_{0} (\Gamma_{\theta_1}^e)^2-2 \Delta_{\pi} (\Gamma_{\theta_1}^e)^2\nonumber\\
	&+2 \Gamma_{\pi-\theta_1}^d (\Gamma_{\theta_1}^e)^2+\Gamma_{\pi-\theta_1}^e (\Gamma_{\theta_1}^e)^2+(\Gamma_{\theta_1}^e)^3-2 \Delta_{\pi} \Delta_{\pi-\theta_1} \Upsilon_{0}-2 \Delta_{0} \Delta_{\theta_1} \Upsilon_{0}-3 \Delta_{\pi} \Delta_{\theta_1} \Upsilon_{0}-2 \Delta_{\theta_1} \Gamma_{\pi-\theta_1}^e \Upsilon_{0}+4 \Delta_{\pi-\theta_1} \Gamma_{\theta_1}^e \Upsilon_{0}\nonumber\\
	&-2 \Delta_{\theta_1} \Gamma_{\theta_1}^e \Upsilon_{0}+2 \Delta_{\pi} (\Upsilon_{0})^2+2 \Gamma_{\theta_1}^e (\Upsilon_{0})^2-2 \Delta_{\pi-\theta_1} \Delta_{\theta_1} \Upsilon_{\theta_1}^d-4 \Delta_{\pi} \Gamma_{\pi-\theta_1}^d \Upsilon_{\theta_1}^d-2 \Delta_{\pi} \Gamma_{\pi-\theta_1}^e \Upsilon_{\theta_1}^d-2 \Delta_{0} \Gamma_{\theta_1}^e \Upsilon_{\theta_1}^d\nonumber\\
	&+4 \Delta_{\pi} \Gamma_{\theta_1}^e \Upsilon_{\theta_1}^d-2 \Gamma_{\pi-\theta_1}^d \Gamma_{\theta_1}^e \Upsilon_{\theta_1}^d+2 \Gamma_{\pi-\theta_1}^e \Gamma_{\theta_1}^e \Upsilon_{\theta_1}^d-2 \Delta_{\theta_1} \Upsilon_{0} \Upsilon_{\theta_1}^d-2 \Delta_{\pi} (\Upsilon_{\theta_1}^d)^2+\Gamma_{\theta_1}^e (\Upsilon_{\theta_1}^d)^2+2 \Delta_{\pi-\theta_1} \Delta_{\theta_1} \Upsilon_{\theta_1}^e\nonumber\\
	&+2 (\Delta_{\theta_1})^2 \Upsilon_{\theta_1}^e-2 \Delta_{0} \Gamma_{\pi-\theta_1}^e \Upsilon_{\theta_1}^e-2 \Delta_{\pi} \Gamma_{\pi-\theta_1}^e \Upsilon_{\theta_1}^e+2 \Gamma_{\pi-\theta_1}^d \Gamma_{\pi-\theta_1}^e \Upsilon_{\theta_1}^e+2 (\Gamma_{\pi-\theta_1}^e)^2 \Upsilon_{\theta_1}^e+2 \Delta_{0} \Gamma_{\theta_1}^e \Upsilon_{\theta_1}^e-6 \Delta_{\pi} \Gamma_{\theta_1}^e \Upsilon_{\theta_1}^e\nonumber\\
	&+\Gamma_{\pi-\theta_1}^d \Gamma_{\theta_1}^e \Upsilon_{\theta_1}^e-2 \Gamma_{\pi-\theta_1}^e \Gamma_{\theta_1}^e \Upsilon_{\theta_1}^e+2 \Delta_{\pi-\theta_1} \Upsilon_{0} \Upsilon_{\theta_1}^e-2 \Delta_{\theta_1} \Upsilon_{0} \Upsilon_{\theta_1}^e+2 (\Upsilon_{0})^2 \Upsilon_{\theta_1}^e-4 \Delta_{0} \Upsilon_{\theta_1}^d \Upsilon_{\theta_1}^e+4 \Delta_{\pi} \Upsilon_{\theta_1}^d \Upsilon_{\theta_1}^e\nonumber\\
	&-4 \Gamma_{\pi-\theta_1}^d \Upsilon_{\theta_1}^d \Upsilon_{\theta_1}^e+2 \Gamma_{\pi-\theta_1}^e \Upsilon_{\theta_1}^d \Upsilon_{\theta_1}^e-2 \Gamma_{\pi-\theta_1}^e (\Upsilon_{\theta_1}^e)^2+2 \Gamma_{\theta_1}^e (\Upsilon_{\theta_1}^e)^2+2 \Upsilon_{\theta_1}^d (\Upsilon_{\theta_1}^e)^2-2 \Gamma_{\theta_1}^d \Big(2 (\Delta_{\theta_1})^2+\Delta_{0} \Gamma_{\theta_1}^e\nonumber\\
	&-4 \Gamma_{\pi-\theta_1}^d \Gamma_{\theta_1}^e-\Gamma_{\pi-\theta_1}^e \Gamma_{\theta_1}^e+(\Gamma_{\theta_1}^e)^2-\Delta_{\theta_1} (2 \Delta_{\pi-\theta_1}+\Upsilon_{0})-\Gamma_{\theta_1}^e \Upsilon_{\theta_1}^d+\Delta_{\pi} (\Gamma_{\theta_1}^e+2 \Upsilon_{\theta_1}^d)-2 \Gamma_{\pi-\theta_1}^e \Upsilon_{\theta_1}^e+\Gamma_{\theta_1}^e \Upsilon_{\theta_1}^e\nonumber\\
	&+\Upsilon_{\theta_1}^d \Upsilon_{\theta_1}^e\Big)+\Gamma_{0} \Big((\Delta_{\theta_1})^2-2 \Delta_{\theta_1} (3 \Delta_{\pi-\theta_1}+2 \Upsilon_{0})-2 [\Gamma_{\theta_1}^d \Gamma_{\theta_1}^e+\Gamma_{\theta_1}^e (\Gamma_{\pi-\theta_1}^d+\Gamma_{\pi-\theta_1}^e+\Gamma_{\theta_1}^e-2 \Upsilon_{\theta_1}^e)+\Upsilon_{\theta_1}^d (\Gamma_{\theta_1}^e+\Upsilon_{\theta_1}^e)]\Big)\Big]\nonumber\\
	&-\frac{1}{64\Gamma_{\theta_1}^e\pi^4N_f^2 (1+v^2)^2}\frac{1}{\epsilon}\Big[3 (\Delta_{0})^2 \Gamma_{\theta_1}^e-2 \Delta_{0} \Delta_{\pi} \Gamma_{\theta_1}^e+2 (\Delta_{0}+\Delta_{\pi}) \Gamma_{0} \Gamma_{\theta_1}^e-(\Gamma_{0})^2 \Gamma_{\theta_1}^e+(\Delta_{\pi})^2 (-2 \Gamma_{\pi-\theta_1}^e+\Gamma_{\theta_1}^e)\Big]
\end{align}

\begin{align}
	A^{(2l)}_{\Gamma_{\theta_2}^d}&=\frac{e^{-\frac{v^2}{v_c^2}}}{64  \pi^4N_f^2\Gamma_{\theta_2}^d}\frac{1}{\epsilon} \Big[-2 (\Delta_{\pi-\theta_1})^2 \Gamma_{\pi/2}^d+2 \Gamma_{\pi/2}^d \Gamma_{\pi-\theta_1}^d \Gamma_{\pi-\theta_1}^e-(\Delta_{\pi-\theta_1})^2 \Gamma_{\pi-\theta_2}^d-2 \Gamma_{\pi-\theta_1}^d \Gamma_{\pi-\theta_1}^e \Gamma_{\pi-\theta_2}^d+3 \Gamma_{\pi/2}^d \Gamma_{\pi-\theta_1}^e \Gamma_{\theta_1}^d\nonumber\\
	&+2 \Gamma_{\pi-\theta_1}^e \Gamma_{\pi-\theta_2}^d \Gamma_{\theta_1}^d+\Gamma_{\pi/2}^d \Gamma_{\pi-\theta_1}^d \Gamma_{\theta_1}^e+2 \Gamma_{\pi-\theta_1}^d \Gamma_{\pi-\theta_2}^d \Gamma_{\theta_1}^e-2 \Gamma_{\pi-\theta_2}^d \Gamma_{\theta_1}^d \Gamma_{\theta_1}^e-(\Delta_{\pi-\theta_1})^2 \Gamma_{\theta_2}^d+4 \Gamma_{\pi-\theta_1}^d \Gamma_{\pi-\theta_1}^e \Gamma_{\theta_2}^d\nonumber\\
	&+4 \Gamma_{\theta_1}^d \Gamma_{\theta_1}^e \Gamma_{\theta_2}^d-(\Delta_{\pi-\theta_1})^2 \Gamma_{\theta_2}^e+(\Gamma_{\pi-\theta_1}^e)^2 \Gamma_{\theta_2}^e+(\Gamma_{\theta_1}^e)^2 \Gamma_{\theta_2}^e-(\Delta_{\theta_1})^2 \Big(-2 \Gamma_{\pi/2}^d-\Gamma_{\pi-\theta_2}^d+\Gamma_{\theta_2}^d+\Gamma_{\theta_2}^e\Big)+\Delta_{\pi-\theta_1} \Gamma_{\pi-\theta_2}^d \Upsilon_{0}\nonumber\\
	&-2 \Delta_{\pi-\theta_1} \Gamma_{\theta_2}^d \Upsilon_{0}-2 \Delta_{\pi-\theta_1} \Gamma_{\theta_2}^e \Upsilon_{0}+2 \Gamma_{\pi/2}^d (\Upsilon_{0})^2+2 \Gamma_{\pi-\theta_2}^d (\Upsilon_{0})^2-\Gamma_{\theta_2}^d (\Upsilon_{0})^2-\Gamma_{\theta_2}^e (\Upsilon_{0})^2+\Delta_{\theta_1} \Big(4 \Delta_{\pi-\theta_1} \Gamma_{\pi/2}^d\nonumber\\
	&+2 \Delta_{\pi-\theta_1} \Gamma_{\pi-\theta_2}^d-2 \Delta_{\pi-\theta_1} \Gamma_{\theta_2}^d+2 \Gamma_{\pi-\theta_2}^d \Upsilon_{0}-\Gamma_{\theta_2}^d \Upsilon_{0}\Big)-\Gamma_{\pi/2}^d \Gamma_{\pi-\theta_1}^d \Upsilon_{\theta_1}^d-2 \Gamma_{\pi-\theta_1}^d \Gamma_{\pi-\theta_2}^d \Upsilon_{\theta_1}^d+2 \Gamma_{\pi/2}^d \Gamma_{\theta_1}^d \Upsilon_{\theta_1}^d\nonumber\\
	&+2 \Gamma_{\pi-\theta_2}^d \Gamma_{\theta_1}^d \Upsilon_{\theta_1}^d+4 \Gamma_{\pi-\theta_1}^d \Gamma_{\theta_2}^d \Upsilon_{\theta_1}^d+2 \Gamma_{\pi-\theta_1}^e \Gamma_{\theta_2}^e \Upsilon_{\theta_1}^d+\Gamma_{\theta_2}^e (\Upsilon_{\theta_1}^d)^2+\Gamma_{\pi-\theta_2}^e\Big((\Delta_{\pi-\theta_1})^2+(\Delta_{\theta_1})^2-(\Gamma_{\pi-\theta_1}^e)^2-(\Gamma_{\theta_1}^e)^2\nonumber\\
	&+2 \Delta_{\pi-\theta_1} \Upsilon_{0}+(\Upsilon_{0})^2-2 \Gamma_{\pi-\theta_1}^e \Upsilon_{\theta_1}^d-(\Upsilon_{\theta_1}^d)^2\Big)-3 \Gamma_{\pi/2}^d \Gamma_{\pi-\theta_1}^e \Upsilon_{\theta_1}^e+2 \Gamma_{\pi/2}^d \Gamma_{\theta_1}^e \Upsilon_{\theta_1}^e+2 \Gamma_{\pi-\theta_1}^e \Gamma_{\theta_2}^d \Upsilon_{\theta_1}^e-2 \Gamma_{\theta_1}^e \Gamma_{\theta_2}^d \Upsilon_{\theta_1}^e\nonumber\\
	&+4 \Gamma_{\pi/2}^d \Upsilon_{\theta_1}^d \Upsilon_{\theta_1}^e+2 \Gamma_{\theta_2}^d \Upsilon_{\theta_1}^d \Upsilon_{\theta_1}^e\Big] +\frac{1}{64  \pi^4N_f^2 \Gamma_{\theta_2}^d(1+v^2)^2}\frac{1}{\epsilon}\Big[(\Delta_{\pi})^2 (\Gamma_{\pi-\theta_2}^d-3 \Gamma_{\theta_2}^d)+(\Delta_{0})^2 (\Gamma_{\pi-\theta_2}^d-\Gamma_{\theta_2}^d)+4 \Delta_{0} \Delta_{\pi} \Gamma_{\theta_2}^d\nonumber\\
	&+\Delta_{\pi} \Gamma_{0} (4 \Gamma_{\pi-\theta_2}^d+\Gamma_{\theta_2}^d)+\Delta_{0} \Gamma_{0} (\Gamma_{\pi-\theta_2}^d+2 \Gamma_{\theta_2}^d)+(\Gamma_{0})^2 (-2 \Gamma_{\pi-\theta_2}^d+3 \Gamma_{\theta_2}^d)\Big]\\
	A^{(2l)}_{\Gamma_{\theta_2}^e}&=-\frac{e^{-\frac{v^2}{v_c^2}}}{64\pi^4N_f^2\Gamma_{\theta_2}^e}\frac{1}{\epsilon} \Big[-2 (\Delta_{\pi-\theta_1})^2 \Gamma_{\pi/2}^e+(\Delta_{\pi-\theta_1})^2 \Gamma_{\theta_2}^e-2 \Gamma_{\pi-\theta_1}^d \Gamma_{\pi-\theta_1}^e \Gamma_{\theta_2}^e+2 \Gamma_{\pi-\theta_1}^e \Gamma_{\theta_1}^d \Gamma_{\theta_2}^e+2 \Gamma_{\pi-\theta_1}^d \Gamma_{\theta_1}^e \Gamma_{\theta_2}^e\nonumber\\
	&-2 \Gamma_{\theta_1}^d \Gamma_{\theta_1}^e \Gamma_{\theta_2}^e+(\Delta_{\theta_1})^2 (2 \Gamma_{\pi/2}^e+\Gamma_{\theta_2}^e)-4 \Delta_{\theta_1} \Gamma_{\pi/2}^e \Upsilon_{0}+2 \Delta_{\pi-\theta_1} \Gamma_{\theta_2}^e \Upsilon_{0}+2 \Gamma_{\pi/2}^e (\Upsilon_{0})^2+\Gamma_{\theta_2}^e (\Upsilon_{0})^2\nonumber\\
	&-2 \Delta_{\theta_1} \Gamma_{\pi-\theta_2}^e (\Delta_{\pi-\theta_1}+\Upsilon_{0})-4 \Gamma_{\pi/2}^e \Gamma_{\pi-\theta_1}^d \Upsilon_{\theta_1}^d-4 \Gamma_{\pi/2}^e \Gamma_{\theta_1}^d \Upsilon_{\theta_1}^d-2 \Gamma_{\pi-\theta_1}^d \Gamma_{\theta_2}^e \Upsilon_{\theta_1}^d+2 \Gamma_{\theta_1}^d \Gamma_{\theta_2}^e \Upsilon_{\theta_1}^d+4 \Gamma_{\pi/2}^e \Upsilon_{\theta_1}^d \Upsilon_{\theta_1}^e\nonumber\\
	&+2 \Gamma_{\pi-\theta_2}^e (\Gamma_{\pi-\theta_1}^e-\Gamma_{\theta_1}^e+\Upsilon_{\theta_1}^d) \Upsilon_{\theta_1}^e\Big] -\frac{1}{64 \Gamma_{\theta_2}^e \pi^4N_f^2 (1+v^2)^2}\Big[-2 (\Delta_{\pi})^2 \Gamma_{\pi-\theta_2}^e+\Big(3 (\Delta_{0})^2-2 \Delta_{0} \Delta_{\pi}+(\Delta_{\pi})^2\nonumber\\
	&+2 (\Delta_{0}+\Delta_{\pi}) \Gamma_{0}-(\Gamma_{0})^2\Big) \Gamma_{\theta_2}^e\Big]
\end{align}

\begin{align}
	A^{(2l)}_{\Gamma_{\pi/2}^d}&=\frac{e^{-\frac{v^2}{v_c^2}}}{128\Gamma_{\pi/2}^d\pi^4N_f^2}\frac{1}{\epsilon}\Big[4 \Gamma_{\pi/2}^d \Gamma_{\pi-\theta_1}^d \Gamma_{\pi-\theta_1}^e+2 \Gamma_{\pi-\theta_1}^d \Gamma_{\pi-\theta_1}^e \Gamma_{\pi-\theta_2}^d+4 \Gamma_{\pi/2}^d \Gamma_{\pi-\theta_1}^e \Gamma_{\theta_1}^d+3 \Gamma_{\pi-\theta_1}^e \Gamma_{\pi-\theta_2}^d \Gamma_{\theta_1}^d+4 \Gamma_{\pi/2}^d \Gamma_{\pi-\theta_1}^d \Gamma_{\theta_1}^e\nonumber\\
	&+\Gamma_{\pi-\theta_1}^d \Gamma_{\pi-\theta_2}^d \Gamma_{\theta_1}^e+4 \Gamma_{\pi/2}^d \Gamma_{\theta_1}^d \Gamma_{\theta_1}^e+2 \Gamma_{\pi-\theta_1}^d \Gamma_{\pi-\theta_1}^e \Gamma_{\theta_2}^d+3 \Gamma_{\pi-\theta_1}^e \Gamma_{\theta_1}^d \Gamma_{\theta_2}^d+\Gamma_{\pi-\theta_1}^d \Gamma_{\theta_1}^e \Gamma_{\theta_2}^d+4 \Delta_{\pi-\theta_1} \Delta_{\theta_1} (\Gamma_{\pi-\theta_2}^d+\Gamma_{\theta_2}^d)\nonumber\\
	&+2 (\Delta_{\theta_1})^2 (\Gamma_{\pi-\theta_2}^d+\Gamma_{\theta_2}^d)-2 (\Delta_{\pi-\theta_1})^2 (2 \Gamma_{\pi/2}^d+\Gamma_{\pi-\theta_2}^d+\Gamma_{\theta_2}^d)-2 \Delta_{\pi-\theta_1} \Gamma_{\pi/2}^d \Upsilon_{0}+2 \Delta_{\theta_1} \Gamma_{\pi/2}^d \Upsilon_{0}+2 \Gamma_{\pi/2}^d (\Upsilon_{0})^2\nonumber\\
	&+2 \Gamma_{\pi-\theta_2}^d (\Upsilon_{0})^2+2 \Gamma_{\theta_2}^d (\Upsilon_{0})^2+4 \Gamma_{\pi/2}^d \Gamma_{\pi-\theta_1}^d \Upsilon_{\theta_1}^d-\Gamma_{\pi-\theta_1}^d \Gamma_{\pi-\theta_2}^d \Upsilon_{\theta_1}^d+4 \Gamma_{\pi/2}^d \Gamma_{\theta_1}^d \Upsilon_{\theta_1}^d+2 \Gamma_{\pi-\theta_2}^d \Gamma_{\theta_1}^d \Upsilon_{\theta_1}^d-\Gamma_{\pi-\theta_1}^d \Gamma_{\theta_2}^d \Upsilon_{\theta_1}^d\nonumber\\
	&+2 \Gamma_{\theta_1}^d \Gamma_{\theta_2}^d \Upsilon_{\theta_1}^d+4 \Gamma_{\pi/2}^d \Gamma_{\pi-\theta_1}^e \Upsilon_{\theta_1}^e-3 \Gamma_{\pi-\theta_1}^e \Gamma_{\pi-\theta_2}^d \Upsilon_{\theta_1}^e-4 \Gamma_{\pi/2}^d \Gamma_{\theta_1}^e \Upsilon_{\theta_1}^e+2 \Gamma_{\pi-\theta_2}^d \Gamma_{\theta_1}^e \Upsilon_{\theta_1}^e-3 \Gamma_{\pi-\theta_1}^e \Gamma_{\theta_2}^d \Upsilon_{\theta_1}^e\nonumber\\
	&+2 \Gamma_{\theta_1}^e \Gamma_{\theta_2}^d \Upsilon_{\theta_1}^e+4 \Gamma_{\pi/2}^d \Upsilon_{\theta_1}^d \Upsilon_{\theta_1}^e+4 \Gamma_{\pi-\theta_2}^d \Upsilon_{\theta_1}^d \Upsilon_{\theta_1}^e+4 \Gamma_{\theta_2}^d \Upsilon_{\theta_1}^d \Upsilon_{\theta_1}^e\Big] +
	\frac{1}{64 \pi^4 N_f^2(1+v^2)^2}\Big[4 \Delta_{0} \Delta_{\pi}-2 (\Delta_{\pi})^2\nonumber\\
	&+3 \Delta_{0} \Gamma_{0}+5 \Delta_{\pi} \Gamma_{0}+(\Gamma_{0})^2\Big]\\
	A^{(2l)}_{\Gamma_{\pi/2}^e}&=-\frac{e^{-\frac{v^2}{v_c^2}}}{64 \Gamma_{\pi/2}^e  \pi^4N_f^2}\frac{1}{\epsilon} \Big[(\Delta_{\pi-\theta_1})^2 \Gamma_{\pi/2}^e-2 \Gamma_{\pi/2}^e \Gamma_{\pi-\theta_1}^d \Gamma_{\pi-\theta_1}^e+2 \Gamma_{\pi/2}^e \Gamma_{\pi-\theta_1}^e \Gamma_{\theta_1}^d+2 \Gamma_{\pi/2}^e \Gamma_{\pi-\theta_1}^d \Gamma_{\theta_1}^e-2 \Gamma_{\pi/2}^e \Gamma_{\theta_1}^d \Gamma_{\theta_1}^e\nonumber\\
	&-(\Delta_{\pi-\theta_1})^2 \Gamma_{\theta_2}^e+(\Delta_{\theta_1})^2 (\Gamma_{\pi/2}^e+\Gamma_{\theta_2}^e)+2 \Delta_{\pi-\theta_1} \Gamma_{\pi/2}^e \Upsilon_{0}+\Gamma_{\pi/2}^e (\Upsilon_{0})^2+\Gamma_{\theta_2}^e (\Upsilon_{0})^2-2 \Delta_{\theta_1} \Big(\Delta_{\pi-\theta_1} \Gamma_{\pi/2}^e\nonumber\\
	&+(\Gamma_{\pi/2}^e+\Gamma_{\theta_2}^e) \Upsilon_{0}\Big)-2 \Gamma_{\pi/2}^e \Gamma_{\pi-\theta_1}^d \Upsilon_{\theta_1}^d+2 \Gamma_{\pi/2}^e \Gamma_{\theta_1}^d \Upsilon_{\theta_1}^d-2 \Gamma_{\pi-\theta_1}^d \Gamma_{\theta_2}^e \Upsilon_{\theta_1}^d-2 \Gamma_{\theta_1}^d \Gamma_{\theta_2}^e \Upsilon_{\theta_1}^d+2 \Gamma_{\pi/2}^e \Gamma_{\pi-\theta_1}^e \Upsilon_{\theta_1}^e\nonumber\\
	&-2 \Gamma_{\pi/2}^e \Gamma_{\theta_1}^e \Upsilon_{\theta_1}^e+2 \Gamma_{\pi/2}^e \Upsilon_{\theta_1}^d \Upsilon_{\theta_1}^e+2 \Gamma_{\theta_2}^e \Upsilon_{\theta_1}^d \Upsilon_{\theta_1}^e+\Gamma_{\pi-\theta_2}^e \Big(-(\Delta_{\pi-\theta_1})^2+(\Delta_{\theta_1})^2-2 \Delta_{\theta_1} \Upsilon_{0}+(\Upsilon_{0})^2-2 \Gamma_{\pi-\theta_1}^d \Upsilon_{\theta_1}^d\nonumber\\
	&-2 \Gamma_{\theta_1}^d \Upsilon_{\theta_1}^d+2 \Upsilon_{\theta_1}^d \Upsilon_{\theta_1}^e\Big)\Big] +\frac{1}{64 \pi^4N_f^2 (1+v^2)^2)}\frac{1}{\epsilon}\Big[-3 (\Delta_{0})^2+2 \Delta_{0} \Delta_{\pi}+(\Delta_{\pi})^2-2 (\Delta_{0}+\Delta_{\pi}) \Gamma_{0}+(\Gamma_{0})^2\Big]
\end{align}

\begin{align}
	A^{(2l)}_{\Gamma_{\pi-\theta_1}^d}&=-\frac{e^{-\frac{v^2}{v_c^2}}}{64 \pi^4N_f^2}\frac{1}{\epsilon} \Big[2 \Delta_{0} (\Delta_{\pi-\theta_1})^2-\Delta_{\pi} (\Delta_{\pi-\theta_1})^2-6 \Delta_{0} \Delta_{\pi-\theta_1} \Delta_{\theta_1}+2 \Delta_{\pi} \Delta_{\pi-\theta_1} \Delta_{\theta_1}+2 \Delta_{0} (\Delta_{\theta_1})^2-\Delta_{\pi} (\Delta_{\theta_1})^2\nonumber\\
	&+2 (\Delta_{\pi-\theta_1})^2 \Gamma_{\pi-\theta_1}^d+2 \Delta_{\pi-\theta_1} \Delta_{\theta_1} \Gamma_{\pi-\theta_1}^d+4 (\Delta_{\theta_1})^2 \Gamma_{\pi-\theta_1}^d+2 (\Gamma_{\pi-\theta_1}^d)^3-(\Delta_{\pi-\theta_1})^2 \Gamma_{\pi-\theta_1}^e+2 \Delta_{\pi-\theta_1} \Delta_{\theta_1} \Gamma_{\pi-\theta_1}^e\nonumber\\
	&+(\Delta_{\theta_1})^2 \Gamma_{\pi-\theta_1}^e-4 (\Gamma_{\pi-\theta_1}^d)^2 \Gamma_{\pi-\theta_1}^e-2 \Delta_{0} (\Gamma_{\pi-\theta_1}^e)^2+\Delta_{\pi} (\Gamma_{\pi-\theta_1}^e)^2+3 \Gamma_{\pi-\theta_1}^d (\Gamma_{\pi-\theta_1}^e)^2-(\Gamma_{\pi-\theta_1}^e)^3-(\Delta_{\pi-\theta_1})^2 \Gamma_{\theta_1}^e\nonumber\\
	&+2 \Delta_{\pi-\theta_1} \Delta_{\theta_1} \Gamma_{\theta_1}^e-3 (\Delta_{\theta_1})^2 \Gamma_{\theta_1}^e-2 \Delta_{0} \Gamma_{\pi-\theta_1}^d \Gamma_{\theta_1}^e-2 \Delta_{\pi} \Gamma_{\pi-\theta_1}^e \Gamma_{\theta_1}^e+(\Gamma_{\pi-\theta_1}^e)^2 \Gamma_{\theta_1}^e+\Delta_{\pi} (\Gamma_{\theta_1}^e)^2+\Gamma_{\pi-\theta_1}^d (\Gamma_{\theta_1}^e)^2\nonumber\\
	&-\Gamma_{\pi-\theta_1}^e (\Gamma_{\theta_1}^e)^2+(\Gamma_{\theta_1}^e)^3-2 \Delta_{0} \Delta_{\pi-\theta_1} \Upsilon_{0}-2 \Delta_{\pi} \Delta_{\pi-\theta_1} \Upsilon_{0}-2 \Delta_{0} \Delta_{\theta_1} \Upsilon_{0}+6 \Delta_{\pi-\theta_1} \Gamma_{\pi-\theta_1}^d \Upsilon_{0}+\Delta_{\theta_1} \Gamma_{\pi-\theta_1}^d \Upsilon_{0}\nonumber\\
	&-2 \Delta_{\pi-\theta_1} \Gamma_{\theta_1}^e \Upsilon_{0}+2 \Delta_{\theta_1} \Gamma_{\theta_1}^e \Upsilon_{0}-4 \Delta_{0} (\Upsilon_{0})^2-\Delta_{\pi} (\Upsilon_{0})^2+\Gamma_{\pi-\theta_1}^e (\Upsilon_{0})^2-\Gamma_{\theta_1}^e (\Upsilon_{0})^2+3 \Delta_{0} \Gamma_{\pi-\theta_1}^d \Upsilon_{\theta_1}^d\nonumber\\
	&-4 (\Gamma_{\pi-\theta_1}^d)^2 \Upsilon_{\theta_1}^d-2 \Delta_{0} \Gamma_{\pi-\theta_1}^e \Upsilon_{\theta_1}^d+2 \Delta_{\pi} \Gamma_{\pi-\theta_1}^e \Upsilon_{\theta_1}^d-2 (\Gamma_{\pi-\theta_1}^e)^2 \Upsilon_{\theta_1}^d+2 \Gamma_{\pi-\theta_1}^e \Gamma_{\theta_1}^e \Upsilon_{\theta_1}^d+2 \Delta_{\pi-\theta_1} \Upsilon_{0} \Upsilon_{\theta_1}^d-2 \Delta_{\theta_1} \Upsilon_{0} \Upsilon_{\theta_1}^d\nonumber\\
	&+2 (\Upsilon_{0})^2 \Upsilon_{\theta_1}^d+\Delta_{\pi} (\Upsilon_{\theta_1}^d)^2+\Gamma_{\pi-\theta_1}^d (\Upsilon_{\theta_1}^d)^2-\Gamma_{\pi-\theta_1}^e (\Upsilon_{\theta_1}^d)^2+\Gamma_{\theta_1}^e (\Upsilon_{\theta_1}^d)^2-2 (\Gamma_{\theta_1}^d)^2 (-\Gamma_{\pi-\theta_1}^d+\Gamma_{\pi-\theta_1}^e-\Gamma_{\theta_1}^e+\Upsilon_{\theta_1}^d)\nonumber\\
	&-2 (\Delta_{\theta_1})^2 \Upsilon_{\theta_1}^e+4 \Delta_{0} \Gamma_{\pi-\theta_1}^e \Upsilon_{\theta_1}^e-2 \Delta_{\pi} \Gamma_{\pi-\theta_1}^e \Upsilon_{\theta_1}^e-2 \Gamma_{\pi-\theta_1}^d \Gamma_{\pi-\theta_1}^e \Upsilon_{\theta_1}^e-2 (\Gamma_{\pi-\theta_1}^e)^2 \Upsilon_{\theta_1}^e+2 \Delta_{0} \Gamma_{\theta_1}^e \Upsilon_{\theta_1}^e+2 \Delta_{\pi} \Gamma_{\theta_1}^e \Upsilon_{\theta_1}^e\nonumber\\
	&+2 \Gamma_{\pi-\theta_1}^d \Gamma_{\theta_1}^e \Upsilon_{\theta_1}^e-2 \Gamma_{\pi-\theta_1}^e \Gamma_{\theta_1}^e \Upsilon_{\theta_1}^e-2 \Delta_{\pi-\theta_1} \Upsilon_{0} \Upsilon_{\theta_1}^e+2 (\Upsilon_{0})^2 \Upsilon_{\theta_1}^e-2 \Delta_{\pi} \Upsilon_{\theta_1}^d \Upsilon_{\theta_1}^e-2 \Gamma_{\pi-\theta_1}^d \Upsilon_{\theta_1}^d \Upsilon_{\theta_1}^e+2 \Gamma_{\pi-\theta_1}^e \Upsilon_{\theta_1}^d \Upsilon_{\theta_1}^e\nonumber\\
	&-2 \Gamma_{\theta_1}^e \Upsilon_{\theta_1}^d \Upsilon_{\theta_1}^e-2 \Delta_{0} (\Upsilon_{\theta_1}^e)^2+2 \Delta_{\pi} (\Upsilon_{\theta_1}^e)^2+\Gamma_{\theta_1}^d \Big((\Delta_{\pi-\theta_1})^2+(\Delta_{\theta_1})^2+4 (\Gamma_{\pi-\theta_1}^d)^2-\Delta_{0} \Gamma_{\pi-\theta_1}^e+2 \Gamma_{\pi-\theta_1}^d \Gamma_{\pi-\theta_1}^e\nonumber\\
	&+2 (\Gamma_{\pi-\theta_1}^e)^2-2 \Delta_{0} \Gamma_{\theta_1}^e-6 \Gamma_{\pi-\theta_1}^d \Gamma_{\theta_1}^e+\Delta_{\pi-\theta_1} \Upsilon_{0}-4 (\Upsilon_{0})^2-2 \Delta_{\theta_1} (\Delta_{\pi-\theta_1}+\Upsilon_{0})+2 \Gamma_{\pi-\theta_1}^d \Upsilon_{\theta_1}^d-2 \Gamma_{\pi-\theta_1}^e \Upsilon_{\theta_1}^d\nonumber\\
	&-2 (\Upsilon_{\theta_1}^e)^2\Big)+\Gamma_{0} \Big((\Delta_{\pi-\theta_1})^2+(\Delta_{\theta_1})^2-2 \Gamma_{\pi-\theta_1}^d \Gamma_{\pi-\theta_1}^e-3 (\Gamma_{\pi-\theta_1}^e)^2-2 \Gamma_{\pi-\theta_1}^e \Gamma_{\theta_1}^d+\Gamma_{\pi-\theta_1}^d \Gamma_{\theta_1}^e+2 \Gamma_{\theta_1}^d \Gamma_{\theta_1}^e-(\Gamma_{\theta_1}^e)^2\nonumber\\
	&-(\Upsilon_{0})^2+2 \Delta_{\theta_1} (\Delta_{\pi-\theta_1}+\Upsilon_{0})-2 \Gamma_{\pi-\theta_1}^d \Upsilon_{\theta_1}^d-4 \Gamma_{\pi-\theta_1}^e \Upsilon_{\theta_1}^d-2 \Gamma_{\theta_1}^d \Upsilon_{\theta_1}^d-(\Upsilon_{\theta_1}^d)^2-\Gamma_{\pi-\theta_1}^e \Upsilon_{\theta_1}^e-4 \Gamma_{\theta_1}^e \Upsilon_{\theta_1}^e-4 \Upsilon_{\theta_1}^d \Upsilon_{\theta_1}^e\nonumber\\
	&-2 (\Upsilon_{\theta_1}^e)^2\Big)\Big] +\frac{1}{64  \pi^4N_f^2 (1+v^2)^2}\frac{1}{\epsilon}\Big[-\Big((\Delta_{0})^2-4 \Delta_{0} \Delta_{\pi}+3 (\Delta_{\pi})^2\Big) \Gamma_{\pi-\theta_1}^d+(2 \Delta_{0}+\Delta_{\pi}) \Gamma_{0} \Gamma_{\pi-\theta_1}^d\nonumber\\
	&+(\Gamma_{0})^2 (3 \Gamma_{\pi-\theta_1}^d-2 \Gamma_{\theta_1}^d)+\Big((\Delta_{0})^2+(\Delta_{\pi})^2\Big) \Gamma_{\theta_1}^d+(\Delta_{0}+4 \Delta_{\pi}) \Gamma_{0} \Gamma_{\theta_1}^d\Big]\\
	A^{(2l)}_{\Gamma_{\pi-\theta_1}^e}&=-\frac{e^{-\frac{v^2}{v_c^2}}}{64 \Gamma_{\pi-\theta_1}^e \pi^4N_f^2}\frac{1}{\epsilon} \Big[-2 \Delta_{0} (\Delta_{\pi-\theta_1})^2+4 \Delta_{0} \Delta_{\pi-\theta_1} \Delta_{\theta_1}-4 (\Delta_{\pi-\theta_1})^2 \Gamma_{\pi-\theta_1}^d+8 \Delta_{\pi-\theta_1} \Delta_{\theta_1} \Gamma_{\pi-\theta_1}^d+4 (\Delta_{\pi-\theta_1})^2 \Gamma_{\pi-\theta_1}^e\nonumber\\
	&-2 \Delta_{\pi-\theta_1} \Delta_{\theta_1} \Gamma_{\pi-\theta_1}^e+2 (\Delta_{\theta_1})^2 \Gamma_{\pi-\theta_1}^e+2 \Delta_{0} \Gamma_{\pi-\theta_1}^d \Gamma_{\pi-\theta_1}^e+6 (\Gamma_{\pi-\theta_1}^d)^2 \Gamma_{\pi-\theta_1}^e+2 \Delta_{0} (\Gamma_{\pi-\theta_1}^e)^2-2 \Gamma_{\pi-\theta_1}^d (\Gamma_{\pi-\theta_1}^e)^2\nonumber\\
	&+(\Gamma_{\pi-\theta_1}^e)^3-2 (\Delta_{\pi-\theta_1})^2 \Gamma_{\theta_1}^d+6 \Delta_{\pi-\theta_1} \Delta_{\theta_1} \Gamma_{\theta_1}^d+2 \Delta_{0} \Gamma_{\pi-\theta_1}^e \Gamma_{\theta_1}^d+8 \Gamma_{\pi-\theta_1}^d \Gamma_{\pi-\theta_1}^e \Gamma_{\theta_1}^d+2 (\Gamma_{\pi-\theta_1}^e)^2 \Gamma_{\theta_1}^d+2 \Gamma_{\pi-\theta_1}^e (\Gamma_{\theta_1}^d)^2\nonumber\\
	&+4 (\Delta_{\pi-\theta_1})^2 \Gamma_{\theta_1}^e-2 \Delta_{\pi-\theta_1} \Delta_{\theta_1} \Gamma_{\theta_1}^e+2 \Delta_{0} \Gamma_{\pi-\theta_1}^e \Gamma_{\theta_1}^e+2 \Gamma_{\pi-\theta_1}^d \Gamma_{\pi-\theta_1}^e \Gamma_{\theta_1}^e+(\Gamma_{\pi-\theta_1}^e)^2 \Gamma_{\theta_1}^e-2 \Gamma_{\pi-\theta_1}^e \Gamma_{\theta_1}^d \Gamma_{\theta_1}^e\nonumber\\
	&+\Gamma_{\pi-\theta_1}^e (\Gamma_{\theta_1}^e)^2-3 \Delta_{0} \Delta_{\pi-\theta_1} \Upsilon_{0}+2 \Delta_{0} \Delta_{\theta_1} \Upsilon_{0}-2 \Delta_{\theta_1} \Gamma_{\pi-\theta_1}^d \Upsilon_{0}+6 \Delta_{\pi-\theta_1} \Gamma_{\pi-\theta_1}^e \Upsilon_{0}+2 \Delta_{\pi-\theta_1} \Gamma_{\theta_1}^e \Upsilon_{0}-4 \Delta_{0} (\Upsilon_{0})^2\nonumber\\
	&+2 \Gamma_{\pi-\theta_1}^e (\Upsilon_{0})^2-2 \Gamma_{\theta_1}^d (\Upsilon_{0})^2-2 \Gamma_{\theta_1}^e (\Upsilon_{0})^2-2 (\Delta_{\pi-\theta_1})^2 \Upsilon_{\theta_1}^d+2 \Delta_{\pi-\theta_1} \Delta_{\theta_1} \Upsilon_{\theta_1}^d-2 \Delta_{0} \Gamma_{\pi-\theta_1}^d \Upsilon_{\theta_1}^d+4 \Delta_{0} \Gamma_{\pi-\theta_1}^e \Upsilon_{\theta_1}^d\nonumber\\
	&-2 \Gamma_{\pi-\theta_1}^d \Gamma_{\pi-\theta_1}^e \Upsilon_{\theta_1}^d+2 \Delta_{0} \Gamma_{\theta_1}^d \Upsilon_{\theta_1}^d-\Gamma_{\pi-\theta_1}^d \Gamma_{\theta_1}^d \Upsilon_{\theta_1}^d+2 \Gamma_{\pi-\theta_1}^e \Gamma_{\theta_1}^d \Upsilon_{\theta_1}^d+(\Gamma_{\theta_1}^d)^2 \Upsilon_{\theta_1}^d-2 \Delta_{0} \Gamma_{\theta_1}^e \Upsilon_{\theta_1}^d-(\Gamma_{\theta_1}^e)^2 \Upsilon_{\theta_1}^d\nonumber\\
	&-2 \Delta_{\pi-\theta_1} \Upsilon_{0} \Upsilon_{\theta_1}^d+2 \Delta_{0} (\Upsilon_{\theta_1}^d)^2-2 \Gamma_{\theta_1}^e (\Upsilon_{\theta_1}^d)^2+(\Upsilon_{\theta_1}^d)^3+2 (\Delta_{\pi-\theta_1})^2 \Upsilon_{\theta_1}^e+2 (\Delta_{\theta_1})^2 \Upsilon_{\theta_1}^e-3 \Gamma_{\pi-\theta_1}^d \Gamma_{\pi-\theta_1}^e \Upsilon_{\theta_1}^e\nonumber\\
	&+2 \Gamma_{\pi-\theta_1}^d \Gamma_{\theta_1}^e \Upsilon_{\theta_1}^e+2 \Gamma_{\pi-\theta_1}^e \Gamma_{\theta_1}^e \Upsilon_{\theta_1}^e-2 (\Gamma_{\theta_1}^e)^2 \Upsilon_{\theta_1}^e-2 \Delta_{\pi-\theta_1} \Upsilon_{0} \Upsilon_{\theta_1}^e-2 \Delta_{0} \Upsilon_{\theta_1}^d \Upsilon_{\theta_1}^e-2 \Gamma_{\pi-\theta_1}^d \Upsilon_{\theta_1}^d \Upsilon_{\theta_1}^e-4 \Gamma_{\theta_1}^d \Upsilon_{\theta_1}^d \Upsilon_{\theta_1}^e\nonumber\\
	&+2 \Gamma_{\theta_1}^e \Upsilon_{\theta_1}^d \Upsilon_{\theta_1}^e+2 \Gamma_{\pi-\theta_1}^e (\Upsilon_{\theta_1}^e)^2-2 \Gamma_{\theta_1}^e (\Upsilon_{\theta_1}^e)^2-\Upsilon_{\theta_1}^d (\Upsilon_{\theta_1}^e)^2+\Delta_{\pi} \Big(2 (\Delta_{\pi-\theta_1})^2+4 (\Delta_{\theta_1})^2+2 \Gamma_{\pi-\theta_1}^d \Gamma_{\pi-\theta_1}^e\nonumber\\
	&+2 (\Gamma_{\pi-\theta_1}^e)^2+2 \Gamma_{\pi-\theta_1}^e \Gamma_{\theta_1}^d-2 \Gamma_{\pi-\theta_1}^e \Gamma_{\theta_1}^e-2 \Delta_{\pi-\theta_1} \Upsilon_{0}-6 \Delta_{\theta_1} \Upsilon_{0}-(\Upsilon_{0})^2+2 \Gamma_{\pi-\theta_1}^d \Upsilon_{\theta_1}^d+2 \Gamma_{\pi-\theta_1}^e \Upsilon_{\theta_1}^d-2 \Gamma_{\theta_1}^d \Upsilon_{\theta_1}^d\nonumber\\
	&+2 \Upsilon_{\theta_1}^d \Upsilon_{\theta_1}^e\Big)-\Gamma_{0} \Big((\Delta_{\pi-\theta_1})^2+2 \Delta_{\pi-\theta_1} \Upsilon_{0}+\Delta_{\theta_1} (-6 \Delta_{\pi-\theta_1}+\Upsilon_{0})+2 (-\Gamma_{\pi-\theta_1}^d \Gamma_{\pi-\theta_1}^e-(\Gamma_{\pi-\theta_1}^e)^2-\Gamma_{\pi-\theta_1}^e \Gamma_{\theta_1}^d\nonumber\\
	&-\Gamma_{\pi-\theta_1}^e \Gamma_{\theta_1}^e+(\Upsilon_{0})^2+3 \Gamma_{\pi-\theta_1}^d \Upsilon_{\theta_1}^d+\Gamma_{\theta_1}^d \Upsilon_{\theta_1}^d-\Gamma_{\theta_1}^e \Upsilon_{\theta_1}^d+(\Upsilon_{\theta_1}^d)^2+\Gamma_{\pi-\theta_1}^e \Upsilon_{\theta_1}^e+\Gamma_{\theta_1}^e \Upsilon_{\theta_1}^e)\Big)\Big]\nonumber\\
	&-\frac{1}{64 \Gamma_{\pi-\theta_1}^e \pi^4N_f^2 (1+v^2)^2}\frac{1}{\epsilon}\Big[2 \Delta_{\pi} (-\Delta_{0}+\Gamma_{0}) \Gamma_{\pi-\theta_1}^e+(3 (\Delta_{0})^2+2 \Delta_{0} \Gamma_{0}-(\Gamma_{0})^2) \Gamma_{\pi-\theta_1}^e+(\Delta_{\pi})^2 \Big(\Gamma_{\pi-\theta_1}^e-2 \Gamma_{\theta_1}^e\Big)\Big]
\end{align}

\begin{align}
	A^{(2l)}_{\Gamma_{\pi-\theta_2}^d}&=-\frac{e^{-\frac{v^2}{v_c^2}}}{64\Gamma_{\pi-\theta_2}^d  \pi^4N_f^2}\frac{1}{\epsilon}\Big[2 (\Delta_{\pi-\theta_1})^2 \Gamma_{\pi/2}^d-2 \Gamma_{\pi/2}^d \Gamma_{\pi-\theta_1}^d \Gamma_{\pi-\theta_1}^e+(\Delta_{\pi-\theta_1})^2 \Gamma_{\pi-\theta_2}^d-4 \Gamma_{\pi-\theta_1}^d \Gamma_{\pi-\theta_1}^e \Gamma_{\pi-\theta_2}^d-3 \Gamma_{\pi/2}^d \Gamma_{\pi-\theta_1}^e \Gamma_{\theta_1}^d\nonumber\\
	&-\Gamma_{\pi/2}^d \Gamma_{\pi-\theta_1}^d \Gamma_{\theta_1}^e-4 \Gamma_{\pi-\theta_2}^d \Gamma_{\theta_1}^d \Gamma_{\theta_1}^e+(\Delta_{\pi-\theta_1})^2 \Gamma_{\theta_2}^d+2 \Gamma_{\pi-\theta_1}^d \Gamma_{\pi-\theta_1}^e \Gamma_{\theta_2}^d-2 \Gamma_{\pi-\theta_1}^e \Gamma_{\theta_1}^d \Gamma_{\theta_2}^d-2 \Gamma_{\pi-\theta_1}^d \Gamma_{\theta_1}^e \Gamma_{\theta_2}^d\nonumber\\
	&+2 \Gamma_{\theta_1}^d \Gamma_{\theta_1}^e \Gamma_{\theta_2}^d-2 \Delta_{\pi-\theta_1} \Delta_{\theta_1}\Big(2 \Gamma_{\pi/2}^d-\Gamma_{\pi-\theta_2}^d+\Gamma_{\theta_2}^d\Big)-(\Delta_{\pi-\theta_1})^2 \Gamma_{\theta_2}^e+(\Gamma_{\pi-\theta_1}^e)^2 \Gamma_{\theta_2}^e+(\Gamma_{\theta_1}^e)^2 \Gamma_{\theta_2}^e-(\Delta_{\theta_1})^2 (2 \Gamma_{\pi/2}^d\nonumber\\
	&-\Gamma_{\pi-\theta_2}^d+\Gamma_{\theta_2}^d+\Gamma_{\theta_2}^e)+2 \Delta_{\pi-\theta_1} \Gamma_{\pi-\theta_2}^d \Upsilon_{0}+\Delta_{\theta_1} (\Gamma_{\pi-\theta_2}^d-2 \Gamma_{\theta_2}^d) \Upsilon_{0}-\Delta_{\pi-\theta_1} \Gamma_{\theta_2}^d \Upsilon_{0}-2 \Delta_{\pi-\theta_1} \Gamma_{\theta_2}^e \Upsilon_{0}-2 \Gamma_{\pi/2}^d (\Upsilon_{0})^2\nonumber\\
	&+\Gamma_{\pi-\theta_2}^d (\Upsilon_{0})^2-2 \Gamma_{\theta_2}^d (\Upsilon_{0})^2-\Gamma_{\theta_2}^e (\Upsilon_{0})^2+\Gamma_{\pi/2}^d \Gamma_{\pi-\theta_1}^d \Upsilon_{\theta_1}^d-4 \Gamma_{\pi-\theta_1}^d \Gamma_{\pi-\theta_2}^d \Upsilon_{\theta_1}^d-2 \Gamma_{\pi/2}^d \Gamma_{\theta_1}^d \Upsilon_{\theta_1}^d+2 \Gamma_{\pi-\theta_1}^d \Gamma_{\theta_2}^d \Upsilon_{\theta_1}^d\nonumber\\
	&-2 \Gamma_{\theta_1}^d \Gamma_{\theta_2}^d \Upsilon_{\theta_1}^d+2 \Gamma_{\pi-\theta_1}^e \Gamma_{\theta_2}^e \Upsilon_{\theta_1}^d+\Gamma_{\theta_2}^e (\Upsilon_{\theta_1}^d)^2+\Gamma_{\pi-\theta_2}^e\Big((\Delta_{\pi-\theta_1})^2+(\Delta_{\theta_1})^2-(\Gamma_{\pi-\theta_1}^e)^2-(\Gamma_{\theta_1}^e)^2+2 \Delta_{\pi-\theta_1} \Upsilon_{0}\nonumber\\
	&+(\Upsilon_{0})^2-2 \Gamma_{\pi-\theta_1}^e \Upsilon_{\theta_1}^d-(\Upsilon_{\theta_1}^d)^2\Big)+3 \Gamma_{\pi/2}^d \Gamma_{\pi-\theta_1}^e \Upsilon_{\theta_1}^e-2 \Gamma_{\pi-\theta_1}^e \Gamma_{\pi-\theta_2}^d \Upsilon_{\theta_1}^e-2 \Gamma_{\pi/2}^d \Gamma_{\theta_1}^e \Upsilon_{\theta_1}^e+2 \Gamma_{\pi-\theta_2}^d \Gamma_{\theta_1}^e \Upsilon_{\theta_1}^e\nonumber\\
	&-4 \Gamma_{\pi/2}^d \Upsilon_{\theta_1}^d \Upsilon_{\theta_1}^e-2 \Gamma_{\pi-\theta_2}^d \Upsilon_{\theta_1}^d \Upsilon_{\theta_1}^e\Big] +\frac{1}{64 \pi^4N_f^2\Gamma_{\pi-\theta_2}^d (1+v^2)^2}\Big[4 \Delta_{0} \Delta_{\pi} \Gamma_{\pi-\theta_2}^d+(\Gamma_{0})^2 (3 \Gamma_{\pi-\theta_2}^d-2 \Gamma_{\theta_2}^d)\nonumber\\
	&+(\Delta_{\pi})^2 (-3 \Gamma_{\pi-\theta_2}^d+\Gamma_{\theta_2}^d)+(\Delta_{0})^2 (-\Gamma_{\pi-\theta_2}^d+\Gamma_{\theta_2}^d)+\Delta_{0} \Gamma_{0} (2 \Gamma_{\pi-\theta_2}^d+\Gamma_{\theta_2}^d)+\Delta_{\pi} \Gamma_{0} (\Gamma_{\pi-\theta_2}^d+4 \Gamma_{\theta_2}^d)\Big]\\
	A^{(2l)}_{\Gamma_{\pi-\theta_2}^e}&=-\frac{e^{-\frac{v^2}{v_c^2}}}{64 \pi^4N_f^2\Gamma_{\pi-\theta_2}^e}\frac{1}{\epsilon} \Big[-2 (\Delta_{\pi-\theta_1})^2 \Gamma_{\pi/2}^e+2 (\Delta_{\theta_1})^2 \Gamma_{\pi/2}^e+2 \Gamma_{\pi/2}^e (\Upsilon_{0})^2-2 \Delta_{\theta_1} (\Delta_{\pi-\theta_1} \Gamma_{\theta_2}^e+(2 \Gamma_{\pi/2}^e+\Gamma_{\theta_2}^e) \Upsilon_{0})\nonumber\\
	&-4 \Gamma_{\pi/2}^e \Gamma_{\pi-\theta_1}^d \Upsilon_{\theta_1}^d-4 \Gamma_{\pi/2}^e \Gamma_{\theta_1}^d \Upsilon_{\theta_1}^d+\Gamma_{\pi-\theta_2}^e \Big((\Delta_{\pi-\theta_1})^2+(\Delta_{\theta_1})^2-2 \Gamma_{\pi-\theta_1}^d \Gamma_{\pi-\theta_1}^e+2 \Gamma_{\pi-\theta_1}^e \Gamma_{\theta_1}^d+2 \Gamma_{\pi-\theta_1}^d \Gamma_{\theta_1}^e\nonumber\\
	&-2 \Gamma_{\theta_1}^d \Gamma_{\theta_1}^e+2 \Delta_{\pi-\theta_1} \Upsilon_{0}+(\Upsilon_{0})^2-2 \Gamma_{\pi-\theta_1}^d \Upsilon_{\theta_1}^d+2 \Gamma_{\theta_1}^d \Upsilon_{\theta_1}^d\Big)+2 \Gamma_{\pi-\theta_1}^e \Gamma_{\theta_2}^e \Upsilon_{\theta_1}^e-2 \Gamma_{\theta_1}^e \Gamma_{\theta_2}^e \Upsilon_{\theta_1}^e+4 \Gamma_{\pi/2}^e \Upsilon_{\theta_1}^d \Upsilon_{\theta_1}^e\nonumber\\
	&+2 \Gamma_{\theta_2}^e \Upsilon_{\theta_1}^d \Upsilon_{\theta_1}^e\Big] -\frac{1}{64 \Gamma_{\pi-\theta_2}^e\pi^4N_f^2 (1+v^2)^2}\Big[\Big(3 (\Delta_{0})^2-2 \Delta_{0} \Delta_{\pi}+(\Delta_{\pi})^2\Big) \Gamma_{\pi-\theta_2}^e+2 (\Delta_{0}+\Delta_{\pi}) \Gamma_{0} \Gamma_{\pi-\theta_2}^e-(\Gamma_{0})^2 \Gamma_{\pi-\theta_2}^e\nonumber\\
	&-2 (\Delta_{\pi})^2 \Gamma_{\theta_2}^e\Big]
\end{align}

\begin{align}
	A_{\Delta_0}^{(2l)}&=-\frac{e^{-\frac{v^2}{v_c^2}}}{64\Delta_0\pi^4N_f^2}\frac{1}{\epsilon} \Big[2 \Delta_{0} (\Delta_{\pi-\theta_1})^2+\Delta_{\pi} (\Delta_{\pi-\theta_1})^2+2 \Delta_{0} \Delta_{\pi-\theta_1} \Delta_{\theta_1}+2 \Delta_{0} (\Delta_{\theta_1})^2+\Delta_{\pi} (\Delta_{\theta_1})^2+2 (\Delta_{\pi-\theta_1})^2 \Gamma_{\pi-\theta_1}^d\nonumber\\
	&-6 \Delta_{\pi-\theta_1} \Delta_{\theta_1} \Gamma_{\pi-\theta_1}^d-3 (\Delta_{\pi-\theta_1})^2 \Gamma_{\pi-\theta_1}^e+2 \Delta_{\pi-\theta_1} \Delta_{\theta_1} \Gamma_{\pi-\theta_1}^e-(\Delta_{\theta_1})^2 \Gamma_{\pi-\theta_1}^e-4 \Delta_{0} \Gamma_{\pi-\theta_1}^d \Gamma_{\pi-\theta_1}^e+\Delta_{0} (\Gamma_{\pi-\theta_1}^e)^2\nonumber\\
	&-\Delta_{\pi} (\Gamma_{\pi-\theta_1}^e)^2+(\Gamma_{\pi-\theta_1}^e)^3+(\Delta_{\pi-\theta_1})^2 \Gamma_{\theta_1}^e-2 \Delta_{\pi-\theta_1} \Delta_{\theta_1} \Gamma_{\theta_1}^e+3 (\Delta_{\theta_1})^2 \Gamma_{\theta_1}^e-2 (\Gamma_{\pi-\theta_1}^d)^2 \Gamma_{\theta_1}^e-2 \Delta_{\pi} \Gamma_{\pi-\theta_1}^e \Gamma_{\theta_1}^e\nonumber\\
	&-(\Gamma_{\pi-\theta_1}^e)^2 \Gamma_{\theta_1}^e+\Delta_{0} (\Gamma_{\theta_1}^e)^2-\Delta_{\pi} (\Gamma_{\theta_1}^e)^2+\Gamma_{\pi-\theta_1}^e (\Gamma_{\theta_1}^e)^2-(\Gamma_{\theta_1}^e)^3+4 \Delta_{0} \Delta_{\pi-\theta_1} \Upsilon_{0}+2 \Delta_{\pi} \Delta_{\pi-\theta_1} \Upsilon_{0}+\Delta_{0} \Delta_{\theta_1} \Upsilon_{0}\nonumber\\
	&-4 \Delta_{\pi-\theta_1} \Gamma_{\pi-\theta_1}^d \Upsilon_{0}-2 \Delta_{\theta_1} \Gamma_{\pi-\theta_1}^d \Upsilon_{0}-4 \Delta_{\pi-\theta_1} \Gamma_{\pi-\theta_1}^e \Upsilon_{0}+2 \Delta_{\pi-\theta_1} \Gamma_{\theta_1}^e \Upsilon_{0}-2 \Delta_{\theta_1} \Gamma_{\theta_1}^e \Upsilon_{0}+2 \Delta_{0} (\Upsilon_{0})^2+\Delta_{\pi} (\Upsilon_{0})^2\nonumber\\
	&-2 \Gamma_{\pi-\theta_1}^d (\Upsilon_{0})^2-\Gamma_{\pi-\theta_1}^e (\Upsilon_{0})^2+\Gamma_{\theta_1}^e (\Upsilon_{0})^2-4 \Delta_{0} \Gamma_{\pi-\theta_1}^d \Upsilon_{\theta_1}^d+3 (\Gamma_{\pi-\theta_1}^d)^2 \Upsilon_{\theta_1}^d+2 \Delta_{0} \Gamma_{\pi-\theta_1}^e \Upsilon_{\theta_1}^d-2 \Delta_{\pi} \Gamma_{\pi-\theta_1}^e \Upsilon_{\theta_1}^d\nonumber\\
	&+2 (\Gamma_{\pi-\theta_1}^e)^2 \Upsilon_{\theta_1}^d-2 \Delta_{\pi} \Gamma_{\theta_1}^e \Upsilon_{\theta_1}^d-2 \Gamma_{\pi-\theta_1}^e \Gamma_{\theta_1}^e \Upsilon_{\theta_1}^d+2 \Delta_{\pi-\theta_1} \Upsilon_{0} \Upsilon_{\theta_1}^d-2 \Delta_{\theta_1} \Upsilon_{0} \Upsilon_{\theta_1}^d+2 (\Upsilon_{0})^2 \Upsilon_{\theta_1}^d+\Delta_{0} (\Upsilon_{\theta_1}^d)^2\nonumber\\
	&-\Delta_{\pi} (\Upsilon_{\theta_1}^d)^2-4 \Gamma_{\pi-\theta_1}^d (\Upsilon_{\theta_1}^d)^2+\Gamma_{\pi-\theta_1}^e (\Upsilon_{\theta_1}^d)^2-\Gamma_{\theta_1}^e (\Upsilon_{\theta_1}^d)^2-2 (\Gamma_{\theta_1}^d)^2 (\Gamma_{\pi-\theta_1}^e-\Gamma_{\theta_1}^e+\Upsilon_{\theta_1}^d)+\Gamma_{0} \Big(-2 (\Delta_{\theta_1})^2\nonumber\\
	&+2 \Gamma_{\pi-\theta_1}^d \Gamma_{\pi-\theta_1}^e+(\Gamma_{\pi-\theta_1}^e)^2-2 \Gamma_{\pi-\theta_1}^e \Gamma_{\theta_1}^d-2 \Gamma_{\pi-\theta_1}^d \Gamma_{\theta_1}^e+2 \Gamma_{\theta_1}^d \Gamma_{\theta_1}^e+(\Gamma_{\theta_1}^e)^2-3 \Delta_{\pi-\theta_1} \Upsilon_{0}-3 (\Upsilon_{0})^2\nonumber\\
	&-2 \Delta_{\theta_1} (\Delta_{\pi-\theta_1}+\Upsilon_{0})+2 \Gamma_{\pi-\theta_1}^d \Upsilon_{\theta_1}^d+2 \Gamma_{\pi-\theta_1}^e \Upsilon_{\theta_1}^d-2 \Gamma_{\theta_1}^d \Upsilon_{\theta_1}^d+(\Upsilon_{\theta_1}^d)^2\Big)-2 (\Delta_{\pi-\theta_1})^2 \Upsilon_{\theta_1}^e+2 (\Delta_{\theta_1})^2 \Upsilon_{\theta_1}^e\nonumber\\
	&-2 \Delta_{0} \Gamma_{\pi-\theta_1}^e \Upsilon_{\theta_1}^e+2 \Delta_{\pi} \Gamma_{\pi-\theta_1}^e \Upsilon_{\theta_1}^e+4 \Gamma_{\pi-\theta_1}^d \Gamma_{\pi-\theta_1}^e \Upsilon_{\theta_1}^e+2 \Delta_{0} \Gamma_{\theta_1}^e \Upsilon_{\theta_1}^e-2 \Delta_{\pi} \Gamma_{\theta_1}^e \Upsilon_{\theta_1}^e+2 \Gamma_{\pi-\theta_1}^d \Gamma_{\theta_1}^e \Upsilon_{\theta_1}^e\nonumber\\
	&+2 \Delta_{\pi-\theta_1} \Upsilon_{0} \Upsilon_{\theta_1}^e+2 \Delta_{\theta_1} \Upsilon_{0} \Upsilon_{\theta_1}^e-2 \Delta_{0} \Upsilon_{\theta_1}^d \Upsilon_{\theta_1}^e+2 \Delta_{\pi} \Upsilon_{\theta_1}^d \Upsilon_{\theta_1}^e-2 \Gamma_{\pi-\theta_1}^e \Upsilon_{\theta_1}^d \Upsilon_{\theta_1}^e-2 \Gamma_{\theta_1}^e \Upsilon_{\theta_1}^d \Upsilon_{\theta_1}^e+2 (\Upsilon_{\theta_1}^d)^2 \Upsilon_{\theta_1}^e\nonumber\\
	&-\Gamma_{\theta_1}^d \Big(2 (\Delta_{\theta_1})^2+3 \Gamma_{\pi-\theta_1}^d \Gamma_{\pi-\theta_1}^e+4 \Delta_{0} \Gamma_{\theta_1}^e+\Gamma_{\pi-\theta_1}^d \Gamma_{\theta_1}^e+4 \Delta_{\pi-\theta_1} \Upsilon_{0}-2 \Delta_{\theta_1} (\Delta_{\pi-\theta_1}+\Upsilon_{0})+2 \Gamma_{\pi-\theta_1}^d \Upsilon_{\theta_1}^d+4 (\Upsilon_{\theta_1}^d)^2\nonumber\\
	&+\Gamma_{\pi-\theta_1}^e \Upsilon_{\theta_1}^e+4 \Gamma_{\theta_1}^e \Upsilon_{\theta_1}^e+4 \Upsilon_{\theta_1}^d \Upsilon_{\theta_1}^e\Big)\Big]-\frac{1}{64\Delta_0\pi^4N_f^2(1+v^2)^2}\frac{1}{\epsilon}\Big[3 (\Delta_{0})^3-4 (\Delta_{0})^2 \Delta_{\pi}+5 \Delta_{0} (\Delta_{\pi})^2-2 (\Delta_{\pi})^3\nonumber\\
	&+\Big((\Delta_{0})^2-\Delta_{0} \Delta_{\pi}-(\Delta_{\pi})^2\Big) \Gamma_{0}-2 (\Delta_{0}+2 \Delta_{\pi}) (\Gamma_{0})^2+2 (\Gamma_{0})^3\Big]
\end{align}

\begin{align}
	A_{\Delta_\pi}^{(2l)}&=-\frac{e^{-\frac{v^2}{ v_c^2}}}{64\pi^4N_f^2}\frac{1}{\epsilon}\Big[2\Delta_{0} \Delta_{\pi-\theta_1} \Delta_{\theta_1}-4 \Delta_{\pi-\theta_1} \Delta_{\theta_1} \Gamma_{\pi-\theta_1}^e+4 (\Delta_{\theta_1})^2 \Gamma_{\pi-\theta_1}^e-4 (\Delta_{\pi-\theta_1})^2 \Gamma_{\theta_1}^e+4 \Delta_{\pi-\theta_1} \Delta_{\theta_1} \Gamma_{\theta_1}^e+2 \Delta_{0} \Delta_{\theta_1} \Upsilon_{0}\nonumber\\
	&-4 \Delta_{\theta_1} \Gamma_{\pi-\theta_1}^d \Upsilon_{0}-4 \Delta_{\theta_1} \Gamma_{\pi-\theta_1}^e \Upsilon_{0}-4 \Delta_{\theta_1} \Gamma_{\theta_1}^d \Upsilon_{0}-2 \Delta_{\pi-\theta_1} \Gamma_{\theta_1}^e \Upsilon_{0}-2 \Delta_{\theta_1} \Gamma_{\theta_1}^e \Upsilon_{0}+2 \Gamma_{\theta_1}^e (\Upsilon_{0})^2-2 \Delta_{\theta_1} \Gamma_{0} (\Delta_{\pi-\theta_1}+\Upsilon_{0})\nonumber\\
	&-2 (\Delta_{\theta_1})^2 \Upsilon_{\theta_1}^d-4 \Gamma_{\pi-\theta_1}^d \Gamma_{\theta_1}^e \Upsilon_{\theta_1}^d-2 \Gamma_{\pi-\theta_1}^e \Gamma_{\theta_1}^e \Upsilon_{\theta_1}^d-4 \Gamma_{\theta_1}^d \Gamma_{\theta_1}^e \Upsilon_{\theta_1}^d-2 \Gamma_{\theta_1}^e (\Upsilon_{\theta_1}^d)^2+2 (\Delta_{\pi-\theta_1})^2 \Upsilon_{\theta_1}^e+2 \Delta_{\pi-\theta_1} \Delta_{\theta_1} \Upsilon_{\theta_1}^e\nonumber\\
	&-4 (\Delta_{\theta_1})^2 \Upsilon_{\theta_1}^e+2 \Delta_{0} \Gamma_{\pi-\theta_1}^e \Upsilon_{\theta_1}^e-2 \Gamma_{\pi-\theta_1}^d \Gamma_{\pi-\theta_1}^e \Upsilon_{\theta_1}^e+2 (\Gamma_{\pi-\theta_1}^e)^2 \Upsilon_{\theta_1}^e-2 \Gamma_{\pi-\theta_1}^e \Gamma_{\theta_1}^d \Upsilon_{\theta_1}^e-2 \Delta_{0} \Gamma_{\theta_1}^e \Upsilon_{\theta_1}^e+\Gamma_{\pi-\theta_1}^d \Gamma_{\theta_1}^e \Upsilon_{\theta_1}^e\nonumber\\
	&+2 \Gamma_{\theta_1}^d \Gamma_{\theta_1}^e \Upsilon_{\theta_1}^e-2 (\Gamma_{\theta_1}^e)^2 \Upsilon_{\theta_1}^e+2 \Delta_{\pi-\theta_1} \Upsilon_{0} \Upsilon_{\theta_1}^e+4 \Delta_{\theta_1} \Upsilon_{0} \Upsilon_{\theta_1}^e+2 \Delta_{0} \Upsilon_{\theta_1}^d \Upsilon_{\theta_1}^e-6 \Gamma_{\pi-\theta_1}^d \Upsilon_{\theta_1}^d \Upsilon_{\theta_1}^e+2 \Gamma_{\pi-\theta_1}^e \Upsilon_{\theta_1}^d \Upsilon_{\theta_1}^e\nonumber\\
	&-6 \Gamma_{\theta_1}^d \Upsilon_{\theta_1}^d \Upsilon_{\theta_1}^e+2 \Gamma_{\theta_1}^e \Upsilon_{\theta_1}^d \Upsilon_{\theta_1}^e+2 \Gamma_{0} (\Gamma_{\pi-\theta_1}^e-\Gamma_{\theta_1}^e+\Upsilon_{\theta_1}^d) \Upsilon_{\theta_1}^e-\Gamma_{\pi-\theta_1}^e (\Upsilon_{\theta_1}^e)^2-4 \Gamma_{\theta_1}^e (\Upsilon_{\theta_1}^e)^2+\Delta_{\pi} \Big(2 (\Delta_{\pi-\theta_1})^2\nonumber\\
	&+2 (\Delta_{\theta_1})^2+(\Gamma_{\pi-\theta_1}^e)^2+4 \Gamma_{\pi-\theta_1}^e \Gamma_{\theta_1}^d+4 \Gamma_{\pi-\theta_1}^d \Gamma_{\theta_1}^e+(\Gamma_{\theta_1}^e)^2+4 \Delta_{\pi-\theta_1} \Upsilon_{0}+\Delta_{\theta_1} \Upsilon_{0}+2 (\Upsilon_{0})^2+2 \Gamma_{\pi-\theta_1}^e \Upsilon_{\theta_1}^d\nonumber\\
	&+4 \Gamma_{\theta_1}^d \Upsilon_{\theta_1}^d+(\Upsilon_{\theta_1}^d)^2-2 \Gamma_{\pi-\theta_1}^e \Upsilon_{\theta_1}^e+2 \Gamma_{\theta_1}^e \Upsilon_{\theta_1}^e-2 \Upsilon_{\theta_1}^d \Upsilon_{\theta_1}^e\Big)\Big] -\frac{1}{64\pi^4N_f^2 (1+v^2)^2}\frac{1}{\epsilon}\Big[11 (\Delta_{0})^2-4 \Delta_{0} \Delta_{\pi}+5 (\Delta_{\pi})^2\nonumber\\
	&+\Big(14 \Delta_{0}+\Delta_{\pi}\Big) \Gamma_{0}+3 (\Gamma_{0})^2 \Big]\\
	A^{(2l)}_{\Delta_{\theta_1}}&=\frac{e^{-\frac{v^2}{v_c^2}}}{64 \Delta_{\theta_1}\pi^4N_f^2}\Big[-2 \Delta_{\pi} \Delta_{\pi-\theta_1} \Gamma_{\pi-\theta_1}^d-4 \Delta_{0} \Delta_{\theta_1} \Gamma_{\pi-\theta_1}^d+6 \Delta_{\pi} \Delta_{\theta_1} \Gamma_{\pi-\theta_1}^d-4 \Delta_{\theta_1} (\Gamma_{\pi-\theta_1}^d)^2+4 \Delta_{\pi} \Delta_{\pi-\theta_1} \Gamma_{\pi-\theta_1}^e\nonumber\\
	&+6 \Delta_{0} \Delta_{\theta_1} \Gamma_{\pi-\theta_1}^e-4 \Delta_{\pi} \Delta_{\theta_1} \Gamma_{\pi-\theta_1}^e-2 \Delta_{\theta_1} \Gamma_{\pi-\theta_1}^d \Gamma_{\pi-\theta_1}^e-3 \Delta_{\theta_1} (\Gamma_{\pi-\theta_1}^e)^2-4 \Delta_{\theta_1} (\Gamma_{\theta_1}^d)^2+6 \Delta_{0} \Delta_{\pi-\theta_1} \Gamma_{\theta_1}^e\nonumber\\
	&-4 \Delta_{0} \Delta_{\theta_1} \Gamma_{\theta_1}^e+4 \Delta_{\pi} \Delta_{\theta_1} \Gamma_{\theta_1}^e-4 \Delta_{\pi-\theta_1} \Gamma_{\pi-\theta_1}^d \Gamma_{\theta_1}^e+6 \Delta_{\theta_1} \Gamma_{\pi-\theta_1}^d \Gamma_{\theta_1}^e+2 \Delta_{\pi-\theta_1} \Gamma_{\pi-\theta_1}^e \Gamma_{\theta_1}^e-\Delta_{\theta_1} \Gamma_{\pi-\theta_1}^e \Gamma_{\theta_1}^e\nonumber\\
	&+2 \Delta_{\pi-\theta_1} (\Gamma_{\theta_1}^e)^2-5 \Delta_{\theta_1} (\Gamma_{\theta_1}^e)^2+2 \Delta_{\pi} \Gamma_{\pi-\theta_1}^d \Upsilon_{0}+4 \Delta_{\pi} \Gamma_{\pi-\theta_1}^e \Upsilon_{0}+4 \Delta_{0} \Gamma_{\theta_1}^e \Upsilon_{0}-\Delta_{\pi} \Gamma_{\theta_1}^e \Upsilon_{0}-2 \Gamma_{\pi-\theta_1}^d \Gamma_{\theta_1}^e \Upsilon_{0}\nonumber\\
	&+2 \Gamma_{\pi-\theta_1}^e \Gamma_{\theta_1}^e \Upsilon_{0}+2 (\Gamma_{\theta_1}^e)^2 \Upsilon_{0}+6 \Delta_{0} \Delta_{\theta_1} \Upsilon_{\theta_1}^d-2 \Delta_{\theta_1} \Gamma_{\pi-\theta_1}^d \Upsilon_{\theta_1}^d-6 \Delta_{\theta_1} \Gamma_{\pi-\theta_1}^e \Upsilon_{\theta_1}^d+2 \Delta_{\pi-\theta_1} \Gamma_{\theta_1}^e \Upsilon_{\theta_1}^d+2 \Gamma_{\theta_1}^e \Upsilon_{0} \Upsilon_{\theta_1}^d\nonumber\\
	&-3 \Delta_{\theta_1} (\Upsilon_{\theta_1}^d)^2+2 \Delta_{\pi} \Delta_{\pi-\theta_1} \Upsilon_{\theta_1}^e-4 \Delta_{0} \Delta_{\theta_1} \Upsilon_{\theta_1}^e+6 \Delta_{\pi} \Delta_{\theta_1} \Upsilon_{\theta_1}^e-2 \Delta_{\pi-\theta_1} \Gamma_{\pi-\theta_1}^d \Upsilon_{\theta_1}^e+\Delta_{\theta_1} \Gamma_{\pi-\theta_1}^d \Upsilon_{\theta_1}^e\nonumber\\
	&-2 \Delta_{\pi-\theta_1} \Gamma_{\pi-\theta_1}^e \Upsilon_{\theta_1}^e-2 \Delta_{\theta_1} \Gamma_{\pi-\theta_1}^e \Upsilon_{\theta_1}^e+2 \Delta_{\pi-\theta_1} \Gamma_{\theta_1}^e \Upsilon_{\theta_1}^e-2 \Delta_{\theta_1} \Gamma_{\theta_1}^e \Upsilon_{\theta_1}^e+2 \Delta_{0} \Upsilon_{0} \Upsilon_{\theta_1}^e-4 \Delta_{\pi} \Upsilon_{0} \Upsilon_{\theta_1}^e+4 \Gamma_{\pi-\theta_1}^d \Upsilon_{0} \Upsilon_{\theta_1}^e\nonumber\\
	&+2 \Gamma_{\pi-\theta_1}^e \Upsilon_{0} \Upsilon_{\theta_1}^e-6 \Delta_{\pi-\theta_1} \Upsilon_{\theta_1}^d \Upsilon_{\theta_1}^e+2 \Delta_{\theta_1} \Upsilon_{\theta_1}^d \Upsilon_{\theta_1}^e-2 \Upsilon_{0} \Upsilon_{\theta_1}^d \Upsilon_{\theta_1}^e+2 \Delta_{\pi-\theta_1} (\Upsilon_{\theta_1}^e)^2-2 \Delta_{\theta_1} (\Upsilon_{\theta_1}^e)^2-2 \Upsilon_{0} (\Upsilon_{\theta_1}^e)^2\nonumber\\
	&+2 \Gamma_{\theta_1}^d \Big(4 \Delta_{0} \Delta_{\theta_1}-4 \Delta_{\theta_1} \Gamma_{\pi-\theta_1}^d+\Delta_{\theta_1} \Gamma_{\pi-\theta_1}^e-\Delta_{\pi-\theta_1} \Gamma_{\theta_1}^e+\Delta_{\pi} (\Delta_{\pi-\theta_1}-\Delta_{\theta_1}+3 \Upsilon_{0})+\Delta_{\theta_1} \Upsilon_{\theta_1}^d-2 \Delta_{\pi-\theta_1} \Upsilon_{\theta_1}^e\nonumber\\
	&+\Upsilon_{0} \Upsilon_{\theta_1}^e\Big)+\Gamma_{0} \Big(8 \Delta_{\theta_1} \Gamma_{\theta_1}^d+7 \Delta_{\theta_1} \Gamma_{\theta_1}^e-2 \Delta_{\theta_1} \Upsilon_{\theta_1}^d-2 \Delta_{\theta_1} (2 \Gamma_{\pi-\theta_1}^d+\Gamma_{\pi-\theta_1}^e+\Upsilon_{\theta_1}^e)+2 \Delta_{\pi-\theta_1} (2 \Gamma_{\theta_1}^e+\Upsilon_{\theta_1}^e)\nonumber\\
	&+2 \Upsilon_{0} (\Gamma_{\theta_1}^e+2 \Upsilon_{\theta_1}^e)\Big)\Big] -\frac{1}{64 \Delta_{\theta_1} \pi^4 N_f^2(1+v^2)^2}\frac{1}{\epsilon}\Big[-(\Delta_{0})^2 \Delta_{\theta_1}+2 \Delta_{0} \Delta_{\pi} \Delta_{\theta_1}+(\Delta_{\pi})^2 (-2 \Delta_{\pi-\theta_1}+\Delta_{\theta_1})\nonumber\\
	&+2 (\Delta_{0}-\Delta_{\pi}) \Delta_{\theta_1} \Gamma_{0}+3 \Delta_{\theta_1} (\Gamma_{0})^2\Big]
\end{align}

\begin{align}
	A^{(2l)}_{\Delta_{\pi-\theta_1}}&=\frac{e^{-\frac{v^2}{v_c^2}}}{64\Delta_{\pi-\theta_1} \pi^4N_f^2}\frac{1}{\epsilon} \Big[-8 \Delta_{0} \Delta_{\pi-\theta_1} \Gamma_{\pi-\theta_1}^d+2 \Delta_{\pi} \Delta_{\pi-\theta_1} \Gamma_{\pi-\theta_1}^d-2 \Delta_{\pi} \Delta_{\theta_1} \Gamma_{\pi-\theta_1}^d-4 \Delta_{\pi-\theta_1} (\Gamma_{\pi-\theta_1}^d)^2+6 \Delta_{0} \Delta_{\pi-\theta_1} \Gamma_{\pi-\theta_1}^e\nonumber\\
	&-4 \Delta_{\pi} \Delta_{\pi-\theta_1} \Gamma_{\pi-\theta_1}^e-6 \Delta_{0} \Delta_{\theta_1} \Gamma_{\pi-\theta_1}^e-2 \Delta_{\pi} \Delta_{\theta_1} \Gamma_{\pi-\theta_1}^e-6 \Delta_{\theta_1} \Gamma_{\pi-\theta_1}^d \Gamma_{\pi-\theta_1}^e-5 \Delta_{\pi-\theta_1} (\Gamma_{\pi-\theta_1}^e)^2+2 \Delta_{\theta_1} (\Gamma_{\pi-\theta_1}^e)^2\nonumber\\
	&-6 \Delta_{0} \Delta_{\pi-\theta_1} \Gamma_{\theta_1}^e+4 \Delta_{\pi} \Delta_{\pi-\theta_1} \Gamma_{\theta_1}^e-4 \Delta_{\pi} \Delta_{\theta_1} \Gamma_{\theta_1}^e+2 \Delta_{\pi-\theta_1} \Gamma_{\pi-\theta_1}^d \Gamma_{\theta_1}^e-5 \Delta_{\pi-\theta_1} \Gamma_{\pi-\theta_1}^e \Gamma_{\theta_1}^e+2 \Delta_{\theta_1} \Gamma_{\pi-\theta_1}^e \Gamma_{\theta_1}^e\nonumber\\
	&-3 \Delta_{\pi-\theta_1} (\Gamma_{\theta_1}^e)^2+4 \Delta_{0} \Gamma_{\pi-\theta_1}^d \Upsilon_{0}-2 \Delta_{\pi} \Gamma_{\pi-\theta_1}^d \Upsilon_{0}-2 (\Gamma_{\pi-\theta_1}^d)^2 \Upsilon_{0}+\Delta_{0} \Gamma_{\pi-\theta_1}^e \Upsilon_{0}-2 \Delta_{\pi} \Gamma_{\pi-\theta_1}^e \Upsilon_{0}+2 \Delta_{0} \Gamma_{\theta_1}^e \Upsilon_{0}\nonumber\\
	&+4 \Delta_{\pi} \Gamma_{\theta_1}^e \Upsilon_{0}-2 \Gamma_{\pi-\theta_1}^e \Gamma_{\theta_1}^e \Upsilon_{0}+(\Gamma_{\theta_1}^e)^2 \Upsilon_{0}+(\Gamma_{\theta_1}^d)^2 (-4 \Delta_{\pi-\theta_1}+\Upsilon_{0})+2 \Delta_{0} \Delta_{\pi-\theta_1} \Upsilon_{\theta_1}^d+2 \Delta_{\pi} \Delta_{\pi-\theta_1} \Upsilon_{\theta_1}^d\nonumber\\
	&+2 \Delta_{\pi} \Delta_{\theta_1} \Upsilon_{\theta_1}^d-2 \Delta_{\pi-\theta_1} \Gamma_{\pi-\theta_1}^d \Upsilon_{\theta_1}^d-4 \Delta_{\pi-\theta_1} \Gamma_{\pi-\theta_1}^e \Upsilon_{\theta_1}^d-2 \Delta_{\theta_1} \Gamma_{\pi-\theta_1}^e \Upsilon_{\theta_1}^d+\Delta_{\pi} \Upsilon_{0} \Upsilon_{\theta_1}^d+2 \Gamma_{\pi-\theta_1}^d \Upsilon_{0} \Upsilon_{\theta_1}^d\nonumber\\
	&+3 \Gamma_{\pi-\theta_1}^e \Upsilon_{0} \Upsilon_{\theta_1}^d+4 \Gamma_{\theta_1}^e \Upsilon_{0} \Upsilon_{\theta_1}^d-3 \Delta_{\pi-\theta_1} (\Upsilon_{\theta_1}^d)^2-\Upsilon_{0} (\Upsilon_{\theta_1}^d)^2+2 \Delta_{0} \Delta_{\pi-\theta_1} \Upsilon_{\theta_1}^e+2 \Delta_{0} \Delta_{\theta_1} \Upsilon_{\theta_1}^e+\Delta_{\pi-\theta_1} \Gamma_{\pi-\theta_1}^d \Upsilon_{\theta_1}^e\nonumber\\
	&-2 \Delta_{\theta_1} \Gamma_{\pi-\theta_1}^d \Upsilon_{\theta_1}^e-2 \Delta_{\pi-\theta_1} \Gamma_{\pi-\theta_1}^e \Upsilon_{\theta_1}^e-2 \Delta_{\theta_1} \Gamma_{\pi-\theta_1}^e \Upsilon_{\theta_1}^e-2 \Delta_{\pi} \Upsilon_{0} \Upsilon_{\theta_1}^e+4 \Gamma_{\pi-\theta_1}^d \Upsilon_{0} \Upsilon_{\theta_1}^e+2 \Gamma_{\pi-\theta_1}^e \Upsilon_{0} \Upsilon_{\theta_1}^e\nonumber\\
	&+2 \Gamma_{\theta_1}^e \Upsilon_{0} \Upsilon_{\theta_1}^e-4 \Delta_{\theta_1} \Upsilon_{\theta_1}^d \Upsilon_{\theta_1}^e-2 \Delta_{\pi-\theta_1} (\Upsilon_{\theta_1}^e)^2+2 \Delta_{\theta_1} (\Upsilon_{\theta_1}^e)^2+\Upsilon_{0} (\Upsilon_{\theta_1}^e)^2+\Gamma_{\theta_1}^d \Big(4 \Delta_{0} \Delta_{\pi-\theta_1}+4 \Delta_{\pi-\theta_1} \Gamma_{0}\nonumber\\
	&-8 \Delta_{\pi-\theta_1} \Gamma_{\pi-\theta_1}^d+6 \Delta_{\pi-\theta_1} \Gamma_{\pi-\theta_1}^e-8 \Delta_{\theta_1} \Gamma_{\pi-\theta_1}^e-2 \Delta_{\pi-\theta_1} \Gamma_{\theta_1}^e+4 \Delta_{0} \Upsilon_{0}+\Gamma_{\pi-\theta_1}^d \Upsilon_{0}+2 \Gamma_{\pi-\theta_1}^e \Upsilon_{0}\nonumber\\
	&+2 \Delta_{\pi} \Big(-3 \Delta_{\pi-\theta_1}+\Delta_{\theta_1}+\Upsilon_{0}\Big)+2 \Delta_{\theta_1} \Upsilon_{\theta_1}^d+2 \Delta_{\pi-\theta_1} \Upsilon_{\theta_1}^e+2 \Upsilon_{0} \Upsilon_{\theta_1}^e\Big)-\Gamma_{0}\Big(\Delta_{\theta_1} (4 \Gamma_{\pi-\theta_1}^e+\Upsilon_{\theta_1}^d)\nonumber\\
	&+\Delta_{\pi-\theta_1} (8 \Gamma_{\pi-\theta_1}^d+7 \Gamma_{\pi-\theta_1}^e-2 \Gamma_{\theta_1}^e+4 \Upsilon_{\theta_1}^d)-2 \Upsilon_{0} (\Gamma_{\pi-\theta_1}^e-\Gamma_{\theta_1}^e+3 \Upsilon_{\theta_1}^d+\Upsilon_{\theta_1}^e)\Big)\Big] \nonumber\\
	&+\frac{1}{64 \epsilon \pi^4 N_f^2(1+v^2)^2}\Big[(\Delta_{\pi})^2 (-\Delta_{\pi-\theta_1}+2 \Delta_{\theta_1})+2 \Delta_{\pi} \Delta_{\pi-\theta_1} (-\Delta_{0}+\Gamma_{0})\nonumber\\
	&+\Delta_{\pi-\theta_1} \Big((\Delta_{0})^2-2 \Delta_{0} \Gamma_{0}-3 (\Gamma_{0})^2\Big)\Big]
\end{align}

\begin{align}
	A^{(2l)}_{\Delta_{\theta_2}}&=\frac{e^{-\frac{v^2}{v_c^2}}}{64\Delta_{\theta_2} \pi^4N_f^2}\frac{1}{\epsilon} \Big[2 \Delta_{\pi/2} \Delta_{\pi-\theta_1} \Gamma_{\pi-\theta_1}^e-2 \Delta_{\theta_2} \Gamma_{\pi-\theta_1}^d \Gamma_{\pi-\theta_1}^e-\Delta_{\theta_2} (\Gamma_{\pi-\theta_1}^e)^2+2 \Delta_{\pi/2} \Delta_{\pi-\theta_1} \Gamma_{\theta_1}^e+2 \Delta_{\theta_2} \Gamma_{\pi-\theta_1}^d \Gamma_{\theta_1}^e\nonumber\\
	&+2 \Delta_{\pi-\theta_2} \Gamma_{\pi-\theta_1}^e \Gamma_{\theta_1}^e-\Delta_{\theta_2} (\Gamma_{\theta_1}^e)^2+4 \Delta_{\pi/2} \Gamma_{\pi-\theta_1}^d \Upsilon_{0}+2 \Delta_{\pi/2} \Gamma_{\pi-\theta_1}^e \Upsilon_{0}+4 \Delta_{\pi/2} \Gamma_{\theta_1}^d \Upsilon_{0}+2 \Delta_{\pi/2} \Gamma_{\theta_1}^e \Upsilon_{0}\nonumber\\
	&-2 \Delta_{\pi/2} \Delta_{\pi-\theta_1} \Upsilon_{\theta_1}^d-2 \Delta_{\theta_2} \Gamma_{\pi-\theta_1}^d \Upsilon_{\theta_1}^d-2 \Delta_{\theta_2} \Gamma_{\pi-\theta_1}^e \Upsilon_{\theta_1}^d+2 \Delta_{\pi-\theta_2} \Gamma_{\theta_1}^e \Upsilon_{\theta_1}^d-2 \Delta_{\pi/2} \Upsilon_{0} \Upsilon_{\theta_1}^d-\Delta_{\theta_2} (\Upsilon_{\theta_1}^d)^2\nonumber\\
	&+2 \Delta_{\pi/2} \Delta_{\theta_1} (-\Gamma_{\pi-\theta_1}^e-\Gamma_{\theta_1}^e+\Upsilon_{\theta_1}^d)+2 \Delta_{\theta_2} \Gamma_{\theta_1}^d (\Gamma_{\pi-\theta_1}^e-\Gamma_{\theta_1}^e+\Upsilon_{\theta_1}^d)-2 \Delta_{\pi-\theta_2} \Gamma_{\pi-\theta_1}^e \Upsilon_{\theta_1}^e+2 \Delta_{\pi-\theta_2} \Gamma_{\theta_1}^e \Upsilon_{\theta_1}^e\nonumber\\
	&-4 \Delta_{\pi/2} \Upsilon_{0} \Upsilon_{\theta_1}^e-2 \Delta_{\pi-\theta_2} \Upsilon_{\theta_1}^d \Upsilon_{\theta_1}^e\Big]-\frac{1}{64 \Delta_{\theta_2}\pi^4 N_f^2(1+v^2)^2}\Big[-(\Delta_{0})^2 \Delta_{\theta_2}+2 \Delta_{0} \Delta_{\pi} \Delta_{\theta_2}+(\Delta_{\pi})^2 (-2 \Delta_{\pi-\theta_2}+\Delta_{\theta_2})\nonumber\\
	&+2 (\Delta_{0}-\Delta_{\pi}) \Delta_{\theta_2} \Gamma_{0}+3 \Delta_{\theta_2} (\Gamma_{0})^2\Big]\\
	A^{(2l)}_{\Delta_{\pi-\theta_2}}&=\frac{e^{-\frac{v^2}{v_c^2}}}{64 \Delta_{\pi-\theta_2} \pi^4N_f^2}\frac{1}{\epsilon} \Big[2 \Delta_{\pi/2} \Delta_{\pi-\theta_1} \Gamma_{\pi-\theta_1}^e-2 \Delta_{\pi-\theta_2} \Gamma_{\pi-\theta_1}^d \Gamma_{\pi-\theta_1}^e-\Delta_{\pi-\theta_2} (\Gamma_{\pi-\theta_1}^e)^2+2 \Delta_{\pi/2} \Delta_{\pi-\theta_1} \Gamma_{\theta_1}^e\nonumber\\
	&+2 \Delta_{\pi-\theta_2} \Gamma_{\pi-\theta_1}^d \Gamma_{\theta_1}^e+2 \Delta_{\theta_2} \Gamma_{\pi-\theta_1}^e \Gamma_{\theta_1}^e-\Delta_{\pi-\theta_2} (\Gamma_{\theta_1}^e)^2+4 \Delta_{\pi/2} \Gamma_{\pi-\theta_1}^d \Upsilon_{0}+2 \Delta_{\pi/2} \Gamma_{\pi-\theta_1}^e \Upsilon_{0}+4 \Delta_{\pi/2} \Gamma_{\theta_1}^d \Upsilon_{0}\nonumber\\
	&+2 \Delta_{\pi/2} \Gamma_{\theta_1}^e \Upsilon_{0}-2 \Delta_{\pi/2} \Delta_{\pi-\theta_1} \Upsilon_{\theta_1}^d-2 \Delta_{\pi-\theta_2} \Gamma_{\pi-\theta_1}^d \Upsilon_{\theta_1}^d-2 \Delta_{\pi-\theta_2} \Gamma_{\pi-\theta_1}^e \Upsilon_{\theta_1}^d+2 \Delta_{\theta_2} \Gamma_{\theta_1}^e \Upsilon_{\theta_1}^d-2 \Delta_{\pi/2} \Upsilon_{0} \Upsilon_{\theta_1}^d\nonumber\\
	&-\Delta_{\pi-\theta_2} (\Upsilon_{\theta_1}^d)^2+2 \Delta_{\pi/2} \Delta_{\theta_1} (-\Gamma_{\pi-\theta_1}^e-\Gamma_{\theta_1}^e+\Upsilon_{\theta_1}^d)+2 \Delta_{\pi-\theta_2} \Gamma_{\theta_1}^d (\Gamma_{\pi-\theta_1}^e-\Gamma_{\theta_1}^e+\Upsilon_{\theta_1}^d)-2 \Delta_{\theta_2} \Gamma_{\pi-\theta_1}^e \Upsilon_{\theta_1}^e\nonumber\\
	&+2 \Delta_{\theta_2} \Gamma_{\theta_1}^e \Upsilon_{\theta_1}^e-4 \Delta_{\pi/2} \Upsilon_{0} \Upsilon_{\theta_1}^e-2 \Delta_{\theta_2} \Upsilon_{\theta_1}^d \Upsilon_{\theta_1}^e\Big]+\frac{1}{64 \Delta_{\pi-\theta_2} \pi^4 N_f^2(1+v^2)^2}\Big[(\Delta_{\pi})^2 (-\Delta_{\pi-\theta_2}+2 \Delta_{\theta_2})\nonumber\\
	&+2 \Delta_{\pi} \Delta_{\pi-\theta_2} (-\Delta_{0}+\Gamma_{0})+\Delta_{\pi-\theta_2} \Big((\Delta_{0})^2-2 \Delta_{0} \Gamma_{0}-3 (\Gamma_{0})^2\Big)\Big]
\end{align}

\begin{align}
	A^{(2l)}_{\Delta_{\pi/2}}&=-\frac{e^{-\frac{v^2}{v_c^2}}}{64 \Delta_{\pi/2} \pi^4N_f^2}\Big[-\Delta_{\pi-\theta_1} \Delta_{\pi-\theta_2} \Gamma_{\pi-\theta_1}^e-\Delta_{\pi-\theta_1} \Delta_{\theta_2} \Gamma_{\pi-\theta_1}^e+2 \Delta_{\pi/2} \Gamma_{\pi-\theta_1}^d \Gamma_{\pi-\theta_1}^e+\Delta_{\pi/2} (\Gamma_{\pi-\theta_1}^e)^2\nonumber\\
	&-\Delta_{\pi-\theta_1} \Delta_{\pi-\theta_2} \Gamma_{\theta_1}^e-\Delta_{\pi-\theta_1} \Delta_{\theta_2} \Gamma_{\theta_1}^e-2 \Delta_{\pi/2} \Gamma_{\pi-\theta_1}^d \Gamma_{\theta_1}^e-2 \Delta_{\pi/2} \Gamma_{\pi-\theta_1}^e \Gamma_{\theta_1}^e+\Delta_{\pi/2} (\Gamma_{\theta_1}^e)^2-2 \Delta_{\pi-\theta_2} \Gamma_{\pi-\theta_1}^d \Upsilon_{0}\nonumber\\
	&-2 \Delta_{\theta_2} \Gamma_{\pi-\theta_1}^d \Upsilon_{0}-\Delta_{\pi-\theta_2} \Gamma_{\pi-\theta_1}^e \Upsilon_{0}-\Delta_{\theta_2} \Gamma_{\pi-\theta_1}^e \Upsilon_{0}-\Delta_{\pi-\theta_2} \Gamma_{\theta_1}^e \Upsilon_{0}-\Delta_{\theta_2} \Gamma_{\theta_1}^e \Upsilon_{0}+\Delta_{\pi-\theta_1} \Delta_{\pi-\theta_2} \Upsilon_{\theta_1}^d\nonumber\\
	&+\Delta_{\pi-\theta_1} \Delta_{\theta_2} \Upsilon_{\theta_1}^d+2 \Delta_{\pi/2} \Gamma_{\pi-\theta_1}^d \Upsilon_{\theta_1}^d+2 \Delta_{\pi/2} \Gamma_{\pi-\theta_1}^e \Upsilon_{\theta_1}^d-2 \Delta_{\pi/2} \Gamma_{\theta_1}^e \Upsilon_{\theta_1}^d+\Delta_{\pi-\theta_2} \Upsilon_{0} \Upsilon_{\theta_1}^d+\Delta_{\theta_2} \Upsilon_{0} \Upsilon_{\theta_1}^d\nonumber\\
	&+\Delta_{\pi/2} (\Upsilon_{\theta_1}^d)^2-\Delta_{\theta_1} (\Delta_{\pi-\theta_2}+\Delta_{\theta_2}) (-\Gamma_{\pi-\theta_1}^e-\Gamma_{\theta_1}^e+\Upsilon_{\theta_1}^d)-2 \Gamma_{\theta_1}^d \Big(\Delta_{\pi-\theta_2} \Upsilon_{0}+\Delta_{\theta_2} \Upsilon_{0}+\Delta_{\pi/2} (\Gamma_{\pi-\theta_1}^e-\Gamma_{\theta_1}^e+\Upsilon_{\theta_1}^d)\Big)\nonumber\\
	&+2 \Delta_{\pi/2} \Gamma_{\pi-\theta_1}^e \Upsilon_{\theta_1}^e-2 \Delta_{\pi/2} \Gamma_{\theta_1}^e \Upsilon_{\theta_1}^e+2 \Delta_{\pi-\theta_2} \Upsilon_{0} \Upsilon_{\theta_1}^e+2 \Delta_{\theta_2} \Upsilon_{0} \Upsilon_{\theta_1}^e+2 \Delta_{\pi/2} \Upsilon_{\theta_1}^d \Upsilon_{\theta_1}^e\Big] \nonumber\\
	&-\frac{1}{64  \pi^4 N_f^2(1+v^2)^2} \frac{1}{\epsilon}\Big[-(\Delta_{0}-\Delta_{\pi})^2+2 (\Delta_{0}-\Delta_{\pi}) \Gamma_{0}+3 (\Gamma_{0})^2\Big]\\
	A^{(2l)}_{\Upsilon_0}&=-\frac{e^{-\frac{v^2}{v_c^2}}}{64 \Upsilon_{0} \pi^4N_f^2}\frac{1}{\epsilon}\Big[-6 \Delta_{0} \Delta_{\pi-\theta_1} \Gamma_{\pi-\theta_1}^d+2 \Delta_{\pi} \Delta_{\pi-\theta_1} \Gamma_{\pi-\theta_1}^d-4 \Delta_{\pi} \Delta_{\theta_1} \Gamma_{\pi-\theta_1}^d-4 \Delta_{\pi} \Delta_{\pi-\theta_1} \Gamma_{\pi-\theta_1}^e-2 \Delta_{0} \Delta_{\theta_1} \Gamma_{\pi-\theta_1}^e\nonumber\\
	&-4 \Delta_{\pi} \Delta_{\theta_1} \Gamma_{\pi-\theta_1}^e-2 \Delta_{\pi-\theta_1} \Gamma_{\pi-\theta_1}^d \Gamma_{\pi-\theta_1}^e-2 \Delta_{\theta_1} \Gamma_{\pi-\theta_1}^d \Gamma_{\pi-\theta_1}^e-2 \Delta_{\pi-\theta_1} (\Gamma_{\pi-\theta_1}^e)^2-2 \Delta_{0} \Delta_{\pi-\theta_1} \Gamma_{\theta_1}^e-4 \Delta_{\pi-\theta_1} \Gamma_{\pi-\theta_1}^e \Gamma_{\theta_1}^e\nonumber\\
	&-\Delta_{\pi-\theta_1} (\Gamma_{\theta_1}^e)^2+2 \Delta_{0} \Gamma_{\pi-\theta_1}^d \Upsilon_{0}-2 \Delta_{\pi} \Gamma_{\pi-\theta_1}^d \Upsilon_{0}-2 (\Gamma_{\pi-\theta_1}^d)^2 \Upsilon_{0}-2 \Delta_{0} \Gamma_{\pi-\theta_1}^e \Upsilon_{0}+\Delta_{\pi} \Gamma_{\pi-\theta_1}^e \Upsilon_{0}+2 \Gamma_{\pi-\theta_1}^d \Gamma_{\pi-\theta_1}^e \Upsilon_{0}\nonumber\\
	&+3 (\Gamma_{\pi-\theta_1}^e)^2 \Upsilon_{0}+6 \Delta_{0} \Gamma_{\theta_1}^e \Upsilon_{0}-2 \Gamma_{\pi-\theta_1}^d \Gamma_{\theta_1}^e \Upsilon_{0}-\Gamma_{\pi-\theta_1}^e \Gamma_{\theta_1}^e \Upsilon_{0}+3 (\Gamma_{\theta_1}^e)^2 \Upsilon_{0}+(\Gamma_{\theta_1}^d)^2 (\Delta_{\pi-\theta_1}+2 \Upsilon_{0})+2 \Delta_{\pi} \Delta_{\pi-\theta_1} \Upsilon_{\theta_1}^d\nonumber\\
	&+\Delta_{\pi-\theta_1} \Gamma_{\pi-\theta_1}^e \Upsilon_{\theta_1}^d-2 \Delta_{\theta_1} \Gamma_{\pi-\theta_1}^e \Upsilon_{\theta_1}^d+2 \Delta_{\pi-\theta_1} \Gamma_{\theta_1}^e \Upsilon_{\theta_1}^d-2 \Delta_{\theta_1} \Gamma_{\theta_1}^e \Upsilon_{\theta_1}^d+\Delta_{0} \Upsilon_{0} \Upsilon_{\theta_1}^d+4 \Delta_{\pi} \Upsilon_{0} \Upsilon_{\theta_1}^d+8 \Gamma_{\pi-\theta_1}^d \Upsilon_{0} \Upsilon_{\theta_1}^d\nonumber\\
	&+8 \Gamma_{\pi-\theta_1}^e \Upsilon_{0} \Upsilon_{\theta_1}^d+6 \Gamma_{\theta_1}^e \Upsilon_{0} \Upsilon_{\theta_1}^d-\Delta_{\pi-\theta_1} (\Upsilon_{\theta_1}^d)^2+2 \Delta_{\theta_1} (\Upsilon_{\theta_1}^d)^2+\Upsilon_{0} (\Upsilon_{\theta_1}^d)^2+2 \Delta_{0} \Delta_{\pi-\theta_1} \Upsilon_{\theta_1}^e+2 \Delta_{\pi} \Delta_{\pi-\theta_1} \Upsilon_{\theta_1}^e\nonumber\\
	&+4 \Delta_{\pi} \Delta_{\theta_1} \Upsilon_{\theta_1}^e-2 \Delta_{\pi-\theta_1} \Gamma_{\pi-\theta_1}^d \Upsilon_{\theta_1}^e-2 \Delta_{\theta_1} \Gamma_{\pi-\theta_1}^d \Upsilon_{\theta_1}^e+2 \Delta_{\theta_1} \Gamma_{\pi-\theta_1}^e \Upsilon_{\theta_1}^e+2 \Delta_{\pi-\theta_1} \Gamma_{\theta_1}^e \Upsilon_{\theta_1}^e-4 \Delta_{\theta_1} \Gamma_{\theta_1}^e \Upsilon_{\theta_1}^e\nonumber\\
	&+4 \Delta_{0} \Upsilon_{0} \Upsilon_{\theta_1}^e+5 \Gamma_{\pi-\theta_1}^d \Upsilon_{0} \Upsilon_{\theta_1}^e+4 \Gamma_{\theta_1}^e \Upsilon_{0} \Upsilon_{\theta_1}^e+2 \Delta_{\pi-\theta_1} \Upsilon_{\theta_1}^d \Upsilon_{\theta_1}^e-2 \Upsilon_{0} \Upsilon_{\theta_1}^d \Upsilon_{\theta_1}^e-\Delta_{\pi-\theta_1} (\Upsilon_{\theta_1}^e)^2+2 \Upsilon_{0} (\Upsilon_{\theta_1}^e)^2\nonumber\\
	&+\Gamma_{\theta_1}^d \Big(-2 \Delta_{\pi} \Delta_{\pi-\theta_1}-\Delta_{\pi-\theta_1} \Gamma_{\pi-\theta_1}^d-4 \Delta_{\theta_1} \Gamma_{\pi-\theta_1}^e+2 \Delta_{0} (\Delta_{\pi-\theta_1}-\Upsilon_{0})+6 \Delta_{\pi} \Upsilon_{0}+2 \Gamma_{\theta_1}^e \Upsilon_{0}-2 \Delta_{\pi-\theta_1} \Upsilon_{\theta_1}^d\nonumber\\
	&+2 \Delta_{\theta_1} \Upsilon_{\theta_1}^d+2 \Upsilon_{0} \Upsilon_{\theta_1}^d+4 \Upsilon_{0} \Upsilon_{\theta_1}^e\Big)+\Gamma_{0}\Big(-6 \Delta_{\pi-\theta_1} \Gamma_{\pi-\theta_1}^d-6 \Delta_{\pi-\theta_1} \Gamma_{\pi-\theta_1}^e+2 \Delta_{\pi-\theta_1} \Gamma_{\theta_1}^e+2 \Gamma_{\theta_1}^d (\Delta_{\pi-\theta_1}-3 \Upsilon_{0})\nonumber\\
	&-2 \Gamma_{\pi-\theta_1}^d \Upsilon_{0}+4 \Gamma_{\pi-\theta_1}^e \Upsilon_{0}-2 \Gamma_{\theta_1}^e \Upsilon_{0}+\Delta_{\pi-\theta_1} \Upsilon_{\theta_1}^d+12 \Upsilon_{0} \Upsilon_{\theta_1}^d+6 \Upsilon_{0} \Upsilon_{\theta_1}^e-\Delta_{\theta_1} (\Gamma_{\pi-\theta_1}^e+2 \Upsilon_{\theta_1}^d+2 \Upsilon_{\theta_1}^e)\Big)\Big] \nonumber\\
	&-\frac{1}{64 \pi^4N_f^2 (1+v^2)^2}\frac{1}{\epsilon}\Big[-(\Delta_{0})^2+2 \Delta_{0} \Delta_{\pi}+(\Delta_{\pi})^2+2 (\Delta_{0}-\Delta_{\pi}) \Gamma_{0}+3 (\Gamma_{0})^2\Big]
\end{align}

\begin{align}
	A^{(2l)}_{\Upsilon_{\theta_1}^d}&=-\frac{e^{-\frac{v^2}{v_c^2}}}{64 \Upsilon_{\theta_1}^d \pi^4N_f^2}\frac{1}{\epsilon} \Big[-4 \Delta_{0} (\Delta_{\pi-\theta_1})^2-4 \Delta_{\pi-\theta_1} \Delta_{\theta_1} \Gamma_{\pi-\theta_1}^d-4 \Delta_{0} \Gamma_{\pi-\theta_1}^d \Gamma_{\pi-\theta_1}^e-2 (\Gamma_{\pi-\theta_1}^d)^2 \Gamma_{\pi-\theta_1}^e+2 \Delta_{0} (\Gamma_{\pi-\theta_1}^e)^2\nonumber\\
	&-2 (\Gamma_{\pi-\theta_1}^e)^3-2 (\Delta_{\pi-\theta_1})^2 \Gamma_{\theta_1}^d-2 \Delta_{\pi-\theta_1} \Delta_{\theta_1} \Gamma_{\theta_1}^d-\Gamma_{\pi-\theta_1}^d \Gamma_{\pi-\theta_1}^e \Gamma_{\theta_1}^d+3 \Gamma_{\pi-\theta_1}^e (\Gamma_{\theta_1}^d)^2-2 (\Delta_{\pi-\theta_1})^2 \Gamma_{\theta_1}^e-2 \Delta_{0} \Gamma_{\pi-\theta_1}^e \Gamma_{\theta_1}^e\nonumber\\
	&-2 (\Gamma_{\pi-\theta_1}^e)^2 \Gamma_{\theta_1}^e-\Gamma_{\pi-\theta_1}^e (\Gamma_{\theta_1}^e)^2-2 \Delta_{0} \Delta_{\pi-\theta_1} \Upsilon_{0}-2 \Delta_{0} \Delta_{\theta_1} \Upsilon_{0}+2 \Delta_{\theta_1} \Gamma_{\pi-\theta_1}^d \Upsilon_{0}+2 \Delta_{\pi-\theta_1} \Gamma_{\pi-\theta_1}^e \Upsilon_{0}-2 \Delta_{\theta_1} \Gamma_{\pi-\theta_1}^e \Upsilon_{0}\nonumber\\
	&+2 \Delta_{\pi-\theta_1} \Gamma_{\theta_1}^e \Upsilon_{0}-2 \Delta_{\theta_1} \Gamma_{\theta_1}^e \Upsilon_{0}+5 \Delta_{0} (\Upsilon_{0})^2+4 \Gamma_{\pi-\theta_1}^d (\Upsilon_{0})^2+2 \Gamma_{\pi-\theta_1}^e (\Upsilon_{0})^2+6 \Gamma_{\theta_1}^d (\Upsilon_{0})^2+4 \Gamma_{\theta_1}^e (\Upsilon_{0})^2+2 (\Delta_{\pi-\theta_1})^2 \Upsilon_{\theta_1}^d\nonumber\\
	&+2 (\Delta_{\theta_1})^2 \Upsilon_{\theta_1}^d-4 \Delta_{0} \Gamma_{\pi-\theta_1}^d \Upsilon_{\theta_1}^d+4 \Delta_{0} \Gamma_{\pi-\theta_1}^e \Upsilon_{\theta_1}^d-2 \Gamma_{\pi-\theta_1}^d \Gamma_{\pi-\theta_1}^e \Upsilon_{\theta_1}^d-4 \Delta_{0} \Gamma_{\theta_1}^d \Upsilon_{\theta_1}^d+2 \Gamma_{\pi-\theta_1}^e \Gamma_{\theta_1}^d \Upsilon_{\theta_1}^d+2 \Delta_{0} \Gamma_{\theta_1}^e \Upsilon_{\theta_1}^d\nonumber\\
	&+2 \Gamma_{\pi-\theta_1}^d \Gamma_{\theta_1}^e \Upsilon_{\theta_1}^d-\Gamma_{\pi-\theta_1}^e \Gamma_{\theta_1}^e \Upsilon_{\theta_1}^d-2 \Gamma_{\theta_1}^d \Gamma_{\theta_1}^e \Upsilon_{\theta_1}^d+(\Gamma_{\theta_1}^e)^2 \Upsilon_{\theta_1}^d+2 \Delta_{\pi-\theta_1} \Upsilon_{0} \Upsilon_{\theta_1}^d+2 \Delta_{\theta_1} \Upsilon_{0} \Upsilon_{\theta_1}^d+2 \Delta_{0} (\Upsilon_{\theta_1}^d)^2\nonumber\\
	&-2 \Gamma_{\pi-\theta_1}^d (\Upsilon_{\theta_1}^d)^2+3 \Gamma_{\pi-\theta_1}^e (\Upsilon_{\theta_1}^d)^2+2 \Gamma_{\theta_1}^d (\Upsilon_{\theta_1}^d)^2+2 \Gamma_{\theta_1}^e (\Upsilon_{\theta_1}^d)^2+(\Upsilon_{\theta_1}^d)^3+4 \Delta_{\pi-\theta_1} \Delta_{\theta_1} \Upsilon_{\theta_1}^e-4 (\Delta_{\theta_1})^2 \Upsilon_{\theta_1}^e\nonumber\\
	&-2 \Gamma_{\pi-\theta_1}^e \Gamma_{\theta_1}^d \Upsilon_{\theta_1}^e-2 \Delta_{0} \Gamma_{\theta_1}^e \Upsilon_{\theta_1}^e-2 \Gamma_{\pi-\theta_1}^d \Gamma_{\theta_1}^e \Upsilon_{\theta_1}^e+2 \Delta_{\pi-\theta_1} \Upsilon_{0} \Upsilon_{\theta_1}^e+4 \Delta_{\theta_1} \Upsilon_{0} \Upsilon_{\theta_1}^e-2 (\Upsilon_{0})^2 \Upsilon_{\theta_1}^e+6 \Delta_{0} \Upsilon_{\theta_1}^d \Upsilon_{\theta_1}^e\nonumber\\
	&+3 \Gamma_{\pi-\theta_1}^d \Upsilon_{\theta_1}^d \Upsilon_{\theta_1}^e+6 \Gamma_{\theta_1}^d \Upsilon_{\theta_1}^d \Upsilon_{\theta_1}^e-\Gamma_{\pi-\theta_1}^e (\Upsilon_{\theta_1}^e)^2+2 \Upsilon_{\theta_1}^d (\Upsilon_{\theta_1}^e)^2+\Delta_{\pi} \Big(-4 (\Delta_{\pi-\theta_1})^2-2 (\Delta_{\theta_1})^2-2 \Delta_{\theta_1} (\Delta_{\pi-\theta_1}-2 \Upsilon_{0})\nonumber\\
	&+\Delta_{\pi-\theta_1} \Upsilon_{0}+2 (\Gamma_{\pi-\theta_1}^d \Gamma_{\pi-\theta_1}^e-\Gamma_{\pi-\theta_1}^d \Gamma_{\theta_1}^e-\Gamma_{\pi-\theta_1}^e \Gamma_{\theta_1}^e+(\Gamma_{\theta_1}^e)^2+(\Upsilon_{0})^2+\Gamma_{\pi-\theta_1}^d \Upsilon_{\theta_1}^d+\Gamma_{\pi-\theta_1}^e \Upsilon_{\theta_1}^d-2 \Gamma_{\theta_1}^e \Upsilon_{\theta_1}^d\nonumber\\
	&+(\Upsilon_{\theta_1}^d)^2+\Gamma_{\theta_1}^d (-\Gamma_{\pi-\theta_1}^e-\Gamma_{\theta_1}^e+\Upsilon_{\theta_1}^d)+\Gamma_{\pi-\theta_1}^e \Upsilon_{\theta_1}^e+2 \Gamma_{\theta_1}^e \Upsilon_{\theta_1}^e)\Big)+\Gamma_{0} \Big(-2 (\Delta_{\pi-\theta_1})^2-3 \Delta_{\pi-\theta_1} \Delta_{\theta_1}+\Delta_{\pi-\theta_1} \Upsilon_{0}\nonumber\\
	&+2 (2 (\Upsilon_{0})^2+(\Upsilon_{\theta_1}^d)^2+\Gamma_{\pi-\theta_1}^e (-2 \Gamma_{\pi-\theta_1}^d-\Gamma_{\pi-\theta_1}^e+\Gamma_{\theta_1}^e+\Upsilon_{\theta_1}^e)+\Upsilon_{\theta_1}^d (\Gamma_{\theta_1}^e+2 \Upsilon_{\theta_1}^e))\Big)\Big]\nonumber\\
	& +\frac{1}{64  \pi^4N_f^2 (1+v^2)^2}\frac{1}{\epsilon}\Big[-3 (\Delta_{0})^2+2 \Delta_{0} \Delta_{\pi}-(\Delta_{\pi})^2-2 (\Delta_{0}+\Delta_{\pi}) \Gamma_{0}+(\Gamma_{0})^2 \Big]
\end{align}

\begin{align}
	A^{(2l)}_{\Upsilon_{\theta_1}^e}&=\frac{e^{-\frac{v^2}{v_c^2}}}{64\Upsilon_{\theta_1}^e \pi^4N_f^2}\frac{1}{\epsilon}\Big[4 \Delta_{0} \Delta_{\pi-\theta_1} \Delta_{\theta_1}-2 \Delta_{\pi-\theta_1} \Delta_{\theta_1} \Gamma_{\pi-\theta_1}^d+2 \Delta_{\pi-\theta_1} \Delta_{\theta_1} \Gamma_{\pi-\theta_1}^e-2 (\Delta_{\theta_1})^2 \Gamma_{\pi-\theta_1}^e+2 (\Delta_{\pi-\theta_1})^2 \Gamma_{\theta_1}^e\nonumber\\
	&-2 \Delta_{\pi-\theta_1} \Delta_{\theta_1} \Gamma_{\theta_1}^e+4 \Delta_{0} \Gamma_{\pi-\theta_1}^e \Gamma_{\theta_1}^e-2 \Gamma_{\pi-\theta_1}^e \Gamma_{\theta_1}^d \Gamma_{\theta_1}^e+2 \Delta_{0} \Delta_{\theta_1} \Upsilon_{0}+2 \Delta_{\theta_1} \Gamma_{\pi-\theta_1}^e \Upsilon_{0}+2 \Delta_{\theta_1} \Gamma_{\theta_1}^d \Upsilon_{0}-2 \Delta_{\pi-\theta_1} \Gamma_{\theta_1}^e \Upsilon_{0}\nonumber\\
	&+4 \Delta_{\theta_1} \Gamma_{\theta_1}^e \Upsilon_{0}-4 \Gamma_{\theta_1}^e (\Upsilon_{0})^2-6 \Delta_{\pi-\theta_1} \Delta_{\theta_1} \Upsilon_{\theta_1}^d+4 (\Delta_{\theta_1})^2 \Upsilon_{\theta_1}^d+2 \Delta_{0} \Gamma_{\theta_1}^e \Upsilon_{\theta_1}^d+2 \Gamma_{\pi-\theta_1}^d \Gamma_{\theta_1}^e \Upsilon_{\theta_1}^d-2 \Gamma_{\pi-\theta_1}^e \Gamma_{\theta_1}^e \Upsilon_{\theta_1}^d\nonumber\\
	&-6 \Delta_{\theta_1} \Upsilon_{0} \Upsilon_{\theta_1}^d-2 \Gamma_{\theta_1}^e (\Upsilon_{\theta_1}^d)^2-2 (\Delta_{\pi-\theta_1})^2 \Upsilon_{\theta_1}^e+2 \Delta_{\pi-\theta_1} \Delta_{\theta_1} \Upsilon_{\theta_1}^e-2 (\Delta_{\theta_1})^2 \Upsilon_{\theta_1}^e+6 \Delta_{0} \Gamma_{\pi-\theta_1}^d \Upsilon_{\theta_1}^e-(\Gamma_{\pi-\theta_1}^e)^2 \Upsilon_{\theta_1}^e\nonumber\\
	&+6 \Delta_{0} \Gamma_{\theta_1}^d \Upsilon_{\theta_1}^e-4 \Gamma_{\pi-\theta_1}^e \Gamma_{\theta_1}^d \Upsilon_{\theta_1}^e-4 \Gamma_{\pi-\theta_1}^d \Gamma_{\theta_1}^e \Upsilon_{\theta_1}^e+2 \Gamma_{\pi-\theta_1}^e \Gamma_{\theta_1}^e \Upsilon_{\theta_1}^e-(\Gamma_{\theta_1}^e)^2 \Upsilon_{\theta_1}^e-4 \Delta_{\pi-\theta_1} \Upsilon_{0} \Upsilon_{\theta_1}^e-\Delta_{\theta_1} \Upsilon_{0} \Upsilon_{\theta_1}^e\nonumber\\
	&-2 (\Upsilon_{0})^2 \Upsilon_{\theta_1}^e-2 \Gamma_{\pi-\theta_1}^e \Upsilon_{\theta_1}^d \Upsilon_{\theta_1}^e-4 \Gamma_{\theta_1}^d \Upsilon_{\theta_1}^d \Upsilon_{\theta_1}^e-(\Upsilon_{\theta_1}^d)^2 \Upsilon_{\theta_1}^e+2 \Gamma_{\pi-\theta_1}^e (\Upsilon_{\theta_1}^e)^2-2 \Gamma_{\theta_1}^e (\Upsilon_{\theta_1}^e)^2+2 \Upsilon_{\theta_1}^d (\Upsilon_{\theta_1}^e)^2-2 (\Upsilon_{\theta_1}^e)^3\nonumber\\
	&+2 \Gamma_{0} \Big(\Delta_{\pi-\theta_1} \Delta_{\theta_1}+\Gamma_{\pi-\theta_1}^e \Gamma_{\theta_1}^e+\Gamma_{\pi-\theta_1}^d \Upsilon_{\theta_1}^e+\Gamma_{\theta_1}^d \Upsilon_{\theta_1}^e\Big)+\Delta_{\pi} \Big(4 (\Delta_{\theta_1})^2+4 \Gamma_{\pi-\theta_1}^d \Gamma_{\pi-\theta_1}^e+4 \Gamma_{\pi-\theta_1}^e \Gamma_{\theta_1}^d-3 \Gamma_{\pi-\theta_1}^d \Gamma_{\theta_1}^e\nonumber\\
	&-4 \Gamma_{\theta_1}^d \Gamma_{\theta_1}^e+2 (\Gamma_{\theta_1}^e)^2-2 \Delta_{\pi-\theta_1} \Upsilon_{0}-2 (\Upsilon_{0})^2-2 \Delta_{\theta_1} (\Delta_{\pi-\theta_1}+\Upsilon_{0})+4 \Gamma_{\pi-\theta_1}^d \Upsilon_{\theta_1}^d-2 \Gamma_{\pi-\theta_1}^e \Upsilon_{\theta_1}^d+4 \Gamma_{\theta_1}^d \Upsilon_{\theta_1}^d-2 (\Upsilon_{\theta_1}^d)^2\nonumber\\
	&-2 \Gamma_{\pi-\theta_1}^d \Upsilon_{\theta_1}^e+\Gamma_{\pi-\theta_1}^e \Upsilon_{\theta_1}^e-2 \Gamma_{\theta_1}^d \Upsilon_{\theta_1}^e+4 \Gamma_{\theta_1}^e \Upsilon_{\theta_1}^e\Big)\Big] -\frac{1}{64 \Upsilon_{\theta_1}^e \pi^4N_f^2 (1+v^2)^2}\frac{1}{\epsilon}\Big[4 \Delta_{0} \Delta_{\pi} \Upsilon_{\theta_1}^d+5 (\Delta_{0})^2 \Upsilon_{\theta_1}^e-2 \Delta_{0} \Delta_{\pi} \Upsilon_{\theta_1}^e\nonumber\\
	&+3 (\Delta_{\pi})^2 \Upsilon_{\theta_1}^e+(\Gamma_{0})^2 \Upsilon_{\theta_1}^e+\Gamma_{0} \Big(4 \Delta_{\pi} \Upsilon_{\theta_1}^d+6 \Delta_{0} \Upsilon_{\theta_1}^e+3 \Delta_{\pi} \Upsilon_{\theta_1}^e
	\Big)\Big]
\end{align}

\begin{align}
	A^{(2l)}_{\Xi_{\theta_1}^d}&=-\frac{e^{-\frac{v^2}{v_c^2}}}{64 \Xi_{\theta_1}^d \pi^4N_f^2}\frac{1}{\epsilon} \Big[(\Delta_{\pi-\theta_1})^2 \Xi_{\theta_1}^d-2 \Delta_{\pi-\theta_1} \Delta_{\theta_1} \Xi_{\theta_1}^d+(\Delta_{\theta_1})^2 \Xi_{\theta_1}^d-2 \Delta_{0} \Gamma_{\pi-\theta_1}^d \Xi_{\theta_1}^d+2 \Delta_{\pi} \Gamma_{\pi-\theta_1}^d \Xi_{\theta_1}^d+(\Gamma_{\pi-\theta_1}^d)^2 \Xi_{\theta_1}^d\nonumber\\
	&+2 \Delta_{0} \Gamma_{\pi-\theta_1}^e \Xi_{\theta_1}^d-2 \Delta_{\pi} \Gamma_{\pi-\theta_1}^e \Xi_{\theta_1}^d+(\Gamma_{\theta_1}^d)^2 \Xi_{\theta_1}^d-2 \Delta_{0} \Gamma_{\theta_1}^e \Xi_{\theta_1}^d+2 \Delta_{\pi} \Gamma_{\theta_1}^e \Xi_{\theta_1}^d+2 \Delta_{\pi-\theta_1} \Upsilon_{0} \Xi_{\theta_1}^d-2 \Delta_{\theta_1} \Upsilon_{0} \Xi_{\theta_1}^d\nonumber\\
	&+(\Upsilon_{0})^2 \Xi_{\theta_1}^d+2 \Delta_{0} \Upsilon_{\theta_1}^d \Xi_{\theta_1}^d-2 \Delta_{\pi} \Upsilon_{\theta_1}^d \Xi_{\theta_1}^d-2 \Delta_{0} \Upsilon_{\theta_1}^e \Xi_{\theta_1}^d+2 \Delta_{\pi} \Upsilon_{\theta_1}^e \Xi_{\theta_1}^d-2 \Gamma_{\pi-\theta_1}^d \Upsilon_{\theta_1}^e \Xi_{\theta_1}^d+(\Upsilon_{\theta_1}^e)^2 \Xi_{\theta_1}^d\nonumber\\
	&-2 (\Delta_{\pi-\theta_1})^2 \Xi_{\theta_1}^e+4 \Delta_{\pi-\theta_1} \Delta_{\theta_1} \Xi_{\theta_1}^e-2 (\Delta_{\theta_1})^2 \Xi_{\theta_1}^e-2 \Delta_{0} \Gamma_{\pi-\theta_1}^e \Xi_{\theta_1}^e+2 \Delta_{\pi} \Gamma_{\pi-\theta_1}^e \Xi_{\theta_1}^e+2 \Gamma_{\pi-\theta_1}^d \Gamma_{\pi-\theta_1}^e \Xi_{\theta_1}^e\nonumber\\
	&+2 \Delta_{0} \Gamma_{\theta_1}^e \Xi_{\theta_1}^e-2 \Delta_{\pi} \Gamma_{\theta_1}^e \Xi_{\theta_1}^e+2 \Gamma_{\pi-\theta_1}^d \Gamma_{\theta_1}^e \Xi_{\theta_1}^e+2 (\Upsilon_{0})^2 \Xi_{\theta_1}^e-2 \Delta_{0} \Upsilon_{\theta_1}^d \Xi_{\theta_1}^e+2 \Delta_{\pi} \Upsilon_{\theta_1}^d \Xi_{\theta_1}^e-2 \Gamma_{\pi-\theta_1}^d \Upsilon_{\theta_1}^d \Xi_{\theta_1}^e\nonumber\\
	&-2 \Gamma_{\pi-\theta_1}^e \Upsilon_{\theta_1}^e \Xi_{\theta_1}^e-2 \Gamma_{\theta_1}^e \Upsilon_{\theta_1}^e \Xi_{\theta_1}^e+2 \Upsilon_{\theta_1}^d \Upsilon_{\theta_1}^e \Xi_{\theta_1}^e+2 \Gamma_{\theta_1}^d (\Delta_{0} \Xi_{\theta_1}^d-\Delta_{\pi} \Xi_{\theta_1}^d+\Gamma_{\pi-\theta_1}^d \Xi_{\theta_1}^d-\Upsilon_{\theta_1}^e \Xi_{\theta_1}^d+\Gamma_{\pi-\theta_1}^e \Xi_{\theta_1}^e\nonumber\\
	&+\Gamma_{\theta_1}^e \Xi_{\theta_1}^e-\Upsilon_{\theta_1}^d \Xi_{\theta_1}^e)+2 \Gamma_{0} \Big(-\Gamma_{\pi-\theta_1}^d \Xi_{\theta_1}^d-\Gamma_{\pi-\theta_1}^e \Xi_{\theta_1}^d+\Gamma_{\theta_1}^d \Xi_{\theta_1}^d+\Gamma_{\theta_1}^e \Xi_{\theta_1}^d-\Upsilon_{\theta_1}^e \Xi_{\theta_1}^d-\Gamma_{\pi-\theta_1}^e \Xi_{\theta_1}^e+\Gamma_{\theta_1}^e \Xi_{\theta_1}^e\nonumber\\
	&-\Upsilon_{\theta_1}^d (\Xi_{\theta_1}^d+\Xi_{\theta_1}^e)\Big)\Big]
\end{align}

\begin{align}
	A^{(2l)}_{\Xi_{\theta_1}^e}&=-\frac{e^{-\frac{v^2}{v_c^2}}}{64 \Xi_{\theta_1}^e \pi^4N_f^2}\frac{1}{ \epsilon} \Big[-2 \Delta_{0} \Gamma_{\pi-\theta_1}^e \Xi_{\theta_1}^d+2 \Delta_{\pi} \Gamma_{\pi-\theta_1}^e \Xi_{\theta_1}^d-2 \Gamma_{0} \Gamma_{\pi-\theta_1}^e \Xi_{\theta_1}^d+2 \Gamma_{\pi-\theta_1}^d \Gamma_{\pi-\theta_1}^e \Xi_{\theta_1}^d+2 \Gamma_{\pi-\theta_1}^e \Gamma_{\theta_1}^d \Xi_{\theta_1}^d\nonumber\\
	&+2 \Delta_{0} \Gamma_{\theta_1}^e \Xi_{\theta_1}^d-2 \Delta_{\pi} \Gamma_{\theta_1}^e \Xi_{\theta_1}^d+2 \Gamma_{0} \Gamma_{\theta_1}^e \Xi_{\theta_1}^d+2 \Gamma_{\pi-\theta_1}^d \Gamma_{\theta_1}^e \Xi_{\theta_1}^d+2 \Gamma_{\theta_1}^d \Gamma_{\theta_1}^e \Xi_{\theta_1}^d+2 (\Upsilon_{0})^2 \Xi_{\theta_1}^d-2 \Delta_{0} \Upsilon_{\theta_1}^d \Xi_{\theta_1}^d\nonumber\\
	&+2 \Delta_{\pi} \Upsilon_{\theta_1}^d \Xi_{\theta_1}^d-2 \Gamma_{0} \Upsilon_{\theta_1}^d \Xi_{\theta_1}^d-2 \Gamma_{\pi-\theta_1}^d \Upsilon_{\theta_1}^d \Xi_{\theta_1}^d-2 \Gamma_{\theta_1}^d \Upsilon_{\theta_1}^d \Xi_{\theta_1}^d-2 \Gamma_{\pi-\theta_1}^e \Upsilon_{\theta_1}^e \Xi_{\theta_1}^d-2 \Gamma_{\theta_1}^e \Upsilon_{\theta_1}^e \Xi_{\theta_1}^d+2 \Upsilon_{\theta_1}^d \Upsilon_{\theta_1}^e \Xi_{\theta_1}^d\nonumber\\
	&-2 \Delta_{0} \Gamma_{\pi-\theta_1}^d \Xi_{\theta_1}^e+2 \Delta_{\pi} \Gamma_{\pi-\theta_1}^d \Xi_{\theta_1}^e-2 \Gamma_{0} \Gamma_{\pi-\theta_1}^d \Xi_{\theta_1}^e+(\Gamma_{\pi-\theta_1}^d)^2 \Xi_{\theta_1}^e+2 \Delta_{0} \Gamma_{\pi-\theta_1}^e \Xi_{\theta_1}^e-2 \Delta_{\pi} \Gamma_{\pi-\theta_1}^e \Xi_{\theta_1}^e-2 \Gamma_{0} \Gamma_{\pi-\theta_1}^e \Xi_{\theta_1}^e\nonumber\\
	&+2 \Delta_{0} \Gamma_{\theta_1}^d \Xi_{\theta_1}^e-2 \Delta_{\pi} \Gamma_{\theta_1}^d \Xi_{\theta_1}^e+2 \Gamma_{0} \Gamma_{\theta_1}^d \Xi_{\theta_1}^e+2 \Gamma_{\pi-\theta_1}^d \Gamma_{\theta_1}^d \Xi_{\theta_1}^e+(\Gamma_{\theta_1}^d)^2 \Xi_{\theta_1}^e-2 \Delta_{0} \Gamma_{\theta_1}^e \Xi_{\theta_1}^e+2 \Delta_{\pi} \Gamma_{\theta_1}^e \Xi_{\theta_1}^e+2 \Gamma_{0} \Gamma_{\theta_1}^e \Xi_{\theta_1}^e\nonumber\\
	&+2 \Delta_{\pi-\theta_1} \Upsilon_{0} \Xi_{\theta_1}^e+(\Upsilon_{0})^2 \Xi_{\theta_1}^e+2 \Delta_{0} \Upsilon_{\theta_1}^d \Xi_{\theta_1}^e-2 \Delta_{\pi} \Upsilon_{\theta_1}^d \Xi_{\theta_1}^e-2 \Gamma_{0} \Upsilon_{\theta_1}^d \Xi_{\theta_1}^e-2 \Delta_{0} \Upsilon_{\theta_1}^e \Xi_{\theta_1}^e+2 \Delta_{\pi} \Upsilon_{\theta_1}^e \Xi_{\theta_1}^e-2 \Gamma_{0} \Upsilon_{\theta_1}^e \Xi_{\theta_1}^e\nonumber\\
	&-2 \Gamma_{\pi-\theta_1}^d \Upsilon_{\theta_1}^e \Xi_{\theta_1}^e-2 \Gamma_{\theta_1}^d \Upsilon_{\theta_1}^e \Xi_{\theta_1}^e+(\Upsilon_{\theta_1}^e)^2 \Xi_{\theta_1}^e+(\Delta_{\pi-\theta_1})^2 (-2 \Xi_{\theta_1}^d+\Xi_{\theta_1}^e)+(\Delta_{\theta_1})^2 (-2 \Xi_{\theta_1}^d+\Xi_{\theta_1}^e)\nonumber\\
	&+\Delta_{\theta_1} \Big(4 \Delta_{\pi-\theta_1} \Xi_{\theta_1}^d-2 \Delta_{\pi-\theta_1} \Xi_{\theta_1}^e-2 \Upsilon_{0} \Xi_{\theta_1}^e\Big)\Big]
\end{align}

\begin{align}
	A^{(2l)}_{\Xi_{\theta_2}^d}&=-\frac{e^{-\frac{v^2}{v_c^2}}}{64  \Xi_{\theta_2}^d \pi^4N_f^2}\frac{1}{\epsilon} \Big[-4 \Delta_{0} \Delta_{\pi-\theta_1} \Xi_{\pi/2}^d-2 \Delta_{\pi} \Delta_{\pi-\theta_1} \Xi_{\pi/2}^d+2 \Delta_{0} \Delta_{\theta_1} \Xi_{\pi/2}^d-2 \Delta_{\pi-\theta_1} \Gamma_{\pi-\theta_1}^d \Xi_{\pi/2}^d+2 \Delta_{\theta_1} \Gamma_{\pi-\theta_1}^d \Xi_{\pi/2}^d\nonumber\\
	&+2 \Delta_{\pi-\theta_1} \Gamma_{\pi-\theta_1}^e \Xi_{\pi/2}^d-2 \Delta_{\theta_1} \Gamma_{\pi-\theta_1}^e \Xi_{\pi/2}^d+2 \Delta_{\pi-\theta_1} \Gamma_{\theta_1}^e \Xi_{\pi/2}^d+2 \Delta_{\theta_1} \Gamma_{\theta_1}^e \Xi_{\pi/2}^d-\Delta_{0} \Upsilon_{0} \Xi_{\pi/2}^d+\Delta_{\pi} \Upsilon_{0} \Xi_{\pi/2}^d\nonumber\\
	&+2 \Gamma_{\pi-\theta_1}^d \Upsilon_{0} \Xi_{\pi/2}^d+2 \Gamma_{\pi-\theta_1}^e \Upsilon_{0} \Xi_{\pi/2}^d+2 \Gamma_{\theta_1}^e \Upsilon_{0} \Xi_{\pi/2}^d-2 \Delta_{\pi-\theta_1} \Upsilon_{\theta_1}^d \Xi_{\pi/2}^d+2 \Delta_{\theta_1} \Upsilon_{\theta_1}^d \Xi_{\pi/2}^d-2 \Upsilon_{0} \Upsilon_{\theta_1}^d \Xi_{\pi/2}^d\nonumber\\
	&+2 \Delta_{\pi-\theta_1} \Upsilon_{\theta_1}^e \Xi_{\pi/2}^d+2 \Delta_{\theta_1} \Upsilon_{\theta_1}^e \Xi_{\pi/2}^d-2 \Upsilon_{0} \Upsilon_{\theta_1}^e \Xi_{\pi/2}^d+2 \Delta_{0} \Gamma_{\pi-\theta_1}^d \Xi_{\theta_2}^d+2 \Delta_{\pi} \Gamma_{\pi-\theta_1}^d \Xi_{\theta_2}^d+(\Gamma_{\pi-\theta_1}^d)^2 \Xi_{\theta_2}^d\nonumber\\
	&+2 \Delta_{0} \Gamma_{\pi-\theta_1}^e \Xi_{\theta_2}^d+2 \Delta_{\pi} \Gamma_{\pi-\theta_1}^e \Xi_{\theta_2}^d-(\Gamma_{\pi-\theta_1}^e)^2 \Xi_{\theta_2}^d+2 (\Gamma_{\theta_1}^d)^2 \Xi_{\theta_2}^d+2 \Delta_{0} \Gamma_{\theta_1}^e \Xi_{\theta_2}^d-2 \Delta_{\pi} \Gamma_{\theta_1}^e \Xi_{\theta_2}^d-\Gamma_{\pi-\theta_1}^e \Gamma_{\theta_1}^e \Xi_{\theta_2}^d\nonumber\\
	&+2 \Delta_{0} \Upsilon_{\theta_1}^d \Xi_{\theta_2}^d+2 \Delta_{\pi} \Upsilon_{\theta_1}^d \Xi_{\theta_2}^d+\Gamma_{\pi-\theta_1}^e \Upsilon_{\theta_1}^d \Xi_{\theta_2}^d+2 (\Upsilon_{\theta_1}^d)^2 \Xi_{\theta_2}^d+2 \Delta_{0} \Upsilon_{\theta_1}^e \Xi_{\theta_2}^d-2 \Delta_{\pi} \Upsilon_{\theta_1}^e \Xi_{\theta_2}^d-\Gamma_{\pi-\theta_1}^d \Upsilon_{\theta_1}^e \Xi_{\theta_2}^d\nonumber\\
	&+\Gamma_{\theta_1}^d (-2 \Delta_{\pi-\theta_1} \Xi_{\pi/2}^d+2 \Delta_{\theta_1} \Xi_{\pi/2}^d+2 \Upsilon_{0} \Xi_{\pi/2}^d+2 \Delta_{0} \Xi_{\theta_2}^d+2 \Delta_{\pi} \Xi_{\theta_2}^d+\Gamma_{\pi-\theta_1}^d \Xi_{\theta_2}^d)+\Gamma_{0} \Big(-\Delta_{\pi-\theta_1} \Xi_{\pi/2}^d+\Delta_{\theta_1} \Xi_{\pi/2}^d\nonumber\\
	&+2 \Gamma_{\pi-\theta_1}^d \Xi_{\theta_2}^d+2 \Gamma_{\pi-\theta_1}^e \Xi_{\theta_2}^d+2 \Gamma_{\theta_1}^d \Xi_{\theta_2}^d+2 \Gamma_{\theta_1}^e \Xi_{\theta_2}^d+2 \Upsilon_{\theta_1}^d \Xi_{\theta_2}^d+2 \Upsilon_{\theta_1}^e \Xi_{\theta_2}^d\Big)\Big]\\
	A^{(2l)}_{\Xi_{\theta_2}^e}&=-\frac{e^{-\frac{v^2}{v_c^2}}}{64 \Xi_{\theta_2}^e \pi^4N_f^2}\frac{1}{\epsilon} \Big[-2 \Delta_{0} \Delta_{\pi-\theta_1} \Xi_{\pi/2}^e+2 \Delta_{\pi} \Delta_{\pi-\theta_1} \Xi_{\pi/2}^e+2 \Delta_{0} \Delta_{\theta_1} \Xi_{\pi/2}^e-2 \Delta_{\pi} \Delta_{\theta_1} \Xi_{\pi/2}^e+2 \Delta_{\pi-\theta_1} \Gamma_{\pi-\theta_1}^d \Xi_{\pi/2}^e\nonumber\\
	&+2 \Delta_{\theta_1} \Gamma_{\pi-\theta_1}^d \Xi_{\pi/2}^e-2 \Delta_{\pi-\theta_1} \Gamma_{\pi-\theta_1}^e \Xi_{\pi/2}^e+2 \Delta_{\theta_1} \Gamma_{\pi-\theta_1}^e \Xi_{\pi/2}^e+2 \Delta_{\pi-\theta_1} \Gamma_{\theta_1}^e \Xi_{\pi/2}^e-2 \Delta_{\theta_1} \Gamma_{\theta_1}^e \Xi_{\pi/2}^e-2 \Delta_{0} \Upsilon_{0} \Xi_{\pi/2}^e\nonumber\\
	&+2 \Delta_{\pi} \Upsilon_{0} \Xi_{\pi/2}^e-2 \Gamma_{\pi-\theta_1}^d \Upsilon_{0} \Xi_{\pi/2}^e-2 \Gamma_{\pi-\theta_1}^e \Upsilon_{0} \Xi_{\pi/2}^e+2 \Gamma_{\theta_1}^e \Upsilon_{0} \Xi_{\pi/2}^e+2 \Delta_{\pi-\theta_1} \Upsilon_{\theta_1}^d \Xi_{\pi/2}^e-2 \Delta_{\theta_1} \Upsilon_{\theta_1}^d \Xi_{\pi/2}^e+2 \Upsilon_{0} \Upsilon_{\theta_1}^d \Xi_{\pi/2}^e\nonumber\\
	&-2 \Delta_{\pi-\theta_1} \Upsilon_{\theta_1}^e \Xi_{\pi/2}^e-2 \Delta_{\theta_1} \Upsilon_{\theta_1}^e \Xi_{\pi/2}^e+2 \Upsilon_{0} \Upsilon_{\theta_1}^e \Xi_{\pi/2}^e+2 \Delta_{0} \Gamma_{\pi-\theta_1}^d \Xi_{\theta_2}^e-2 \Delta_{\pi} \Gamma_{\pi-\theta_1}^d \Xi_{\theta_2}^e+(\Gamma_{\pi-\theta_1}^d)^2 \Xi_{\theta_2}^e-2 \Delta_{0} \Gamma_{\pi-\theta_1}^e \Xi_{\theta_2}^e\nonumber\\
	&-2 \Delta_{\pi} \Gamma_{\pi-\theta_1}^e \Xi_{\theta_2}^e+(\Gamma_{\pi-\theta_1}^e)^2 \Xi_{\theta_2}^e+(\Gamma_{\theta_1}^d)^2 \Xi_{\theta_2}^e+2 \Delta_{0} \Gamma_{\theta_1}^e \Xi_{\theta_2}^e+2 \Delta_{\pi} \Gamma_{\theta_1}^e \Xi_{\theta_2}^e-2 \Gamma_{\pi-\theta_1}^e \Gamma_{\theta_1}^e \Xi_{\theta_2}^e+(\Gamma_{\theta_1}^e)^2 \Xi_{\theta_2}^e-2 \Delta_{0} \Upsilon_{\theta_1}^d \Xi_{\theta_2}^e\nonumber\\
	&-2 \Delta_{\pi} \Upsilon_{\theta_1}^d \Xi_{\theta_2}^e+2 \Gamma_{\pi-\theta_1}^e \Upsilon_{\theta_1}^d \Xi_{\theta_2}^e-2 \Gamma_{\theta_1}^e \Upsilon_{\theta_1}^d \Xi_{\theta_2}^e+(\Upsilon_{\theta_1}^d)^2 \Xi_{\theta_2}^e-2 \Delta_{0} \Upsilon_{\theta_1}^e \Xi_{\theta_2}^e+2 \Delta_{\pi} \Upsilon_{\theta_1}^e \Xi_{\theta_2}^e-2 \Gamma_{\pi-\theta_1}^d \Upsilon_{\theta_1}^e \Xi_{\theta_2}^e\nonumber\\
	&+(\Upsilon_{\theta_1}^e)^2 \Xi_{\theta_2}^e-2 \Gamma_{\theta_1}^d \Big(-\Delta_{\pi-\theta_1} \Xi_{\pi/2}^e-\Delta_{\theta_1} \Xi_{\pi/2}^e+\Upsilon_{0} \Xi_{\pi/2}^e+\Delta_{0} \Xi_{\theta_2}^e-\Delta_{\pi} \Xi_{\theta_2}^e-\Gamma_{\pi-\theta_1}^d \Xi_{\theta_2}^e+\Upsilon_{\theta_1}^e \Xi_{\theta_2}^e\Big)\nonumber\\
	&-2 \Gamma_{0} \Big(\Delta_{\pi-\theta_1} \Xi_{\pi/2}^e-\Delta_{\theta_1} \Xi_{\pi/2}^e+\Upsilon_{0} \Xi_{\pi/2}^e-\Gamma_{\pi-\theta_1}^d \Xi_{\theta_2}^e-\Gamma_{\pi-\theta_1}^e \Xi_{\theta_2}^e+\Gamma_{\theta_1}^d \Xi_{\theta_2}^e+\Gamma_{\theta_1}^e \Xi_{\theta_2}^e-\Upsilon_{\theta_1}^d \Xi_{\theta_2}^e+\Upsilon_{\theta_1}^e \Xi_{\theta_2}^e\Big)\Big]
\end{align}

\begin{align}
	A^{(2l)}_{\Xi_{\pi/2}^d}&=-\frac{e^{-\frac{v^2}{v_c^2}}}{64  \Xi_{\pi/2}^d \pi^4N_f^2}\frac{1}{\epsilon} \Big[2 \Delta_{0} \Gamma_{\pi-\theta_1}^d \Xi_{\pi/2}^d+2 \Delta_{\pi} \Gamma_{\pi-\theta_1}^d \Xi_{\pi/2}^d+(\Gamma_{\pi-\theta_1}^d)^2 \Xi_{\pi/2}^d+2 \Delta_{0} \Gamma_{\pi-\theta_1}^e \Xi_{\pi/2}^d+2 \Delta_{\pi} \Gamma_{\pi-\theta_1}^e \Xi_{\pi/2}^d\nonumber\\
	&-(\Gamma_{\pi-\theta_1}^e)^2 \Xi_{\pi/2}^d+2 (\Gamma_{\theta_1}^d)^2 \Xi_{\pi/2}^d+2 \Delta_{0} \Gamma_{\theta_1}^e \Xi_{\pi/2}^d-2 \Delta_{\pi} \Gamma_{\theta_1}^e \Xi_{\pi/2}^d-\Gamma_{\pi-\theta_1}^e \Gamma_{\theta_1}^e \Xi_{\pi/2}^d+2 \Delta_{0} \Upsilon_{\theta_1}^d \Xi_{\pi/2}^d+2 \Delta_{\pi} \Upsilon_{\theta_1}^d \Xi_{\pi/2}^d\nonumber\\
	&+\Gamma_{\pi-\theta_1}^e \Upsilon_{\theta_1}^d \Xi_{\pi/2}^d+2 (\Upsilon_{\theta_1}^d)^2 \Xi_{\pi/2}^d+2 \Delta_{0} \Upsilon_{\theta_1}^e \Xi_{\pi/2}^d-2 \Delta_{\pi} \Upsilon_{\theta_1}^e \Xi_{\pi/2}^d-\Gamma_{\pi-\theta_1}^d \Upsilon_{\theta_1}^e \Xi_{\pi/2}^d-4 \Delta_{0} \Delta_{\pi-\theta_1} \Xi_{\theta_2}^d-2 \Delta_{\pi} \Delta_{\pi-\theta_1} \Xi_{\theta_2}^d\nonumber\\
	&+2 \Delta_{0} \Delta_{\theta_1} \Xi_{\theta_2}^d-2 \Delta_{\pi-\theta_1} \Gamma_{\pi-\theta_1}^d \Xi_{\theta_2}^d+2 \Delta_{\theta_1} \Gamma_{\pi-\theta_1}^d \Xi_{\theta_2}^d+2 \Delta_{\pi-\theta_1} \Gamma_{\pi-\theta_1}^e \Xi_{\theta_2}^d-2 \Delta_{\theta_1} \Gamma_{\pi-\theta_1}^e \Xi_{\theta_2}^d+2 \Delta_{\pi-\theta_1} \Gamma_{\theta_1}^e \Xi_{\theta_2}^d\nonumber\\
	&+2 \Delta_{\theta_1} \Gamma_{\theta_1}^e \Xi_{\theta_2}^d-\Delta_{0} \Upsilon_{0} \Xi_{\theta_2}^d+\Delta_{\pi} \Upsilon_{0} \Xi_{\theta_2}^d+2 \Gamma_{\pi-\theta_1}^d \Upsilon_{0} \Xi_{\theta_2}^d+2 \Gamma_{\pi-\theta_1}^e \Upsilon_{0} \Xi_{\theta_2}^d+2 \Gamma_{\theta_1}^e \Upsilon_{0} \Xi_{\theta_2}^d-2 \Delta_{\pi-\theta_1} \Upsilon_{\theta_1}^d \Xi_{\theta_2}^d\nonumber\\
	&+2 \Delta_{\theta_1} \Upsilon_{\theta_1}^d \Xi_{\theta_2}^d-2 \Upsilon_{0} \Upsilon_{\theta_1}^d \Xi_{\theta_2}^d+2 \Delta_{\pi-\theta_1} \Upsilon_{\theta_1}^e \Xi_{\theta_2}^d+2 \Delta_{\theta_1} \Upsilon_{\theta_1}^e \Xi_{\theta_2}^d-2 \Upsilon_{0} \Upsilon_{\theta_1}^e \Xi_{\theta_2}^d+\Gamma_{0} \Big(2 (\Gamma_{\pi-\theta_1}^d+\Gamma_{\pi-\theta_1}^e+\Gamma_{\theta_1}^d+\Gamma_{\theta_1}^e\nonumber\\
	&+\Upsilon_{\theta_1}^d+\Upsilon_{\theta_1}^e) \Xi_{\pi/2}^d-\Delta_{\pi-\theta_1} \Xi_{\theta_2}^d+\Delta_{\theta_1} \Xi_{\theta_2}^d\Big)+\Gamma_{\theta_1}^d \Big(2 \Delta_{0} \Xi_{\pi/2}^d+2 \Delta_{\pi} \Xi_{\pi/2}^d+\Gamma_{\pi-\theta_1}^d \Xi_{\pi/2}^d-2 \Delta_{\pi-\theta_1} \Xi_{\theta_2}^d+2 \Delta_{\theta_1} \Xi_{\theta_2}^d\nonumber\\
	&+2 \Upsilon_{0} \Xi_{\theta_2}^d\Big)\Big]\\
	A^{(2l)}_{\Xi_{\pi/2}^e}&=-\frac{e^{-\frac{v^2}{v_c^2}}}{64  \Xi_{\pi/2}^e \pi^4N_f^2}\frac{1}{\epsilon} \Big[2 \Delta_{0} \Gamma_{\pi-\theta_1}^d \Xi_{\pi/2}^e-2 \Delta_{\pi} \Gamma_{\pi-\theta_1}^d \Xi_{\pi/2}^e+(\Gamma_{\pi-\theta_1}^d)^2 \Xi_{\pi/2}^e-2 \Delta_{0} \Gamma_{\pi-\theta_1}^e \Xi_{\pi/2}^e-2 \Delta_{\pi} \Gamma_{\pi-\theta_1}^e \Xi_{\pi/2}^e\nonumber\\
	&+(\Gamma_{\pi-\theta_1}^e)^2 \Xi_{\pi/2}^e+(\Gamma_{\theta_1}^d)^2 \Xi_{\pi/2}^e+2 \Delta_{0} \Gamma_{\theta_1}^e \Xi_{\pi/2}^e+2 \Delta_{\pi} \Gamma_{\theta_1}^e \Xi_{\pi/2}^e-2 \Gamma_{\pi-\theta_1}^e \Gamma_{\theta_1}^e \Xi_{\pi/2}^e+(\Gamma_{\theta_1}^e)^2 \Xi_{\pi/2}^e-2 \Delta_{0} \Upsilon_{\theta_1}^d \Xi_{\pi/2}^e\nonumber\\
	&-2 \Delta_{\pi} \Upsilon_{\theta_1}^d \Xi_{\pi/2}^e+2 \Gamma_{\pi-\theta_1}^e \Upsilon_{\theta_1}^d \Xi_{\pi/2}^e-2 \Gamma_{\theta_1}^e \Upsilon_{\theta_1}^d \Xi_{\pi/2}^e+(\Upsilon_{\theta_1}^d)^2 \Xi_{\pi/2}^e-2 \Delta_{0} \Upsilon_{\theta_1}^e \Xi_{\pi/2}^e+2 \Delta_{\pi} \Upsilon_{\theta_1}^e \Xi_{\pi/2}^e-2 \Gamma_{\pi-\theta_1}^d \Upsilon_{\theta_1}^e \Xi_{\pi/2}^e\nonumber\\
	&+(\Upsilon_{\theta_1}^e)^2 \Xi_{\pi/2}^e-2 \Delta_{0} \Delta_{\pi-\theta_1} \Xi_{\theta_2}^e+2 \Delta_{\pi} \Delta_{\pi-\theta_1} \Xi_{\theta_2}^e+2 \Delta_{0} \Delta_{\theta_1} \Xi_{\theta_2}^e-2 \Delta_{\pi} \Delta_{\theta_1} \Xi_{\theta_2}^e+2 \Delta_{\pi-\theta_1} \Gamma_{\pi-\theta_1}^d \Xi_{\theta_2}^e+2 \Delta_{\theta_1} \Gamma_{\pi-\theta_1}^d \Xi_{\theta_2}^e\nonumber\\
	&-2 \Delta_{\pi-\theta_1} \Gamma_{\pi-\theta_1}^e \Xi_{\theta_2}^e+2 \Delta_{\theta_1} \Gamma_{\pi-\theta_1}^e \Xi_{\theta_2}^e+2 \Delta_{\pi-\theta_1} \Gamma_{\theta_1}^e \Xi_{\theta_2}^e-2 \Delta_{\theta_1} \Gamma_{\theta_1}^e \Xi_{\theta_2}^e-2 \Delta_{0} \Upsilon_{0} \Xi_{\theta_2}^e+2 \Delta_{\pi} \Upsilon_{0} \Xi_{\theta_2}^e-2 \Gamma_{\pi-\theta_1}^d \Upsilon_{0} \Xi_{\theta_2}^e\nonumber\\
	&-2 \Gamma_{\pi-\theta_1}^e \Upsilon_{0} \Xi_{\theta_2}^e+2 \Gamma_{\theta_1}^e \Upsilon_{0} \Xi_{\theta_2}^e+2 \Delta_{\pi-\theta_1} \Upsilon_{\theta_1}^d \Xi_{\theta_2}^e-2 \Delta_{\theta_1} \Upsilon_{\theta_1}^d \Xi_{\theta_2}^e+2 \Upsilon_{0} \Upsilon_{\theta_1}^d \Xi_{\theta_2}^e-2 \Delta_{\pi-\theta_1} \Upsilon_{\theta_1}^e \Xi_{\theta_2}^e-2 \Delta_{\theta_1} \Upsilon_{\theta_1}^e \Xi_{\theta_2}^e\nonumber\\
	&+2 \Upsilon_{0} \Upsilon_{\theta_1}^e \Xi_{\theta_2}^e+2 \Gamma_{0} (\Gamma_{\pi-\theta_1}^d \Xi_{\pi/2}^e+\Gamma_{\pi-\theta_1}^e \Xi_{\pi/2}^e-\Gamma_{\theta_1}^d \Xi_{\pi/2}^e-\Gamma_{\theta_1}^e \Xi_{\pi/2}^e+\Upsilon_{\theta_1}^d \Xi_{\pi/2}^e-\Upsilon_{\theta_1}^e \Xi_{\pi/2}^e-\Delta_{\pi-\theta_1} \Xi_{\theta_2}^e+\Delta_{\theta_1} \Xi_{\theta_2}^e\nonumber\\
	&-\Upsilon_{0} \Xi_{\theta_2}^e)+2 \Gamma_{\theta_1}^d (-\Delta_{0} \Xi_{\pi/2}^e+\Delta_{\pi} \Xi_{\pi/2}^e+\Gamma_{\pi-\theta_1}^d \Xi_{\pi/2}^e-\Upsilon_{\theta_1}^e \Xi_{\pi/2}^e+\Delta_{\pi-\theta_1} \Xi_{\theta_2}^e+\Delta_{\theta_1} \Xi_{\theta_2}^e-\Upsilon_{0} \Xi_{\theta_2}^e)\Big]
\end{align}